\documentclass[a4paper,twoside,openright]{cernmono}

\newif\ifRelease 

\newif\ifYellowReport 

\Releasetrue 
\YellowReporttrue 
\ifRelease
\else
	\usepackage{draftwatermark,color,ulem}
	\SetWatermarkText{CLIC PIP draft}  
\fi

\ifYellowReport
\usepackage[
     plainpages        = false,
     bookmarks         = true,
     bookmarksnumbered = true,
     bookmarksopen     = true,
     pdfstartview      = FitH,
     pdfpagelayout     = SinglePage,
     colorlinks        = true,
     citecolor         = black,
     linkcolor         = black,
     urlcolor          = black, 
     pdftitle={CLIC Project Implementation Plan},
     pdfauthor={CLIC},
     pdfkeywords={CLIC, physics performances, detector concepts},
     pdfcreator={latex},
     pdfproducer={pdfLaTex}]{hyperref}

\else
\usepackage[
     plainpages        = false,
     bookmarks         = true,
     bookmarksnumbered = true,
     bookmarksopen     = true,
     pdfstartview      = FitH,
     pdfpagelayout     = SinglePage,
     colorlinks        = true,
     citecolor         = red,
     linkcolor         = violet,
     urlcolor          = blue, 
     pdftitle={CLIC Project Implementation Plan},
     pdfauthor={CLIC},
     pdfkeywords={CLIC, physics performances, detector concepts},
     pdfcreator={latex},
     pdfproducer={pdfLaTex}]{hyperref}
\fi

\usepackage{CLIC} 
\usepackage[T1]{fontenc} 
\usepackage{epic,eepic,pspicture,mathptmx,times,graphpap,epsf,titleref,wasysym} 
\usepackage[utf8]{inputenc} 
\usepackage{lmodern} 
\usepackage[english]{babel}
\usepackage{booktabs}
\usepackage{mathtools}
\usepackage{doi}
\usepackage{subfigure}
\usepackage{csquotes}
\usepackage{makecell}
\usepackage{adjustbox}
\usepackage{tikz}
\usepackage{pgfplots}
   \pgfplotsset{compat=1.13}
\usepackage{pgfplotstable}
\usepackage{sansmath}
\usepackage{siunitx} 
\usepackage{multirow}
\usepackage{etoolbox}
\usepackage{rotating}
\usepackage{ulem}
\usepackage{pdfpages}
\usepackage[sorting=none, bibstyle=ieee, citestyle=numeric-comp]{biblatex}
\robustify\bfseries	
\renewcommand{\bibname}{Bibliography}

\begin{document}
\title{CLIC-PIP}

\pagenumbering{roman}
\setcounter{page}{1}

%
%
%
%
%
\includepdf{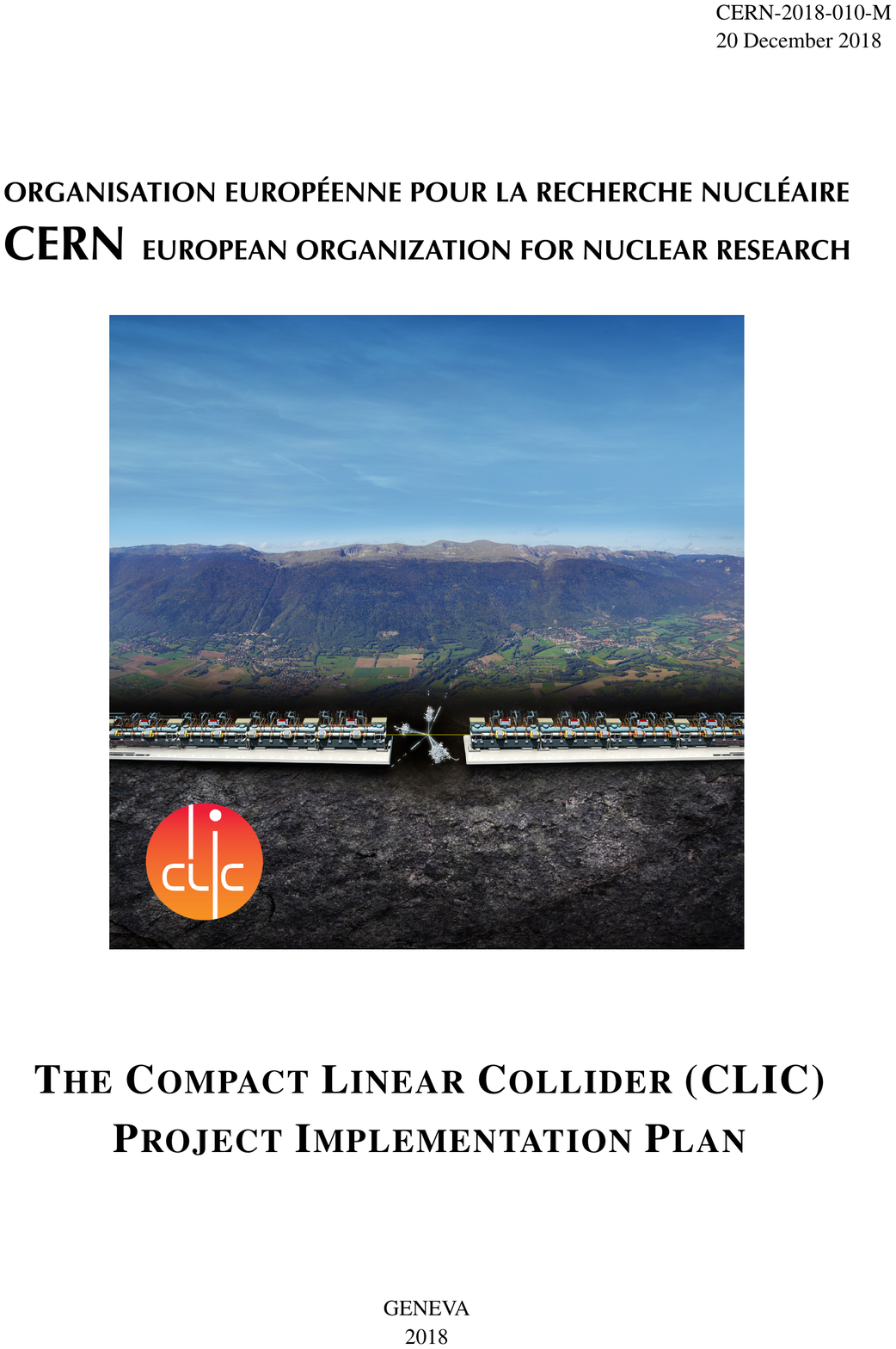}
\begin{titlepage}
\thispagestyle{empty}
\mbox{}
\vfill

\begin{flushleft}
CERN Yellow Reports: Monographs\\
Published by CERN, CH-1211 Geneva 23, Switzerland\\[3mm]

\begin{tabular}{@{}l@{~}l}
  ISBN & 978--92--9083--514--1 (paperback) \\
  ISBN & 978--92--9083--515--8 (PDF) \\
  ISSN & 2519-8068 (Print)\\ 
  ISSN & 2519-8076 (Online)\\ 
DOI & \url{https://doi.org/10.23731/CYRM-2018-004}\\

\end{tabular}\\[3mm]
Accepted for publication by the CERN Report Editorial Board (CREB) on 20 December 2018\\
Available online at \url{http://publishing.cern.ch/} and \url{http://cds.cern.ch/}\\[3mm]

Copyright \copyright{} CERN, 2018\\[1mm]
\raisebox{-1mm}{\includegraphics[height=12pt]{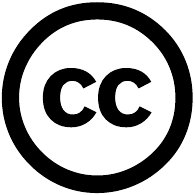}}
 Creative Commons Attribution 4.0\\[1mm]
Knowledge transfer is an integral part of CERN's mission.\\[1mm]
CERN publishes this volume Open Access under the Creative Commons Attribution 4.0 license\\
(\url{http://creativecommons.org/licenses/by/4.0/}) in order to permit its wide dissemination and use.\\
The submission of a contribution to a CERN Yellow Report series shall be deemed to constitute the contributor's agreement to this copyright and license statement. Contributors are requested to obtain any clearances that may be necessary for this purpose.\\[5mm]

This volume is indexed in: CERN Document Server (CDS), INSPIRE.\\[5mm]

This volume should be cited as:\\[1mm]

The Compact Linear Collider (CLIC) -- Project Implementation Plan,\\
edited by M.~Aicheler, P.N.~Burrows, N.~Catalan, R.~Corsini, M.~Draper,  J.~Osborne, D.~Schulte,  S.~Stapnes, M.J.~Stuart,  CERN-2018-010-M,\\
CERN Yellow Reports: Monographs, Vol. 4/2018, CERN--2018--010--M (CERN, Geneva, 2018). https://doi.org/10.23731/CYRM-2018-004

\end{flushleft}
\end{titlepage}

\cleardoublepage

\thispagestyle{empty}
\vspace*{3cm}
\begin{center}
  \large{\bfseries\sffamily Abstract}
\end{center}
\begin{quotation}
\noindent
The Compact Linear Collider (CLIC) is a TeV-scale high-luminosity linear e$^{+}$e$^{-}$
collider under development by international collaborations hosted by CERN. This document provides an overview of the design, technology, and implementation aspects of the CLIC accelerator. For an optimal exploitation of its physics potential, CLIC is foreseen to be built and operated in stages, at centre-of-mass energies of 380\,GeV, 1.5\,TeV and 3\,TeV, for a site length ranging between 11\,km and 50\,km. CLIC uses a Two-Beam acceleration scheme, in which normal-conducting high-gradient 12\,GHz accelerating structures are powered via a high-current Drive Beam. For the first stage, an alternative with X-band klystron powering is also considered. CLIC accelerator optimisation, technical developments, and system tests have resulted in significant progress in recent years. Moreover, this has led to an increased energy efficiency and reduced power consumption of around 170\,MW for the 380\,GeV stage, together with a reduced cost estimate of approximately 6~billion~CHF. The construction of the first CLIC energy stage could start as early as 2026 and first beams would be available by 2035, marking the beginning of a physics programme spanning 25--30 years and providing excellent sensitivity to Beyond Standard Model physics, through direct searches and via a broad set of precision measurements of Standard Model processes, particularly in the Higgs and top-quark sectors.
\end{quotation}
\vfill

\begin{center}
{\large {\bfseries\sffamily Corresponding editors}} 
\vspace*{0.25cm}

{
Markus Aicheler (Helsinki Institute of Physics), Philip N.\ Burrows (University of Oxford), \\ Nuria Catalan Lasheras (CERN), Mick Draper (CERN),  John Andrew Osborne (CERN), \\ 
Daniel Schulte (CERN), Steinar Stapnes (CERN), Matthew James Stuart (CERN)
}
\end{center}

\cleardoublepage

\setcounter{tocdepth}{1}
\tableofcontents
\cleardoublepage

\collName{The CLIC accelerator collaboration}
\addauthor{
T.K.~Charles, 
P.J.~Giansiracusa,
T.G.~Lucas,
R.P.~Rassool,
M.~Volpi$^{1}$\\ 
\emph{University of Melbourne, Melbourne, Australia}
}

\addauthor{
C.~Balazs\\
\emph{Monash University, Melbourne, Australia}
}

\addauthor{
A.~Patapenka, 
I.~Zhuk\\
\emph{Joint Institute for Power and Nuclear Research - Sosny, Minsk, Belarus}
}

\addauthor{
C.~Collette\\
\emph{Universit\'e libre de Bruxelles, Brussels, Belgium}
}

\addauthor{
M.J.~Boland\\ 
\emph{University of Saskatchewan, Saskatoon, Canada}
% ...
}

\addauthor{
\begin{flushleft}
{$^{1}$}Also at CERN, Geneva, Switzerland\\
{$^{2}$}Also at Department of Physics, University of Helsinki, Helsinki, Finland\\
{$^{3}$}Now at CERN, Geneva, Switzerland\\
{$^{4}$}Now at Johannes-Gutenberg University, Mainz, Germany\\
{$^{5}$}Also at Department of Informatics, Athens University of Business and Economics, Athens, Greece\\
{$^{6}$}Also at University of Patras, Patras, Greece\\
{$^{7}$}Also at University of Antananarivo, Antananarivo, Madagascar\\
{$^{8}$}Also at IFIC, Valencia, Spain\\
{$^{9}$}Now at PSI, Villigen, Switzerland\\
{$^{10}$}Now at CEA, Gif-sur-Yvette, France\\
{$^{11}$}Also at LAL, Orsay, France\\
{$^{12}$}Also at Uppsala University, Uppsala, Sweden\\
{$^{13}$}Now at LBNL, Berkeley CA, USA\\
{$^{14}$}Now at DESY, Zeuthen, Germany\\
{$^{15}$}Also at Shandong University, Jinan, China\\
{$^{16}$}Also at EPFL, Lausanne, Switzerland\\
{$^{17}$}Also at Omer Halis Demir University, Nigde, Turkey\\
{$^{18}$}Also at The Cockcroft Institute, Daresbury, United Kingdom\\
{$^{19}$}Now at Diamond Light Source, Harwell, United Kingdom\\

\end{flushleft}
}

\begin{center}%
  \Large\bfseries\sffamily\MyCollName%
\end{center}%

\setcounter{footnote}{0}\def\@currentlabel{}%
\begingroup\def\thefootnote{\arabic{footnote}}
\def\@makefnmark{\hbox{$^{\@thefnmark)}$}}
\large
{
\MyAuthors
\par}
\endgroup

\vspace{-5mm}
\clearpage
\normalsize
\pagebreak
\pagenumbering{arabic}
\setcounter{page}{1}

\refsection 
\chapter{Project Implementation Plan}
\label{sec:introduction}
\section{Introduction}
The Compact Linear Collider (CLIC) is a multi-TeV high-luminosity linear collider under development by the CLIC accelerator collaboration. CLIC uses a novel Two-Beam acceleration technique, with normal conducting accelerating structures operating in the range of 70-100\,MeV/m. The CLIC Conceptual Design Report (CDR) was published in 2012 \cite{Aicheler2012}.  The main focus of the CDR was to demonstrate the feasibility of the CLIC accelerator at high energy (3\,TeV) and to confirm that high-precision physics measurements can be performed, despite the luminosity spectrum and the presence of particles from beam-induced background. 

Following the completion of the CDR, detailed studies on Higgs~\cite{Abramowicz:2016zbo} and top-quark physics~\cite{Abramowicz:2018rjq}, with particular focus on the first energy stage of CLIC, concluded that the optimal centre-of-mass energy for the first stage is 380\,GeV. As a result, a comprehensive optimisation study of the CLIC accelerator complex was performed,  by scanning the full parameter space for the accelerating structures, and by using the CLIC performance,  cost and energy consumption as a gauge for operation at 380\,GeV and 3\,TeV. The results led to optimised accelerator design parameters for the proposed CLIC staging scenario, with operation at 380\,GeV, 1.5\,TeV and 3\,TeV, as reported in~\cite{Boland2016}.

This report summarises progress and results of the CLIC studies at the time of submitting input to the update of the European Strategy for Particle Physics, in December 2018.  The report describes recent achievements in accelerator design, technology development, system tests and beam tests. Large-scale CLIC-specific beam tests have taken place, for example, at the CLIC Test Facility CTF3 at CERN~\cite{Geschonke2002}, at the Accelerator Test Facility ATF2 at KEK~\cite{Kuroda2016,Okugi2016}, at the FACET facility at SLAC~\cite{FACET} and at the FERMI facility in Trieste~\cite{FERMI}. 
An overview of the CLIC studies including also the detector design and physics case was published as separate Yellow Report~\cite{Charles:2018vfv}. Input from the theory community on the CLIC potential for new physics is collected in a separate publication~\cite{deBlas:2018mhx}.

Crucial experience also emanates from the expanding field of Free Electron Laser (FEL) linacs and recent-generation light sources.  Together they provide the demonstration that all implications of the CLIC design parameters are well understood and reproduced in beam tests. Therefore the CLIC performance goals are realistic. An alternative CLIC scenario for the first stage, where the accelerating structures are powered by X-band klystrons, is also under study. The implementation of CLIC near CERN has been investigated. Principally focusing on the 380\,GeV stage, this includes civil engineering aspects, electrical networks, cooling and ventilation, installation scheduling, transport, and safety aspects.  All CLIC studies have put emphasis on optimising cost and energy efficiency, and the resulting power and cost estimates are reported.  

Chapter~\ref{Chapter:Base_Design} and Chapter~\ref{Chapter:Kly_Design} provide an overview of the CLIC accelerator design and performance at 380\,GeV for both the Two-Beam baseline design and the Klystron-based option. Chapter~\ref{Chapter:HE_Design} describes the path to the higher energies, 1.5\,TeV and 3\,TeV, and gives an overview of the key CLIC technology developments.  Chapter~\ref{Chapter:TECH} describes in detail the technologies and systems used for the accelerator depicting technological choices and development since the CDR. Chapter~\ref{Chapter:CEIS} reports on civil engineering, infrastructure and siting.  Chapter~\ref{Chapter:IMP} describes the present plans for the implementation of CLIC, with emphasis on the 380\,GeV stage, and provides estimates of the energy consumption and of the cost for construction and operation.  Chapter~\ref{Chapter:PERF} describes system tests incorporating key technologies and provides evidence that the CLIC performance goals can be met.

\printbibliography[heading=subbibintoc]
\endrefsection

\refsection 
\chapter{Baseline Collider Design}
\label{Chapter:Base_Design}
\section{Introduction}
\begin{figure}[ht!]
\centering
\includegraphics[scale=0.7]{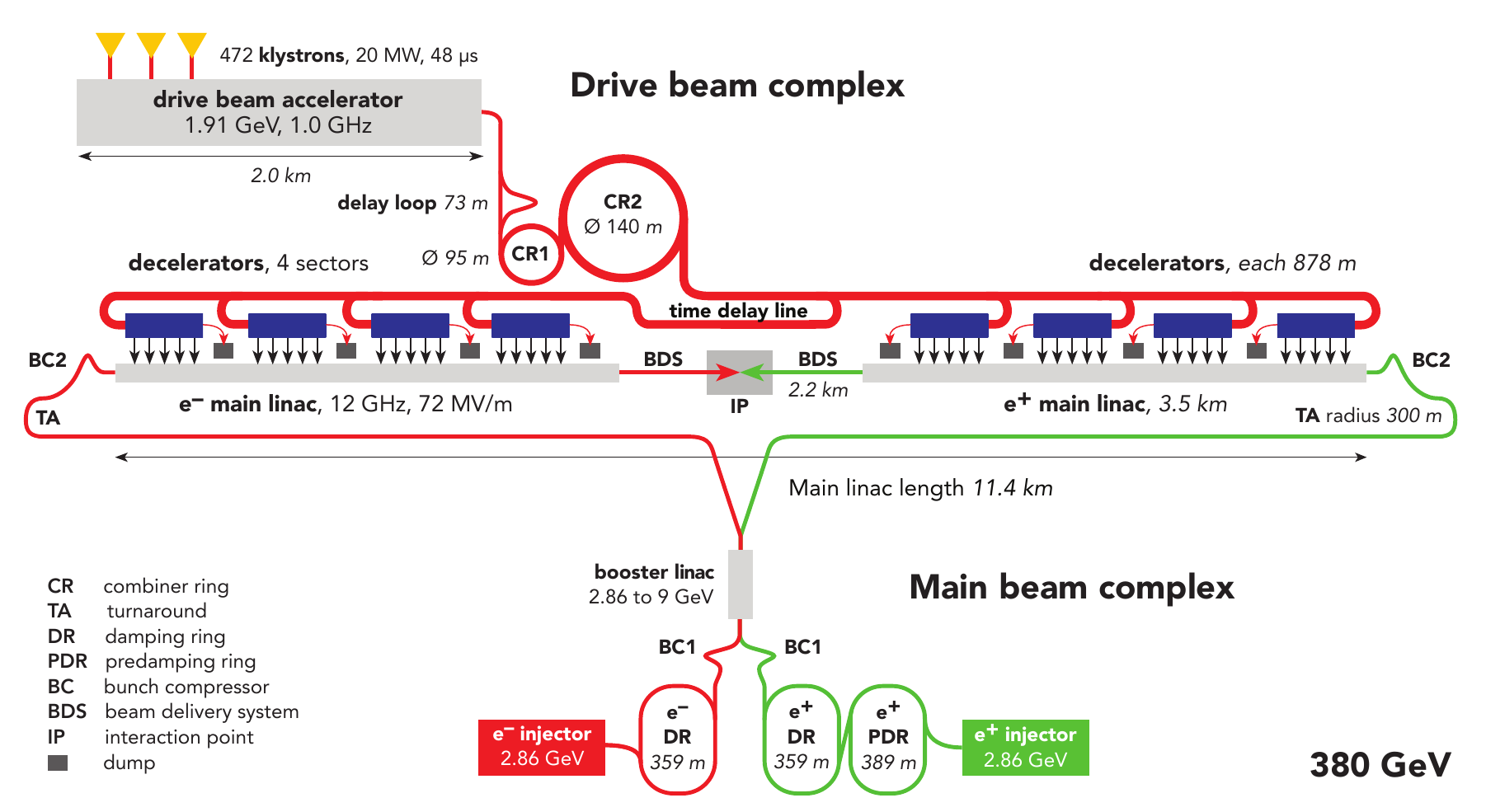}
\caption{Overview of the CLIC layout at $\sqrt{s}$~=~380\,GeV.}
\label{scd:clic_layout}
\end{figure}

The schematic layout of the CLIC complex can be seen in Fig.~\ref{scd:clic_layout} and the key parameters can be found in Table~\ref{t:scdup1}.

The main electron beam is produced in a conventional RF source and accelerated to 2.86\,GeV. The beam emittance is then reduced in a Damping Ring. To produce the positron beam, an electron beam is accelerated to 5\,GeV and sent into a crystal to produce energetic photons, which hit a second target and produce positrons. The positrons are captured and accelerated to 2.86\,GeV. Their beam emittance is reduced first in a Pre-Damping Ring and then in a Damping Ring. The Ring To Main Linac system accelerates the beams to 9\,GeV and compresses their length. The Main Linacs accelerate the beams to the collision energy of 190\,GeV. The Beam Delivery System removes transverse tails and off-energy particles with collimators and compresses the beams to the required small sizes at the collision point. After the collision the beams are transported by the Post Collision Line to the Beam Dump.

The RF power for each Main Linac is provided by a high-current, low-energy Drive Beam that runs parallel to the colliding beam through a sequence of Power Extraction and Transfer Structures (PETS). The Drive Beam generates RF power in the PETS that is then transferred to the accelerating structures via waveguides.

The Drive Beam is generated in a central complex with a fundamental frequency of 1\,GHz. A 46\,$\mu$s long beam pulse is produced in the injector and fills every other bucket, i.e. with a bunch spacing of 0.6\,m. Every 244\,ns, the injector switches from filling even buckets to filling odd buckets and vice versa creating 244\,ns long sub-pulses. The beam is accelerated in the Drive-Beam linac to 2\,GeV. A 0.5\,GHz, resonant RF deflector sends half the sub-pulses through the Delay Loop such that the bunches can be interleaved with those of the following sub-pulses that are not delayed. This generates a sequence of 244\,ns trains in which every bucket is filled separated by gaps of the same length. In a similar fashion three of these trains are merged in the first Combiner Ring. Groups of four of the trains, now with 0.1\,m bunch separation, are then merged in the second Combiner Ring. The final pulses are thus 244\,ns long and have a bunch spacing of 2.5\,cm, i.e. 24 times the initial beam current. The distance between the pulses has increased to 24$\times$244\,ns, which corresponds to twice the length of a decelerator. The first four sub-pulses are transported through a delay line before they are used to power one of the linacs while the next four sub-pulses are used to power the other linac directly. The first sub-pulse in each linac powers the first Drive-Beam decelerator, running in parallel to the colliding beam. When this sub-pulse reaches the decelerator end, the second sub-pulse has reached the beginning of the second Drive-Beam decelerator and will power it running in parallel to the colliding beam. The scheme continues with the other sub-pulses.

\begin{table}[htb!]
\caption{Key parameters of 380\,GeV baseline collider.}
\label{t:scdup1}
\centering
\begin{tabular}{l l l l }
\toprule
\textbf{Parameter}                  & \textbf{Symbol}         & \textbf{Unit}& \textbf{}  \\
\midrule
Centre-of-mass energy               & $\sqrt{s}$              &GeV           & 380 \\
Repetition frequency                & $f_{\text{rep}}$        &Hz            & 50  \\
Number of bunches per train         & $n_{b}$                 &              & 352 \\
Bunch separation                    & $\Delta\,t$             &ns            & 0.5 \\
Pulse length                        & $\tau_{\text{RF}}$      &ns            & 244 \\
\midrule
Accelerating gradient               & $G$                     &MV/m          & 72 \\
\midrule
Total luminosity                    & $\mathcal{L}$           &$10^{34}\;\text{cm}^{-2}\text{s}^{-1}$     & 1.5 \\
Luminosity above 99\% of $\sqrt{s}$ & $\mathcal{L}_{0.01}$    &$10^{34}\;\text{cm}^{-2}\text{s}^{-1}$     & 0.9 \\
\midrule
Main tunnel length                  &                         &km            & 11.4 \\
Number of particles per bunch       & $N$                     &$10^9$        & 5.2 \\
Bunch length                        & $\sigma_z$              &$\mu$m        & 70 \\
IP beam size                        & $\sigma_x/\sigma_y$     &nm            & 149/2.9\\
Normalised emittance (end of linac) & $\epsilon_x/\epsilon_y$ &nm            & 900/20 \\
\bottomrule
\end{tabular}
\end{table}

\section{Main-Beam Injectors}
\label{sect:MBI}
\begin{figure}[ht!]
\centering
\includegraphics[scale=0.45]{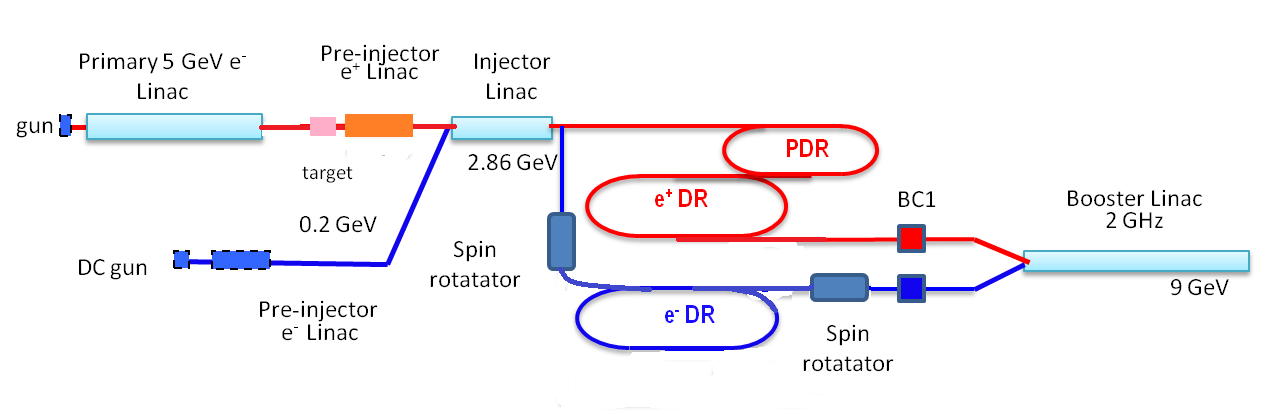}
\caption{\label{fig:MBI_1}Schematic layout of the Main-Beam Injector complex.}
\end{figure}

The CLIC Main-Beam Injectors (MBIs) consist of a polarized electron source and a conventional un-polarized positron source. Both particle species are pre-accelerated up to 200\,MeV in individual linacs before they are injected into a common injector linac which increases their energy up to 2.86\,GeV. The beams are then injected into the Damping Rings (DR) and finally accelerated to 9\,GeV in the Booster Linac. A schematic view of the injector complex can be seen in Fig.\,\ref{fig:MBI_1}. It is situated in a central complex on the surface parallel to the Main Linac tunnel, this allows the injector installations to be retained for upgrades to higher energies.

The common injector linac beam dynamics design is based on the positron beam since its emittance constrains the physical aperture and beta-function requirements. A new detailed beam dynamics design has been performed and is described in detail in \cite{Bayar2017}. The injector linac is 250\,m long and uses a 2\,GHz RF system and a FODO lattice. Two compressed RF pulses spaced by 3.4\,$\mu$s are used to accelerate the positron and the electron beam allowing individual beam loading compensation.

The CLIC polarized electron source uses a DC-photo injector followed by a 2\,GHz bunching and accelerating system. The spin-polarized electrons are generated using a polarized laser impinging on a strained GaAs cathode. Such cathodes have been used at several accelerator laboratories and demonstrated the CLIC requirements in terms of lifetime and charge extraction \cite{Zhou2009}. The electrons are accelerated in a pre-injector up to 200\,MeV before being injected into the common injector linac which boosts their energy to 2.86\,GeV. A spin rotator orients the spin vertically before injection into the DR. 

\begin{table}[htb!]
\caption{Beam parameters at the entrance of the Pre-Damping Ring for polarized electrons and for positrons at 2.86\,GeV.}
\label{tab:MBI_1}
\centering
\begin{tabular}{l l c c} 
\toprule
Parameter & Unit & Polarized electrons & Positrons \\ \hline 
\midrule
E 										& GeV 		& 2.86 	& 2.86 \\  
N 										& 10$^{9}$ 	& 6 	& 6 \\  
n${}_{b}$ 								& - 		& 352 	& 352 \\  
$\Delta$t$_{b}$ 						& ns 		& 0.5 	& 0.5 \\  
t${}_{pulse}$ 							& ns 		& 176 	& 176 \\  
$\epsilon_{x,y}$ 						& mm~mrad 	& < 25 	& 7071, 7577 \\  
$\sigma _{z}$ 							& mm 		& < 4 	& 3.3 \\  
$\sigma _{E}$ 							& \% 		& < 1 	& 1.63 \\  
Charge stability shot-to-shot 			& \% 		& 1 	& 1 \\  
Charge stability flatness on flat top 	& \% 		& 1 	& 1 \\  
f${}_{rep}$ 							& Hz 		& 50 	& 50 \\  
P & kW & 45 & 45 \\ 
\bottomrule 
\end{tabular}
\end{table}

The CLIC positron source consists of a 5\,GeV electron beam impinging on a tungsten hybrid target (one thin crystal target plus one thick amorphous target) taking advantage of photon enhancement via the channelling process. The positrons are captured after the target with an adiabatic matching device and a 2\,GHz capture linac accelerating the positrons up to 200\,MeV. One target station is sufficient because recent simulations and updates to the system showed that the positron yield from the target to the Pre-Damping Ring (PDR) could be increased by a factor 2.5 compared to the CDR value of $\approx$~1~e$^+$/e$^-$ \cite{Bayar2017}. The beam current of the primary electron linac was therefore reduced by the same factor leading to significantly less peak power deposition density in the amorphous target. More details can be found in Chapter~\ref{Chapter:PERF}.

Several changes have been implemented in the DR complex relative to the systems described in the CDR \cite{Aicheler2012}. The CDR employed a PDR for both the electron and positron beams, operated at an RF frequency of 1\,GHz, preceding the main DR, operated at the same RF frequency. A delay and recombination loop, downstream of the main rings, would provide a unique train with the required 2\,GHz bunch structure. The beams would spend 20\,ms in each DR. The updated scheme consists of changing the RF frequency from 1 to 2\,GHz throughout, removing the need for the Delay Loop. The PDR for electrons is also no longer needed because the electron injector is capable of delivering an emittance smaller than 25\,mm\,mrad and can be injected directly into the DR. In contrast, with its high incoming emittance, the positron beam continues to require a PDR. Thus, the positron bunch train is delayed 20\,ms relative to the corresponding electron beam, and is therefore synchronous with the subsequent electron bunch train given the 50\,Hz train repetition frequency. Finally the yield in the positron production chain has been increased significantly and a single positron production target is sufficient instead of two parallel targets as in the CDR.

The beam parameters at the entrance to the rings have changed slightly relative to the CDR and can be found in Table~\ref{tab:MBI_1}. The bunch charge is increased and the number of bunches is slightly higher. The bunch spacing is now 0.5\,ns and therefore the train length is shorter. The injectors have been designed to be able to provide about 15\,\% more intensity than needed at the input of the DR. The electron and positron beams are accelerated on different parts of the same RF pulse both in the Injector Linac and in the Booster Linac. The time interval between the first electron and positron bunch, in both the Injector and Booster Linacs, is 3.4\,$\mu$s, matching the geometry of the collider.

\section{Damping Rings}
\label{sect:DR}

The Damping Rings (DRs) are a fundamental part of the CLIC injector complex and are required to damp the large emittance of the injector linac beams, particularly for positrons, in all three dimensions to obtain the desired luminosity. The normalized transverse emittance of the incoming beam is to be reduced to 500\,nm and 5\,nm in horizontal and vertical directions respectively at an energy of 2.86\,GeV, corresponding to geometric emittances of around 90\,pm\,rad and 0.9\,pm\,rad. These transverse emittances, combined with the longitudinal emittance specification of 6\,keV\,m (i.e. ultra-short bunch lengths of below 2\,mm), while individually achieved in light sources in operation and under construction, are unprecedented in combination with a high bunch population of $5.7\times 10^9$. The generation and preservation of this high bunch brightness is subject to a number of collective effects (intra-beam scattering, space-charge, coherent instabilities including e-cloud for positrons and fast-ion for electrons) and drives both the optics of the rings and the various mitigation measures for vacuum design and feedbacks~\cite{Papaphilippou2012,Antoniou2012,Antoniou2014}. After the publication of the CDR~\cite{Aicheler2012} and, in view of a CLIC staged approach~\cite{Boland2016}, an effort was made to revise the DR system, adapting its performance to the different stages of the collider and in particular to the 380\,GeV first stage.

A schematic view of the DR complex with the e$^-$ and e$^+$ DR and the positron PDR (red) is shown in Fig.~\ref{fig:MBI_1}. In the original design, four rings were foreseen, a PDR and a main DR for each particle species. The PDRs were found to be necessary due to the large input emittance coming from the positron source, necessitating dynamic and momentum acceptances incompatible with the design of a high-focusing ultra-low emittance main DR. In addition, the storing time of 20\,ms corresponding to the high repetition collider rate of 50\,Hz is not long enough to allow the large injected emittances to damp to the required output value. However, in the case of the electrons, a combination of a high-brightness source and a careful emittance preservation in the injector linac could allow for a transverse input normalised emittance value of around 10\,$\mu$m, which is lower in the horizontal but higher in the vertical plane, with respect to the actual performance of the PDR. A careful Dynamic Aperture (DA) optimisation of the DRs in the vertical plane confirmed that such an emittance could be accommodated in the main rings, thereby making the electron PDR obsolete. There is the future possibility of replacing the positron PDR with a booster ring which could simultaneously damp and accelerate the beam, thereby reducing the positron linac cost.

In the CDR~\cite{Aicheler2012} the injected bunch train structure in the DRs was composed of two trains with twice the nominal separation (1\,vs\,0.5\,ns), separated by half the DR circumference. The two-train structure of 1\,GHz was chosen in order to reduce the transient beam loading effects in the RF cavities of the DRs. These trains were damped simultaneously and then extracted in a single turn from the main DR. A delay and recombination loop was located downstream of the rings, and served to combine the two trains into one, through an RF deflector, thus providing the required 2\,GHz bunch structure. However, it was apparent~\cite{Grudiev2012} that an RF system with a frequency of 2\,GHz has no higher technological difficulties than a system with an RF frequency of 1\,GHz. In addition, using 2\,GHz avoids the complexity of the RF deflector for the train recombination and reduces the impedance that might affect the transverse emittances. It was thus decided to choose as baseline the RF option with a frequency of 2\,GHz.

\begin{table}[!htb]
	\begin{center}
		\caption{ Design parameters for the improved design of the CLIC DRs, for the case of $f_{RF}~=~2$\,GHz and $N_b~=~5.7 \times 10^9$. The magnetic field is varying along the dipoles.}
		\label{tab:DRparams9}
		\centering
		{\small \begin{tabular}{lc}
        \toprule
				
				\multirow{1}{*}{Parameters, Symbol [Unit]} & \multirow{1}{*}{Variable dipole}\\ 
				\midrule
				Energy, $E$ [GeV]         		    				& 	{2.86}  \\     
				Bunch population, $N_b$	[$10^9$]      				&   {5.7}	  \\                          
				Circumference, $C$ [m]             					& 359.4 \\   
				Number of arc cells/wigglers,	$N_d/N_{w}$			& 90/40\\
				RF Voltage,	$V_{RF}$ [MV]				        	& 6.50\\
				RF Stationary phase [$^o$]  						& 63.0\\
				Harmonic number,	$h$        						& 2398\\
				Momentum compaction,	$\alpha_c$ [$10^{-4}$]		& 1.2 \\
				Damping times, ($\tau_x, \tau_y, \tau_{l}$) [ms] 	&(1.15, 1.18, 0.60)\\
				Energy loss/turn, $U$ [MeV]							& 5.8 \\
				Horizontal and vertical tune, ($Q_x$, $Q_y$) 	   & (45.61, 13.55)\\
				Horizontal and vertical chromaticity, ($\xi_x$, $\xi_y$) & (-169, -51)\\  		
				Wiggler peak field, $B_w$ [T]                       &  3.5\\  
				Wiggler length, $L_{w}$ [m]                         & 2 \\  
				Wiggler period, $\lambda_w$ [cm]					& 4.9 \\
				Normalized horiz. emittance with IBS, $\gamma\epsilon_{x}$ [nm-rad] & 535.9  \\
				Normalized horiz. emittance with IBS, $\gamma\epsilon_{x}$ [nm-rad] & 6.5  \\
				Longitudinal emittance with IBS, $\epsilon_l$ [keVm] 			    &  4.8\\
				IBS factors hor./ver./long. 					   	& 1.22/1.96/1.05\\
				\bottomrule
			\end{tabular}
		}
	\end{center}
\end{table}

The DRs have a racetrack shape composed of Theoretical Minimum Emittance (TME) arc cells and FODO cells straight sections and Super-Conducting (SC) damping wigglers. A revision of the DR design was undertaken based on recent design developments in the low emittance rings community for both beam dynamics and technology. This resulted in a new DR arc cell employing a novel concept of a TME with a longitudinally varying bend~\cite{Papadopoulou2018}. This dipole with varying field along the longitudinal position and including a small transverse gradient can further reduce the horizontal emittance while keeping the same space constraints of the TME arc cell. These magnets  are currently under design and prototyping at CIEMAT~\cite{Martinez2018}.  Furthermore, by employing a novel wire technology with Nb$_3$Sn, higher fields can be reached in the SC wigglers~\cite{Fajardo2016,Bernhard2016} for roughly the same period length. These two considerations enabled the reduction of the DR circumference by roughly 20\,\%, while maintaining their performance. The parameters for the new tentative design are found in Table~\ref{tab:DRparams9}. Some further optimisation will be applied in order to reduce the impact of Intra-Beam Scattering (IBS) in the vertical emittance, in particular by allowing a larger initial longitudinal emittance. Additional work was undertaken in collaboration with the light source community for beam tests in critical hardware, such as the SC wigglers~\cite{Schoerling2012} and kicker system~\cite{Iriso2018}.

\section{Ring To Main Linac Sections}
\label{sect:RTML}
\subsection{System Description}
\begin{figure}[!ht]
\begin{center}
\includegraphics[width = 0.9 \columnwidth]{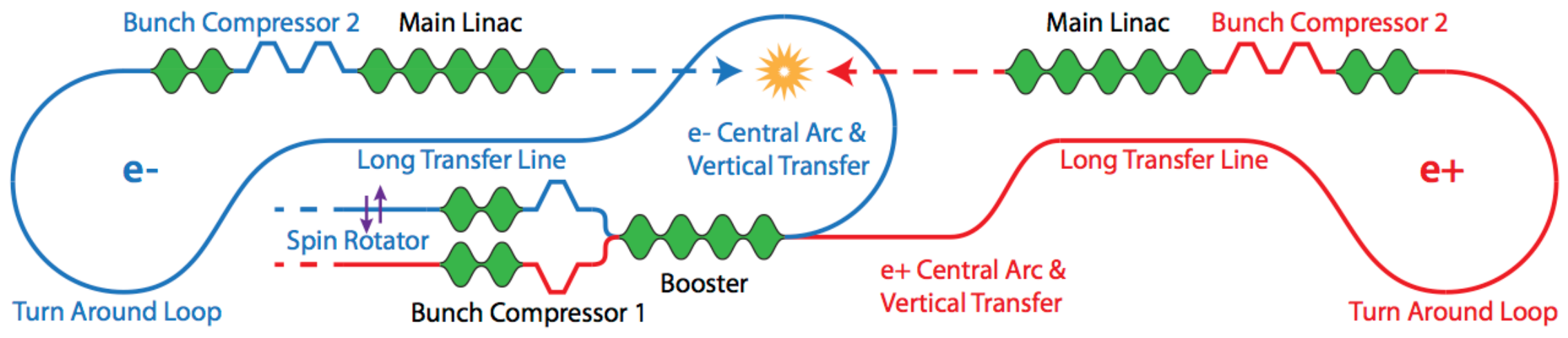}
\caption{Sketch RTML section, all subsystems are visible (dimensions not-to-scale).}
\label{fig:RTML_1} 
\end{center}
\end{figure}

The Ring To Main Linac (RTML) sections transport the electron and the positron beams from their respective damping ring, at ground level, to the main linacs start points, underground. While transporting the beam and matching the geometric layout of the beam lines, the RTML must accomplish three tasks: preserve the ultra-low beam transverse emittances from the damping rings, increase the beam energy from 2.86 GeV to 9 GeV, and compress the bunch length from 1.8\,mm to $\approx 70~\mu$m. The total lengths of the two RTMLs are slightly different, as the two lines have to accommodate different beam line layouts and guarantee the correct arrival time of the beams at the Interaction Point (IP). The electron line comprises eight subsystems: the Spin Rotator (SR), the two Bunch Compressors (BC1 and BC2), the booster linac (shared between electrons and positrons), the Central Arc (CA), the Vertical Transfer (VT), the Long Transfer Line (LTL) and the Turn-Around Loop (TAL). The positron RTML is composed of the same subsystems as the electron line, with the exception of the SR which is absent. A sketch of the whole RTML lines is visible in Fig.~\ref{fig:RTML_1}. The RF system of the booster linac, which accelerates both electrons and positron trains, determines the position and the length of the two arms of the RTML. An optimisation of the booster RF system aimed at minimising cost while maximising efficiency, lead to having the two electron / positron pulses separated by $\approx{}$3.4\,$\mu$s. This imposes a path-length difference from the booster to the IP of about 1020\,m.

\subsection{Design Choices}
The spin rotator in the electron line, located at the beginning of the RTML, offers full control of the beam polarisation.  Its design is unchanged with respect to the CDR and it is documented in \cite{Latina2010}. The electron beam depolarisation between the SR and the ML's end is kept minimal by design, having the SR run parallel to the ML. The small bending angle introduced by the Beam Delivery System (BDS) leads to a negligible depolarisation. The bunch compressors 1 and 2, respectively located at the start and at the end of each RTML, have been readjusted from the 3\,TeV CLIC design to the new bunch parameters. Given the flexibility of the original design, it has been sufficient to adjust the magnetic strengths of the chicanes and the operational parameters of the RF systems \cite{Han2017}. The RF systems of the two bunch compressors, 2\,GHz for BC1 and 12\,GHz for BC2, are unchanged with respect to the CDR, with the exception of the BC2's RF iris apertures, which have needed to be increased by a factor $<1.5$ to accommodate the larger bunch charge \cite{Han2017}. The bunch length is compressed from 1.8\,mm to 235\,$\mu$m in BC1, and from 235\,$\mu$m to 70\,$\mu$m in BC2.

\subsection{Beam Performance}
\begin{table}[!htb]
\caption{\label{tab:RTML_1}Normalised emittance budgets in the RTML.}
\begin{center}
\begin{tabular}{ccccc}
\toprule
\multirow{2}{*}{} & \multirow{2}{*}{Initial } & \multicolumn{3}{c}{Final emittance$^{(\star)}$}\tabularnewline
 &  & by Design & with Static Imperfections & with Dynamic Imperfections\tabularnewline
\midrule
$\epsilon_{x}$ {[}nm{]} & 700 & $<800$ & $<820$ & $<850$\tabularnewline

$\epsilon_{y}$ {[}nm{]} & 5 & $<6$ & $<8$ & $<10$\tabularnewline
\bottomrule
\end{tabular}
\end{center}
$^{(\star)}$ $90^\text{th}$ percentile.
\end{table}

One of the main challenges of the RTML is to fulfil the tight emittance budgets, summarised in Table~\ref{tab:RTML_1}. With respect to the 3\,TeV bunch parameters, the increased bunch charge and length induce the two competing effects that drive the RTML design. The larger charge induces stronger synchrotron radiation emission in all bending magnets, and magnifies the wakefield effects in the accelerating structures. The increased bunch length reduces the coherent synchrotron radiation emission in the magnetic chicanes, but enhances the effect of the transverse wakefields. All these effects, which induce the so-called ``design'' emittance growth, are taken into account. The design of the arc cells is the same in both the CA and the TAL, as presented in the CDR. It features strong-focusing double-bend achromat cells, with sextupoles for chromaticity correction, which make the arcs particularly sensitive to imperfections. 

The complexity of the design, together with the challenges due to the relatively high charge and increased length of the bunches, make the RTML extremely sensitive to both static and dynamic imperfections. 
Beam-based alignment (BBA) techniques are therefore mandatory to preserve the ultra-low emittances from the DR to the Main-Linac entrance and then to the IP, under the effects of misalignments and other static imperfections. Table~\ref{rtml.static_errors} summarises the static imperfections considered, which therefore constitute pre-alignment requirements. All the tabulated numbers have already been achieved and exceeded in existing machines, e.g. in light sources.

\begin{table}[!htb]
\caption{\label{rtml.static_errors} Static imperfections considered.}
\begin{center}
\begin{tabular}{ccc}
\toprule
Imperfection & RTML w/o CA and TAL & CA and TAL\tabularnewline
\midrule
R.M.S. position error & 100 $\mu$m & 30 $\mu$m\tabularnewline
R.M.S. tilt error & 100 $\mu$rad & 30 $\mu$rad\tabularnewline
R.M.S. roll error & 100 $\mu$rad & 30 $\mu$rad\tabularnewline
$\Delta B/B$ quadrupoles & $10^{-3}$ & $10^{-4}$\tabularnewline
$\Delta B/B$ other magnets & \multicolumn{2}{c}{$10^{-3}$}\tabularnewline
Magnetic-center shift w/strength & \multicolumn{2}{c}{0.35 $\mu$m / $5\%$}\tabularnewline
BPM resolution  & \multicolumn{2}{c}{1 $\mu$m}\tabularnewline
Sextupole movers step size & - & 1 $\mu$m\tabularnewline
\bottomrule
\end{tabular}
\end{center}
\end{table}

\begin{figure}[!htb]
\begin{center}
\includegraphics[width=0.45\textwidth]{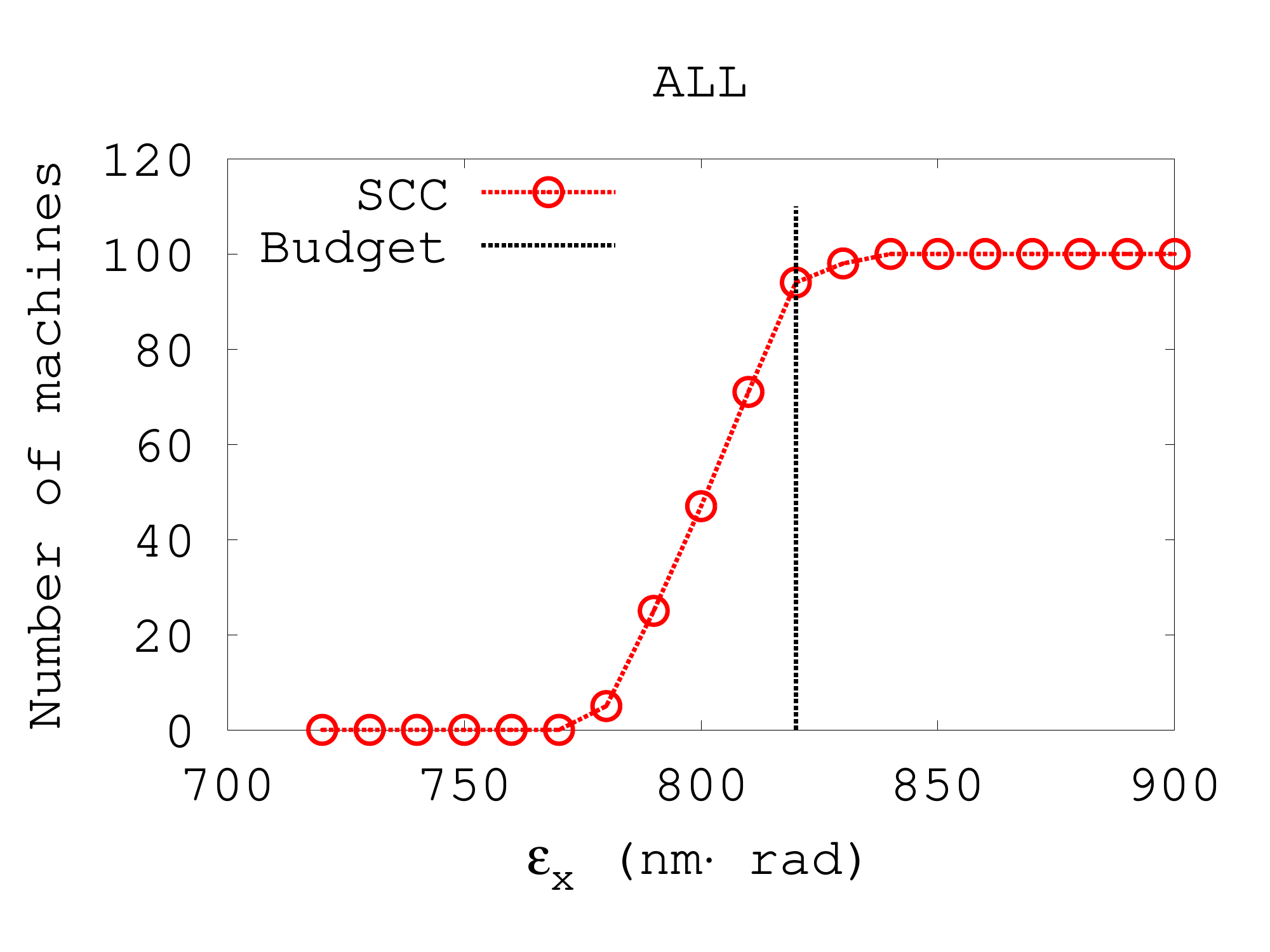}
\includegraphics[width=0.45\textwidth]{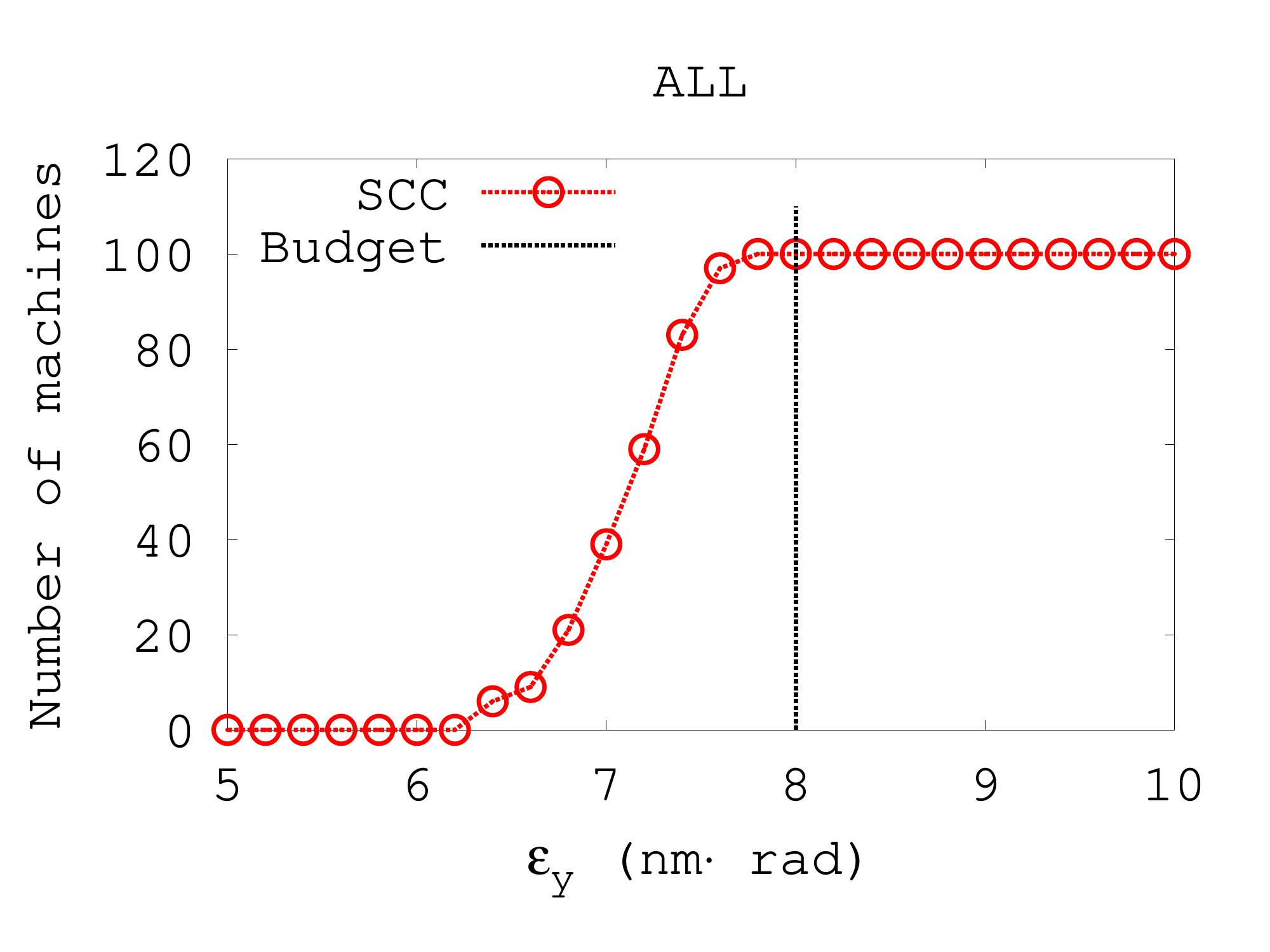}
\caption{\label{fig:RTML_2}Emittance at the end of the electron RTML after beam-based alignment and tuning. In the horizontal plane just 6 out of 100 machines exceed the budget; in the vertical plane all machines are below the budget. The acronym SCC means Sextupole Coupling Correction, i.e. the final step of the tuning procedure.}
\end{center}
\end{figure}

The BBA procedure designed for the RTML consists of applying dispersion-free steering (DFS) in all subsystems, followed by an iteration of emittance tuning bumps. In the CA, DFS is applied using a test beam which is prepared by decreasing the gradient of the booster linac by 5\% and adjusting its optics accordingly. In the TAL, the residual dispersion is measured by decreasing the magnetic strength of the magnets by 5\,\%. The emittance tuning bumps are implemented using the first five sextupoles of both the CA and the TAL, targeting the beam size at the Main-Linac entrance to remove the effects of coupling and optics mismatches. Detailed studies of BBA performance taking into account element pre-alignment errors, angle, magnetic strength errors, and dynamic magnetic-centre shifts (Table~\ref{rtml.static_errors}), showed that the emittance budgets can be fulfilled both for the 380\,GeV and the 3\,TeV stages of CLIC \cite{Han2017,Han2017A}. In the vertical plane, which is the most critical, all seeds are corrected to be within budget. The distribution of final emittances of 100 random misalignment seeds is shown in Fig.~\ref{fig:RTML_2}. These new studies showed increased robustness with respect to the CDR results, and enabled the relaxation of the pre-alignment tolerances for the component installation in the tunnel.  

The emittance growth in the RTML is sensitive also to dynamic effects, such as transverse orbit jitter at  injection, vibrations, and stray magnetic fields. The transverse orbit jitter seems the most critical among the dynamic effects, and results in stringent stability requirements of the DR extraction elements. This led to dedicated R\&D for the extraction kickers from the DR as documented in \cite{Belver-Aguilar2014}. To further minimise the impact of incoming beam jitter on the RTML, an option for a feed-forward solution via hardware is presented in \cite{Lienart2012,Apsimon2014}.

\section{Main Linacs}
\label{sect:ML}
\subsection{Overview}
The two Main Linacs (MLs), one for positrons and one for electrons, accelerate the beams from an initial energy of 9\,GeV to the final value of 190\,GeV using normal conducting accelerating structures with an RF frequency of 12\,GHz and a gradient of 72\,MV/m. This choice of frequency and gradient is based on an optimisation of the total accelerator cost. The linac design is identical for electrons and positrons and the linacs are each about 3.5\,km long. This includes a total energy overhead of 10\,\% to allow for different operational margins. A key design goal is the preservation of the ultra-low transverse emittances during beam transport. This goal is achieved by a combination of careful lattice design, precise pre-alignment of the beam line components, stabilisation of the beam-guiding quadrupoles against vibrations and use of beam-based correction methods. The Main-Linac tunnel and the beam line are laser straight. This avoids the complications that would result from a linac that follows the curvature of the Earth~\cite{Latina2006}.

\subsection{Beam Parameters}
Table~\ref{tab:ML_1} shows the key beam parameters for the Main Linacs. In the linac the bunch length remains constant, while the transverse emittances increase due to machine imperfections. The beam is accelerated at an average RF phase of $12^\circ$ in order to limit the final energy spread; this results in an effective gradient reduction of about 2\,\%.

\begin{table}[!ht]
\caption{Key beam parameters in the ML.}
\label{tab:ML_1}
\begin{center}
\begin{tabular}{l c l c}
\toprule
Particles per bunch				&$5.2\times10^9$&
Bunches per pulse				&$352$\\
Bunch spacing					&$15\,$cm&
Bunch length					&$70\,\mu$m\\
Initial R.M.S. energy spread	&$\le2\,\%$&
Final R.M.S. energy spread		&$0.35\,\%$\\
Initial horizontal emittance	&$\le850\,$nm&
Final horizontal emittance		&$\le900\,$nm\\
Initial vertical emittance		&$\le10\,$nm&
Final vertical emittance		&$\le20\,$nm\\
\bottomrule
\hline
\end{tabular}
\end{center}
\end{table}

\subsection{Linac Layout and Optics}
The Main Linac consists of a sequence of 2.343\,m-long modules. Three different types exist: the most common supports eight accelerating structures (type T0). The other two (T1 and T2) support six or four accelerating structures and a quadrupole of 0.43\,m or 1.01\,m length, respectively. Each pair of accelerating structures is fed by a PETS in the Drive Beam line.

The ML consists of five lattice sectors each using FODO optics~\cite{Schulte2009}. The phase advance is about $72^\circ$ per cell throughout the ML. The quadrupole spacing is constant in any particular sector but varies from sector to sector following an approximate scaling with $\sqrt{E}$ (see Table~\ref{tab:ML_2}). There is a quadrupole on every module in the first sector (2.343\,m spacing). The quadrupole spacing increases along the linac until there is one quadrupole for every four accelerating modules in the last sector (9.372\,m spacing). This quadrupole spacing balances the contributions to emittance growth from dispersive and wakefield effects along the linac. The total length of quadrupoles is roughly the same in every sector resulting in an almost constant fill factor (the ratio of the length of the accelerating structures to the total length). The lattice functions between sectors are matched using the last four quadrupole of the lower energy sector and first three quadrupoles of the higher energy sector.

At the end of each Drive-Beam sector, four of the Main-Beam modules are not equipped with accelerating structures but only with drifts. In this section the used Drive Beam is bent into the dump line and the new Drive Beam is brought in. The total number of PETS per decelerator is 1,273.

\begin{table}[!htb]
\caption{The main parameters of the different ML sectors.}
\label{tab:ML_2}
\begin{center}
\begin{tabular}{l *5{c}}
\toprule
\textbf{sector number}& \textbf{1} & \textbf{2}& \textbf{3}& \textbf{4}& \textbf{5} \\
\midrule
quadrupole number& 120 & 150 & 86 & 62 & 156 \\
quadrupole length [m] & 0.43& 0.43& 0.43& 1.01& 1.01\\
quadrupole spacing [m] & 2.343& 4.686& 7.029& 7.029 & 9.372\\
\bottomrule
\end{tabular}
\end{center}
\end{table}

\subsection{Accelerator Physics Issues}
\begin{figure}[!htb]
\centering
\includegraphics[scale=0.66]{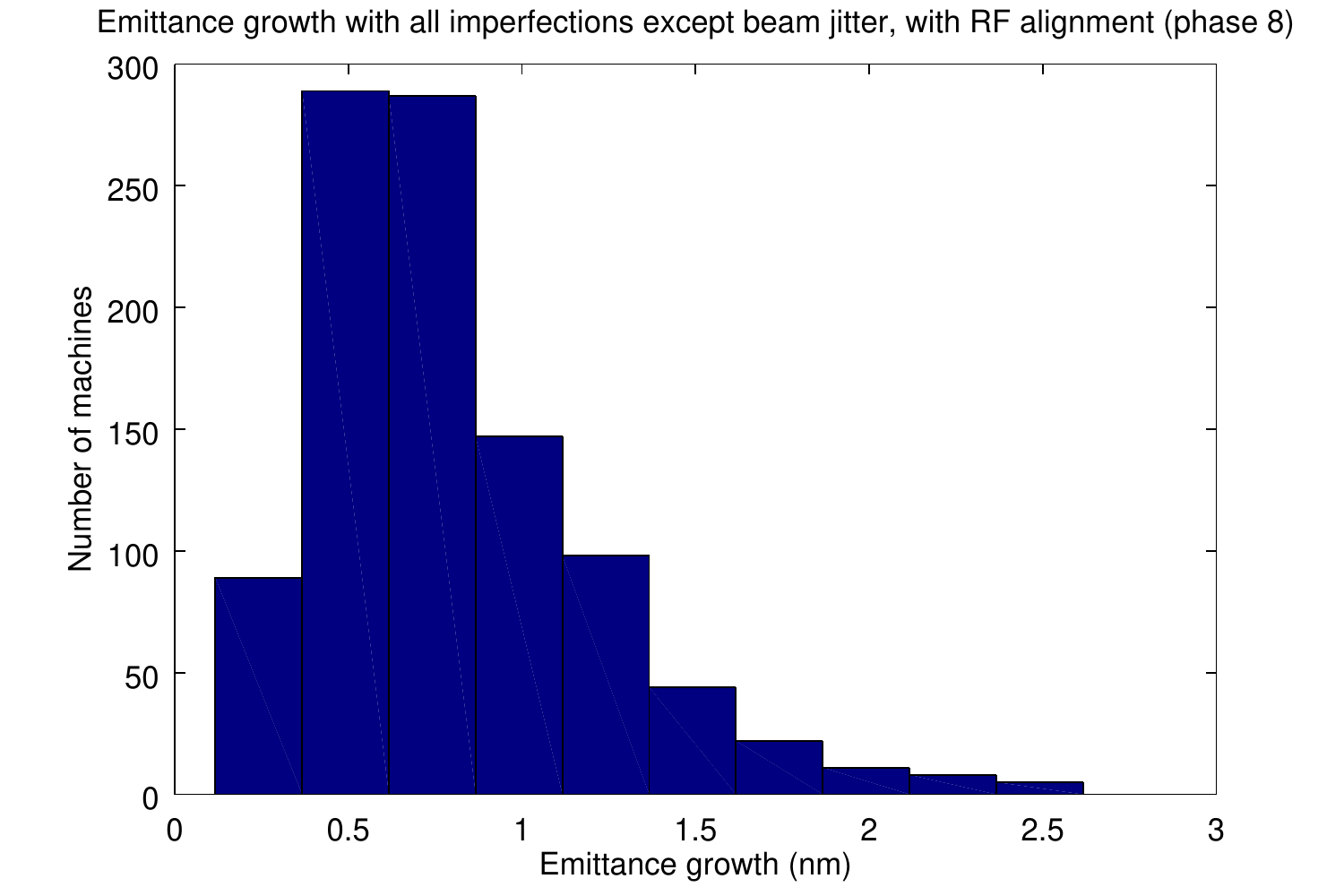}
\caption{Probability distribution of the emittance growth for static imperfections for a ML average RF phase of $12^\circ$.}
\label{f:dsstatic}
\end{figure}

The accelerator physics issues for the 380\,GeV stage are the same as those for the 3\,TeV stage. Of particular importance is the preservation of the vertical emittance in the presence of static misalignments of the accelerator components. The goal is to achieve an emittance growth below 5\,nm with a 90\,\% likelihood, the same goal as for the 3\,TeV design. In principle, several of the tolerances could be reduced by  a factor of about two at 380\,GeV compared with 3\,TeV, since the ML is shorter. However, this would require upgrading the systems for better performance when the energy is upgraded. Hence the system design and the specifications remain unchanged. The specifications for the imperfections and the resulting values of the emittance growths are given in Table~\ref{tab:ML_3}~\cite{c:MLperf}. The expected value for the total growth is about 1\,nm. The emittance growth has a stochastic probability distribution (Fig.~\ref{f:dsstatic}). With a probability that 90\,\% of the machines remain below 1.5\,nm emittance growth, which is well within the budget.

\begin{table}[!htb]
\caption{Key alignment specifications for the ML components and the resulting emittance growth. The values after simple steering (1-2-1), Dispersion Free Steering (DFS) and realignment of the accelerating structures using the wakefield monitors (RF) are shown.}
\label{tab:ML_3}
\begin{center}
\begin{tabular}{cccccc}
\toprule
                 &                    &              & \multicolumn{3}{c}{$\Delta \epsilon_y$ [nm]} \\
Imperfection     & With respect to    & Value        & 1-2-1  & DFS    & RF     \\
\midrule
Girder end point  & Wire reference     & 12 $\mu$m    & 12.91  & 12.81  & 0.07   \\
Girder end point  & Articulation point & 5 $\mu$m     & 1.31   & 1.30   & 0.02   \\
Quadrupole roll   & Longitudinal axis  & 100 $\mu$rad & 0.05   & 0.05   & 0.05   \\
BPM offset        & Wire reference     & 14 $\mu$m    & 188.99 & 7.12   & 0.06   \\
Cavity offset     & Girder axis        & 14 $\mu$m    & 5.39   & 5.35   & 0.03   \\
Cavity tilt       & Girder axis        & 141 $\mu$rad & 0.12   & 0.40   & 0.27   \\
BPM resolution    &                    & 0.1 $\mu$m   & 0.01   & 0.76   & 0.03   \\
Wake monitor      & Structure centre   & 3.5 $\mu$m   & 0.01   & 0.01   & 0.35   \\
\midrule
All               &                    &              & 204.53 & 25.88  & 0.83   \\
\bottomrule
\end{tabular}
\end{center}
\end{table}

Good vacuum quality prevents fast beam ion instability and keeps tail generation, due to beam-gas scattering, low. A quality similar to the 3\,TeV design \cite{Aicheler2012} is foreseen; studies with a more refined simulation code indicate that the requirements could be relaxed by a factor of a few~\cite{c:lotta}.

\subsection{Components}
The total number of components is listed in Table~\ref{tab:ML_4}. The alignment tolerances are listed in Table~\ref{tab:ML_3}.

\begin{table}[!htb]
\caption{Key components of the Main Linac.}
\label{tab:ML_4}
\centering
\begin{tabular}{{l}{r}}
\toprule
Structures T0 & 914 \\
Quadrupoles T1 & 356 \\
Quadrupoles T2 & 218 \\
Accelerating structures & 10,184\\
PETS & 5,092\\
\bottomrule
\end{tabular}
\end{table}

\section{Beam Delivery System}
\label{sect:BDS}
\noindent The Beam Delivery System (BDS) transports the e$^{+}$ and e$^{-}$ beams from the exit of the linacs to the IP and performs the critical functions required to meet the CLIC luminosity goal. First the beam is cleaned in the collimation section and then it is focused with the Final Focus System (FFS) while correcting higher order transport aberrations in order to deliver the design IP beam sizes. The BDS new baseline foresees an $L^{*}$ of 6~metres with final quadrupoles mounted outside the detector volume~\cite{clicdet2017,Plassard2018a}, directly on the tunnel floor. The Final Focus lattice with $L^{*}$~=~6~m has been lengthened by a factor 6/4.3 compared with the previous $\sqrt{s}$~=~500\,GeV design~\cite{Lebrun2012}.

\subsection{BDS Optics Design}
\begin{table}[!htb]
\centering
\caption{\label{beam_param} CLIC 380\,GeV beam parameters.}
\begin{tabular}{lcc} 
\toprule
Norm. emittance (end of linac) $\gamma\epsilon _{x}/ \gamma\epsilon _{y}$ & [nm]     &     900 / 20          \\ 
Norm. emittance (IP) $\gamma\epsilon _{x}/ \gamma\epsilon _{y}$ & [nm]     &     950 / 30          \\ 
Nominal beta function (IP) $\beta_{x}^{*}/ \beta_{y}^{*}$ &  [mm]   &         8.0 / 0.1       \\ 
Target IP beam size $\sigma_{x}^{*}/ \sigma_{y}^{*}$ &  [nm]      &     149 / 2.9     \\ 
Bunch length $\sigma_{z}$  &  [$\mu$m]           &        70      \\ 
R.M.S. energy spread $\delta_{p}$  &   [\%]      &       0.35        \\ 
Bunch population $N_{e}$  &   [$10^{9}$]      &       5.2        \\ 
Number of bunches  $n_{\text{b}}$  &      &         352        \\
Repetition rate $f_{\text{rep}}$ &   [Hz]         &         50         \\ 
Luminosity $\mathcal{L}_0$  & [$10^{34} \text{cm}^{-2}\text{s}^{-1}$]      &     1.5     \\ 
Peak luminosity $\mathcal{L}_{0.01}$  &   [$10^{34} \text{cm}^{-2}\text{s}^{-1}$]        &      0.9        \\ 
\bottomrule
\end{tabular}
\end{table}

\begin{figure}[!htb]
\centering
\includegraphics[scale=0.35, angle=-90]{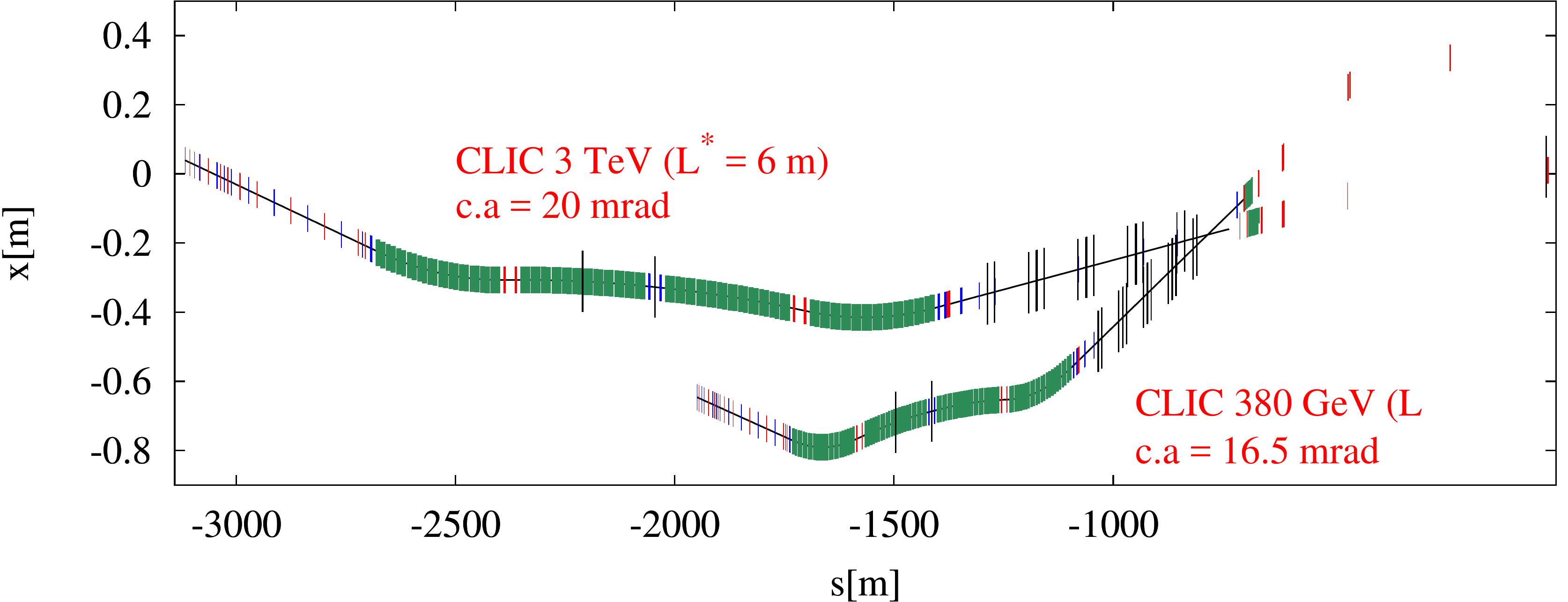}
\caption{\label{fig_bds_tunnel} CLIC 3\,TeV and 380\,GeV BDS layouts with $L^{*}$~=~6~m.}
\end{figure}
\begin{figure}[htb!]
\centering
\includegraphics[scale=0.33, angle=-90]{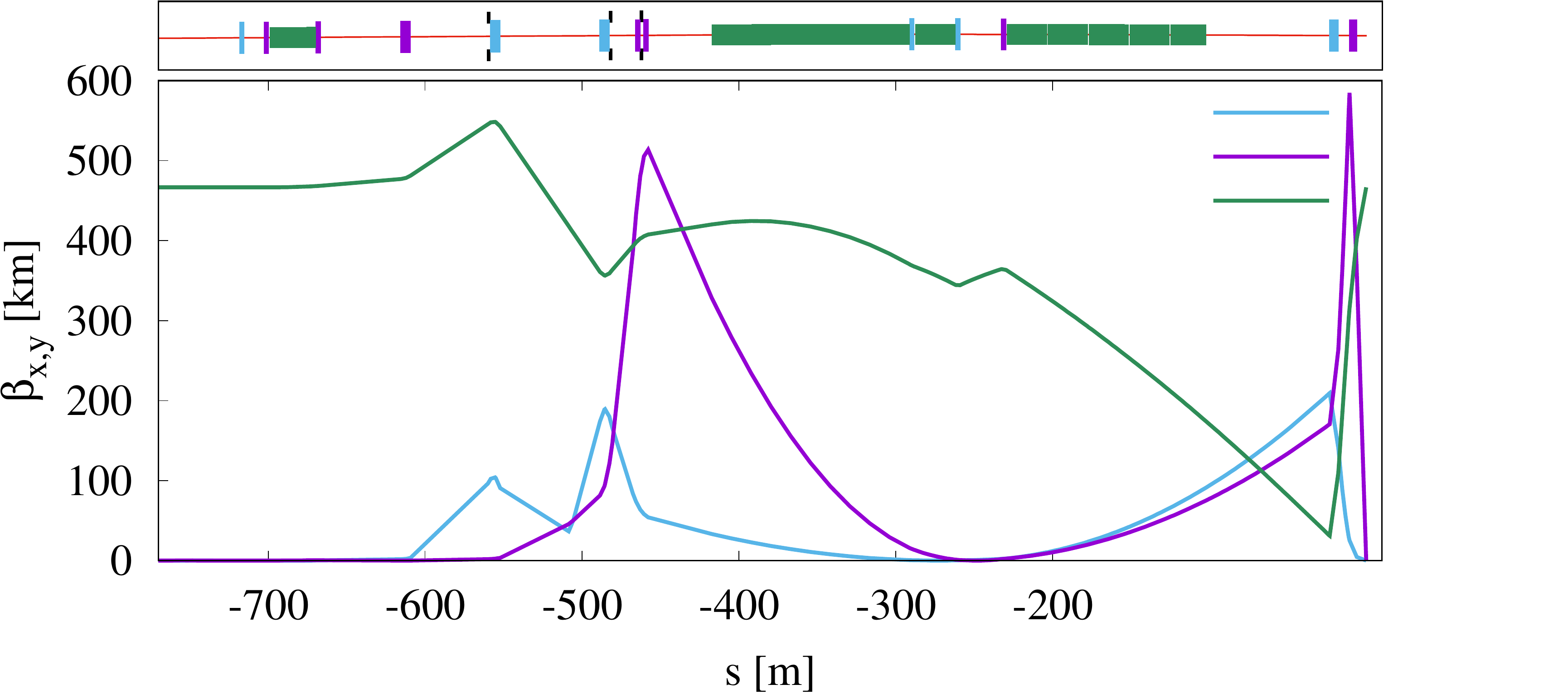}
\caption{\label{twiss} Optical functions through the FFS with $L^{*}$~=~6\,m.  }
\end{figure}

\noindent The BDS has been optimized with respect to the beam parameters foreseen for the first CLIC energy stage given in Table~\ref{beam_param}. The length of the entire BDS is $\approx 1950$\,m and the FFS length is 770\,m. The BDS, extended to 2,222\,m by means of a drift, is 878\,m shorter than the 3,100\,m BDS for the 3\,TeV design. In order to allow the energy upgrade inside the CLIC tunnel, the end of the 380\,GeV and 3\,TeV BDS beamlines have been matched such that the axis along which the ML is located is unchanged by the upgrade. The crossing angle for the CLIC 3\,TeV BDS is 20\,mrad and the required crossing angle for the CLIC~380\,GeV BDS is 16.5\,mrad for the $L^{*}$~=~6\,m BDS (see Fig.~\ref{fig_bds_tunnel}). The FFS dipoles, quadrupoles and sextupoles have been optimized to match the desired beam parameters at the IP while locally correcting the chromaticity generated by the Final Doublet~\cite{Raimondi2001} (QF1 and QD0 quadrupoles). The main optical functions are shown in Fig.~\ref{twiss}. A pair of octupoles has been introduced in the lattice to correct the remaining 3$^{\text{rd}}$ order chromatic and geometric aberrations. To benchmark the optics of the system a simplified assessment is made by tracking a beam from the end of the ML with emittances (950\,/\,30\,nm) throughout the perfect BDS. This corresponds to the case where the full emittance budget was used and the results are close to the CLIC requirements (see Table~\ref{lumi_perf}). Due to the non-uniform energy spread and highly nonlinear properties of the FFS the R.M.S. values of the beam sizes get artificially inflated by the bunch tails and therefore the quoted beam sizes are computed from Gaussian fits to the core of the beam at IP. The nominal horizontal beam size is somewhat smaller than the target which gives some overhead but lowers $\mathcal{L}_{0.01}$.

\begin{table}
\begin{center}
\caption{\label{lumi_perf}Performances of both $L^{*}$ options for the CLIC 380\,GeV BDS.}
\begin{tabular}{lcccc}
\toprule
Design &  $\sigma_{x}^{*}$ & $\sigma_{y}^{*}$ & $\mathcal{L}_0$  & $\mathcal{L}_{0.01}$   \\ 
  & [nm] & [nm] & [$10^{34} \text{cm}^{-2}\text{s}^{-1}$] & [$10^{34} \text{cm}^{-2}\text{s}^{-1}$] \\ 
\midrule
 $L^{*}$~=~6~m  & 144.7 & 2.95 & 1.58 & 0.89\\ 
$L^{*}$~=~4.3~m & 141.3 & 2.81 & 1.63 & 0.92 \\
\bottomrule
\end{tabular}
\end{center}
\end{table}

\subsection{FFS Tuning}
Tuning of the FFS aims to bring the system to its design performance under realistic beamline imperfections~\cite{Dalena2012,Marin2018a}. The procedure consists of beam-based alignment (BBA) techniques for correcting trajectory and dispersion, and orthogonal sextupole knobs that aim to correct chosen aberrations at the IP~\cite{Nosochkov2002}, independently. The following steps are included in the tuning procedure \cite{c:BDSperf}:
\begin{enumerate}
\item Beam-based alignment (BBA) with all multipoles switched OFF
\item Pre-alignment of sextupoles by powering them one-by-one and monitoring luminosity
\item Sextupole linear knobs (transverse position)
\item Octupole linear knobs (transverse position)
\item Sextupole linear knobs (transverse position), second iteration
\end{enumerate}
The evaluation of the tuning efficiency is estimated over 500 randomly misaligned machines and the figure of merit of the tuning procedure is the luminosity. Imperfections assumed for the FFS tuning are given in Table~\ref{imperfections}. The tracking simulations are done in PLACET with synchrotron radiation included and luminosities are computed from full beam-beam simulations in GUINEA-PIG, which includes a random noise (order of 1\,\%) to the luminosity signal. The effect of ground motion is not included in the simulations and a single beam is simulated and then mirrored to evaluate the luminosity. 

\begin{table}[htb!]
\caption{\label{imperfections} Imperfections applied to the FFS lattice.}
\centering
\begin{tabular}{lc}
\toprule
$\sigma_{\text{offset}}$ (Quadrupoles, Sextupoles, Octupoles and BPMs) & 10~$\mu\text{m}$ \\
BPM resolution &  20 $\text{nm}$ \\
$\sigma_{\text{roll}}$ (Quadrupoles, Sextupoles, Octupoles and BPMs) & 100~$\mu\text{rad}$ \\
Strength error (Quadrupoles, Sextupoles and Octupoles) & 0.01~\% \\
\bottomrule
\end{tabular}
\end{table}
\par
\begin{figure}[htb!]
\centering
\includegraphics[scale=0.55]{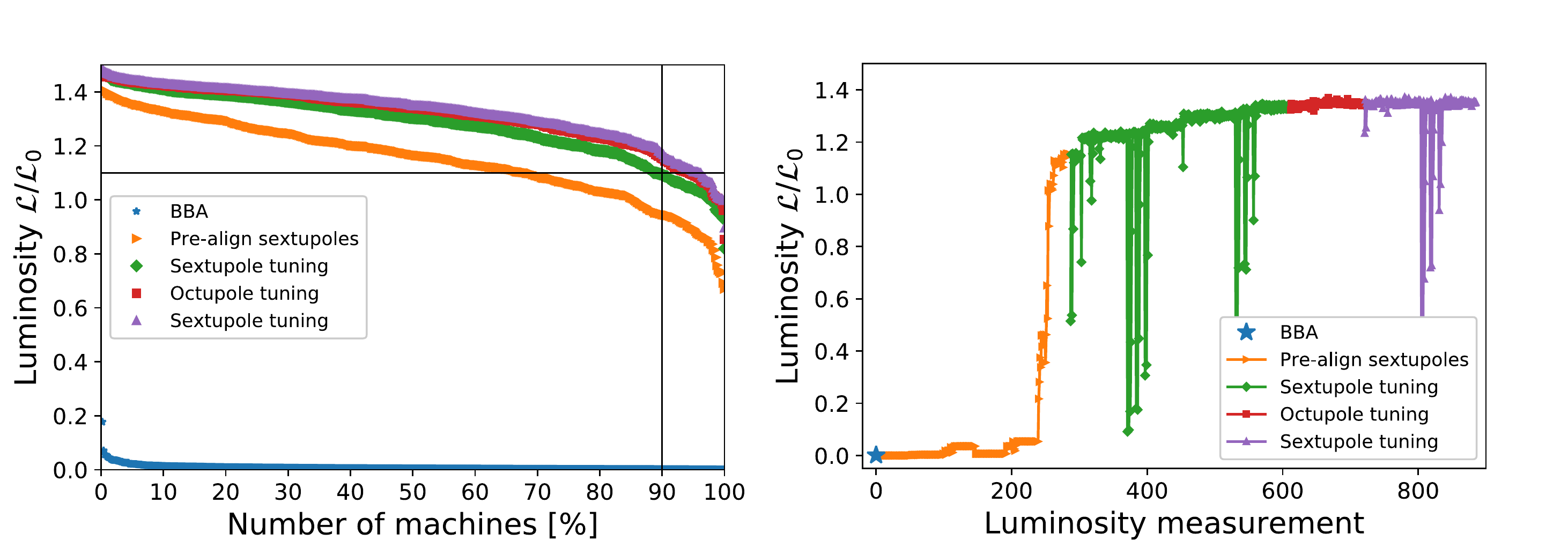}
\caption{\label{fig:tuning} Left: luminosity after different tuning stages for 500 machines with randomly distributed static imperfections. After the final stage (about 900 luminosity measurements) 90\,\% of the machines reach 117\,\% of the nominal luminosity. Right: the luminosity evolution for the average machine. }
\end{figure}

The tuning performance goal is that 90\,\% of the machines should reach a luminosity of 110\,\% of the design luminosity $\mathcal{L}_{0}~=~1.5\times 10^{34}\text{cm}^{-2}\text{s}^{-1}$. The 110\,\% is to ensure a budget for dynamic imperfections. In the simulations, a realistic beam is used which is obtained from tracking studies from the DR to the end of the ML. In the current studies, imperfections and tuning of the collimation part of the BDS has not been included, but only for the final focus system, which is also the most challenging part. Figure~\ref{fig:tuning} shows the simulation results for the $L^{*}$~=~6\,m system where, after BBA, sextupole pre-alignment and tuning of sextupoles and octupoles, 90\,\% of the machines reach a luminosity of 117\,\% of $\mathcal{L}_{0}$. Approximately 900 luminosity measurements were needed for the full tuning procedure. The results look promising but the study should be extended to include Two-Beam tuning and dynamic imperfections. Great progress has already been made for the 3\,TeV CLIC FFS on Two-Beam tuning~\cite{Marin2018a} and tuning with dynamic imperfections~\cite{Marin2018}. Furthermore, the influence of measured beam energy and BPM calibration errors on the BBA and its consequences on the tuning should also be further studied.

\section{Machine Detector Interface}
\label{sect:MDI}

The Machine Detector Interface (MDI) is the region within the detector cavern where the beam lines of the accelerator pass through the detector. Key issues are the support of the final beam line components within the detector, luminosity monitoring and feedback, background suppression, and radiation shielding. The MDI includes the positioning of the final focus quadrupoles, QD0, which have very stringent alignment and stabilisation requirements. Luminosity monitoring is integrated within the detector. The spent beams must be transported cleanly away through the experiment onto two beam dumps, via the post-collision lines. Collimators and masking must suppress backgrounds from the incoming beams, from the beam-beam interaction and from the beam dumps. The cavern layout must minimize the exposure of equipment and personnel to radiation.

Since the CDR \cite{Aicheler2012}, the final quadrupole QD0 has been moved from inside the detector (with $L^*$~=~3.5\,m at 3\,TeV or $L^*$~=~4.3\,m at 500\,GeV) to the tunnel floor outside the detector (using $L^*$~=~6\,m for both 380\,GeV and 3\,TeV designs). The angular acceptance of the detector in the forward region becomes significantly larger and a number of technical systems become simpler. The peak luminosity decreases only marginally.

The detector community has decided to concentrate on a single detector, designed for opening in a garage position only, away from the IP. At 3\,TeV the crossing angle between the beam lines is 20\,mrad. At 380\,GeV it is 16.5\,mrad in order to geometrically match the BDS in the same tunnel. This imposes some changes to the vacuum layout and modifications to calorimeter closest to the beam, the BeamCal, which measures very small angle particles. Lower backgrounds from incoherent pairs at 380\,GeV allow for a central vacuum pipe with a smaller diameter, and thus a smaller radius of the innermost vertex detector layer. However, most of the detector elements - in particular the heavy and expensive calorimeters - will remain unchanged when moving from the lower energy stages to 3\,TeV.

As many of the design constraints, such as beam-beam backgrounds and radiation effects, as well as required fields and gradients are more stringent at 3\,TeV, most of the new MDI layout features are based on the 3\,TeV parameters.

In the new design, the QD0 quadrupoles are mounted on the tunnel floor. Measurements at the CMS cavern have shown that the tunnel floor movements are almost two orders of magnitudes smaller than vibrations inside the detector. In the tunnel sufficient space is available to comfortably house the stabilisation and alignment systems. As the quadrupoles are now no longer embedded inside the detector, the alignment can be based on the same technology as the rest of the linacs and Beam Delivery Systems. The baseline layout no longer foresees a pre-absorber, as the cantilever supporting the QD0 magnets has been suppressed. It has been shown that the luminosity does not change appreciably if the QD0 magnets are split into two or three shorter parts. The QD0 quadrupoles have a much smaller gradient, 25\,T/m, and a larger aperture radius (25\,mm). Shorter magnets are lighter, easier to produce and to stabilise. The overall design features could remain the same, but with the lower gradient an alternative solution with classical electromagnets is now under consideration. No anti-solenoid is needed, as the QD0 magnets are outside the detector solenoid field.

The new layout simplifies the Machine Detector Interface inside the detectors considerably. The IP feedback Beam Position Monitors (BPMs) and kicker can stay roughly in their CDR positions. The larger aperture minimises the pumping time and venting of the QD0 is no longer required before moving the detector to its garage position. Therefore, the vacuum sectorisation can be significantly simplified. 

\begin{figure}[ht!]
\centering
\includegraphics[scale=0.95]{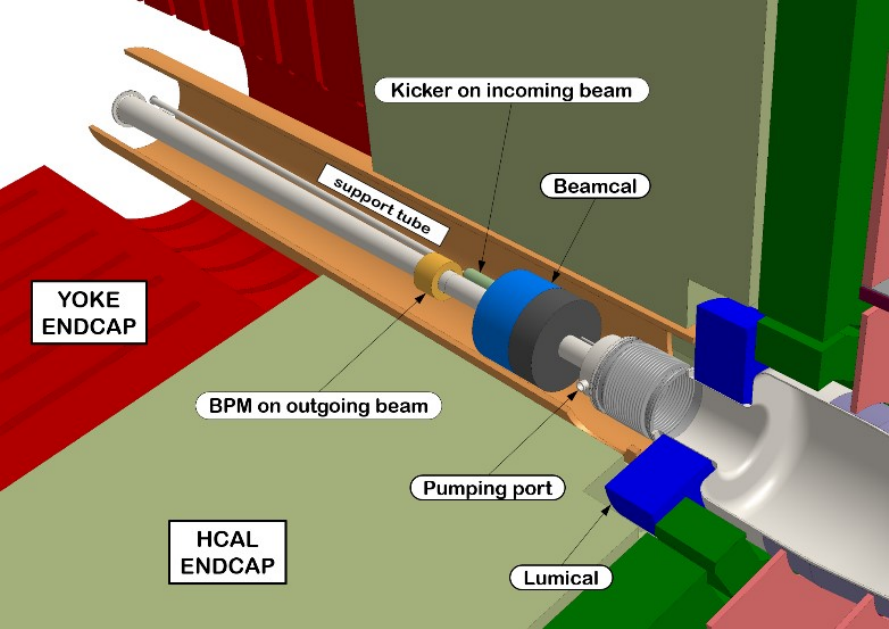}
\caption{\label{fig:MDI_2}A blown-up view of the MDI region inside the detector.}
\end{figure}

\begin{figure}[h!]
\centering
\includegraphics[scale=0.95]{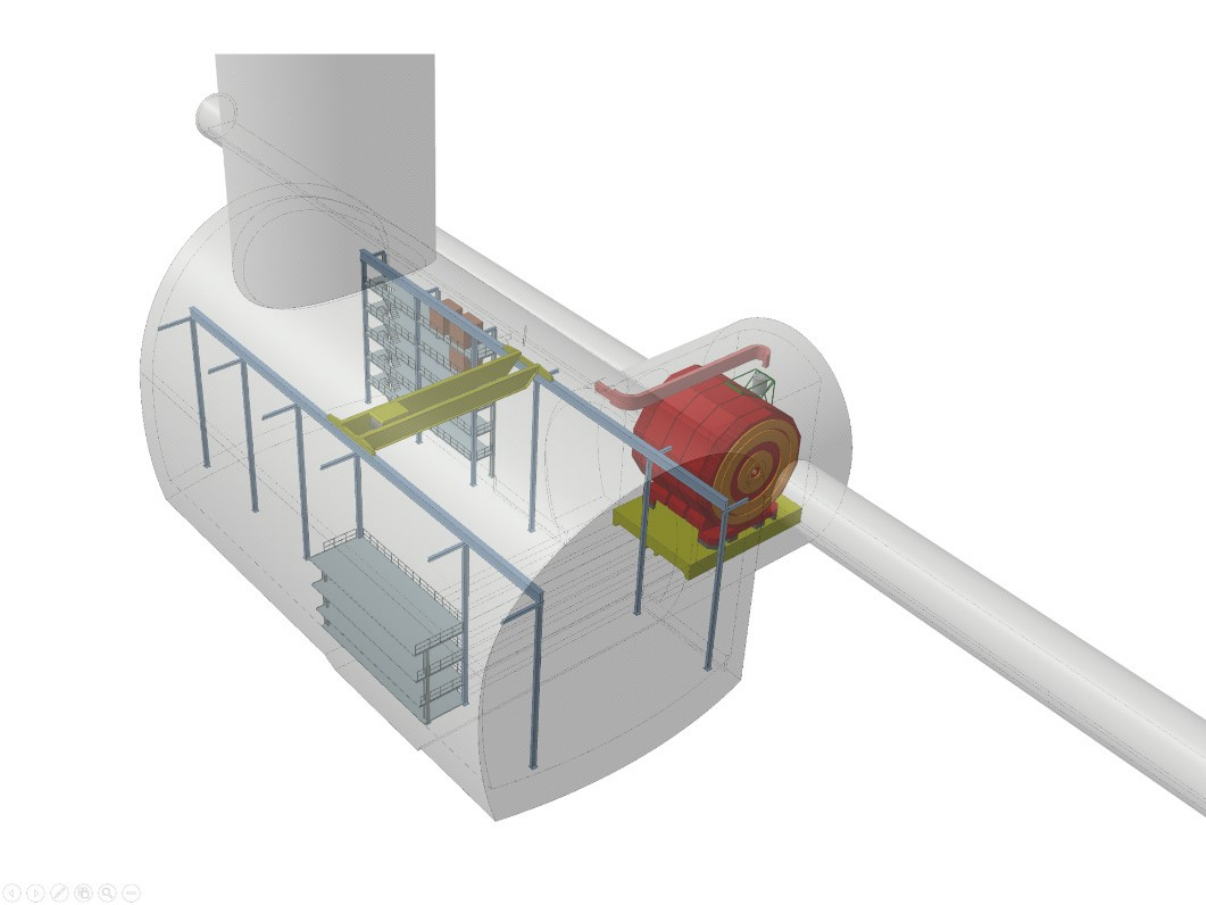}
\caption{\label{fig:MDI_3}The new cavern layout.}
\end{figure}

A blown-up view of the MDI region is shown in Fig.~\ref{fig:MDI_2}. The suppression of the second detector reduces the cavern excavation volume. The new layout is shown in Fig.~\ref{fig:MDI_3}.

\section{Post-Collision Line}
\label{sect:PCL}
The CLIC Post-Collision Line (PCL) is described in detail in the CDR \cite{Aicheler2012}. This description focused on the 3\,TeV design. Accounting for changes and improvements \cite{Ferrari2009,Angal-Kalinin2013,Gschwendtner2010}, the PCL has been re-investigated for the 380\,GeV design. The expectation is that the lower-energy design can more easily meet the requirements described in the CDR. The investigation confirms that by scaling the dipole magnet strengths down from 0.8\,T to approximately 0.1\,T and maintaining the other key aspects of the 3\,TeV design, the 380\,GeV PCL is better capable of meeting the design requirements than the 3\,TeV PCL.

The general layout remains essentially the same as that described in the CDR, with the exception of the first dipole magnet pair and the final drift length. Looking at the magnet designs described in \cite{Vorozhtsov2010}, the lengths of the pair of magnets were kept at 2\,m each, instead of 0.5\,m and 3.5\,m, as previously stated. This change caused no impact on the overall performance of the PCL and followed the design parameters described in \cite{Vorozhtsov2010}. The final drift is no longer 210\,m, but 50\,m, as described in \cite{Ferrari2009}.

\begin{figure}[!htb]
  \centering
    \includegraphics[width=1.0\textwidth]{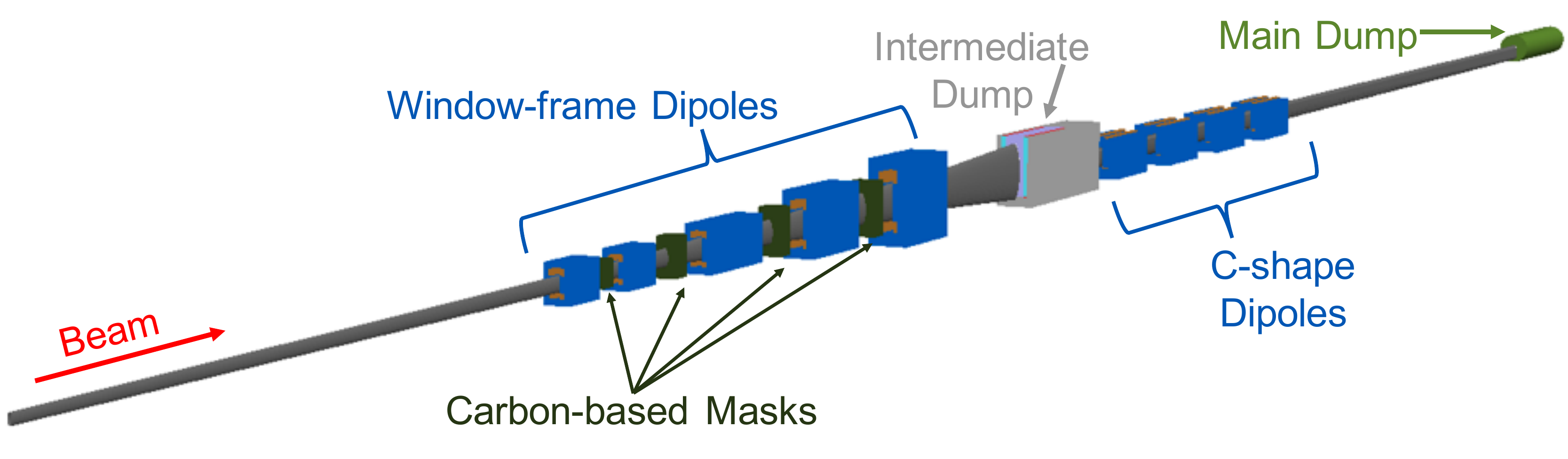}
    \caption{Post-collision line layout from BDSIM viewer.}
    \label{fig:PCL_Overview}
\end{figure}

To be sure that the re-investigation provided similar results to those shown in the CDR and previous studies, the model was rebuilt from the ground-up using BDSIM \cite{Nevay2018} and its geometry utility, pyg4ometry \cite{pyg4ometry2018}. The geometries of the beampipes, the intermediate dump, and the carbon-based masks were based upon the designs listed in \cite{Ferrari2009,Angal-Kalinin2013,Gschwendtner2010}. For the beam distributions, the data were provided by the CLIC beam-beam interactions studies for both the 3\,TeV and the 380\,GeV machines \cite{beambeam2018}. The main dump was based upon the design detailed in \cite{Mereghetti2011}. An overview of the full PCL design as visualized in BDSIM can be seen in Fig.\,\ref{fig:PCL_Overview}.

After confirming that this re-built 3\,TeV design achieves similar results to those previously reported, investigation of the 380\,GeV design began by scaling down the magnet strengths to provide the same angular kick to the nominal 190\,GeV beam. The dipole magnets for the 3\,TeV design are all set to 0.8\,T, which provides a kick angle of 0.64\,mrad to the 1.5\,TeV electron beam at each dipole. For a 190\,GeV electron beam, the same kick angle is provided by a magnet strength of 0.1\,T.

\begin{figure}[!htb]
  \centering
    \includegraphics[width=1.0\textwidth]{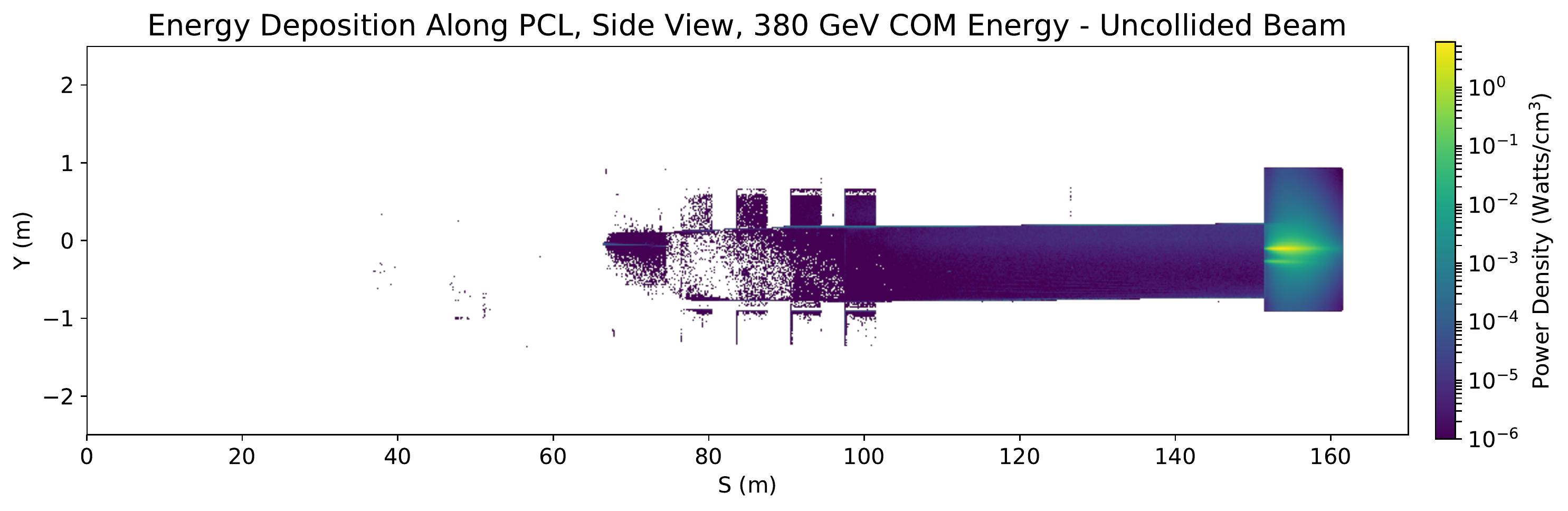}
    \caption{Energy deposition along the PCL for 190\,GeV uncollided electron beam.}
    \label{fig:380sideuncollided}
\end{figure}

\begin{figure}[!htb]
  \centering
    \includegraphics[width=1.0\textwidth]{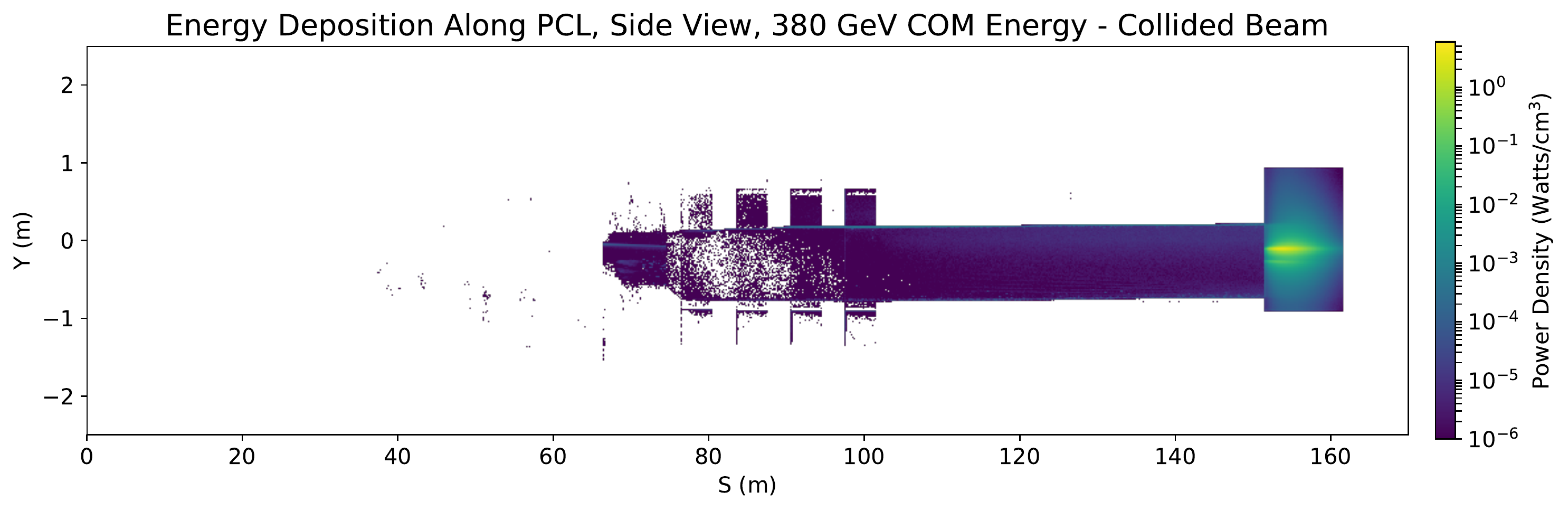}
    \caption{Energy deposition along the PCL for 190\,GeV collided electron beam.}
    \label{fig:380sidecollided}
\end{figure}

Investigation of the energy deposition along the PCL and in the main dump shows that there are no unexpected ``hot spots'' along the beamline. Figs.~\ref{fig:380sideuncollided} and \ref{fig:380sidecollided} show the energy deposited along the PCL for the uncollided and collided 190\,GeV electron beams, respectively. The colors correspond to power density per unit volume, including the power from deposited secondary particles. The main dump receives the vast majority of the power deposition, and less power is deposited along the PCL than for the 3\,TeV design. Table~\ref{tab:Uncollided_Power} summarizes the power deposition in several key elements of the PCL for both the 3\,TeV and 380\,GeV designs, including both collided and uncollided electron beams. Since the main dump is designed to handle 14\,MW of power from the 1.5\,TeV electron beam, the 2.91\,MW provided by the uncollided 190\,GeV electron beam is easily managed, as expected. Furthermore, the majority of the energy deposition into the main dump occurs below the vertical center, around $-10$\,cm. This should allow for luminosity monitoring by detecting Beamstrahlung photons, as described in the CDR.

The PCL is capable of achieving the requirements stated in the CDR for both the 3\,TeV and 380\,GeV designs. Studies indicate that there are no unexpected consequences of scaling down the dipole magnet strengths from 0.8\,T to 0.1\,T for the 190\,GeV electron beam. No major changes to the PCL design are required. Minor changes to the carbon-based masks and the intermediate dump are being investigated to further reduce deposition on the magnets, but these changes should not affect the overall performance or cost of the PCL. Instrumentation studies for the PCL are in progress.

\begin{table}[htb]
    \caption{Power deposited (MW) in key PCL elements.}
    \label{tab:Uncollided_Power}
    \centering
    \begin{tabular}{lccc}
    \toprule
         & \textbf{Intermediate Dump} & \textbf{Final Drift} & \textbf{Main Dump} \\
    \midrule
        \textbf{3\,TeV Uncollided} & $2.10\times10^{-4}$ & $1.97\times10^{-2}$ & 13.6\\
        \textbf{3\,TeV Collided} & $3.67\times10^{-2}$ & $2.96\times10^{-2}$ & 10.2\\
        \textbf{380\,GeV Uncollided} & $5.19\times10^{-5}$ & $4.08\times10^{-3}$ & 2.91\\
        \textbf{380\,GeV Collided}  & $7.77\times10^{-5}$ & $4.23\times10^{-3}$ & 2.70\\
    \bottomrule
\end{tabular}
\end{table}

\section{Drive-Beam Accelerator}
\label{sect:DBA}
\subsection{Introduction}

\begin{figure}[htb!]
\centering
\includegraphics[scale=0.75]{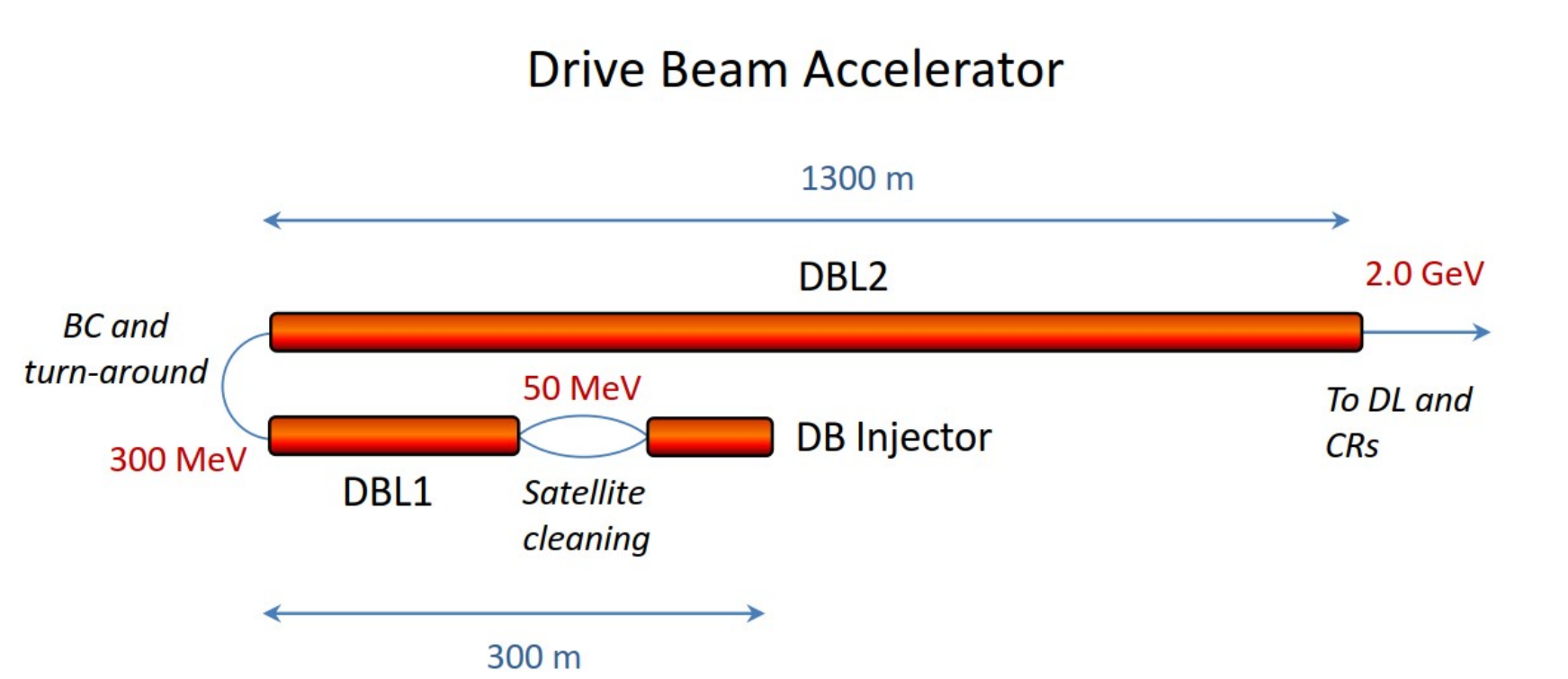}
\caption{\label{fig:DBA_1} Schematic layout of the DB Injector and Accelerator.}
\end{figure}

The Drive-Beam Accelerator (DBA) accelerates the Drive Beam (DB). It consists of an injector generating a 48\,$\mu$s-long Drive-Beam bunch train with an average current of 4.2\,A and a bunch repetition frequency of 499.75\,MHz. The two Drive-Beam Linacs (DBL1 and DBL2) accelerate the beam up to 1.91\,GeV, and are separated by a bunch compressor. The injector generates electron bunches with a charge of $q{}_{b}$~=~8.4\,nC, an R.M.S. bunch length of 3\,mm and a normalised emittance of 100\,$\mu$m at an energy of 50\,MeV. The bunch compressor between DBL1 and DBL2 compresses the bunches to a final length of 1\,mm, in a 180$^\circ$ turn-around which reverses the direction of propagation of the beam and minimises the overall footprint.  The length of the whole accelerator complex is 1.3\,km.  The schematic layout is shown in Fig.~\ref{fig:DBA_1}. 

Since it is the biggest power consumer in the entire CLIC complex the DBA has been designed to maximise efficiency. The accelerating structures are operated in a fully loaded mode reaching an RF-to-Beam efficiency of 95\,\%. A multi-beam klystron and a solid state modulator have been developed together with industry, efficiencies above 70\,\% and 90\,\% respectively have been obtained. In addition, the focus of the design was on beam stability. The charge along the bunch train and shot-to-shot has to be stable to 0.1\,\%. The beam energy and phase have to be stable to within 0.2\,\% and 0.05$^\circ$ respectively. These specifications drive the technology choice of the electron source, the modulators, and the RF power source.

The CLIC 380\,GeV stage compared with the CDR design~ \cite{Aicheler2012} has a 20\,\% reduced Drive-Beam energy and a bunch train length which is a factor 3 shorter. In addition, the power per klystron has been increased to 20\,MW which reduces the overall length further due to a higher gradient in the fully loaded accelerating structures.

The baseline emittance of the drive beam at the end of the DBA is $\varepsilon_x=\varepsilon_y=50\,\mu$m. This is smaller than in the CDR, since recent studies of the Drive-Beam Injector suggest that a small emittance growth can be expected~\cite{Kelisani2017}. This is also consistent with the experience at CTF3 and LEETCHEE (Chapter~\ref{Chapter:PERF}).

\subsection{Drive-Beam Injector}

\begin{figure}[htb!]
\centering
\includegraphics[scale=0.95]{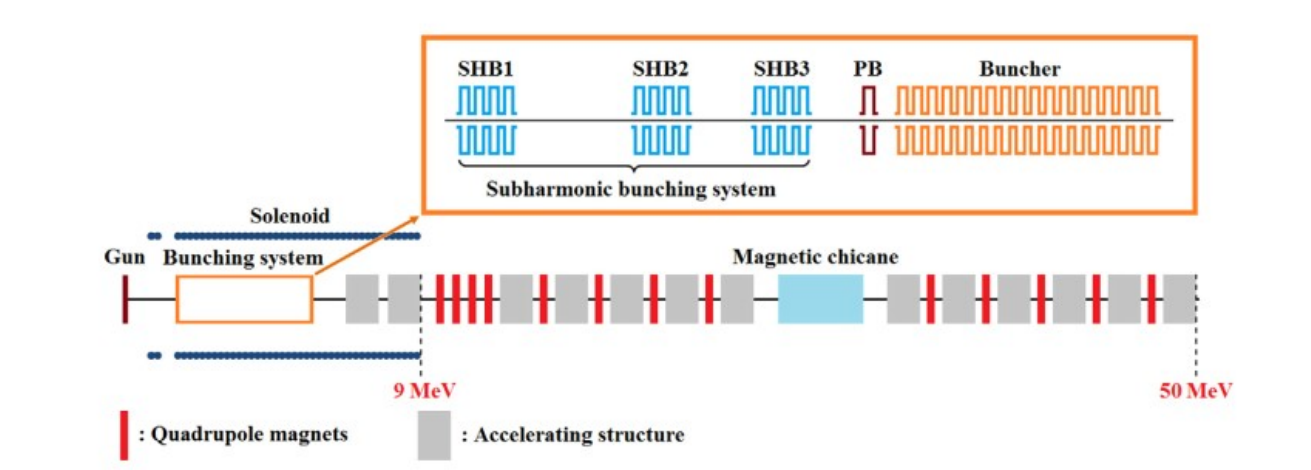}
\caption{\label{fig_DBA_2} Schematic layout of the Drive-Beam injector.}
\end{figure}

The Drive-Beam Injector (DBI) has been completely re-designed and optimised relative to the description in the CDR. In particular the capture efficiency has been increased, and the satellite population drastically reduced. The new design is documented in \cite{Hajari2015,Shaker2013,Shaker2015} and a schematic layout is shown in Fig.~\ref{fig_DBA_2}. The electrons are produced by a thermionic gun at 140\,keV. The gun is followed by a sub-harmonic bunching system consisting of three standing wave bunchers at a frequency of 500\,MHz (SHB), a 1\,GHz Pre-Buncher (PB) and a 1\,GHz Travelling Wave Buncher (TWB). Since the beam energy after the TWB is 2.3\,MeV two more accelerating structures are embedded in a continuous solenoidal magnetic field to preserve the emittance. At this stage the focusing is done by quadrupoles and 11 more accelerating structures bring the beam energy up to the desired 50\,MeV. A magnetic chicane is used to clean up the phase space at low energy. The injector design fulfills all requirements for the DB Linac. At 50\,MeV the energy spread is 0.95\,\%, the bunch length is 3\,mm and the emittance is below 30\,$\mu$m. The beam losses in the cleaning chicane are 3.5\,\% and the satellite population is reduced to 2.7\,\%. This low value significantly eases the load of the cleaning system for the satellite bunches described in the CDR.

In the last few years prototypes of the Drive-Beam electron source, sub-harmonic buncher and the RF systems have been built and tested. Results are described in Chapter~\ref{Chapter:PERF}.

\subsection{Drive-Beam Accelerator}

The DBA increases the beam energy from 50 to 300\,MeV. It uses a FODO lattice with one accelerating structure between quadrupoles and a phase advance of 106$^\circ$. 

\begin{figure}[!htb]
\centering
\includegraphics[scale=0.95]{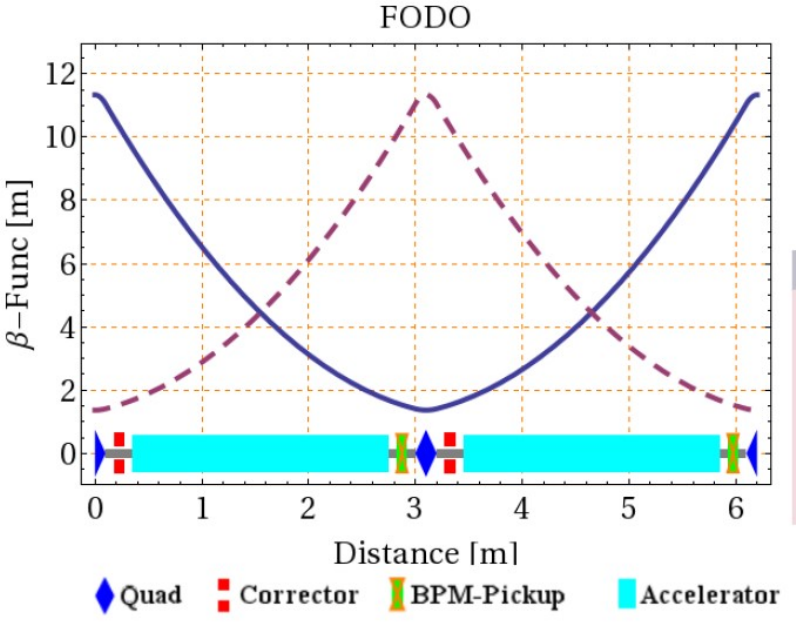}
\caption{\label{fig_DBA_3} FODO-lattice for the first DL with one accelerating structure between quadrupoles.}
\end{figure}

Each FODO-cell is equipped with a Beam Position Monitor (BPM) and corrector magnets for beam-based alignment as shown in Fig.~\ref{fig_DBA_3}. As mentioned, each accelerating structure is operating in fully loaded mode. The input power is 18\,MW per structure corresponding to an energy gain of 4.1\,MeV. The RF-to-beam efficiency is 95\,\%.

After the first linac a 180$^\circ$ turn-around bends the beam into the second DBA. This construction allows the full DBA complex to be located on the CERN Prevessin site. The bend compresses the bunches to a length of 1\,mm. Here the electrons are accelerated up to 2\,GeV using a FODO cell with 3 accelerating structures between quadrupoles and a phase advance of 101$^\circ$. A total of 200\,quadrupoles and 460\,structures are used in the DBA.

Triplet, Doublet and FODO lattices were studied to optimise the beam stability and emittance growth before the lattice described above was adopted as the baseline \cite{Aksoy2018}. For an RMS misalignment of 100\,$\mu$m of critical components, such as quadrupoles, accelerating structures and BPMs, the normalised emittance growth remains below 5\,$\mu$m after beam-based correction, and the total emittance well below the goal of 50\,$\mu$m. Beam jitter is not amplified significantly.

\section{Drive-Beam Recombination Complex}
\label{sect:DBRC}

\subsection{Overview}
In order to power efficiently the accelerating structures to their nominal gradient of 72\,MV/m, the bunches coming out of the DB Linac~\cite{Aksoy2011} at a frequency of 500\,MHz must be recombined by a factor 24, compressing a 5.86\,$\mu$s sub-train into a 244\,ns pulse with a frequency of 12\,GHz. This operation is the function of the Drive Beam Recombination Complex (DBRC), which consists of a Delay Loop (DL), Combiner~Ring~1~(CR1), and Combiner~Ring~2~(CR2). Each of them recombines 2, 3, and 4 sub-pulses respectively, thereby increasing the bunch frequency by the same amount.  Additionally, the DBRC has three transfer lines connecting the DL, CR1, CR2, and the transfer turnaround (see Fig.~\ref{fig:DBRC_1}). Further details regarding operation of the DBRC can be found in~\cite{Biscari2009,Aicheler2012,Costa2018}.

\subsection{System Description}

\begin{figure}[ht!]

\centering{\includegraphics[scale=.70,trim=10 10 0 10,clip]{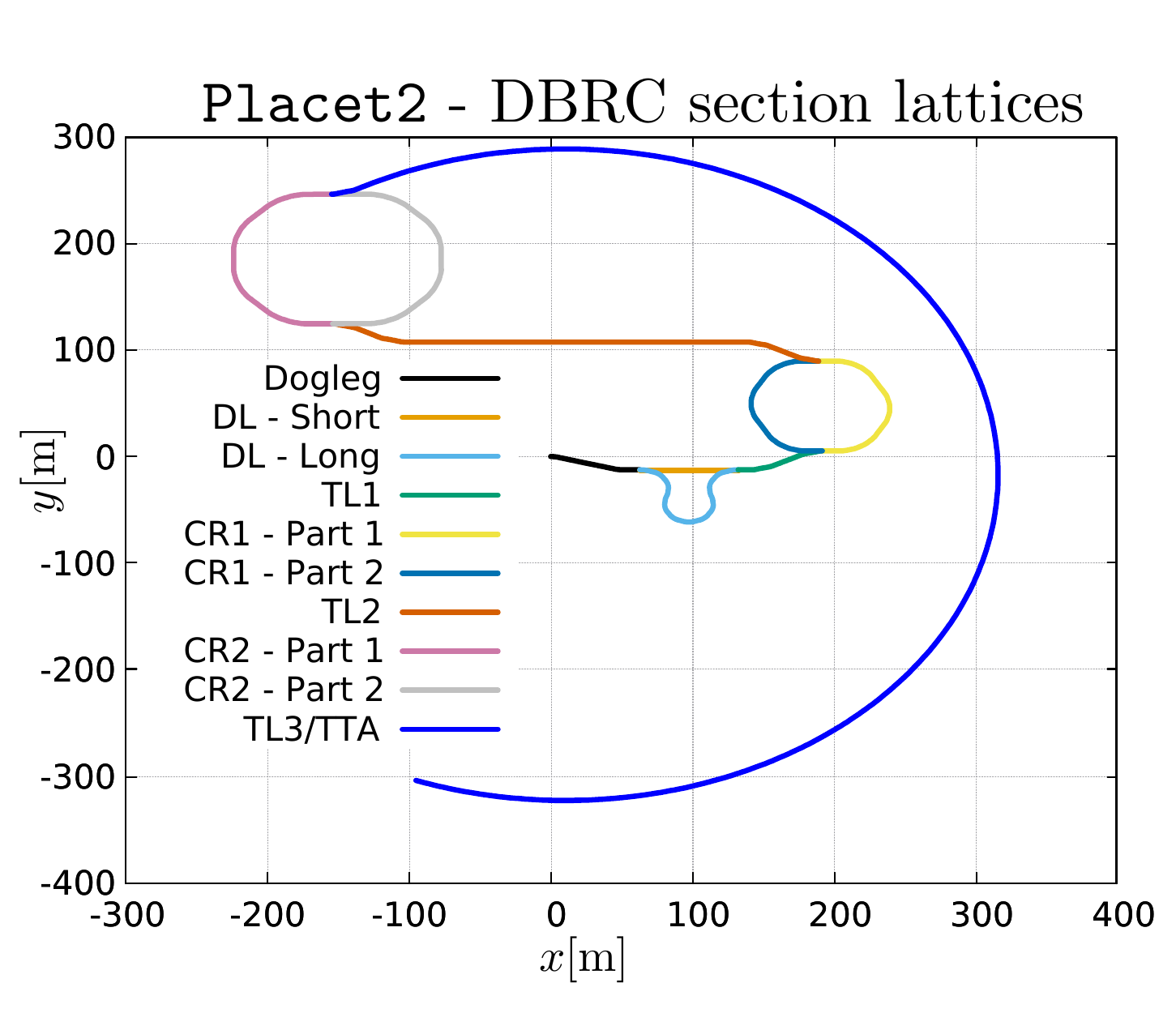}}
\caption{Schematic view of the full Drive-Deam Recombination Complex, the beamlines that composed the DBRC are denoted by different colours. Figure generated with \texttt{Placet2}~\cite{Pellegrini2015}.}
\label{fig:DBRC_1}
\end{figure}

\begin{figure}[h!]
\centering{\includegraphics[width=0.9\textwidth]{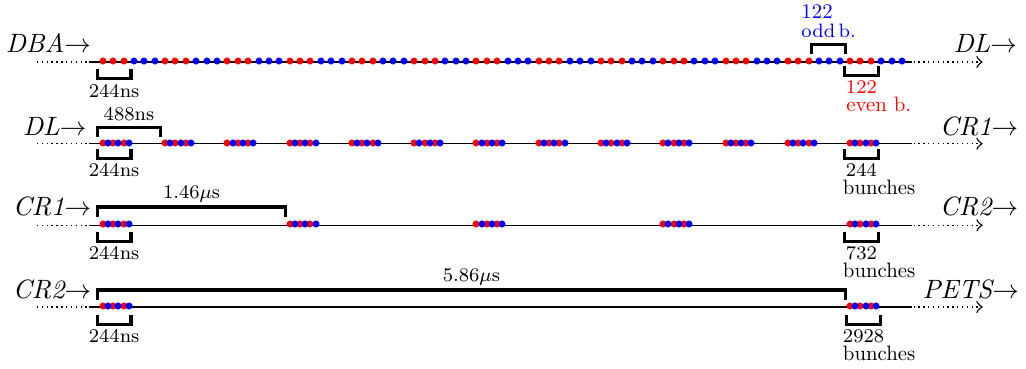}}
\caption{Train structure in the various stages of recombination \cite{Costa2018}.}
\label{fig:DBRC_recombination}
\end{figure}

The first recombination step separates two 244\,ns Drive-Beam sub-trains and interleaves them into a single pulse (see Fig.~\ref{fig:DBRC_recombination}). This is achieved by separating even and odd sub-trains using an RF deflector; the even sub-train is sent through the Delay Loop (DL), the odd sub-train through the straight transfer line. After recombining the two sub-trains, the DB bunch frequency is increased by a factor 2, forming a single 1\,GHz pulse.

After the DL, the 244\,ns separation between pulses allows the use of combiner rings to further multiply the bunch frequency by a factor 3. In these rings, the time taken by a pulse to complete a full turn is $n\times 244 \text{[ns]}+\Delta t_\text{Rec}$, where $\Delta t_\text{Rec}$ is the product of the time between bunches at injection and the ring's frequency multiplication factor. This allows us to interlock several pulses into a single one, multiplying both pulse frequency and current.
As shown in Fig.~\ref{fig:DBRC_1}, the DBRC has two combiner rings (CR1 and CR2). The frequency multiplication factor of CR1 and CR2 is $3$ and $4$, respectively. The bunch frequency at CR2 extraction is $12$ GHz, which corresponds to the frequency required in the Power Extraction and Transfer Structures (PETS). The injection and the extraction of the recombined pulses into and from CR1 and CR2 rely on transverse RF deflectors and striplines.

\subsection{Performance}

The CLIC Test Facility (CTF3) was essential in demonstrating the feasibility of the beam recombination scheme on which the CLIC DBRC is based (see Chapter~\ref{Chapter:PERF}). The new design of the DBRC is an improved version of the design presented in~\cite{Adli2009,Aicheler2012}. The changes in the lattice design allowed an increase in the energy acceptance of the entire complex, while respecting the transverse emittance budgets, taking account of the effects of Coherent Synchrotron Radiation (CSR) emission~\cite{Esberg2014}. The goal has been set to be close to 1\,\% R.M.S. relative energy spread.

\subsubsection{Design Updates}

Increasing the bunch length at the exit of the Drive-Beam Linac from 1\,mm to 2\,mm significantly reduces the impact of CSR when the beam circulates through the DBRC. One dogleg has been inserted before the DBRC to lengthen the bunch~\cite{Marin2016} from 1\,mm to 2\,mm, and one chicane has been added after the Transfer Turn Around (TTA)  to recompress the bunch length from 2\,mm to 1\,mm (see Section~\ref{sect:DBA}). This compression of the longitudinal phase space is needed for CSR compensation, as well as a longitudinal energy chirp (also known as correlated energy spread) for the phase feed forward to preserve the Drive-Beam-to-Main-Beam synchronization.

In order to increase the energy bandwidth of the system, and accommodate such a correlated energy spread, the design of the CR arc cells has been modified and the original triple-bend achromatic cells have been replaced by a double Chassman-Green cell~\cite{Marin2016}, which had already been adopted for the DL in the CDR design~\cite{Aicheler2012}.

\subsubsection{Longitudinal Phase Space Effects}

The development of the tracking code PLACET2~\cite{Pellegrini2015} enabled the study of the performance of the DBRC in a fully realistic manner~\cite{Costa2018}, including RF-deflectors (and their impact on the path-length), synchrotron radiation emission and associated energy loss and bunch train recombination.

\begin{figure}[ht!]
\includegraphics[scale=.28,trim=0 0 10 10,clip]{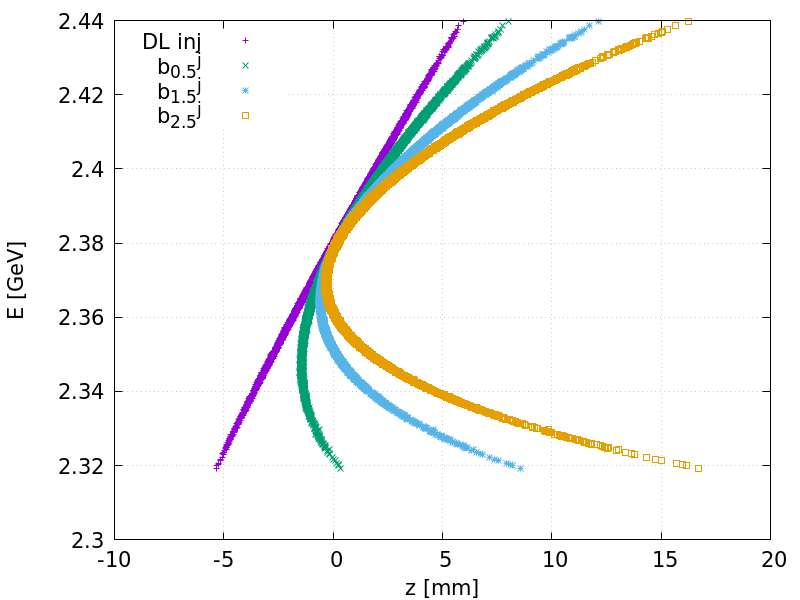}
\includegraphics[scale=.28,trim=0 0 10 10,clip]{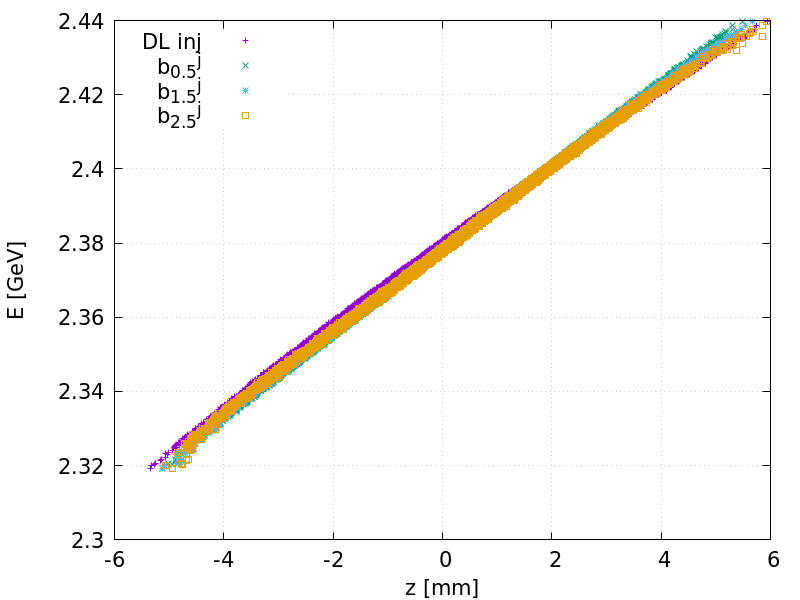}
\caption{Longitudinal energy profile before (left panel) and after (right panel) sextupole optimisation. The first layer shows the bunch at the DL injection point (target profile) and the following layers show the bunch after $0.5$ $\left(b_{0.5}^{\protect\hphantom{aa}j}\right)$, $1.5$ $\left(b_{1.5}^{\protect\hphantom{aa}j}\right)$ and $2.5$ $\left(b_{2.5}^{\protect\hphantom{aa}j}\right)$ turns in CR1.}
\label{fig:DBRC_2}
\end{figure}

One of the main issues of the recombination process is a strong parabolic correlation between particle momentum and longitudinal position (see Fig.~\ref{fig:DBRC_2}-left). This effect can be corrected by tuning the sextupoles located in dispersive regions (see Fig.~\ref{fig:DBRC_2}-right). However, one must consider potential detrimental effects introduced by the sextupoles on the transverse emittances. Optimisation studies to address the issue were performed in~\cite{Costa2018}.
The initial beam emittance is 50\,$\mu$m.
The goal is to reach an emittance after the DBRC of less than 150\,$\mu$m, as required by the decelerators. Simulations of the full DBRC show that this goal can be reached even with initial emittances of up to 80\,$\mu$m, while fully correcting the beam's longitudinal aberration \cite{Costa2018a}.

\subsubsection{Imperfections}

The results of beam-based alignment schemes, like Dispersion-Free Steering (DFS) in CTF3's CR, indicate that misalignment of the element impacts the performance~\cite{Gamba2016}. These studies confirmed that DFS is a promising technique and dedicated studies of its use in the DBRC should be performed. No major problems are expected.

\section{Decelerators}
\label{sect:DB_Decelerator}

\subsection{Overview}
The decelerator lines run in parallel with the two Main-Beam lines for the full length of the Main Linacs. Each linac has four decelerator sectors. In each sector a Drive-Beam train of 101\,A, 1.91\,GeV, and 244\,ns length is decelerated and the kinetic energy of the beam particles is converted to RF power by the Power Extraction and Transfer Structures (PETS) \cite{Syratchev2007}.

Four turn-around loops bring the Drive-Beam trains to the start of each decelerator sector; the spacing between them is 878\,m. The lengths of the decelerating sectors vary slightly from 900\,m to 840\,m, such that each decelerator feeds the same number of accelerating structures. This length variation is a result of the variation of the filling factor in the Main Linac due to the limited choices of quadrupole spacings. At the end of each sector an insertion of four module lengths (about 10\,m) extracts the used DB into a dump and brings in the new DB with a chicane.

\subsection{Beam Parameters}

Table~\ref{t:dec_parameters} summarizes the main parameters for the decelerator lattice and beam for nominal operation.  During a tune-up period, uncombined (0.5\,GHz bunch spacing) low-current beams may be used with a shorter train length (as short as required for precise enough BPM resolution).
  
\begin{table}[!hbt]
\caption{DB Parameters for Nominal Operation}
\label{t:dec_parameters}
\begin{centering}
\begin{tabular}{l c r}
\toprule 
\textbf{Drive-Beam parameters} [units] & \textbf{Symbol} & \textbf{Value }\tabularnewline
\midrule
Average beam current [A] & $I$  & 101\tabularnewline
Initial energy [GeV] & $E_{0}$  & 1.91\tabularnewline
Minimal final particle energy [GeV]& $E_{min}$ & 0.191\tabularnewline
Train length [ns]& $t_{train}$  & 244\tabularnewline
Bunch frequency [GHz]& $f_{bunch}$ & 11.994\tabularnewline
Bunch-to-bunch distance [mm]& $z_{bb}$ & 25.0\tabularnewline
Bunch length [mm] & $\sigma_{z}$  & 1\tabularnewline
Bunch form factor [\%] & $F(\lambda(\sigma_{z}))$  & 96.9\tabularnewline
Initial normalized emittance [$\mu$m] & $\epsilon_{Nx,Ny}$  & $150$ \tabularnewline
\bottomrule
\end{tabular}
\par\end{centering}
\end{table}

\subsection{Decelerator Layout and Optics}
Each decelerator sector consists of a sequence of 2.343\,m-long modules. Each module supports two quadrupoles and up to four PETS, each of which feeds one pair of Main Linac accelerating structures. If the accelerating structure in the Main Linac is replaced with a lattice quadrupole, the corresponding PETS in the decelerator is replaced by a drift. In total, each decelerator contains 1,273~PETS (see Table~\ref{tab:dece_sectors}. The inner radius of the PETS, and that of the vacuum chamber, is $a_0 = 11.5$\,mm.  The initial normalized emittance is assumed to be $\epsilon_{Nx,Ny} = 150$\,$\mu$m. Strong focusing with a FODO lattice design is used to keep the beam size limited and to mitigate the transverse wakefield effects from the PETS. This configuration also has the required large energy acceptance. The phase advance is 92$^\circ$ per cell, the maximum beta is $\hat{\beta}~=~4$\,m, and the average beta function is ${\langle}\beta{\rangle}~=~1.5$\,m. The initial maximal R.M.S. beam size is $\sigma_{x,y}~=~0.4$\,mm.

\begin{table}[!htb]
\caption{\label{tab:dece_sectors}The decelerator sectors.}
\begin{center}
\begin{tabular}{*5{c}}
\toprule
\textbf{sector number}& \textbf{1} & \textbf{2}& \textbf{3}& \textbf{4} \\
\midrule
number of modules & 381 & 355 & 370 & 364 \\
number of PETS & 1273 & 1273 & 1273 & 1273 \\
\bottomrule
\end{tabular}
\end{center}
\end{table}

The Drive-Beam particles are decelerated as a function of their bunch number and their longitudinal position within the bunch. The first bunches experience little deceleration, while later bunches reach up to 90\,\% deceleration.

The strengths of quadrupole magnets are scaled so that the most decelerated particles maintain a constant phase advance per cell, as proposed in \cite{Riche1994}.  The quadrupole gradient, therefore, has to decrease to 10\,\% of the initial value along each decelerator.  For a perfect machine and injection, the high-energy beam slices in the transient are contained within the envelope of the lower energy slices. 

\subsection{Accelerator Physics Issues}
The accelerator physics of the 380\,GeV energy stage is comparable to the one at 3\,TeV. The slightly larger beta-function and lower beam energy make the beam more sensitive to wakefield effects. The lower PETS impedance and the smaller number of PETS in each decelerator recover most of this.

\subsection{Component Specifications}
\subsubsection{Quadrupoles}
A total of about 6,000 quadrupoles are required for the 8~decelerator sectors. For nominal operation, the integrated gradient decreases, in proportion to the beam deceleration, from 10.5\,T at the start of each decelerator to 1.05\,T at the end. To accommodate the requirements of physics energy scans (see Section~\ref{sect:Energy_Scan}) and operational flexibility during system tuning, the strongest quadrupole magnets need to operate at 7.3--12.5\,T, and the weakest quadrupoles at 0.73--4.2\,T \cite{Adli:1239173}.   

Two magnet design schemes have been investigated; one based on electromagnets and the other on mechanically tunable permanent magnets, as discussed in Section~\ref{sect:Magnets}. For electromagnets, a special powering strategy has been developed to reduce power dissipation in the cables and to reduce the power cost \cite{Adli2010}. The requirement on the relative field accuracy of each individual magnet is $1\times10^{-3}$ (R.M.S.), while the requirement on the pulse-to-pulse field stability is $5 \times 10^{-4}$ (R.M.S.).  The latter is tighter than the former, since the effect of the resulting orbit jitter cannot be taken out by static corrections.

\subsubsection{Correctors}
The movers of the decelerator girders are used to steer the beam in the baseline scenario. This requires precise quadrupole alignment to the girder of better than $20\,\mu$m R.M.S. Alternatively, the quadrupoles could be equipped with movers or corrector dipoles, which would yield better performance and remove the requirement for precise quadrupole alignment but at a higher cost.
\subsubsection{Beam Position Monitors}
The BPM accuracy and precision is key for the orbit correction~\cite{Adli:1239173}. One BPM is installed at each quadrupole with an accuracy of 20\,$\mu$m and a resolution of 2\,$\mu$m. However, a significant number could fail before the correction is compromised in a significant fashion.

\subsubsection{Component Tolerances}
Table~\ref{tab:adli_dec_tolerances} summarizes the tolerances of the beam line components in the decelerator. The quadrupole offset tolerance corresponds to what needs to be achieved by  machine pre-alignment, which would allow successful beam-based tuning and robust performance. The requirement on the BPM accuracy is to ensure sufficient performance of 1-to-1 steering in order to be able to proceed with Dispersion-Free Steering. The requirement on the BPM resolution is for Dispersion-Free Correction, in order to produce a beam envelope close to the minimum achievable for robust decelerator operation.

\begin{table}[!hbt]
\begin{centering}
\caption{Decelerator component tolerances}
\label{tab:adli_dec_tolerances}
\begin{tabular}{ l c r}
\toprule
\textbf{Component tolerances} [units R.M.S.]& \textbf{Symbol} & \textbf{Value}\tabularnewline
\midrule
Quadrupole offsets [$\mu$m ] 					& $\sigma_{quad}$				& 20	\tabularnewline
PETS offset [$\mu$m]							& $\sigma_{PETS}$				& 100	\tabularnewline
Pitch/roll [mrad] 								& $\sigma_{\theta,\phi}$	& 1	\tabularnewline
BPM accuracy (mechanical + electrical) [$\mu$m] & $\sigma_{acc}$				& 20	\tabularnewline
BPM resolution [$\mu$m] 						& $\sigma_{res}$				& 2	\tabularnewline
\bottomrule
\end{tabular}
\par\end{centering}
\end{table}

\subsubsection{Vacuum System Considerations}
To prevent the onset of resistive-wall instabilities the vacuum chambers of the decelerator have to be built with copper on the inner walls so as to have a conductivity of the order of 5.9\,x\,10$^7$~$\Omega^{-1}$\,m$^{-1}$. To prevent the onset of fast beam-ion instabilities the vacuum pressure in the decelerator is required to be 40~$\times$~10$^{-9}$\,mbar or less.

\section{Drive-Beam Dump Lines}
\label{sect:DBDL}

\begin{figure}[!htb]
\begin{center}
\includegraphics[width=0.45\textwidth]{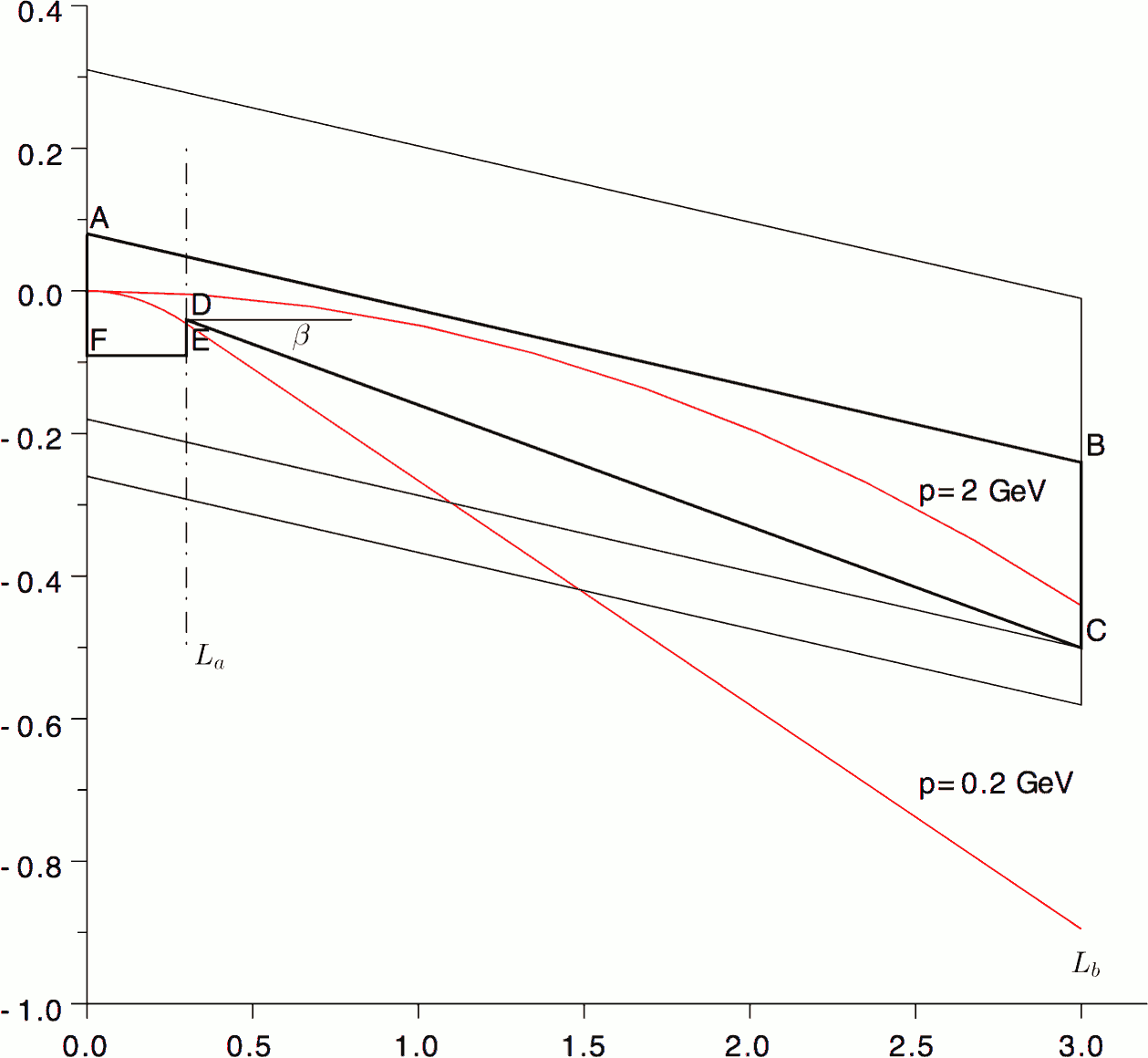}
\end{center}
\caption{\label{fig:DBDL.yoke}A schematic top view of the magnet which extracts the beam from the decelerator line and spreads the beam for adequate beam density on the dump. The abscissa represents the longitudinal $s$ coordinate which is parallel to the decelerator axis; the ordinate is the horizontal coordinate $x$. The beam enters the magnet along $s$ at $x=s=0$. The area of nominal constant magnetic field $B$ is  delimited  by  the  thick  black  line. The  thin  black  external line delimits the  yoke (on  the positive $x$-side) and the coil (negative $x$-side). The beam trajectories of the two extreme momenta are drawn in red.}
\end{figure}

After leaving $\approx$\,90\,\% of its power in the PETS,  the spent Drive Beam at the end of each decelerator has to be bent away from the decelerator axis and disposed of at a beam dump. This needs to be done in a way that leaves a sufficient amount of space for injecting the fresh Drive Beam into the next sector, which starts 8\,m downstream. The spent beam is characterised by an energy spread of a factor ten, since the bunch is composed of a 2\,GeV electron head followed by a long tail of $\approx$\,200\,MeV electrons. Such a large energy spread requires a dedicated magnet design to bend the spent beam to the 2\,m wide dump window located about 15\,m away. The shape of the magnet and a description of the yoke's footprint are shown in Fig.~\ref{fig:DBDL.yoke} and Table~\ref{tab:DBDL.yoke_params}.

The peculiar shape of the magnet does not pose a problem from a magnet design point of view. Exotic pole shapes can be manufactured, in particular if the magnet is operated in DC and the yoke is of solid iron. A permanent magnet solution could also be envisaged. The drive beam dump system is described in the CDR, the extraction magnet is found in \cite{DBDL.Jeanneret}. 

\begin{table}[!htb]
\begin{center}
\caption{\label{tab:DBDL.yoke_params}The geometry of the field area of the extraction magnet.}
\begin{tabular}{ccc}
\toprule
Vertex & $s$ {[}m{]} & $x$ {[}m{]}\tabularnewline
\midrule
A & 0 & 0.080\tabularnewline
B & 3 & -0.241\tabularnewline
C & 3 & -0.501\tabularnewline
D & 0.3 & -0.041\tabularnewline
E & 0.3 & -0.091\tabularnewline
F & 0 & -0.091\tabularnewline
\midrule
$\beta$ & 0.169 & rad\tabularnewline
\bottomrule
 &  & \tabularnewline
\end{tabular}
\end{center}
\end{table}

\section{Integrated Performance}
\label{sect:INT_PERF}
The ability of CLIC to reach its energy and luminosity target has been thoroughly studied in the CDR for the 3\,TeV stage~\cite{Aicheler2012}. Reaching the energy goal requires the target gradient in the accelerating structures to be achieved. This in turn requires that the structures can sustain the gradient and that the Drive Beam provides enough power. Both are feasible and  discussed in Section~\ref{sect:PERF_RF} and Section~\ref{sect:PERF_DB}, respectively. Reaching the luminosity goal requires control of the beam quality. Its feasibility is established through a combination of hardware development and prototyping as well as experiments at beam test facilities, see Chapter~\ref{Chapter:TECH} and Chapter~\ref{Chapter:PERF}. However, theoretical studies are required to translate this into luminosity performance predictions and are discussed below.

The total integrated luminosity goal for CLIC is $180\,\text{fb}^{-1}$
per year at 380\,GeV. The tentative operation plan \cite{Bordry2018}
 includes a yearly shutdown of 120 days, 30 days of commissioning, 20 days for machine development, and 10 days for planned technical stops.
This leaves 185 days of operation per year for the experiments. Considering the availability goal of 75\,\%, this results in an integrated luminosity per year that is equivalent to operating at full luminosity of $\mathcal{L}=1.5\times10^{34}\;\text{cm}^{-2}\text{s}^{-1}$ for $1.2\times10^7\,\text{s}$. 

\subsection{Luminosity and Parameters}
The CLIC target luminosity at 380\,GeV is $\mathcal{L}=1.5\times10^{34}\;\text{cm}^{-2}\text{s}^{-1}$. The nominal beam parameters at the IP
are given in Table~\ref{t:scdup1}. The key considerations are:
\begin{itemize}
\item The choice of bunch charge and length ensures stable transport of the beam. The main limitation arises from short-range wakefields in the Main Linac.
\item The spacing between subsequent bunches ensures that the long-range wakefields in the Main Linac can be sufficiently damped to avoid beam break-up instabilities.
\item The horizontal beam size at the collision point ensures that the beamstrahlung caused by the high beam brightness is kept to an
acceptable level for the given bunch charge. This ensures a luminosity spectrum consistent with the requirements of the physics experiments.
\item The horizontal emittance is dominated by single particle and collective effects in the Damping Rings and includes some additional contributions from the RTML.
\item The vertical emittance is given mainly by the Damping Ring and additional contributions from imperfections of the machine implementation.
The target parameters take into account budgets for detrimental effects from static and dynamic imperfections such as component misalignments and jitter.
\item The vertical beta-function is the optimum choice for luminosity. The horizontal beta-function is determined by the combination of required beam size and horizontal emittance.
\end{itemize}
In summary, the parameters are largely determined by fundamental beam physics and machine design with the exception of the vertical emittance which is determined by imperfections. Without them a
luminosity of $\mathcal{L}=4.3\times10^{34}\;\text{cm}^{-2}\text{s}^{-1}$ would be achieved.

\subsection{Luminosity and Imperfections}
Emittance growth budgets have been defined to account for static and dynamic imperfections, see Table~\ref{tab:INT_PERF_1}. For static imperfections it is required that the systems achieve a smaller emittance with a likelihood of more than 90\%. For the dynamic budgets the average value has to remain below the target. The vertical budgets are similar to the 3\,TeV design, as specified in the CDR~\cite{Aicheler2012}, but are, in most cases, easier to meet because of the shorter Main Linac and the larger beam size at the interaction point. The horizontal targets are larger at 380\,GeV than at 3\,TeV because the bunch charge is higher.

\begin{table}[htb!]
\centering
\caption{
all the different sub-systems.
The evolution of the transverse emittances (horizontal/vertical), in nm, along the CLIC complex if the full emittance budgets are used.
After the BDS, the effective emittance is given.}
\label{tab:INT_PERF_1}
\begin{tabular}{*5{c}}
\toprule
System & Initial & Design & Static Imperfections &Dynamic Imperfections\\
\midrule
RTML & 700/5 & 800/6 & 820/8 & 850/10 \\
Main Linac & 850/10 & 850/10 & 875/15 & 900/20 \\
BDS & 900/20 & 900/20 & 925/25 & 950/30\\
\bottomrule
\end{tabular}
\end{table}

Because of its smaller budgets, the focus in the following is on the vertical emittance. It is also important to note that if the
actual horizontal emittance is smaller than the target, 
the horizontal beta-function will be increased to achieve the nominal beam size and hence nominal level of beamstrahlung.
In contrast, if the vertical emittance is smaller than the target, one can profit from the higher luminosity as long as the detectors can tolerate it. Otherwise one can level the luminosity by increasing vertical or horizontal beta-function (the latter also improving the luminosity spectrum).

Technical systems have been developed that reduce the imperfections to a level that meets the performance goals even for the 3\,TeV stage with margin~\cite{Aicheler2012}. These include the pre-alignment system, the wakefield monitors in the accelerating structures, the quadrupole vibration suppression system and the phase control system. In addition, beam-based mitigation methods such as dispersion-free steering and feedback are used. Further improvements can be obtained with emittance or luminosity tuning bumps. More detail on the hardware development and prototyping as well as the beam studies in facilities is given in Chapter~\ref{Chapter:PERF}.

For an efficient acceleration, the Main Beam should experience a correct RF phase and amplitude~\cite{Aicheler2012}. A 1\,\% luminosity loss budget sets a maximum allowed Drive Beam R.M.S. phase jitter of $0.20^{\circ}$ at 12\,GHz~\cite{bib:scd_rfjitter_a}. The use of a phase-feedforward prototype at the CTF3 drive beam test facility has successfully stabilized the pulse-to-pulse phase jitter to $0.20^{\circ} \pm 0.01^{\circ}$ at 12 GHz~\cite{Roberts2018}, meeting the CLIC requirement.

For the static imperfections, the hardware performance specifications are the same for the 380\,GeV and 3\,TeV stages and corresponds to the CDR~\cite{Aicheler2012}. Also the technical systems for mitigating them are the same. Most tolerances could be relaxed for the lower energy stage, typically by about a factor of two. However, the original, better performances are required for the energy upgrade. No substantial cost saving has been identified by relaxing the specifications for the first stage. Therefore, it has been decided to make the systems consistent with the final energy from the very beginning, thus avoiding upgrades. This also provides additional margin to achieve the luminosity.
In addition, the tuning procedures for the RTML and the BDS have been improved relative to the CDR.

At 380\,GeV, detailed studies of the RTML show that the emittance growth goal can be achieved and that the average vertical emittance growth is only 1\,nm~\cite{Han2017}. In the Main Linac an average growth of 1\,nm is found, with 90\,\% of the machines remaining below 1.5\,nm, also well within the budget~\cite{c:MLperf}.
Similarly, the BDS tuning achieves its performance goal, which is to
reach 110\,\% of the nominal luminosity with 90\% likelihood, when the beam starts with the emittances given in Table~\ref{tab:INT_PERF_1}.
This corresponds to the situation that all systems use up the full budgets and provides the budget for dynamic imperfections in the BDS.
The procedures achieve the 110\,\% luminosity level with 95\,\% likelihood and with 90\,\% likelihood 117\,\% of the nominal luminosity~\cite{c:BDSperf}. They also require a few hundred luminosity measurements instead of a few thousand and hence are significantly faster.

Integrated studies of RTML, Main Linac and BDS using only static imperfections find an average luminosity of $\mathcal{L}=3\times10^{34}\;\text{cm}^{-2}\text{s}^{-1}$, about twice the target value. Hence, one can expect that the luminosity is significantly higher than the target.

For the dynamic imperfections and their mitigation systems, the hardware specifications are also the same at 380\,GeV and 3\,TeV, following the same rationale as for the static imperfections. The impact of dynamic imperfections on the luminosity has been studied in detail for 3\,TeV. It has been found to be well within the target. The impact on luminosity is more important at 3\,TeV than at 380\,GeV due to the longer Main Linac and smaller beam sizes in the BDS.
Since the budgets for dynamic emittance growth are the same at the lower energy, one can thus conclude that dynamic imperfections will generally not pose a danger to the luminosity target. 

An exception to this is the effect of time varying magnetic fields. These deflect the colliding beams, leading to trajectory jitter and emittance growth, thus reducing the luminosity. Their impact is particularly large in the RTML and the BDS. In the latter they are more important at 380\,GeV than in the 3\,TeV design due to the lower beam energy; the field variation at the position of the beam needs to remain below O(0.1\,nT)~\cite{Gohil2019b}.

A study in collaboration with experts from the Hungarian Geophysics Institute has commenced to investigate the typical power spectrum of magnetic fields from different sources: natural sources, such as geomagnetic storms; environmental sources, such as railway trains and power lines and technical sources, i.e. the collider itself.
The study concluded that natural sources are not expected to significantly impact the integrated luminosity~\cite{Heilig2018}.
As part of the collaboration a measurement station has been established in the Jura mountains (close to CERN) to monitor the power spectrum in the region over long timescales.

The study of the environmental and technical sources has started but is not yet complete. Preliminary estimates have been performed using the magnetic field variations that
were measured in the LHC tunnel and show a 100\,nT power spectrum amplitude above 1\,Hz, i.e. three orders of magnitude above the tolerance. However, a number of mitigation methods bring the impact of these fields down to an acceptable level. They include:
\begin{itemize}
\item Field variations at the grid frequency of 50\,Hz are the largest environmental source. They are mitigated by operating at the same beam-pulse frequency, locked to the grid.
The dynamic fields thus appear static to the beam.
\item Beam-based trajectory feedback mitigates slow field variations.
\item Passive shielding reduces the fields at the location of the beam. In part this is obtained by the collider components such as accelerating structures, magnets and to a lesser extent the
beam pipe. In addition shielding using mu-metal will be added in sensitive areas, such as the long drifts in the RTML and the BDS.
\item The components of the collider will be designed to produce only acceptable dynamic magnetic fields.
\item In addition, measurement of the field variations will allow active correction.
\end{itemize}
Studies of the first three methods show that they reduce the impact of the fields on the trajectory by four orders of magnitude, i.e. one order of magnitude below the tolerance~\cite{Gohil2019b}.
Once can thus be confident that the magnetic field effects can be mitigated with careful design of the collider.
More detail is given in Section~\ref{sect:PERF_Stray}.

The mitigation of ground motion has also changed compared to the old 3\,TeV design. The magnets have been moved further away from the interaction point such that they are now outside of the detector. This allows their supports to be mounted directly to the tunnel floor and avoids the previous cantilever system. Due to the resulting increase in stability, it is possible to remove the pre-isolator that supported the cantilever.

Simulation studies have been performed to estimate the impact of the new configuration on luminosity. They use a conservative model of the ground motion, based on measurements in the rather noisy CMS detector hall \cite{Kuzmin2009}; measurements in the LEP tunnel \cite{Juravlev1993} showed significantly lower values. The studies assumed that the static imperfections use the full emittance budget attributed to them, which leads to a luminosity of 
$\mathcal{L}=2.0\times10^{34}\,$cm$^{-2}$s$^{-1}$, i.e. 33\,\% more than nominal. The ground motion reduced this value by 3\,\%, i.e.  well within the budget~\cite{Gohil2019a}.

Further development of the foreseen technical and beam-based imperfection mitigation systems should allow a reduction in the emittance budgets and an increase in the luminosity target. Also new systems could be devised to this end. As an example, higher frequency accelerating structures could allow reduction of the energy spread of the colliding beams, which can improve the luminosity and also the luminosity spectrum for specific measurements such as the top threshold scan.

\subsection{Operational Considerations}

The operational considerations at 380\,GeV are similar to those at 3\,TeV, described in the CDR \cite{Aicheler2012}. The main operational action is to safely bring the machine from a state with no beam to a state where beams of nominal intensity are available for further luminosity optimisation. Besides the obvious initial start-up after a long shutdown, several possible events can potentially affect the nominal beam operation and are discussed below (ordered by increasing impact on operation and decreasing frequency).

\begin{itemize}
\item {\bf Hardware failures that do not require an interlock.}

Many failures are transparent for luminosity production. For example a repeated RF breakdown in a Main Linac accelerating structure can be avoided by switching the corresponding PETS off. The resulting energy loss and transverse deflection of the beam is much smaller than the capture range of the trajectory and the energy feedback.

\item {\bf Short beam interruptions due to spurious interlocks.}

Beam interruptions due to spurious hardware interlocks will inevitably happen and their number increases with system complexity. There might be many cases during daily operation where the interlock system, after detecting beam losses, may trigger a so called beam quality interlock. Most notably, coinciding RF breakdowns in several accelerating structures may cause beam losses that eventually could trigger such an interlock. In the case of a beam quality interlock, the system is envisaged to automatically go through a partial or full beam validating intensity ramp-up as described in \cite{Aicheler2012}. In case the error was incidental only and the basic machine performance was intrinsically not affected, such a beam validating energy ramp-up will be relatively quick; i.e. of the order of one second. A very conservative upper limit for the resulting unavailability can be obtained assuming that two simultaneous breakdowns in one linac are sufficient to trigger the interlock. This would occur every 400\,s and thus contribute an unavailability of about 0.25\,\%.

\item {\bf  Beam interruptions due to short hardware failure recovered by automated procedure.}

This case is almost identical to the previous one, except that minor changes of the machine configuration may need to take place for recovery. For example, after simultaneous RF breakdowns of a number of structures, it can automatically be decided to switch off the PETS in question and to compensate using neighbouring equipment. This procedure can be applied within a few pulses. Similar situations can occur for BPM, corrector, and even for certain quadrupole failures. Detailed studies, experience and, in this case, a short time might be required to reach full luminosity again, as certain machine parameters will have to be optimized. Detailed studies, experience and `machine learning' will allow automatic detection of these failures and to achieve short recovery times.

\item {\bf Beam interruption due to a failure requiring a hardware intervention.}

In this case the interruption of the operation will be more significant and the recovery will take longer. Moreover, failure could also cause blindness to the beam behaviour in affected parts. Whilst under normal conditions the machine state is continuously monitored and corrected for e.g. ground motions and thermal drifts, if such ``beam-blindness'' lasts for a period longer than the characteristic time of beam deterioration, then there will be a substantially longer period to assure safe beam in the machine and to optimise the luminosity performance. This is similar to the beam start after a planned technical stop.

Temperature variations due to abrupt changes in thermal loads are important. The temperature regulation system and the beam control have to be able to cope with them. Different strategies can be employed. Fast regulation of the temperature allows delay of beam operation until the temperature has stabilised. Slower temperature variations can be handled by the beam feedback systems.

In the BDS and the Main Linac, ground motion can induce misalignments that can degrade the performance when the beam is switched on again. However, the components can be tracked by the alignment system and fast recovery procedures with beam can be used to mitigate this.

The above arguments only apply to those parts of the machine that are affected by the failure. In other parts of the machine not affected by the failure  the beam can be maintained, hence minimizing the need for beam recommissioning in those parts.

\item {\bf Beam interruption due to a failure requiring a hardware intervention under controlled access in some areas of the machine.}

This situation is like the previous one. However, to minimize operational cost, it may be decided to switch off other operational parts of the machine, or to run them with a reduced intensity/and or duty-cycle. Consequently, the start-up period will become longer. A precise recipe for optimization cannot be given as this will depend on many parameters, such as cost to keep the machine ready, and speed of which performance can be recovered. Specific measures can reduce the time of the start-up. For example, heating the klystron cathodes during the stop will allow beam start up in a few RF pulses.

\item {\bf Scheduled beam interruption imposed by power consumption constraints.}

A study is being performed to investigate if the machine could be switched off during times of peak power demand in order to save cost.
This case is almost identical to the previous one. Also here the optimal strategy will depend on a trade of the cost to keep the machine ready, and speed (and cost) by which performance can be recovered.
\end{itemize}

The ultimate improvement in performance recovery time will come through experience. When the control system has followed and analysed enough intensity ramp-ups and luminosity optimisations, performance recovery can be improved based on experience. Even in the presence of so called moving targets that may take place over long time periods (e.g. during temperature stabilisation), a fast optimisation response time and anticipation of corrections based on experience, the control system should be able to achieve a respectable performance.

\subsection{Availability}
\label{sect:AVAIL}

A dedicated availability study is in progress. The principal goals of this study are to:
\begin{itemize}
\item  Demonstrate that CLIC availability requirement of 75\,\% can be reached.
\item  Identify the key accelerator systems and components that drive availability.
\item  Optimise the design with the best balance between availability and cost.
\item Find the optimal technical stops and operational schedule that maximizes availability. 
\item  Provide guidelines for availability-driven improvements of system and component designs.
\item Contribute to the research and development road map.
\end{itemize}
To achieve this, a two-pronged strategy is used consisting of a top-down and a bottom-up approach. 

\subsubsection{Availability Requirements for CLIC Subsystems}
\begin{figure}[ht!]
\centering
\includegraphics[width=\textwidth]{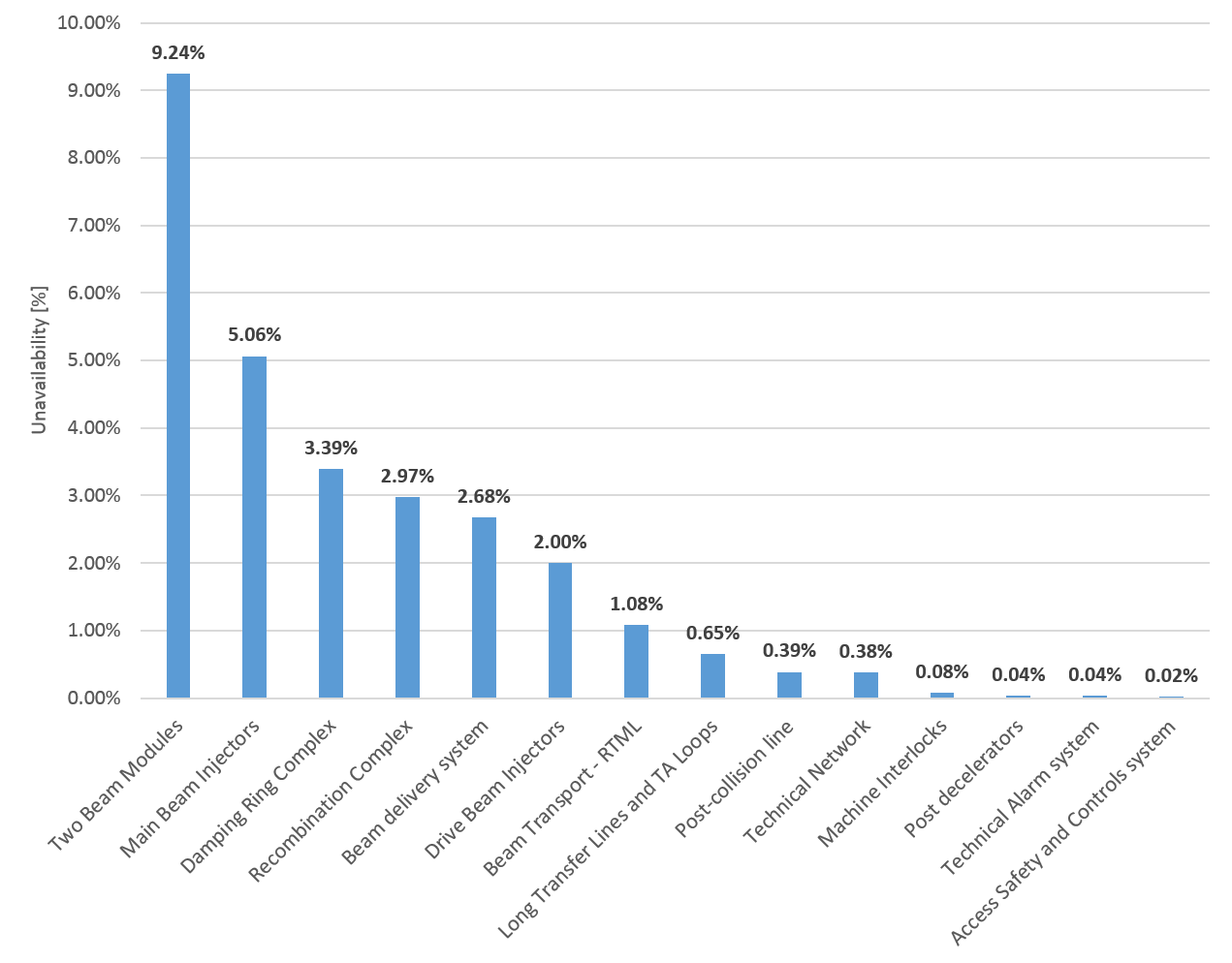}
\caption{Availability allocation to CLIC subsystems based on complexity criteria. The target availability for the overall system is 75\,\%.}
\label{fig:AVAIL_1}
\end{figure}

The top-down approach translates the overall CLIC availability goal into availability requirements for each system based on an evaluation of the complexity \cite{Orozco2018a}.
This is particularly useful for cases in which the detailed design is not known, or where new technologies are developed and no failure data are available for a detailed availability assessment.
During the complexity definition process, many factors are considered to obtain so-called complexity weights. These complexity weights are then used to allocate availabilities to the individual systems in relation to their weights. Based on the analysis of factors affecting CLIC availability, six criteria have been defined for complexity and availability allocation. To this end system experts provided the estimates of chosen factors. Considering only the major systems and a target availability of 75\% \cite{Bordry2018}, Fig.~\ref{fig:AVAIL_1} shows the required availability of the systems. These 'unavailability budgets' for each system include the downtime induced in other systems and the time needed to recover after a failure. For example, the downtime associated to beam losses that are triggered by a system failure is assigned to the failed system, i.e. to the root cause of the problem.  The chain of two beam modules in the Main Linac is considered the most complex subsystem and therefore has the highest allocated unavailability budget of 9\,\%.

The next step will consist of an availability distribution onto sub-systems, such as magnets, power converters and vacuum system. This optimisation should also consider cost as a boundary condition.

\subsubsection{Availability Prediction}
The bottom-up approach uses component reliability data to estimate the overall machine availability using Monte Carlo simulations. This will allow verification of the overall availability and  identification of the need for improvements of components or machine configuration in case the predictions will exceed the availability budgets.
A model of the CLIC systems and their components is being developed, including the impact of component failures on the machine
and models of the possible mitigation measures and recovery strategies.
In the current study, the model is implemented in two different software packages to validate results \cite{Orozco2017, Orozco2018}.
A fundamental ingredient of the study is the reliability data for the individual components,
such as Mean Time To Failures (MTTF) and Mean Time To Repair (MTTR).

A sensitivity analysis allows ranking of the components according to their impact on machine availability. In particular this analysis determines the expected impact of changes in component reliability on the overall availability.
Using Differential Importance Measures (DIMS) of first, second and total order,
also allows investigation of the effects of interactions among components \cite{Zio2004} and to determination of the system design
that leads to the best performance at the lowest cost. This will guide the component specifications and the optimum
scheduling of technical short stops.

As a first step, this analysis focuses on the most critical CLIC systems;
the final goal is to cover all subsystems and merge them into a unique availability model.
The key components that are expected to have the highest impact on the availability are: 1) the RF systems, i.e. klystrons, modulators, low-level RF systems, waveguides, the structures and their cooling system, and 2) the magnet system, in particular the power converters that supply the magnets. The associated failures can be grouped into two classes:
1. Those failures that require an immediate beam stop to perform a system repair
2. Those failures that rely on so-called hot-spare modules to compensate for the failure occurrence (fault tolerance). Due to a finite number of hot spare modules, beam operation might nevertheless be stopped after a given number of failures of this class occurs.
Power converter failures cause the magnets to fail and fall into the first class. They typically require a beam stop and are therefore mitigated by use of redundant power converters and by reducing their number. If one considers only this class, the combined availability of several systems is the product of the individual system availabilities.
The RF system failures fall into the second class. During planned or unplanned technical stops failed equipment belonging to the second class will be replaced; the reserves of hot spares are therefore dimensioned to be sufficient to bridge the time between subsequent stops. For a typical stop duration of 24\,h and the planned availability of 75\,\%, the average operation time between stops would be about 3 days. However, the tentative goal is to have reserves for 500\,h, corresponding to a stop on average every three weeks. The resulting unavailability would be about 5\,\%.
If one considers only this class, the combined availability of several systems is larger than the product of the individual system availabilities and can be quite close to the individual worst system.
The developed availability models allow to verify that the overall availability goals can be met, to refine the specifications for the components and to optimise the strategy (fault tolerance and frequency of technical stops) to provide high availability. Table~\ref{tab:mainfail} shows key values of MTTF and MTTR used in the simulation.

Studies are on going for the Two Beam Modules,
the most complex system according to the top-down method. A first estimate is obtained by considering the most important components.

Failures of the power converters that supply the magnets require a technical stop for repair have been studied for the 3\,TeV case, which is more difficult than the 380\,GeV stage due to the larger number of components.
In this case, the modules of the Drive Beam decelerator complex contain 41,000 quadrupoles, which are powered in groups with redundant power converters; small trims are used to finely adjust their strengths \cite{Siemaszko2012}. This reduces the number of power converters and the number of associated failures strongly.
Failures of the trims can be tolerated to a large extent. An availability of 99.8\,\% is reached with this scheme.

At 3\,TeV, the two Main Linacs contain a total of 4,000 quadrupoles. Most magnets are grouped in a similar way as in the drive beam and only seven magnets per sector are powered individually. First estimates show that an availability of 99\,\% can be reached.

The main linac modules also contain 4000 BPMs and 4000 correctors,
which are used to mitigate the impact of ground motion on the beam.
A first tentative study found that if 10\,\% of each fail,
the quality of the ground motion mitigation is compromised by 14\,\%, which is acceptable. A very modest MTTF of only 5,000\,h would already be sufficient to ensure a time of 500\,h until the exhaustion of the hot spares. Therefore, these systems can be considered highly available but further studies remain to be done.

The basic RF units of the modules consist of one PETS that feeds two accelerating structures, the associated waveguides and loads. If any of these components fails, the PETS has to be switched off. The Main Linac includes a 5\,\% reserve of hot-spare RF units. Hence to achieve the required 500\,h mean time to exhaustion of the reserve, a MTTF of 10,000\,h is required for the basic unit.

\begin{table}[!tbp]
    \centering
    \caption{Failure parameters of critical systems.}
    \begin{tabular}{lrr}
    \toprule
    System          					& MTTF [h] 	& MTTR [h]  \\ 
    \midrule
    Klystron, Drive-Beam Accelerator 	& 50,000   	& 24        \\
    Klystron, Main Linac (alt.)        	& 60,000   	& 24        \\
    Power Converter 					& 300,000  	& 4         \\
    Modulator      						& 100,000   & 12        \\
    LLRF            					& 26,300   	& 3         \\ 
    \bottomrule
    \end{tabular}
    \label{tab:mainfail}
\end{table}

The Drive Beam Accelerator contains 472\,RF units, including 12 hot-spare units. The study includes failures of klystrons, modulators, the low-level RF systems, waveguides, the structures and the cooling system. Klystrons, modulators and low-level RF systems can be replaced without stopping the beam. The mean time until the reserve is exhausted is expected to be 524\,h, meeting the target.

A klystron powered Main Linac for 380\,GeV c.o.m. energy requires about 1,500 pairs of klystrons. Considering an individual MTTF of the klystrons of 60,000 hours, a klystron failure every 20~hours is to be statistically expected. Adding all the component failures and with a 5\,\%  reserve, the availability models estimate the expected mean time to reserve exhaustion to be 933\,h, which is above the target. The sensitivity analysis shows that, of all the components, the modulators have the highest impact on the system performance. 

The availability study is still in progress. The investigated key failures
do not compromise the operation of CLIC thanks to the implemented redundancy and reserve. Further studies are required to include a more complete list of failures and to optimise the design for robustness and cost. A maintenance schedule will be developed with optimised length and frequency of technical short stops. Of particular interest will be the integration of stops that are introduced to reduce the power consumption during times of peak demand in the general power grid. This will allow minimisation of the impact of maintenance on the machine operation.

\section{Centre-of-Mass Energy Flexibility}
\label{sect:Energy_Scan}

CLIC can be operated at different centre-of-mass energies so as to perform scans. This can be achieved by operating the Main Linacs with reduced accelerating gradients. Currently, the only request from the physics community is to scan the top threshold, i.e. around 350\,GeV, with an integrated luminosity of 100\,fb$^{-1}$. This is illustrated below, but other operating energies would also be possible.

At the top threshold, a small intra-bunch energy spread tends to be beneficial since it allows one to resolve the onset of the cross section better. Therefore, two scenarios were explored. The first uses the same bunch charge and length as at 380\,GeV and the second a bunch that is 10\% longer and has only 90\% of the nominal charge. This choice ensures the same beam stability but allows one to reduce the beam energy spread.

\begin{figure}[ht!]
\begin{center}
\includegraphics[width=14cm]{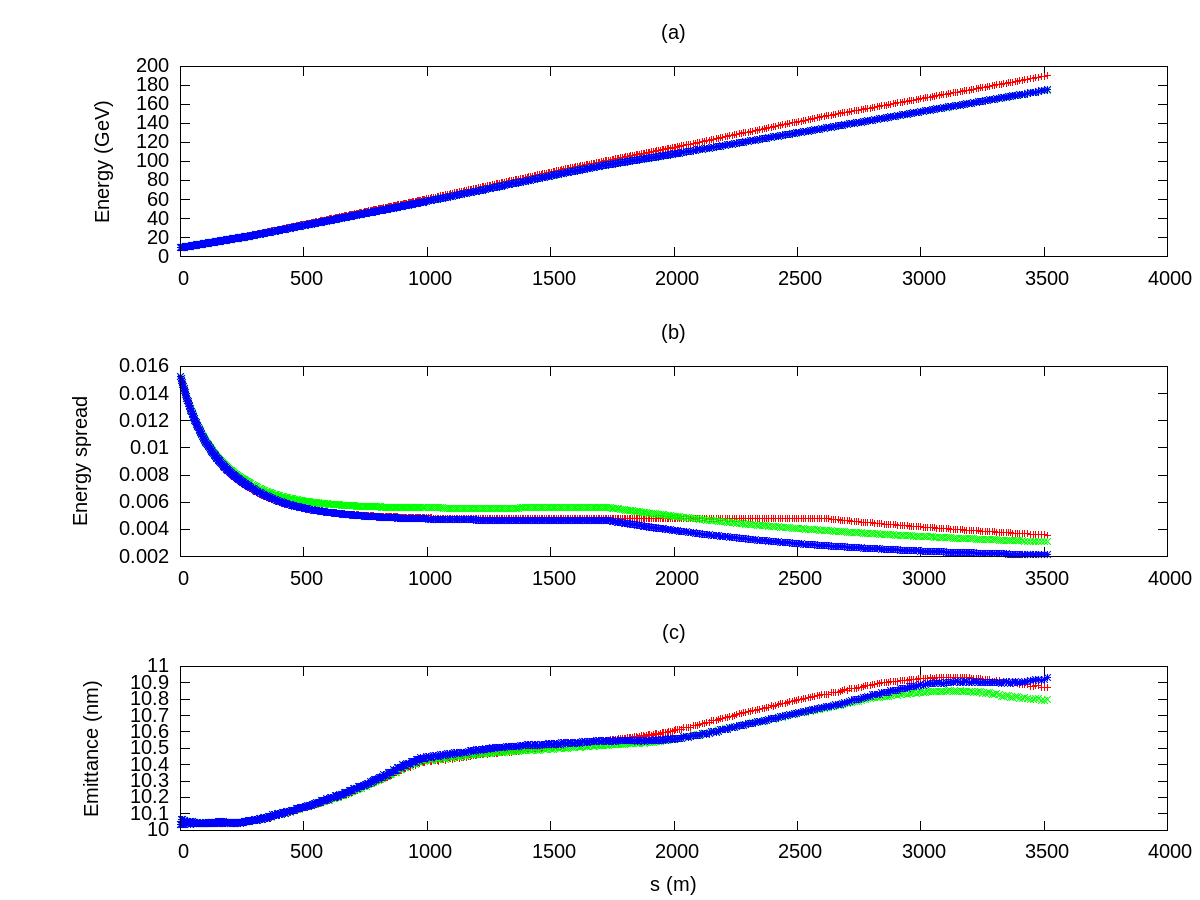}
\caption{(a) Energy, (b) R.M.S. fractional energy spread and (c) vertical emittance versus length $s$ of the Main Linac, operating with nominal bunch length and charge at 380\,GeV (red) and 350\,GeV (green), and at 110\% of the nominal bunch length and 90\% of the nominal bunch charge at 350\,GeV (blue).}
\label{fig:OP_TOP_1}
\end{center}
\end{figure}

The beam energy spread is determined by the short-range longitudinal wakefields of the bunch and the choice of RF phases in each of the four Drive Beam sectors. At the beginning of the linac the phase $\phi_1$ is set to introduce an increasing correlated energy spread while the uncorrelated part is naturally decreasing. At the end the phase $\phi_2$ is set to reduce the energy spread. This choice takes advantage of the so-called BNS damping \cite{Balakin1983}. The correlated energy spread counteracts the transverse wakefields induced by a jittering beam. This suppresses beam break-up.

Figure~\ref{fig:OP_TOP_1} shows the energy and energy spread along the Main Linac for operation at the nominal 380\,GeV and for the two options at 350\,GeV. The incoming R.M.S. energy spread from the RTML is 1.6\%. and an R.M.S. vertical quadrupole position error of 10\,nm is assumed throughout the Main Linac. The emittance growth due to an artificially large quadrupole jitter is also shown and demonstrates that the beam stability is the same in the different cases.

For the 380\,GeV case, $\Phi_1 = 8^{\circ}$ is used for the first three-quarters of accelerating structures and $\Phi_2 = 29.6^{\circ}$ for the last quarter. This gives an R.M.S. energy spread of 0.35\%, which is acceptable for the BDS \cite{Aicheler2012}.

For the first 350\,GeV case, $\Phi_1~=~6^{\circ}$ is used for the first half and $\Phi_2~=~30^{\circ}$ for the second half of accelerating structures, allowing a larger energy spread in the middle of the linac, hence producing a more stable low-emittance beam; this energy spread is removed by the end of the Main Linac. The resulting luminosity is $1.52 \times 10^{34}$~cm$^{-2}$~s$^{-1}$ and the R.M.S. energy spread is 0.30\% in which 60\% of particles are above 99\% of the design energy.

\begin{figure}[!ht]
\begin{center}
\includegraphics[width=10cm]{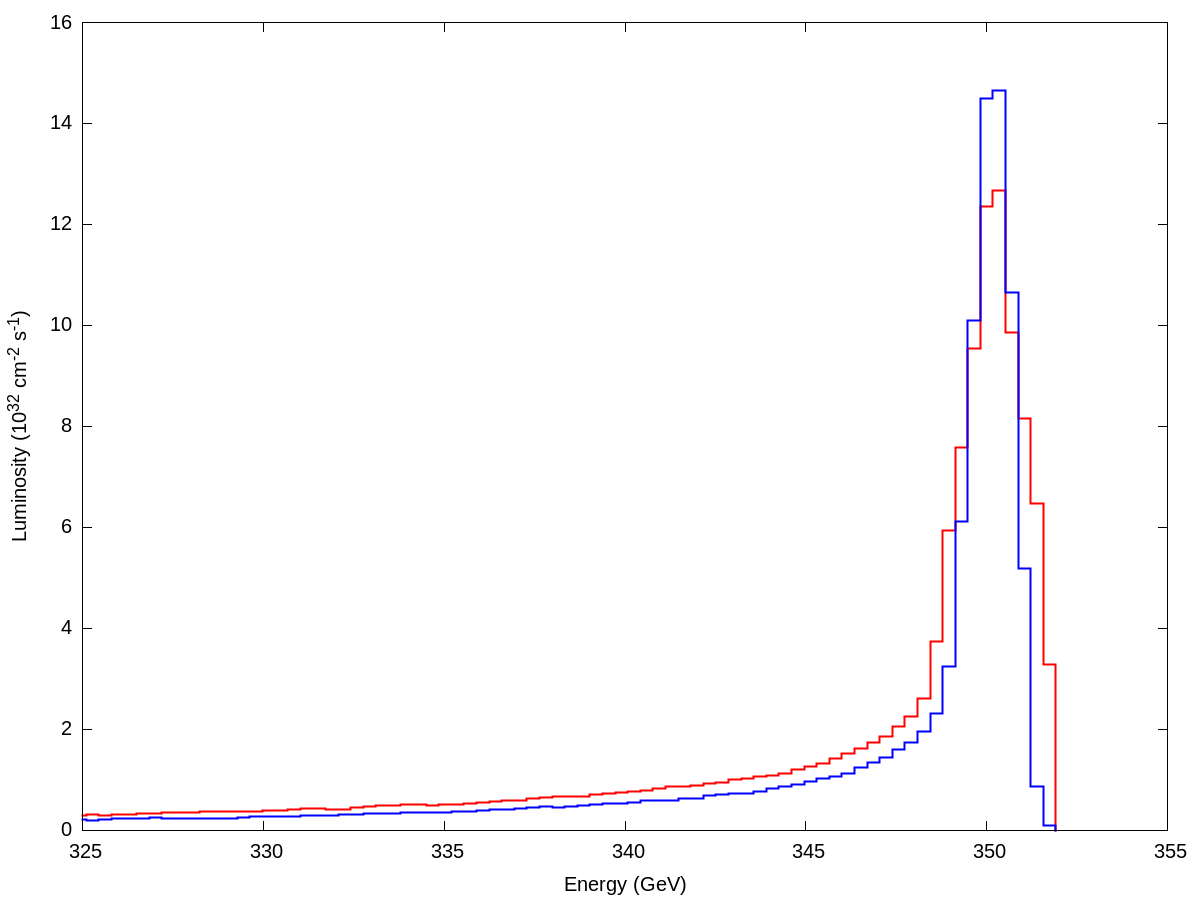}
\caption{Luminosity versus collision energy, operating with nominal bunch length and charge (red), and at 110\% of the nominal bunch length and 90\% of the nominal bunch charge (blue).}
\label{fig:OP_TOP_2}
\end{center}
\end{figure}

For the second 350\,GeV case, the same phases are used but the charge is reduced to 90\% and the bunch length increased to 110\%. This reduces the energy spread to 0.20\%, whilst preserving the same beam quality.
The luminosity achieved is $1.18 \times 10^{34}$~cm$^{-2}$~s$^{-1}$, 64\% of particles are above 99\% of the design energy.
Fig.~\ref{fig:OP_TOP_2} shows the luminosity spectrum for both cases.
Further  optimisation will be done.

\printbibliography[heading=subbibintoc]
\endrefsection

\refsection 
\chapter{Klystron-based Alternative Design}
\label{Chapter:Kly_Design}
\section{Introduction}
An alternative design for the 380\,GeV stage of CLIC is based on the use of X-band klystrons to produce the RF power for the Main Linac. On the one hand, this solution increases the cost of the Main Linac because the klystrons and modulators are more expensive than the Drive-Beam Decelerator and also because a larger tunnel is needed to house the additional equipment. On the other hand, it avoids the substantial cost of the construction of the Drive-Beam Complex and makes the linac more modular. One can therefore expect a competitive cost at low energies while the Drive Beam solution leads to lower cost at high energies. The upgrade of the complex is cheaper with a Drive Beam-based design, since the additional cost to upgrade the Drive-Beam Complex to feed a longer linac is relatively modest. However, an important advantage of the Klystron-based design is that the Main Linac modules can easily be fully tested for performance when they are received. In contrast, the Drive Beam option requires the construction of a substantial complex that can produce a 100\,A Drive Beam before modules can be fully tested. The Klystron-based option could therefore be implemented more rapidly than the Drive Beam-based solution.

\section{Design Choice}
The Klystron-based alternative design is based on a study~\cite{StagingBaseline} that used the same optimisation tools as for the Drive Beam-based option. The required changes were made in the algorithms and a cost model included for the klystrons and modulators. Based on the conclusions of the study a tentative structure and parameter set have been chosen for this design. The optimum structure differs from the Drive Beam-based design, the parameters are given in Table~\ref{t:scdkl0}. If one were to use the same structure as for the Drive Beam-based design, the expected cost would be about 330\,MCHF higher.

The basic unit of the Main Linac consists of a pair of X-band klystrons, a pulse-compressor, an RF distribution system and a number of accelerating structures, see Fig.~\ref{fig:k-module}. The klystrons produce an RF pulse, which is then compressed in time to match the required RF pulse length. This compression increases the power of the RF pulse and hence the number of accelerating structures that can be powered per pair of klystrons. However, a significant amount of power is lost in the compressors. This limits the useful klystron pulse length. The final RF pulse is then split to feed a number of accelerating structures. For each design of the accelerating structure, the RF input power and pulse length were used to determine the pulse compressor parameters.

As can be seen in Table~\ref{t:scdkl1}, the beam emittance, energy spread and charge of the Klystron-based design are very similar to the 3\,TeV parameters, while the bunch is somewhat longer. The vertical emittance is also the same as for the Drive Beam-based design, while the horizontal emittance is smaller and proportional to the bunch charge. The number of bunches per train is significantly higher in order to produce the required luminosity.

The evolution of the vertical emittance along the collider is similar to the Drive Beam-based design, while the horizontal emittance corresponds to the 3\,TeV design. The horizontal and vertical emittances remain below 500\,nm and 5\,nm at extraction from the Damping Ring, below 600\,nm and 10\,nm at injection into the Main Linac and below 630\,nm and 20\,nm at the end of the Main Linac.

\begin{table}[!htb]
\caption{Key parameters of the CLIC accelerating structures.}
\label{t:scdkl0}
\centering
\begin{tabular}{l r r r r}
\toprule
\textbf{Parameter}                  &\textbf{Symbol}        &\textbf{Unit} &\textbf{Klystron} &\textbf{Drive Beam}\\
\midrule
Frequency                           & $f$                   & GHz          & 12         & 12    \\
Acceleration gradient               & $G$                   & MV/m            & 75        & 72           \\
\midrule
RF phase advance per cell           & $\Delta \phi$         & $^{\circ}$   & 120        & 120        \\
Number of cells                     & $N_{\text{c}}$        &          & 28        & 33          \\
First iris radius / RF wavelength   & $a_1/\lambda$         &              & 0.145     & 0.1625       \\
Last iris radius / RF wavelength    & $a_2/\lambda$         &               & 0.09      & 0.104         \\
First iris thickness / cell length  & $d_1/L_{\text{c}}$    &              & 0.25      & 0.303        \\
Last iris thickness / cell length   & $d_2/L_{\text{c}}$    &              & 0.134     & 0.172        \\
\midrule
Number of particles per bunch       & $N$                   & $10^9$       & 3.87      & 5.2      \\
Number of bunches per train         & $n_{\text{b}}$        &              & 485       & 352       \\
Pulse length                        & $\tau_{\text{RF}}$    & ns           & 325       & 244          \\
Peak input power into the structure & $P_{\text{in}}$       & MW           & 42.5      & 59.5         \\
\bottomrule
\end{tabular}
\end{table}

\begin{table}[!htb]
\caption{Key beam parameters at the collision point.}
\label{t:scdkl1}
\begin{center}
\begin{tabular}{l c l c}
\toprule
Particles per bunch&$3.87\times10^9$&
bunches per pulse&$485$\\
bunch spacing&$15$\,cm&
bunch length&$60\,\mu$m\\
final r.m.s. energy spread&$0.35\,\%$\\
horizontal/vertical beam size&119\,nm/2.9\,nm\\
\bottomrule
\end{tabular}
\end{center}
\end{table}

\section{Design Implications}
For the Klystron-based alternative the Main Linac has been redesigned to evaluate the cost and verify the beam dynamics. No design optimisation has been performed for the other systems. We expect this to have a minor impact on cost and system performance as detailed below.

The Klystron-based injector RF design has to differ from the Drive Beam-based case, in order to accommodate the longer bunch train. However, the total charge per pulse is the same in both cases. Hence, to first order the same amount of installed RF is required. However, the choice of accelerating structure and pulse compressor in the injection system would be slightly different to obtain optimum efficiency.

The single bunch parameters at the entrance of the linac are very similar to the Drive Beam-based design. The main difference is the lower bunch charge, which helps the emittance preservation, and the smaller horizontal emittance, which needs to be achieved. Both parameters are very similar to the 3\,TeV design and can thus be achieved with the corresponding design. The bunch length at the start of the Main Linac and afterwards is larger in the Klystron-based design, which requires less compression in the RTML and eases the system requirements.

The BDS has to provide the same beta-functions in the Drive Beam and klystron-based designs. Hence the Drive Beam-based design fully achieves the required performance for both options.

\section{Main Linac Layout and Optics}
The Main Linac consists of a sequence of accelerating modules interleaved with modules that support quadrupoles detailed in Table~\ref{t:ml2}. The accelerating modules are 2.01\,m long and support eight structures. Two different types of quadrupole modules exist, with a length of 0.6\,m and 0.85\,m. Each of them supports a BPM and a quadrupole with a length of either 0.4\,m or 0.65\,m, respectively.

\begin{table}[!htb]
\caption{The main parameters of the different lattice sectors.}
\label{t:ml2}
\begin{center}
\begin{tabular}{l c c c c c}
\toprule
\textbf{Sector Number}& \textbf{1} & \textbf{2}& \textbf{3}& \textbf{4}& \textbf{5} \\
\midrule
Number of quadrupoles & 148 & 138 & 76 & 100 & 126 \\
Quadrupole length [m] & 0.4& 0.4& 0.4& 0.65& 0.65\\
Quadrupole spacing [m] & 2.613& 4.62& 6.63& 6.88 & 8.89\\
\bottomrule
\end{tabular}
\end{center}
\end{table}

\subsection{Accelerator Physics Issues}
\begin{figure}[!htb]
\centering
\includegraphics[scale=0.66]{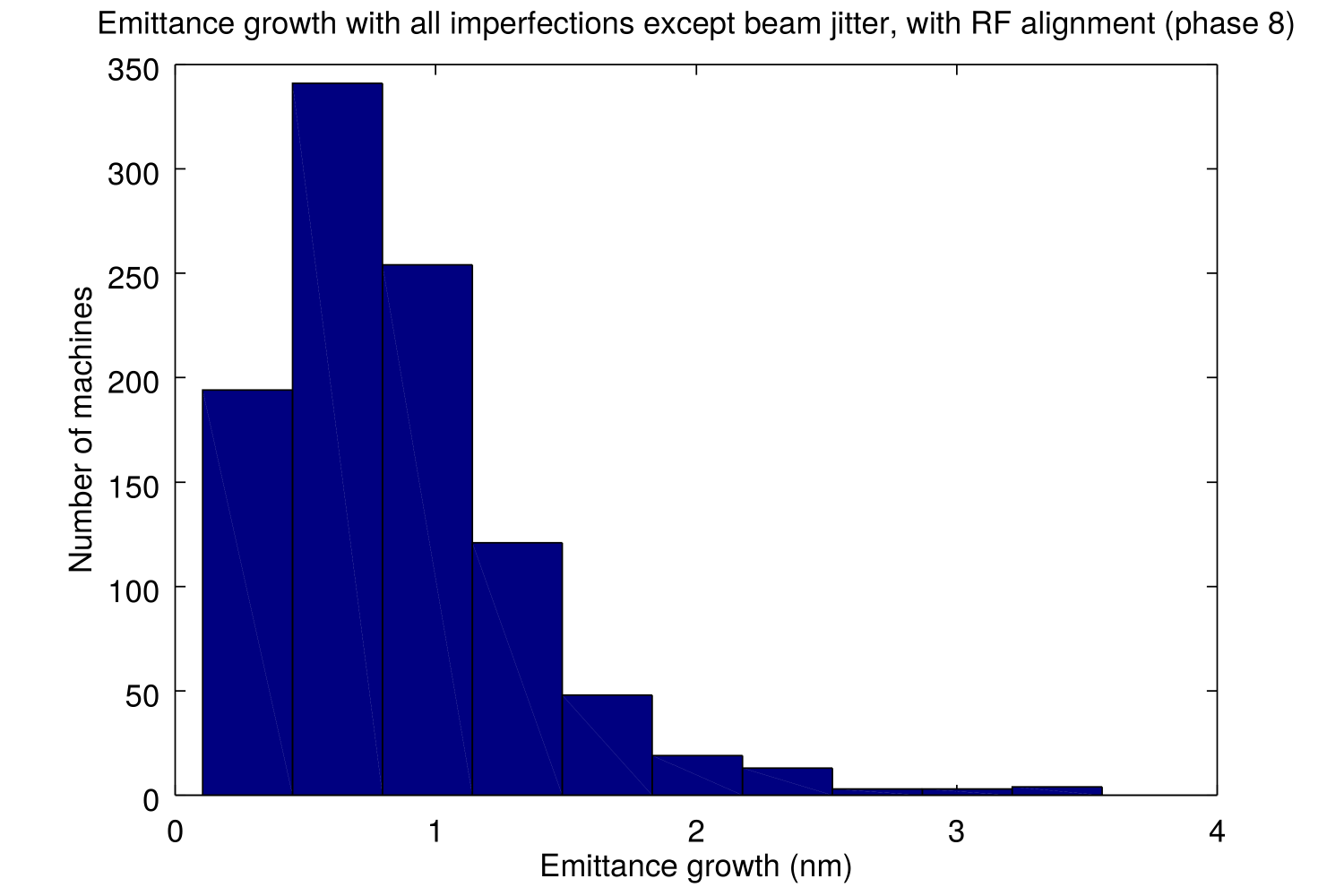}
\caption{Probability distribution of the emittance growth for static imperfections for a Main Linac average RF phase of $12^\circ$.}
\label{f:dsstatic_kl}
\end{figure}

The beam dynamics of the Klystron-based design do not differ from the Drive Beam-based design. The increase of the wakefields due to the smaller accelerating structure aperture is, by design, compensated by the reduced bunch charge and length. The expected emittance growth is shown in Table~\ref{t:align_ml_kl} and the probability distribution in Fig.~\ref{f:dsstatic_kl}~\cite{c:MLperf}.

\begin{table}[!htb]
\caption{Key alignment specifications for the Main-Linac components and the resulting emittance growth. The values after simple steering (1-2-1), Dispersion Free Steering (DFS) and realignment of the accelerating structures using the wakefield monitors (RF) are shown.}
\label{t:align_ml_kl}
\begin{center}
\begin{tabular}{cccccc}
\toprule
                  &                    &              & \multicolumn{3}{c}{$\Delta \epsilon_y$ [nm]} \\ 
Imperfection      & With respect to    & Value        & 1-2-1  & DFS    & RF     \\
\midrule
Girder end point  & Wire reference     & 12 $\mu$m    & 11.37  & 11.31  & 0.07  \\
Girder end point  & Articulation point & 5 $\mu$m     & 1.45   & 1.45   & 0.02  \\
Quadrupole roll   & Longitudinal axis  & 100 $\mu$rad & 0.04   & 0.04   & 0.04  \\
BPM offset        & Wire reference     & 14 $\mu$m    & 154.54 & 14.01  & 0.10  \\
Cavity offset     & Girder axis        & 14 $\mu$m    & 5.51   & 5.50   & 0.04  \\
Cavity tilt       & Girder axis        & 141 $\mu$rad & 0.10   & 0.47   & 0.25  \\
BPM resolution    &                    & 0.1 $\mu$m   & 0.01   & 1.03   & 0.02  \\
Wake monitor      & Structure centre   & 3.5 $\mu$m   & 0.01   & 0.01   & 0.40  \\
\midrule
All               &                    &              & 176.68 & 32.72  & 0.84  \\
\bottomrule
\end{tabular}
\end{center}
\end{table}

\subsection{Components}
The total number of components is listed in Table~\ref{t:scdkl3}. The alignment tolerances are the same as for the Drive Beam-based design.

\begin{table}[!htb]
\caption{Key components for each Main Linac.}
\label{t:scdkl3}
\begin{center}
\begin{tabular}{*2{c}}
\toprule
Quadrupoles T0 & 1456 \\
Quadrupoles T1 & 362 \\
Quadrupoles T2 & 226 \\
BPMs & 588 \\
Accelerating structures & 11648 \\
\bottomrule
\end{tabular}
\end{center}
\end{table}

\section{Main Linac RF Unit}
\label{sect:KLY_ML}

\begin{figure}[!hbtp]
\begin{center}
\includegraphics[width=15 cm]{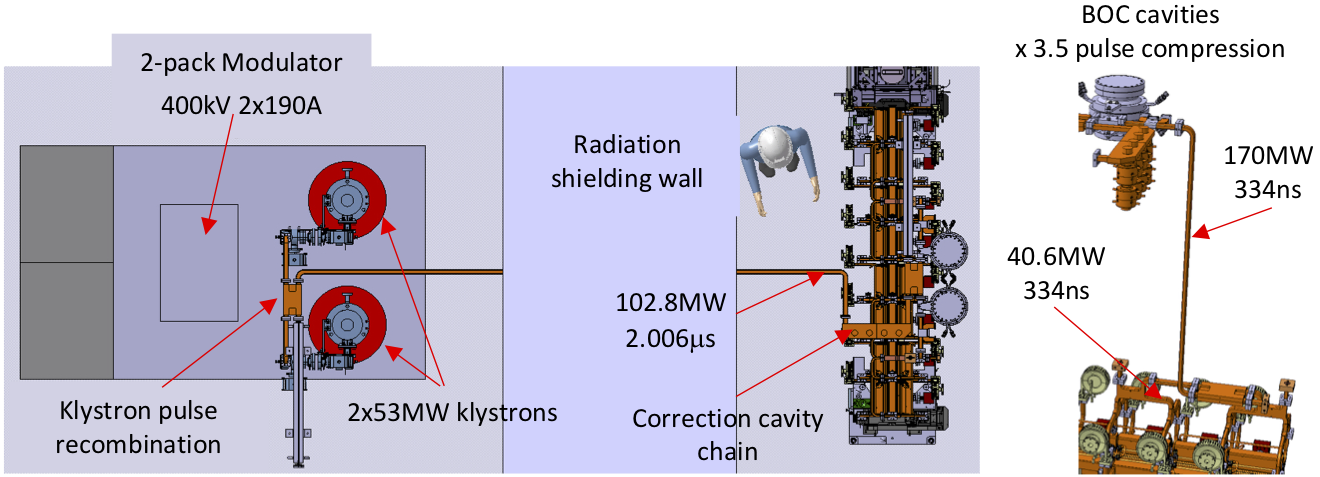}
\caption{The RF unit: top view of one RF unit in the klystron and Main-Linac tunnels (left) and detailed view of the distribution network (right).}
\label{fig:RF-distr}
\end{center}
\end{figure}

Figure~\ref{fig:RF-distr} shows the top view of one RF unit installed in the klystron and Main-Linac tunnels, with the detailed view of the distribution network from the pulse compression onwards. A solid-state modulator delivering 400\,kV and 2~$\times$~190\,A drives two 53\,MW X-band klystrons (see Section~\ref{sect:Klystrons}). The RF power from the two klystrons is recombined and sent to the Main-Linac tunnel beyond the shielding wall, which has been dimensioned to allow access to the klystron gallery under certain conditions during operation. After having been processed by an 8-cavity linearisation system, the RF signal is split in two and delivered to two pulse compression devices, one on each half of the module (Section~\ref{sect:Pulse_Comp}). The compressed RF pulse is then distributed through two hybrid splitters in cascade to four of the accelerating cavities by a double-height waveguide system, so to minimise RF losses and the risk of breakdown.

The numbers in the drawing assume a klystron efficiency of 70\%, which is the minimum goal for the development of high efficiency klystrons \cite{Constable2017}; any additional gain in klystron efficiency could then be reflected in a reduction of the HV level of the modulator pulse and in an increase of the operational margins, with benefits for cost and reliability. This scheme also can foresee a pulse compression factor of 3.5, obtained by means of two separate compression cavities of the BOC type, one for each side of the RF module; a well known compression scheme already used in S-band and C-band RF systems. The RF pulse, compressed in time, reaches 170\,MW at the entrance of the first hybrid splitter and 40.6\,MW are delivered to each of the four accelerating structures connected to each BOC pulse compressor. Some of the fundamental RF devices that have been specifically developed to be integrated into the RF power delivery system include the correction cavity chain, for the linearisation of the compressed pulse \cite{Wang2017}, compact spiral loads and high-power hybrid splitters.

The klystron feeding would give an advantage in terms of flexibility for operation, by providing the possibility to independently set the amplitude and phase of the accelerating field for each module; this flexibility would not only be restricted to the 380\,GeV stage, but could be extended to the higher energy stages by maintaining the klystron-powered section in front of the BDS, making the control of the colliding beam energy spread more straightforward.

\printbibliography[heading=subbibintoc]
\endrefsection

\refsection 
\chapter{Upgrade to Higher Energies}
\label{Chapter:HE_Design}
\section{Introduction}
The CLIC 380\,GeV energy stage can be efficiently upgraded to higher energies; viz. 1.5\,TeV and 3\,TeV. This flexibility has been an integral part of the design choices for the first energy stage. The highest energy stage corresponds to the design described in the  CDR~\cite{Aicheler2012}, with minor modifications due to the first energy stages, as described below. The only important difference to the CDR design is a new final focus system that has an increased distance between the	last quadrupole	of the machine and the interaction point. This allows the magnet to be installed in the tunnel and outside of the detector.

\section{Sample Parameters}
\label{sect:HE_Intro}

\begin{figure}
\begin{center}
\includegraphics[width=12cm]{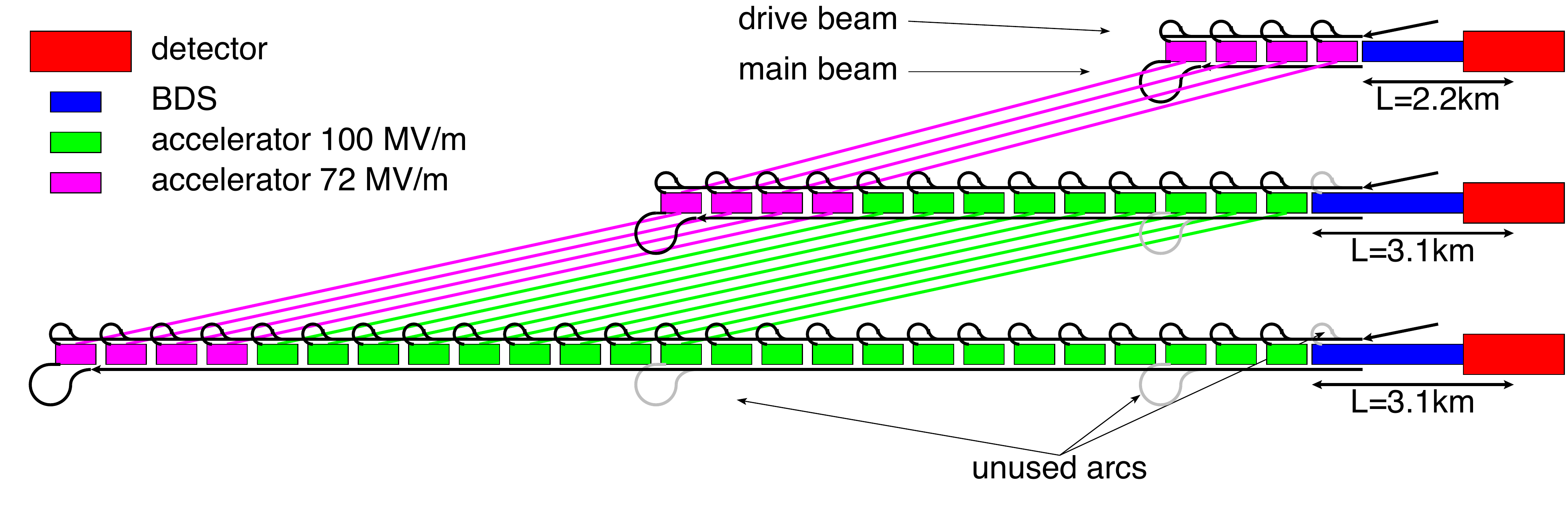}
\end{center}
\caption{Main Linac upgrade concept for the Drive-Beam-based first energy stage.}
\label{f:scdup1}
\end{figure}

The key parameters for the three energy stages of CLIC are given in Table~\ref{t:scdup2}. The concept of the staging implementation is illustrated in Fig.~\ref{f:scdup1}. In the first stage, the linac consists of modules that contain accelerating structures that are optimised for this energy range. At higher energies these modules are reused and new modules are added to the linac. First,  the linac tunnel is extended and a new Main-Beam turn-around is constructed at its new end. The technical installations in the old turn-around and the subsequent bunch compressor are then moved to this new location. Similarly, the Main Linac installation is moved to the beginning of the new tunnel. Finally, the new modules that are optimised for the new energy are added to the Main Linac. The BDS has to be modified by installing magnets that are suited for the higher energy and it will be extended in length. The beam extraction line also has to be modified to accept the larger beam energy but the dump remains untouched.

The baseline scenario to reuse the old structures requires that the modules of the first energy stage are capable of accelerating the beam pulses of the later stages. This constraint has been taken into account in the design of the first energy stage. In addition to this baseline, alternative scenarios also exist. First, one can distribute the old modules along the new linac. Secondly, one can replace the existing modules with new ones; however, this would increase the cost of the upgrade. 

In order to minimise modifications to the Drive-Beam Complex, the RF pulse length of the first stage is chosen to be the same as the subsequent energy stages. This is important since the lengths of the Delay Loop and the Combiner Rings, as well
as the spacings of the turn-around loops in the Main Linac, are directly proportional to the RF pulse length. Hence, the constant RF pulse length allows the reuse of the whole Drive-Beam Complex and the turn-arounds in the tunnel. In addition, it is optimum if the fill time of the accelerating structures in the Drive Beam Accelerator correspond to the RF pulse length as this minimises the sensitivity to Drive-Beam klystron RF amplitude and phase jitters.

In order to upgrade from 380\,GeV to 1.5\,TeV some modifications are required for the Drive-Beam production complex. The Drive-Beam Accelerator pulse length has to increase to be able to feed all of the new decelerators and also the beam energy has to be increased by 20\,\%. The energy increase is achieved by adding more Drive-Beam modules. The pulse length increase is achieved by increasing the stored energy in the modulators to produce longer pulses. The remainder of the Drive-Beam Complex remains unchanged except that all magnets after the linac need to operate at a 20\% larger field. The upgrade to 3\,TeV requires the construction of a second Drive-Beam generation complex.

Within the Main-Beam pulses the bunches have the same spacing at all energy stages to minimise the impact on the Main-Beam production complex.  To be able to accelerate the full train of the final stage, the fill time of the first-stage structures must be shorter and the bunch charge limit higher than in the final stage. This constraint has been taken into account in the first energy stage design.
For the target value 
at 380\,GeV, this resulted in a cost increase of 50\,MCHF. The gradient of the structures for the first stage is 72\,MV/m.  Consequently, four decelerator stages are required per Main Linac in the first stage.  The upgrade to 3\,TeV requires an additional 21 decelerator stages.

\begin{table}
\caption{Key parameters of the three energy stages.}
\label{t:scdup2}
\centering
\begin{tabular}{l l l l l l}
\toprule
\textbf{Parameter}                  & \textbf{Symbol}         & \textbf{Unit}& \textbf{Stage 1} & \textbf{Stage 2} & \textbf{Stage 3} \\
\midrule
Centre-of-mass energy               & $\sqrt{s}$              &GeV                                        & 380 & 1500 & 3000\\
Repetition frequency                & $f_{\text{rep}}$        &Hz                                         & 50 & 50 & 50\\
Number of bunches per train         & $n_{b}$                 &                                           & 352 & 312 & 312\\
Bunch separation                    & $\Delta\,t$             &ns                                         & 0.5 & 0.5 & 0.5\\
Pulse length                        & $\tau_{\text{RF}}$      &ns                                         &244 &244 &244\\
\midrule
Accelerating gradient               & $G$                     &MV/m                                       & 72 & 72/100 & 72/100\\
\midrule
Total luminosity                    & $\mathcal{L}$           &$10^{34}\;\text{cm}^{-2}\text{s}^{-1}$     & 1.5 & 3.7 & 5.9 \\
Luminosity above 99\% of $\sqrt{s}$ & $\mathcal{L}_{0.01}$    &$10^{34}\;\text{cm}^{-2}\text{s}^{-1}$     & 0.9 & 1.4 & 2\\
\midrule
Main tunnel length                  &                         &km                                         & 11.4 & 29.0 & 50.1\\
Number of particles per bunch                    & $N$                     &$10^9$                                     & 5.2 & 3.7 & 3.7\\
Bunch length                        & $\sigma_z$              &$\mu$m                                     & 70 & 44 & 44\\
IP beam size                        & $\sigma_x/\sigma_y$     &nm                                         & 149/2.9 & $\sim$ 60/1.5 & $\sim$ 40/1\\
Normalised emittance (end of linac) & $\epsilon_x/\epsilon_y$ &nm                                         & 900/20 & 660/20 & 660/20\\
\bottomrule
\end{tabular}
\end{table}

\begin{figure}
\begin{center}
\includegraphics[width=12cm]{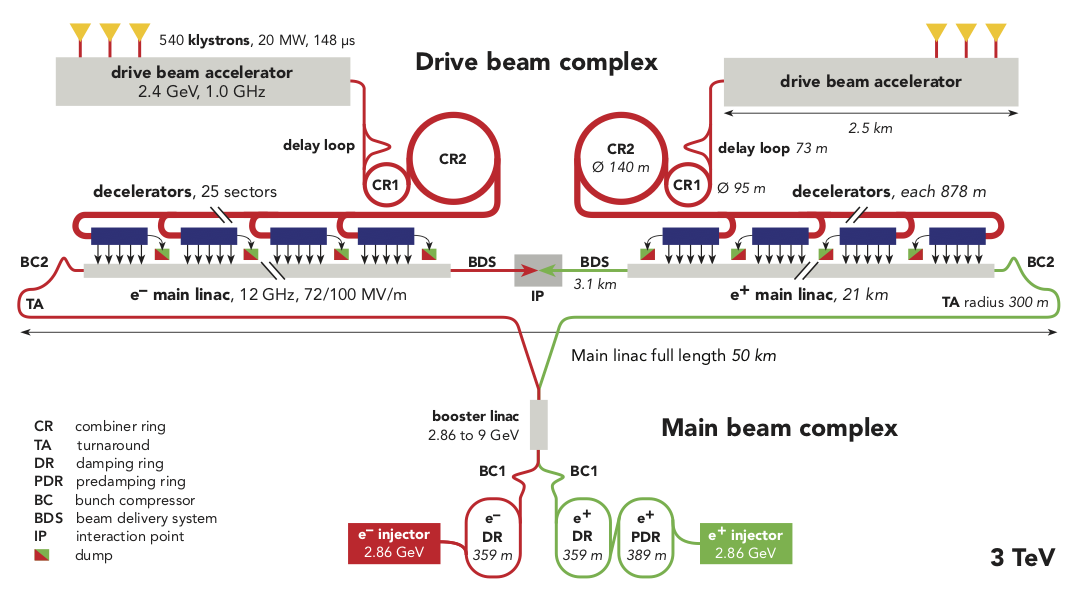}
\end{center}
\caption{Overview of the CLIC layout at $\sqrt{s}$~=~3\,TeV.}
\label{f:scduplayout}
\end{figure}

\section{Impact on Systems}
\label{sect:HE_Impact}
The 1.5\,TeV beam parameters at injection into the Main Linac are the same as for 3\,TeV. The preservation of the beam emittance will be easier due to the shorter linac and the beam cleaning and focusing into the collision point are also less challenging due to the lower beam energy. Therefore the 1.5\,TeV case is not discussed separately. The beam parameters of the 3\,TeV stage are the same as in the CLIC CDR~\cite{Aicheler2012} where the feasibility of the 3\,TeV parameters has been established. Therefore all the beam line and component designs from this 3\,TeV can be used for the energy upgrade.

\subsection{Main-Beam Injector Complex}
Some modifications of the 380\,GeV complex are required to accommodate the high energy parameters. The injectors have to be modified to deal with smaller bunch charges and number of bunches, while the BDS and spent beam system have to be modified to accommodate the higher energy.

At injection into the Main Linac the bunch charge at 3\,TeV is smaller than at 380\,GeV and the trains contain fewer bunches while the vertical emittance and the beam energy remain the same. This reduces the impact of wakefields in the injector and booster linacs leading to better beam stability and more relaxed tolerances for misalignments. The smaller bunch charge and number also reduces the beam loading in the injector, so that the klystrons can be operated at lower output power. Only two beam parameters are more ambitious than at 380\,GeV, the horizontal emittance is smaller and the bunch length is also smaller. Both values depend strongly on the damping ring. The reduced bunch produces a smaller horizontal emittance because it leads to reduced intra-beam scattering. It also leads to a reduced longitudinal emittance which means shorter bunch length in the compressors of the RTML transport system. The difficulty in reaching these parameters remains roughly the same independent of energy.

\subsection{Main Linac}

The preservation of the beam quality in the Main Linac is slightly more challenging at the high energy than at 380\,GeV. However, the specifications for the performance of alignment and stabilisation systems for the 380\,GeV stage are based on the requirements for the 3\,TeV stage. They are therefore sufficient for the high energy stage and no upgrades of these systems are required.

\subsection{Beam Delivery System and Extraction Line}
In the BDS the magnets will have to be replaced with stronger ones for the higher energy, in particular for the final doublet. The overall system has also to be increased in length. The main reason is the required increase in the length of the energy collimation system. This system protects the machine from the most important failure, which is the failure of one of the kickers that dispatches the Drive-Beam pulses into the decelerators. This will prevent one of the Drive-Beam sectors to be powered resulting in a large energy error of the Main Beam. The energy collimation will protect the machine from this beam and requires more length at the higher beam energy to avoid collimator damage.

The final focus system is longer than at lower energies. This reduces the bending of the beam required for the chromatic correction, which reduces the impact of synchrotron radiation of the bends which is more critical at higher energies.
The layout of the BDS at 380\,GeV takes the upgrade into account, see Section~\ref{sect:BDS}.

Finally, the extraction line has to be rebuilt to deal with the larger beam energy. The 3\,TeV design can deal with the large energy spread in the beam from the strong beam-beam interaction and the 14\,MW of beam power. The 380\,GeV design can take advantage of the smaller beam power of 2.8\,MW and a much smaller energy spread. This makes instrumentation possible in the extraction line that is not possible at 3\,TeV. The beam dump will be dimensioned from the beginning to handle 14\,MW in order not to have to replace it.

\subsection{Drive-Beam Complex}
For the upgrade the Drive-Beam production complex has to produce more Drive-Beam pulses. The beam current and the pulse length remain constant but the beam energy has to increase by about 20\,\%. This minimises the changes that are required for the upgrade.

With the same Drive-Beam current, the PETS of the old modules will continue to produce the RF power required for the old accelerating structures. The PETS of the new modules use a modified design that produces the RF power required for the new structures, which is quite similar but not exactly the same. The choice of constant Drive-Beam current implies that the existing Drive-Beam Accelerator RF units are perfectly matched for the new energy stage and achieve the efficiency as before.

Additional RF units in the Drive-Beam Accelerator are used to achieve the higher drive beam energy. Their design corresponds exactly to the already existing ones and they can be appended at the end of the Drive-Beam Accelerator Complex in a new building.

An upgrade of the Drive-Beam Accelerator modules allows the production of longer RF pulses. Increasing the stored energy in them allows the klystron to produce longer RF pulses and accelerate a longer Drive-Beam train that is then combined into a larger number of final pulses. The klystrons are already dimensioned to be able to handle the longer pulses.

The constant RF pulse length allows the reuse of the Drive-Beam Recombination Complex. The circumferences of the delay loop and the combiner rings are directly proportional to the RF pulse length. As a consequence the lattice of the combination complex remains unchanged and only the field in the magnets has to be increased in proportion to the beam energy, which is taken into account in the magnet design. The distance between the Drive-Beam turn-arounds in the Main Linac is also proportional to the RF pulse length. Therefore the existing turn-arounds can be reused for the higher energy stages. The only exception is the last turn-around, which will not be used since this sector is taken from the Main Linac and added to the beam delivery system. As in the combination complex all of the magnets in the Drive-Beam distribution system need to be operated at higher field for the upgrade.

The Drive-Beam Decelerator sectors with the new modules have more accelerating structures since they are shorter. Hence more PETS will extract power from the Drive Beam and decelerate it more. The energy of the Drive Beam is therefore increased by about 20\%. This ensures that the Drive Beam can be safely dumped at the end of the decelerator. In the decelerators with the old modules the Drive-Beam deceleration remains unchanged so that the beam will be dumped at higher energy, which leads to a small inefficiency but poses no issue for the machine operation.

\section{Upgrade from the Klystron-Based Option}
\label{sect:HE_K_Upgrade}
The upgrade from a klystron-based first stage to higher energies is also possible by reusing the klystron-driven structures and the klystrons and by adding new Drive-Beam powered structures. In the klystron-based first energy stage, the single bunch parameters are the same as for the high energy stages, only the bunch charge is slightly larger. At 380\,GeV more bunches per train are accelerated than at higher energies, so for the upgrade the klystron pulse length can be slightly shortened.

\begin{figure}
\begin{center}
\includegraphics[width=12cm]{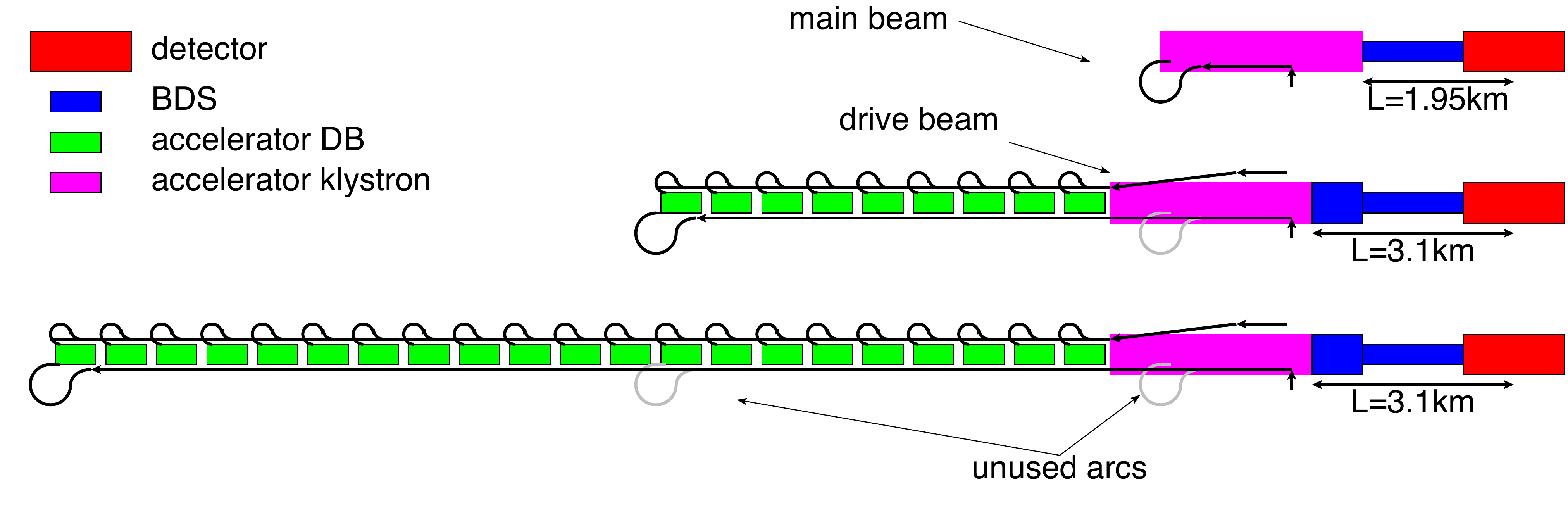}
\end{center}
\caption{Main Linac upgrade concept for the Klystron-based first energy stage.}
\label{f:scdkl1}
\end{figure}

An important difference with respect to the Drive-Beam powered first energy stage is the placement of modules. The klystron-powered Main Linac has to be larger in radius than the part housing beam-driven acceleration in order to provide the space for klystrons and modulators. Therefore it appears best to extend the Main Linac for 1,200\,m with a large tunnel and then continue with a smaller tunnel, see Fig.~\ref{f:scdkl1}. The Drive-Beam powered modules are then placed in the smaller tunnel. The klystron-powered structures remain in the large tunnel. They need to be moved longitudinally slightly in order to adjust the lattice for the high energy, which requires longer quadrupoles with a wider spacing. The last 1,200\,m of the linac is moved to the beginning of the large tunnel to provide the space for the high energy BDS.

The impact of the energy upgrade on the Main-Beam Injectors and Damping Rings is quite small. The bunch charge at 3\,TeV is smaller than at 380\,GeV; the difference is at the 4\%-level, significantly smaller than for the upgrade of the Drive-Beam-based machine. At higher energy, the number of bunches per beam pulse is also smaller, which is straightforward to accommodate. The BDS for klystron and Drive-Beam-based design are the same; hence the upgrade path is also the same.

\section{Beam Delivery System}
\label{sect:HE_BDS}

The design and tuning of the BDS has been further optimised. A design with a distance of 6\,m between the collision point and the last focusing magnet has been developed, hence the magnets are not inside of the detector any more. Also the tuning procedures have been improved in speed and performance. 

The BDS transports the high energy $e^-$ and $e^+$ beams from the exit of the linacs, focusing them by the Final Focus Systems (FFS) to the small beam sizes required at the Interaction Point (IP) to meet the CLIC luminosity ($\mathcal{L}_0=\;5.9\cdot 10^{34}cm^{-2}s^{-1}$) goal. In addition the BDS is responsible for measuring, collimating and protecting the downstream systems as explained in~\cite{Aicheler2012}.

\begin{figure}[!htb]
\begin{center}
\includegraphics[width = 0.45 \columnwidth]{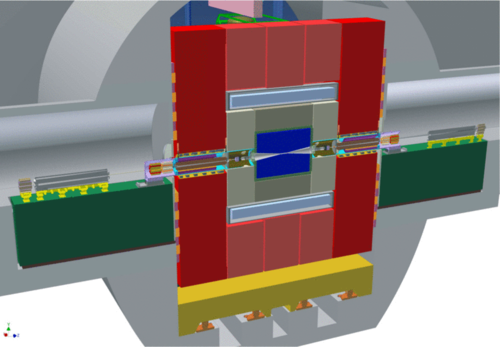}
\includegraphics[width = 0.41 \columnwidth]{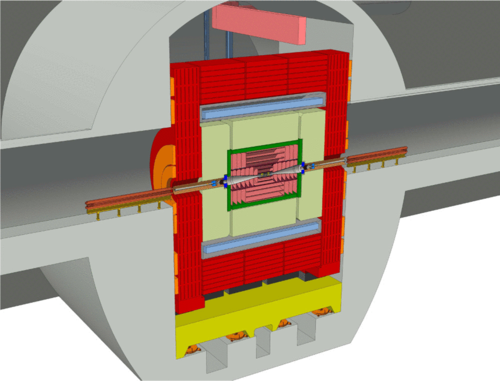}
\caption{MDI for the case with L$^*=\;3.5$~m (left) and L$^*=\;6$~m (right). }
\label{fig:cantilever}
\end{center}
\end{figure}

The FFS presented here is based on the local chromaticity correction scheme proposed in~\cite{Raimondi2001}. The key parameters of the FFS are summarized in Table~\ref{beam_param2}. With regard to the distance (L$^*$) between the Final Doublet (FD), the last pair of quadrupoles, and the IP, two designs are envisaged, the baseline design with L$^*$=~3.5\,m and the quadrupole free detector design with L$^*$=~6\,m. Increasing L$^*$ to 6\,m simplifies enormously the Machine Detector Interface (MDI), removes interplay between the solenoid field and QD0 field, reduces QD0 vibration, and also increases the forward acceptance and the feasibility of the system without compromising the luminosity delivered to the experiments. Figure~\ref{fig:cantilever} shows the adopted MDI solutions in both cases. 

\begin{table}[htb!]
\begin{center}
\caption{\label{beam_param2} CLIC 3\,TeV design parameters.}
\begin{tabular}{lccc} 
\toprule
L$^{*}$    &  [m]        &     3.5       &  6  \\ 
\midrule
FFS length  &  [m]      &       450       &  770\\ 
Norm. emittance (IP) $\gamma\epsilon _{x}/ \gamma\epsilon _{y}$ &   [nm]     &     660 / 20         &  660 / 20      \\ 
Beta function (IP) $\beta_{x}^{*}/ \beta_{y}^{*}$ &   [mm]   &         7 / 0.068       &   7 / 0.12   \\ 
IP beam size $\sigma_{x}^{*}/ \sigma_{y}^{*}$ &  [nm]      &     40 / 0.7      &  40 / 0.9    \\ 
Bunch length $\sigma_{z}$   &   [$\mu$m]           &        44       &    44    \\ 
R.M.S. energy spread $\delta_{p}$  &  [\%]      &       0.3        &  0.3     \\ 
Bunch population $N_{e}$  &  [$\times 10^{9}$]      &       3.72        &    3.72     \\ 
Number of bunches  $n_{b}$        &   &      312        &    312      \\
Repetition rate $f_{rep}$  &   [Hz]         &         50       &    50       \\ 
Luminosity $\mathcal{L}_{total}$ ($\mathcal{L}_{1\%}$) &  [$10^{34}\,cm^{-2}s^{-1}$]      &     5.9 (2.0)       &    5.9 (2.0) \\ 
\bottomrule
\end{tabular}
\end{center}
\end{table}

\subsection{CLIC-FFS 3\,TeV with L$^*=3.5$\,m}
Since the CDR~\cite{Aicheler2012} the tuning study of the baseline design of the CLIC-FFS has made remarkable progress in terms of considered machine imperfections and tuning performance. The new tuning procedure based on linear and non-linear knobs made it possible to address the feasibility of the CLIC-FFS. Moreover the new procedure treats the e$^-$ and e$^+$ systems independently as opposed to the CDR study, where only one system was considered for tuning. Table~\ref{tab:imperfections} compares the list of machine imperfections considered in the CDR and the present tuning study. 
\begin{table}
   \caption{List of machine imperfections included in the tuning studies for the L$^*$ cases 3.5\,m and 6\,m.}
	\begin{center}
	 \begin{tabular}{l|c|c|cc}
     \toprule
           \textbf{Imperfection}    & \textbf{Unit} & \multicolumn{3}{c}{\textbf{$\sigma_{error}$}} \\
     \midrule
       								&			   	    &		CDR		&   \multicolumn{2}{c}{Present study} \\
									&					&		L$^*$=3.5~m	& L$^*$=3.5\,m	&	L$^*$=6\,m \\
      \midrule
           BPM Transverse Alignment     & [$\mu$m]     & 		10		&	10     &    10   \\
           BPM Roll          	    	& [$\mu rad$]   & 		-		&	300    &    300    \\ 
           BPM Resolution         		& [nm]              & 		10		&	20     &    10    \\ 
           Magnet Transverse Alignment  & [$\mu$m]     & 		10		&	10     &    10   \\ 
           Magnet Roll          	    & [$\mu rad$]   & 		-		&	300    &    300   \\ 
           Magnet Strength				& [$\%$]        	&		-		&	0.01	&	0.01	    \\
           Ground Motion				&           		&		-		&	Model-B10 & -	    \\
           \bottomrule
     \end{tabular}
     \end{center}
   \label{tab:imperfections}
\end{table}

The inclusion of additional imperfections brings the study into a more realistic scenario,but it also adds new aberrations at the IP. To target these, a set of 4 skew sextupole magnets have been included at the proper phase advances~\cite{Okugi2014} into both beamlines. Non-linear knobs based on sextupole strength variations have been designed to target the most relevant $2^{nd}$-order aberrations.

Treating independently the e$^-$ and e$^+$ systems slows the tuning process by 40\,\%. In reality when considering the single beam case (e.g. tracking only one beam and then mirroring the obtained IP particle distribution to compute the luminosity) there is a perfect overlap between e$^-$ and e$^+$ beams at the IP, which speeds up the tuning convergence by a factor $\sqrt{2}$, since both $\sigma_{e^-}$ and $\sigma_{e^+}$ are equals, as explained in~\cite{Marin2018a}. 
The results obtained with the new procedure including only static imperfections show that 90\,\% of the machines reach a luminosity above 100\,\% of $\mathcal{L}_0$ at the cost of 6000 luminosity measurements, which is a factor 3 faster to deliver 10\,\% more luminosity with respect to the CDR results. Nevertheless the goal has been set to 110\,\% of $\mathcal{L}_0$, the additional 10\,\% accounting for dynamic imperfections one of the most important being ground motion.

Several models exist for ground motion~\cite{Shiltsev2010}. In this study the model B10 is chosen, as it contains measurements of the power spectra density of the seismic motion and also the contribution from the noisy environment of a particle detector, the ATLAS detector of the LHC.

\begin{figure}[!htb]
\begin{center}
\includegraphics[width = 0.99 \columnwidth]{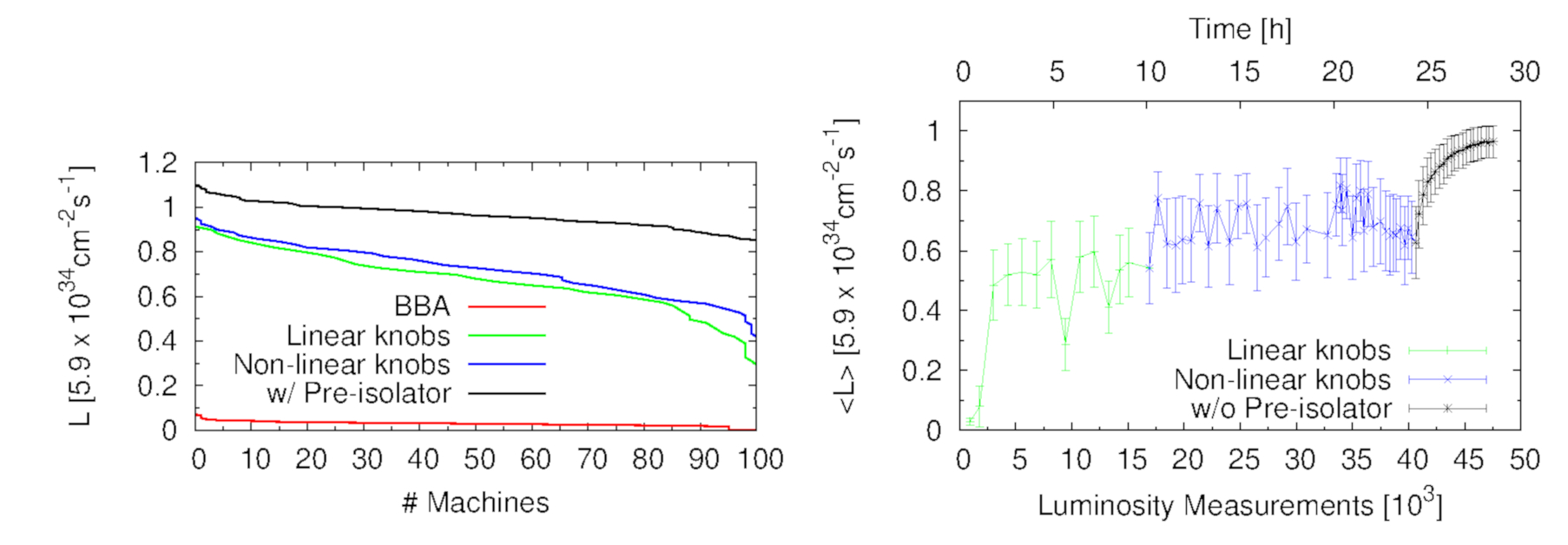}
\caption{Left plot: Accumulated histograms obtained after applying the beam-based alignment (red), after applying the linear and non-linear knobs (green and blue, respectively) and after scanning linear and non-linear knobs without the pre-isolator (black). Right plot: Evolution of the mean value of the luminosity and its standard deviation of the 100 machines versus luminosity measurement, or equivalently in time (top horizontal axis), assuming that each $\mathcal{L}$ measurement takes 2\,s.}
\label{fig:histograms}
\end{center}
\end{figure}

QD0 sits inside the detector for the L$^*~=~3.5$\,m design, thus a cantilever is foreseen to isolate it from the vibration environment of the detector. Moreover a large so-called pre-isolator mass, supports QD0, through the cantilever, and QF1 as shown in Fig.~\ref{fig:cantilever}-left. Under this configuration the stability of the systems against ground motion was studied and the sub-nm tolerances specified for the FD were satisfied \cite{Balik2012}. Nevertheless this solution limits the tuning performance to 60\% of $\mathcal{L}_0$, as shown by the blue curves in Fig.~\ref{fig:histograms}. This is due to the uncorrelated motion of the FD with respect to the rest of the beamline, which leads to a luminosity signal that jitters by about 6\%, see~\cite{Snuverink2011}. If instead, the pre-isolator is removed from the systems, the tuning procedure can improve the performance of all machines, bringing 90\% of the machines to $\mathcal{L}\geq\;90\%\;\mathcal{L}_0$, as shown in Fig.~\ref{fig:histograms}, see~\cite{Marin2018} for more details.

In conclusion, the static imperfections do not present an obstacle to tune the CLIC-FFS 3\,TeV baseline design. However due to the complex MDI designed, the tuning feasibility is compromised. It is expected that the longer FFS design with L$^*$~=~6\,m would not suffer from this condition since the FDs sit directly on the tunnel floor and no pre-isolator is required.

\section{CLIC-FFS 3\,TeV with L$^{*}$~=~6\,m}
\begin{figure}[h!]
\centering
\includegraphics[scale=0.3, angle=-90]{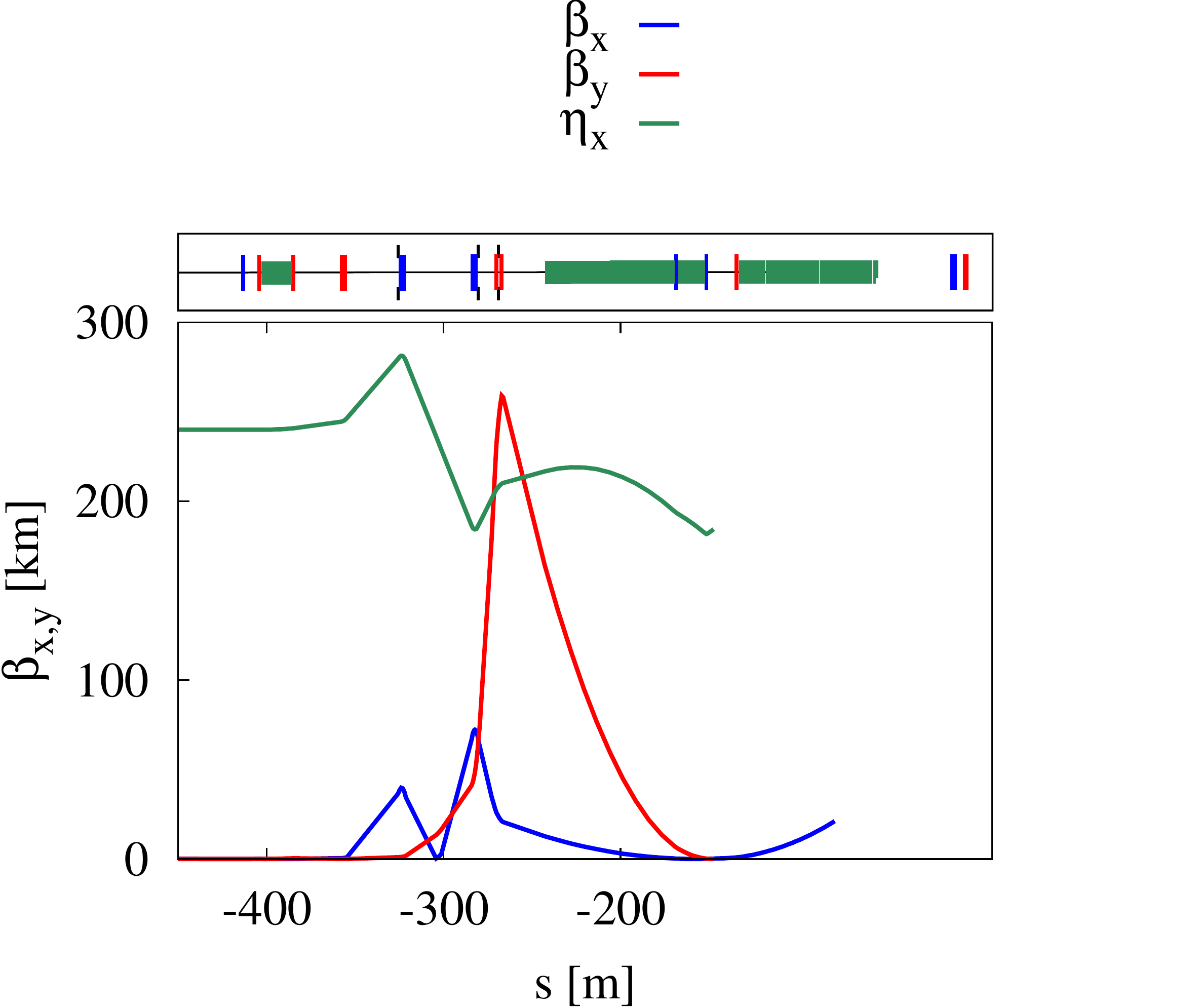}
\includegraphics[scale=0.3, angle=-90]{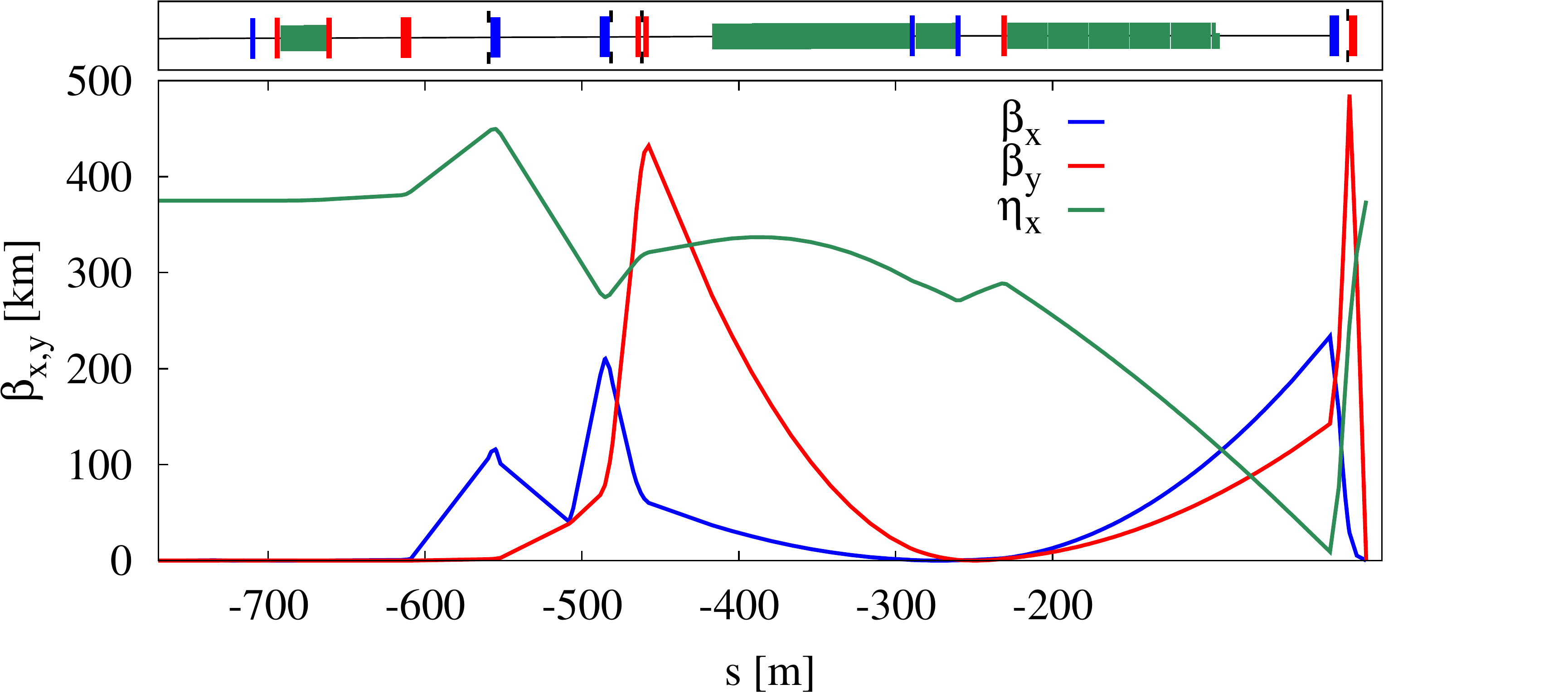}
\caption{\label{twiss2} Beta ($\beta_{x,y}(s)$) and dispersion ($\eta_{x}(s)$) functions through the FFS for L$^{*}$~=~3.5\,m (top) and L$^{*}$~=~6\,m (bottom). The lattice for L$^{*}$~=~6\,m has been lengthened with respect to the increase of L$^{*}$ from the CDR~\cite{Aicheler2012} design.}
\end{figure}

The long L$^{*}$ CLIC-FFS design~\cite{Plassard2018a} is obtained by scaling up the original lattice~\cite{Garcia2014}. The quadrupoles have been retuned in order to match the design optics parameters at the IP (see Table~\ref{beam_param2}). Figure~\ref{twiss2} compares the optical functions of the new FFS with L$^{*}$~=~6\,m lattice with respect to the ones of the baseline design.\\

The optics of the FFS has been optimized in order to achieve luminosities above $\mathcal{L}_0$ in order to allow for some budget for static and dynamic imperfections. For both L$^{*}$ options octupoles and decapoles have been inserted and optimized in order to correct the remaining 3$^{rd}$ and 4$^{th}$ order chromatic and geometric aberrations, minimizing their contribution to the vertical beam size growth. The resulting total luminosities ($\mathcal{L}$) after optimization is $7.6\cdot 10^{34}\,cm^{-2}\,s^{-1}$ and $6.4 \cdot 10^{34}\,cm^{-2}\,s^{-1}$ for the baseline and longer L$^*$ designs respectively. In terms of peak luminosity ($\mathcal{L}_{1\%}$) the baseline and longer designs reach $2.4\cdot 10^{34}\,cm^{-2}\,s^{-1}$ and $2.3\cdot 10^{34}\,cm^{-2}\,s^{-1}$, respectively.

Both $\mathcal{L}$ and $\mathcal{L}_{1\%}$ exceed $\mathcal{L}_0$ by 10\% and 15\% respectively. In terms of energy bandwidth both systems show comparable performances~\cite{Plassard2018a}.

The tuning efficiency of the L$^{*}$~=~6\,m FFS design was studied assuming realistic static error conditions (see Table~\ref{tab:imperfections}-right column). In the present simulations, the luminosity is computed assuming that both the electron and positron FFS are identical, so the same beam distribution at the IP is assigned for both e$^{-}$ and e$^{+}$ beam lines. The tuning simulations show that 85\% of the machines reach more than 110\% of the design luminosity $\mathcal{L}_{0}$ in approximately 6300 luminosity measurements, falling close to the tuning goal. The luminosity obtained for 90\% of the machines, after scanning linear knobs is $\geq$~97\% of $\mathcal{L}_{0}$ which is a 10\%  improvement with respect to the baseline design, when only scanning linear knobs~\cite{Plassard2017}. It is expected that the implementation of 2$^{nd}$-order knobs would further improve the results as observed in the L$^*$=3.5\,m tuning study.

\section{Energy Upgrades with Future Technologies}
\label{sect:Energy_Upgrades}
\subsection{Introduction}

Future novel acceleration technologies have either demonstrated high gradients of 1\,GV/m or more, or have experimental programs aimed at achieving such gradients \cite{Cros2017}. A study has been carried out to ensure that  CLIC is consistent with upgrades using these technologies. In this section ideas are discussed for such upgrades. While there is currently no defined collision energy target for such a future upgrade, in the long term new technology may aid in attaining e$^-$~e$^+$ collisions at tens of TeV centre of mass energy. Preliminary physics studies indicate that the luminosity goals for a 10\,TeV machine may be similar, or higher, than for the 3\,TeV CLIC \cite{Roloff2018}. Therefore the novel technology used for upgrades must meet stringent requirements for efficiency and beam quality preservation.

The novel technologies being developed are either plasma-based or dielectric-based. Both media can be driven either by a particle drive beam or by a laser drive beam \cite{Cros2017}. None of the technologies are sufficiently mature to develop consistent designs for an electron-positron collider, and a number of challenges have to be addressed before they can be considered for such machines, including better understanding of efficiencies and instabilities. Novel accelerator technology for focusing and injection could also potentially improve a collider, and is  briefly discussed below. 

\subsection{Dielectric Structures}

Dielectric structures, with operating frequencies ranging from X-band to THz, are being developed at multiple laboratories \cite{Cros2017}. The structures could in principle be made as hollow dielectric cylinders, potentially easier and cheaper to manufacture than copper structures. One development branch, headed by Argonne National Laboratory, concerns X-band dielectric structures driven by a parallel Drive Beam, thus their mode of operation is similar to CLIC. Such structures could provide a path for an upgrade of the CLIC linac, if superior high-gradient behaviour can be demonstrated, and sufficient control of the beam break up instability can be demonstrated. An experimental program would be very important to complete the understanding of such structures, and a collaboration between CLIC and Argonne is planned to this end. 

It has also been proposed to build colliders using direct laser acceleration in silicon chips - the ``system on a chip'' type technology \cite{England2014}. Although this technology holds the promise for very compact accelerators, the concept and the suggested beam parameters are very different from those of CLIC, and a performance comparison is therefore challenging at this point in time.

\subsection{Plasma Wakefield Acceleration}

Beam driven Plasma Wakefield Accelerators (PWFA) have demonstrated very high gradients, >50\,GV/m \cite{Blumenfeld2007} and good wake-to-witness efficiencies, >30\% \cite{Litos2014}. Open questions remain concerning the Beam-Break Up instability (BBU), in PWFA often referred to as the hosing instability, and its mitigation possibilities. The above gradient and efficiency results were obtained in the so-called PWFA blow-out regime, where the driver is sufficiently dense as to expel all plasma electrons from the axis, and form an ion cavity trailing the Drive Beam. The ion cavity provides very strong, linear focusing fields (MT/m) for electrons, which helps to mitigate the BBU. However, the presence of strong focusing fields implies that the dynamic tolerances, in particular drive-beam jitter, become very tight. For example, assuming the plasma stage parameters in \cite{Adli2013} it is found that the transverse position jitter of the Drive Beam must be less than 1.4\,nm \cite{Schulte2016}. Angular jitter tolerances may be even more severe. Various mitigation techniques, including adjusting cell lengths to compensate kicks, and plasma density ramps \cite{Floettmann2014}, may improve the tolerances by some factor. 

Due to the promise of very high gradient, PWFA may provide a path towards very high energies in an upgraded CLIC linac. Therefore the CLIC project is contributing to understanding and addressing the instabilities and quantifying tolerances for this technology, which will allow an optimized and consistent parameter set for a PWFA-collider to be developed, providing ground for a better assessment of the technology. 

Laser driven plasma wakefield accelerators \cite{Leemans2014} may also be an interesting future technology, however, this would require that the efficiency from the wall plug to the laser beam is increased by an order of magnitude from today's few \%. 

\subsection{Positron Acceleration and $\gamma\gamma$~Colliders}

In dielectric structures, positrons can be accelerated with the same performance as electrons. Plasmas, on the other hand, respond differently to a positron beam than to an electron beam. While high gradient acceleration of positrons has been demonstrated (about 4\,GV/m \cite{Corde2015}), positron acceleration in plasmas has been much less studied than electron acceleration. Currently it is not clear how to perform high-gradient, high-efficiency positron acceleration while also preserving good emittance \cite{Cros2017}. In particular, the PWFA blow-out regime, as discussed above, seems unsuitable for positron acceleration. Another approach is to use hollow plasma channels \cite{Gessner2016} which does not have the strong ion focusing on axis, and can in principle be used to accelerate both positrons and electrons. However, the lack of ion focusing implies a very strong BBU in hollow channels \cite{Lindstrom2018}, and it is not clear how the instability can be mitigated. 

One could imagine a scenario where RF-technology is used to build the first stage of CLIC, and later, plasma technology for electron acceleration has matured sufficiently in time to be used for an energy upgrade, whereas technology for positron acceleration has not. In such a scenario, one could envisage a multi-TeV $\gamma\gamma$-collider as an upgrade. A $\gamma\gamma$-collider requires the acceleration of two electron beams (for example using the PWFA blow-out regime) with two laser beams shone onto the electron beams, just ahead of the collision point. Through inverse Compton-scattering, about 80\% of the initial electron energy can be transferred to laser photons \cite{Telnov1998}. Although $\gamma\gamma$~colliders have been studied in some details for the last decades, there have not been conclusive studies of the physics opportunities in the assumption that the $\gamma\gamma$~collider will be instead of, and not in addition to, a multi-TeV e$^+$e$^-$ collider. Studies for the physics potential of a 3-30\,TeV $\gamma\gamma$~collider, as the only multi-TeV linear collider, are currently underway. 

\subsection{Novel Focusing Technology}

The BDS constitutes a significant fraction of the CLIC 380\,GeV length, 4\,km out of an 11\,km footprint. Future technologies may also be considered for upgrades of this part of the machine. Technologies which are presently being studied are crystal focusing, which can provide focusing in one plane without defocusing in the other plane \cite{Scandale2018}, or active plasma lenses \cite{Tilborg2015}, which can provide focusing in both planes simultaneously. Both technologies have, therefore, the potential for reducing the length, the chromatic errors \cite{Lindstrom2016} and/or the complexity of the BDS.

\subsection{Plasma-Based Injectors}

Plasma-based injectors, producing GeV-range beams from a short plasma cell, have been proposed as injectors into electron linacs, for example the ``plasma photocathode'' \cite{Hidding2012}. According to simulations, GeV-range beams with very low emittance could be produced from such injectors. If beams with sufficient charge and quality can be demonstrated, plasmas could be considered as a compact injector for a future technology machine, replacing the e$^-$ damping ring and booster linacs. 

\subsection{Plasma Afterburners}

It has been proposed to keep the Main Linac of an existing linear collider, and add a plasma stage to boost the energy of the beams exiting the RF Main Linac \cite{Lee2002}. Afterburner concepts as proposed in \cite{Lee2002} imply the use of part of the original main bunch as a driver, likely leading to a significant reduction in luminosity. Keeping this limitation in mind, afterburners could still be interesting to investigate further as a limited-cost upgrade option.

\subsection{Reuse Opportunities of CLIC}

The following features of the CLIC accelerator may be of immediate benefit for a machine upgraded with future technology:
\begin{itemize}
\item The tunnel 
\item The crossing angle of 20\,mrad 
\item The Drive Beam
\item Injection complex
\item Alignment and stability developments
\end{itemize}

\subsubsection{Main Linac Tunnels}

After competition of the physics program at 380\,GeV, the Main Linacs of 2\,x\,3.5\,km could be replaced with a dielectric or a plasma-based Main Linac. Assuming an effective gradient of 1\,GV/m for the future technology \cite{Cros2017}, beams of up to 3.5\,TeV could be produced in the Main Linac, allowing for collisions of up to 7\,TeV, assuming the BDS is upgraded to handle the energy increase.

\subsubsection{Crossing Angle}

The full length of the CLIC tunnels, hosting the Main Linacs and the BDS, are laser straight, crossing at an angle of 20\,mrad \cite{Schulte2001}. The crossing angle is optimal for 3\,TeV centre-of-mass energy collisions, and also likely to be a good choice for higher centre-of-mass energy collisions. Furthermore, the CLIC crossing angle is also compatible with high-energy $\gamma\gamma$ collisions \cite{Telnov2005}. 

\subsubsection{Drive Beam Re-use}

Beam driven plasma wakefield accelerators require high-power electron Drive Beams. For a plasma stage with parameters as suggested in \cite{Adli2013}, with a pulsed time structure of 2\,ns spacing between drive bunches delivered in pulses at 50\,Hz repetition rate it could be possible to reuse parts of the CLIC Drive-Beam Complex to produce appropriately spaced Drive Beams, with limited modifications \cite{Schulte2017}. The Drive Beam Accelerator could be arranged in a recirculating configuration with new structures but using the existing RF power production systems, as an example a 25\,GeV Drive Beam is suggested \cite{Adli2013}. For a lower Drive-Beam energy, for example 10\,GeV, even the structures could be reused.

\subsubsection{Injector Complex}

The CLIC injectors for a 380\,GeV CLIC, providing 9\,GeV low emittance electron and positron beams, could be reused directly to inject into a Main Linac based on future technology, possibly with some modifications to adapt to a time structure and bunch length optimized for the future technology. 

\subsubsection{Alignment and Stability Studies}

Tolerances on alignment and stability, both transverse and longitudinal, will likely be at least an order of magnitude more stringent for linacs based on future technology than for CLIC~\cite{Schulte2016}. Significant effort and resources have been put into developing methods and technology in this domain. A few examples are demonstration of stabilisation of the CLIC final quadrupole to below the nm scale \cite{Janssens2015}, demonstration of phase feedback on the 50\,fs level \cite{Roberts2018}, development of beam-based alignment techniques \cite{Latina2014,Pfingstner2014} and metrology and static alignment techniques \cite{Pacman2014}. These examples highlight that studies for the implementation of CLIC, will advance the state-of-the-art of accelerator stability and alignment, and will also be directly useful for the development of an advanced linear collider based on novel accelerator technology.

\printbibliography[heading=subbibintoc]
\endrefsection
\addtocontents{toc}{\vspace{\normalbaselineskip}}
\newrefsection 
\chapter{Technologies}
\label{Chapter:TECH}
\section{Introduction}
This chapter describes the technological challenges and the solutions implemented in the current design of the accelerator. It starts with the the production of the particles and their journey along the accelerator including the various power production schemes and related technologies. Other essential systems are highlighted and discussed, such as Beam Instrumentation, Vacuum, Survey and Stabilisation as well as general machine governing technologies such as Controls, Timing and Machine Protection. Where significant progress has been made with respect to the CDR \cite{Aicheler2012} these differences are pointed out.  

\section{Sources and Injectors}
\label{sect:Sources}
The injector of the Drive-Beam Accelerator is described together with the Drive-Beam Accelerator in Section~\ref{sect:DBA}. Here we focus on the Main-Beam Injectors which provide polarized electrons and non-polarized positrons. 

\subsection{Polarized Electron Source and Pre-injector Linac}

The CLIC polarized electron source is based on a DC-photo injector followed by a 2\,GHz bunching system and a 200\,MeV accelerator. After the common injector linac a spin rotator at 2.86\,GeV will be used to orient the spin vertically for injection into the damping ring. 
The electron injector can deliver an emittance smaller than 25\,mm\,mrad.
The source is based on demonstrated technologies and installations. The SLC electron source and the ILC design are used as references. The gun test facility at SLAC has shown the parameters needed for the CLIC source using a commercial strained multi-layered GaAs-cathode \cite{Zhou2009}. The cathode needs to be operated in a vacuum environment of the order of 10${}^{-11}$\,mbar. Under such conditions several thousand hours of lifetime have been achieved. The gun has to be equipped with a load-lock system to allow cathode changes and activation under vacuum. Additionally, the gun optics have to be carefully designed to minimize particle losses close to the cathode. 

A polarized laser based on flash-lamp pumped TiSa technology is used to illuminate the Ga-As cathodes. The laser has to be tuneable in frequency to be able to tune each cathode individually for maximum polarization. This tuneable bandwidth of approx. 0.7\,nm is achieved with a quartz plate inside the laser cavity. The system produces a long pulse of several $\mu$s of which a 176\,ns long flat pulse is cut out via a Pockels-cell. Experimental results indicate that a polarization above 80\,\% can typically be achieved. Details of the proposed laser system can be found in the CDR and the main parameters of the CLIC polarized electron source can be found in Table~\ref{Tab_SRC_1}.

\begin{table}[htb!]
\centering
\caption{Laser and electron parameters for the polarized electron source and comparison with experimentally achieved results.}
\label{Tab_SRC_1}
\begin{tabular}{l c c} 
\toprule
~ &  CLIC 1\,GHz & CLIC DC/SLAC Demo \\ 
\midrule
\textbf{Electrons} \\ 
\midrule
Number of electrons per bunch (10$^{9}$)  & 6 & 1365 \\  
Charge/single bunch (nC)  & 0.96 & NA \\  
Charge/macrobunch (nC)  & 338 & 300 \\  
Bunch spacing (ns)  & 0.5 & DC \\  
RF frequeny (GHz)  & 2 & DC \\  
Bunch length at cathode (ps)  & 100 & DC \\ 
Number of bunches  & 352 & NA \\ 
Repetition rate (Hz) & 50 & 50 \\ 
QE (\%)  & 0.3 & 0.3 \\ 
Polarization  & $>$80\,\% & $>$80\,\% \\  
Circular polarization  & $>$99\,\% & $>$99\,\% \\ 
\midrule 
\textbf{Laser} \\ 
\midrule
Laser wavelength (nm) & 780-880 & 865 \\ 
Energy/micropulse on cathode (nJ) & 320 & --- \\ 
Energy/macropulse on cathode ($\mu$J) & 113 & 190 \\  
Energy/micropulse laser room (nJ) & 1000 & --- \\ 
Energy/macrop. Laser room ($\mu$J) & 350 & 633 \\  
Peak power  per pulse (kW) & 2 & 2 \\  
Average power at cathode wavelength (mW) & 8 & 9.5 \\ 
\bottomrule 
\end{tabular}
\end{table}

\begin{figure}[htb!]
\centering
\includegraphics[width=0.98\textwidth]{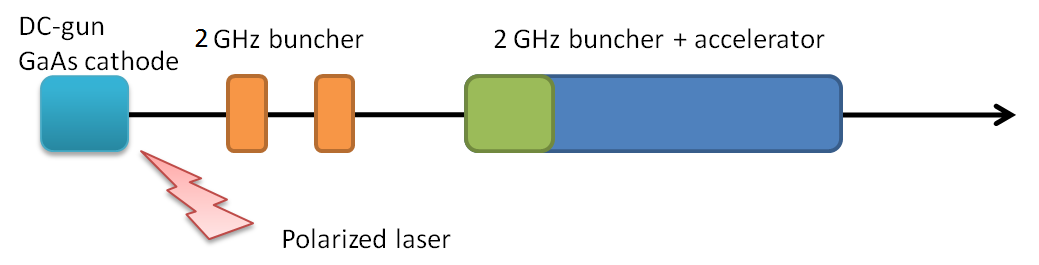}
\caption{\label{fig_SRC_1}Schematic view of the polarized electron source and pre-injector.}
\end{figure}

The 176\,ns-long electron pulse produced at the cathode will be pre-bunched with a pair of 2\,GHz standing wave buncher cavities followed by a five-cell travelling wave buncher. After that, 2\,GHz accelerating structures are used to bring up the energy to 200\,MeV for injection into the common injector linac. The performance of such a system was simulated and showed an 88\% capture efficiency with an emittance of 22\,mm\,mrad \cite{Zhou2009a}. A schematic of the pre-injector for the polarized electrons can be seen in Fig.~\ref{fig_SRC_1}.

\subsection{Positron Source}

\begin{figure}[htb!]
\centering
\includegraphics[width=0.6\textwidth]{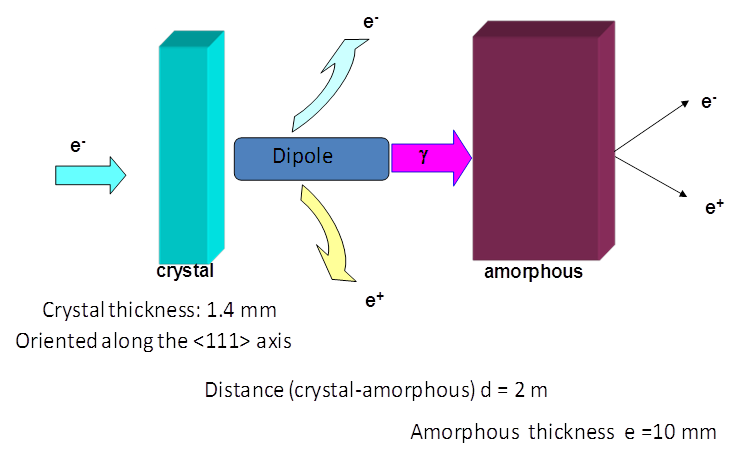}
\caption{\label{fig_SRC_2}The CLIC positron source based on a hybrid target.}
\end{figure}

The CLIC positron source consists of a 5\,GeV electron linac, a hybrid target followed by an adiabatic matching device and a capture pre-injector linac. The hybrid target uses a thin tungsten crystal to generate an enhanced photon flux taking advantage of the channelling effect. After the first target, a dipole magnet sweeps away charged particles and ideally only the $\gamma$-ray beam hits the amorphous target which generates positrons via pair-production. The advantage of this scheme is that it minimizes the peak power deposition density onto the production target enhancing its lifetime. Past experience at SLAC determined a destruction threshold of 35\,J/g which should not be exceeded in order to guarantee a lifetime of 1000 days. The newly optimised design for 380\,GeV has a peak power deposition density of 23\,J/g. Therefore a second parallel target station, as described in the CDR for the 500\,GeV case, is not needed. The thickness of the targets and their separation have been optimized together with the incoming beam parameters to maximise the positron yield \cite{Han2018} and can be seen in a schematic view in Fig.~\ref{fig_SRC_2}. Downstream of the target an adiabatic matching device (AMD) focuses the positrons into the capture linac. The simulations assume a 20\,cm long device with an aperture of 20\,mm radius and with a magnetic field decaying from 6\,T to 0.5\,T (B\,=\,B${}_{0}$\,/\,(1\,+\,$\alpha$z), with B${}_{0}$\,=\,6\,T and $\alpha$\,=\,55\,m${}^{-1}$).

The positron pre-injector linac and the common injector linac have been completely redesigned relative to the CDR. The yield of positrons fitting into the pre-damping ring energy acceptance has been increased by a factor 2.5 and the longitudinal phase space has been improved \cite{Bayar2017}. The positron yield after the target is 8\,e${}^{+}$/e${}^{-}$ and the yield at the entrance of the pre-damping ring reaches 1\,e${}^{+}$/e${}^{-}$.

These results allow for lowering the current of the 5\,GeV linac leading to substantial cost savings. In addition, it is possible now to work with a single positron production target. The positron driver linac has a length of 350\,m and uses a 2\,GHz RF system with an accelerating gradient of 20\,MV/m.

The beam optics of the common injector linac has been designed for the large emittance positron beam. A sequence of FODO lattices have been used with increasing beta-function and quadrupole spacing to minimize the number of magnets used. This linac is approximately 250\,m long and uses a 2\,GHz accelerating structure with a  loaded gradient of 18\,MV/m.

\subsection{Injector RF system}

The whole injector complex use a 2\,GHz RF system for the linear accelerators. The generic unit consists of a pair of klystrons with a peak power of 50\,MW and a pulse length of 8\,$\mu$s at 50\,Hz repetition rate and a cavity-type pulse compressor. This combination allows the generation of one or a pair of RF pulses with a flat top and amplitude modulation for beam loading compensation. The pair of klystrons provides power for three or four accelerating structures installed on a common girder together with focusing and beam diagnostics elements. The injector linac and the booster linac use a double RF pulse spaced by 3.4\,$\mu$s to accelerate positrons and electrons.

An accelerating structure has been designed with a length of 1.5\,m which will be operated at a gradient between 15\,MV/m and 20\,MV/m depending on the specific configuration for each linac. A shorter accelerating structure is needed due to the strong beam loading in the positron capture section. More details about the RF system can be found in the corresponding sections of this document.

\section{The Klystron and Drive-Beam Modules}
\label{sect:Technology_Modules}
\subsection{Introduction}

The two Main Linacs take the electron and positron beams from 9\,GeV and accelerate them to the collision energy. In the 380\,GeV stage the opportunity is given to study two different accelerator configurations, based on two quite different technologies to feed the Main Linacs with the RF power required for the beam acceleration. Both have advantages in different areas at roughly similar cost. The optimisation work that has been performed for the two configurations independently has produced two very different layouts \cite{Vamvakas2018}, the differences going down to the level of the single accelerating structures. The main components of the accelerators are the RF accelerating structures, which are mechanically assembled two-by-two to form the super-accelerating structures (SAS). They are used to accelerate the beams and are assembled on supporting girders to form what we call the ``modules''. Modules will be pre-assembled on the surface and then installed in the machine tunnel, so they represent the fundamental building block for the construction of the accelerator, thousands of these are required even at the lowest collision energy.

The implications coming from the differences between the two configurations are evident in the module architecture and are reflected in the accelerator infrastructure. The module nomenclature uses the terms
T0, T1 etc. to refer to the number of SAS that are installed on the Main-Beam girder. The Main Beam Quadrupole (MBQ) is not part of the CLIC module and it requires a dedicated support and stabilization system. It will be treated separately (see Section~\ref{sect:Magnets}).

\subsection{The Drive-Beam-based Module}

\begin{figure}[hbt!]
\centering
\includegraphics[width=0.9\textwidth]{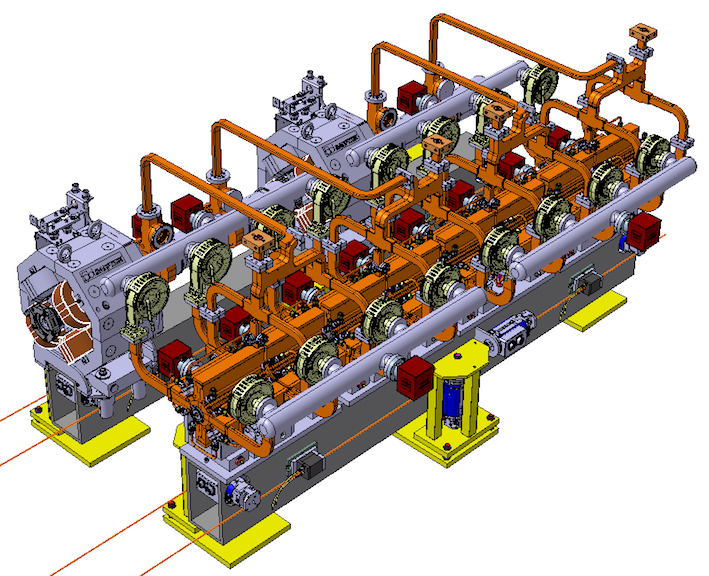}
\caption{The Module in the Drive Beam based configuration: the Main Beam girder (right) and the Drive Beam girder (left).}
\label{fig:DB_Mod}
\end{figure}

In the Drive Beam (DB) configuration, the RF power is transferred to the Main Linac by the Power Extraction and Transfer Structures (PETS) that are aligned in the Drive-Beam Decelerator, which runs parallel, at a distance of a few cm, to the Main Linac. Each PETS supplies one SAS and there can be up to four PETS in one module. In the case of the DB configuration, we consider as one module the assembly of the Main Linac and Drive-Beam Linac girders, since the two are interconnected by the waveguides transferring the RF power from the Drive beam to the Main beam.
The DB-based module is represented in Fig.~\ref{fig:DB_Mod}. 

The optimisation of the accelerating structure geometry, from the point of view of the RF efficiency and of the beam wakefield tolerance, has generated RF structures that are made of 33 cells each and have a length of 0.271\,m (see Section~\ref{sect:PETS}). In the case of the DB-based accelerator the module length is defined by the periodicity of the Drive Beam lattice, so by the interval between Drive Beam Quadrupoles (DBQ). The length of a T0 module including four SAS is 2.343\,m. By considering the fraction of RF power entering the accelerating structure that goes into the beam acceleration the RF-to-beam efficiency of the accelerating structures is 39.8\,\%; 59.7\,MW peak RF power is required in a pulse length of 244\,ns to accelerate the beam with an average gradient of 72\,MV/m. In a T0 module with four SAS the average energy gain per module is 156\,MeV.

A total number of 2,976 modules are required for the two Main Linacs, which will be filled with 10,184 RF structures; the SAS are on the Main Beam side and the PETS on the Drive Beam side. In the  fabrication tests it has been shown that connecting two accelerating structures to form a SAS represents a challenge for the correct alignment of the two structures, which must be within 14\,$\mu$m R.M.S. with respect to the beam axis. It is our intention to directly assemble the SAS during the fabrication phase of the accelerating structures, when the final alignment can be better kept under control.

The main characteristics of the module in the DB configuration have been summarised in Table~\ref{tab:DBMod}.

\begin{table}[htb!]
\centering
\caption{A summary of the module characteristics in the Klystron-based accelerator}
\label{tab:DBMod}
\begin{tabular}{lc}
\toprule
\textbf{Module and RF Structure Characteristics} & \textbf{Values}\\
\midrule
Cells / structure                       & 33 \\
SAS length [m]                          & 0.542 \\
SAS / T0 module                         & 4 \\
SAS / T1 module                         & 3 \\
SAS / T2 module                         & 2 \\
Average accelerating gradient [MV/m]    & 72 \\
Input power / structure [MW]            & 59.7 \\
RF Frequency [GHz]                      & 12 \\
Module length [m]                       &  2.343 \\
DBQ / module                            & 2 \\
\bottomrule
\end{tabular}
\end{table}

A detailed description of the DB Quadrupoles and their powering architecture can be found in Section~\ref{sect:Magnets}. The alignment of the SAS on the supporting girder relies on adjustable supports, the overall alignment strategy for the DB-based accelerator is described in Section~\ref{sect:Survey}. The supporting girder must provide sufficient rigidity and stability. Two materials are being considered, structural steel and Epument, a mineral cast material. The measured static deformation under maximum load is about 15\,$\mu$m in the case of Epument compared to about 35\,$\mu$m for structural steel; a choice between the two materials has not been made yet, however cost considerations and simplicity of fabrication would indicate a preference for the structural steel option.

\begin{figure}[hbt!]
\centering
\includegraphics[width=0.7\textwidth]{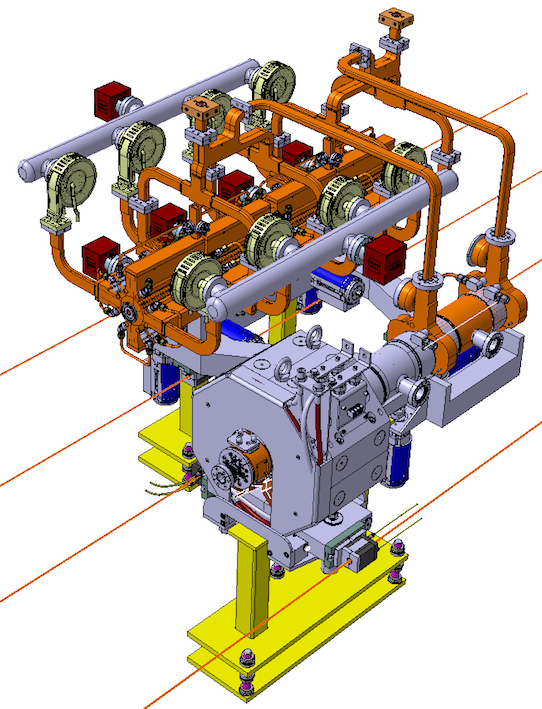}
\caption{The RF Unit assembly with PETS and SAS connected to a common supporting frame. The DBQs sit on independent supports.}
\label{fig:RF-Unit}
\end{figure}

The fact that the PETS and the SAS are tightly connected by waveguides for the RF power transfer has also led to the idea of considering the two components as a single unit with a lighter way of assembling the DB-based accelerator, as shown in Fig.\,\ref{fig:RF-Unit}. In this scheme the DBQs would be supported separately, also in consideration of the importance that their alignment has on the DB performance. For the time being this solution remains an option that will be further explored.

As opposed to what was established in the CDR, the vacuum system will  rely on individual mini-pumps connected to each PETS and SAS, providing ion and getter pumping in the same device; additional pumping will be provided by vacuum pumps connected to the compact spiral loads and to the waveguides. The choice of suppressing the common pumping manifold between the MB and DB sides of the module \cite{Aicheler2012} descends from the need to reduce the mechanical coupling and the forces that are exerted between the two girders. A detailed description of the vacuum system is provided in Section~\ref{sect:Vacuum}.

Due to the tight mechanical tolerances that strongly impact on the Main Beam quality, the thermal stabilisation of the module assumes a critical operational role; most of the power that is dissipated in the module comes from the RF structures and from the RF loads. An extensive study had already been performed at the time of the CDR and here power figures have been reviewed in light of improved efficiencies. In Table~\ref{tab:DBPow} a summary of the power dissipated in a T0 module is provided, in the most unfavourable condition of unloaded operation.

\begin{table}[htb!]
\centering
\caption{Power dissipation in the DB module under unloaded conditions (case of a T0 module)}
\label{tab:DBPow}
\begin{tabular}{p{5cm}rrr}
\hline
Module Component & Number & Item Dissipation [W]   & Total Dissipation [W]\\
\hline
SAS    &  4      &  772     & 3,088 \\
PETS    &  4      &  11     & 42 \\
SAS RF Loads    & 16       & 168      & 2,690 \\
Waveguides    &  4      &  60     & 241 \\
DBQs    &  2      & 171      & 342 \\
\hline
Total per module      &        &      & 6,403 \\
\hline
\end{tabular}
\end{table}

A compact spiral load has been specifically developed to dissipate the power transmitted through the accelerating structures. It consists of a 3D\,printed spiral waveguide built by Electron Beam Melting (EBM) additive manufacturing using pure titanium. It provides -40\,dB attenuation at 12\,GHz, It can manage up to 200\,W average power dissipation and withstand up to 16\,MW peak power \cite{DAlessandro2015}.

\subsection{The Klystron-based Module}

If a Klystron-based design is adopted for the 380\,GeV stage, the acceleration of the two colliding beams, from the injection energy of 9\,GeV to 190\,GeV, is provided by a sequence of 1,456 modules for each linac, each module being fed by a dedicated RF power station. The modules are assembled in a FODO lattice with quadrupoles of two different lengths, 0.4\,m and 0.65\,m, following the beam energy increase along the length of the accelerator. A total of 724 short quadrupoles and 452 long quadrupoles are required for the two linacs.

In the Klystron-based design, a single type of module is considered; the module includes a common support for eight RF accelerating structures, which are coupled two-by-two to form SAS. The total number of SAS required for the two linacs is 11,648. The quadrupoles in the focusing channel are mechanically independent of the module itself, being equipped with a dedicated support that, in addition, integrates a stabilisation system capable of assuring the compensation of ground vibrations. Since those vibrations can strongly affect the beam luminosity at the IP, the stabilisation system must guarantee a quadrupole stability of the order of 1\,nm.

\begin{figure}[hbtp]
\begin{center}
\includegraphics[width=10 cm]{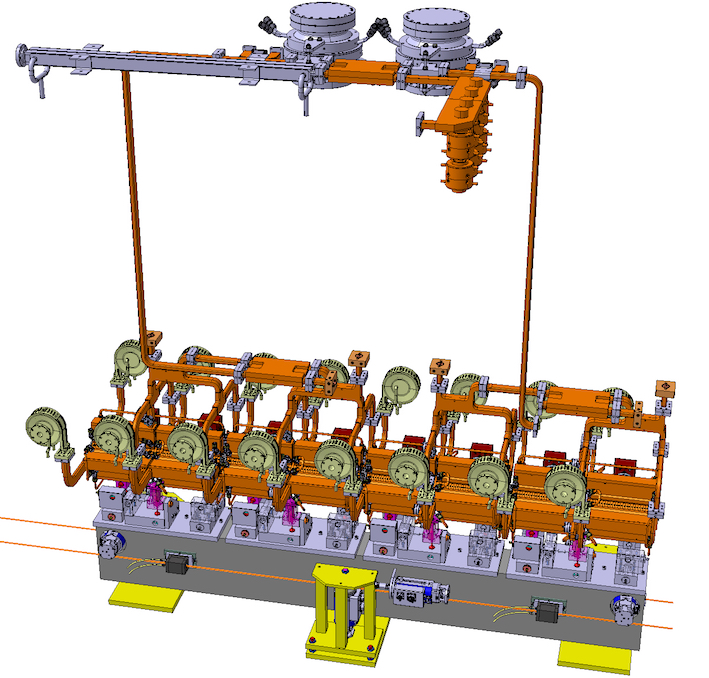}
\caption{The klystron-based module with pulse compression and linearisation system}
\label{fig:k-module}
\end{center}
\end{figure}

An image of the Klystron-based module is shown in Fig.~\ref{fig:k-module}; most RF components used for this module have been built and are being tested in the X-band test facilities under nominal power conditions.

The accelerating structures are 0.23\,m long and require a peak RF power 40.6\,MW in the pulse length of 334\,ns (see Section~\ref{sect:PETS}). The choice of the module length is defined by the RF power available from a modulator station and how that power can be distributed to the accelerating structures. A two-pack modulator equipped with two 53\,MW klystrons can feed one module made of eight accelerating structures. If the design accelerating gradient of 75\,MV/m is reached in the RF structures, the average energy gain per module is about 135\,MeV. The module length is 2.01\,m. The powering system efficiency, from plug to RF delivered to the accelerating structures, is 22.2\,$\%$.

In a machine where the main source of power consumption is the RF losses in the copper walls of the accelerating structures, the quest for efficiency represents a crucial task in the design exercise. This starts from the design of the RF structure cells, which has increased the RF efficiency to 38.2\,\%. The same attention has been given to the design of the RF network. This has led to the adoption of double-height waveguides, which are effective in reducing the attenuation by 40\,$\%$ and electric and magnetic fields by 30\,\% with respect to standard-height waveguides. Table~\ref{tab:BRF} gives a summary of the beam characteristics and the RF structures for the klystron-based Main Linac.

\begin{table}[h]
\begin{center}
\caption{Summary of beam and RF structure characteristics in the klystron-based accelerator.}
\label{tab:BRF}
\begin{tabular}{lcc}
\toprule
\textbf{Beam Parameters}                & \textbf{Value}\\
\midrule
Particles / bunch [x10$^{9}$]           & 3.87 \\
Bunches / train                         & 485 \\
Train length [ns]                       & 242 \\
Bunch separation [ns]                   & 0.5 \\
\midrule
\textbf{RF Structure Characteristics}   &\textbf{Value}\\
\midrule
Cells / structure                       & 28 \\
Structures / module                     & 8 \\
Accelerating gradient [MV/m]            & 75 \\
Structure input power [MW]              & 40.6 \\
Module input power [MW]                 & 340 \\
RF Frequency [GHz]                      & 12 \\
\bottomrule
\end{tabular}
\end{center}
\end{table}

Due to the intrinsic inefficiency of the pulse compression process the power dissipation in the Main-Linac tunnel is increased in the klystron-based module configuration, with respect to the DB accelerator, and is summarised in Table~\ref{tab:KPow}.

\begin{table}[htb!]
\centering
\caption{Power dissipation in the klystron-based module under loaded conditions}
\label{tab:KPow}
\begin{tabular}{lccc}
\toprule
Module Component & Number & Item Dissipation [W]   & Total Dissipation [W]\\
\midrule
SAS    &  4      &  1,100     & 4,400 \\
SAS RF Loads    & 16       & 178      & 2,842 \\
Waveguides    &  1      &  1,034     & 1,034 \\
\hline
Total per module      &        &      & 8,276 \\
\bottomrule
\end{tabular}
\end{table}

For the Klystron-based configuration, a technical complication and an important contribution to the total cost of the accelerator arises from the need to have a parallel gallery to the Main-Linac tunnel hosting the klystrons and modulators. An extensive study compared the two configurations and the technical details of this study can be found in Section~\ref{ssect:Main_Tunnel_xsect}. 

\section{Pulse Compression System for the Klystron-Based Option}
\label{sect:Pulse_Comp}

\begin{figure}[htb!]
\centering
\includegraphics[width=0.5\textwidth]{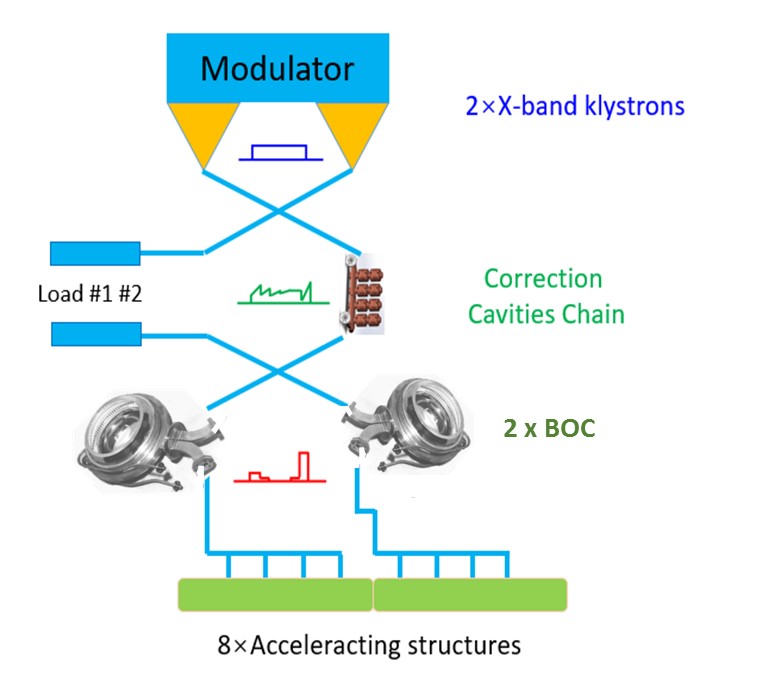}
\caption{\label{fig_Pulse_1} The layout of RF unit of the klystron-based CLIC Main Linac.}
\end{figure}

The layout of an RF unit of the klystron-based Main Linac is shown in Fig.~\ref{fig_Pulse_1}. The two X-band klystrons, mounted on a single HV modulator, produce long RF pulses. These pulses are combined on the 3\,dB hybrid into a single pulse with doubled RF power. The combined RF power is conveyed via waveguide to the Correction Cavities Chain (CCC). The CCC pre-modulates the shape of RF pulse and then splits it in two with each sharing half the power. These pulses feed the two BOC-type RF pulse compressors \cite{Yu1992}. Each BOC shortens the pulse and amplifies the peak RF power. This arrangement is chosen in order to reduce the maximal peak power that has to be transported in a single waveguide system. Finally, these pulses are split and distributed to the eight accelerating structures.

\begin{figure}[htb!]
\centering
\includegraphics[width=0.9\textwidth]{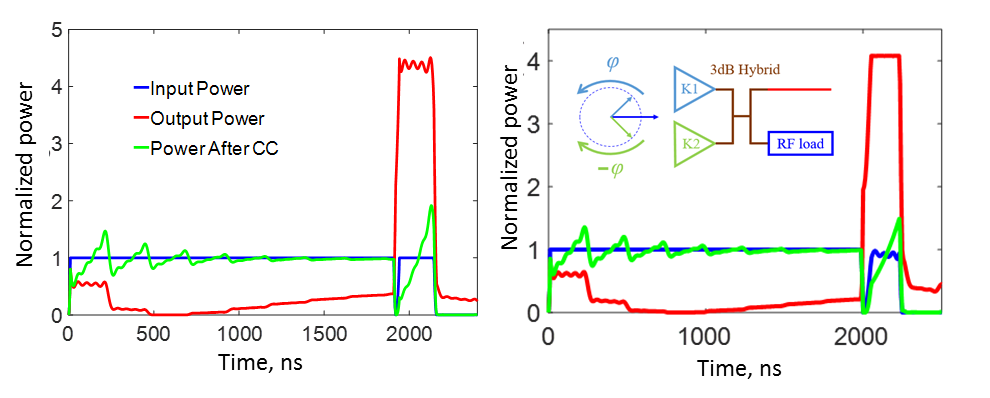}
\caption{\label{fig_Pulse_2} The RF pulse envelopes in a system without (left) and with (right) special RF phase modulation of the klystron pair.}
\end{figure}

The combination of the BOC single cavity pulse compressor and CCC allows one to generate efficiently the RF pulses with a flat top \cite{Zennaro2013}. This system is very compact compared to the SLED\,II, as it does not requires the long delay lines for the RF pulse-forming network. The use of the CCC with its limited number of elements, however, provides the pulses with a small residual modulation on the flat top (see Fig.~\ref{fig_Pulse_2}, left). To mitigate this effect, and to enable the required initial ramping of the RF pulse for the beam loading compensation, special RF phase modulation is applied to the klystron pair (see Fig.~\ref{fig_Pulse_2}, right).

\begin{figure}[htb!]
\centering
\includegraphics[width=0.9\textwidth]{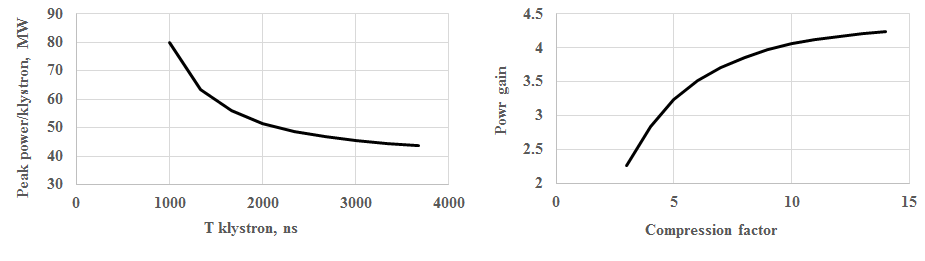}
\caption{\label{fig_Pulse_3} Power gain vs. compression factor (left) and klystron power / pulse length curve (right).}
\end{figure}

The power gain in such a system depends on the frequency, on the quality factor of the BOC storage cavity, on the duration of the output pulse and on the compression factor -- the ratio between input and output pulse lengths. In Fig.~\ref{fig_Pulse_3} (left), the power gain curve simulated for the system with Q${}_{BOC}$~=~2.0~x~10${}^{5}$ and Q${}_{CCC}$~=~4.5~x~10${}^{4}$ for 334\,ns compressed pulses as a function of the compression factor is shown. For four accelerating structure per klystron and 40.6\,MW per structure, anticipating 90\,\% efficiency of the waveguide RF transmission circuit, this curve can be translated into  one which connects the required peak power and pulse length of the individual klystron (Fig.~\ref{fig_Pulse_3} right and Table~\ref{tab_pulse1}.). Following these simulations, the most comfortable parameters for the high-power klystrons are 51.4\,MW peak RF power and 2\,$\mu$s pulse length.

\begin{table}
\centering
\caption{The klystron peak power and pulse length adopted for the CLIC accelerator unit.}
\label{tab_pulse1}
\begin{tabular}{lccccccc} 
\toprule
Pulse\,(ns) & 1,002 & 1,336 & 1,670 & 2,004 & 2,338 & 2,672 & 3,006 \\
Power\,(MW) & 79.8 & 63.5 & 55.9 & 51.4 & 48.6 & 46.7 & 45.5 \\ 
\bottomrule
\end{tabular}
\end{table}

\begin{figure}[htb!]
\centering
\includegraphics[width=0.8\textwidth]{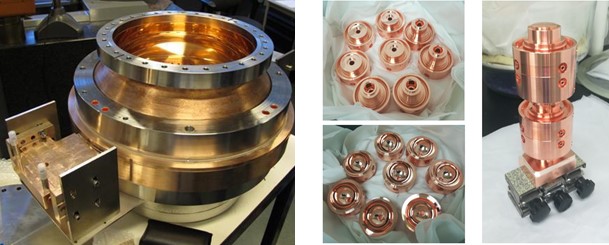}
\caption{\label{fig_Pulse_4} C-band BOC fabricated at PSI shown is left. CCC cavities half shells and one of four assembled CCC cavities doublet are shown right.}
\end{figure}

The X-band version of BOC RF pulse compressor is now under development through a collaboration between PSI and CERN. At PSI a 5.7\,GHz version of such a system has already been built (see Fig.\,\ref{fig_Pulse_4}, left) and tested \cite{Wang2017}. An X-band CCC unit has been built at Tsinghua University in China. In Fig.\,\ref{fig_Pulse_4} (right) the CCC cavities half shells and one of four assembled CCC cavity doublets are shown. The complete CCC unit will be installed and tested at CERN at high RF power early in 2019.

\section{Klystrons and Modulators}
\label{sect:Klystrons}
\subsection{Drive Beam RF Power}

The Drive-Beam based solutions for CLIC require High Voltage klystron modulators to feed several hundred klystrons. For the CDR \cite{Aicheler2012} different options of RF power sources and klystron configurations were studied. It was found that the most reliable and inexpensive solution would be a multi-beam klystron (MBK) with an output power in the range of 15 to 20\,MW \cite{Aicheler2012}. Specifications for the klystron and modulator are summarized in Table~\ref{Tab:specs}.

In addition, the amplitude and phase of the RF output power has to be controlled to $\pm$0.2\,\% and $\pm$0.05\,degrees respectively, putting constraints on all equipment involved.

\begin{table}[htb!]
\centering
\caption{\label{Tab:specs} Specifications for the klystron and modulator for the Drive Beam.}
\begin{tabular}{lccc} 
\toprule 
\textbf{Main parameter} & \textbf{380 GeV} & \textbf{1.5 TeV} & \textbf{3 TeV} \\
\midrule
Frequency [MHz]                 & 999.516       & 999.516   & 999.516\\ 
Number of Klystrons             & 446           & 540       & 1080 \\  
RF Peak Power {MW}              & 20            & 20        & 20\\ 
RF Average Power [kW]           & 48            & 148       & 148\\ 
Pulse Length [$\mu$s]           & 48            & 148       & 148\\
Pulse Repetition Rate [Hz]      & 50            &  50       & 50 \\ 
Klystron Efficiency [\%]        & 70            & 70        & 70\\ 
Klystron Perveance [$\mu$perve] & 2.8           & 2.8       & 2.8\\
Modulator Voltage [kV]          &160-170        & 160-170   & 160-170\\
Peak Pulse Current [A]          & 180           & 180       & 180\\ 
Modulator Average Power [kW]    & 84            & 244       & 244\\
Klystron Solenoid Average Power [kW] & 5        & 5         & 5\\ 
Klystron Heater Power [kW]      & 1             & 1         & 1\\ 
Average Power Consumption [MW]  & 40.1          & 135       & 270\\
Flat-Top Stability [\%]   & 0.1           & 0.1       & 0.1\\ 
Pulse-to-Pulse Repeatability [ppm]& 50    & 50        & 50\\
Rise and fall times (t${}_{rise}$, t${}_{fall}$) [$\mu$s] & 3,3 & 3,3 & 3,3\\ 
Voltage Settling Time (t${}_{set}$) [$\mu$s]    & 5  & 5 & 5\\
\bottomrule
\end{tabular}
\end{table}

\subsubsection{Klystrons for the Drive Beam}

In the Two-Beam version, the Drive Beam is the main power consumer of the facility; therefore, the efficiency of the RF power source is extremely important. For 380\,GeV more than 9\,GW of peak power and an average power of 22\,MW are needed at an RF frequency of 1\,GHz. Two high efficiency klystron prototypes have been developed together with industry with the specifications shown in Table~\ref{Tab:specs}.

\begin{figure}[!htb]
  \centering
  \includegraphics[width=0.4\textwidth]{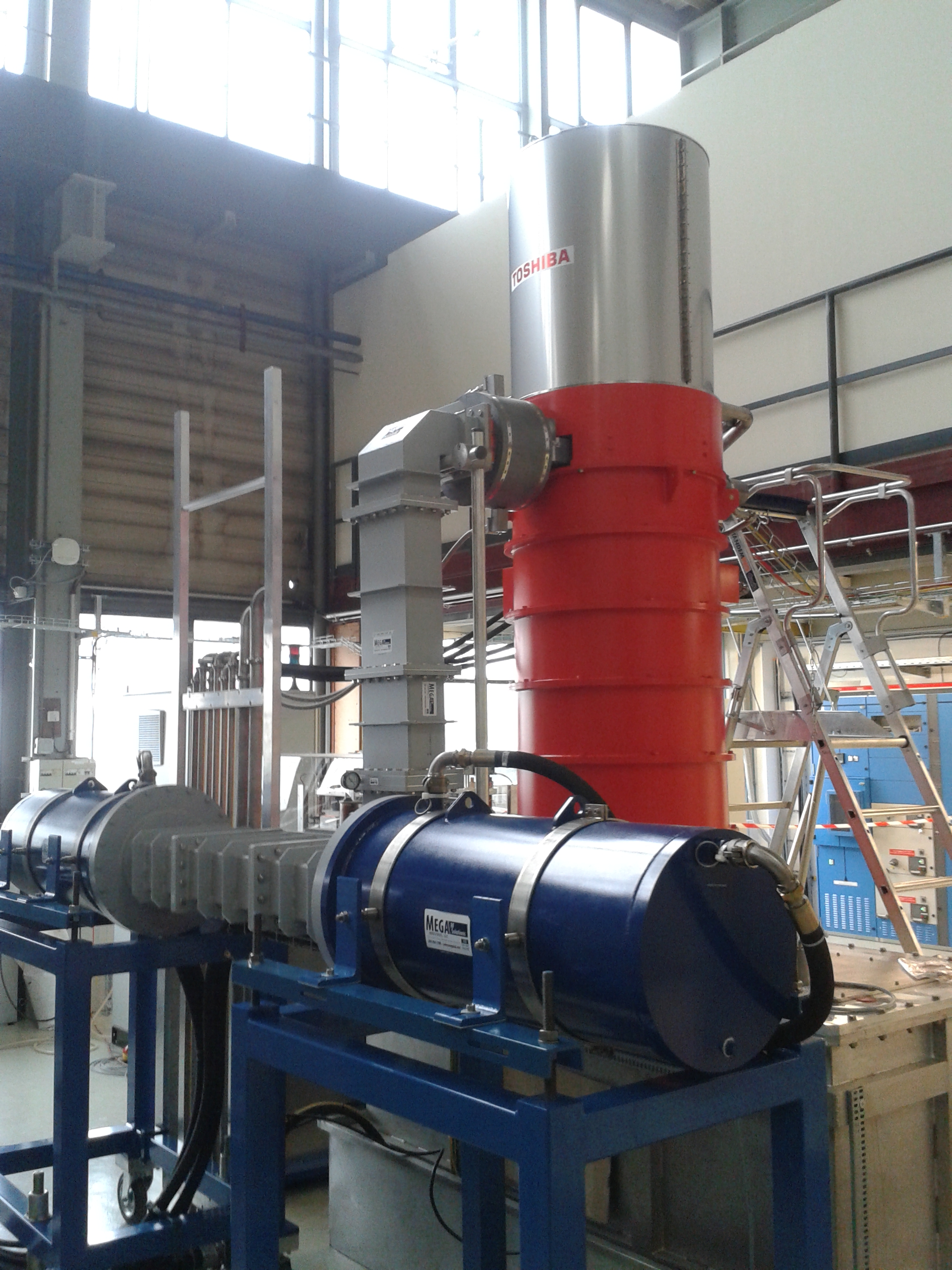}
  \includegraphics[width=0.58\textwidth]{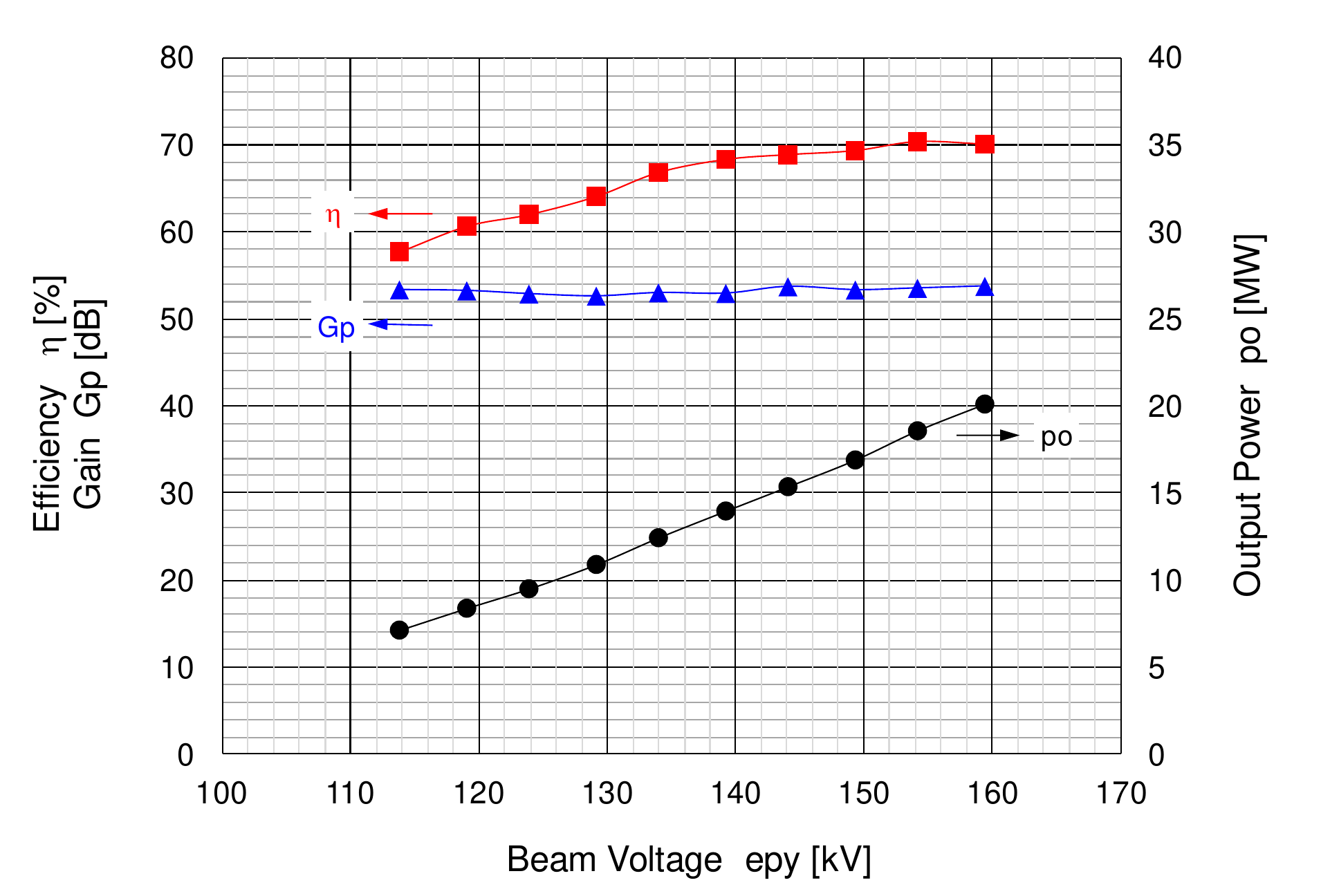}  \caption{Left: The 1\,GHz klystron test stand at CERN with the Toshiba klystron installed; Right: Efficiency, Output power and gain as a function of voltage for the Toshiba MBK.}
  \label{fig:Klystrons_1}
\end{figure}

The first prototype was made by Toshiba, using a six-beam Multi Beam Klystron (MBK). The tube reached in factory tests 21\,MW output power and an efficiency of 71.5\,\% fulfilling all the specifications. Fig.~\ref{fig:Klystrons_1} (right) shows the measurements of the klystron gain, efficiency and output power as a function of voltage. The efficiency remains remarkably high for a wide range of output power. A second prototype has been built by Thales. Here the choice of a ten-beam MBK was made. This tube also reached the required peak power and an efficiency of 73\,\%  but could not yet fulfil the average power and stability requirements during factory tests. Extensive testing of both tubes will continue in a dedicated test stand at CERN which can be seen in Fig.~\ref{fig:Klystrons_1} (left).

These two innovative prototype MBK commissioned by the CLIC study represent the state of the art of what is available in industry and reach record efficiencies. Based on this experience a large-scale production of such klystrons can be mapped out. Detailed testing of the complete RF power unit including the modulator and the low-level RF system will have to demonstrate the required stability.

\subsubsection{Modulator for the Drive Beam}

\begin{figure}[!htb]
  \centering
  \includegraphics[width=0.98\textwidth]{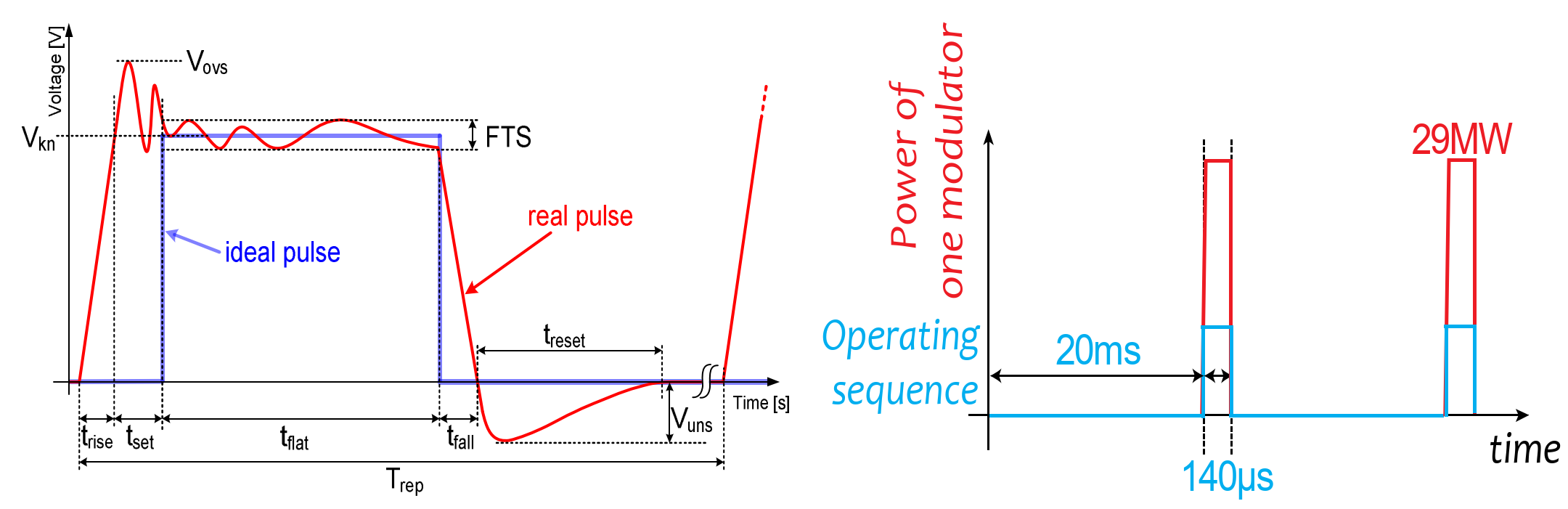}
  \caption{Left: voltage pulse definition; Right: 3\,TeV pulsed power illustration (one modulator).}
  \label{fig:Klystrons_2}
\end{figure}

Ther modulator requirements shown in Table~\ref{Tab:specs} fall in an unexplored range where specifications of the fast pulse modulators (voltage fast rise and fall times to minimize power losses) and long pulse modulators (long voltage flat-top) have to be merged. See Fig.~\ref{fig:Klystrons_2} (right) for pulse definitions \cite{Aicheler2012}.

From the global modulator efficiency of 95\,\%, which includes the electrical efficiency and ``pulse efficiency'', defined as the ratio between the ideal and real voltage pulses energies (Fig.~\ref{fig:Klystrons_2} (left)) pulse rise, fall and settling times have been defined, which shall be minimized to reduce klystron losses. The total power fluctuation (all modulators synchronously operated) is 38\,GW (for the 3\,TeV case). To obtain a stable voltage distribution network it has been decided to limit the AC power fluctuation to 1\,\%. Details on design challenges are summarized in \cite{Aguglia2011}.

\subsection{General Klystron Modulator configuration}

The search for a suitable klystron modulator configuration has been approached by considering the high power electrical distribution over an approx. 2\,km long DB. Global optimization studies \cite{Jankovic2013}, considering overall infrastructure cost, operability (modularity), and efficiency, clearly show the need for a medium voltage distribution. The final solution for the 3\,TeV option consists of six powering sectors, each with a powering substation, equally distributed along the DB length.  Each substation is composed of a set of AC power transformers, AC switchgears and AC bus bars to provide a 10\,kV AC network with n+1 redundancy. The 10\,kV AC network supplies a modular AC to DC converter, a so-called Modular Multilevel Converter (MMC), which provides 20\,kV DC voltage \cite{Jankovic2014}. A novel MMC control has been developed to respect the 1\,\% power fluctuation without the need to synchronize the CLIC RF to the utility grid. All voltage levels result from the global optimization approach in the 3\,TeV stage. Each of the six centralized AC to DC converters supplies approx.~180 modulator-klystron sets.

\begin{figure}[!htb]
  \centering
  \includegraphics[width=0.8\textwidth]{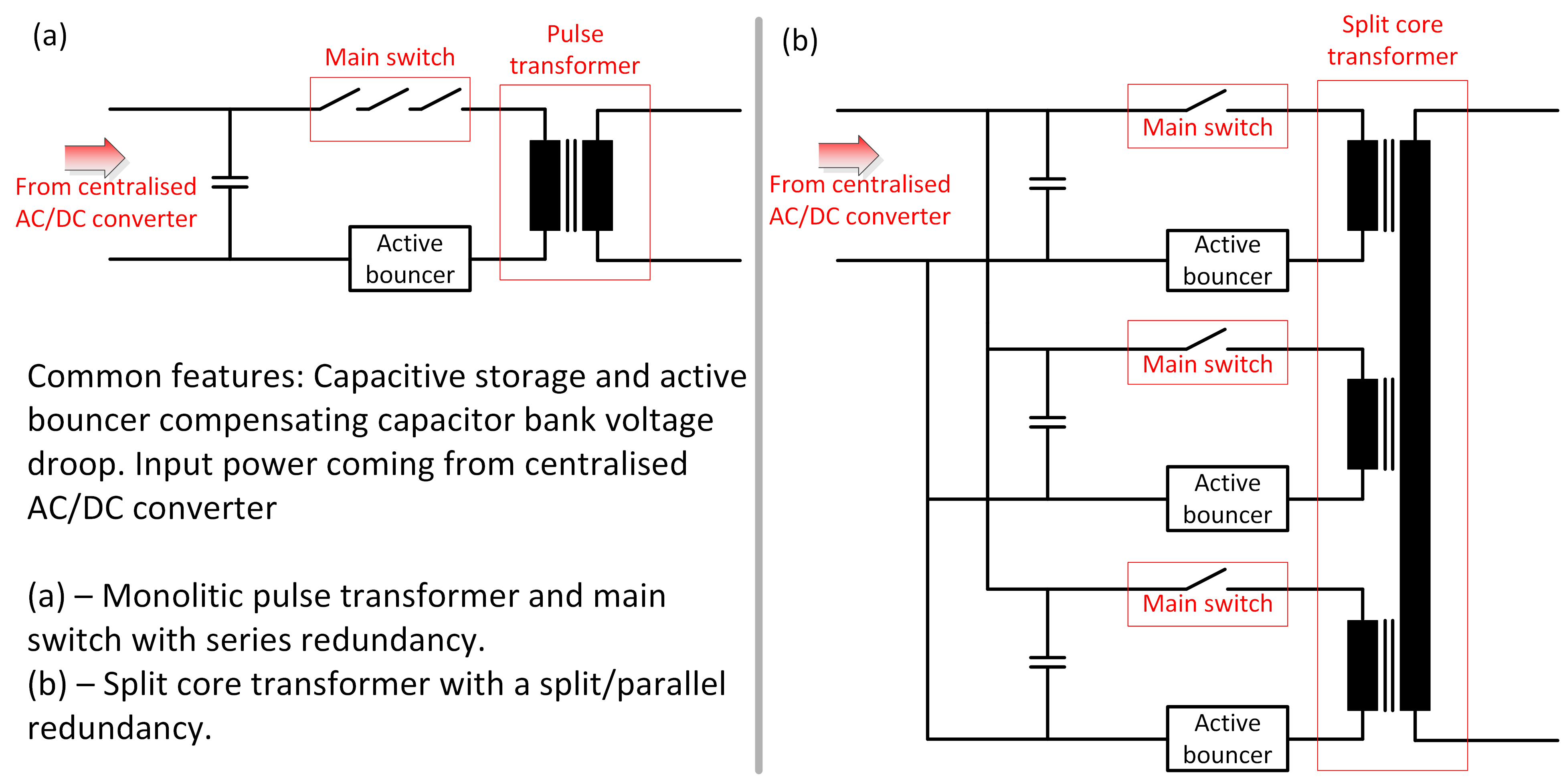}
  \caption{Principle topology for: (a) serial redundancy and (b) parallel redundancy.}
  \label{fig:Klystrons_3}
\end{figure}

The results of the global electrical distribution and AC to DC conversion imposes the modulator configuration, with a medium voltage DC stage and a voltage step-up pulse transformer. Two basic configurations, with series and with parallel redundancies (Fig.~\ref{fig:Klystrons_3} (a) \& (b)), have been studied \cite{Magallanes2015,Candolfi2014,Blume2013,Blume2015}. 

The monolithic pulse-transformer-based configuration (Fig.~\ref{fig:Klystrons_3} (a)) has been designed and validated via small-scale prototypes \cite{Magallanes2018}, whereas a full-scale split-core based configuration prototype has been built and tested. The stringent pulse-to-pulse repeatability requirement \cite{Gobbo2014} drove the choices of the active bouncer configuration (together with the fast rise time) and the design of special acquisition systems for high rate voltage measurements \cite{Baccigalupi2015}.

Preliminary studies show that the proposed 3\,TeV powering layout is also an optimal solution for the 380\,GeV and 1.5\,TeV stages. Since all the equipment up to the Medium Voltage DC distribution is modular, a staged approach minimally influences the cost of each stage; however, an initial extra investment in the modulators (no more than 20\,\% ) for 380\,GeV is necessary. This allows the modulators to be dimensioned correctly (pulse transformer) for the larger pulse width requirements in the 3\,TeV stage.

\begin{figure}[!htb]
  \centering
  \includegraphics[width=0.98\textwidth]{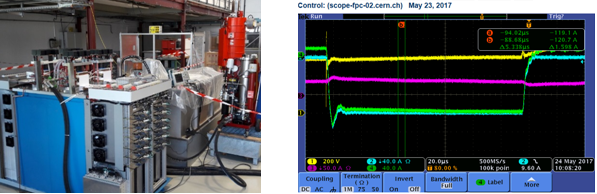}
  \caption{Left: Photograph of the full-scale modulator prototype installed at the CERN RF test stand. Right: First measured nominal voltage pulses demonstrating the dynamical capabilities.}
  \label{fig:Klystrons_4}
\end{figure}

A full scale modulator prototype based on configuration of parallel redundancy (as seen in Fig.~\ref{fig:Klystrons_3}~(b)) has been designed and delivered to CERN from ETH Z\"{u}rich (Fig.~\ref{fig:Klystrons_4}~left). First tests on an electrical dummy load demonstrated the feasibility of the voltage pulse dynamics up to 180\,kV. This modulator represents the new state of the art in fast-pulsed modulators with flat-top medium voltage input.  

Following the results (in 2016) of  global powering optimisation, the configuration of serial redundancy (as seen in Fig.\,\ref{fig:Klystrons_3}~(a)) is more adapted for CLIC. Therefore a second full-scale modulator prototype based on a configuration of serial redundancy is being constructed and first tests are expected in 2019.

\subsection{RF Power Source for 380\,GeV Klystron Option}

The klystron-driven 380\,GeV option will require a high-efficiency RF power source comprising 5,800 klystrons and 2,900 pulsed modulators. Each klystron will provide a peak RF power of 53\,MW at a frequency of 11.9942\,GHz with a pulse width of 2\,$\mu$s and a pulse repetition rate of 50\,Hz.The klystrons will be combined in pairs to increase the peak power available to the pulse compression system, feeding eight structures in the Main-Beam Linac and allowing control of the amplitude and phase stability. With such a large number of RF units required, efficiency is imperative in the design of both the klystron and modulator.

An RF unit consists of the following components, which all contribute to the overall operational performance and cost:

\begin{enumerate}
	\item  Modulator
	\item  Klystron (+ Klystron Heater Power Supply)
 	\item  Solenoid Magnet (+ Solenoid Power Supplies)
	\item  Low-Level RF (LLRF) (+ Klystron pre amp driver)
\end{enumerate}

\begin{figure}[!htb]
  \centering
  \includegraphics[width=0.8\textwidth]{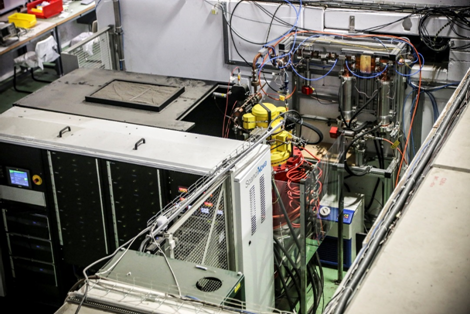}
  \caption{High-Gradient Test Stand facility at CERN, showing modulator, klystron and pulse compressor.}
  \label{fig:Klystrons_5}
\end{figure}

At present, the high-gradient test stands at CERN \cite{Catalan2014,Catalan2017} are operating on a 24/7 basis showing that the peak power and reliability requirements are achievable using technology already available from industry. This includes a commercially available modulator and a high-power X-band klystron, based on the original SLAC XL4 concept \cite{Sprehn2010}. Figure~\ref{fig:Klystrons_5} shows a picture of a high-power X-band RF source at CERN.

\subsection{X-band Klystrons}

With recent developments on higher efficiency klystrons \cite{Baikov2015}, a study is ongoing to improve the existing design of the klystron to achieve an efficiency of 70\,\% while maintaining the required peak power. This will have a significant impact on reducing the average power required and the physical size of the modulators needed to fit in the tunnel.

The operational parameters for the existing klystron modulator and the future high-efficiency, dual klystron modulator option are shown in Table~\ref{Tab:params_dual}.

\begin{table}[htb!]
\centering
\caption{\label{Tab:params_dual} Parameters for dual X-band klystron modulator.}
\begin{tabular}{|p{5cm}|p{4cm}|p{4cm}|} \hline 
\textbf{Main Parameters} & \textbf{Existing klystron modulator} & \textbf{Future high-efficiency, dual klystron modulator} \\ \hline 
RF Peak Power & 50\,MW & 2\,x\,53\,MW \\ \hline 
RF Average Power & 3.75\,kW & 2\,x\,5.3\,kW \\ \hline 
Pulse Length & 1.5\,$\mu$S & 2\,$\mu$S \\ \hline 
Pulse Repetition Rate & 50\,Hz & 50\,Hz \\ \hline 
Klystron Efficiency & 44\,\% & 70\,\% \\ \hline 
Perveance & 1.15\,$\mu$perv & 0.75\,$\mu$perv \\ \hline 
Modulator Voltage  & 420\,kV & 400\,kV \\ \hline 
Peak Pulse Current & 2\,x\,310\,A & 2\,x\,190\,A \\ \hline 
Modulator Average Power  & 24.597\,kW & 21.85\,kW \\ \hline 
Klystron Solenoid Average Power (Super Conducting Solenoid Average power) & 20\,kW & 20\,kW (1.6\,kW) \\ \hline 
Klystron Heater Power & 1\,kW & 1\,kW \\ \hline 
Average Power Consumption (5,800 klystrons, 2900 modulators) & 193.131\,MW & 185.169.4\,MW (78.469\,MW) SC Solenoids \\ \hline 
Flat-Top Stability (FTS) & +/- 0.25\,\% & +/- 0.25\,\% \\ \hline 
Pulse to pulse repeatability (PPR) & 0.01\,\% & 0.01\,\% \\ \hline 
Rise and fall times (t${}_{rise}$, t${}_{fall}$) & $<$1\,$\mu$s & $<$1.2\,$\mu$s \\ \hline 
\end{tabular}
\end{table}

\subsection{X-band Modulators}

\begin{figure}[!htb]
  \centering
  \includegraphics[width=0.49\textwidth]{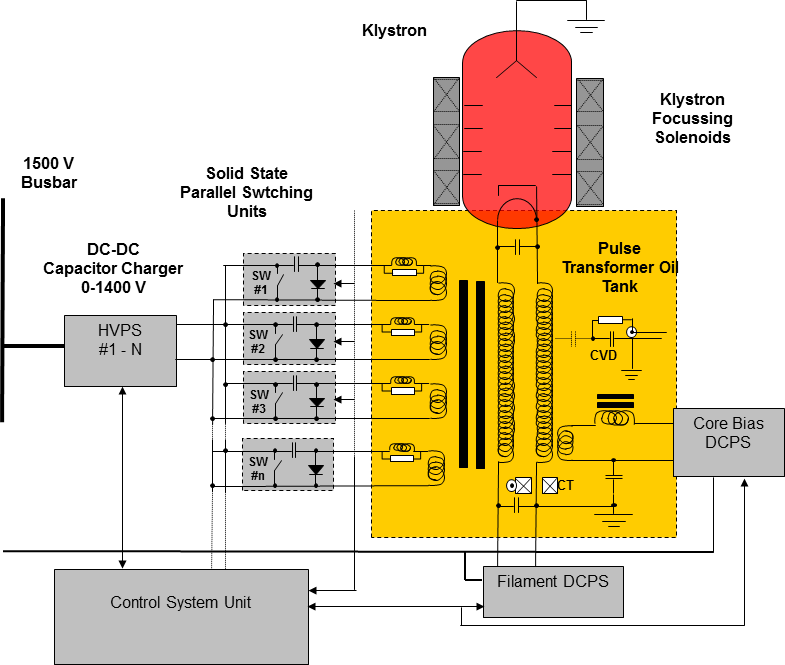}
  \includegraphics[width=0.49\textwidth]{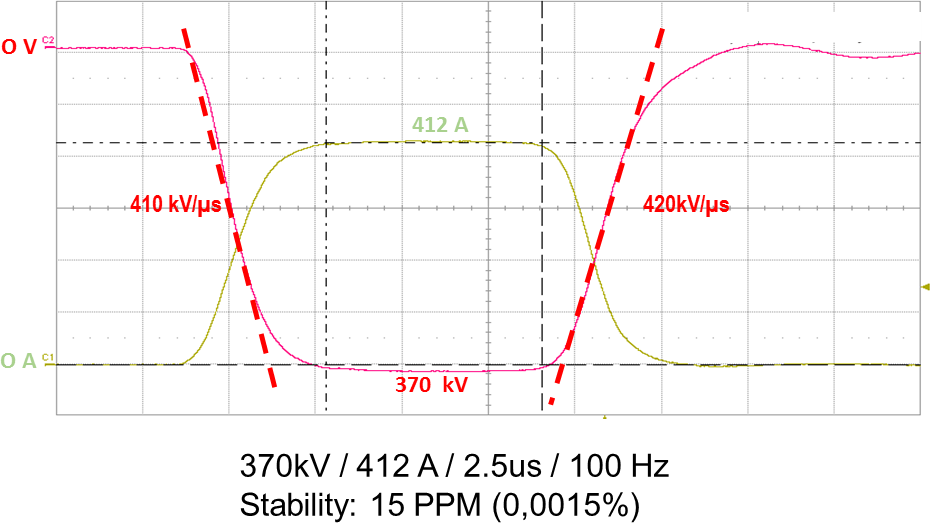}  \caption{Left: Schematic of solid-state modulator and klystron; Right: Typical voltage and current waveforms achieved in state-of-the-art modulators.}
  \label{fig:Klystrons_6}
\end{figure}

State of the art modulators using solid-state switching technology have shown that the pulse-to-pulse stability and pulse flatness requirements for CLIC are already achievable today. The high voltage pulse in the modulator is generated by charging and discharging many capacitors in parallel using solid-state switches, reducing the DC voltage requirements on the primary side of the pulse step-up transformer to less than 1.4\,kV. This results in a significant reduction in the volume and footprint of the modulator compared to the traditional 50\,kV line-type modulators, easing the constraints for installation in the tunnel. Fig.~\ref{fig:Klystrons_6} shows a simplified functionality diagram of the modulator and klystron system.

Using a pulse transformer implies rising and falling edges on the voltage pulses that are normally not usable for the RF pulse thus reducing the efficiency of the electrical energy available to 77\,\%. A study is ongoing to see if the efficiency increases by applying RF during the rise time of the voltage pulse, to fill the pulse compressor cavities.

\subsection{Solenoid Focusing Magnets}

The klystron focusing solenoids are another source of high average power consumption, contributing to approximately 50\,\% of the electrical energy consumed. To make a significant reduction in this power consumption a study to replace the normal-conducting solenoid with a superconducting solenoid has started, where the average power per solenoid is expected to be 1.6\,kW, resulting in an overall power saving of 106.7\,MW. This will reduce the average power consumption for the 380\,GeV stage RF to 78.5\,MW; this figure is in the baseline design.

\subsection{Low-Level RF (LLRF)}

Recent developments in commercially available solid state pre-amplifiers now being used in the high-gradient Test Facilities at CERN show a phase stability of  better than 0.05\,degrees which is more than sufficient to reach the CLIC requirements. Performances in the operation of the high-gradient Tests Stands and recent X-band linearizers in the Paul Scherrer Institute \cite{Gough2017} and in Synchrotron Trieste \cite{DAuria2010} show that the LLRF used to drive these amplifiers can also achieve the required specifications necessary for CLIC.

\section{PETS and Accelerating Structures}
\label{sect:PETS}

There are ten RF systems in the accelerator complex including the Main and Drive Beam generation, and the Main Linac. Each of these uses different accelerating/decelerating structures that will be described below.  

\subsection{Drive-Beam Generation Structures}

\subsubsection{Drive-Beam Linac Accelerating Structure}

This structure has been designed to maximise efficiency since it is the biggest power consumer in the entire CLIC complex. Therefore the accelerating structures are operated in a fully-loaded mode reaching an RF-to-Beam efficiency of 95\,\%. The main challenge of the system is the handling of long-range wakefields generated by the high current (4.2\,A) Drive Beam. The chosen baseline is a disc-loaded travelling wave structure of SICA type (Slotted Iris--Constant Aperture). The required damping is achieved through four slots in each iris to couple to the azimuthal currents generated by transverse modes. The HOM absorbers (SiC loads) are located directly in the iris slot. Other modes are detuned by changing the length of the nose cones in every cell. The structure operates at 1\,GHz fed directly by klystrons and is composed of 21 accelerating cells plus 2 coupling cells, giving a total structure length of 2.6\,m. The CLIC structure is a scaled version of the 3\,GHz structure built and successfully tested in CTF3. Fabrication of the final structure can be done via conventional precision machining as geometrical tolerances required are of the order of 0.1\,mm.

For the 380\,GeV stage, a higher gradient for the fully-loaded accelerating structures has been adopted, resulting in a reduced overall length.

\subsubsection{RF Deflectors}

The bunch train compression scheme for CLIC relies on the use of fast RF Deflectors (RFDs) for injection into the Delay Line (DL) and in the Combiner Rings (CRs). The three types of deflectors are all based on the travelling wave concept and have different operation frequencies depending on the recombination factor of the rings. The frequencies are 0.5\,GHz, 2\,GHz, and 3\,GHz for the DL, CR1 and CR2, respectively. The RF and mechanical designs of a generic RFD exist and have been tested in CTF3. The various parameters of the above-mentioned structures are summarized in Table~\ref{Tab:PETS_1}.

\begin{table}[htb!]
\caption{Overview of parameters for various RF structures for the Drive Beam generation for 3\,TeV (380\,GeV numbers in brackets)}
\label{Tab:PETS_1}
\begin{tabular}{l c c c c} \toprule
							 			& DBA & RFD DL & RFD CR1 & RFD CR2 \\
\midrule
Operating frequency [GHz] 	 			& 999.5 & 0.5 & 2 & 3 \\ 
Number of structures 		 			& 1638 (459) & 2 & 2 & 6 \\  
Active structure length [mm] 			& 240 & 600 & 450 & 150 \\ 
Number of cells per structure 			& 23 &  &  &  \\  
Pulse length [$\mu$s] 					& 150 (48) & 140 & 140 & 140 \\ 
Aperture diameter [mm] 					& 98 & 50 & 40 & 40 \\
Filling time [ns] 						& 245 & 110 & 80 & 16 \\ 
Input peak power [MW] 					& 18 & 50 & 50 & 50 \\ 
Accelerating gradient unloaded [MV/m] 	& 6.6 & 11 & 13 & 14 \\
Accelerating gradient loaded [MV/m] 	& 3.4 & - & - & - \\ 
\bottomrule
\end{tabular}
\end{table}

\subsection{Main-Beam Generation Structures}

\subsubsection{Pre-Damping Ring RF Systems}

The description of the Pre-Damping Ring RF system in the CLIC CDR \cite{Aicheler2012} is still valid and no further developments have been done. As a reminder, it is based on the Single Cell Cavity developed and used in KEK-B high-energy ring (HER) scaled from 0.509 to 1\,GHz. In total ten of these superconducting cavities would be required. 

\subsubsection{Damping Ring RF Systems}

In order to compensate the energy lost per turn in the wigglers of the damping ring, the RF system has to work in a very high beam loading regime. The required RF system is close to the standard RF system for high beam current storage rings such as KEK-B low energy ring (LER). It is proposed to use the same type of cavities (ARES-type) which provide the low R/Q necessary to mitigate the strong beam-loading effects. The proposed solution consists of scaling the accelerating and coupling cavity to 2\,GHz while keeping the storage cavity almost at the same dimensions. In this way, R/Q is reduced and the voltage per cavity stays within reasonable limits. However, the final implementation of the structure geometry is still unknown. Studies are ongoing to optimize the overall parameters of the whole RF system including feedbacks to satisfy beam constraints on one hand, and minimize the power and size of the structure on the other. The use of a scaled version of the accelerating structure used in light sources like the ALBA synchrotron is also under investigation \cite{Bravo2017}.

\subsection{Main-Linac RF Structures}
\subsubsection{Power Extraction and Transfer Structure (PETS)}

These passive microwave devices interact with the Drive Beam to generate RF power along a constant-impedance periodic structure. The power is collected downstream in an RF power extractor that converts the TM01 mode in a 23\,mm diameter circular waveguide into a TE10 mode in a rectangular WR90 waveguide. In its final configuration, the PETS comprises eight octants (bars) separated by the 2.2\,mm wide damping slots. Each of the bars is equipped with HOM damping loads. A total of sixteen PETS have been manufactured and tested in the two beam lines of CTF3. The performance of the PETS in CTF3 is described in Section~\ref{sect:PERF_DB}.

\begin{figure}[!htb]
  \centering
  \includegraphics[width=0.6\textwidth]{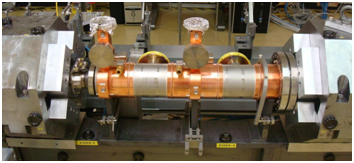}
\caption{Photograph of PETS}
  \label{fig:PETS_1}
\end{figure}

\subsubsection{Main-Beam Accelerating Structure} 

The compactness of CLIC is based on the high gradient achieved in the Main-Beam Accelerator in order to reach 3\,TeV in a reasonable tunnel length. The main challenge for these accelerating structures is producing extremely elevated electric fields without excessive vacuum arcing during operation. In this case, we use disk-loaded travelling-wave structures working at 12\,GHz and with constant gradient to increase RF-to-beam efficiency. The very small aperture of the iris generates considerable transverse short-range wakefields which translate into very demanding alignment tolerances of the order of 14\,$\mu$m per super-structure (as described in Section~\ref{sect:Survey}). Damping of the long-range transverse wakefields is provided by four radial waveguides ending in SiC loads. The continual decrease of the iris radii through the structure ensures as a side effect detuning of the remaining modes. The final optimization of the length, gradient and overall geometry of the structure has been made using a general costing model which takes into account also power and civil engineering costs. Therefore, we have a different structure for each stage of CLIC, as well as for the klystron-based option. The main parameters of all three structures can be found in Table~\ref{Tab:PETS_2}. A large number of prototypes corresponding to the 3\,TeV stage have been manufactured over the last years. They have been successfully tested in high power conditions in the X-band test facilities at CERN, KEK and SLAC (see Chapter~\ref{Chapter:PERF}). Prototypes for the 380\,GeV stage are now under fabrication. 

\begin{table}[htb!]
\centering
\caption{RF parameters for the Main Linac RF structures working at 11.994\,GHz. The Main-Beam accelerating structure has been optimized for the 380\,GeV initial stage both Drive Beam \cite{DarvishRoknabadi2016} and klystron based \cite{Huang2018} as well as for the 3\,TeVstage \cite{Liu2018}.}
\label{Tab:PETS_2}
\begin{tabular} {l c c c c c c} 
\toprule 
 & \multicolumn{3}{c}{Main Beam Accelerator } & \multicolumn{2}{c}{PETS} & Crab \\  
 & \multicolumn{2}{c}{380 GeV} & 3 TeV & 380 GeV & 3 TeV &  \\
 & DB & Klystron &  &  &  &  \\ 
 \midrule 
Number of structures & 20,592 & 23,296 & 143,232 & 10,296 & 71,616 & 2 \\ 
Active structure length [mm] & 272 & 230 & 230 &  &  &  \\
Number of cells & 33 & 28 & 28 & 33 & 34 & 12 \\
Pulse length [us] & \multicolumn{3}{c}{0.244} & \multicolumn{2}{c}{0.244} & $\sim$200 \\ 
Aperture diameter [mm] & 8.2-5.2 & 7.25-4.5 & 6.3-4.7 & 23 & 23 & 10 \\ 
Filling time [ns] & 55.75 & 63.75 & 66.27 & 1.52 & 1.55 &  \\
Input peak power [MW] & 59.2 & 40.6 & 61.1 & 123.3 & 127.3 & 20 \\
Average Q factor & 5504 & 5846 & 5843 & 7200 & 7200 &  \\ 
Accelerating voltage unloaded [MV] & 92.2 & 94.9 & 27.8 & - & - & 2.55 \\ 
Accelerating voltage loaded [MV] & 72 & 75 & 100 & - & - &  \\ 
\bottomrule
\end{tabular}
\end{table}

In all cases, the fabrication process is critical to achieve the highest gradient. The structure is made of OFE copper disks machined by turning and milling using single-crystal diamond tools and with tolerances in the micron range. The stack of disks is then bonded in a furnace at 1040\,$^\circ$C with a hydrogen protective atmosphere. Ancillary systems like cooling blocks, manifolds with loads, and the vacuum chamber are brazed separately after this. After RF measurements and bead-pulling, the final unit is then baked for 1-2 days to desorb the Hydrogen and finish cleaning the surface. 

\begin{figure}[!htb]
  \centering
  \includegraphics[width=0.4\textwidth]{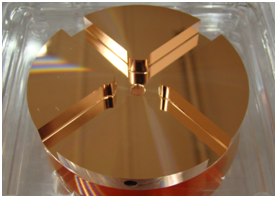}
  \includegraphics[width=0.4\textwidth]{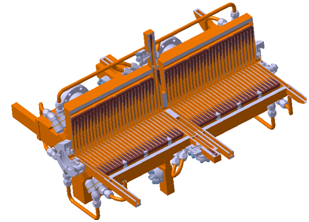}
  \caption{Photograph of a single disc (left) and an engineering model of a full super-accelerating structure (right)}
  \label{fig:PETS_2}
\end{figure}

Even if the final CLIC prototype accelerating structure has been manufactured and successfully tested, the complexity of assembly and the need for ultra-precision machining made the structure costly and difficult to produce in large numbers. Alternative manufacturing techniques based on vacuum brazing and electron-beam welding of structures made in halves instead of disks are under investigation. In parallel, studies are being launched together with industry to estimate the necessary modifications in the production flow, production yields, quality assurance needs, costs and learning coefficients for the full CLIC production. 

One of the remaining challenges with respect to the Main-Linac accelerating structure is the accuracy required for the integrated wakefield monitor of only 3.5\,microns. Laboratory and beam studies in CLEAR are focusing on verifying the geometric accuracy of the cells and wakefield monitors as well as minimizing errors in the electronics, acquisition and signal to noise ratio. 

\section{Beam Instrumentation}
\label{sect:BI}
\subsection{Overview of the CLIC beam instrumentation requirements}

Beam dynamic considerations dictate most of the requirements for beam instrumentation and CLIC is expected to operate with extremely tight tolerances on most beam parameters. Extremely low-emittance beams are generated in the damping rings and must be conserved over kilometres of beam lines requiring a precise control of the beam position over such long distances. Before entering the Main Linac, the bunch length must be shortened and controlled at the femtosecond level. At the interaction point, the beam is finally focused to only a few nanometres in size. After collisions, the highly disrupted beam must be dumped in clean conditions, making sure that the 2.91\,MW of power carried by the particles are safely absorbed.

An overview of CLIC beam instrumentation was presented in the CDR \cite{Aicheler2012} with a collection of technical specifications for the different parts of the 3\,TeV stage, together with a description of the possible technologies in use and their expected performance. In the context of a 380\,GeV stage, all those specifications and challenges remain valid, and only the total number of devices required is reduced as it scales almost linearly with the beam energy and the overall length of the accelerator complex. 
The total number of instruments required on the Drive and Main beams for 380\,GeV is presented in Table~\ref{tab:Tab_INS_1}.

\begin{table}[htb!]
\centering
\caption{Number of beam instruments for the Drive and Main Beams}
\label{tab:Tab_INS_1}
\begin{tabular}{l c c} \toprule
Instrument 			& Drive Beam 	& Main Beam \\ 
\midrule
Intensity 			& 50 			& 130 \\ 
Position 			& 7,875 		& 4,165 \\  
Beam Size 			& 80 			& 110 \\  
Energy 				& 50 			& 30 \\  
Energy Spread 		& 50 			& 30 \\  
Bunch Length 		& 60 			& 30 \\  
Beam Loss /Halo 	& 7,790 		& 4,950 \\  
Beam Polarization 	&  				& 20 \\  
Tune 				&  				& 6 \\  
Luminosity 			&  				& 2 \\  
\midrule
Total 				& 15,955 		& 9,475 \\ 
\bottomrule
\end{tabular}
\end{table}

\subsection{Main-Beam Instrumentation}
\subsubsection{Main-Beam Cavity Beam Position Monitor}

\begin{figure}[!htb]
  \centering
  \includegraphics[width=0.8
  \textwidth]{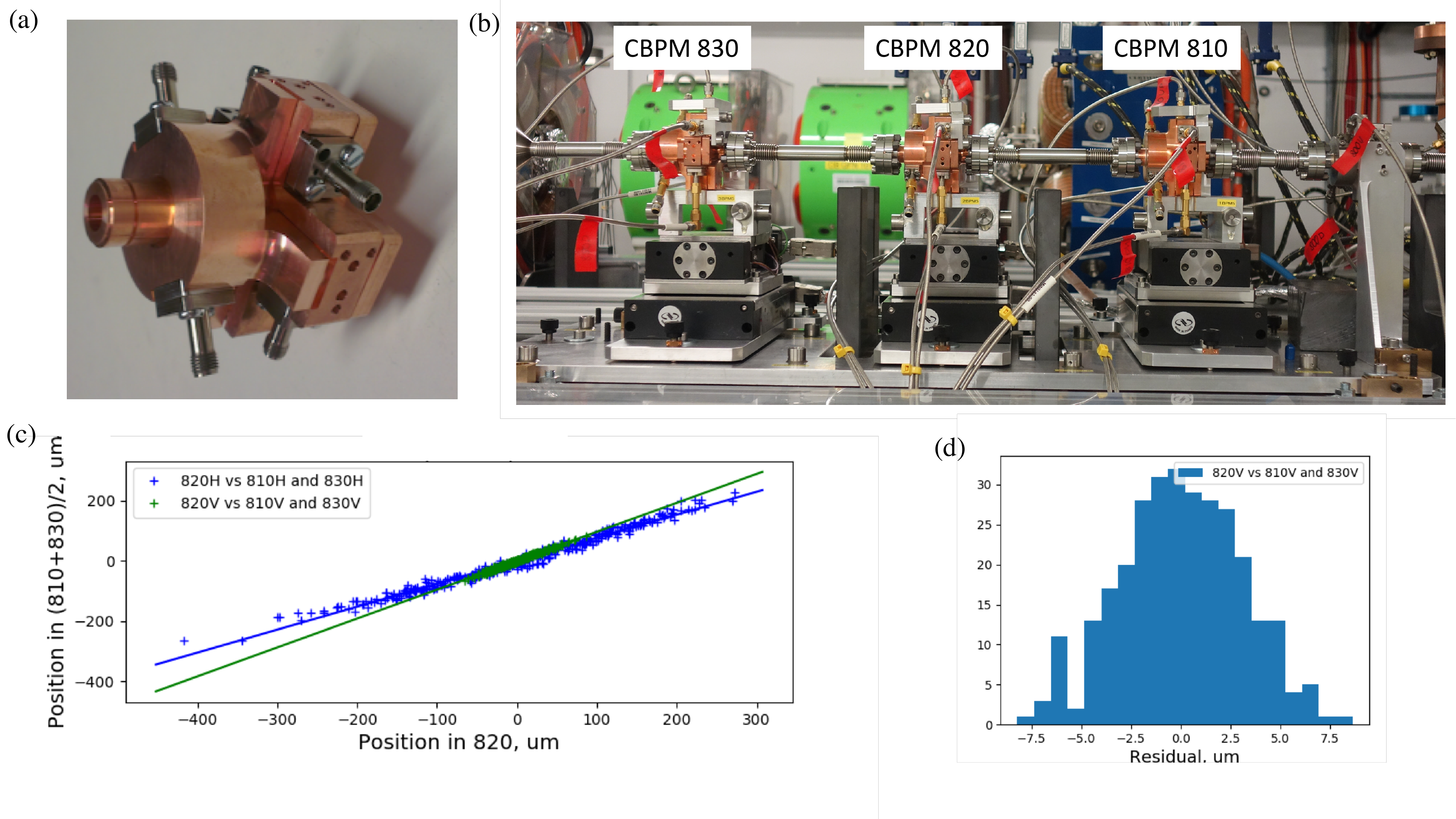}
  \caption{(a) A picture of the CLIC Cavity BPM; (b) Experimental test-stand installed on the CTF3 beam line with 3 consecutive cavity BPMs, each BPM being installed on movers for remote alignment; (c) Correlation plot of the beam position measured by the central BPM (820) with respect to the positions measured by the two others; (d) The corresponding residual error measured in the vertical plane.}
  \label{fig:CBPM}
\end{figure}

The ability of Cavity Beam Position Monitors to achieve nanometre-level resolution was already demonstrated in \cite{Walstong2007}, but an additional challenge in CLIC is the use of dispersion free-steering along the Main Linac, which would require both high spatial resolution and time resolution better than 50\,ns. This has launched the development of a lower-Q cavity BPM for the Main Beam, which has been constructed and tested in CTF3 and has demonstrated its capability to measure the beam position, with a 200\,ns long train of bunches, with a time resolution better than 20\,ns \cite{Cullinan2015}. A picture of the CLIC cavity BPM installed on the CTF3 beam line is shown in Fig.\,\ref{fig:CBPM}. 

\subsubsection{Non-invasive Short Bunch Length Monitor}

Bunches as short as 150\,fs will be produced and their longitudinal profile must remain under control up to the collision point to guarantee maximum luminosity. The bunch length needs to be measured and controlled accurately with time resolution better than 20\,fs. Radio-frequency deflectors \cite{Maxson2016} have already demonstrated their capability to provide fs resolution but in a destructive way. An R\&D program was thus pursued since 2009 to design and test non-invasive bunch length monitors using laser pulses and bi-refringent electro-optical crystal \cite{Berden2004,Berden2007}. Based on an EO encoding process described in \cite{Jamison2006} which explains the encoding in terms of optical sideband generation via sum and difference frequency mixing of the optical and terahertz fields in the crystal, a new technique called Spectral Upconversion \cite{Jamison2010} has been developed. The technique directly measures the Fourier spectrum of the electron bunch, through first upconverting the far-IR-mid-IR spectrum to the optical region, followed by optical spectral imagining.  The technique uses a long-pulse ($>$10\,ps) laser probe, for which laser transport in optical fibre becomes relatively trivial. The laser system can be significantly simpler than the ultrafast amplified systems of other EO decoding schemes. Combined with a time-explicit Frequency Resolved Optical Gating (FROG) system, the temporal profile of the beam would be measurable \cite{Walsh2015}.
A study is still on-going in order to test such a system with short electron pulses. In addition, it was also proposed to investigate how a multiple-crystal detection arrangement may provide larger response bandwidth, sufficient to characterize bunches with 20\,fs and potentially even higher resolution, . 

\subsubsection{Measuring very small beam sizes}

The extremely small transverse emittance generated in the damping rings puts tight requirements on the measurement of transverse beam size, which should achieve micron resolution. In rings, synchrotron radiation \cite{Hofmann2004} provides a great opportunity to measure the small beam size. The 3${}^{rd}$ generation light source community has been very active over the last 20 years in developing better and better transverse beam diagnostics based on x-ray optics \cite{Takano2006} or optical light interferometric techniques \cite{Mitsuhashi2006}. Recently studies using randomly distributed interferometric targets have been performed in an attempt to improve further the performance of SR-based techniques \cite{Siano2017}.  

Other methods were studied in the early 2000's and Laser Wire Scanner (LWS) technology has demonstrated excellent performance in measuring sub-micron beams non-invasively \cite{Boogert2010}. However, the system's complexity and the need for an expensive high-power laser triggered the study of alternative solutions aiming for better reliability, simplicity and cost saving. A breakthrough was achieved in 2011, with the experimental measurement of the point-spread function of optical transition radiation \cite{Karataev2011}. This then led to the development of beam size measurement techniques with sub-micron resolution using a simple, cheap and compact optical imaging system \cite{Bolzon2015}. Theoretical considerations on how to use this technique practically for highly relativistic beams was also studied and verified experimentally \cite{Kieffer2018a}. 

\subsection{Drive-Beam Instrumentation}

During the last 13 years, in the framework of CTF3, a full suite of beam instruments has been successfully developed to fulfil the requirements in the Drive Beam complex. These are longitudinal beam diagnostics for  measuring the bunch frequency multiplication in the combiner rings \cite{Welsch2006,Micheler2010,Lekomtsev2012,Dabrowski2010}, and also transverse profile monitors for high energy spread beams in the Drive-Beam Decelerator \cite{Welsch2007,Welsch2006a,Olvegard2012}. 

\subsubsection{Non-invasive Transverse Beam Profile for Low-Energy, High-Charge Beams}

Compared to CTF3, the CLIC Drive Beam has a considerably higher total charge,  corresponding to an average current of 4.2\,A over a 48\,$\mu$s pulse length. This has severe consequences on the technological choices for beam instrumentation. Any intercepting devices would be limited to the observation of a small fraction of the Drive Beam, most likely by reducing the beam pulse length or current. We have launched an R\&D program to develop non-invasive transverse beam profile monitors based on the interaction of the particles with a supersonic gas-jet. With encouraging preliminary results \cite{Jeff2014,Jeff2013,Tzoganis2017}, this work is now being followed up by a large scientific community as it may also be of interest to many other accelerators. 

\subsubsection{Reliable and Maintainable Beam Position Monitor System for the Drive-Beam Decelerator}

With more than 6,000 devices, the BPMs in the Drive-Beam Decelerator will be the largest under-vacuum beam instrumentation system in CLIC. It will also be the biggest cost driver and a critical system that needs to be  highly reliable and easily maintainable. Over the last 10 years, we have tested different BPM technologies, such as inductive pick-ups \cite{Gasior2003} or stripline pick-ups. We  did not find a big difference in cost and performance between these two designs. However inductive pick-ups present several advantages compared to striplines. First, in terms of complexity, inductive pick-ups, based on the measurement of the wall current, have no element under vacuum, whereas striplines sit inside the beam pipe and need to be carefully matched to 50~ohms impedance, which would require longer tuning procedures that would increase the production cost of the system. Moreover, a recent development made for LHC \cite{Krupa2016} has shown that inductive pick-ups can be designed and manufactured such that they can be dismounted without breaking the vacuum, which would also enable maintenance and repair to be carried out with minimum downtime and impact on other services. 

\subsection{Technology Developments Valid for Both Beams}
\subsubsection{Non-invasive Beam-size Monitoring using Polarisation Radiation}

Non-invasive transverse beam size monitoring is key to the operation and the optimization of high-beam-charge accelerators. In 2010 an effort was launched to develop techniques based on polarisation radiation, such as diffraction radiation \cite{Aumeyr2015}. After the first promising measurements in 2004 \cite{Karataev2004}, we launched a series of experimental investigations both on the Cornell electron storage ring \cite{Bobb2018} as well as on the Advanced Test Facility 2 (ATF2) at KEK, and we have recently demonstrated the possibility to measure beams with a spot size of a few microns using UV light and very small ($<$100\,$\mu$m aperture) diffraction slits. This would find application, as an alternative to laser wire scanner in the Main Beam RTML as well as in the Drive Beam complex where the electron beam energy is in excess of 1\,GeV. 

\begin{figure}[!htb]
  \centering
  \includegraphics[width=0.8\textwidth]{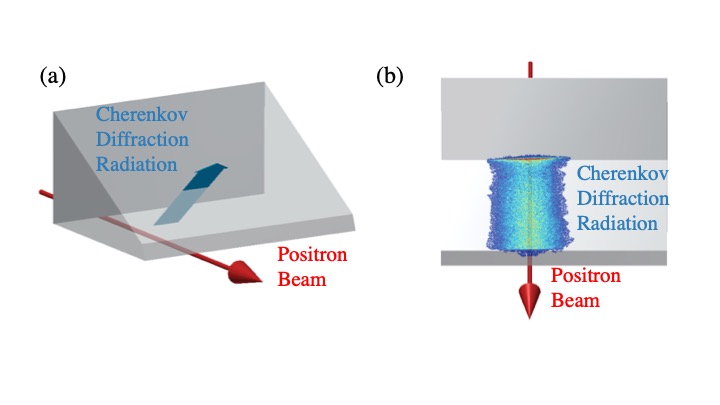}
  \caption{First detection of Cherenkov Diffraction radiation emitted by a 5.3\,GeV positron circulating in the Cornell storage ring;(a): Schematic view of the emission of Cherenkov diffraction radiation in a prismatic dielectric radiator; (b): Corresponding Beam image showing an horizontal beam size of 4\,mm.}
  \label{fig:ChDR}
\end{figure}

Very recently, looking for a way to overcome background limitation in diffraction radiation from small aperture slits, we investigated the Cherenkov Diffraction radiation emitted in longer dielectric material. First measurements performed in 2017 \cite{Kieffer2018} on CESR have shown very promising results, as depicted in Fig.~\ref{fig:ChDR}. A more extensive study of the effect has been recently initiated in order to measure precisely the resolution of this technique as a non-invasive beam-imaging system. 

\subsubsection{A Distributed Beam Loss Monitoring System along the CLIC Main Linac}

Along the decelerator, the Drive-Beam energy will be continually transformed into 12\,GHz RF power over some hundreds of meters. The beam energy spread rises up linearly to reach a value of 90\,\% at the end of a decelerator section. The beam optics must be adapted to ensure constant RF power production without consequent beam losses and beam operation relies on the continual monitoring of the beam properties. This is achieved using a beam loss monitoring system with a large number of classical ionization chambers \cite{Holzer2005} distributed along the main linac. Pursuing an alternative solution providing a better measurement of longitudinal distribution of the losses, simulations and experimental validations have led to the development of a high performance and cost-efficient BLM system \cite{Kastriotou2016} based on optical fibre measuring Cherenkov light induced by lost charged particles. In particularl, the study has addressed several key features of the BLM system such as the position resolution of optical fibre detection system when using long electron pulses (i.e. 200\,ns) \cite{Nebot2015} and the crosstalk between losses from the Main and Drive Beams \cite{Kastriotou2013}.

\section{Vacuum System}
\label{sect:Vacuum}
\subsection{Introduction}

Apart from its large size, the vacuum system for the CLIC complex contains a multitude of challenges: synchrotron radiation, electron cloud, sparking and degassing induced by high electrical field, conductance limited system, cryogenic vacuum system, etc. Most of the CLIC complex requires high or ultra-high vacuum.
From a general point of view, different technical solutions are chosen: 

\begin{enumerate}
\item  \textbf{For bakeable vacuum systems operating at room temperature}, a solution based on a Ti-Zr-V film coating is used. The thickness of this non-evaporable getter (NEG) coating can be reduced below 1\,$\mu$m if required.  After in-situ activation by heating, it provides a distributed pumping speed for the most abundant gas species released in the vacuum chambers. Also, the desorption yields induced by photon and electron bombardment and the secondary electron yield are reduced drastically when compared to traditional materials. A limited number of lump ion pumps are required to remove noble gas and methane not absorbed by the NEG material.

\item  \textbf{For non-bakeable vacuum systems operating at room temperature}, vacuum performance is driven by water vapour outgassing. High vacuum can be obtained after a long pumping period. Lumped pumps are used. A combination of NEG pumps with large pumping speed and sputter-ion pumps is recommended to limit the cost of cabling. If low secondary electron yield is required, amorphous carbon (aC) coating can be applied. 

\item  \textbf{For vacuum systems operating at cryogenic temperature}, the cold surface acts as a cryopump for the impinging gas molecules sticking to it. The pumping speed and saturated vapour pressure of the different gas species depend on the temperature of the vacuum chamber. To reduce secondary electron yield, and therefore the heat load to the cryogenic system, amorphous carbon coating can be applied. Measurements of aC outgassing rates are ongoing in the framework of the HL-LHC project for an implementation in the LHC long straight sections. 
\end{enumerate}

\subsection{Update of the Main-Linac Vacuum System}

The vacuum system for the Main Linac is integrated in the Two-Beam Module design. It is a non-baked system characterized by low vacuum conductance and large surface areas. Its concept was based on a main pumping speed applied to a central tank and redistributed to the AS and PETS by means of flexible connections. Even if the system performance is limited by the conductance, this solution provided enough pumping speed to each element while minimizing the number and cost of the pumping components. Vacuum tests have been carried out on a dedicated set-up. As expected for a metallic non-baked system, the pressure decreases with time (1/t dependence) in the Main-Beam chamber and reaches around 7\,x\,10${}^{-9}$\,mbar after 100\,h pumping \cite{Garion2011}. Several geometries of the vacuum manifold of the AS have been tested and it turns out that it does not have a significant impact on the vacuum performance. Nevertheless, transvere forces due to the application of the vacuum on the main and drive beam lines generate displacements which were not compatible with the CLIC requirements. 

A new concept based on local compact pumps has been proposed. It consists of NEG cartridge pumps with 100\,l.s${}^{-1}$ pumping speed installed directly on the AS vacuum manifold or the PETS vacuum enclosure. The conductance limitation of the flexible connection between the tank and the AS and PETS is removed. Therefore, the vacuum performance is given only by the AS or PETS geometry. Noble gas and methane are pumped by small ion sputter pumps that are directly integrated with the NEG cartridge pump. Vacuum tests have been carried out on the dedicated AS with a NEG pump with nominal pumping speed 100\,l.s${}^{-1}$  combined with a sputter ion pump providing 5\,l.s${}^{-1}$ pumping speed. After activation of the NEG cartridge, the pressure decreased with time (1/t) and reached 3\,x\,10${}^{-9}$\,mbar along the beam axis after 100\,h pumping. To reduce the number of sputter ion pumps, a configuration of alternating pumps with and without sputter ion pumps is proposed; it has been successfully tested \cite{Garion2011}. The architecture of the vacuum system is therefore based on a combination of NEG cartridge pumps combined with sputter ion pumps (100\,+\,5\,l.s${}^{-1}$), and NEG cartridge pumps (100\,l.s${}^{-1}$). 

In addition, a T-shape connector has to be installed on each line. Installed on the pumping port of the AS or PETS, it is equipped with a NEG pump on one side and a manual all-metal right angle valve on the other side allowing the rough pumping and leak detection of the modules. 

Vacuum gauges, a set of Pirani and Penning gauges, are installed on each beam line in each module. This allows pressure measurement from atmospheric pressure down to 10${}^{-10}$\,mbar. This is needed during operation and  commissioning of the vacuum system. Vacuum gauges provide also signals for interlocks of vacuum sector valves and machine protection systems. 

Unless radiation-hard electronics is developed, vacuum gauge controllers and power supplies have to be installed in a protected area, requiring long high voltage cables. Cost reduction can be achieved by a single power supply used for multiple ion pumps equipped with dispatch boxes and small local cables. 

\subsection{Development of vacuum Technologies of Interest to CLIC}

Some developments in vacuum technologies have been recently initiated at CERN which might be of interest to the CLIC project and are presented here.

\subsubsection{Low aperture copper vacuum chamber with NEG coating}

\begin{figure}[htb!]
\centering
\includegraphics[width=0.7\textwidth]{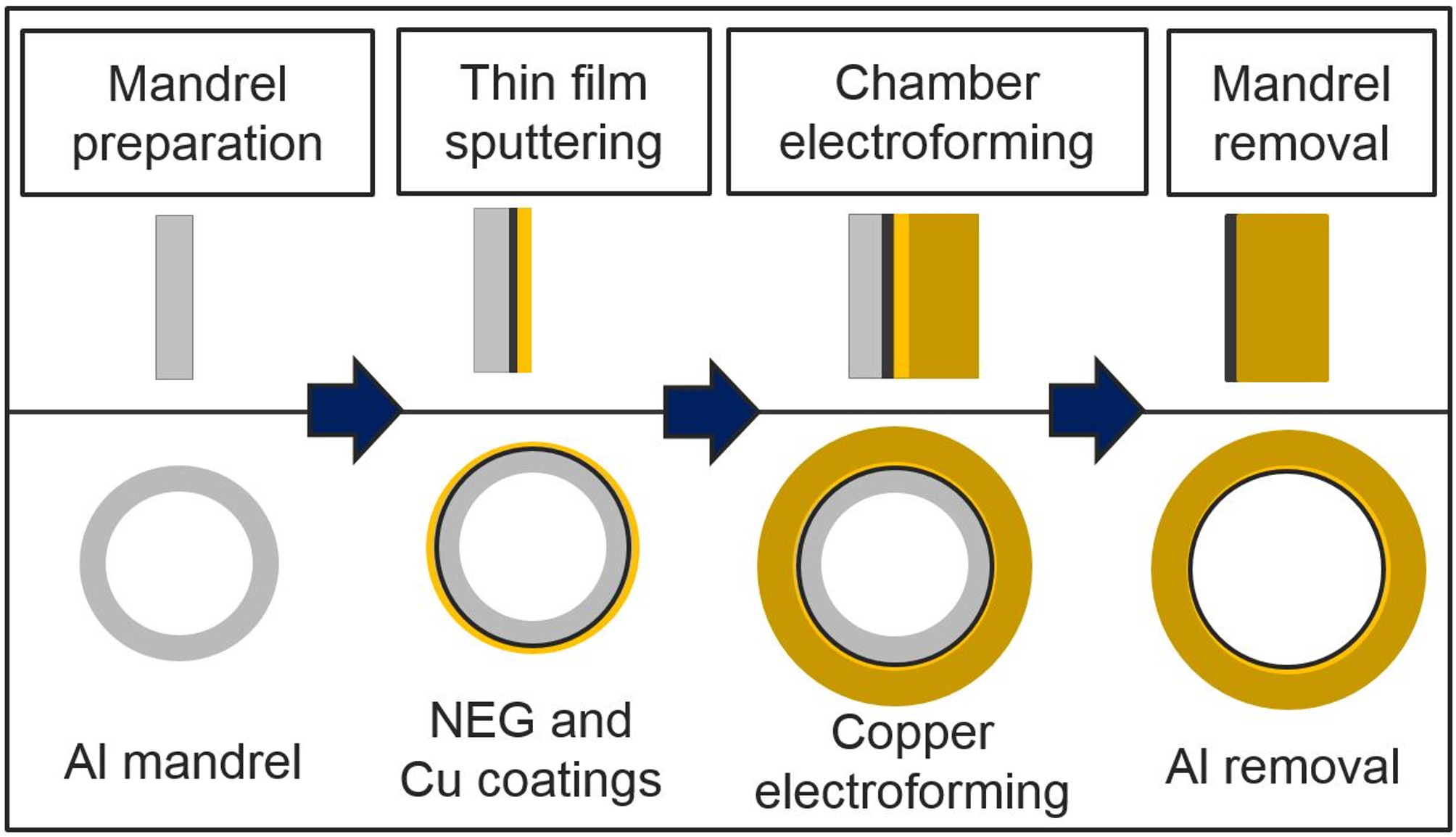}
\caption{\label{fig_VAC_1} Electroforming process of low aperture copper vacuum chamber.}
\end{figure}

\begin{figure}[htb!]
\centering
\includegraphics[width=0.7\textwidth]{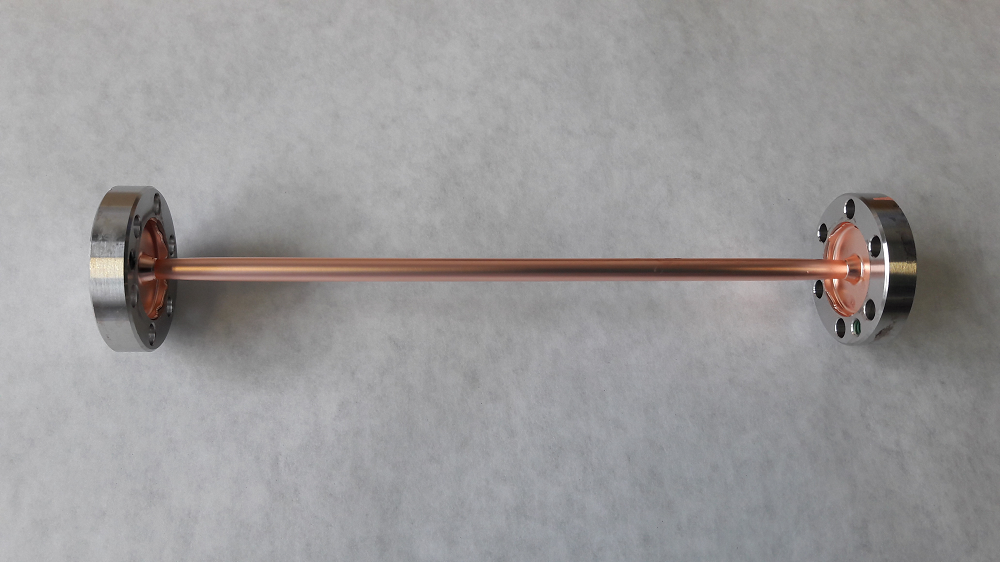}
\caption{\label{fig_VAC_2} Electroformed copper chamber integrating stainless steel flanges.}
\end{figure}

Currently CERN develops copper electroformed vacuum chambers with integrated NEG film coatings. This new method is based on a sacrificial aluminium mandrel with the required shape on which a thin NEG coating is applied. The vacuum chamber wall in copper is then built up by electrodeposition on top of the NEG film. Finally, the aluminium mandrel is removed by chemical etching. A schematic of the process is shown in Fig.~\ref{fig_VAC_1}. This innovative manufacturing method allows the production of vacuum chambers with very low aperture (down to few millimetres) while integrating NEG coating, used as a distributed pump and/or low secondary electron yield surface. This procedure can also integrate and join the stainless steel flanges during the electroforming step (neither electron beam welding nor brazing is needed). Performance of the NEG produced with this method is being assessed. Last results have shown a good H2 sticking coefficient when the chambers are activated around 260${}^\circ$ -- 280${}^\circ$\,C. For a standard activation at 230${}^\circ$\,C for 24\,hours the H2 pumping speed exhibits five to ten times lower values than expected for standard NEG. The studies focus on the evaluation of the effect on the performance of the aluminium mandrel characteristics (e.g. thickness and alloy) and the optimization of the copper electro formed layer. As low as 3\,mm diameter coated vacuum chambers were successfully produced using this method (Fig.\,\ref{fig_VAC_2}) \cite{LainAmador2018}. 

\subsubsection{Permanent Radiation Hard Bake-out System}

\begin{figure}[htb!]
\centering
\includegraphics[width=0.35\textwidth]{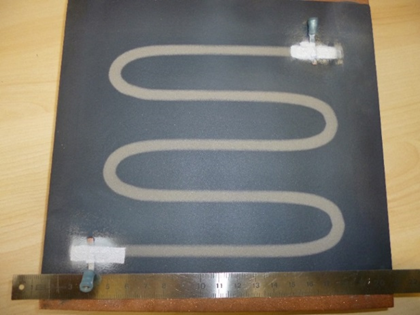}
\includegraphics[width=0.45\textwidth]{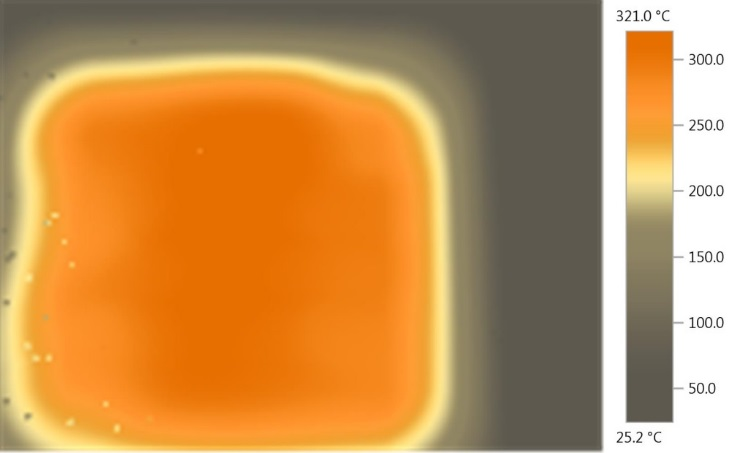}
\caption{\label{fig_VAC_3} Permanent radiation hard heating system.}
\end{figure}

A new permanent radiation-hard bake-out system is being developed (Fig.~\ref{fig_VAC_3}). It is based on a ceramic layer deposited on the vacuum chamber or the element to be baked. On top of this insulating coating, a conductive track, preferably made of high electrical resistivity materials, is deposited. Heating is obtained by the Joule effect. The pattern of the heating track is defined to minimize cold spots that may occur during the bake out. An additional insulating protective layer can be applied on the track to avoid damage or short circuit of the heating track. A first test has been successfully performed on a copper element. The thickness of the heating system is around half a millimetre. This solution is suitable for large-scale series, for example for the vacuum chambers of the main and drive beam transfer lines. In addition, based on metallic and ceramic materials, it is expected to be applicable to in-vacuum bake out systems that could be used, e.g. for collimator jaws. 

\subsubsection{Compact Temperature-Controlled UHV Connector}

\begin{figure}[htb!]
\centering
\includegraphics[width=0.4\textwidth]{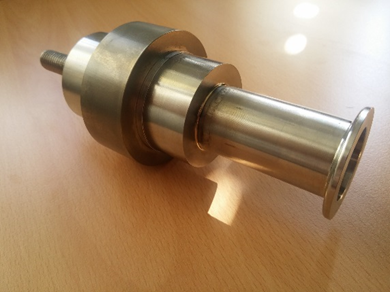}
\includegraphics[width=0.4\textwidth]{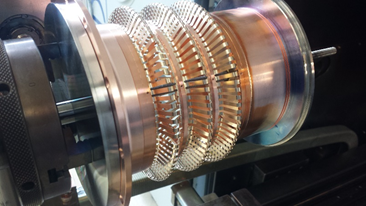}
\caption{\label{fig_VAC_4} Left: SMA UHV connector.Right: Deformable RF bridge.}
\end{figure}

Beam-pipe coupling in particle accelerators is nowadays provided by metallic flanges that are tightly connected by several screws or heavy collars to deform plastically a gasket made from soft material which is placed placed between the two flanges. A new concept based on a Shape Memory Alloy (SMA) connector has been proposed at CERN. It consists in a SMA ring that is clamped by heating above 100${}^\circ$\,C around the beam pipe extremities with a soft material sleeve placed in between (Fig.\,\ref{fig_VAC_4} left). This generates a high contact pressure and the leak tightness between the vacuum chamber and the gasket. The assembly remains tight at operating (room) temperature and is disconnected by cooling below -40${}^\circ$\,C using the two way effect of the SMA. The ring offers a reliable, compact and light solution. Connection between pipes can be obtained with reduced space requirement. This should simplify the integration in very compact accelerators such as CLIC. The SMA rings can significantly reduce the intervention time and can be operated remotely. Consequently, the radioactive doses received by personnel during interventions is strongly reduced. Tests have proved the robustness of the design. Suitability for applications in accelerator environment, in particular the behaviour under irradiation, is being assessed \cite{Niccoli2017,Niccoli2017a}. 

\subsubsection{Deformable RF Bridge}

Initially proposed for the Drive-Beam interconnections in the Main Linac, the deformable RF bridges have been further developed for the LHC (modules adjacent to collimators) and HL-LHC (interconnection in the triplet areas). It is based on a deformable thin-walled structure in copper beryllium (Fig.~\ref{fig_VAC_4} right), which fulfils different requirements without the need for sliding contacts: longitudinal, angular and transversal movements due to both thermal effects (during bake-out or cool-down) and mechanical misalignments (during assembly, alignment, commissioning and operation phases). Extensive mechanical tests have been done. They show the robustness of the bridge, in particular when subjected to large transverse offset \cite{Garion2012,Perez-Espinos2016}. 

\section{Survey and Alignment}
\label{sect:Survey}
\begin{figure}[htb!]
\centering
\includegraphics[width=0.9\textwidth]{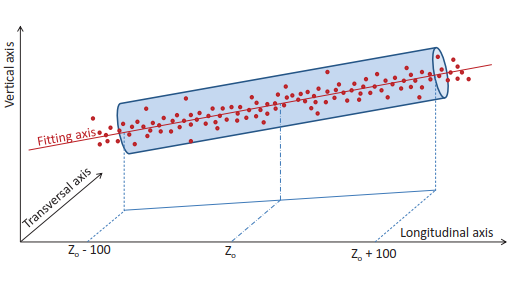}
\caption{\label{fig_SURV_1}Error budget for the absolute positioning of the reference axes of components.}
\end{figure}

In the updated baseline of CLIC, survey and alignment requirements for 380\,GeV are the same than for the 3\,TeV layout: an active, remotely controlled pre-alignment for the components of the Main Linac (ML) and Beam Delivery System (BDS); a classical pre-alignment using standard means from large scale metrology, manual jacks and the intervention of technicians in all other areas. The total error budget allocated to the absolute positioning of the reference axes of the major accelerator components (magnetic axis of quadrupoles, electrical zero of BPMs or electro-magnetic axis of RF structures) can be represented by points inside a cylinder over a sliding window of 200\,m. Its radius is equal in the ML to 14\,$\mu$m R.M.S. for MB BPM and RF structures, 17\,$\mu$m for MB quadrupole and 20\,$\mu$mfor DB quadrupoles. Further downstream in the BDS its radius is equal to 10\,$\mu$m R.M.S. over a sliding window of 500\,m (see Fig.~\ref{fig_SURV_1}).

Since the CDR \cite{Aicheler2012} deeper studies have been carried out on fiducialisation and initial alignment of components on their common support, leading to a change of strategy for this step; to improve accuracy, efficiency and flexibility. We now have a better knowledge of the sensor configuration needed for each module, following the results obtained on the Two-Beam Test Modules. Two different solutions of supporting and micrometric adjusting have been validated successfully on dedicated test setups. We have also a better knowledge of the accuracy that can be reached on the Metrological Reference Network (MRN) as simulations were confirmed by results on a 140\,m long facility.

\subsection{Determination of MRN}

The Metrological Reference Network (MRN) will be installed in the ML and BDS as soon as the tunnel floor is available. It consists of parallel, overlapping stretched wires to provide a straight reference of alignment along the whole length of the tunnel. Wire Positioning Sensors (WPS) are fixed on the same support plate to very accurately define wire-to-wire distances, combined with Hydrostatic Levelling Sensors (HLS) to model the catenary of the wires. This will represent the straight reference of alignment in the tunnel. Simulations have been carried out for such a MRN \cite{Durand2017}, considering wires with a length of 200\,m and an accuracy of alignment sensors (WPS and HLS sensors) of 5\,$\mu$m. It was shown that the standard deviation for the position of each component w.r.t. a straight line was included in a cylinder with a radius below 7\,$\mu$m. This was confirmed experimentally on a 140\,m long facility, for the radial position.

\subsection{Fiducialisation and Initial Alignment of the Components on a Common Support}

In order to facilitate the alignment process in the tunnel, several components are aligned on the same support assembly. In the CDR \cite{Aicheler2012}, this was achieved by high-precision manufacturing of the supports and outer surface of the components. As an example, both the outer diameter of RF structures and the V-shaped supports in the girders supporting them, were machined with micrometric accuracy. We initially considered determining the position of the external targets (fiducials) w.r.t. the mechanical axis of the components. A new strategy, however, has been proposed for this fiducialisation and initial alignment of the components on their common support assembly \cite{Durand2018}, based on results obtained from the PACMAN project \cite{Caiazza2017,Pacman2014} and on the development of a 5\,DOF adjustment platform \cite{Durand2018}.

The following sequence is proposed (see Fig.\,\ref{fig_SURV_2}):

\begin{figure}[htb!]
\centering
\includegraphics[width=0.8\textwidth]{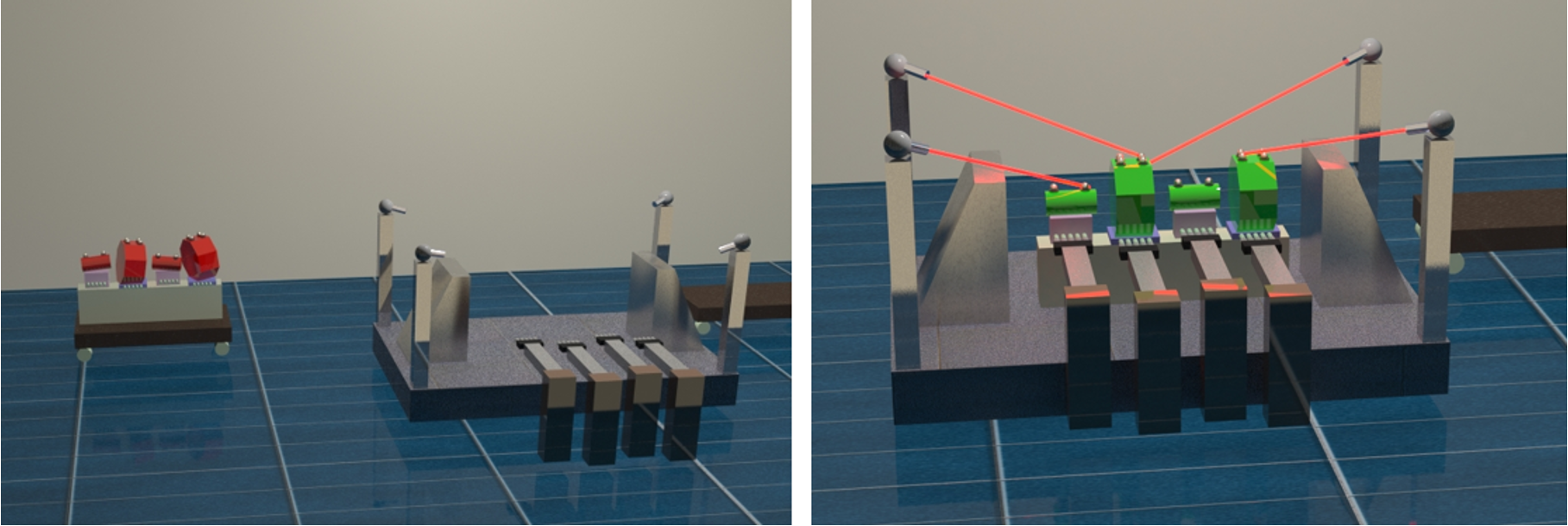}
\caption{\label{fig_SURV_2}Automatic initial alignment of the components on their girder assembly.}
\end{figure}

\begin{enumerate}
\item  Individual fiducialisation of each component, using techniques developed in the PACMAN project, using a stretched wire which now represents the active access of the component.

\item  A 5\,DoF adjustment platform is inserted between each component and the girder assembly. All components are roughly pre-aligned on the same girder.

\item  The girder is transferred to a measurement marble equipped with Frequency Scanning Interferometry (FSI) heads, to determine with micrometric accuracy the position of the alignment targets of each component in the girder assembly referential frame.

\item  Plug-in motors are temporarily connected to the 5\,DoF adjustment platforms. Once the position of components is known, the adjustment of each component can take place using these plug-in motors. If all the components are at their theoretical position on the girder, the plug-in motors are disconnected and the girder is stored, being ready for installation. Otherwise, there is an additional iteration of adjustment and position determination.
\end{enumerate}

Such a sequence can be completely automated; it can be performed at the manufacturer's premises, or at CERN in the Metrology lab, or even in the tunnel, provided the FSI heads are installed on a rigid and portable structure. It allows a gain in accuracy and efficiency (an initial alignment can be performed within a few minutes) and allows the possibility of performing alignment checks after transport in the tunnel.

\subsection{The Support Pre-alignment Network (SPN)}

The Two-Beam Test Modules, a full scale mock-up of four CLIC modules, offered the possibility to perform tests on actuators and sensors \cite{Pacman2014}. This led to a series of modifications for this updated baseline.

\begin{figure}[htb!]
\centering
\includegraphics[width=0.8\textwidth]{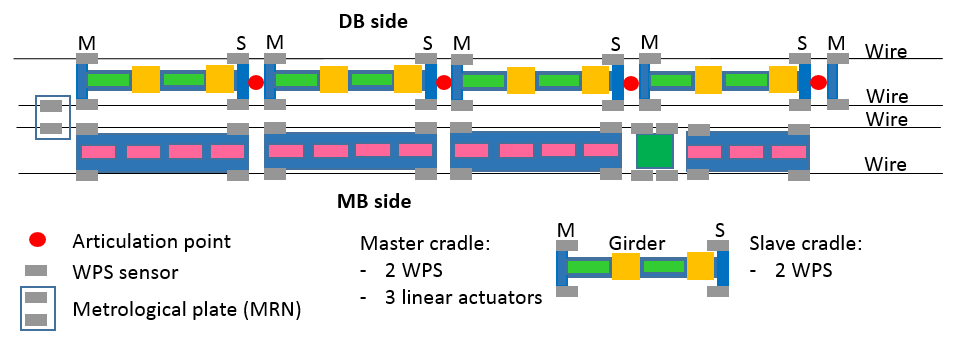}
\caption{\label{fig_SURV_3}Sensors configuration for 380\,GeV DB option.}
\end{figure}

Adjustment of girders:

\begin{enumerate}
\item  The snake configuration, with an articulation point linking two girders, will be kept for the DB side, when there are no discontinuities between girders due to MB quadrupoles (see Fig.~\ref{fig_SURV_3}). In such a configuration, two adjacent girders are interlinked by an articulation point allowing a natural smoothing of the girder and limiting the DoF between girders to three (vertical and horizontal translations and roll rotation). The articulation point will be adjustable, its adjustment being controlled by FSI measurements to an accuracy of less than 5\,$\mu$m. For such a configuration, three linear actuators, supporting the master cradle, will perform the adjustment \cite{Mainaud-Durand2018}.

\item  For the MB girders and MB quadrupoles, aligned independently, we propose to use cam movers. A configuration of five cam movers was validated for two lengths of quadrupoles: 2\,m and 0.5\,m, where we showed that the positioning requirements (sensor offsets below 1\,$\mu$m and roll below 5\,$\mu$rad) can be met in one movement using feedback from alignment sensors \cite{Sosin2016,Kemppinen2016,Kostka2017}. To fulfil the new requirement of micrometric adjustment in the longitudinal direction, we propose to add a sixth cam mover to the system and reorganize the layout of cam movers. The previous prototypes did not meet the MBQ pre-alignment stage stiffness requirement. A new prototype, taking into consideration both the positioning accuracy and stiffness, is being built. Results from the new prototype, together with more general discussion about precise multi-axis machines with pre-load, will be published in a doctoral thesis and articles later in 2019.
\end{enumerate}

\begin{figure}[htb!]
\centering
\includegraphics[width=0.8\textwidth]{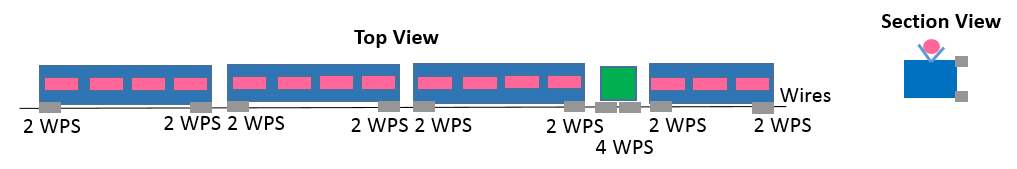}
\caption{\label{fig_SURV_4}Sensors configuration for 380\,GeV Klystrons option.}
\end{figure}

Position determination:

\begin{enumerate}
\item  Alignment sensors will be installed on girder cradles; cradles and girders being one block, of the same material.

\item  The tests performed showed that it was difficult to develop, install and calibrate an absolute inclinometer within a uncertainty of measurement, and that it was far more accurate to use two parallel wires, stretched on both sides of the module (or vertically on the same side as proposed for the klystron option in Fig.~\ref{fig_SURV_4}), measured by two WPS (one per side of cradle).

\item  A redundancy is needed in the configuration of alignment sensors to increase the accuracy of position determination and detect errors. Two additional WPS will be added on the slave cradle of the girder, in case of a snake configuration.  
\end{enumerate}

\subsection{Summary}

The recent studies undertaken at several facilities confirm that pre-alignment requirements can be fulfilled in the ML using a combination of WPS and HLS sensors for the determination of the position, linear actuators or cam movers for the adjustment of the component assembly and with a new scenario for the fiducialisation and initial alignment of the components on their assembly, based on the techniques developed in the PACMAN project. The global error budget is summarized in Table~~\ref{tab:GlobErrBudget}.

\begin{table}[htb!]
\centering
\caption{Sensors configuration for 380\,GeV DB option.}
\label{tab:GlobErrBudget}
\small
\begin{tabular}{|p{3.5cm}|p{1.5cm}|p{1.5cm}|p{1.5cm}|p{1.5cm}|p{1.5cm}|p{1.5cm}|} \hline 
 & \multicolumn{2}{p{3cm}|}{AS, BPM in $\mu$m} & \multicolumn{2}{p{3cm}|}{MB quad in $\mu$m} & \multicolumn{2}{p{3cm}|}{DB quad in $\mu$m} \\ \cline{2-7}
 &  (2012) & (2018) &  (2012) & (2018) & (2012)    & (2018) \\ \hline 
Fiducialisation & 5 (TBC) & & 10 (TBC) &  10 (TBC) &   &  \\ \hline 
Fiducials to pre-ali. sensor interface & 5 &  & 5 &  &  &  \\ \hline 
Pre-alignment sensor accuracy & 5 & 5 & 5 & 5 & 5 & 5 \\ \hline 
Sensor linearity & 5 & 5 & 5 & 5 & 5 & 5 \\ \hline 
Straight reference & 10 (TBC) & 7 (rad., vert.\,TBC) & 10 (TBC) & 7 (rad., vert.\,TBC) & 10 (TBC) & 7 (rad., vert.\,TBC) \\ \hline 
\textbf{Total error budget} & \textbf{14} & \textbf{11} & \textbf{17} & \textbf{11} & \textbf{20} & \textbf{11} \\ \hline 
\end{tabular}
\end{table}

\section{Ground Motion}
\label{sect:Ground_Motion}
\subsection{Introduction}
To obtain the desired luminosity of CLIC, very stringent specifications have to be satisfied in term of vibrations \cite{Redaelli2003,Janssens2015}. Indeed, a beam with very small emittance is sensitive to all imperfections of the Beam Delivery System (BDS) and Final Focus System. The Ground Motion (GM) / structural vibrations effects are one of the most critical causes affecting beam brightness and position stability at the Interaction Point (IP). In this section, the GM, its content and its influence on a collider like CLIC are described. The required specific instrumentations and the measurement methods are also detailed.

\subsection{Ground Motion Content}

Ground motion stems mainly from two sources: the seismic activities (or natural earth motion) and the cultural noise.
The seismic activities are basically composed of earthquakes and seismic waves. There are different kinds of seismic waves that move in different ways. The two main types of waves are body waves (can travel through the earth's inner layers) and surface waves (can only move along the surface). Earthquakes radiate seismic energy as both body and surface waves. Seismic activity is a coherent motion and dominates in low frequency range, typically lower than 2\,Hz.
The second contribution is the cultural noise that results from human activities, whether it is the outside environment like roads, air traffic or internal disturbances like cooling system or vacuum pumps. It is dominant above a few Hertz and is usually not coherent.

\begin{figure}[htb!]
\centering
\includegraphics[width=0.7\textwidth]{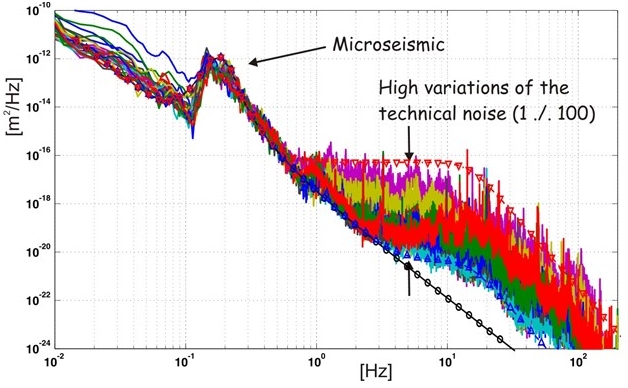}
\caption{\label{fig_GM_1}PSD measurements in the LHC tunnel at different locations \cite{Collette2010}.}
\end{figure}

Because of the widespread GM issues for large experiments \cite{Stochino2009}, GM collider measurements were already performed worldwide \cite{Amirikas2005}. To consider an environment similar to the foreseen CLIC tunnel, the LHC experiment was selected with measurements at different locations in the tunnel \cite{Collette2010} (Fig.~\ref{fig_GM_1}).

The GM Power Spectral Density (PSD) is basically a steep function of frequency which falls off as $\frac{1}{f4}$. Above 1\,Hz, the cultural noise level depends on the proximity to internal systems or cryogenic pumps for example. Below 1\,Hz vibrations are dominated by the earth motion like the micro-seismic peak at 0.17\,Hz which is due to incoming sea waves.

At the end, GM induces absolute displacements of about a few dozens of nanometers integrated r.m.s. at low frequency, which is already higher than the requirements for CLIC MBQs and BDS. However, to evaluate the disturbances consequences and the relative motion between the different elements, GM coherence has to be considered. It could be expressed like in equation\,\ref{equ:GrindEQ1} and a good coherence is equal to: 

\begin{equation}
C_{xy}\left(f\right)=\frac{{\left|P_{xy}(f)\right|}^2}{P_{xx}\left(f\right)P_{yy}\left(f\right)} 
\label{equ:GrindEQ1}
\end{equation}

\begin{figure}[htb!]
\centering
\includegraphics[width=0.8\textwidth]{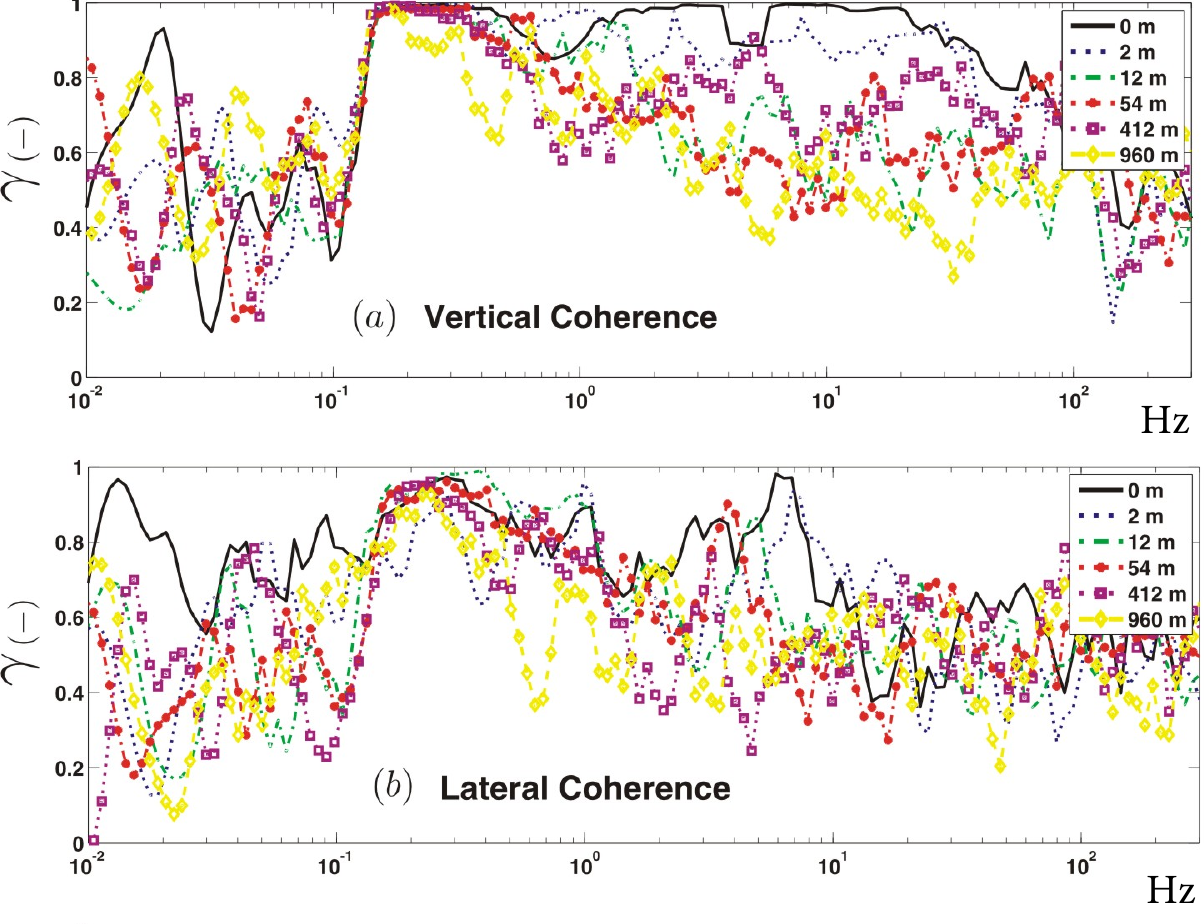}
\caption{\label{fig_GM_2}Coherence measured in vertical and lateral directions \cite{Collette2010}.}
\end{figure}

Measurements of Fig.~\ref{fig_GM_2} show that the motion is coherent over a long distance only in the narrow frequency range around the frequency of the micro-seismic wave (0.17\,Hz). Above that frequency, the coherence fades out rapidly and confirms the necessity to manage such vibrations in the mitigation strategy.

From measurements, GM can be simulated and several models are already tuned for different experiments \cite{Seyi2000} and which could be integrated into a beam dynamic simulation tool like PLACET (Program for Linear Accelerator Correction and Efficiency Tests). The prediction of the GM influence reveals that many accelerator elements, like the quadrupoles, will move independently on a large bandwidth all along the accelerator reducing the performance of the collider \cite{Balik2012,Pfingstner2013}. Various control systems and beam controls will be necessary as described in the control section, and all the mechanical design / infrastructure have to be optimized.

\subsection{Available Sensors}

Given these constraints, it is necessary to be able to measure very low amplitudes of GM and structural vibrations. Consequentially, specific instrumentation is required \cite{Bolzon2007}. 

To compare the performances of various sensors, the main properties to evaluate are the sensitivity, the resolution and the dynamic range. The sensitivity is the ratio between the real motion \textit{(M)} and its measurement \textit{(V)}, for example the sensor voltage, and could be expressed as $S=\frac{M}V$. The noise \textit{(N)} is the part of the signal \textit{(V)} which is not representative of the motion \textit{(M)} and allows to obtain the main critical criteria named resolution \textit{(R)} which is the smallest motion that the sensor is able to measure $R=\frac{N}{S}$. An efficient sensor has very small \textit{(R)} and \textit{(N)} but on a sufficient dynamic range \textit{(DR)} which is a function of the maximal measurable motion taking into account the noise.

In order to evaluate the absolute measurement of the GM (displacement, velocity or acceleration), inertial sensors are the most relevant. The working principle is based on a mass \textit{m}, linked in the axis direction that one would like to measure via a spring with a stiffness \textit{k} and a damping \textit{c}.
Among inertial sensors, the geophone is a reliable sensor. In this case, a coil is added around the mass and the motion of the mass inside the coil provides a signal representative of the velocity of the support from a few Hertz to a few hundred Hertz. The limitations of the bandwidth are the characteristics of the mechanical system: the spurious frequencies (high limit) and the fundamental frequency (low limit) \cite{Havskov2004}.
To optimise the measurement in low frequencies, these sensors are managed in a feedback loop: force-balanced accelerometers or broad-band seismometers (Table~\ref{tab:Tab_GM_1}). In this case, the relative displacement of the mass is measured and allows to evaluate the needed coil command to generate a force which compensates the internal mass motion.

\begin{table}[htb!]
\centering
\caption{Characteristics of common inertial sensors \cite{STS,GU,PMD}}
\label{tab:Tab_GM_1}
\begin{tabular}{lccccc}
\toprule
\textbf{Name} & \textbf{Range} & \textbf{Freq. res.} & \textbf{Sensitivity} & \textbf{Susp. mass} & \textbf{Size} \\ 
 & \textbf{[Hz]} & \textbf{[Hz]} & \textbf{[V/(m/s)]} & \textbf{[g]} & \textbf{[mm]} \\ 

\midrule 
GS1 & $\ge$1 & 1 & 40 & 700 & 160x75 \\ 
GS11D & $\ge$4,5 & 4.5 & 23 & 23 & 33x31 \\ 
L4C & $\ge$1 & 1 & 276 & 1000 & 76x130 \\  
STS-2 & 0.008-50 & 0.0083 & 1500 &  & 230x235 \\  
CMG-3ESP & 0.03-500 & 0.03 & 1000 & 8 & 168x258 \\  
CMG-6T & 0.03-100 & 0.03 & 2000 & 2.5 & 154x205 \\  
SP500 & 0.016-75 &  & 2000 & 0.75 & 50$^2$x100 \\ 
\bottomrule 
\end{tabular}
\end{table}

The last type of sensors are classical broadband accelerometers in which the mass is fixed on piezoelectric transducers \cite{WI}. They are well adapted to mechanical structures and infrastructure vibration measurements and could complete in high frequencies the performances of the previously described sensors.

All these sensors are very efficient for GM and / or structural vibrations, but they present two main disadvantages: they are sensitive to radiation and are limited / not dedicated to be integrated into vibration control systems. For these reasons, many specific developments are proceeding to overcome these issues.

\subsection{Dedicated Sensor R\&D}

Many R\&D projects were dedicated to inertial sensors over the past years. The majority were focused on the improvement of  commercial geophones for seismic issues \cite{Brazilai2000}, but some of them were focused on the application of geophones to the collider environment and its specific aspects \cite{Frisch2003}. For CLIC, the main recent studies were done at CERN, ULB and LAPP.

\begin{figure}[htb!]
\centering
\includegraphics[width=0.7\textwidth]{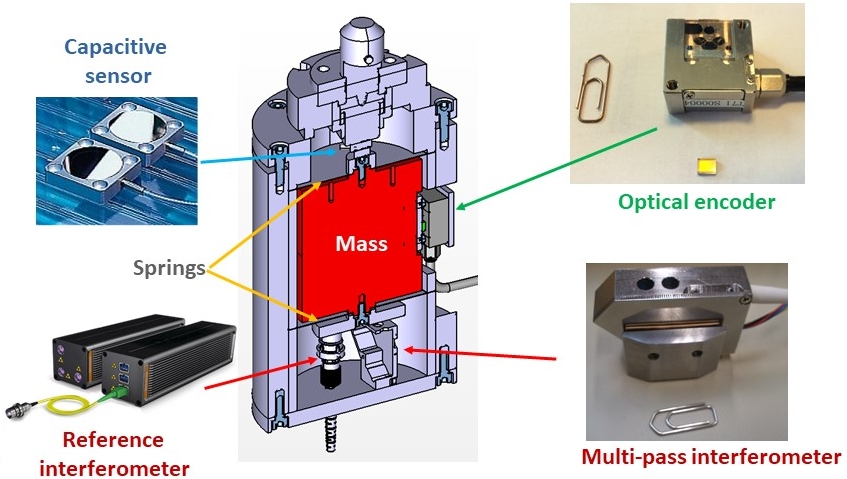}
\caption{\label{fig:fig_GM_3}Comparison measurement technology for the relative motion of the mass \cite{Novotny2017}.}
\end{figure}

The first study was conducted in the innovative doctoral program named PACMAN (Particle Accelerator Components' Metrology and Alignment to the Nanometre scale). One of the topics was to design a new vibration sensor including a comparison of the differential measurement technologies for the internal mass motion \cite{Novotny2017}. The compared transducers are capacitive sensors \cite{PI}, optics encoder \cite{MA} and interferometers \cite{AT} with a multi-pass one \cite{Pisani2009}, as shown in Fig.~\ref{fig:fig_GM_3}.

\begin{figure}[htb!]
\centering
\includegraphics[width=0.9\textwidth]{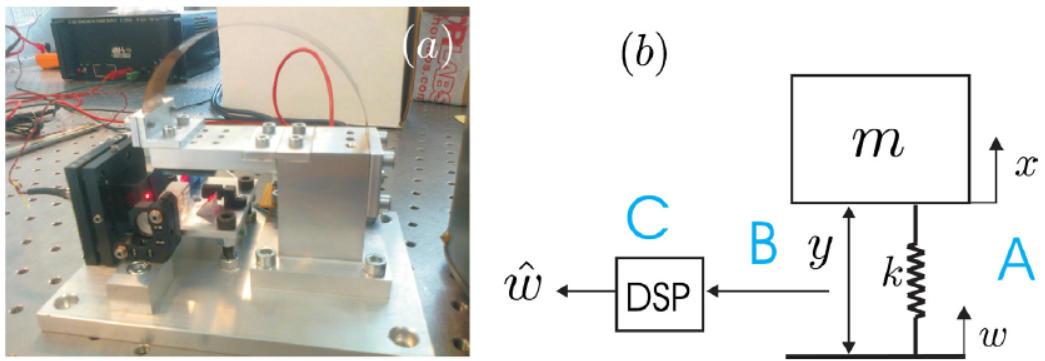}
\caption{\label{fig:fig_GM_4} Picture (a) and schematic (b) of the prototype of the interferometric inertial sensor named NOSE \cite{Collette2012a}.}
\end{figure}

\begin{figure}[htb!]
\centering
\includegraphics[width=0.9\textwidth]{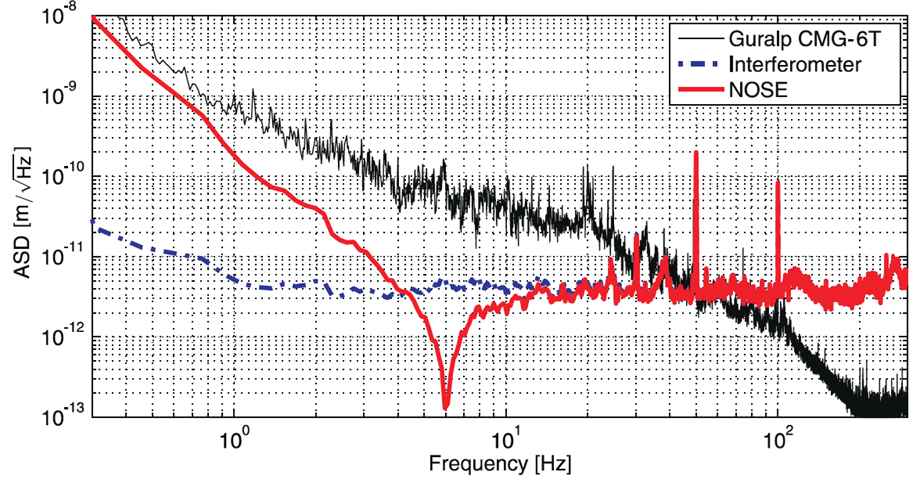}
\caption{\label{fig:fig_GM_5} Measured resolution from experimental sensors: Guralp CMG-6T (two sensors side by side in a quiet environment); estimated resolution of the NOSE prototype, measured by blocking the inertial mass \cite{Collette2012a}.}
\end{figure}

Another approach is being carried out by the Universit\'{e} Libre de Bruxelles (ULB) in collaboration with CERN \cite{Collette2012a,Hellegouarch2016}. The classical spring mass of the inertial sensor is replaced by an internal beam - pendulum in cantilever mode and the beam relative motion is measured by interferometry, Fig.~\ref{fig:fig_GM_4}. Such a setup allows for optimising the Eigen-Frequencies (avoiding as much as possible the spurious ones, increasing the resolution and being efficient in a vibration control system), decreasing the thermal noise and improving the resilience in magnetic environment.
In this setup the instrumentation noise is reduced with respect to classic inertial sensors (Fig.~\ref{fig:fig_GM_5}) and satisfies the vibration control requirements.

\begin{figure}[h!]
\centering
\includegraphics[height=5cm]{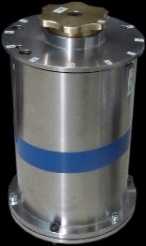}
\includegraphics[height=5cm]{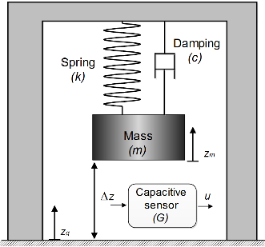}
\caption{\label{fig:fig_GM_6} LAPP sensor image and layout \cite{Caron2012}.}
\end{figure}

Yet another approach has been conducted at LAPP. A specific vibration sensor (patent n${}^\circ$\,FR\,13\,59336) has been designed (Fig.\,\ref{fig:fig_GM_6}) \cite{Caron2012}. It is based on an internal mass-spring-damper system and a capacitive sensor, which gives the relative motion between this mass and the GM. The GM can be deduced by the sensor's dynamic $L\left(s\right)=\frac{u(s)}{z_g(s)}$, through Equation~\ref{equ:DynamicL}.

\begin{equation}
\label{equ:DynamicL}
L\left(s\right)=G\frac{\frac{1}{{{\omega }_0}^2}s^2}{1+\frac{2\xi }{{\omega }_0}s+\frac{1}{{{\omega }_0}^2}s^2}{,{\omega }_0}^2=\frac{k}{m},\ \xi =\frac{c}{2}\sqrt{\frac{1}{m.k}},\ f_0=\frac{\omega_0}{2\pi}
\end{equation}

\begin{figure}[htb!]
\centering
\includegraphics[width=0.98\textwidth]{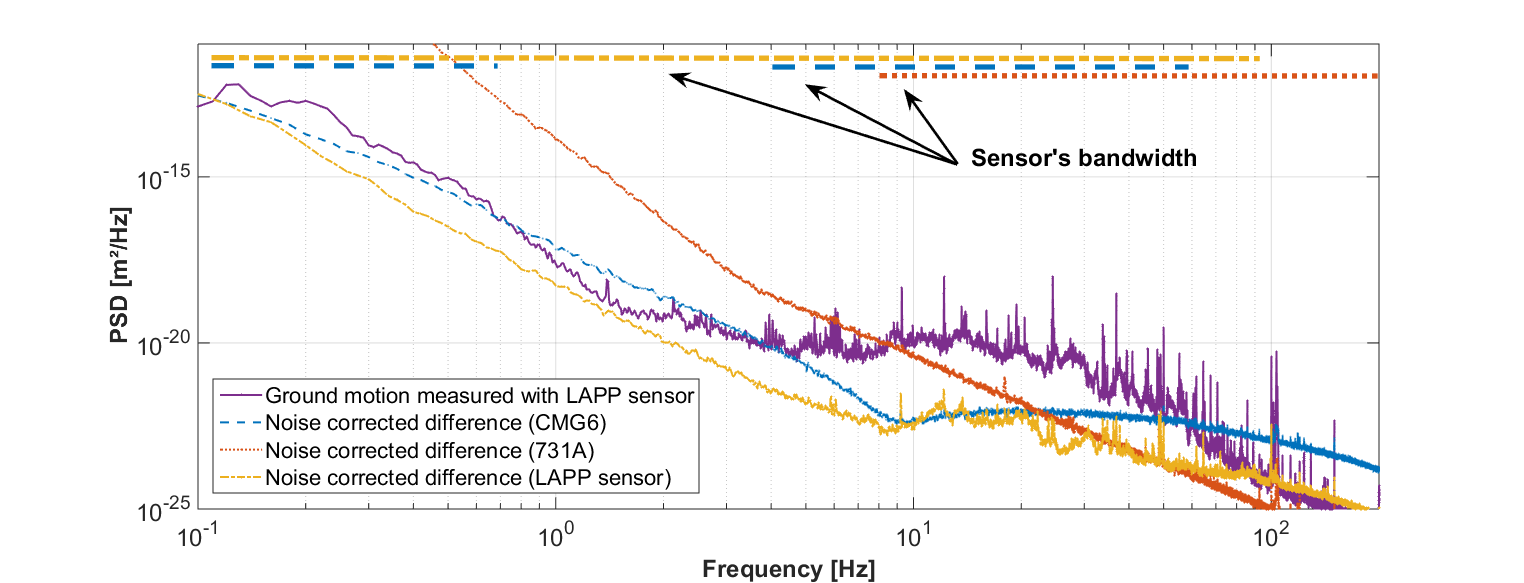}
\caption{\label{fig_GM_7} PSD of the seismic motion measured with LAPP sensor - Experimental measure of the noise of CMG-6T, 731-A, and LAPP sensor \cite{Caron2012}.}
\end{figure}

Comparative measurements were performed simultaneously with inertial sensors in a very well adapted environment, which guarantees the quality of the GM coherence. Each sensor's noise, calculated by using the corrected difference method \cite{NLC1996}, has been characterized by measuring the seismic motion with two sensors of the same model placed side-by-side. The effective bandwidth of each sensor (i.e. the ability of the sensors to measure the seismic level in a specific environment) measured at LAPP is shown in Table~\ref{Tab_GM_2} and Fig.~\ref{fig_GM_7}.

\begin{table}[htb!]
\centering
\label{Tab_GM_2}
\caption{Sensor Bandwidth and Noise Level \cite{Caron2012}}
\begin{tabular}{lccc} 
\toprule 
Sensor & Bandwidth [Hz] & Effective bandwidth [Hz]  & Noise level R.M.S. \\ 
& (constructor data) & (measured) &  @\,4\,Hz [nm] \\
\midrule
CMG-6T & 0.03-100 & 0.1-0.8 U 4-60 & 0.1 \\
731-A & 0.01-500 & 8-150 & 0.5 \\ 
LAPP & 0.1-3000 & 0.1-100 & 0.04 \\ 
\bottomrule 
\end{tabular}
\end{table}

Note that this sensor was processed in a dedicated active vibration control system \cite{Breton2013}. The control strategy has been optimized by using only one LAPP sensor in a FB loop instead of using up to four sensors for feedback and feedforward. A damping ratio of 8.5 has been achieved at 4\,Hz, leading to a r.m.s. displacement of the support of 0.25\,nm \cite{Balik2017}, very close to Final Focus CLIC specification (0,2\,nm\,@4\,Hz).

\subsection{Conclusions}

The worldwide GM measurements confirm that the GM mitigation is a critical aspect to the CLIC feasibility. However, the GM spectrum is relatively well known and could be simulated in order to design and develop the control system / beam. Additionally, the outcome of various sensor developments allow for overcoming the limitations of the current commercial sensors in an accelerator environment.

\section{Stabilisation}
\label{sect:Stab}
\subsection{Introduction}

Dynamic imperfections in the Main Linac, Beam Delivery System (BDS) and Final-Focus (FF) magnets lead to luminosity loss. Ground motion and vibrations that generate motion of the magnets and their field centre create such imperfections. The integrated R.M.S. $\sigma$${}_{x}$(f) \cite{Seryi1996} of the spectral density $\Phi$${}_{x}$(f) of absolute displacements of the magnetic field center of the quadrupoles is the first quantity to be observed. As a first approximation, the integrated R.M.S. displacement above 1\,Hz shall stay below 1.5\,nm in the vertical direction and below 5\,nm in the horizontal direction for all Main-Beam Quadrupoles (MBQ). For the FF magnets, the integrated displacement above 4\,Hz shall remain below 0.14\,nm in the vertical plane. 

The magnetic field displacements can be created by ground motion transmitted through the magnet supports and by forces arriving directly on the magnet by e.g. water cooling. Information from several measurement campaigns and models of ground motion are available in Section~\ref{sect:Ground_Motion} and in literature \cite{Seryi1996,Collette2010,Artoos2009,DESY,Peterson1993}. This first quantity $\sigma{}_{x}$(f) is only a measurement at one point over the length of an accelerator. The relative motion between quadrupole magnets at different points is also a significant factor quantified by the correlation or coherence between the different points \cite{Seryi1996,Collette2010}. The lowest frequency seismic vibrations with a long spatial wavelength move the whole accelerator lattice as a rigid body with little effect on performance if the magnet supports are sufficiently stiff. The higher frequency vibrations in ground motion will have a shorter wavelength that will create relative motion between the lattice components. The propagation from local technical vibrations and their attenuation over distance is also rather well known and was measured near the LHC in operation \cite{Guinchard2018}. An inventory of possible vibration sources acting as direct forces were studied and measured \cite{Artoos2010,Artoos2011,Bolzon2009}. 

\subsection{Overall Strategy to Reach the Required Stability}

The approach to reach the required vibrational stability combines the selection of the site with appropriate geological characteristics and seismic background, the minimization of technical noise sources in and around the CLIC tunnel and finally the use of a mechanical active vibration stabilization system. 

By adapting the civil engineering of the tunnel and cavern by, for instance, placing machinery on a separate floor structure \cite{PSI_Building} or the use of adapted technical concrete, the length of the vibration path and hence the attenuation of the vibrations will be improved. Vibration dampers can be used under the machinery. The depth of the tunnel will reduce the vibrations created at the surface.

Resonant frequencies of poorly damped mechanical structures will be excited by any broad band excitation (ground motion, water cooling, acoustic noise and ventilation) and this can create significant local vibration sources.  This was nicely demonstrated at the accelerator test facility ATF2 where beam jitter was reduced significantly after identification of two vibration sources created by poorly supported water pipes \cite{Pfingstner2014a}. Another demonstration can be found in \cite{Guinchard2018} where oscillations of the beam orbit were created by exciting support modes of the Final-Focus magnets. Therefore, the design of all components and supports in the CLIC tunnel shall have natural frequencies as high as possible by careful design. Also a verification with seismometers during the construction phase will help to remove unwanted vibration sources. The most significant consequence is the beam height from the floor that shall be as small as possible. Where possible, the mass of the components shall be reduced and also the position of support points can increase the stiffness of components (Airy points). 

\subsection{Vibration Stabilization System Studies}

From the different measurement campaigns in different particle accelerators \cite{Collette2010,Artoos2009,DESY} and as described in the CDR \cite{Aicheler2012}, a vibration isolation system is needed to reach the specified stability levels. A survey of existing vibration stabilization or isolation systems in various fields of precision engineering was performed \cite{Collette2011,Collette2015}. Studies were carried out at CERN and at LAPP (Annecy) and ULB (Brussels) and prototypes have been constructed. 

\begin{figure}[htb!]
\centering
\includegraphics[width=0.5\textwidth]{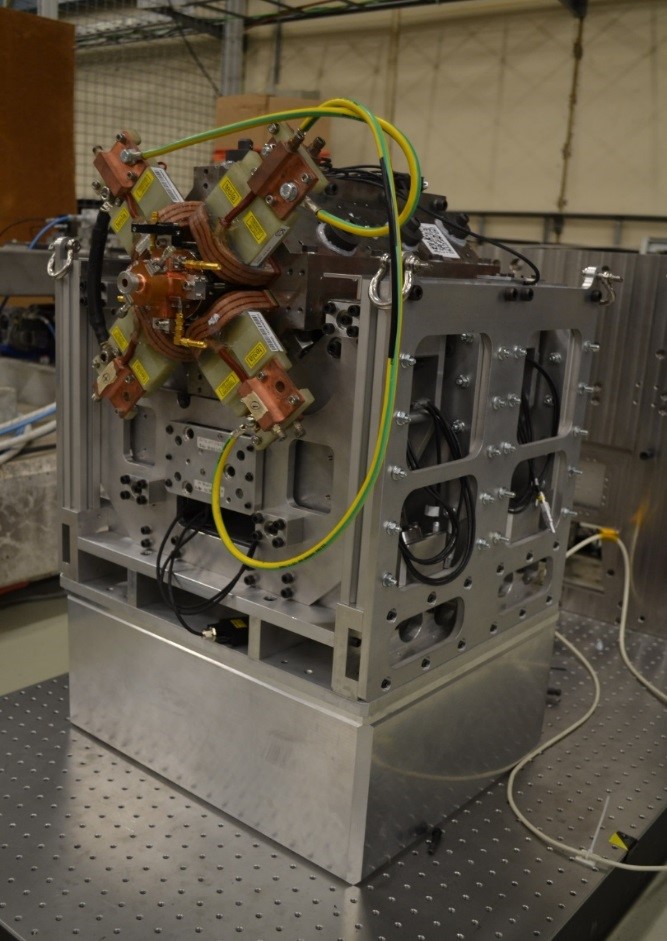}
\caption{\label{fig_STAB_1}Main Beam Quad type~1 stabilization prototype.}
\end{figure}

The very stringent alignment requirements and the presence of forces acting directly on the MBQ (water-cooling, acoustic noise, vacuum pipes and electrical cabling) as well as the low frequencies for stability, resulted in the choice of active stabilization \cite{Collette2010a,Collette2010b} for the 4,000 Main-Beam Quadrupoles (MBQ). Active stabilization is based on a stiff actuating support that reduces the compliance for the direct acting external forces mentioned above. The use of stiff piezo actuators gives an additional possibility to make fast repositioning of the magnets possible in between beam pulses (every 20\,ms) with a high resolution; so-called nano-positioning \cite{Esposito2012,Collette2011a}. Five prototypes were built with increasing complexity, mass and degrees of freedom. The first four prototypes in Table~\ref{tab:Tab-STAB_1} (with references for complete information) all reached the requirements with commercially available piezo actuators and seismic sensors. Prototype~4 is a MBQ (type~1) prototype magnet with nominal magnetic field and water cooling that reached the requirements from a higher vibration back ground (R.M.S. ratio). The last MBQ prototype (Fig.~\ref{fig_STAB_1}) is a complete, fully integrated stabilization system with type~1 MBQ with the same measured transmissibility as prototype~4 but that was not yet tested in a low vibration background (not in Table~\ref{tab:Tab-STAB_1}).

\begin{table}[htb!]
\centering
\small
\caption{List of prototypes that reached the requirements}
\label{tab:Tab-STAB_1}
\begin{tabular}{p{1.6in}p{0.8in}p{0.7in}p{0.6in}p{1.2in}} 
\toprule
\textbf{Prototype} & \textbf{Integrated R.M.S. (nm)} & \textbf{R.M.S.\newline Ratio} & \textbf{Set-up} & \textbf{Remark} \\ 
\midrule
1. Membrane one d.o.f \cite{Janssens2015} & 0.3 (1\,Hz)\newline 0.6 (1\,Hz)\newline 0.2 (4\,Hz) & Ratio 6\newline Ratio 10\newline Ratio 8.5 & FF+FB\newline Analogue & Down scaled \\ 
2. Tripod \cite{Collette2011a} & 0.9 (1\,Hz)\newline 0.7 (4\,Hz) & Ratio 2-2.5 & FB\newline Digital & Down scaled \\  
3. Two d.o.f. x-y \cite{Janssens2015} & 0.5 (1\,Hz) & Ratio 9 & FF + FB Analogue Hybrid & Real scale \\  
4. Magnet tripod \cite{Janssens2015} & 0.45 (1\,Hz)\newline 0.35 (4\,Hz) & Ratio 13.3 & FF + FB Analogue Hybrid & Real scale, Measured with nominal magnet field and cooling water flow \\ 
5. LAPP \cite{Balik2017,Balik2018} & 0.25 (4\,Hz) & Ratio 8.5 & FB\newline Digital & Down scaled \\ 
\bottomrule 
\end{tabular}
\end{table}

The FF magnets are supported by a cantilevered structure in the particle detector and their stabilization require a higher level of integration with the surrounding infrastructure. Different approaches were studied \cite{Caron2012,Collette2012} with or without passive components. Prototypes were built with specially developed sensors \cite{Balik2017,Balik2018,Watchi2018} or using stiff carbon tie rods for active damping \cite{Collette2013}. Two prototypes were measured with a low vibration back ground and came very close to the requirements (number 1 and 5 in Table~\ref{tab:Tab-STAB_1}). 

The success of studies and prototype with respect to integrated r.m.s. displacement will depend on the resolution/noise level and controller stability that can be reached with the combination of sensors, actuating structures, controller and electronics. Some of the studies not reaching the integrated r.m.s objective have however also demonstrated more stable and hence more reliable controllers with for instance the use of collocated force sensors \cite{Collette2013,Collette2015} or systems that have components better adapted to the accelerator environment (radiation). 

The measured transmissibility or transfer functions of the prototypes were used to estimate the luminosity that can be obtained with the respective stabilization strategies \cite{Janssens2011,Snuverink2011,Balik2012}. These studies indicated that the shape of the transmissibility function is more important for the luminosity than the obtained integrated r.m.s. displacement. This understanding has triggered the development of more adapted sensors for the stabilization (see Section~\ref{sect:Ground_Motion}) and defined a new luminosity objective for the combined control systems (stabilization and beam).  

\subsubsection{The QD0 Stabilisation System}

One of the most critical elements in terms of stabilization is the final focusing quadrupole QD0, which is positioned on the tunnel floor, just outside the detector. It focuses the vertical beam size down to about 2.9\,nm r.m.s. The distance L* of the downstream end of this quadrupole to the IP has been chosen to be 6\,m, the shortest distance compatible with detector space requirements. Any movement of the quadrupole in the transverse plane would affect the transverse position of the beam at the IP by a comparable amount. Therefore, the quadrupole has to be mechanically stabilised to 0.3\,nm r.m.s. in the vertical plane for frequencies above 4\,Hz. The IP feedback system complements the stabilisation system at lower frequencies. It measures the position of the outgoing beam and applies a calculated kick to the incoming beam in order to optimise  luminosity. Bunch-to-bunch correction is not possible, but the latency loop of the system is short enough to allow several iterations during the 156\,ns bunch train. An active pre-alignment system ensures that the average position of the QD0 is correct to within 10\,$mu$m with respect to the other BDS elements.

\begin{figure}[htb!]
\centering
\includegraphics[height=5.5cm]{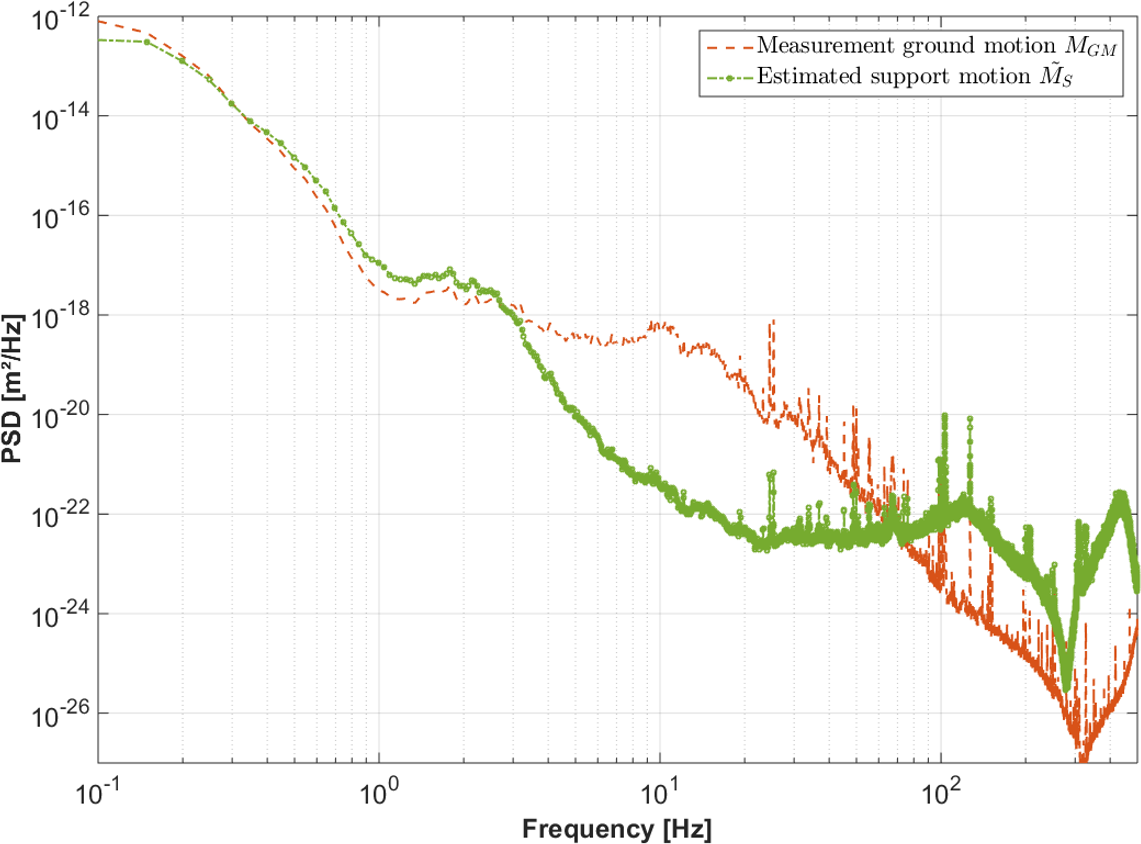}
\includegraphics[height=5.5cm]{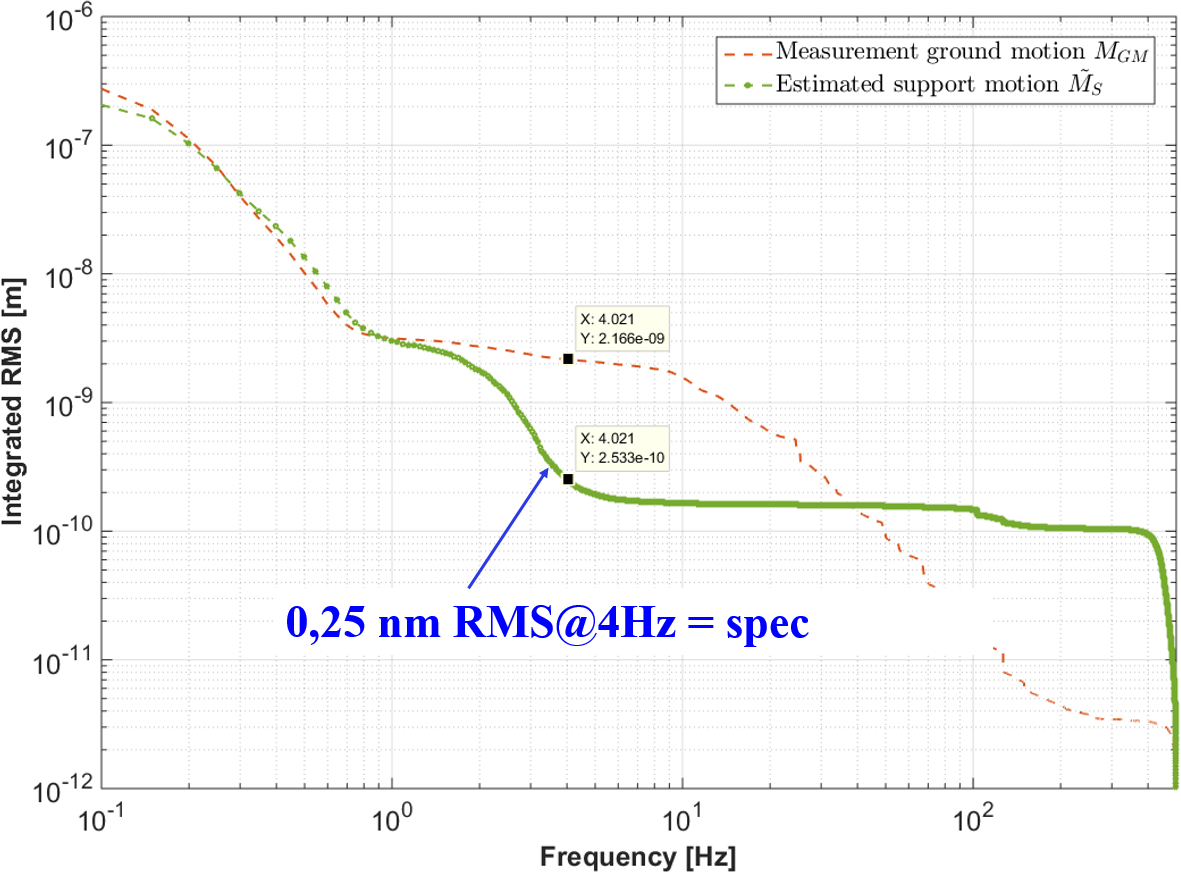}
\caption{\label{fig_MDI_2}The measurements of the new stabilisation with and without the QD0 stabilisation system. On the left hand side the PSD, on the right hand side the integrated r.m.s. achieved.}
\end{figure}

An intense R\&D program, including the development of custom-made new sensors and an active support, has allowed to reach stabilisation at the 0.25\,nm r.m.s. at 4\,Hz and above \cite{Balik2018}. The result of the measurements is shown in Fig.\,\ref{fig_MDI_2}. Therefore, a pre-absorber is no longer needed (or could in the worst case be implemented with existing commercial solutions rather than with a specific development).

\subsection{Readiness for CLIC}

The work described above shows that the required level of magnet field centre stability can be reached on scaled or real size prototypes in a laboratory environment.  Most equipment used was commercially available or the components that were developed in house are technologically well within reach. Also the integration with other CLIC systems seems technically possible. The studies and prototypes built by different teams also brought forward the following common factors: 

\begin{itemize}
\item Piezo actuators combine the required stiffness with the needed sub nano-metric resolution and suitable band width. It is the technology that is commercially most available. Piezo actuators are non linear actuators with hysteresis, this is linearized by special piezo amplifiers.

\item High resolution vibration sensors are based on the measurement of an inertial mass with a coil, capacitance measurement, laser interferometer or optical encoders. 

\item Controller hardware: digital or analogue (hybrid) circuits requiring special care for phase margin in the control (ADC delay, impedance problems).

\item Low frequency input and output signals with a high dynamic range.
\end{itemize}

Most of these factors call for short cable lengths, i.e. the controller hardware (amplifiers, conditioners and controller) shall be near  the stabilization system. This is not evident in an accelerator tunnel with radiation. Complete shielding against radiation in high energy particle accelerators is difficult and would be expensive. It seems, therefore, that we will be required to develop and test radiation hard components and at the same time to develop components that can deal with longer cable lengths.

\section{Beam Transfer}
\label{sect:BT}
\subsection{Main Beam: Damping Ring Kicker Systems}

To achieve high luminosity at the interaction point, it is essential that the beams have very low transverse emittance: the Pre-Damping Ring (PDR) and Damping Ring (DR) damp the beam emittance to extremely low values in all three planes. In order to limit the beam emittance blow-up, due to oscillations at extraction from the DR, the combined flat-top ripple and droop of the field pulse must be less than $\pm$\,0.02\,\%. In addition, the total allowable beam coupling impedance, in each ring, is also very low: 1\,$\Omega$\,$\times$\,\textit{n} longitudinally and 10\,M$\Omega$/m transversally. This section discusses means for achieving the demanding requirements for the DR kickers. Table~\ref{tab:BT_table1} shows the specifications for the DR extraction kicker system.

\begin{table}[htb!]
\centering
\caption{DR Extraction Kicker Specifications}
\label{tab:BT_table1}
\begin{tabular}{lc} 
\toprule
\textbf{Parameter} 								& \textbf{DR} \\ 
\midrule
Beam Energy (GeV) 								& 2.86 \\  
Deflection Angle (mrad) 						& 1.5 \\ 
Aperture (mm) 									& 20 \\ 
Field maximum rise and fall time (ns) 			& 1000 \\  
Pulse flat-top duration (ns) 					& $\sim$160 or 900 \\ 
Flat-top reproducibility 						& $\pm$1x10${}^{-4}$ \\ 
Stability 										& $\pm$2x10${}^{-4}$ \\ 
Field inhomogeneity (\%), over 1~mm radius		& $\pm$0.01${}^{*}$ \\ 
Repetition rate (Hz) 							& 50 \\ 
Available length (m) 							& $\sim$1.7 \\ 
Vacuum (mbar) 									& 10${}^-10$ \\ 
Pulse voltage per stripline (kV) 				& $\pm$12.5 \\  
Stripline pulse current (A) 					& $\pm$309 \\ 
\bottomrule
\end{tabular}
\end{table}

To achieve the demanding specifications for low beam coupling impedance, striplines will be used for the DR extraction kickers \cite{Belver-Aguilar2014,Belver-Aguilar2015}. Striplines have both an odd and an even mode impedance: the odd mode is when both electrodes are driven to opposite polarity voltages, to extract beam from the DR, whereas the even mode is when the electrodes are at the same potential, e.g. not driven by pulse generators. The characteristic impedance of both odd and even modes should ideally be optimized to 50\,$\Omega$. However, for coupled electrodes this is not possible to achieve \cite{Belver-Aguilar2014,Belver-Aguilar2015}. Existing designs of stripline electrodes, e.g. \cite{Garcia2009,Alesini2010,Naito2011}, could not achieve the demanding specifications. Hence a novel electrode shape has been developed for the CLIC DR extraction kicker that achieves the excellent field homogeneity, good matching of both odd and even mode characteristic impedances, and a decrease in the beam-coupling impedance at low frequencies \cite{Belver-Aguilar2015}. 

To achieve the aforementioned requirements, various shapes of electrodes were studied and optimized. In addition the electrode supports, feedthroughs, and manufacturing tolerances were studied. A novel shape of electrode, called a ``half-moon electrode'', was selected as the optimum shape \cite{Belver-Aguilar2015}. For the DR extraction striplines, manufactured with the optimized half-moon electrodes, the odd mode characteristic impedance is 40.9\,$\Omega$. Since an inductive adder will be connected to each electrode with commercial coaxial cable of 50\,$\Omega$ impedance, and each electrode will be terminated with 50\,$\Omega$, there is an impedance mismatch for the odd mode at both the input and output of the electrodes. A mismatched odd mode impedance can significantly influence the striplines performance: predictions for the influence of the odd mode characteristic impedance upon the contribution of each field component, electric and magnetic, to the deflection angle, have been presented in \cite{Belver-Aguilar2016}. A new idea has been proposed to match the load side characteristic impedance for both the odd and even modes of excitation of the striplines \cite{Belver-Aguilar2015}.

The maximum field inhomogeneity allowed is $\pm$\,0.01\,\%, over 1\,mm radius (Table~\ref{tab:BT_table1}), although a radius of 0.5\,mm has also been accepted from beam optics considerations \cite{Apsimon2013}. The magnetic field homogeneity for different frequencies, for the optimized stripline geometry, has also been studied \cite{Belver-Aguilar2016}.

Stripline kickers are generally assumed to have equal contributions from the electric and magnetic field to the total deflection angle, for ultra-relativistic beams. Hence, the deflection angle is usually determined by simulating the striplines from an electrostatic perspective. However, recent studies show that, when exciting the striplines with a trapezoidal current pulse containing high frequency components, the magnetic field changes during the flat-top of the pulse, due to eddy currents induced in the stripline electrodes, and this can have a significant effect upon the striplines performances \cite{Belver-Aguilar2016}. The variation of the magnetic field during the pulse flat-top, and its effect upon the deflection angle, has been presented in \cite{Belver-Aguilar2016a}. The solution proposed to compensate for this variation is to modulate the pulses created by the inductive adders \cite{Holma2018} which will power the striplines.

A first prototype of the extraction kicker for the CLIC DR has been designed and built. The beam coupling impedance of the striplines has been studied analytically, as well as numerically with CST Particle Studio followed by measurements in the laboratory. A new approach for understanding the dipolar component of the horizontal impedance has been derived, when considering both odd and even operating modes of the striplines. This new approach, presented in \cite{Belver-Aguilar2017}, has been used to understand the differences found between the predicted transverse impedance and the two wire measurements carried out in the laboratory for the prototype CLIC DR striplines.

In order to complete its characterization, the prototype stripline kicker has been installed in one of the medium straight sections, of the ALBA Synchrotron, to be tested with beam \cite{Iriso2018}. The main purposes of these tests is to measure the beam impedance and DC electric field homogeneity for comparison with predictions from simulations. However, during conditioning of the electrodes, with circulating beam, pressure increased with beam current: this unexpected behaviour of the pressure might be related to ion instabilities and requires further studies. Since the performance of the striplines under these conditions is not compatible with user operation, the striplines have to be removed from the ring during user operation and are only installed for measurements on the striplines -- this has limited the time available for measurements, but some initial measurements have been carried out nevertheless \cite{Iriso2018}.

The DR extraction kicker has very demanding specifications: the combined flat-top ripple and droop of the field pulse must be within $\pm$\,0.02\,\%. In addition, the flat-top repeatability must be within $\pm$\,0.01\,\% (Table~\ref{tab:BT_table1}). An inductive adder has been selected as a contender for achieving the demanding specifications for the DR extraction kicker modulator \cite{Holma2018,Holma2011,Holma2015,Holma2018a,Holma2018b}. The inductive adder is a solid-state modulator, which can provide relatively short and precise pulses: it is modular and thus the design can be adapted to various requirements.

An inductive adder consists of multiple layers, each of which has a transformer: the transformer usually has a 1:1 turn ratio in order to ensure that it is suitable for fast pulses. The single turn primary totally encloses a magnetic core; hence, the leakage inductance of this geometry is negligible \cite{Takayama2010}. The secondary winding of each of these transformers is connected in series: hence a step-up voltage ratio of 1:\textit{N} is achieved by using \textit{N}-layers, with adequate voltage isolation. The primary circuit typically has many parallel branches: each branch generally contains a single capacitor and a single power semiconductor switch. All the power semiconductor switches and gate drive circuits are referenced to ground and there are no electronics referenced directly to the high-voltage output pulse. In general, the capacitors of all the layers can be charged with a single power supply -- this is not, however, necessary and in some applications the capacitors are deliberately charged to different voltages to allow modulation of the output pulse \cite{Cook2002}. Each layer has an array of clamp diodes which carry magnetizing current from the core following turn-off of the power semiconductor switches in the layer \cite{Holma2015}.

An inductive adder has good built-in fault tolerance and redundancy, because if one or more solid-state switches of a layer of the inductive adder stack fails to turn-on, the magnitude of the output pulse of the inductive adder is reduced by, at most, only the voltage of a single layer. Since the inductive adder is of a modular design, extra layers can be added to improve redundancy and/or increase voltage rating. In addition, the modular construction allows for a good scalability, adaptation of individual components to the particular application, and a path for future upgrades. 

Considerable research and development has been carried out on the inductive adder for the CLIC DR \cite{Holma2018,Holma2011,Holma2015,Holma2018a,Holma2018b}. A significant advantage of the inductive adder over some alternative technologies is the ability to modulate the output \cite{Takayama2010}. Recent measurements on a prototype inductive adder show that the flat-top stability achieved by applying modulation was $\pm$\,0.02\,\% ($\pm$\,2.2\,V) over 900\,ns at 10.2\,kV output voltage. This pulse meets the stability specifications for the 2\,GHz (baseline) and 1\,GHz specifications for the DR extraction kicker.

\begin{figure}[htb!]
\centering
\includegraphics[width=0.95\textwidth]{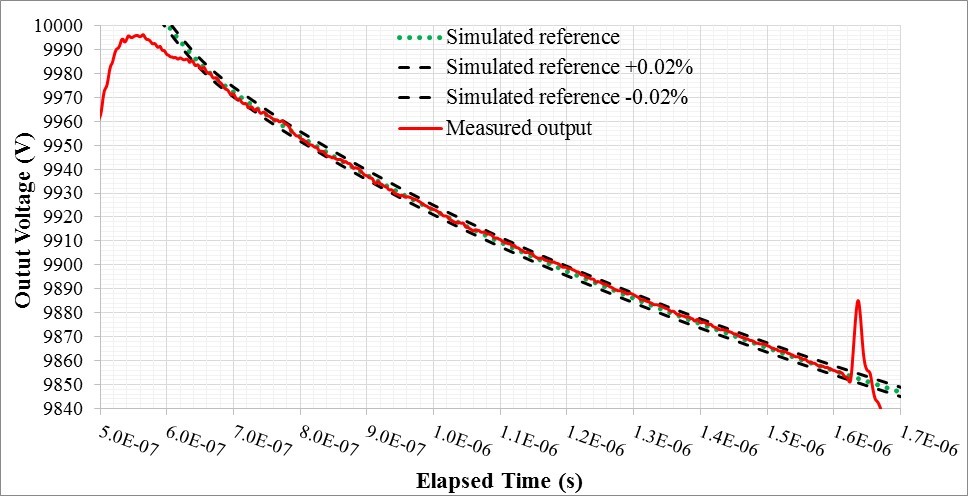}
\caption{\label{fig_BT_1}Simulated optimum waveform for the CLIC DR extraction kicker (green) with $\boldsymbol{\pm}$\,0.02\,\% stability margins (black), and an average of 100 measured pulses (red) of the prototype inductive adder.}
\end{figure}

As mentioned above, exciting the striplines with a trapezoidal current pulse results in induced eddy current in the electrodes. This causes the magnetic field to increase in amplitude during the ``flat-top'' of the current pulse \cite{Belver-Aguilar2016a}. The solution proposed, to compensate for this variation, is to modulate the pulse created by the inductive adder: a controlled decay-waveform is required \cite{Holma2018,Holma2018b}. Fig.\,\ref{fig_BT_1} shows a simulated reference controlled decay waveform (green) with error margins for stability of $\pm$\,0.02\,\% (black): the measured waveform (red) is within the $\pm$\,0.02\,\%, for more than 900\,ns, with respect to the simulated reference waveform. This fulfils the 1 and 2\,GHz specifications for the CLIC DR extraction kicker system.

As a result of the low source impedance, the source voltage of an inductive adder does not need to be doubled, unlike a matched impedance Pulse Forming Line (PFL) or Pulse Forming Network (PFN). The solid-state switches typically used in the inductive adder are either metal-oxide-semiconductor field effect transistors (MOSFETs) or insulated-gate bipolar transistors (IGBTs). These solid-state switches can be opened when conducting full load current, hence only a portion of the stored energy is delivered to the load during the pulse: therefore a PFL or PFN is not required. In addition, opening of the solid-state switches potentially limits the duration of fault current in the event of a magnet (load) electrical breakdown. Hence the inductive adder is also seen as a promising technology for use in existing CERN kicker systems \cite{Barnes2015,Woog2018}, where a suitable replacement for PFL cable is challenging to purchase these days, as well as for possible future kicker systems at CERN, e.g. for injection systems of the Future Circular Collider \cite{Kramer2016,Woog2018a}. As a result of the use of magnetic cores for the transformer \cite{Holma2015}, a drawback of the inductive adder is its limitations concerning the available pulse duration: however many kicker magnet applications at CERN require pulse durations of only up to 2.6\,$\mu$s, which is achievable with the discussed design.

\subsection{Conclusions and Future Work}

A novel shape of stripline electrodes has been designed to meet the demanding CLIC DR extraction kicker requirements for field homogeneity and low beam coupling impedance: this shape also results in reasonably close values for odd and even mode impedance. Nevertheless, to match the odd mode impedance an additional resistor is proposed to be connected on the output of the striplines, between the electrodes. Detailed studies show that, during field rise and fall, eddy currents are induced in the electrodes, which modifies the deflecting magnetic field. To compensate for this the output pulse of the inductive adder can be modulated -- this modulation has been demonstrated. The striplines have been prototyped, and beam coupling impedance measurements carried out. The prototype electrodes have recently been installed in the ALBA synchrotron and initial measurements carried out to characterize the field homogeneity and beam coupling impedance. Detailed characterization of the striplines, in an accelerator, is required to validate the predictions for the longitudinal and transverse beam coupling impedance and field uniformity.

Measurements of the output pulse of the inductive adder show a stability for a flat-top of $\pm$0.02\,\%, over 900\,ns, for a 10.2\,kV pulse. For the controlled decay waveform, the measured stability was $\pm$0.02\,\%, with respect to the simulated reference waveform, for 900\,ns duration at 10\,kV: this stability meets the requirements for the CLIC DR extraction kicker, at approximately 80\,\% of nominal voltage. Tests and measurements are ongoing and the output pulse voltage will be increased to 12.5\,kV in the near future. In addition, a second inductive adder has been constructed: the polarity of the output pulse is changed by moving the output connector from one end of the inductive adder to the other. Furthermore, work is ongoing to develop the necessary control system, and an automated way to modify the analogue modulation to achieve the required shape and stability of the deflection waveform. Commercial terminating resistors have proved to be unreliable for the pulse loading. Hence, suitable terminating resistors, for use with the inductive adder, have been designed: these will be manufactured in the near future and tested with the striplines and inductive adder.

The striplines will be tested in the laboratory with pulses of up to $\pm$12.5\,kV, using both inductive adders: long-term testing is required to verify the reliability and response during fault conditions. In addition, it is highly desirable that the striplines and inductive adders are fully tested together with beam: a single bunch could be used to scan through the flat-top and verify the deflection stability and repeatability of the flat-top. Such a flat-top measurement is considered essential to confirm the predictions for the required controlled decay waveform, to achieve a flat-top deflection pulse, and addition as it is not possible to guarantee the absolute precision of the electrical measurements.

\section{Normal Conducting Electro-Magnets and Permanent Magnets}
\label{sect:Magnets}

\subsection{Introduction}

Even though the size of the initial stage of CLIC at 380\,GeV is substantially smaller compared to the 3\,TeV final machine, the number of normal-conducting Electro-Magnets (EM) and Permanent Magnets (PM) to be produced will stay well beyond the quantities produced so far for particle accelerators. Therefore the industrialization, cost optimization, quality control, assembly, and installation of such large numbers of magnets will be one of the major challenges of the project.
Latter stages concern the Main Linacs and their long transfer lines, but the global layout and beam output energy of the Drive Beam and Main-Beam Injector systems are required for the initial phase.  During an intensive R\&D design campaign, various magnet types have been designed with numerical methods to assess their feasibility and evaluate their dimensions, power consumption, and cost. More than 15 prototype magnets have been manufactured for the technically most challenging configurations. The outcome of this R\&D program is summarized in a magnet catalogue which has been used to select most of the magnetic elements of the new lattices for the various parts of the 380\,GeV machine.

\subsection{Drive-Beam Magnets}

\subsubsection{Quadrupoles in the Main Decelerator}

Two versions of quadrupoles have been developed in parallel for the Drive Beam: a conventional electro-magnet (EM) version, and a tuneable permanent magnet (PM) version. 

The parameters and performance of the EM version have already been presented in the CDR \cite{Aicheler2012} as well as in several other papers \cite{Modena:2014}. Since the CDR, this design is the baseline and several prototype units have been manufactured, which confirmed the design parameters and achievable tolerances. Some of these magnets have been installed in the  CLIC Test Facility (CTF3) for further studies with beam.

\begin{figure}[htb!]
\centering
\includegraphics[width=0.6\textwidth]{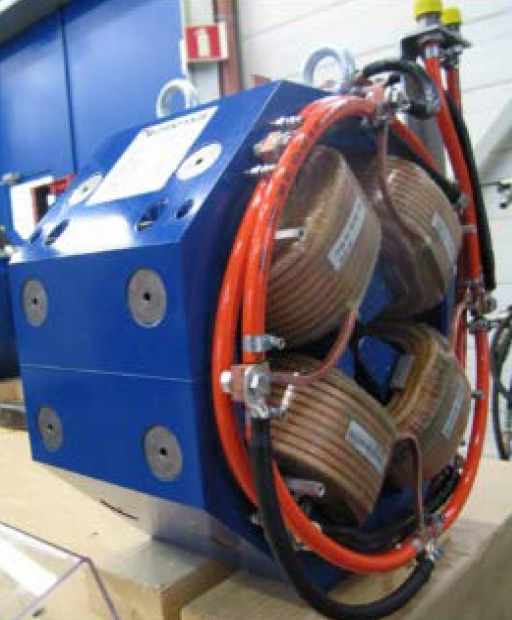}
\caption{\label{fig_Magnets_1}An electro-magnet DBQ prototype.}
\end{figure}

As an alternative to the EM version, a PM version has been extensively developed in collaboration with the STFC Daresbury Laboratory in the UK \cite{Shepherd:2013,Shepherd:2014,Shepherd:2012}. The obvious advantage of this solution is the major savings in electrical power and cooling requirements. However, as the field gradient scales after each Two-Beam module all along the decelerator, up to a factor of 10 for the 3\,TeV machine,two types of mechanically tuneable permanent magnets are needed. The first, the High Gradient PMQ, covers the first 60\,\% of the energy range, while the Low Gradient PMQ covers the last 40\,\% of the energy range. The mechanical adjustment of the gradient is piloted with a stepper motor which controls the positioning of the permanent magnet blocks via transfer gearboxes. The challenge resides in the accuracy and stability of the mechanical parts, as well as in the stability of the permanent magnet field. 

\begin{figure}[htb!]
\centering
\includegraphics[width=0.5\textwidth]{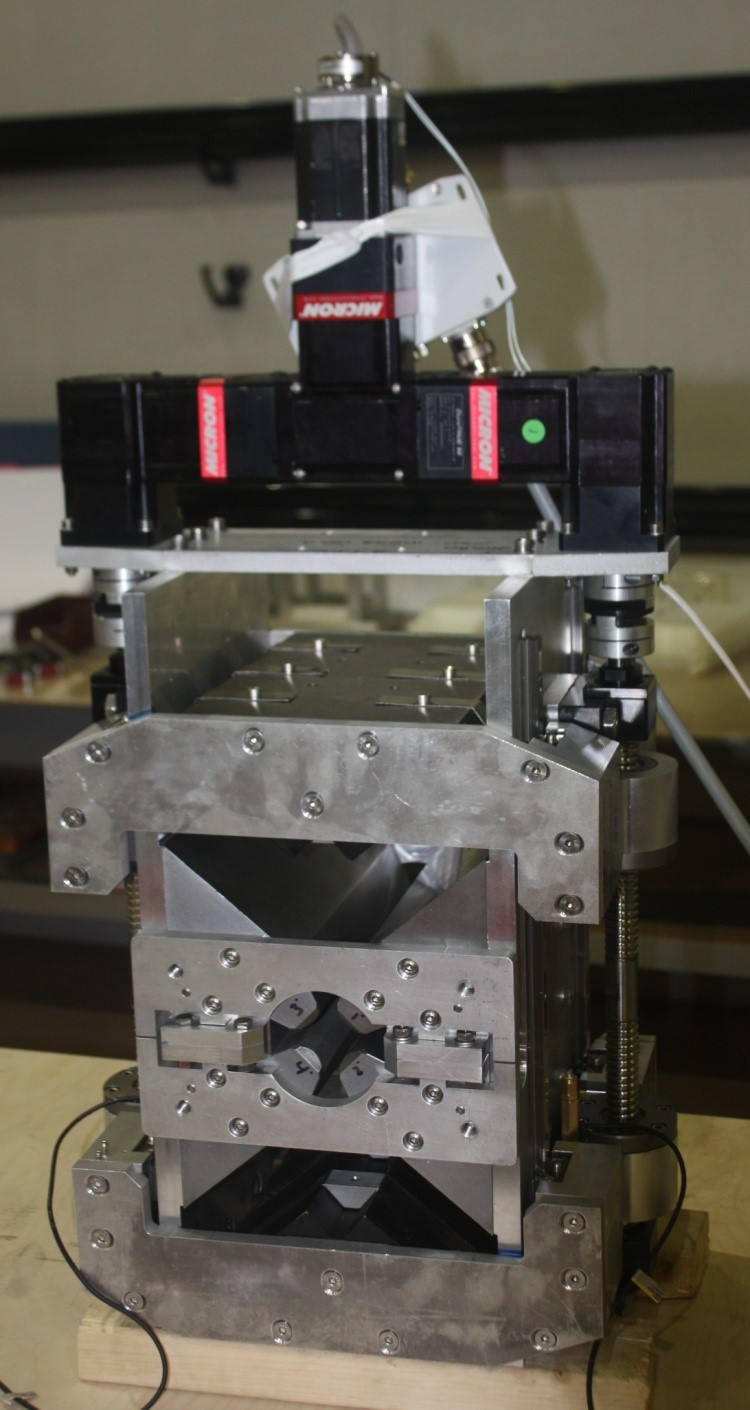}
\caption{\label{fig_Magnets_2}Prototype of the latest version of the PM DBQ, with optimized design.}
\end{figure}

Prototypes of both High and Low Gradient magnets have been built and measured at STFC Daresbury Laboratory (see Fig.~\ref{fig_Magnets_2}, confirming the feasibility of this concept. For the 380\,GeV version, the High Gradient type will be sufficient to cover the whole energy range of the decelerator. Since 2014, the STFC Daresbury Laboratory has continued the development of the PMQ designs to further improve their performance, stability while reducing the manufacturing costs by revisiting in particular the motion system, permanent magnet blocks and by reducing the number of components. With these improvements, the design is now considered mature and cost optimized. It has been proposed to change the baseline from the EM version to the PM version for the construction of these magnets. Such a decision is, however, subject to the assessment of the PM material stability in the radiation conditions of CLIC. A survey of results from studies on the radiation damage to PM materials has been carried out \cite{Shepherd:2018} and will be used to analyse the criticality of this decision.

\subsubsection{Beam Transport Magnets}

The lattices of the DB transport complex -- which include the combiner rings, delay loop and transfer lines have been reworked to allow standardization of the magnets. About 90\% of the new layout is using magnetic elements from the original magnet catalogue. The rest corresponds to some quadrupoles and sextupoles which exceed the current capabilities of the magnets and whose design needs updating. This is also the case for the dipoles in the Turn-Arounds (TA) for which the optics have now adopted combined function magnets. A bespoke magnet design will be developed once the specifications are frozen.

\subsection{Main-Beam Magnets}

\subsubsection{Main-Beam Linacs}

The Main-Beam Linacs' magnetic lattice foresees the focusing of the e$^+$ and e$^-$ beams with two types of MBQs electro-magnets which have a similar cross section but different magnetic lengths (see Table~\ref{Tab_Magnets_6}).  The performances in terms of precision of pole shape and quadrants assembly has been achieved in a 2\,m long prototype magnet developed for the final energy \cite{Modena:1662725}. Therefore the only  remaining challenge expected for this type of magnet is the optimization of the production process considering the large quantity of units to be produced. 

\begin{table}[htb!]
\centering
\small
\caption{\label{Tab_Magnets_6}  List of magnet types for both 380\,GeV and 3\,TeV.}
\begin{tabular}{l c c c c}  
\toprule 
 & \multicolumn{2}{c} {380\,GeV} & \multicolumn{2}{c} {3\,TeV} \\
 & No. variants & No. units & No. variants & No. units \\ \midrule
\textbf{\textit{Dipoles}} \\
Combined function 	&3 	&160 	&- 	&- \\
Bending dipole		&3	&152	&4	&1028 \\
Quadrupole			&4	&796	&4	&5451 \\
Sextupole			&3	&644	&3	&1804 \\
Corrector dipole	&1	&796	&3	&3813 \\
\midrule
\textbf{\textit{Main-Beam Linac}} \\				
MBQ Quadrupole		&2	&1148	&4	&4020 \\
Corrector dipole	&2	&1148	&4	&4020 \\
DBQ Quadrupole		&1	&5952	&1	&41400 \\
\midrule
\textbf{\textit{Damping Rings}} \\				
Combined function	&1	&180 \\		
Bending dipole		&1	&76 \\		
Quadrupole			&6	&1240 \\		
Sextupole			&2	&636 \\		
Skew quadrupole		&2	&580 \\		
Corrector dipole	&2	&1228 \\		
\midrule
\textbf{\textit{Ring to Main Linac Transport}} \\				
Bending dipole		&5	&718	&5	&708 \\
Quadrupole			&9	&1674	&10	&1747 \\
Sextupole			&2	&541	&2	&536 \\
\midrule
\textbf{\textit{Beam Delivery System}} \\				
Bending dipole		&12	&206	&12	&206 \\
Quadrupole			&63	&134	&63	&134 \\
Sextupole			&8	&30		&8	&30 \\
Octupole			&2	&2		&2	&2 \\
\midrule
\textbf{\textit{Post Collision Line}} \\	 	 	 	 
Bending dipole		&5	&18		&5	&18 \\
\midrule
\textbf{\textit{Total}}	&139	&18059	&130	&64917 \\
\bottomrule 
\end{tabular}
\end{table}

At the end of each quadrupole a short laminated electro-magnet dipole is attached,and is capable to steer the beam in one plane. A similar approach as for the MBQs has been adopted, with an identical cross section but two different lengths to cope with the increasing energy of the beam along the linac. The evaluation of possible coupling effect between the two magnets is in progress.

\subsubsection{Main-Beam Transport}

\begin{figure}[htb!]
\centering
\includegraphics[width=0.6\textwidth]{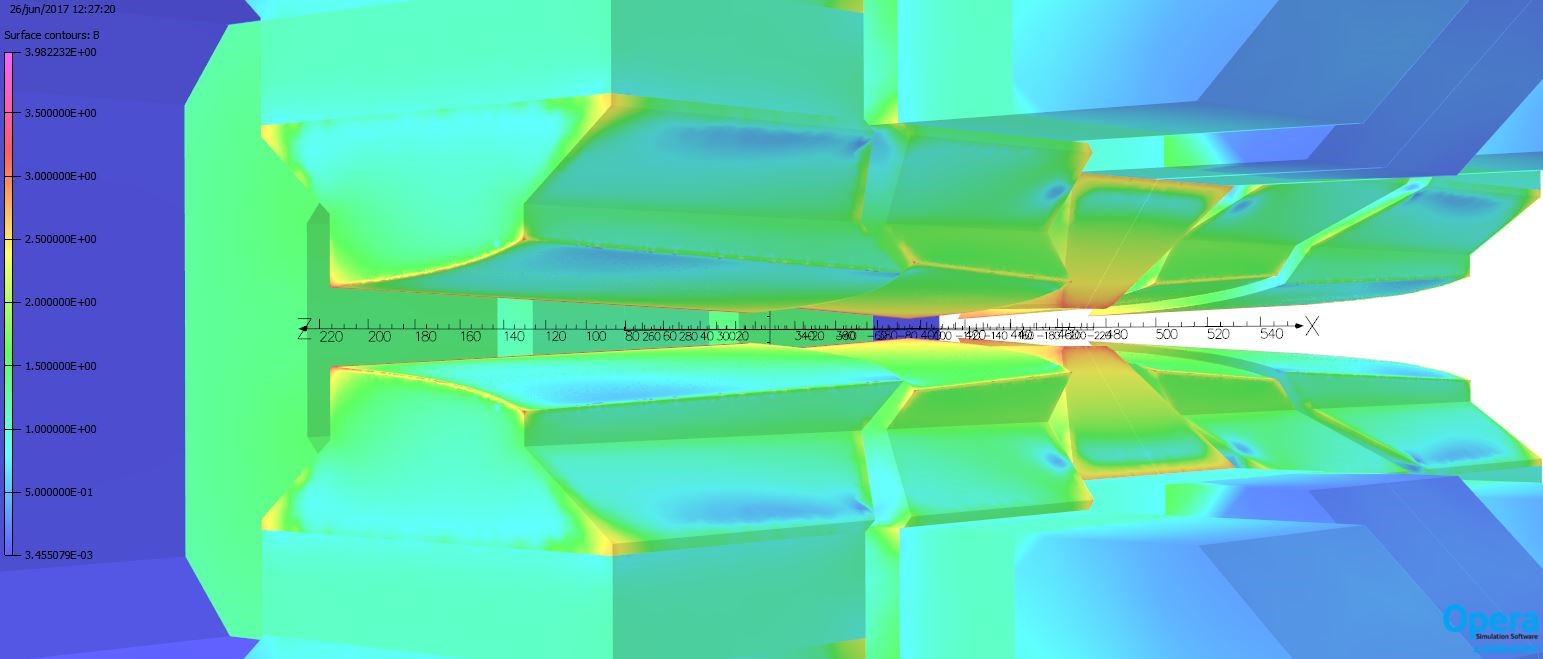}
\caption{\label{fig_MDI_1a}View of the poles of the longitudinal variable combined magnet.}
\end{figure}

The Pre-Damping Ring (PDR) and Damping Ring (DR) lattices use mostly electro-magnetic magnets. A special type of dipole with varying aperture along its length is needed in the damping rings to decrease beam emittance. A prototype of this magnet is presently under development with CIEMAT in Spain. This magnet will allow a reduction in the total circumference of the damping ring of 13\% for the same performance as the previous version. The concept is being applied to the upgrade of light sources like the ESRF, but the CLIC prototype goes beyond, as it is a tuneable permanent magnet combining dipole and quadrupole components, with an ultra-high field of 2.3\,T at its centre (see Fig.~\ref{fig_MDI_1a}).

For the Ring to Main Linacs (RTML) beam transport, the main change concerns the number of quadrupoles in the Long Transfer Lines (LTL). 

\subsubsection{Beam Delivery System (BDS)}

The baseline QD0 magnet design is based on an hybrid design, combining coils and permanent magnets and described in detail in the CDR \cite{Aicheler2012}. It has in the meantime been validated in the lab by a short, 100\,mm long, prototype. It reached a gradient of 531\,T, as expected slightly lower than the nominal gradient as the rising fields at the edges are not yet saturated at the centre of this very short magnet. The field quality was within specification as documented in \cite{Modena2012,Vorozhtsov2012,Modena2014}.

\begin{figure}[htb!]
\centering
\includegraphics[width=0.6\textwidth]{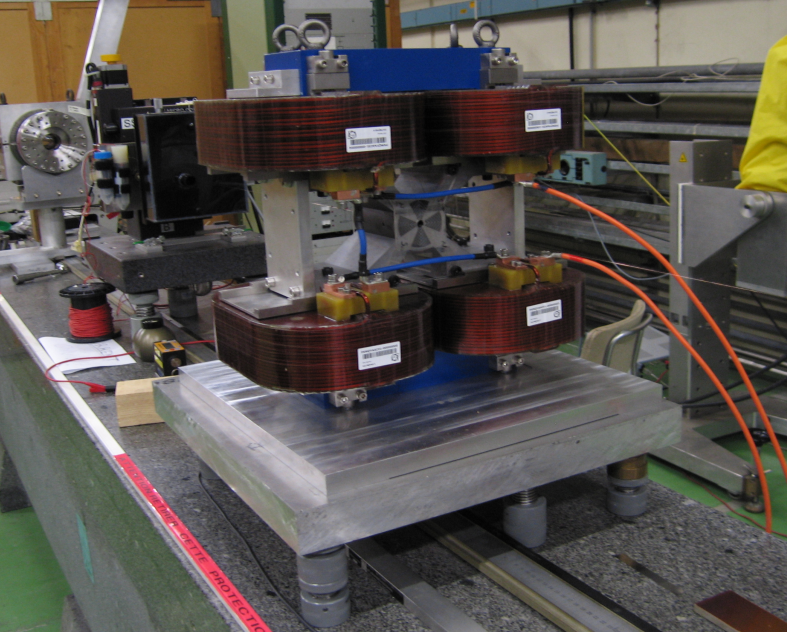}
\caption{\label{fig_MDI_1}The short QD0 prototype.}
\end{figure}

A photo of the prototype is shown in Fig.~\ref{fig_MDI_1}.  The optics studies for the Beam Delivery System have shown that the approx. 5\,m long quadrupole can be split into two or three shorter magnets without significant loss of luminosity, as long as the coil ends can be kept short.

After the CLIC rebaselining,the installation of the QD0 magnets on the tunnel floor rather than inside the detector, considerably simplifies many aspects of the machine detector interface. Also, the maximum gradient has been reduced from 575\,T to 25\,T for 3\,TeV and considerably smaller for the 380\,GeV phase and it is now also conceivable to design it optionally as a conventional quadrupole. The main constraint on the mechanical design of the QD0 magnet stems from the lateral space required for the vacuum pipe of the outgoing beam.

\begin{figure}[htb!]
\centering
\includegraphics[width=0.8\textwidth]{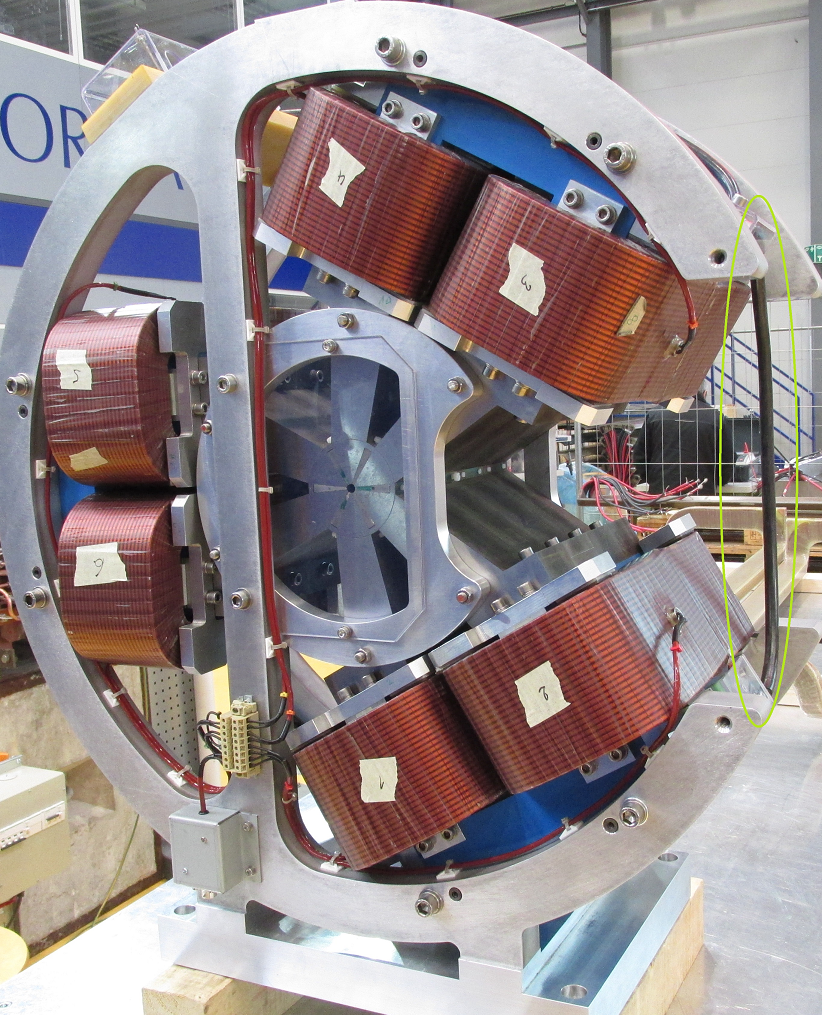}
\caption{\label{fig_Magnets_3}Prototype of the hybrid SD0, with PM blocks, Fe-Co core and air-cooled copper coils.}
\end{figure}

In 2017-2018, a 254\,mm long prototype of the SD0 has been built and is presently under magnetic measurements. The magnet, shown in Fig.~\ref{fig_Magnets_3}, is assembled with a core made of cobalt--iron laminates which allows reducing the saturation while carrying the large flux density needed to achieve the required sextupolar gradient 219403\,T/m${}^{2}$. The open frame allows to accommodate the beam pipe of the opposite beam.
Further studies on the mechanical design of a long magnet have been performed on a hybrid design of the final sextupole, SD0.

\subsubsection{Post-Collision Lines}

The baseline Post-collision Lines (PCL) lattice is composed of five types of dipoles all of which are very large and heavy (up to 90\,tons) due to the large aperture required (close to 500\,mm for the largest ones).

\section{Super-Conducting Damping Wiggler}
\label{sect:Wiggler}
\subsection{Introduction}

As shown in Chapter~\ref{Chapter:Base_Design}, to achieve the required luminosity at the collision point of CLIC, the normalized horizontal emittance has to be drastically reduced by means of damping rings. The positron beam enters first a Pre-Damping Ring and afterwards the main Damping Rings. The damping rings will be racetrack shaped rings with 26 wigglers placed in each straight section. The horizontal equilibrium emittance is designed to be around one order of magnitude smaller than in other planned or existing rings, which can be only achieved with superconducting damping wiggler magnets.

The baseline design foresees Nb-Ti superconducting wigglers with the parameters presented in Table~\ref{Tab-WIG_1}. To test the wiggler system with beam, one prototype superconducting wiggler magnet has been installed in the ANKA storage ring. The operational parameters for this prototype magnet were slightly different from the today's baseline parameters: B${}_{w}$\,=\,3.0\,T, 35\,periods, 51\,mm period length $\lambda$${}_{w}$ and a vacuum cold gap of 13\,mm. A detailed analysis was performed and showed that today's baseline parameters are of similar complexity as the parameters of the prototype magnet and are within reach. No large-scale cryogenic infrastructure is foreseen for CLIC, therefore, cooling each wiggler magnet individually with cryo-coolers has been chosen as baseline \cite{Schoerling2012}. To minimize the He inventory, indirect cooling has been chosen and implemented in the prototype magnet.

\begin{table}[htb!]
\centering
\caption{Main baseline parameters of Nb-Ti wiggler for CLIC}
\label{Tab-WIG_1}
\begin{tabular}{l c c}
\toprule
Period  							& 49 				& mm \\  
Magnet length 						& 1864 				& mm \\ 
Number of full field poles 			& 72 				&  \\ 
Magnetic Field  					&  3.5 				& T \\  
K 									& $\sim$ 16 		&  \\  
Vacuum gap cold 					& 10 				& mm \\  
Magnetic gap cold 					& 12 				& mm \\  
Length flange to flange  			& 2590 				& mm \\ 
Maximum ramping time  				& $<$ 5 			& min \\  
Power supply stability  			& $<$ 10${}^{-4}$ 	&  \\ 
Beam heat load 						& 50 				& W \\  
Period for LHe refill with beam  	& $>$ 6				& month \\ 
Field stability for two weeks 		& $\pm$10${}^{-4}$ 	&  \\
\bottomrule
\end{tabular}
\end{table}

\begin{figure}[htb!]
\centering
\includegraphics[width=0.6\textwidth]{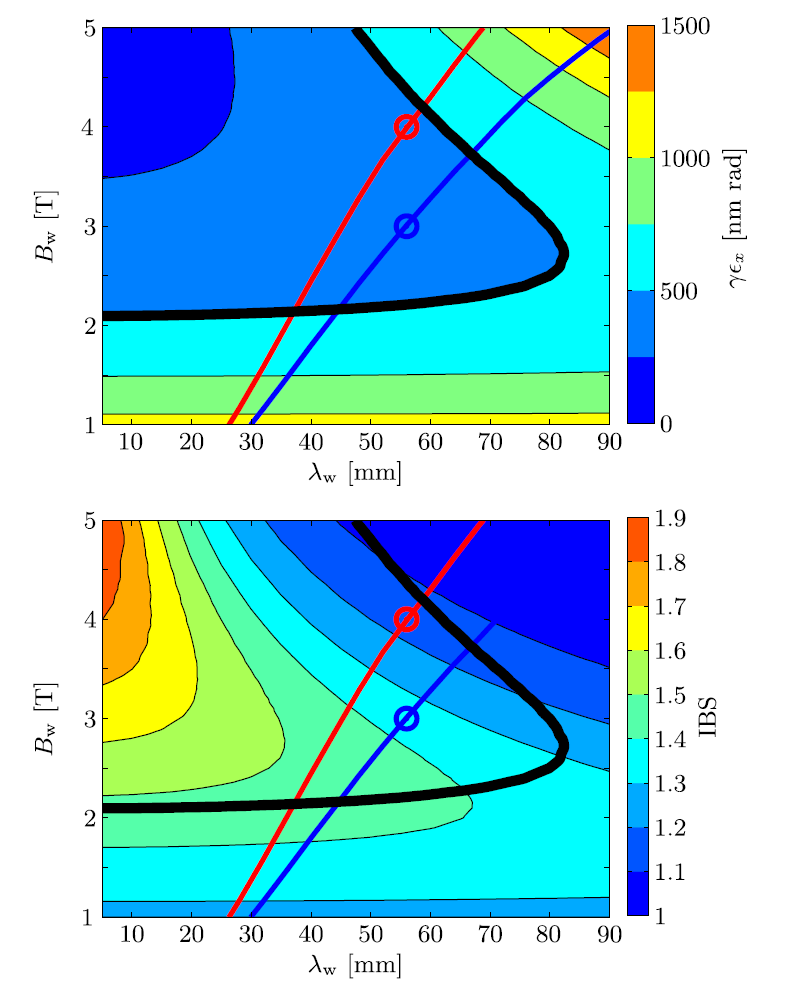}
\caption{\label{fig_WIG_1} Equilibrium normalized horizontal emittance \textit{YE}${}_{x}$ (top) and the effect of IBS (\textit{YE}${}_{x}$=\textit{YE}${}_{x}$;0). The red and the blue curves show the maximum achievable magnetic flux density for superconducting wiggler magnets with Nb${}_{3}$Sn and Nb-Ti wire technology, respectively \cite{Schoerling2012}.}
\end{figure}

In the framework of this programme, wiggler magnets employing Nb${}_{3}$Sn conductor technology have been designed, built and tested \cite{Schoerling2012a,Fajardo2015}. Nb${}_{3}$Sn wiggler magnets have the potential to reach smaller period lengths in combination with larger magnetic fields and margins. This very interesting combination of parameters (see Fig.~\ref{fig_WIG_1}) has triggered the construction of small prototypes. The results of tests of these prototypes are promising in terms of magnetic field but the magnets suffered from insulation issues. Therefore, more work would be required to be able to fully exploit the potential of Nb${}_{3}$Sn wiggler magnets.   

\subsection{Magnet Prototype}

In a collaboration between CERN, BINP and KIT, a prototype of a superconducting damping wiggler for the CLIC damping rings has been designed, manufactured and installed at the ANKA synchrotron light source \cite{Bernhard2013,Bernhard2016}.

\begin{figure}[htb!]
\centering
\includegraphics[width=0.4\textwidth]{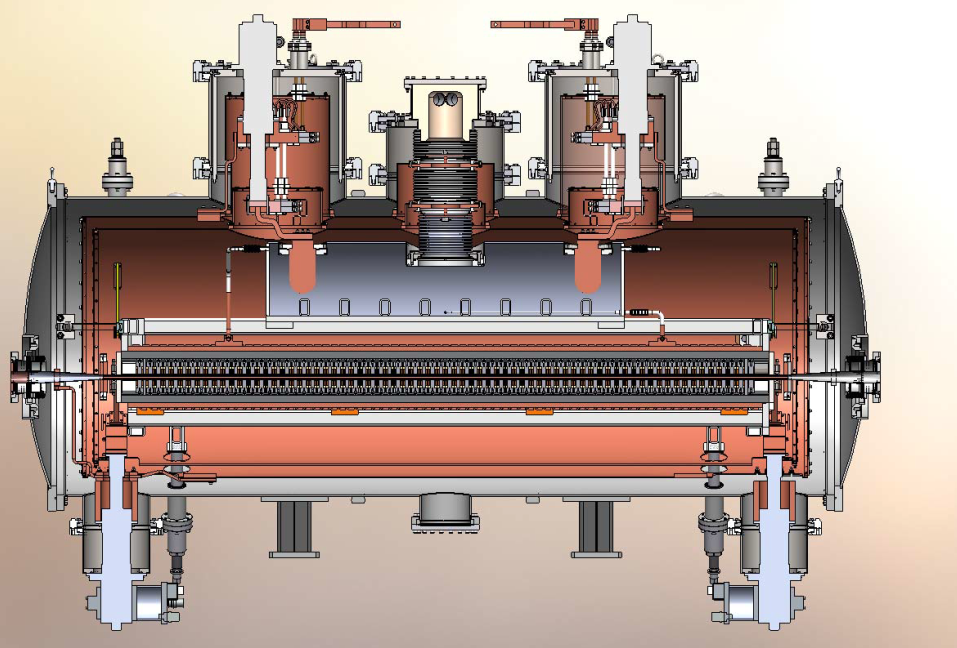}
\includegraphics[width=0.4\textwidth]{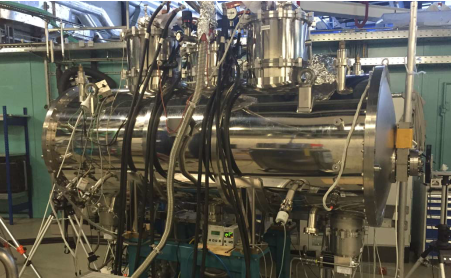}
\caption{\label{fig_WIG_2} Cross section of the CLIC damping wiggler prototype (left) and CLIC damping wiggler during testing at BINP (right).}
\end{figure}

The prototype magnet (see Fig.~\ref{fig_WIG_2}) was used to validate the technical design of the wiggler, particularly the conduction cooling concept applied in its cryostat design, in a long-term study. In this study the expected heat load from the up-stream wiggler's synchrotron radiation in the order of several tens of Watts is also studied by heating the vacuum pipe with a dedicated electrical heater. 

The wiggler's influence on the beam dynamics, particularly in the presence of collective effects, is planned to be investigated. ANKA's low-alpha, short-bunch operation mode will serve as a model system for these studies on collective effects. To simulate these effects and to make verifiable predictions an accurate model of the ANKA storage ring in low-alpha mode, including the insertion devices is under development.

\subsection{Magnet Test Results}

\begin{figure}[htb!]
\centering
\includegraphics[width=0.7\textwidth]{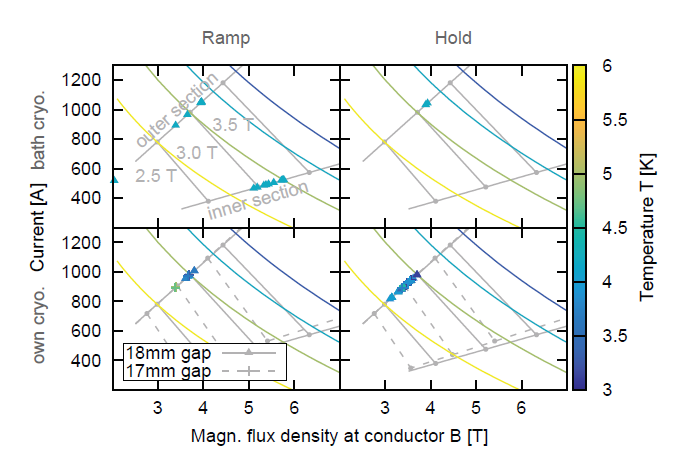}
\caption{\label{fig_WIG_3} Quenches of the wiggler marked on the load lines of inner and outer coil sections during current ramps (left) and at constant current (right), in liquid Helium bath (top) and in own cryostat (bottom). Each point marks a quench, colour-coded according to the temperature immediately before the quench. The coloured lines represent the critical condition for different temperature levels, the labelled grey lines depict the corresponding on-axis field for the original (18\,mm gap) and the final (17\,mm gap) magnet configuration, respectively.}
\end{figure}

The cryostat design, first tested with the superconducting undulator at APS \cite{Ivanyushenkov2015}, is based on the continuous recondensation of Helium vapour on plates cooled with the 2$^{nd}$ stage of cryocoolers. This approach led to an under-pressure inside the Helium tank and in turn to temperatures in equilibrium conditions of around 3\,K, even during beam operation. Despite this lower than expected temperature, the maximum stable current reached in the magnet coils turned out to be slightly lower than expected from the magnet tests in a liquid Helium bath at 4.2\,K \cite{Bragin2016}. Fig.~\ref{fig_WIG_3} summarizes the quenches of the magnet. The maximum on-axis field amplitude reached during ramping was 3.2\,T, both in the bath cryostat and in the wiggler's own cryostat. In the case of indirect cooling, however, holding quenches occur after periods of seconds to several hours in the outer, high current coil sections and basically uncorrelated with the magnet temperature. The physical origin of this instability is not yet satisfactorily explained. The stable field is limited to 2.9\,T by this effect, however, the wiggler can routinely be operated at this field level also in the storage ring under beam operation conditions.

\subsection{Beam Test Results}

Before the installation in the storage ring the field integrals of the wiggler were measured with a stretched wire set-up and minimized by adjusting the current distribution in the wiggler's matching coils. A quite significant horizontal field integral was corrected by additional vertical corrector magnets outside the cryostat. From the closed-orbit measurements a slight readjustment of the field integral compensation settings turned out to be necessary.

The first basic experiments on beam dynamics focused on the wiggler's influence on the closed orbit and betatron tunes in order to confirm the field integral compensation settings, the alignment of the wiggler and the results of the online magnetic characterization. A betatron tune shift due to the vertical focusing strength of the wiggler agrees with the model calculations. The origin of the (unexpected) horizontal tune shift is still a matter of investigation.

\section{Controls}
\label{sect:Controls}

\subsection{Introduction}
The CLIC control system will be similar to the control system of the LHC and its injectors (see \cite{bib:LHC_controls}). Some special consideration has to be given to the Front-End Tier. For all CLIC accelerators excluding the Main Linac tunnel it is proposed to use the ``LHC approach'' where controls electronics are housed in racks close to the accelerators and connected to front-end computers, via dedicated cabling. These are in turn connected to the CERN Technical Network which is based on gigabit Ethernet. This is the basis of the control system for CTF3 and although the scale of the CLIC project means that the distances and the number of components to be controlled are larger than in CTF3, there should be no fundamental difficulty in controlling the CLIC injector complex.

The equipment in the Main-Linac tunnel can not be controlled in the same way. Here there is high control-signal density (about 100 signals/m). Building the acquisition system for this number of channels in the classical modular approach would simply not be possible in terms of available space and also in terms of cost. In addition, limitations on heat dissipation and the radiation environment in the tunnel, impose a new front-end architecture. 

In the scope of the HL-LHC project \cite{Apollinari:2116337}, which is also very demanding in terms of  accelerator controls, a project has been started; ``Distributed I/O Tier and Radiation-Tolerant Fieldbus Project'' \cite{Daniluk:THPHA071}. This is a variation of what was proposed in the CLIC CDR; where the development of a compact acquisition module, to serve as a dedicated acquisition and control module (ACM) was proposed. For the rest of this discussion, we will discuss only the parts of the CLIC control system which differ significantly from that described in the CDR.

The control system has 3 hierarchical layers of equipment communicating
through the CERN Technical Network, a flat Gigabit Ethernet network using the TCP-IP protocol. We will consider only the lowest layer, the  hardware layer.  

\subsection{Hardware Layer}
The Hardware layer is where the interaction with the machine components takes place. A Distributed I/O Tier platform (DIOT) and a new high-speed, radiation-tolerant fieldbus, are the areas where the control system as described in the CDR \cite{Aicheler2012} differs substantially from that proposed here.

\begin{figure}[!htb]
  \centering
  \includegraphics[width=0.8\textwidth]{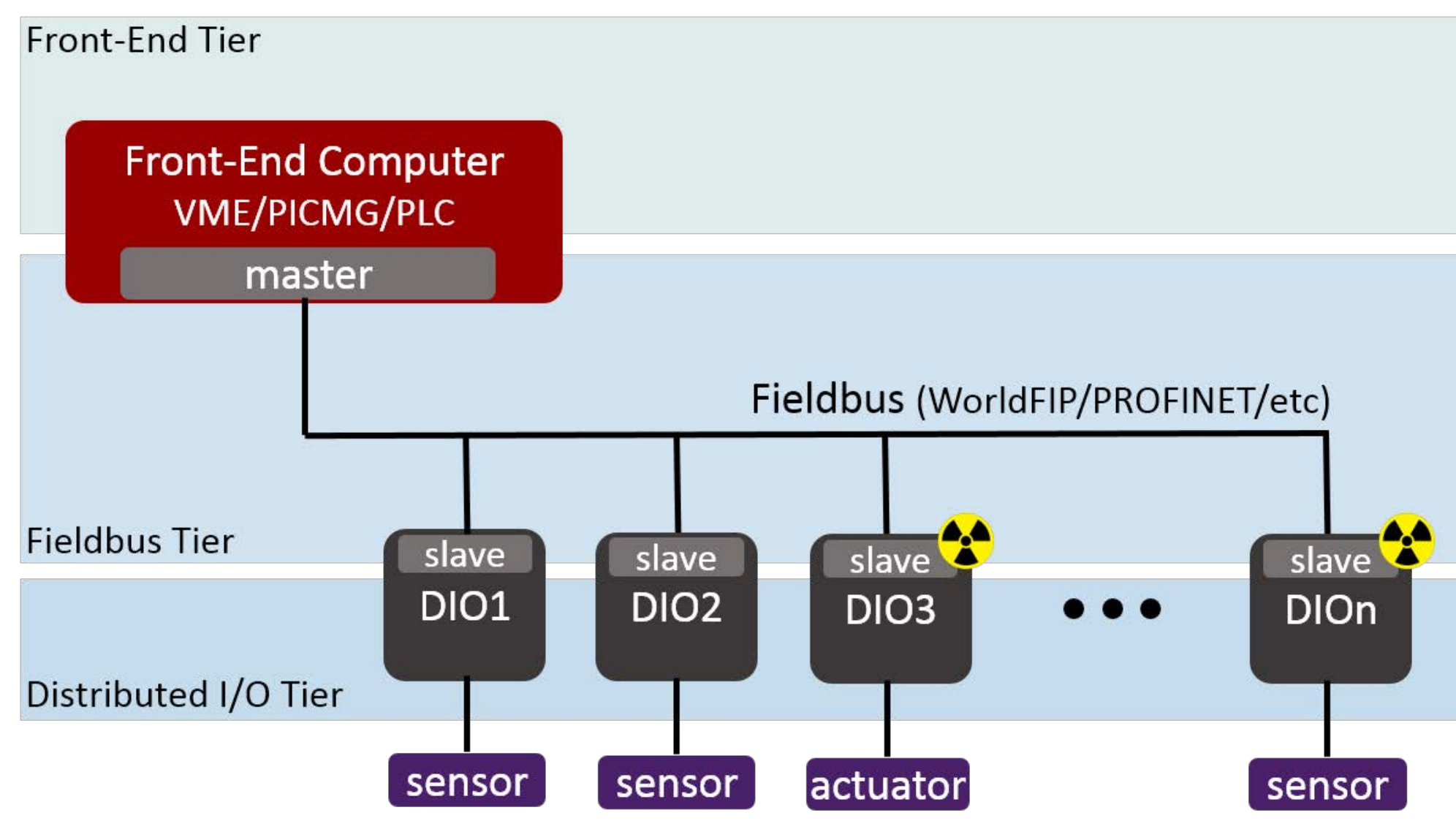}
  \caption{Hardware layers}
  \label{fig:Controls_2}
\end{figure}

As Fig~\ref{fig:Controls_2} shows, within the hardware layer, one can distinguish three tiers:

\begin{itemize}
\item Front-end Tier – a PLC or powerful computer in various form factors (VME, PICMG 1.3, MTCA.4, etc.) running a set of user applications controlling Distributed I/O Tier over a fieldbus.
\item Fieldbus Tier – communication link between the Master in the Front-end Tier and a set of slaves in the Distributed I/O Tier
\item Distributed I/O Tier – electronics modules that interface directly with the accelerator components in radiation-exposed or radiation-free areas, controlled over the fieldbus. These are usually FPGA-based boards sampling digital and analog inputs, driving outputs and performing safety-critical operations.
\end{itemize}

Depending on the needs of a given application, equipment groups can either use off the shelf systems (e.g. PLC-based), design custom electronics, or a combination of the two. The specialised needs of the CERN accelerators often demand the development of custom electronics. For these, there is a centralised service in the Front-end Tier in the form of VME crates and PICMG1.3 computers that can host a modular FMC (FPGA Mezzanine Card) kit. 

In the Fieldbus Tier the current offering for custom electronics is mainly built around the radiation-tolerant WorldFIP bus. However, the limited (2.5\,Mb/s) bandwidth will inflict prohibitive delays on some equipment groups as they try to gather enough diagnostics to find the reasons for a given failure. It is planned to expand the offering by developing a modern fieldbus communication based on 100\,Mb/s Ethernet.

\begin{figure}[!htb]
  \centering
  \includegraphics[width=0.8\textwidth]{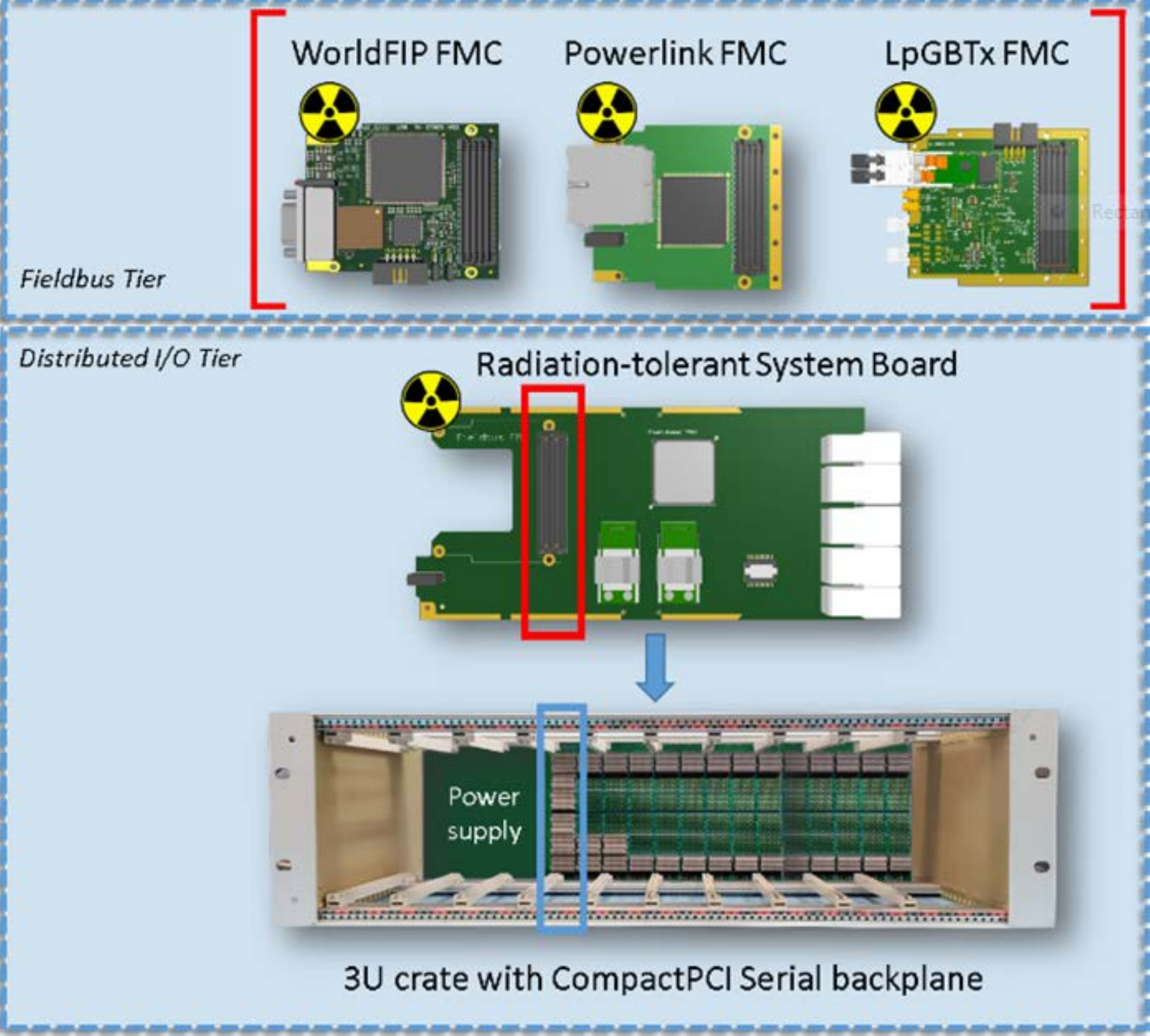}
  \caption{Radiation-tolerant variant of the Distributed I/O Tier kit}
  \label{fig:Controls_3}
\end{figure}

A modular standardised hardware kit that can be customized to suit various applications will be developed as part of the DIOT project. Two important principles of the project are staying close to the standards used in industry and designing the modules together with the future users – to benefit from the scrutiny and review of many developers. While the complete kit consists of modules for radiation-exposed and radiation-free areas, in the scope of the HL-LHC and CLIC the focus will be on radiation-tolerant components (Fig.~\ref{fig:Controls_3}):
\begin{itemize}
\item The 3U crate with a fully passive, standard, off-the-shelf backplane and power supply that can be exposed to radiation.
\item The generic system board that serves as the crate controller, implements crate  diagnostics, interfaces with application-specific peripheral boards plugged into other slots of a 3U crate, and features an FPGA that can be programmed by each equipment group.
\item A set of communication mezzanines that implement various fieldbus protocols: WorldFIP (2.5\,Mbps), Powerlink (100\,Mbps), LpGBTx (10\,Gbps).
\end{itemize}

This modular approach will satisfy the needs of the different subsystems. In particular, specific acquisition systems can adopt the whole kit, or use only some of its elements according to their needs. The customisation of the kit is done by developing application-specific add-in modules and implementing FPGA firmware for the system board.

The WorldFIP FMC has been designed \cite{Daniluk2018} and work is ongoing to develop the Powerlink FMC mezzanine. A first DIOT hardware demonstrator which features an off-the-shelf crate has also been developed. The system board for this first prototype is a modified GEFE board designed for control and acquisition of Beam Instrumentation. This kit is being used as a prototyping platform to work on a radiation-tolerant power supply and a set of FPGA cores for uniform diagnostics. These crates will be available to equipment groups for evaluation and development of their application-specific add-in boards [ref DIOT]. 

\subsubsection{Acquisition and Control for the Main Linac}

For each type of CLIC module approximately 200 acquisition channels and 170 control channels are needed to measure and control the properties of the beam, the RF parameters, the module alignment and stabilization, vacuum and the cooling systems. The data rate for acquisition is dominated by the signals coming from beam instrumentation and from RF and is around 6\,Mbits/s per ``DIO Slave''. Given the capacity of current optical fibres, there should be no data transmission bottleneck to and from the ``Slaves'' to the alcoves.

Owing to the long distance (878\,m) between the underground alcoves (see Chapter~\ref{Chapter:CEIS}), it is not foreseeable to locate the acquisition and control electronics for the 436 modules of one ``sector'' in these alcoves. The required space for the electronics for such a large number of modules and the cost for the cabling to and from such a number of modules to the alcoves would be prohibitive.

The proposed solution, therefore, consists of one or two local ''DIO Slaves'' (DIOS) per CLIC module, located as close as possible to each module. The position of the ``DIOS'' relative to the module has yet to be decided and depends on the form factors of the chosen hardware platforms. The DIOS should, however, be positioned as far as possible from the beam pipes to limit the amount of radiation to which they are subjected. 

The DIOS is responsible for the acquisition and transmission of control signals under the constraints on the radiation hardness and power consumption. In the light of these constraints, a generic approach must be found in order to provide an ``open'' CLIC module acquisition and control solution that would accommodate the controls requirements for the different CLIC sub-systems.

\subsubsection{Digitization and Data Transmission}

The DIOS should perform the following tasks:
\begin{itemize}
\item The prompt digitizing of the incoming signals from the CLIC sub-systems
\item The execution of commands and configuration actions, as received from the higher level(s) of the CLIC control system
\item The execution, for some systems, of pulse-to-pulse feedback loops
\item The synchronization of acquisitions and commands with the CLIC timing system
\item The time stamping of the acquisitions
\item The pulsing of hardware components of the system, in synchronization with the CLIC timing system
\item The bi-directional transmission of data via optical link(s) to the next-higher layer of control system.
\end{itemize}

\subsubsection{Remote Configuration and Diagnostic Facilities}
The quantities and the underground location of the DIOS represent a unique challenge in terms  of operations. Specific aspects should be addressed in terms of operational availability and long-term maintenance and evolution of these systems. The following mandatory services shall be implemented at the level of the DIOS:

\begin{itemize}
\item Monitor in real-time, through the communication link, the correct functioning of any internal hardware component of the system.
\item In case of malfunction, disable parts of the system in order to allow the operation of CLIC in degraded mode.
\item Remotely upgrade firmware and configuration parameters.
\end{itemize}

\subsubsection{Front-End Computers}
In each alcove there is one, or more, dedicated Front-End Computer (FEC) for each module sub-system: beam instrumentation, RF, cooling, alignment, stabilization, vacuum, and power converters. In addition, there is a dedicated FEC for timing and another FEC dedicated to Machine Protection. The FECs will be the fastest real-time computer available at the time of their installation. For the slower signals (e.g., vacuum and cooling, where real-time processing speed is not so important), Programmable Logic Controllers (PLCs), or their equivalent, may be used. For the FECs to send information, via the middleware services, to the CLIC Control Centre the alcoves must have a connection to the surface. As there is only surface access every fourth alcove, it is necessary to link four alcoves via separate fibres. In the alcove which has access to the surface, dedicated switches take care of the connection to the surface.

\subsection{Data Management Services}
Data management is fundamental to the operation of any accelerator complex throughout its full life-cycle from design to dismantling [321]. 

The data that will be gathered and stored for the operation of CLIC can be categorized as configuration or logging data.As mentioned above, we deal only with data logging here.

The current “CERN Accelerator Logging Service” (CALS) subscribes to 20,000 different devices and logs data for some 1.5 million different signals. It has around 1,000 users, which generate 5 million queries per day (coming mainly from automated applications). CALS stores 71 billion records/day that occupy around 2\,TB / day of unfiltered data. Since this amount is quite significant for storage, heavy filtering is actively applied and 95\% of data is filtered out after 3 months. That leaves the long-term storage with around 1\,PB of important data stored long term since 2003.

These numbers mean that a new solution is required for future accelerators. A new project “Next CALS” (NXCALS) \cite{Wosniak2018} has been started and after prototyping with various tools and techniques, Apache Spark was selected as the best tool for extraction and analysis for controls data backed up by the synergy of Apache HBase and Apache Parquet files based storage in Hadoop.

For the visualization, there is a widespread adoption of Python and Jupyter notebooks happening at CERN and other institutes involved with data science and scientific computing. This study truly set the scene, showing directions for how the Controls data could be presented to its users. The scalability of the new system relies on the CERN on-premise cloud services based on OpenStack with 250,000 cores currently available.

\subsection{Conclusions}
The CLIC project comprises several different accelerators, each of which needs a solution for controls. This complex of accelerators resembles the present scheme at CERN for the LHC and its injector chain. Hence a control system as a scaled version of the present LHC system can be envisaged. The basic functionality of a three-tier system has to be provided and the solutions for the front-end computers following the industrial standards available at the time of construction. An exception to this general solution is the acquisition and control of the CLIC modules in the Main Linac and the logging of data coming from the accelerator sub-systems. For this several challenges will have to be addressed as technology develops over the next few years:

\begin{itemize}
\item The scale of the CLIC machine requires an unprecedented amount of controls equipment to be installed within a limited space and under radiation constraints. For each of the 2,976 Main-Beam  modules (380\,GeV stage) one or more DIOS chassis will be installed, each of which holds 10–20 electronics modules. This gives a total of between 30,000 and 60,000 individual electronics modules. A large industrialization and procurement effort will be required to produce and deliver such a quantity of electronic cards.
\item The tunnel of the CLIC Main Beam is a very hostile environment for electronics. The DIOS chassis will be installed as close as possible to the modules. This is a new and yet un-tested, concept for an accelerator control system front-end.
\item From simulations, the radiation levels in the CLIC Main Beam tunnel are one or two orders of magnitude higher than in the LHC. The DIOT Project, from which comes the DIOS concept, has addressed radiation hardness from the beginning and so this issue should be less limiting than in the CDR.
\item Another major challenge is to limit the amount of heat given off by the DIOS into the tunnel. Over the next few years, various low-power solutions will be studied as will the possibility of switching off the electronics between beam-pulses.
\end{itemize}

\section{Fine Time Generation and Distribution}
\label{sect:TECH_fine_timing}

\subsection{Background}
The goal of a timing system is to provide a common  time reference in a distributed environment. This reference is usually the result of counting ticks of a clock signal from an arbitrary instant. The clock signal, $a(t)$, is ideally of perfect periodicity and stability. Real-world clocks, however, present imperfections \cite{bib:Serrano:rubiola09} in both amplitude and phase as expressed in Eq.~\ref{eq:imperfect}.

\begin{equation}
  \label{eq:imperfect}
  a(t)=A \left( 1+ \alpha (t) \right) \sin(\omega t +\varphi (t)) 
\end{equation}
where $\omega$ is the angular frequency of the clock signal.

\begin{figure}[!htb]
  \centering
  \includegraphics[width=0.5\textwidth]{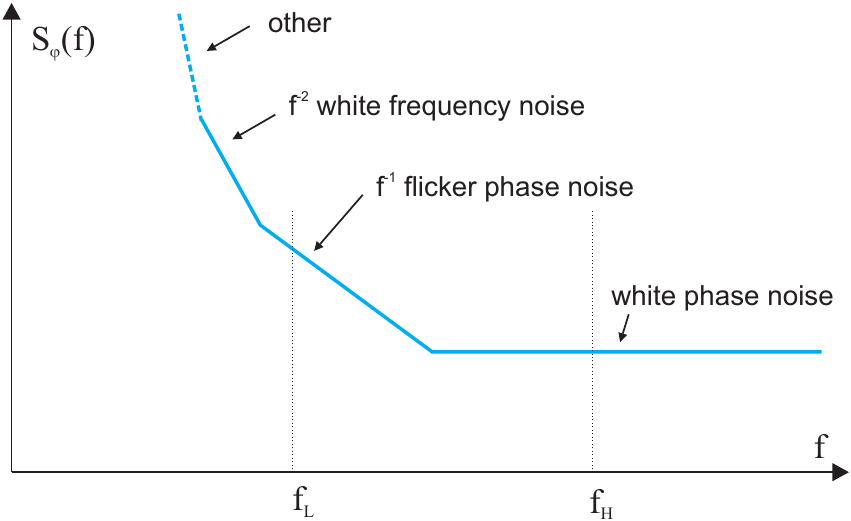}
  \caption{One-sided PSD of phase noise for a typical oscillator.}
  \label{fig:phase_noise}
\end{figure}

It is common practice to use the Power Spectral Density (PSD) of signal $\varphi(t)$ to specify jitter requirements. The integral under the PSD curve is the total jitter. In real life, an application is only sensitive to jitter generated between two finite integration limits. Figure~\ref{fig:phase_noise} shows a typical plot of a one-sided PSD ($S_{\varphi}(f)$) of the phase noise for an oscillator. Integration limits are set between $f_L$ and $f_H$. For a machine with a repetition rate of 50\,Hz such as CLIC, it is estimated that all perturbations below 5\,Hz in Fourier frequency can be dealt with by appropriate inter-pulse feedback strategies. Reasons for establishing an upper limit in integration stem mainly from the inability of some systems to react to such fast variations. These limitations can be in electronics, such as the
bandwidth of the input stage of a digital gate, or in electromechanical systems such as an RF accelerating cavity.

\subsection{CLIC Timing Requirements}
\subsubsection{Drive-Beam RF System}

In Ref.~\cite{bib:Serrano:jensen09} the jitter of the 1\,GHz field in the accelerating cavities of the Drive Beam is specified as 50\,fs integrating between 5\,kHz and 20\,MHz. It is also said that with appropriate feed-forward control in the Main Beam, this figure could be relaxed by a factor of 10. However, the reference phase noise fed to the LLRF system is only responsible for a small percentage of the final jitter in the electromagnetic field. Taking this contribution to be 10\% results in a specification of 50\,fs for the jitter of the
reference clock signal distribution to each one of the 326 accelerating structures in each linac.

\subsubsection{Beam Instrumentation}

Longitudinal profile monitors could be based on distributed lasers and changes in optical properties of bi-refringent materials induced by the Main Beam. These monitors have the task of measuring 150\,fs bunches with a resolution of 20\,fs \cite{bib:Serrano:lefevre10}. The precision required from the clock signal, which allows the synchronisation of the lasers with the beam, would therefore be in the few tens of femtoseconds, integrated between 5\,Hz and a few hundreds of kHz (the bandwidth of the Phase-Locked Loop locking the laser to the reference).

\subsubsection{Two-Beam Acceleration System}

\begin{figure}[!htb]
  \centering
  \includegraphics[width=0.5\textwidth]{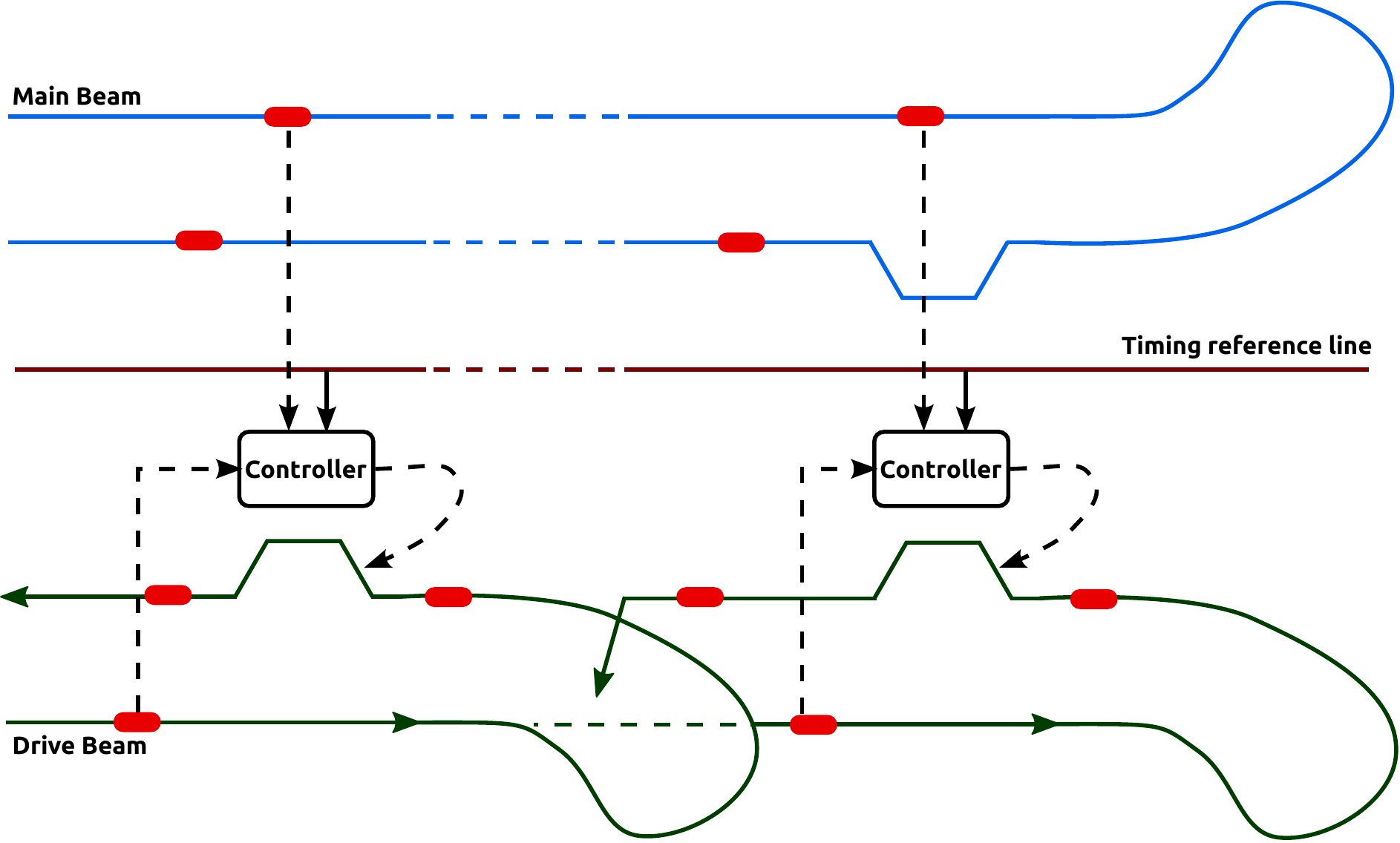}
  \caption{Timing reference line usage for inter-beam
    synchronisation (adapted from Ref.~\cite{bib:Serrano:schulte10}).}
  \label{fig:timing_ref}
\end{figure}

The Two-Beam acceleration scheme in CLIC requires a very precise synchronisation between the Drive Beam and the Main Beam. Figure~\ref{fig:timing_ref} depicts a possible synchronisation solution. A controller measures the phases of the Drive and the Main Beam with respect to a reference line, and uses that information to control the amplitude of a kicker pulse which modifies the trajectory of the Drive Beam in order to keep it well synchronised with the Main Beam.  The required precision of this alignment is around 40\,fs \cite{bib:Serrano:schulte10}, so the timing reference precision clearly needs to be better than that. The bandwidth of the kickers, in the several MHz region, would set a natural upper limit for integration of phase noise.

\subsection{Technical Solutions}

There are two main types of solutions currently developed and tested for the distribution of precise timing signals in the femtosecond realm \cite{bib:Serrano:loehl10}. Both types have achieved synchronisations better than 20\,fs over distances of several hundreds of metres.

In the Continuous-Wave (CW) variant, a laser modulated in amplitude by the RF or microwave signal to be transmitted. A fraction of the light reaching the destination bounces back from a Faraday Rotator Mirror (FRM) and is, in the process, shifted in frequency using a Frequency Shifter (FS). Another FRM reflects a sample of the light signal as it leaves the source. These two signals, upon mixing together, produce a beat whose phase can be easily measured. The key feature of this system is that the heterodyning process preserves phases, so a phase shift induced in the optical frequency by a change of length in the fibre will show up as exactly the same phase shift at the RF frequency, which is much easier to measure. Once the phase of the beat signal is detected, it can be used to digitally shift the phase of the recovered RF signal at the receiving end of the link. It is important to note that what is really measured in this method is the phase delay, not the group delay of the modulation which is ultimately what we are interested in. In order to compensate for group delay, a first-order correction --- based on actual measurements of a given fibre type for a range of temperatures --- is applied to the raw phase delay measurements.

In the pulsed variant, a mode-locked laser is synchronised with the external RF signal, which determines its repetition rate. Narrow light pulses come out of the laser and travel through a dispersion-compensated fibre to the receiver, where a fraction of the light is reflected back towards the emitter. An optical cross-correlator measures the degree of coincidence of the two pulse trains and the result is used to control a piezo actuator that changes the fibre length so as to keep a constant group delay. This mechanical actuator is unavoidable because the cross-correlator needs the pulses to overlap, at least partially, in order to give a meaningful reading. On the other hand, the pulsed system controls group delay directly, so no \textit{ad hoc} conversion between phase delay and group delay is needed.

\subsection{Technical Issues}

Both types of systems have been successfully deployed over distances of several hundred metres, so the 6\,km needed by CLIC Stage 1 is unknown territory. In the case of the CW system, one potential source of concern is the need for a model of phase-group delay corrections vs. temperature. If the temperature is not uniform over the complete length of the fibre, as can easily be the case in CLIC, a solution will need to be found.

Another potential issue is Brillouin scattering, a non-linear effect which is especially strong in optical fibres, due to the large optical intensities in the fibre core. Brillouin scattering in fibres leads to a limit of the optical power which can be transmitted, since above a certain power threshold, most of the light is scattered or reflected. This is especially a problem for narrow-band optical signals, and for long fibre lengths. In the pulsed system, Brillouin scattering is not much of an issue, since the power spectral density, which is the important quantity, is much smaller than for narrowband CW systems.

For the pulsed system, a major difficulty will be the dispersion control of the optical fibres, as the fibre length increases. Additional problems due to non-linearities are not expected to be any more significant than with shorter fibre links. 

Finally, one very important difference between the solutions currently deployed and the CLIC scenario is the scale of the project. Current implementations do not scale very well beyond some tens of destinations. This is mainly due to the number of ports typically available in optical components, such as splitters/combiners. For
CLIC, one of the challenges will be to explore in detail the real needs of each destination and come up with a strategy for partitioning the system in such a way that a reasonable compromise between performance and cost can be found.

\section{Machine Protection}
\label{sect:MachineProtection}
\subsection{introduction}

In the CDR \cite{Aicheler2012} there is a detailed discussion of the destructive capacity of the CLIC beams. Although the beam power is impressive (Main Beam: 14\,MW and Drive Beam for a single decelerator: 2.9\,MW) the energy per pulse 2.8\,MJ and 0.58\,MJ respectively, is more than two orders of magnitude lower compared to the LHC beam of 300\,MJ. Nevertheless, the energy density due to the extremely low beam emittance and spot sizes is very critical. Both beam power and energy density have to be addressed with the machine protection system to allow for safe and reliable operation.

As described in the CDR, the CLIC Machine Protection policy deploys different strategies depending on the time domain of the failures. In the following subsections, the status and implications for each of these strategies is discussed in more detail. The description will start from the interlock protection, which provides real-time protection for equipment failures up about a millisecond before beam passage, and then continues with the strategies to mitigate failures occurring at shorter time scales. Then we discuss the strategies for failures occurring over longer time scales.

\subsection{Machine Interlock System}

The CLIC machine interlock system will assure that all equipment failures, detected in time before the next beam is committed, revoke the ``next beam permit'' and inhibit delivery of otherwise ill-controlled beams.

At CERN, and most notably for the LHC, a lot of experience with interlock systems has been build up. Although there may be many similarities, there is also a substantial number of differences to the CLIC machine. Firstly, whilst the LHC interlock surveys the circulating beams and may demand a beam dump on the next beam passage, in CLIC there is only a single beam passage. Hence, the interlock system only controls the permission for the next pulse. Technically speaking, the CLIC interlock system should be compared with the LHC injection interlock. A second difference is that the CLIC cycle time (20\,ms) is by far not comparable with the LHC inter-cycle time (over 20 minutes). Hence, the CLIC interlock system must be able to rearm much faster and rely on automated analysis to determine whether the next cycle can safely be committed. Finally, the CLIC interlock system has to deal with multiple beams (Main Beam, Drive Beam) of various possibly beam intensities (i.e. during the intensity ramp up), that are traversing a multitude of accelerators and transfer lines.

Due to this complexity, the CLIC interlock system must be closely integrated with the beam scheduling system. The beam scheduling system will send out a token with the next pulse configuration that circulates through the CLIC complex. All interlock managing stations in the CLIC accelerator chain have to add their acceptance signature to the token for the next pulse to be executed.

\subsection{Safe by Design}

The interlock system intercepts any equipment failure that occurs during the inter-beam period. However, as the interlock system has a finite signal-handling time, the interlock system will only cope with equipment errors that are detected up to 1$\sim$2\,ms before the next pulse permit is given. This implies that all the equipment circuits must be designed such that there is enough inertia in the system for the equipment settings to stay within the tolerance required for safe beam passage during 2\,ms following the onset of a failure.

This strategy is not new, and is already deployed in many of the CERN transfer lines which carry high-intensity beams.

\subsection{In-flight Protection}

The very short pulse length (156\,ns for the Main Beam and 244\,ns for the Drive Beam) makes it practically impossible to correct real-time errors. Once a pulse has been committed, there are practically no means to intercept safely the beam in the event that beam losses are detected (note that the CDR discusses a few exceptions where theoretically an interlock system may act on the beam in progress). Hence, for so-called ``in-flight'' failures, the machine protection system has to rely on fixed protections by masks and collimators. Most notably, masks are needed to protect extraction and transfer channels that follow any fast extraction or injection element. The design of the mask will be a challenging task as the charge density of the extracted beam could be high enough to damage the mask.

The interesting concept of reusable surfaces (i.e. where, in case of damage, the collimator surface can be displaced parallel to the beam) which exists in the LHC should also be deployed in CLIC. Furthermore, special attention is given to improving the reliability of the kicker systems through the deployment of inductive adders to pulse the kicker systems.
 
\subsection{Beam Quality Checks and Next Pulse Permit}

The strategy to deal with medium term failures is based on the concept of a ``next pulse permit''. Upon completion of every cycle, the hardware permit of the next pulse will automatically be revoked. Only after a number of predefined automated beam quality checks on the actual pulse have successfully passed, will the permit for the next pulse be re-established.

Conceptually this strategy is not new, beam quality checks are also applied in LHC. However, while in the LHC many post-cycle checks are performed under operator/expert supervision, in CLIC this decision has to be executed within the 20\,ms cycle, imposing further constraints on the data collection and analysis of the beam instrumentation equipment.

The most important beam quality checks are based on beam losses, beam trajectory information and beam-based feedback/feedforward performance.

The beam quality check will work in close conjunction with the beam scheduling system, which orchestrates the beam intensity ramp-up after a beam interlock. Hence, the next pulse permit decision is not just a go/no-go, but it also encompasses the decision on the beam intensities.
 
\subsection{Long-Term Protection}

Long-term protection has to ensure that the machine stays in a healthy state over longer periods. The most important ageing factor of many components is radiation related. Continuous machine optimisation with the aim to reduce radiation losses is important in this context. Radiation levels will be monitored and, where needed, the impact can be reduced by radiation shielding.

\section{Beam Interception Devices}
\label{sect:BID}

\subsection{Main-Beam Dump}

In order to cope with the challenge of absorbing the CLIC Main Beam, a water Beam Dump at the end of the CLIC post-collision line is proposed. 
It consists of a 10\,m long cylinder, filled with water, pressurized at 10\,bar, and with a diameter of 1.8\,m, surrounded by a 15\,mm thick titanium vessel of the same shape. For the present study a circular window has been considered, 30\,cm in diameter, 1\,mm thick, made of a titanium alloy  (Ti-6Al-4V, i.e., ASTM G5, or UNS R56400), and directly cooled by the circulating water inside the dump.

The energy/power deposition by the primary beam on the water dump was calculated using the FLUKA Monte Carlo simulation code \cite{bib:BID:ref4}.  Two different beam scenarios were considered: the 1.5\,TeV uncollided electron beam with a transverse Gaussian beam profile centred on the dump axis, with $\sigma$$_{hor}$\,=\,1.79\,mm and $\sigma$$_{ver}$\,=\,3.15\,mm. For the collided beam the secondary particles produced in the electron--positron interactions at 3\,TeV centre-of-mass energy were simulated with the GUINEA-PIG package \cite{bib:BID:ref6} and were transported along the post-collision line up to the dump with the DIMAD tracking code \cite{bib:BID:ref8}.

Table~\ref{tab:BID_2} shows the peak energy and total power on the water dump and on the titanium vessel/window for the uncollided and the collided beam scenarios. As expected, the case of uncollided beam is more severe than the one of collided beam. 

\begin{table}[!htb]
\begin{center}
\caption{Peak energy and total power on the water dump and on the titanium vessel/window for the un-collided and the collided beam scenarios. Statistical uncertainties on the total values are below 0.1\%.}
\label{tab:BID_2}
\begin{tabular}{l c c c c}
\toprule
 & \multicolumn{2}{c}{\textbf{Max}} & \multicolumn{2}{c}{\textbf{Total power}} \\ 
 & \multicolumn{2}{c}{\textbf{[J cm$^{-3}$ per bunch train]}} & \multicolumn{2}{c}{\textbf{[W]}} \\ 
 & \textbf{uncollided}  & \textbf{collided} & \textbf{uncollided}  & \textbf{collided} \\
\midrule 
H$_{2}$O 									& 230\,$\pm$\,1 						& 9.10\,$\pm$\,0.01 		& 13.8\,M	& 13.4\,M \\ 
Ti window 								& 4.35\,$\pm$\,0.36 				& 83.8\,$\pm$\,1.0\,m 	& 6.24 		& 4.91 \\ 
Ti vessel (side) 					& 569\,$\pm$\,18\,$\mu$ 		& 903\,$\pm$\,29\,$\mu$ & 8.45\,k & 9.07\,k \\ 
Ti vessel (upstr. face) 	& 32.0\,$\pm$\,13.2$\,\mu$	& 2.15\,$\pm$\,0.13\,m 	& 9.07\,	& 45.7 \\ 
Ti vessel (dwnstr. face) 	& 245\,$\pm$\,6\,m 					& 39.1\,$\pm$\,6.2\,m 	& 1.12\,k & 944 \\ 
\bottomrule
\end{tabular}
\end{center}
\end{table}

The induced nuclide production in the dump was estimated; in case of uncollided beam, the short lived radio-nuclides are (with half-lives in brackets) $^{15}$O (2 minutes), $^{13}$N (10 minutes) and $^{11}$C (20 minutes) are produced with rates of 1.19$\times$10$^{15}$\,s${}^{-1}$, 5.51$\times$10$^{13}$\,s$^{-1}$ and 3.39$\times$10$^{14}$\,s$^{-1}$, respectively. The long-lived nuclides $^{7}$Be (53.6\,days) and $^{3}$H (12.3\,years) have a production rate of 1.14$\times$10$^{14}$\,s$^{-1}$ and 3.12$\times$10$^{14}$\,s$^{-1}$, respectively. 

Water circulates in the dump in a closed loop, externally cooled.
A continuous water flow of at least 25--30\,litre/s, at an average speed of 1.5\,m/s, is required to remove the power deposited in the innermost part of the dump and maintain the peak temperature of water slightly below its boiling point. A safety factor of at least two should be applied to these values. 

In case of inefficient heat removal, the sudden increase of the water temperature is about $\Delta T$$\sim$55\,K per pulse~\cite{Mereghetti2011}.  Water reaches the boiling point after few bunch trains and the beam must be interlocked. A safety system must be activated to evacuate the vapour pressure generated in the tank. The initial pressurization at 10\,bar moves the boiling point of water to 180${}^\circ$C. allowing a larger margin of manoeuvre, but AUTODYN\textregistered \cite{bib:BID:ref11} simulations show that the 
pressure wave causes an overstress on the dump walls and window at each bunch train, to be added to the stresses due to the 10\,bar hydrostatic pressure. 
Stiffeners are required, and possibly some type of shock absorbers in critical locations, in order to guarantee the dump structural integrity. The use of gas--water mixtures for a more effective damping of the pressure wave is not \textit{a priori} excluded.

\subsection{Collimation System}

The CLIC collimation system 
consists of pairs of spoilers and absorbers in order to remove the unwanted halo of the beam. The CLIC collimation system is driven by their cleaning capabilities and its ability to survive a possible accident scenario.  
The  system was simulated by means of FLUKA \cite{bib:BID:ref4} for a possible beam accident scenario, as well as for normal cleaning operation.

For the accident scenario it was assumed that the full train of 312\,pulses with intensity of 3.72$\times$10$^{9}$ electrons at 1.5\,TeV hits the front face of the energy spoiler(ESP) and, with lower probability, the horizontal or vertical spoilers (XSP/YSP). For the simulation, a shallow depth of the beam was used with an impact parameter of 2\,mm assuming a beam spread of 780$\times$22\,$\mu$m$^{2}$ in horizontal $\times$ vertical for the energy spoiler ESP and 8$\times$1\,$\mu$m$^{2}$ for the transverse spoiler XSP.

During normal operation all spoilers should see only a fraction of beam (of the order of 10$^{-6}$). For the sake of simulation, 
we assumed a beam spread similar to that the accident case.

\begin{figure}[!htb]
  \centering
  \includegraphics[width=0.45\textwidth]{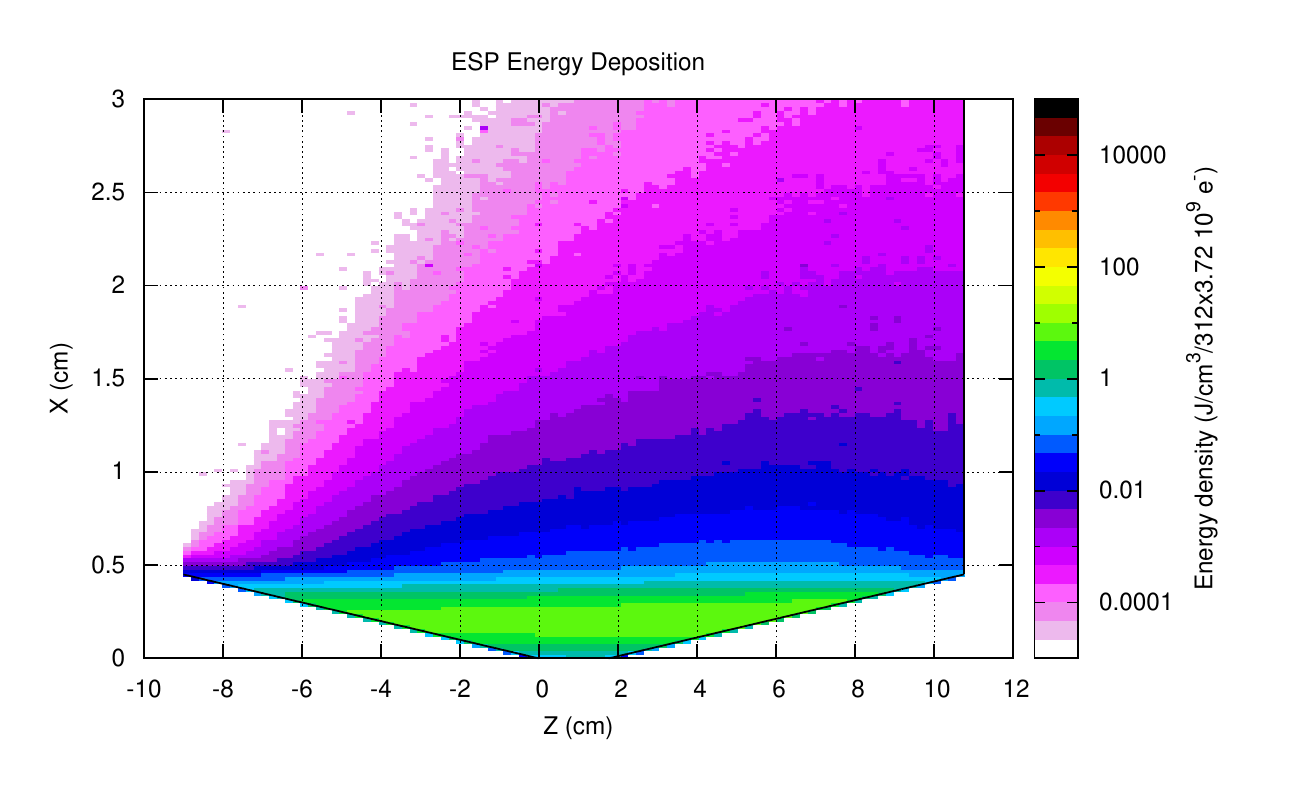}
  \includegraphics[width=0.45\textwidth]{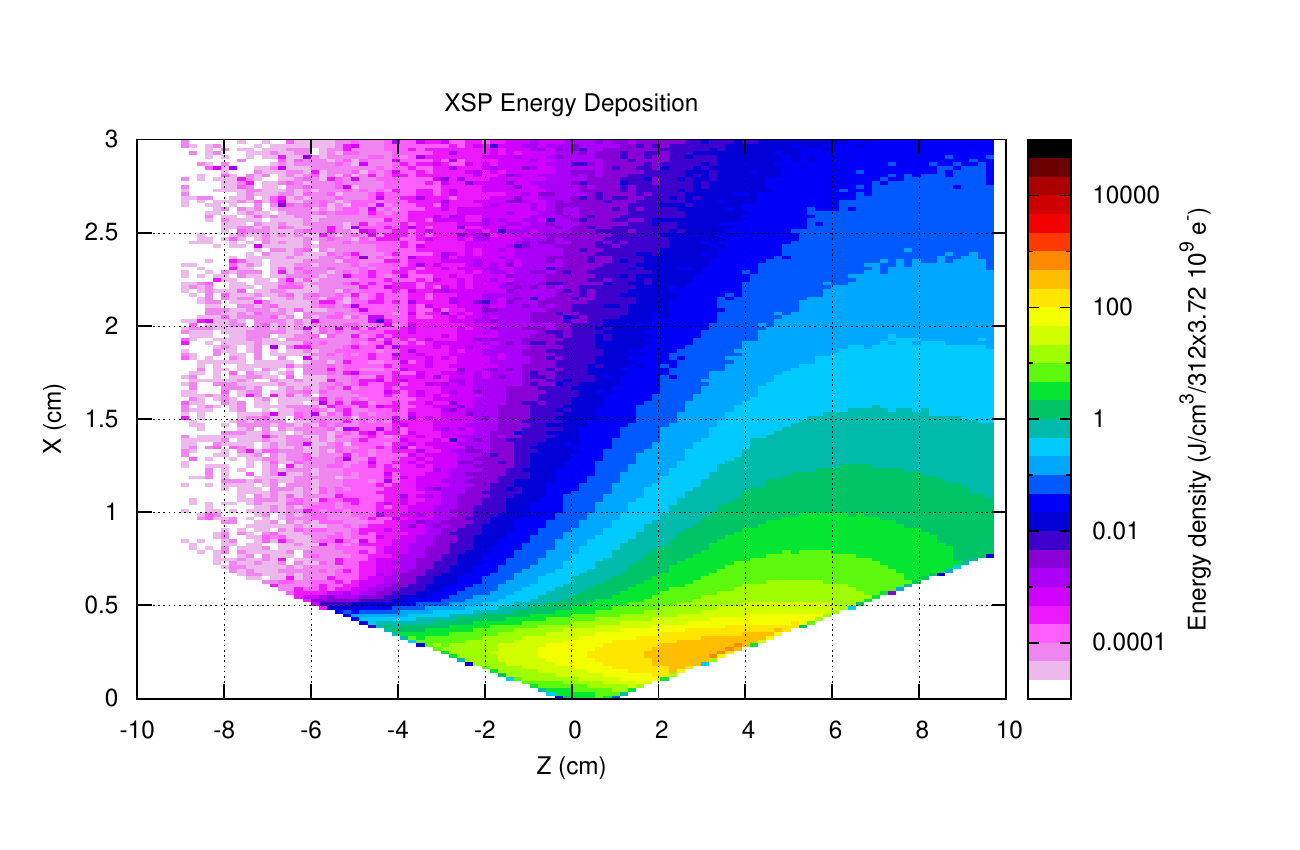}
  \caption{Energy density map in the case of the accident scenario with 2\,mm impact parameter. Left plot shows ESP collimator, right plot XSP collimator}
  \label{fig:BID_3}
\end{figure}

Figure~\ref{fig:BID_3} shows the energy deposition map on a beryllium ESP spoiler for the accident scenario with 2\,mm impact parameter. Although the total beam energy is equivalent to 280\,kJ, the spoiler stops only a tiny fraction of the beam close to 6\,J, about 36.5\,MeV out of 1.5 \,TeV for every primary electron hitting the spoiler. This energy deposition produces a peak energy density of the order of 600\,J/cm$^{3}$, which will generate, assuming adiabatic conditions, an instantaneous temperature rise of 450\,K. The ESP spoiler will survive the accident scenario, since it is practically transparent to the beam, and only a tiny fraction of 2.4$\times$10$^{-5}$ will be stopped on the jaws. The remaining energy will continue downstream, therefore a dedicated study will be required to ensure the protection of the downstream elements that will be exposed on the accidental beam.

In the case of the XSP (Figs.~\ref{fig:BID_3} the considerably smaller beam spread as well as the heavier materials used (titanium for the jaws and copper for coating), results in a very high energy deposition with a peak-energy density of 30\,kJ/cm$^{3}$. Obviously, the jaw material and the coating will not survive this huge energy density therefore a replacement of the spoiler after such a failure scenario would be required.

The main mechanism of secondary particle production on the collimator is the indirect interaction of the electron beam with the spoiler through photo-nuclear interactions from the bremsstrahlung photons. 
Muons, being minimum-ionizing particles, are those which have a higher probability of reaching the experiments at the interaction point. 
The correct evaluation of the muon spectrum reaching the experiments will strongly depend on the material present in the flight path of the muons. A possible remedy would be additional shielding after the spoiler in order to allow the charged mesons to interact before they decay.

According to the design parameters, the ESP should be exposed to fractions of 10$^{-6}$ of the full beam, with a very small impact parameter. Assuming an operation of 200 days per year with a pulse train of 312$\times$3.72$\times$10$^{9}$ electrons every 20\,ms, and a damage threshold for beryllium of 31\,eV, the maximum DPA obtained on the ESP spoiler is of the order of 2.5$\times$10$^{-6}$/year.
Using the same parameters, the core of the ESP spoiler will reach a maximum activity of 200\,mSv/h during operation, which very quickly will drop to values below 0.1\,mSv/h after one hour of cooling due to the very light and short-lived residual nuclei produced.

\subsection{Photon Absorbers in the Damping Rings}

\begin{figure}[!htb]
  \centering
  \includegraphics[width=0.6\textwidth]{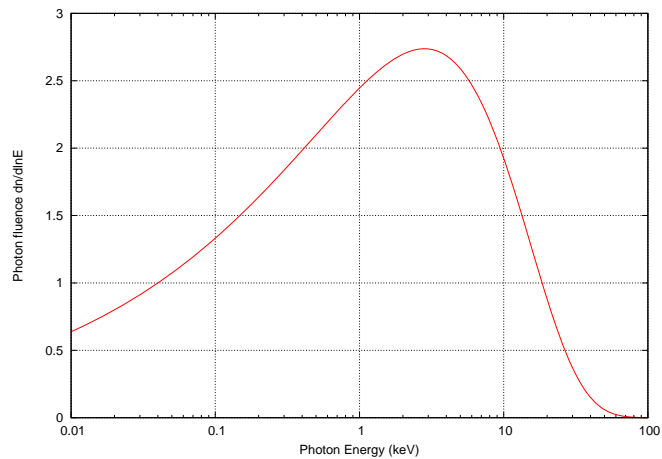}
  \caption{Synchrotron radiation photon spectrum of 2.424\,GeV electrons with a 2.5\,T bending magnet}
  \label{fig:BID_11}
\end{figure}

The twenty six wigglers installed in the CLIC damping rings will generate synchrotron radiation with a critical energy of 9.62\,keV with a spectrum given in Fig.~\ref{fig:BID_11} and emitted in a concentrated light cone with a small opening angle of $\sim$3.2\,mrad. As shown on the plot the maximum energy of these photons is of the order of 50\,keV, resulting in an attenuation length of about 19\,mm on copper. The total power to be absorbed per wiggler is of the order of 8\,kW.

Vertical and horizontal absorbers made of copper are foreseen after each wiggler. Ray tracing simulations \cite{bib:BID:ref13} were performed to estimate the energy density seen by the wiggler absorbers' This gave a total power load of 8\,kW with 95\,W/mm$^{2}$ maximum power density. Owing to the high surface density of deposited power a water cooling system is foreseen on the upper and lower plates of the absorber.

The line of absorbers in the focusing and defocusing quadrupoles of the wiggler section intercepts around 210\,kW of radiation power. The remaining 90\,kW (for the ideal on-axis trajectory) should be stopped by a copper water-cooled absorber downstream of the first bending magnet after the wiggler section. The optimization shows that the most advantageous shape of the final absorber is trapezoidal: in this case we can reduce the power density and improve heat transfer from the illuminated strip on the absorber body to the cooling tubes. The total length of the final absorber is about 5\,m.

\printbibliography[heading=subbibintoc]
\endrefsection
\addtocontents{toc}{\vspace{\normalbaselineskip}}
\refsection 
\chapter{Civil Engineering, Infrastructure and Siting}
\label{Chapter:CEIS}
\section{Civil Engineering}
\label{sect:Civil_Eng}
\subsection{Overview}

\begin{figure}[htb!]
\centering
\includegraphics[width=0.9\textheight, angle=90]{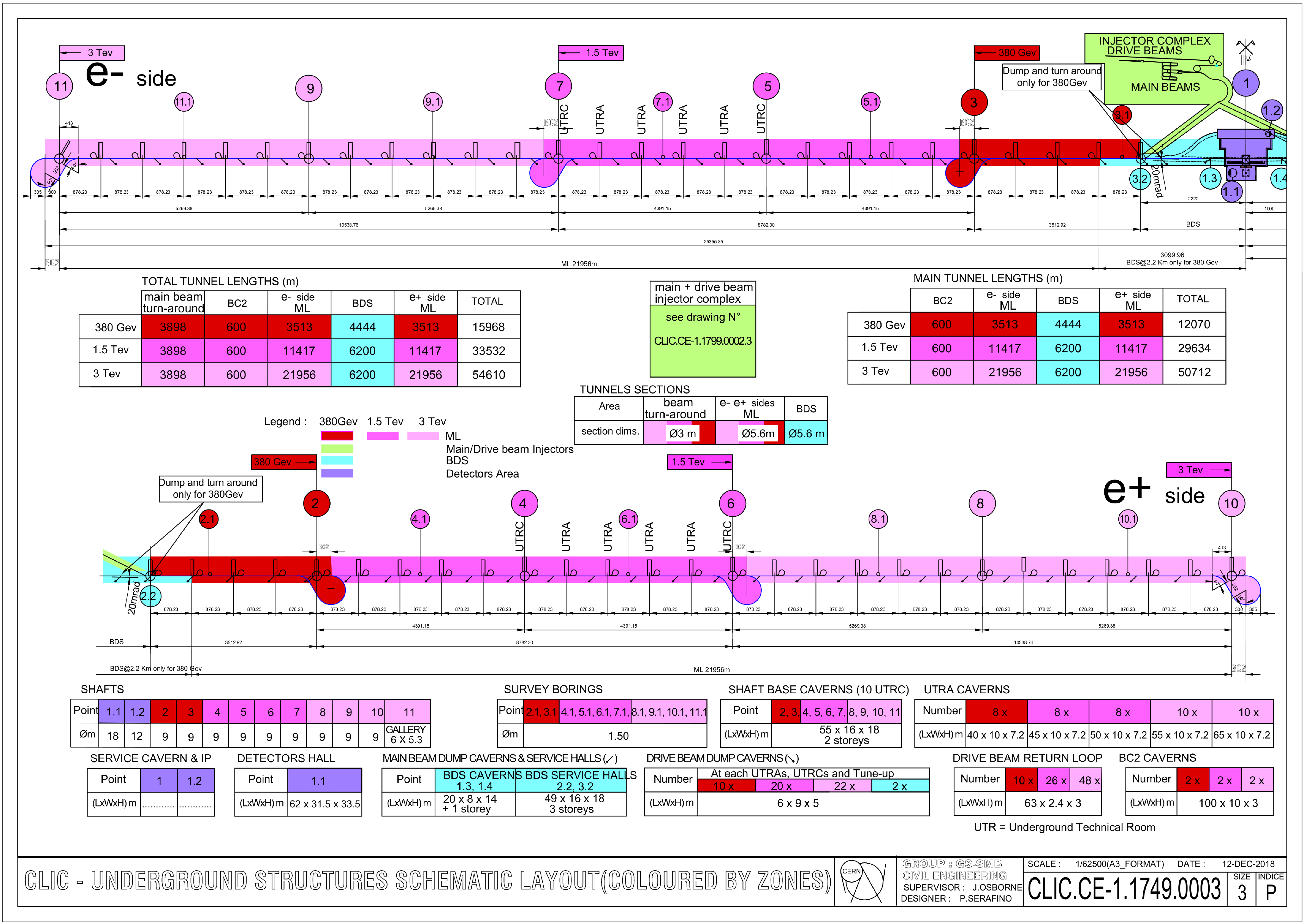}
\caption{\label{fig_CEIS_1} Schematic layout of the three-stage civil engineering complex.}
\end{figure}

Since the CLIC CDR was published in 2012 a number of significant changes have been introduced to the design. The most noteworthy change is the introduction of a new 380\,GeV energy stage for two different machine options, the Drive-Beam option and the Klystron option. The civil engineering design has been optimised to account for this new energy stage, including: a reduced tunnel length, an optimised injector complex, an increased internal tunnel diameter for the Klystron option and site optimisation for access shafts and their associated structures. Figure~\ref{fig_CEIS_1} shows the layout of the civil engineering works for the CLIC project with the three proposed energy stages.

The key features of this layout, noting key differences with respect to the CDR, are:
\begin{itemize}
\item  380\,GeV Drive-Beam machine with a main tunnel length of 12.1\,km.
\item  380\,GeV Klystron machine with a main tunnel length of 11.5\,km.
\item  1.5\,TeV machine with a main tunnel length of 29.6\,km.
\item  3\,TeV machine with a main tunnel length of 50.7\,km.
\item  Inclined access tunnel of 640\,m in place of shaft 11.
\item  One detector cavern for detector assembly and maintenance with a passageway leading to a smaller interaction region (IR) cavern.
\end{itemize}

\subsection{Study Area and Siting}
\begin{figure}[htb!]
\centering
\includegraphics[width=0.6\textwidth]{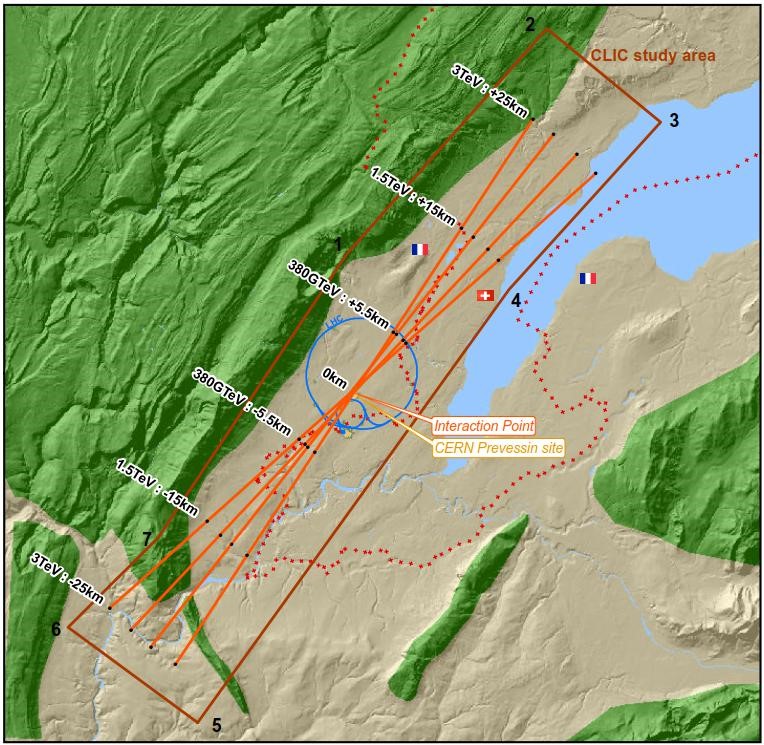}
\caption{\label{fig_CEIS_2} Boundary conditions for the CLIC siting Study area.}
\end{figure}

The location of the CLIC 380\,GeV machine has been optimised whilst still considering the requirements for higher energy stages. This optimisation took into account the availability of existing CERN sites, the regional geology and local environment. Previous experience from the construction of LEP and the LHC has shown that the sedimentary rock in the Geneva basin, known as molasse, provides suitable conditions for tunnelling. Therefore, boundary conditions were established so as to avoid the karstic limestone of the Jura mountain range and to avoid siting the tunnels below Lake Geneva (see Fig.\,\ref{fig_CEIS_2}) whilst maximising the portion of tunnel located in the molasse.

\begin{figure}[htb!]
\centering
\includegraphics[width=0.9\textheight, angle=90]{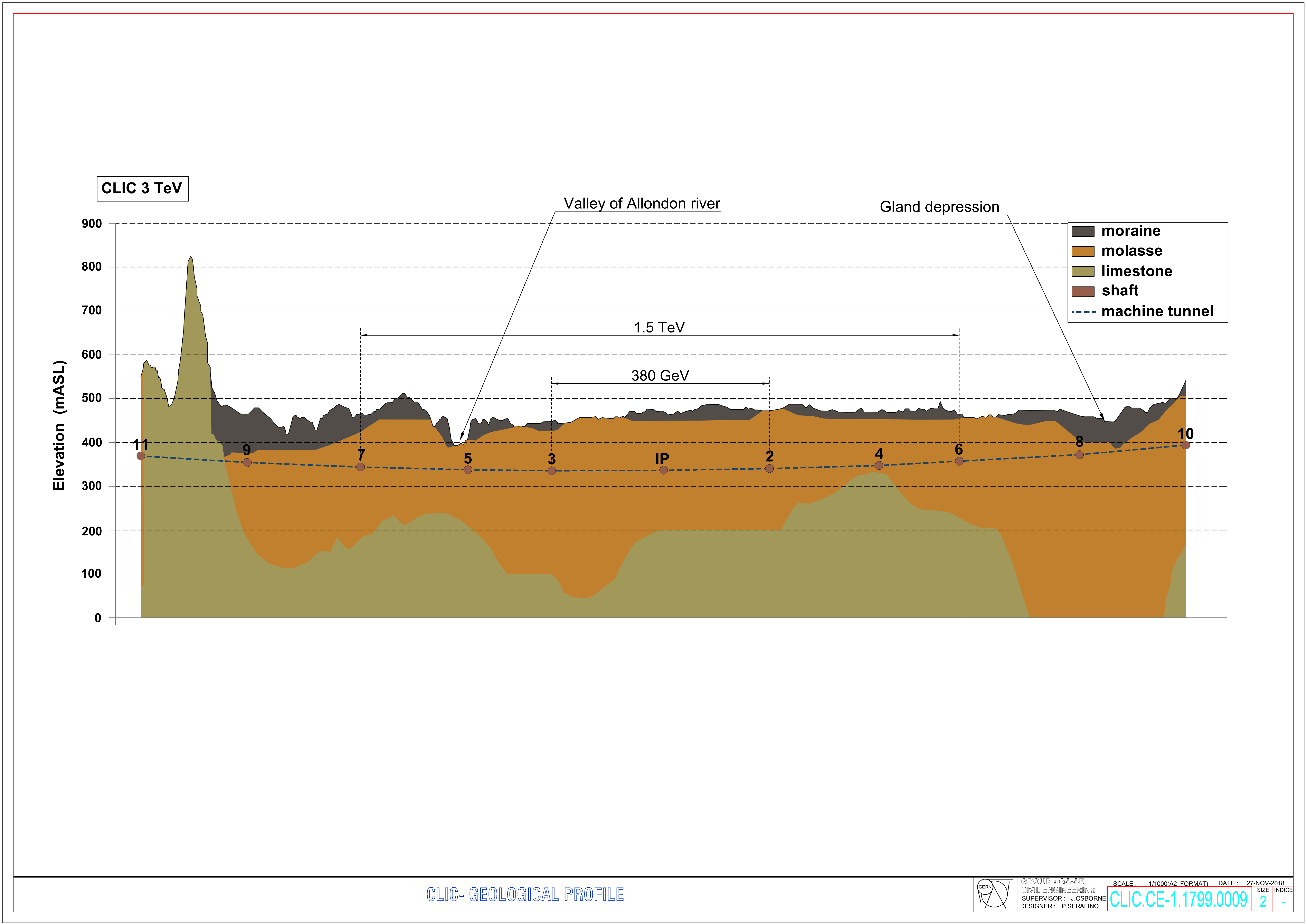}
\caption{\label{fig_CEIS_3} Geological profile of the three-stage linear collider.}
\end{figure}

Based on the regional geological and surface data, and using a bespoke digital modelling tunnel optimisation tool (TOT) developed specifically for CLIC
\cite{Stuart2018}, a 380\,GeV solution has been found that can be feasibly upgraded to the higher energy stages at 1.5\,TeV and 3\,TeV. Figure~\ref{fig_CEIS_3} shows the simplified geological profile of the CLIC machine stages. The 380\,GeV stage is located entirely in molasse rock and avoids the complex Allondon and Gland depressions. This solution is both optimised for 380\,GeV and provides an excellent upgrade possibility to the 1.5\,TeV and 3\,TeV stages. A key advantage of this solution is that the interaction point and injection complex are located on the CERN Pr\'{e}vessin site.

\eject

\subsection{Drive Beam Option}
\subsubsection{Injector Complex}
\begin{figure}[htb!]
\centering
\includegraphics[width=\textwidth]{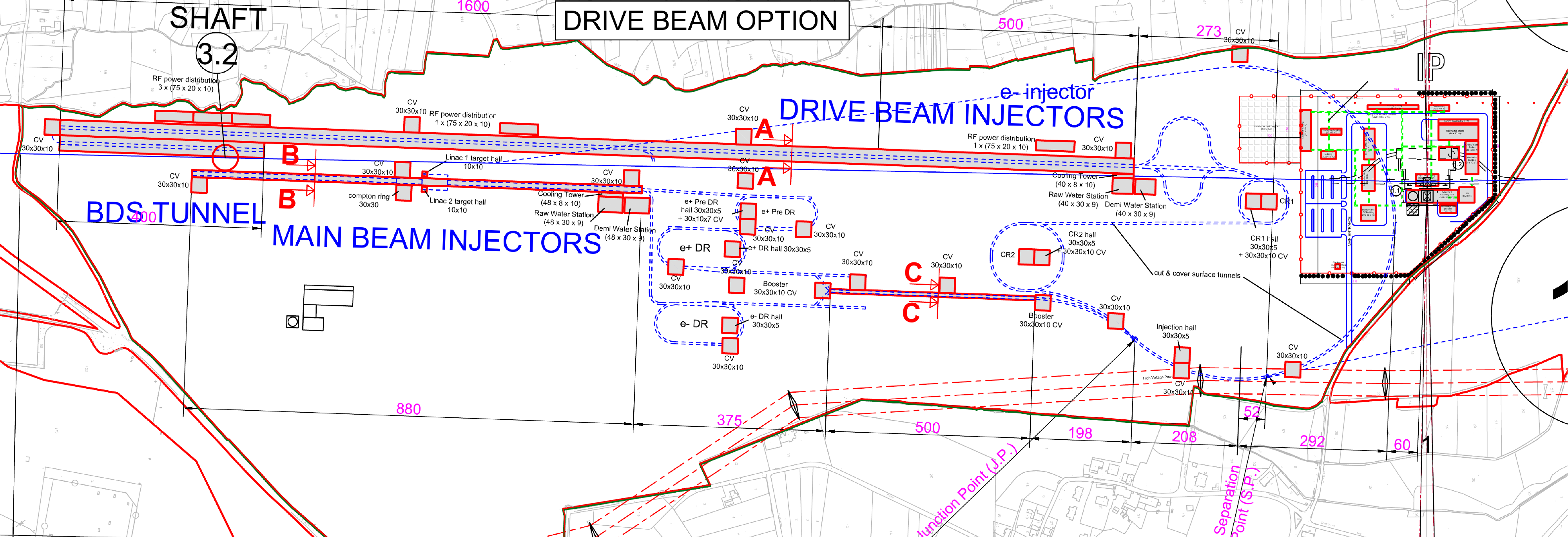}
\caption{\label{fig_CEIS_4} Schematic layout of the Injector Complex.}
\end{figure}

The layout of the Injector and Experimental complexes have been optimised to be completely located on CERN land (see Fig.~\ref{fig_CEIS_4}). At 380\,GeV the two buildings housing the klystrons and modulators of the Drive-Beam Injector complex have a width of 17\,m and a combined length of 2\,km. For the upgrade to 1.5\,TeV the second building is extended by 500\,m in length to allow for increasing the Drive-Beam energy from 1.9 to 2.4\,GeV. To house the RF for the second Drive-Beam Injector for the upgrade to 3\,TeV, a second set of buildings will be constructed running parallel to the initial Drive-Beam Injector buildings. The second accelerator will be integrated into the same tunnel as the first.

The Main-Beam Injectors are also located on the CERN site; they consists of several linacs and damping rings. The layout is similar to that presented in the CDR, except that, the electron Pre-Damping Ring and one experimental shaft are no longer required.

\subsubsection{Linac Tunnel Cross Section}

\begin{figure}[h!]
\centering
\includegraphics[width=\textwidth]{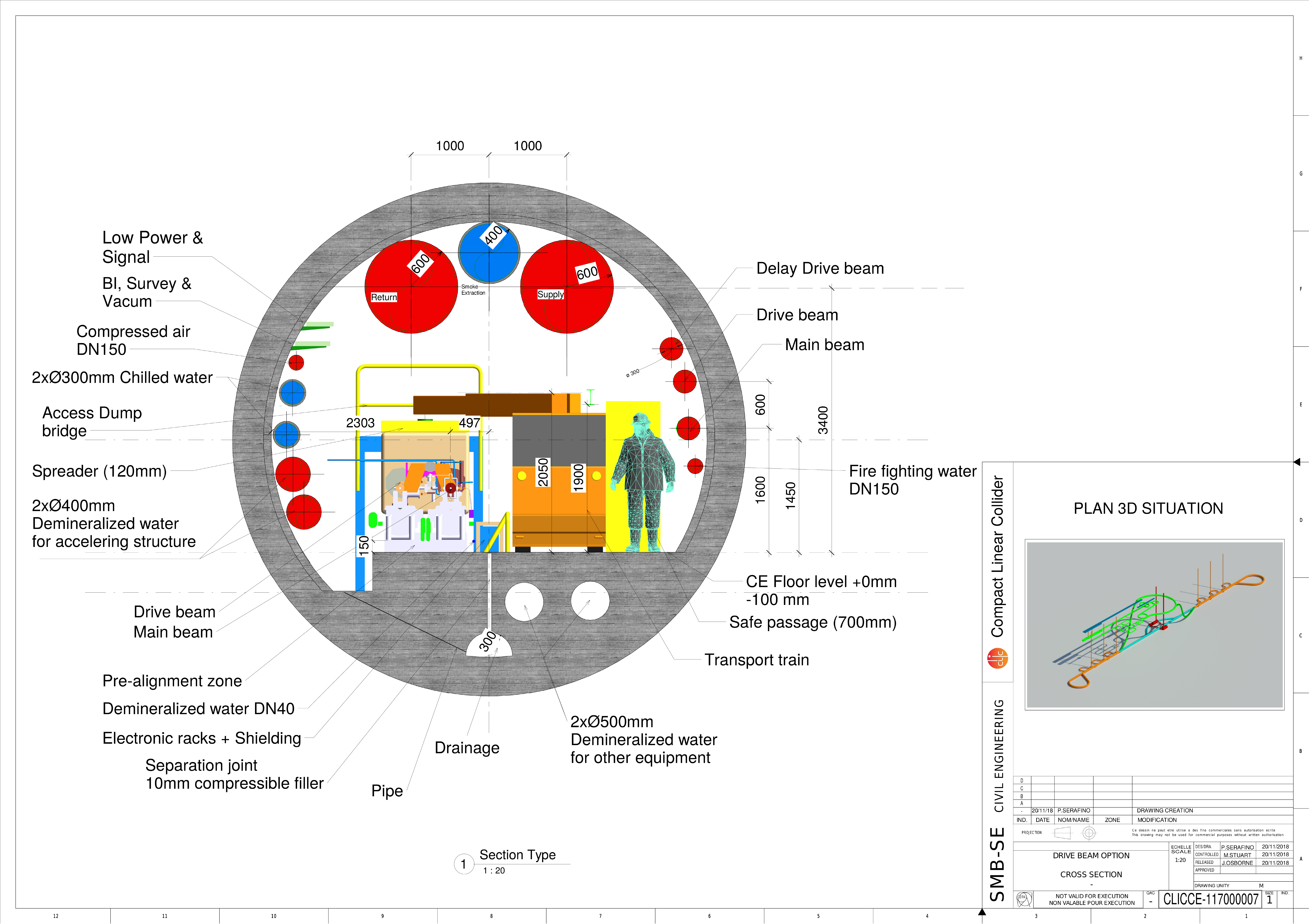}
\caption{\label{fig_CEIS_17} Cross section of the underground tunnel (Drive Beam option).}
\end{figure}

A 5.6\,m internal diameter tunnel is required to house the accelerating structures and all the necessary services (Fig.~\ref{fig_CEIS_17}). This tunnel section has remained largely the same since the CDR. The most significant alteration is the introduction of large ventilation ducts that must be routed through the access shaft caverns (UTRCs) and the service alcoves (UTRAs) located every 878.23\,m (Fig.~\ref{fig_CEIS_1}).

Due to a significant change in the cooling and ventilation solution, the UTRCs and UTRAs have increased in size. It is necessary to construct caverns that allow large CV ducts to pass from the main tunnel to these auxilliary structures which will be required to house the air handling units; the UTRCs are expected to be 60\,m in length and 16\,m in width and the UTRAs for the 380\,GeV energy stage is expected to be 45\,m in length and 15\,m in width.

It is anticipated that the main tunnels and bypass tunnels will be constructed using Tunnel Boring Machines (TBMs). For TBM excavation in a sector with ``good'' conditions, a single pass pre-cast lining is adopted.

\subsubsection{BDS and Interaction Region}
\begin{figure}[htb!]
\centering
\includegraphics[width=0.9\textheight, angle=90]{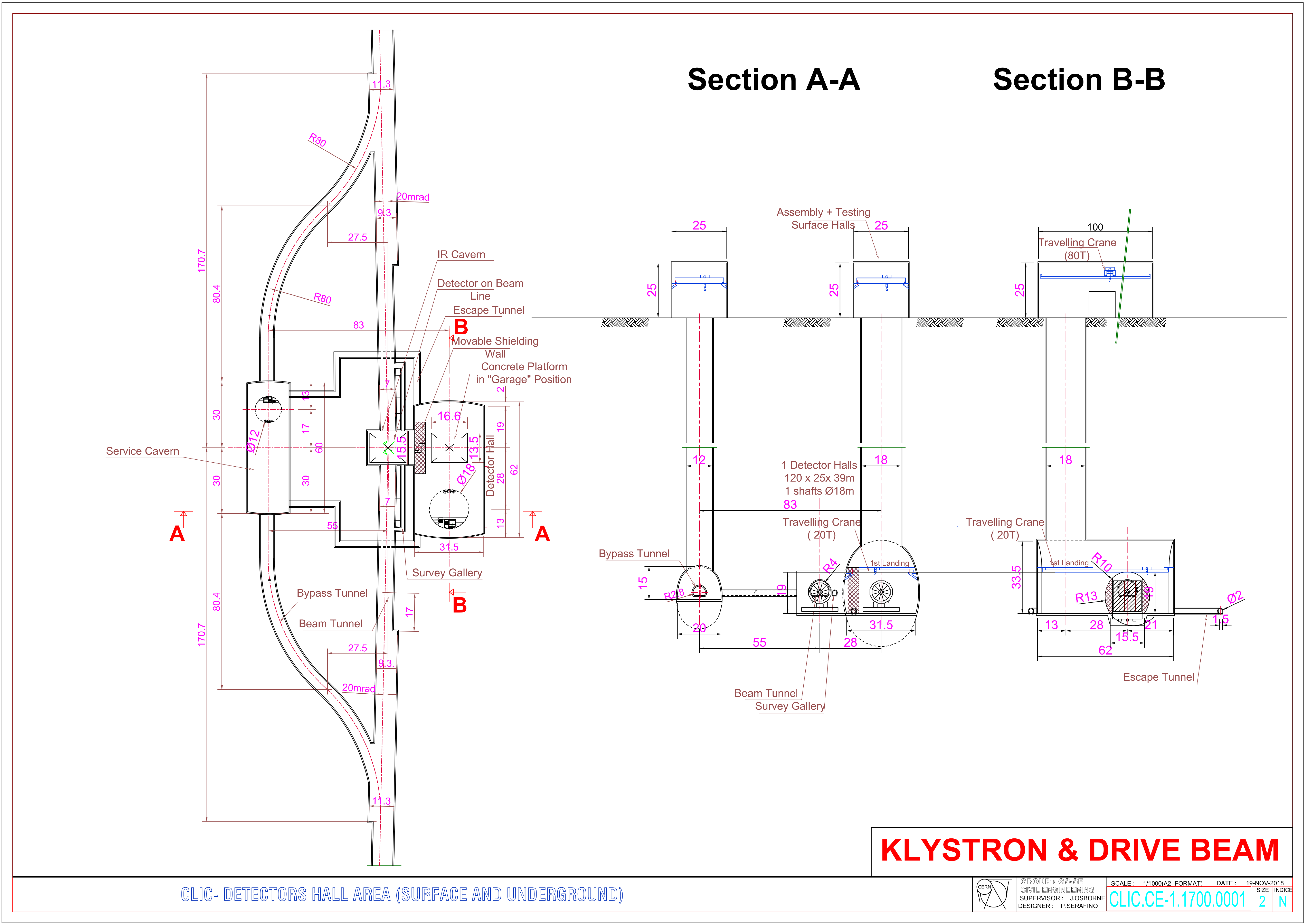}
\caption{\label{fig_CEIS_5} Layout of the CLIC Interaction Region.}
\end{figure}

\begin{figure}[h!]
\centering
\includegraphics[width=\textwidth]{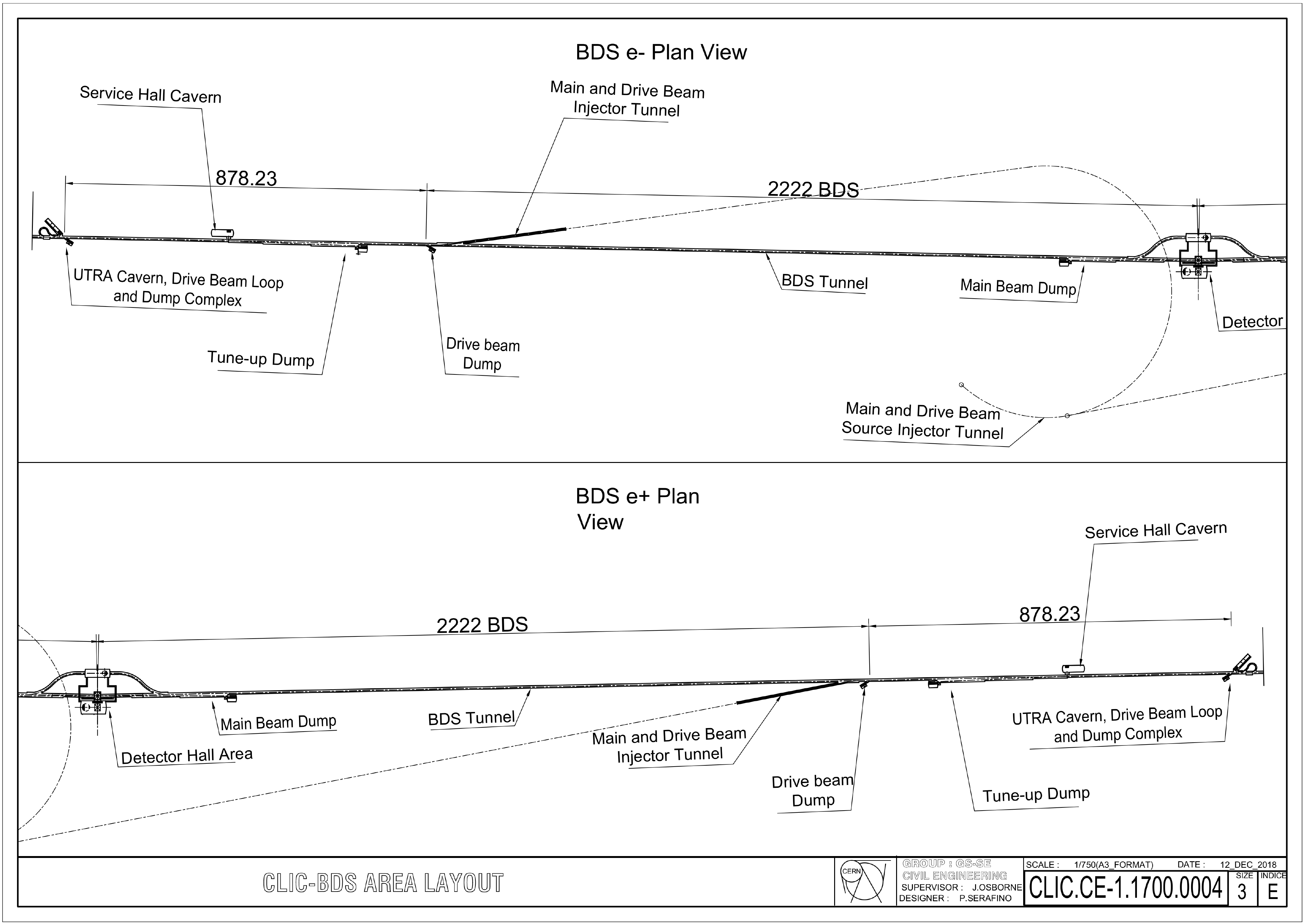}
\caption{\label{fig_CEIS_21} CLIC 380\,GeV BDS Layout}
\end{figure}

A layout of the interaction region is shown in Fig.~\ref{fig_CEIS_5}. It contains one detector and one service cavern, which are interconnected by an escape tunnel that leads to a safe zone in each of the caverns. The service cavern is accessible via a 12\,m internal diameter shaft. The two tunnels that contain the Beam Delivery Systems, shown in Fig.~\ref{fig_CEIS_21}, cross with a total angle of 20\,mrad. This allows effective extraction of the beams after collision and the separation of the post collision line from the incoming beam line. The design is similar to that of the CDR, with one of the experiments removed and a service cavern added. 

\subsection{Klystron Option}

\subsubsection{Injector Complex}
\begin{figure}[h!]
\centering
\includegraphics[width=\textwidth]{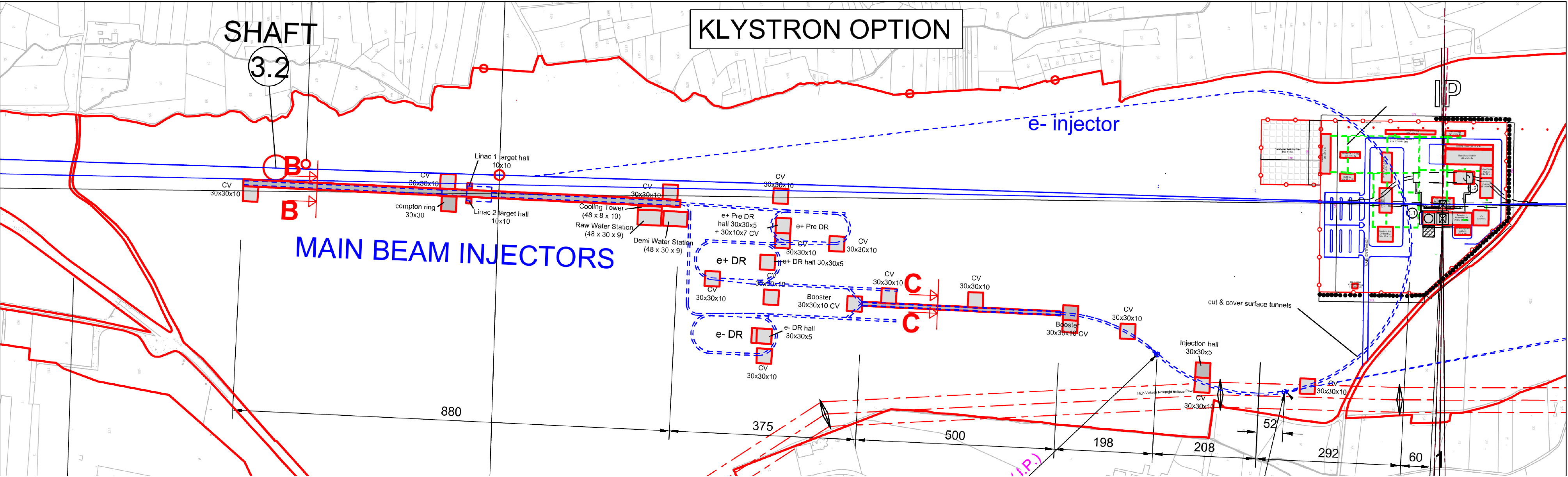}
\caption{\label{fig_CEIS_6} Schematic layout of the Klystron option injection complex.}
\end{figure}

No Drive-Beam Injector is required for the Klystron-based option which simplifies the complex significantly and reduces cost. The Main-Beam Injector complex is similar to the Drive-Beam based design (Fig.\,\ref{fig_CEIS_6}).

\subsubsection{Main Tunnel Layout and Cross Section}
\label{ssect:Main_Tunnel_xsect}
\begin{figure}[h!]
\centering
\includegraphics[width=\textwidth]{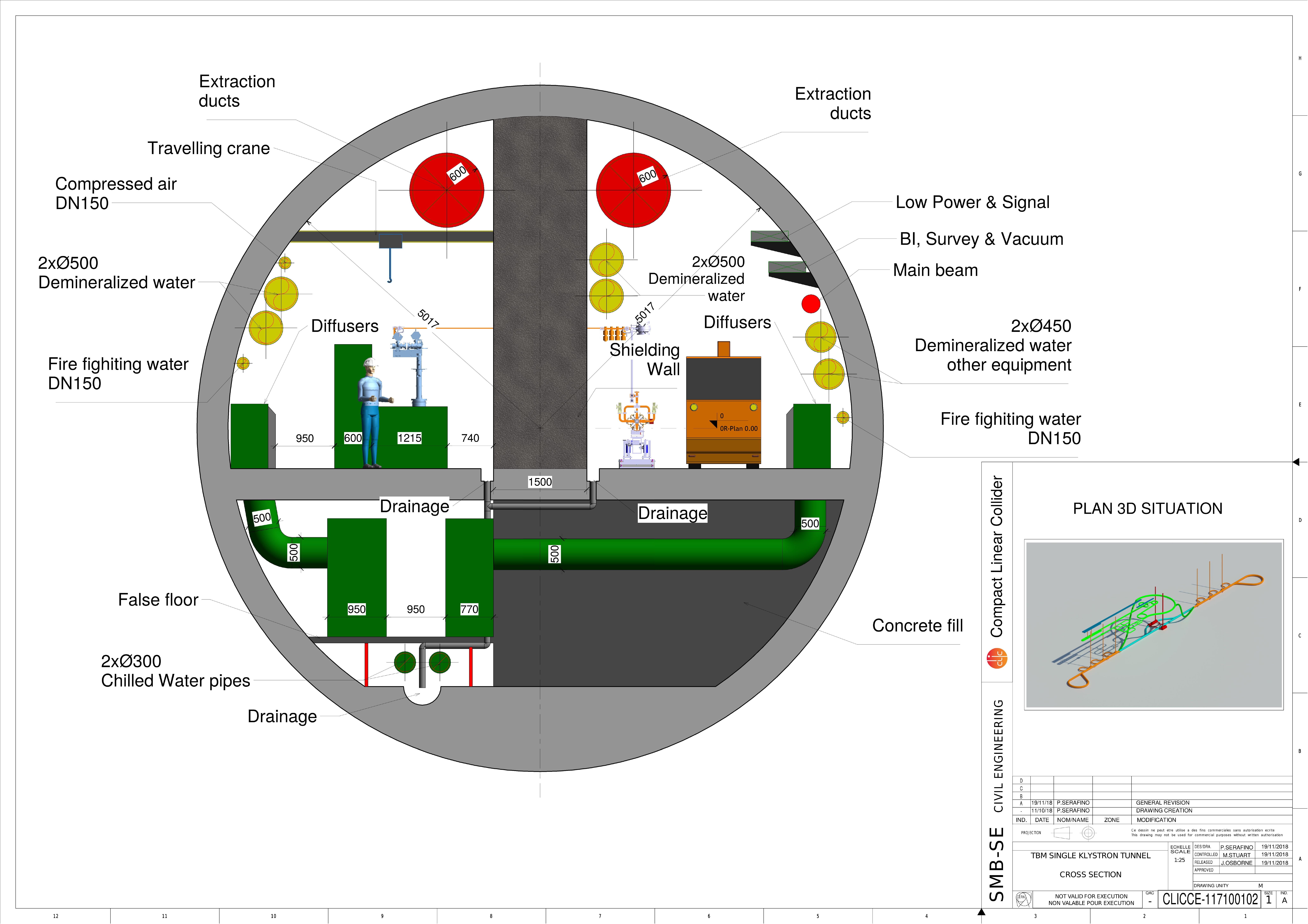}
\caption{\label{fig_CEIS_7} Main Linac cross section for the klystron option.}
\end{figure}

The length of the Main-Linac tunnel is similar to the Drive-Beam option, however, it does not contain the turn-arounds for the Drive Beam. A larger 10\,m internal diameter tunnel is required to house both the accelerating structures and the klystron modules, separated by a 1.5\,m thick shielding wall, which are connected by waveguides (Fig.~\ref{fig_CEIS_7}). In order to minimise the impact of vibrations on the accelerating modules, a services compartment will be located below the less sensitive klystron modules. At each UTRC along the tunnel length the 1.5\,m thick shielding wall will contain a contractible section that can be slid out during long shutdowns to allow vehicle access between the two tunnel compartments.  

The air supply ducts and refrigeration units will be located within this service compartment, which is accessible from the UTRC and UTRA caverns. Access will be provided to these compartments at intervals of 880m via the UTRAs and UTRCs for general maintenance of services. Provisions are required to allow diffusers to be located every 20\,m on both sides of the tunnel. A typical cross-section of the klystron Main-Linac tunnel with access to a UTRA is shown in Fig.~\ref{fig_CEIS_7}. \textit{Smoke extraction duct and sprinkler integration still required.}

\subsubsection{BDS and Interaction Region}
This region is basically the same for both the Drive Beam and Klystron options. However, for efficiency, the cross-section of the BDS tunnel for the klystron design will have a larger internal diameter of 10\,m. This allows the use of the same TBMs for the Main Linacs and the BDS. 

\subsubsection{Construction and Installation Methods}
It is envisaged that the construction method for the Klystron option will be the same as that used for the Main-Linac tunnel, however, once the excavation and tunnel lining is complete the installation of the smaller civil engineering works will differ. As seen in Fig.~\ref{fig_CEIS_7} the klystron option will require a larger concrete invert with a service compartment cast in. The concrete invert will be separated into two sections with a compressive filler between each. The construction of the shielding wall will be staggered with the construction of the concrete invert to allow time for the concrete to cure. This will then be followed by the installation of any internal structural steelwork required for the cooling and ventilation services.

\subsection{Cost Consideration}
The overall cost has been minimised for the 380\,GeV Drive Beam and Klystron options. The cost estimates are based on those used for other ongoing CERN projects and the full cost analysis includes all aspects of construction, including, but not limited to, labour costs, material costs and consumables as well as design, consultancy, and site investigation works.

The major differences between the Klystron-based option in comparison to the Drive Beam option are; larger cost for the Main-Linac tunnel  for the former due to the increased tunnel diameter and a lower cost for the Injector Complex due to the removal of Drive-Beam Injector infrastructure. In comparison to the CDR the cost differs because the e$^{-}$ Pre-Damping Ring and one of the detector shafts along with its cavern and surface buildings are no longer required. It should be noted that multiple cooling and ventilation buildings have been added to both the Drive-Beam and Klystron option Injector Complex. Finally, for the Drive Beam option the Drive-Beam Injector building has been reduced in size, providing significant cost savings. 

\section{Electrical Network}
\label{sect:CEIS_ELec}

The design of the CLIC electrical network is driven by three main factors:

\begin{itemize}
\item  the estimated electrical power requirements (Fig.~\ref{fig_IMP_11}); \item  the location and type of equipment to be supplied and;
\item  the expected level of electrical network availability and operability. 
\end{itemize}

The electrical network is composed of a transmission and a distribution level. The transmission level transmits power from the European Grid to the CLIC sites and between the eleven CLIC surface locations and finally to the underground infrastructure. This network typically operates at high voltage levels of 400\,kV, 135\,kV and 63\,kV. The distribution level distributes the power from the transmission level to the end users at low and medium voltage levels in the range of 400\,V and 36\,kV.

\subsection{Source of Electrical Energy}

\begin{figure}[h!]
\centering
\includegraphics[width=0.75\textwidth]{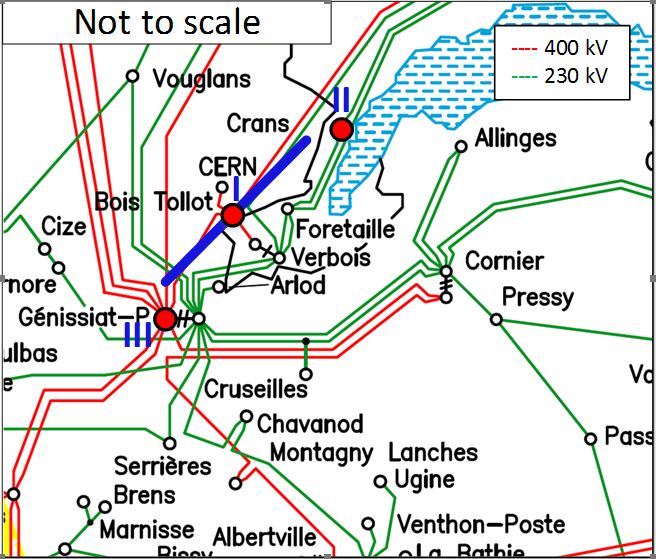}
\caption{\label{fig_CEIS_8} Location of existing power sources near CERN.}
\end{figure}

The electrical power for the 380\,GeV CLIC Drive Beam and Klystron option is supplied from the European Grid. There are two 400\,kV sources and one 230\,kV source located within close proximity to the proposed CLIC location (Fig.~\ref{fig_CEIS_8}). According to RTE (Reseau Transport Electricit\'{e}) on the time horizon of 2035, each of the French supplies is capable of providing 200\,MW of power in addition to their present load. The long-term availability of power from the Swiss supply is unknown. Assuming that 200\,MW of power is available from each of the three sources, the necessary power supply for all four CLIC configurations is currently available. For all four CLIC configurations, roughly 70\% of the total power required is for the injection infrastructure located on the main campus (point 1 in Fig.~\ref{fig_CEIS_8}). The proposed transmission and distribution network considers the upgrade of the European Grid to make available the total CLIC power requirements at point 1.

\subsection{Transmission Network}

The transmission network configuration for 380\,GeV is identical for both the Drive Beam and Klystron options. The network will expand progressively for the 1.5\,TeV and 3\,TeV stages. The proposed voltage levels and network are optimised to allow the use of the existing electrical infrastructure when expanding to the higher CLIC energy stages.

\subsubsection{Transmission Network for the 380\,GeV Stage}

\begin{figure}[h!]
\centering
\includegraphics[width=\textwidth]{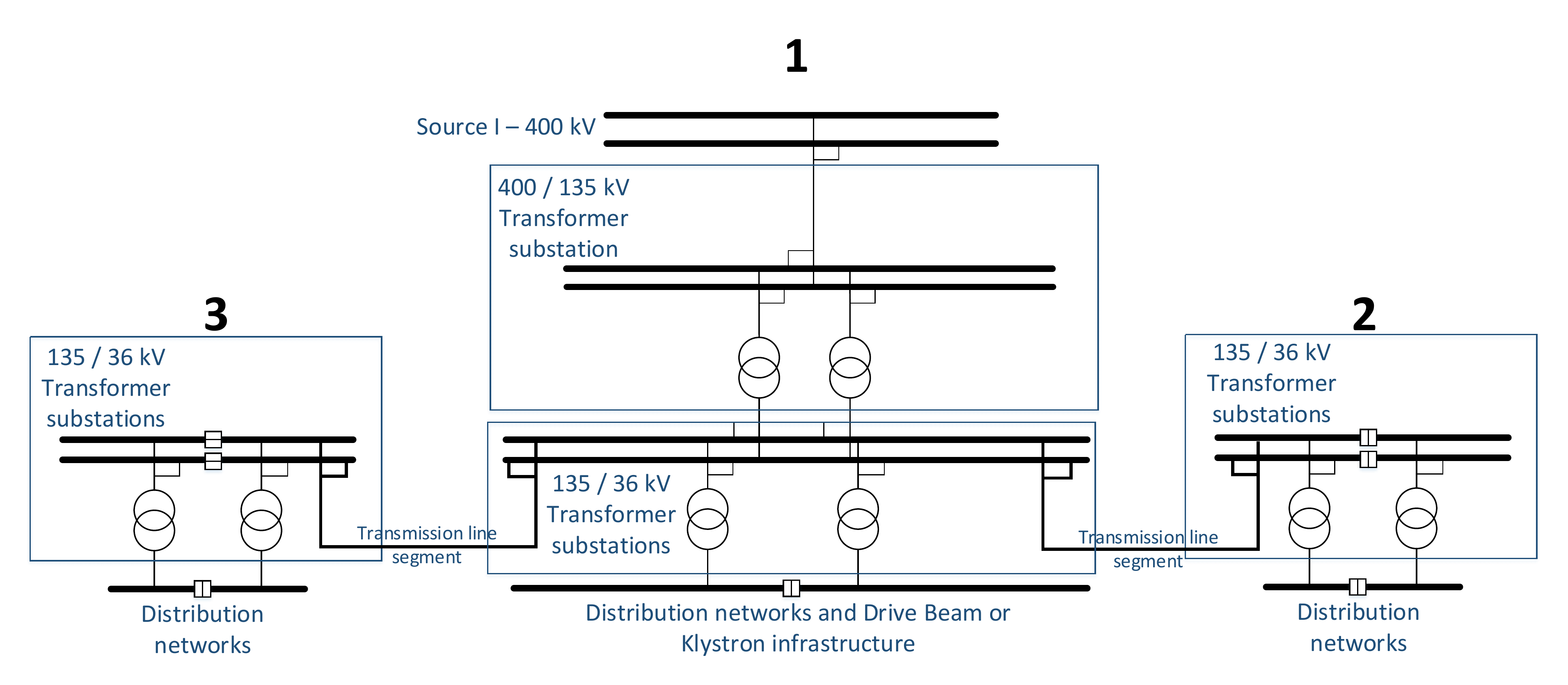}
\caption{\label{fig_CEIS_9} Transmission network scheme for 380\,GeV Drive Beam and Klystron configurations.}
\end{figure}

The transmission network scheme for the 380\,GeV stage (Fig.~\ref{fig_CEIS_9}) includes a primary transformer substation at point 1. From the 400\,kV European Grid, a 400/135\,kV step-down transformer substation supplies a transmission network that connects to points 2 and 3 through a 135\,kV transmission line. From the 135 kV substations at points, 1, 2 and 3 the required 135/36\,kV step-down transformer substations will supply the distribution network. The design includes redundant bus bars at 400\,kV and 135\,kV levels, and redundant power transformers in order to provide the required level of availability, operability and maintainability.

\subsubsection{Transmission Network for the 1.5\,TeV Stage}

\begin{figure}[h!]
\centering
\includegraphics[width=\textwidth]{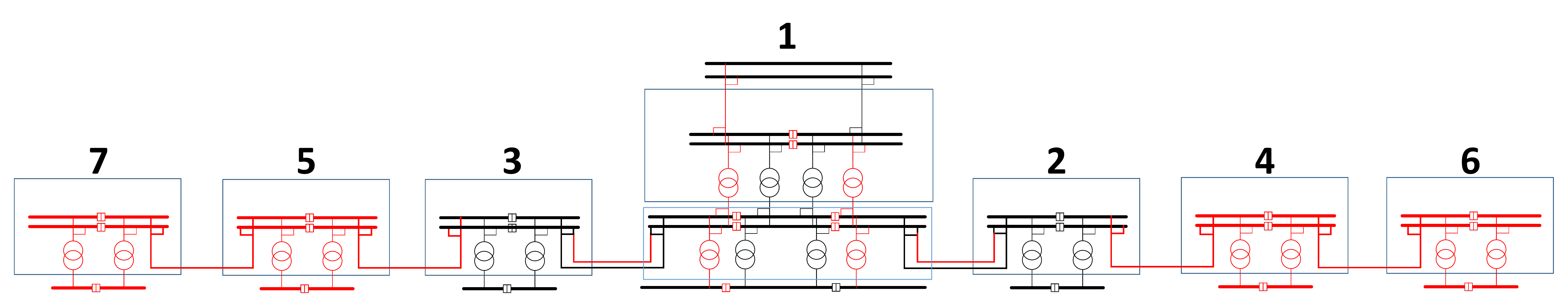}
\caption{\label{fig_CEIS_10} Transmission network baseline for 1.5\,TeV. The extension from the 380\,GeV stage to the 1.5\,TeV stage is shown in red.}
\end{figure}

The upgrade of the transmission network from the 380\,GeV stage to the 1.5\,TeV stage includes the extension of the transmission network to the adjacent points 4, 5, 6 and 7 (Fig.~\ref{fig_CEIS_10}). On each of these four points, a 135/36\,kV step-down transformer substation is constructed and interconnected to the existing substations at points 2 and 3 through new 135 kV transmission line segments. The ampacity of the existing transmission line segments between points 1, 2 and 3 is increased by adding a second line in parallel to the existing ones. At point 1 the existing 400/135\,kV and 135/36\,kV step-down transformer substations are extended to provide the necessary power requirements for the supply of the Drive-Beam infrastructure. 

\subsubsection{Transmission Network for the 3.0\,TeV Stage}

\begin{figure}[h!]
\centering
\includegraphics[width=\textwidth]{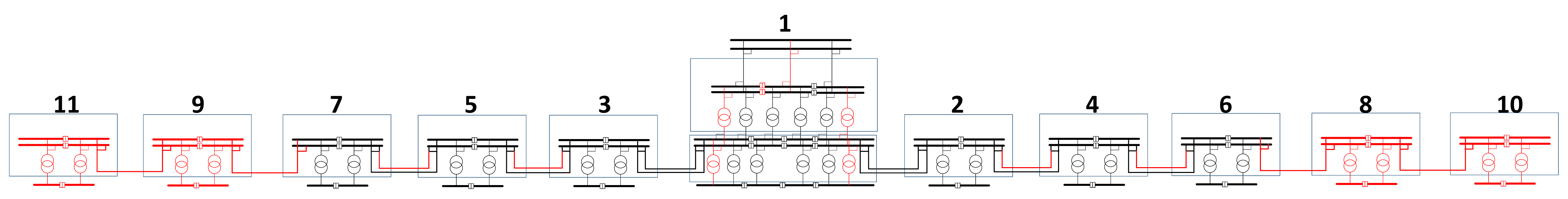}
\caption{\label{fig_CEIS_11} Transmission network baseline for 3\,TeV. The planned extension from the 1.5\,GeV stage to the 3\,TeV Stage is shown in red.}
\end{figure}

The upgrade of the transmission network from the 1.5\,TeV stage to the 3.0\,TeV stage includes the extension of the transmission network to points 8, 9, 10 and 11 (Fig.~\ref{fig_CEIS_11}). At each of these four surface sites, a 135/36\,kV step-down transformer substation is constructed and interconnected to the existing substations, at points 6 and 7, through new 135\,kV transmission line segments. The ampacity of the existing transmission line segments between points 3. 5, 7 and 2, 4, 6 is increased by adding a second line in parallel to the existing ones. At point 1 the existing 400/135\,kV and 135/36\,kV step-down transformer substations are further extended to provide the necessary power required for the supply of the Drive-Beam infrastructure.

\subsection{Distribution Network Topology}

\begin{figure}[hbt]
\centering
\includegraphics[width=\textwidth]{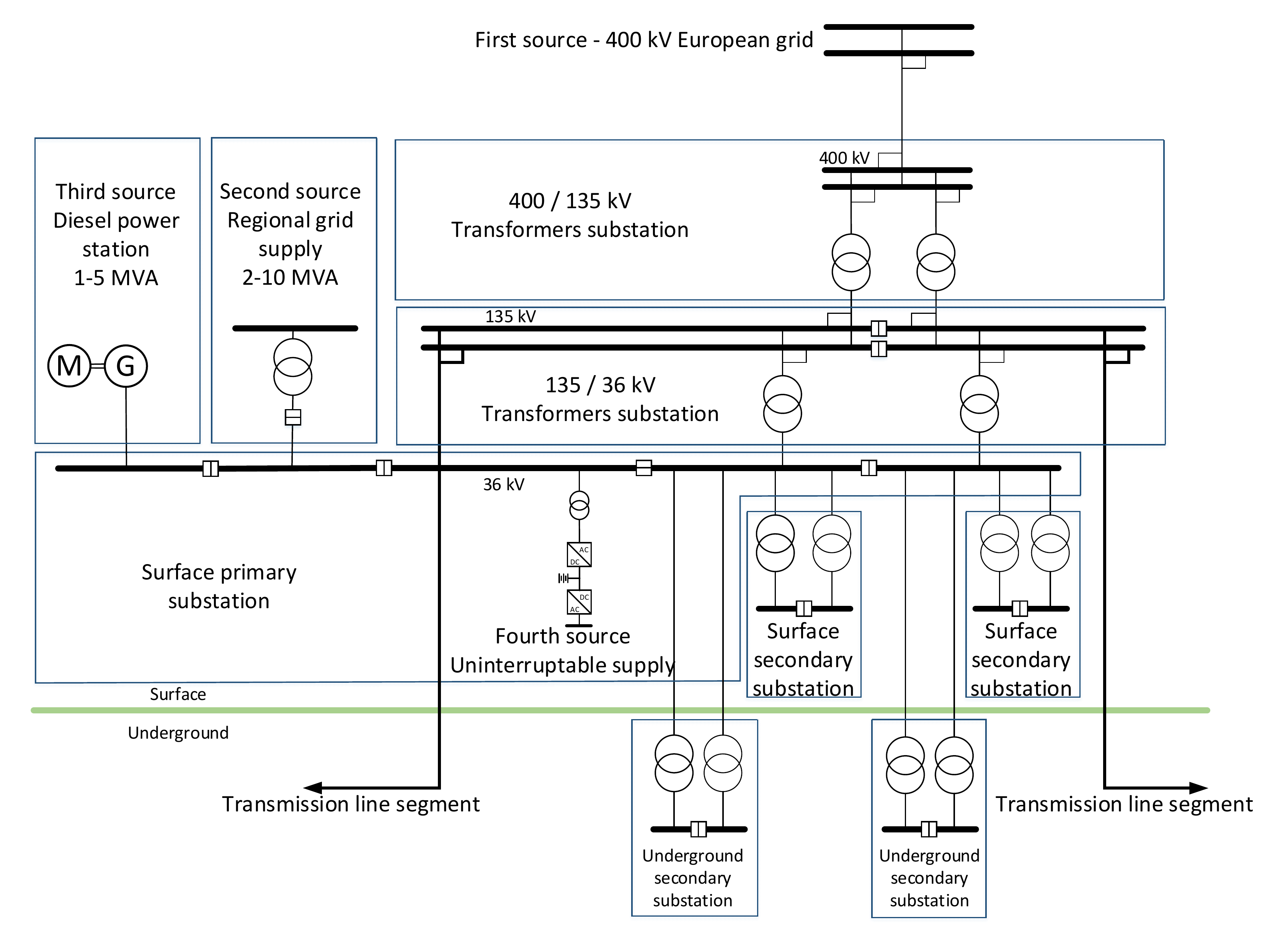}
\caption{\label{fig_CEIS_12} Diagram of the distribution network of one CLIC point including the alternative power sources.}
\end{figure}

The distribution network connects the transmission network to equipment and systems installed on the surface and underground. During nominal operation, the transmission network supplies the distribution network. Alternative sources of supply are required in order to achieve the required level of network availability and to cope with degraded scenarios such as general or local power supply disruption. Therefore, the distribution network includes a second source of supply, rated between 2 and 10\,MVA, fed from a regional grid node, a third source of supply rated from 1 to 5\,MVA of local diesel power stations and a fourth source of supply which provides uninterruptable power. Fig.~\ref{fig_CEIS_12} shows the single line diagram of the distribution network of an individual CLIC surface point including the alternative power sources. 

The distribution network consists of a primary surface indoor substation composed of several bus bars. The incoming feeders are the two redundant 135/36\,kV transformers on the transmission network, that is, the second supply from a regional source and the third supply from a local diesel power station. The outgoing feeders supply the secondary substations. These are located on either the surface or underground, near the load. The operating voltage of the distribution network is typically 36\,kV for the power distribution over distances greater than 750\,m. Voltage step-down transformers feed end users from the secondary substations over a maximum cable length of 750\,m. End users are supplied from the secondary substations at voltage levels between 400\,V for wall plug equipment and 3.3\,kV for high power motors for cooling, ventilation and cryogenic systems. 

\subsection{Emergency Power}

The emergency power concept is based on the requirement to keep essential parts of the accelerator infrastructure operational if the normal power source fails. Particular emphasis is put on loads related to personnel and machine safety during degraded situations. The various load classes and types can be characterized as shown in Table~\ref{CEIS_Table1}. The main ranking parameters are the acceptable duration of the power interruption and whether the load is part of a personnel or accelerator safety system.

\begin{table}[!t]
\centering
\begin{tabular}{|p{1.3in}|p{3.1in}|p{1.3in}|} \hline 
\textbf{Load class} & \textbf{Loads type\newline (non-exhaustive list)} & \textbf{Power unavailability duration in case of degraded scenario} \\ \hline \hline
Machine & Power converters, cooling and ventilation motors, radio frequency & Until return of main supply \\ \hline
General\newline Services & Lighting, pumps, vacuum, wall plugs & Until return of main or secondary supply \\  \hline
Secured & \textbf{Personnel safety:} Lighting, pumps, wall plugs, elevators & 10 -- 30 seconds \\ \hline 
Uninterruptable & \textbf{Personnel safety:} evacuation and anti-panic lighting, fire-fighting system, oxygen deficiency, evacuation \newline \textbf{Machine safety:} sensitive processing and monitoring, beam loss, beam monitoring, machine protection & Interruptions not allowed, continuous service mandatory \\ \hline
\end{tabular}
\caption{Load classes and main characteristics.}
\label{CEIS_Table1}
\end{table}

\begin{figure}[hbt]
\centering
\includegraphics[width=\textwidth]{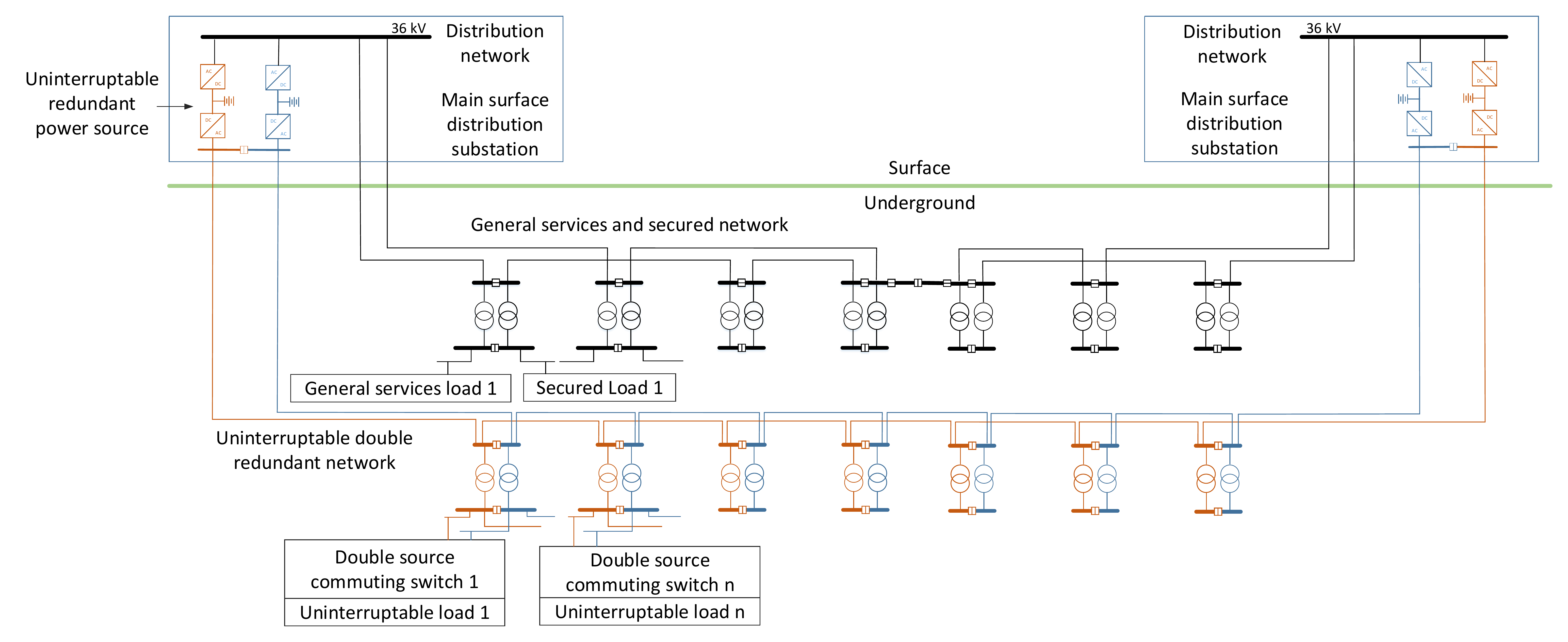}
\caption{\label{fig_CEIS_13} Functional scheme of the general services loads network and the double redundant uninterruptable loads network.}
\end{figure}

Machine loads are energised from the transmission network through the distribution network and do not have a second source of supply. The general services loads typically accept power cuts between several minutes and several hours, sufficiently long to commute to the second source or to wait until the main source is restored. In contrast, a more robust supply is required for secured loads, they include personnel and machine safety equipment or systems that can sustain short power cuts up to a duration of 30\,s. This is provided using three sources of supply. In a degraded situation, the first level backup is provided by the diesel power station, which typically starts up within 10\,s. If the diesel power station is unavailable, the second level back-up supply comes from the regional grid. Finally, the most robust supply is provided to uninterruptable loads, which include personnel and machine safety equipment or systems that require continuous and stable power supply. A specific distribution scheme is used, the uninterruptable network scheme, which is composed of two redundant uninterruptable power supply (UPS) systems supplied from the distribution network in the two adjacent points. Downstream of the redundant UPS systems, a double redundant network delivers two independent sources, each coming from an adjacent point to the end-user plug. Each piece of end-user equipment has two entries and will manage the double source of supply. To meet safety and access requirements, UPS and batteries are located outside the tunnel and above ground. Fig.~\ref{fig_CEIS_13} represents the functional scheme of the general services network and the double redundant uninterruptable loads network.

\section{Cooling and Ventilation}
\label{sect:CEIS_CV}
\subsection{Introduction}

This section is composed of two parts, the first detailing the piped utilities and the second outlining the ventilation systems for the CLIC complex. Throughout this chapter we introduce cooling equipment that has been selected according to the currently available heat loads. No safety margin has been applied to account for uncertainties, possible upgrades, or extensions.

This document reflects the current study for which most of the details are given for the 380\,GeV scenario. Several uncertainties are yet to be clarified and some technical solutions will be revised as the project progresses through the technical design stage; however, the main architecture of the infrastructure is defined and will not change.

\subsection{Piped Utilities}
\subsubsection{Introduction}

The piping systems are mainly dedicated to the cooling of the accelerator equipment and the related infrastructure such as power converters, electronic racks, cables, etc. In addition, specific systems are foreseen to cover other general needs such as fire extinguishing plants, drainage and sump installations (for both surface and underground areas) and compressed air. The main systems presented within this section are: 

\begin{enumerate}
\item  Industrial and demineralized water: for the cooling of accelerator equipment and infrastructure;  
\item  chilled water: for ventilation plants;
\item  drinking water: for sanitary purposes and make up of industrial water circuits as well as production of demineralized water;
\item  industrial water for firefighting systems;
\item  waste water: reject and clear water from underground and surface premises;
\item  compressed air.
\end{enumerate}

\subsubsection{Water Cooling Circuits}

The two main typologies of water cooling circuits are defined according to their working temperatures: circuits cooled by cooling towers and working at temperatures higher than 27°C and circuits using chillers to produce water at 6°C.

The cooling plants have been divided into four independent sectors considering the working parameters, thermal loads and operability constraints. The sectors are as follows:

\begin{enumerate}
\item Drive-Beam Injector complex
\item Main-Linac tunnel
\item Experimental complex
\item Main-Beam Injector complex
\end{enumerate}

The cooling towers for sectors 1, 3 and 4 are centralised at a single location set to minimise the distances to the different premises requiring cooling. For sector 2, a set of cooling towers are located at the surface “odd” shaft sites and will serve the two adjacent sectors via secondary circuits. For the 380\,GeV machine, sector 2 will be exclusively served by the cooling towers placed centrally, close to the interaction point on the Pr\'{e}vessin site.

Where possible, the cooling plants will be located on the surface for accessibility and to facilitate operation and maintenance related activities. 

In order to ensure the smooth running and operation of these systems and considering the accessibility conditions mentioned above, the redundancy has been set to N+1 for electromechanical systems. It is not necessary to ensure the same level of redundancy for the electrical cubicles and the control cubicles. In addition, a secured power supply for the cooling plants is not planned: in case of power failure, all accelerator related equipment will stop and therefore, will not require cooling.

\subsubsubsection{Primary Circuits}

The primary industrial water circuits, cooled by open wet cooling towers, are configured in closed loops to minimise the water consumption. These circuits will be used to cool demineralised secondary circuits, refrigerate chillers and direct expansion units.

Considering a tolerance of 0.5$^\circ$C,the design parameters are the following:

\begin{enumerate}
\item  Cooling tower inlet: 33$^\circ$C 

\item  Cooling tower outlet: 25$^\circ$C
\end{enumerate}

Primary circuits will use raw industrial water and the make-up is made with drinking water; a continuous water treatment against legionellae, scaling and proliferation of algae is foreseen.

Primary circuits are located at the surface. The only exceptions are the circuits connected to the underground caverns, where the distribution systems serving the equipment are installed. For most cases, this distribution system constitutes the secondary circuits.

\subsubsubsection{Secondary Circuits}

Considering a tolerance of 0,5$^\circ$C, the working parameters are:

\begin{enumerate}
\item  Supply from the station: 27$^\circ$C 

\item  Heat exchanger inlet: 35$^\circ$C
\end{enumerate}

Secondary circuits will generally use demineralised water with a maximum conductivity of 0.5\,$\mu$S/cm in a closed loop. 

The demineralized water will partially be produced in a new central station for the entire complex. However, given the long distances between the central area and the equipment, it will not be possible to have an automatic refill pipework without degrading the quality of the water. Therefore, the refill can also be made by transporting the required volume of water in tanks. 

Tables~\ref{tab:Primary_Circuit} and \ref{tab:Secondary_Circuit}, outline the main cooling circuits, the associated cooling powers and flow rates for each sector.

\begin{table}[!htb]
\small
  \centering
  \caption{Primary circuit specifications.}
  \label{tab:Primary_Circuit}
    \begin{tabular}{rlcccr}
    \toprule
    \multicolumn{6}{c}{Primary Circuits} \\
    \midrule
          & \multicolumn{1}{c|}{Structure} & \multicolumn{2}{c|}{Two Beam Machine} & \multicolumn{2}{c}{Klystron Machine} \\
    \midrule
    \multicolumn{1}{c|}{\multirow{2}[4]{*}{\begin{sideways}Sector\end{sideways}}} & \multicolumn{1}{c|}{\multirow{2}[4]{*}{Name}} & \multicolumn{1}{p{4.5em}|}{Cooling Power} & \multicolumn{1}{p{4em}|}{Flow Rate} & \multicolumn{1}{p{4.5em}|}{Cooling Power} & \multicolumn{1}{p{4em}}{Flow Rate} \\
\cmidrule{3-6}    \multicolumn{1}{c|}{} & \multicolumn{1}{c|}{} & \multicolumn{1}{c|}{MW} & \multicolumn{1}{c|}{m3/h} & \multicolumn{1}{c|}{MW} & \multicolumn{1}{c}{m3/h} \\
    \midrule
    \multicolumn{1}{c|}{\multirow{6}[1]{*}{\begin{sideways}1\end{sideways}}} & \multicolumn{1}{p{22.57em}|}{Drive Beam Injector U} & \multicolumn{1}{c|}{5.4} & \multicolumn{1}{c|}{580} & \multicolumn{1}{c|}{\multirow{6}[1]{*}{\textit{---}}} & \multicolumn{1}{c}{\multirow{6}[1]{*}{\textit{---}}} \\
    \multicolumn{1}{c|}{} & \multicolumn{1}{p{22.57em}|}{Drive Beam Injector S} & \multicolumn{1}{c|}{14.2} & \multicolumn{1}{c|}{1530} & \multicolumn{1}{c|}{} &  \\
    \multicolumn{1}{c|}{} & \multicolumn{1}{p{22.57em}|}{Frequency Multiplication Circuit a)} & \multicolumn{1}{c|}{3.2} & \multicolumn{1}{c|}{350} & \multicolumn{1}{c|}{} &  \\
    \multicolumn{1}{c|}{} & \multicolumn{1}{p{22.57em}|}{Frequency Multiplication Circuit b), CR1 S, CR2 S and Transfer Line - CR2 to J.P.} & \multicolumn{1}{c|}{16.8} & \multicolumn{1}{c|}{1820} & \multicolumn{1}{c|}{} &  \\
    \multicolumn{1}{c|}{} & \multicolumn{1}{p{22.57em}|}{Chillers Refrigeration - Drive Beam Injector S/U, Frequency Multiplication Circuit 1, RF Distribution Circuit 1,2 and 3} & \multicolumn{1}{c|}{9.9} & \multicolumn{1}{c|}{1070} & \multicolumn{1}{c|}{} &  \\
    \multicolumn{1}{c|}{} & \multicolumn{1}{p{22.57em}|}{Chillers Refrigeration - CR1 S, CR2 S, Frequency Multiplication Circuit 2, 3, 4 and Transfer Line - CR2 to J.P.} & \multicolumn{1}{c|}{1.8} & \multicolumn{1}{c|}{200} & \multicolumn{1}{c|}{} &  \\
      \midrule
    \multicolumn{1}{c|}{\multirow{10}[1]{*}{\begin{sideways}2/3\end{sideways}}} & \multicolumn{1}{p{22.57em}|}{Accelerator  - Klystron} & \multicolumn{1}{c|}{\textit{---}} & \multicolumn{1}{c|}{\textit{---}} & \multicolumn{1}{c|}{25} & \multicolumn{1}{c}{2670} \\
    \multicolumn{1}{c|}{} & \multicolumn{1}{p{22.57em}|}{Accelerator  - LINAC} & \multicolumn{1}{c|}{18.6} & \multicolumn{1}{c|}{2000} & \multicolumn{1}{c|}{27} & \multicolumn{1}{c}{2930} \\
    \multicolumn{1}{c|}{} & \multicolumn{1}{p{22.57em}|}{Main Tunnel (other equipment)} & \multicolumn{1}{c|}{34.3} & \multicolumn{1}{c|}{3700} & \multicolumn{1}{c|}{24} & \multicolumn{1}{c}{2610} \\
    \multicolumn{1}{c|}{} & \multicolumn{1}{p{22.57em}|}{Injection Hall and Transfer Lines - e+/e- , Loop, J.P to S.P (P\&ID Circuit B)} & \multicolumn{1}{c|}{5.9} & \multicolumn{1}{c|}{640} & \multicolumn{1}{c|}{5.9} & \multicolumn{1}{c}{640} \\
    \multicolumn{1}{c|}{} & \multicolumn{1}{p{22.57em}|}{Detectors S} & \multicolumn{1}{c|}{0.9} & \multicolumn{1}{c|}{100} & \multicolumn{1}{c|}{0.9} & \multicolumn{1}{c}{100} \\
    \multicolumn{1}{c|}{} & \multicolumn{1}{p{22.57em}|}{Detectors U} & \multicolumn{1}{c|}{2.0} & \multicolumn{1}{c|}{230} & \multicolumn{1}{c|}{2.0} & \multicolumn{1}{c}{230} \\
    \multicolumn{1}{c|}{} & \multicolumn{1}{p{22.57em}|}{Chillers Refrigeration - Buildings IP} & \multicolumn{1}{c|}{1.7} & \multicolumn{1}{c|}{190} & \multicolumn{1}{c|}{1.7} & \multicolumn{1}{c}{190} \\
    \multicolumn{1}{c|}{} & \multicolumn{1}{p{22.57em}|}{Chillers Refrigeration - Main tunnel} & \multicolumn{1}{c|}{5.3} & \multicolumn{1}{c|}{580} & \multicolumn{1}{c|}{7.0} & \multicolumn{1}{c}{750} \\
    \multicolumn{1}{c|}{} & \multicolumn{1}{p{22.57em}|}{Chillers Refrigeration - Detectors Hall S/U, Injection Hall,  Transfer Lines - Loop and J.P. to S.P. (P\&ID Circuit A)} & \multicolumn{1}{c|}{2.2} & \multicolumn{1}{c|}{240} & \multicolumn{1}{c|}{2.2} & \multicolumn{1}{c}{240} \\
    \multicolumn{1}{c|}{} & \multicolumn{1}{p{22.57em}|}{Main Tunnel Purge} & \multicolumn{1}{c|}{1.2} & \multicolumn{1}{c|}{130} & \multicolumn{1}{c|}{3.6} & \multicolumn{1}{c}{390} \\
    \midrule
    \multicolumn{1}{c|}{\multirow{6}[2]{*}{\begin{sideways}4\end{sideways}}} & \multicolumn{1}{p{22.57em}|}{Main Beam Injector U} & \multicolumn{1}{c|}{3.9} & \multicolumn{1}{c|}{420} & \multicolumn{1}{c|}{3.9} & \multicolumn{1}{c}{420} \\
    \multicolumn{1}{c|}{} & \multicolumn{1}{p{22.57em}|}{Main Beam Injector S} & \multicolumn{1}{c|}{5.1} & \multicolumn{1}{c|}{560} & \multicolumn{1}{c|}{5.1} & \multicolumn{1}{c}{560} \\
    \multicolumn{1}{c|}{} & \multicolumn{1}{p{22.57em}|}{Booster S/U, Damping Ring e- S/U,  and Transfer Line - Booster to J.P.} & \multicolumn{1}{c|}{11.2} & \multicolumn{1}{c|}{1210} & \multicolumn{1}{c|}{11.2} & \multicolumn{1}{c}{1210} \\
    \multicolumn{1}{c|}{} & \multicolumn{1}{p{22.57em}|}{Pre Damping Ring S/U, Damping Ring e+ S/U} & \multicolumn{1}{c|}{8.3} & \multicolumn{1}{c|}{900} & \multicolumn{1}{c|}{8.3} & \multicolumn{1}{c}{900} \\
    \multicolumn{1}{c|}{} & \multicolumn{1}{p{22.57em}|}{Chillers Refrigeration - Main Beam Injector S/U,  Compton Ring S, Traget Hall S} & \multicolumn{1}{c|}{2.4} & \multicolumn{1}{c|}{260} & \multicolumn{1}{c|}{2.4} & \multicolumn{1}{c}{260} \\
    \multicolumn{1}{c|}{} & \multicolumn{1}{p{22.57em}|}{Chillers Refrigeration - Pre Damping Ring S/U, Damping Rings e+, e- S/U, Booster S/U and Transfer Line - Booster to J.P.} & \multicolumn{1}{c|}{4.4} & \multicolumn{1}{c|}{480} & \multicolumn{1}{c|}{4.4} & \multicolumn{1}{c}{480} \\
    \midrule
          & Total Cooling & 159   &       & 135   &  \\
    \bottomrule
    \end{tabular}%
\end{table}%


\begin{table}[!htb]
  \centering
\caption{Secondary circuit specifications.}
\label{tab:Secondary_Circuit}
    \begin{tabular}{rlcrcr}
    \toprule
    \multicolumn{6}{c}{Secondary Circuits, Demineralized Water} \\
    \midrule
    \multicolumn{1}{c|}{\multirow{3}[5]{*}{\begin{sideways}Sector\end{sideways}}} & \multicolumn{1}{c|}{Structure} & \multicolumn{2}{c|}{Two Beam Machine} & \multicolumn{2}{c}{Klystron Machine} \\
\cmidrule{2-6}    \multicolumn{1}{c|}{} & \multicolumn{1}{c|}{\multirow{2}[3]{*}{Name}} & \multicolumn{1}{p{4.5em}|}{Cooling Power} & \multicolumn{1}{c|}{Flow Rate} & \multicolumn{1}{p{4.5em}|}{Cooling Power} & \multicolumn{1}{c}{Flow Rate} \\
\cmidrule{3-6}    \multicolumn{1}{c|}{} & \multicolumn{1}{c|}{} & \multicolumn{1}{c|}{MW} & \multicolumn{1}{c|}{m3/h} & \multicolumn{1}{c|}{MW} & \multicolumn{1}{c}{m3/h} \\
    \midrule
    \multicolumn{1}{c|}{\multirow{4}[1]{*}{\begin{sideways}1\end{sideways}}} & \multicolumn{1}{p{15em}|}{Drive Beam Injector U} & \multicolumn{1}{c|}{5.4} & \multicolumn{1}{c|}{580} & \multicolumn{1}{c|}{\textit{---}} & \multicolumn{1}{c}{\textit{---}} \\
    \multicolumn{1}{c|}{} & \multicolumn{1}{p{15em}|}{Drive Beam Injector S} & \multicolumn{1}{c|}{14.2} & \multicolumn{1}{c|}{1530} & \multicolumn{1}{c|}{\textit{---}} & \multicolumn{1}{c}{\textit{---}} \\
    \multicolumn{1}{c|}{} & \multicolumn{1}{p{15em}|}{Frequency Multiplication Circuit a)} & \multicolumn{1}{c|}{3.2} & \multicolumn{1}{c|}{350} & \multicolumn{1}{c|}{\textit{---}} & \multicolumn{1}{c}{\textit{---}} \\
    \multicolumn{1}{c|}{} & \multicolumn{1}{p{15em}|}{Frequency Multiplication Circuit b), CR1 S, CR2 S and Transfer Line - CR2 to J.P.} & \multicolumn{1}{c|}{16.8} & \multicolumn{1}{c|}{1820} & \multicolumn{1}{c|}{\textit{---}} & \multicolumn{1}{c}{\textit{---}} \\
    \midrule
    \multicolumn{1}{c|}{\multirow{6}[2]{*}{\begin{sideways}2/3\end{sideways}}} & \multicolumn{1}{p{15em}|}{Accelerator  - Klystron} & \multicolumn{1}{c|}{\textit{---}} & \multicolumn{1}{c|}{\textit{---}} & \multicolumn{1}{c|}{24.8} & \multicolumn{1}{c}{2670} \\
    \multicolumn{1}{c|}{} & \multicolumn{1}{p{15em}|}{Accelerator  - LINAC} & \multicolumn{1}{c|}{18.6} & \multicolumn{1}{c|}{2000} & \multicolumn{1}{c|}{27.1} & \multicolumn{1}{c}{2930} \\
    \multicolumn{1}{c|}{} & \multicolumn{1}{p{15em}|}{Main Tunnel (other equipment)} & \multicolumn{1}{c|}{34.3} & \multicolumn{1}{c|}{3700} & \multicolumn{1}{c|}{24.1} & \multicolumn{1}{c}{2610} \\
    \multicolumn{1}{c|}{} & \multicolumn{1}{p{15em}|}{Injection Hall and Transfer Lines - e+/e- , Loop, J.P to S.P} & \multicolumn{1}{c|}{5.9} & \multicolumn{1}{c|}{640} & \multicolumn{1}{c|}{5.9} & \multicolumn{1}{c}{640} \\
    \multicolumn{1}{c|}{} & \multicolumn{1}{p{15em}|}{Detectors S} & \multicolumn{1}{c|}{0.9} & \multicolumn{1}{c|}{100} & \multicolumn{1}{c|}{0.9} & \multicolumn{1}{c}{100} \\
    \multicolumn{1}{c|}{} & \multicolumn{1}{p{15em}|}{Detectors U} & \multicolumn{1}{c|}{2.0} & \multicolumn{1}{c|}{230} & \multicolumn{1}{c|}{2.0} & \multicolumn{1}{c}{230} \\
    \midrule
    \multicolumn{1}{c|}{\multirow{4}[2]{*}{\begin{sideways}4\end{sideways}}} & \multicolumn{1}{p{15em}|}{Main Beam Injector U} & \multicolumn{1}{c|}{3.9} & \multicolumn{1}{c|}{420} & \multicolumn{1}{c|}{3.9} & \multicolumn{1}{c}{420} \\
    \multicolumn{1}{c|}{} & \multicolumn{1}{p{15em}|}{Main Beam Injector S} & \multicolumn{1}{c|}{5.1} & \multicolumn{1}{c|}{560} & \multicolumn{1}{c|}{5.1} & \multicolumn{1}{c}{560} \\
    \multicolumn{1}{c|}{} & \multicolumn{1}{p{15em}|}{Pre Damping Ring S/U, Damping Ring e+ S/U} & \multicolumn{1}{c|}{8.3} & \multicolumn{1}{c|}{900} & \multicolumn{1}{c|}{8.3} & \multicolumn{1}{c}{900} \\
    \multicolumn{1}{c|}{} & \multicolumn{1}{p{15em}|}{Booster S/U, Damping Ring e- S/U,  and Transfer Line - Booster to J.P.} & \multicolumn{1}{c|}{11.2} & \multicolumn{1}{c|}{1210} & \multicolumn{1}{c|}{11.2} & \multicolumn{1}{c}{1210} \\
    \midrule
          & Total Cooling & 130   &       & 113   &  \\
    \bottomrule
    \end{tabular}%
\end{table}%

\subsubsubsection{Chilled Water Circuits}

Chilled water is used in the cooling batteries of air-handling units to cool the air; the working temperatures are:

\begin{enumerate}
\item  Outlet from the station: 6$^\circ$C if dehumidification is required, if not: 12$^\circ$C 

\item  Return to the station: 12$^\circ$C if dehumidification is required, if not: 18$^\circ$C
\end{enumerate}

Chilled water production plants are as close as possible to the air-handling units to which they are connected. Smaller stations, that are closer to the equipment, are preferred over one single central station.

Most of the chillers will be water-cooled. Air-cooled chillers will only be employed if there are no cooling towers in the vicinity.

Air-handling units are located at the surface and in the caverns; therefore, there are chilled water circuits located in the main tunnels. The possibility of having direct expansion units instead of installing air-handling units is to be studied in the future. 

The redundancy level follows the same principle as the primary circuits.

The chilled water circuit specifications are provided in Table~\ref{tab:Chilled_water}.

\begin{table}[!htb]
  \centering
\caption{Chilled water circuits specifications.}
\label{tab:Chilled_water}
    \begin{tabular}{rlcccr}
    \toprule
    \multicolumn{6}{c}{Chilled Water Circuits} \\
    \midrule
    \multicolumn{1}{c|}{\multirow{3}[6]{*}{\begin{sideways}Sector\end{sideways}}} & \multicolumn{1}{c|}{Structure} & \multicolumn{2}{c|}{Two Beam Machine} & \multicolumn{2}{c}{Klystron Machine} \\
\cmidrule{2-6}    \multicolumn{1}{c|}{} & \multicolumn{1}{c|}{\multirow{2}[4]{*}{Name}} & \multicolumn{1}{p{4.5em}|}{Cooling Power} & \multicolumn{1}{p{4em}|}{Flow Rate} & \multicolumn{1}{p{4.5em}|}{Cooling Power} & \multicolumn{1}{p{4em}}{Flow Rate} \\
\cmidrule{3-6}    \multicolumn{1}{c|}{} & \multicolumn{1}{c|}{} & \multicolumn{1}{c|}{MW} & \multicolumn{1}{c|}{m3/h} & \multicolumn{1}{c|}{MW} & \multicolumn{1}{c}{m3/h} \\
    \midrule
    \multicolumn{1}{c|}{\multirow{9}[2]{*}{\begin{sideways}1\end{sideways}}} & \multicolumn{1}{p{20.07em}|}{Drive Beam Circuit 1} & \multicolumn{1}{c|}{2.0} & \multicolumn{1}{c|}{340} & \multicolumn{1}{c|}{\multirow{9}[2]{*}{\textit{---}}} & \multicolumn{1}{c}{\multirow{9}[2]{*}{\textit{---}}} \\
    \multicolumn{1}{c|}{} & \multicolumn{1}{p{20.07em}|}{Drive Beam Circuit 2} & \multicolumn{1}{c|}{1.4} & \multicolumn{1}{c|}{250} & \multicolumn{1}{c|}{} &  \\
    \multicolumn{1}{c|}{} & \multicolumn{1}{p{20.07em}|}{Drive Beam Circuit 3} & \multicolumn{1}{c|}{1.4} & \multicolumn{1}{c|}{250} & \multicolumn{1}{c|}{} &  \\
    \multicolumn{1}{c|}{} & \multicolumn{1}{p{20.07em}|}{Drive Beam Injector S and  Frequency Multiplication Circuit 1} & \multicolumn{1}{c|}{1.5} & \multicolumn{1}{c|}{260} & \multicolumn{1}{c|}{} &  \\
    \multicolumn{1}{c|}{} & \multicolumn{1}{p{20.07em}|}{CR2 S, Frequency Multiplication Circuit 4 and Transfer Line - CR2 to J.P.} & \multicolumn{1}{c|}{0.8} & \multicolumn{1}{c|}{150} & \multicolumn{1}{c|}{} &  \\
    \multicolumn{1}{c|}{} & \multicolumn{1}{p{20.07em}|}{RF Power Distribution Circuit 1} & \multicolumn{1}{c|}{0.5} & \multicolumn{1}{c|}{80} & \multicolumn{1}{c|}{} &  \\
    \multicolumn{1}{c|}{} & \multicolumn{1}{p{20.07em}|}{RF Power Distribution Circuit 2} & \multicolumn{1}{c|}{0.5} & \multicolumn{1}{c|}{80} & \multicolumn{1}{c|}{} &  \\
    \multicolumn{1}{c|}{} & \multicolumn{1}{p{20.07em}|}{RF Power Distribution Circuit 3} & \multicolumn{1}{c|}{0.5} & \multicolumn{1}{c|}{80} & \multicolumn{1}{c|}{} &  \\
    \multicolumn{1}{c|}{} & \multicolumn{1}{p{20.07em}|}{CR1 S and Frequency Multiplication Circuits 2,3} & \multicolumn{1}{c|}{0.6} & \multicolumn{1}{c|}{100} & \multicolumn{1}{c|}{} &  \\
    \midrule
    \multicolumn{1}{c|}{\multirow{6}[2]{*}{\begin{sideways}2/3\end{sideways}}} & \multicolumn{1}{p{20.07em}|}{Injection Hall, Transfer Line - Loop Circuit 1, J.P. to S.P. and e+} & \multicolumn{1}{c|}{0.7} & \multicolumn{1}{c|}{120} & \multicolumn{1}{c|}{0.7} & \multicolumn{1}{c}{120} \\
    \multicolumn{1}{c|}{} & \multicolumn{1}{p{20.07em}|}{Transfer Line - Loop Circuit 2 and e-} & \multicolumn{1}{c|}{0.3} & \multicolumn{1}{c|}{60} & \multicolumn{1}{c|}{0.3} & \multicolumn{1}{c}{60} \\
    \multicolumn{1}{c|}{} & \multicolumn{1}{p{20.07em}|}{Detectors Hall U and Detectors Hall S} & \multicolumn{1}{c|}{0.7} & \multicolumn{1}{c|}{120} & \multicolumn{1}{c|}{0.7} & \multicolumn{1}{c}{120} \\
    \multicolumn{1}{c|}{} & \multicolumn{1}{p{20.07em}|}{Main Tunnel} & \multicolumn{1}{c|}{4.1} & \multicolumn{1}{c|}{710} & \multicolumn{1}{c|}{5.4} & \multicolumn{1}{c}{930} \\
    \multicolumn{1}{c|}{} & \multicolumn{1}{p{20.07em}|}{Buildings IP} & \multicolumn{1}{c|}{1.3} & \multicolumn{1}{c|}{240} & \multicolumn{1}{c|}{1.3} & \multicolumn{1}{c}{240} \\
    \multicolumn{1}{c|}{} & \multicolumn{1}{p{20.07em}|}{Main Tunnel Purge} &\multicolumn{1}{c|}{0.9} & \multicolumn{1}{c|}{160} & \multicolumn{1}{c|}{2.7} & \multicolumn{1}{c}{160} \\
    \midrule
    \multicolumn{1}{c|}{\multirow{8}[2]{*}{\begin{sideways}4\end{sideways}}} & \multicolumn{1}{p{20.07em}|}{Main Beam S/U Circuit 1} & \multicolumn{1}{c|}{0.8} & \multicolumn{1}{c|}{150} & \multicolumn{1}{c|}{0.8} & \multicolumn{1}{c}{150} \\
    \multicolumn{1}{c|}{} & \multicolumn{1}{p{20.07em}|}{Main Beam S/U Circuit 2} & \multicolumn{1}{c|}{0.4} & \multicolumn{1}{c|}{80} & \multicolumn{1}{c|}{0.4} & \multicolumn{1}{c}{80} \\
    \multicolumn{1}{c|}{} & \multicolumn{1}{p{20.07em}|}{Main Beam S/U Circuit 3} & \multicolumn{1}{c|}{0.4} & \multicolumn{1}{c|}{80} & \multicolumn{1}{c|}{0.4} & \multicolumn{1}{c}{80} \\
    \multicolumn{1}{c|}{} & \multicolumn{1}{p{20.07em}|}{Booster S/U and Transfer Line - Booster to J.P.} & \multicolumn{1}{c|}{0.9} & \multicolumn{1}{c|}{170} & \multicolumn{1}{c|}{0.9} & \multicolumn{1}{c}{170} \\
    \multicolumn{1}{c|}{} & \multicolumn{1}{p{20.07em}|}{Damping Rings S/U and Pre Damping Rings S/U} & \multicolumn{1}{c|}{2.4} & \multicolumn{1}{c|}{420} & \multicolumn{1}{c|}{2.4} & \multicolumn{1}{c}{420} \\
    \multicolumn{1}{c|}{} & \multicolumn{1}{p{20.07em}|}{Buildings Shaft 2} & \multicolumn{1}{c|}{0.3} & \multicolumn{1}{c|}{50} & \multicolumn{1}{c|}{0.3} & \multicolumn{1}{c}{50} \\
    \multicolumn{1}{c|}{} & \multicolumn{1}{p{20.07em}|}{Buildings Shaft 3} & \multicolumn{1}{c|}{0.3} & \multicolumn{1}{c|}{50} & \multicolumn{1}{c|}{0.3} & \multicolumn{1}{c}{50} \\
    \multicolumn{1}{c|}{} & \multicolumn{1}{p{20.07em}|}{Compton Ring S, Traget Hall S} & \multicolumn{1}{c|}{0.2} & \multicolumn{1}{c|}{30} & \multicolumn{1}{c|}{0.2} & \multicolumn{1}{c}{30} \\
    \midrule
          & Total Cooling & 23    &       & 17    &  \\
    \bottomrule
    \end{tabular}%
\end{table}%

\begin{figure}[htb!]
\centering
\includegraphics[width=0.8\textwidth]{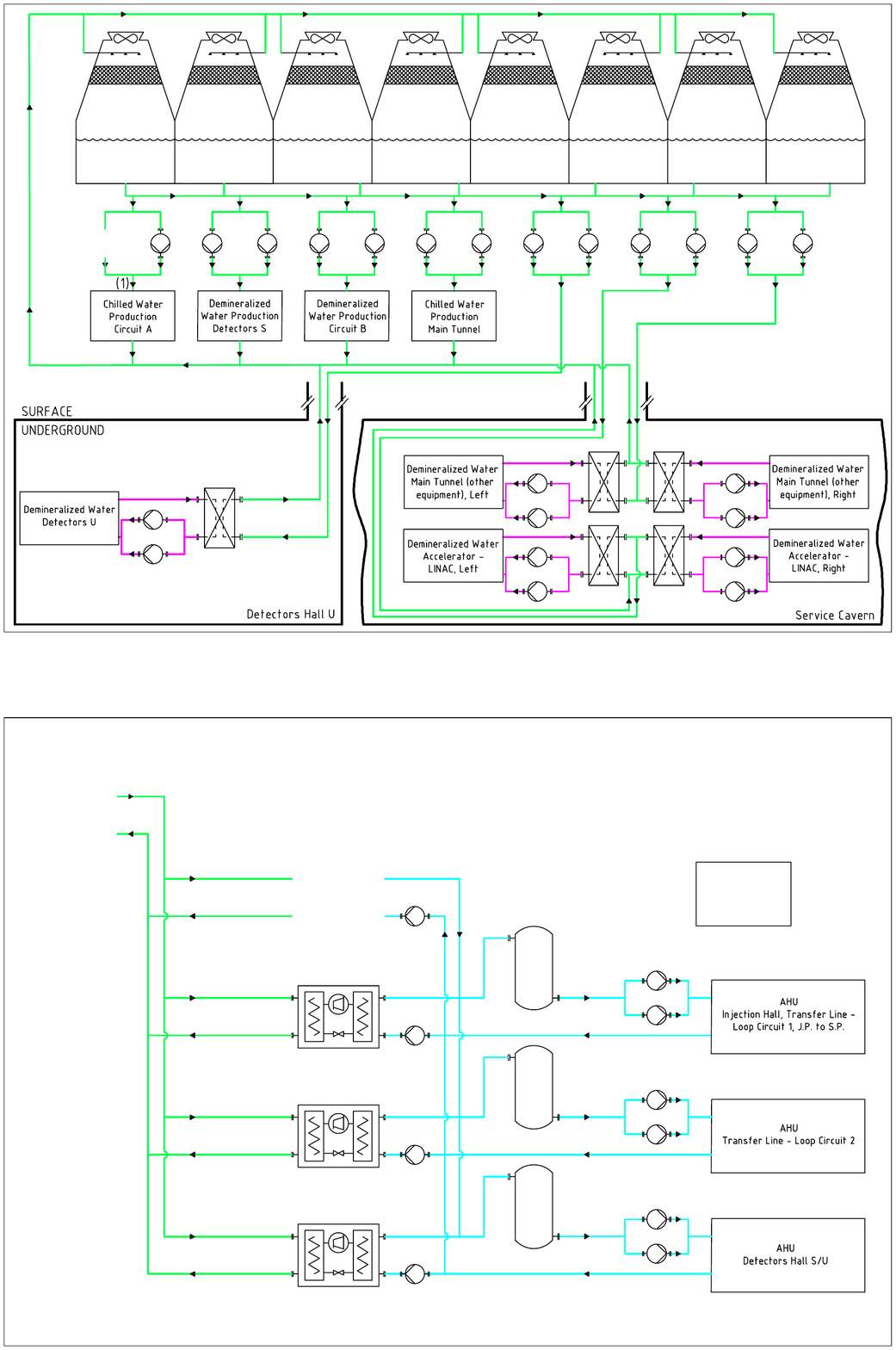}
\caption{\label{fig_CEIS_20} Simplified P\&ID concerning the cooling plants for sector 2 and 3.}
\end{figure}

See Fig.~\ref{fig_CEIS_20} for the process and instrumentation diagrams (P\&ID) concerning the cooling plants for sectors 2 and 3.

\subsubsection{Drinking Water}

Drinking water is used for sanitary purposes and as make-up water for the cooling towers. At the central part of the accelerator and the injector complex it will be supplied via an extension of the existing drinking water network at CERN, the local public network will supply all the other surface sites.

\subsubsection{Firefighting Water Network}

A firefighting water network will be implemented in all underground and surface areas: fire hydrants are foreseen at external areas, while flexible hoses are foreseen inside all premises.

These networks are to be connected to existing firefighting lines that are close to each surface point. For the injector complex they are connected to CERN’s existing line.

\subsubsection{Reject Water}

Two separate networks, one for clear and one for sewage water will be installed in all underground and surface areas; these networks will then be connected to the corresponding existing systems at each area or surface Point.

At the release point in the surface, the pH and temperature of the rejected water will be monitored and corresponding alarms installed; if the quality of the water does not comply with the required level, retention basins or other measures will be taken to improve the quality before discharge.

\subsubsection{Compressed Air}

Compressed air production stations are foreseen on the surface, to serve both surface and underground areas. Their number and location will depend on future needs; it is likely that it will be necessary to install a station for the central area and another one for each surface point. An N+1 redundancy is foreseen for the air compressors of each station.

\subsection{Heating, Ventilation and Air Conditioning}
\subsubsection{Introduction}

The heating, ventilation and air conditioning plants are designed to:

\begin{enumerate}
\item  Supply fresh air for personnel,
\item  Filter the supply and exhaust air,
\item  Sustain a given temperature and/or humidity in each area,
\item  Extract smoke and gas if necessary,
\item  Purge the tunnels if necessary.
\end{enumerate}

\subsubsection{Indoor Conditions}

Additional requirements for the temperature stability along the accelerator tunnel, operational, energetic and economical aspects have been taken into account. However, a detailed study considering all these points is still required.

The defined design ambient temperatures are:

\begin{enumerate}
\item  Main-Linac tunnel, BDS, detector hall, caverns, dumps and turnarounds : 28$^\circ$C
\item  Injectors, Booster, Damping rings, Transfer lines: 22$^\circ$C
\item  Surface Buildings: 18$^\circ$C during winter, 25$^\circ$C during summer
\end{enumerate}

The dew point is kept below 12$^\circ$C within the underground areas, no other regulation for the humidity is foreseen.

The general outdoor conditions for the Geneva region, used to specify the air handling equipment, are 32$^\circ$C dry bulb and 40\% for RH during summer and -12$^\circ$C and 90\% during winter. Free cooling and air recycling principles will be adopted whenever possible.

\subsubsection{Underground Areas}

All underground areas are ventilated by air handling units located at the surface or in the caverns; a redundancy level of N+1 is foreseen, where required, to allow constant operation of the accelerator in case of a breakdown. 

Smoke extraction shall be ensured by extraction units that are not equipped with filters to avoid clogging.

\subsubsubsection{Operational Modes}

A number of different modes are foreseen depending on the operating conditions. The fans have variable speed drives in order to adjust to the different operating conditions (table \ref{CEIS_Table2}).

\begin{table}[!h]
\centering
\caption{Operational modes}
\label{CEIS_Table2}
\begin{tabular}{l c}
\toprule
\textbf{Mode} & \textbf{Conditions} \\ \midrule
Run & No access, machines running, maximum air recycling \\
Purge & Before access where it is necessary, accelerator stopped, only fresh air \\ 
Shutdown & Open access, accelerator stopped, fresh air supply for people \\ \bottomrule
\end{tabular}
\end{table}

\subsubsubsection{Accelerating Gallery, Two-Beam Machine}

Two different ventilation systems are foreseen for the main tunnel according to the operational mode of the accelerator: 

\begin{itemize}
\item  Shutdown and Purge mode: longitudinal ventilation is ensured by air handling units at the surface;
\item  Run mode: local cooling is ensured by units sited in the caverns, fresh air intake is not required.
\end{itemize}

During Shutdown and Purge mode, air handling units will supply air at a certain surface point and a second unit will extract the air at the adjacent surface point. For higher energy stages (1.5\,TeV and 3\,TeV) air is then recycled from the extraction unit to the supply unit of the adjacent point.

Additionally, it may be necessary to heat the air. For this propose, a heating coil is included in the air handling units.

\begin{figure}[htb!]
\centering
\includegraphics[width=0.8\textwidth]{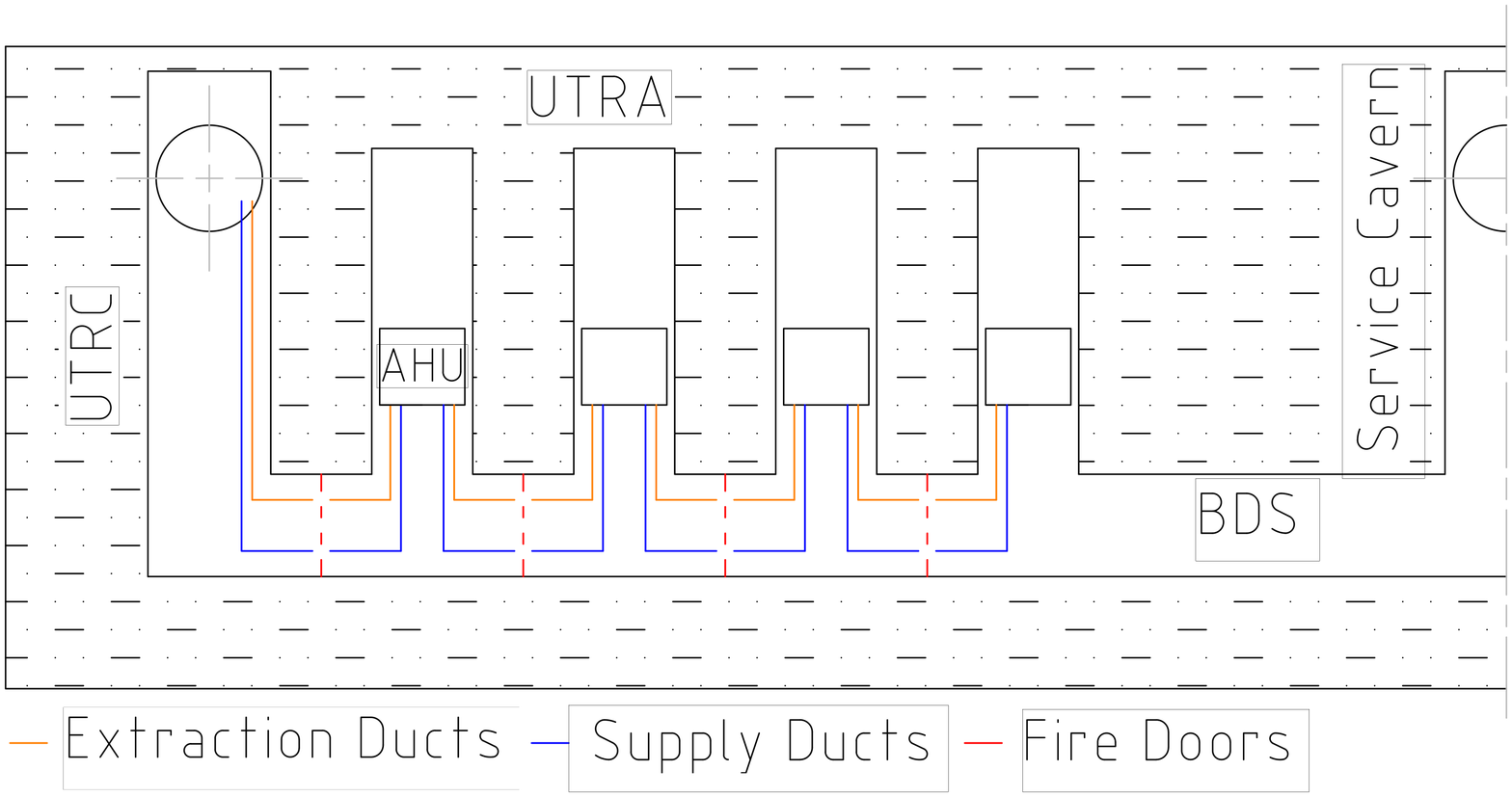}
\caption{\label{fig_CEIS_19} Drive-Beam option, HVAC schematic.}
\end{figure}

In Run Mode, air is treated by air handling units located inside the UTRAs and UTRCs; each of these units are connected to ducts (supply and return) see Fig.~\ref{fig_CEIS_19}. These units will be sized to absorb the heat loads in the tunnel. Presently, these units are not redundant due to space issues.

The supply and return grills at each of the ducts are longitudinally offset to ensure a better air distribution throughout the tunnel and to avoid bypass.

\subsubsubsection{Accelerator Gallery, Klystron Machine}

The ventilation concepts for the klystron and the Two-Beam machines are similar: different systems cover the operational modes as mentioned above.

During shutdown, air-handling units (AHUs) placed at the surface force a push and pull arrangement, similarly to the system described for the Two-Beam machine. The local air handling units are not running.

\begin{figure}[h!]
\centering
\includegraphics[width=0.8\textwidth]{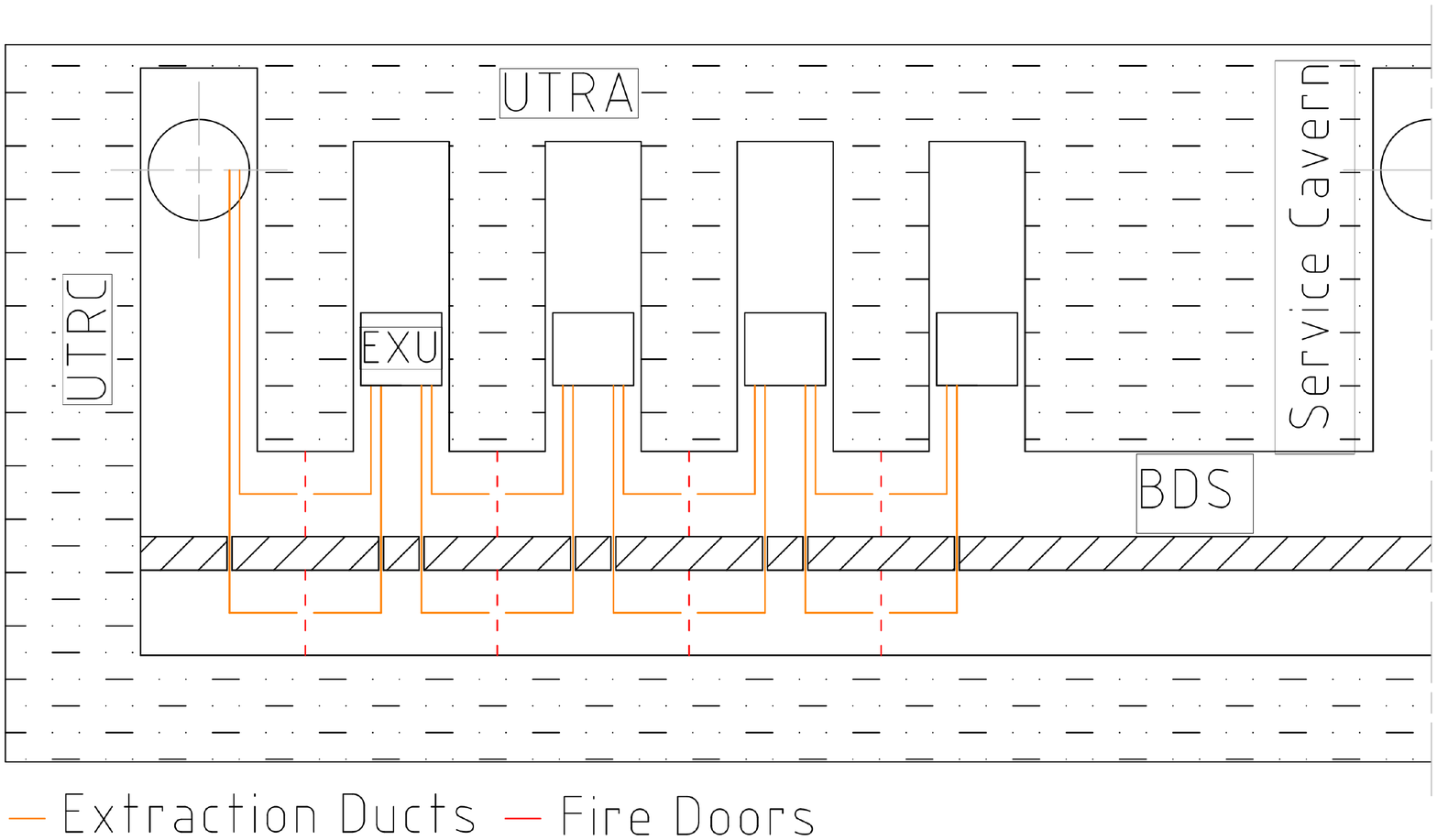}
\caption{\label{fig_CEIS_14} Klystron machine, HVAC schematic.}
\end{figure}

During run mode, air is treated by a number of AHUs cooled by chilled water and located in the service compartment below the tunnel floor. These units are connected to diffusers located at the tunnel invert level in the klystron and accelerator tunnels. Extraction units (EXU) located in the UTRAs and shaft surface buildings drive the air from the tunnel to the service compartment, where it is discharged. A schematic is presented in Fig.~\ref{fig_CEIS_14}. Presently, the units in the UTRAs are not redundant due to space issues.

Table~\ref{tab:Ventilation} shows the main parameters of the ventilation systems for both the Two-Beam and klystron machines during run mode.

\begin{table}[!ht]
\centering
\caption{Ventilation infrastructure for the accelerator gallery.}
\label{tab:Ventilation}
\begin{tabular}{|p{4cm}|p{0.5cm}|p{0.5cm}|p{1.4cm}|p{1.4cm}|p{0.5cm}|p{1.4cm}|p{1.4cm}|} \hline 
\multicolumn{2}{|p{1in}|}{Structure} & \multicolumn{3}{p{3.4cm}|}{Two Beam Machine} & \multicolumn{3}{p{3.4cm}|}{Klystron Machine} \\ \hline 
\multicolumn{2}{|p{1in}|}{} & \multicolumn{3}{p{3.4cm}|}{Air-Handling Units} & \multicolumn{3}{p{3.4cm}|}{Air-Handling Units} \\ \hline 
Name & N$^\circ$ & N$^\circ$ & Cooling Power & Flow Rate & N$^\circ$ & Cooling Power & Flow Rate \\ \cline{2-8} 
 &  & --- & kW & m${}^{3}$/h & --- & kW & m${}^{3}$/h \\ \hline 
\multirow{2}{4cm}{Accelerator Gallery - LINAC side} & 1 & 6 & 190 & 100540 & 8 & 200 & 104550 \\ \cline{3-5} 
 &  & 4 & 100 & 50270 &  &  &  \\ \hline 
Accelerator Gallery - Klystron side & 1 & --- & --- & --- & 16 & 90 & 45780 \\ \hline 
\end{tabular}
\end{table}

\subsubsubsection{Other Underground Areas}

The remaining underground areas will use different ventilation schemes depending on heat loads, geometry and operational requirements.

The BDS is conditioned by air handling units placed at the surface, the detector hall and the experimental caverns will be ventilated by units located on the surface, while the beam dump caverns and UTRAs are ventilated by units installed within each structure. Ducts will provide proper air diffusion at these locations. 

The Drive-Beam turnarounds are spatially close to the UTRAs and UTRCs. Hence, ducts will extend from an AHU, in the caverns, to the turnarounds. Due to space constraints there will be no redundancy for the AHUs serving the Drive-Beam turnarounds.

A longitudinal duct-free HVAC system is foreseen for the shallow underground tunnels: injectors, damping and combiner rings and transfer lines. The air-handling units are placed at the surface, close to the shafts, and supply the required ventilation. 

The operational parameters for the underground areas are listed in Table~\ref{CEIS_Table3}, for the particular case of run mode.

\begin{table}[!htb]
  \centering
  \tiny
\caption{Ventilation infrastructure for the underground facilities, excluding the accelerator gallery and redundant units.}
\label{CEIS_Table3}
    \begin{tabular}{c|l|c|c|c|c|c|c|c|c|c|c}
    \toprule
    \multirow{4}[8]{*}{\begin{sideways}Sector\end{sideways}} & \multicolumn{1}{c|}{\multirow{2}[4]{*}{Structure}} & \multicolumn{5}{c|}{Two Beam Machine} & \multicolumn{5}{c}{Klystron Machine} \\
\cmidrule{3-12}          &       & \multicolumn{3}{c|}{Air-Handling Units} & \multicolumn{2}{c|}{Extraction Units} & \multicolumn{3}{c|}{Air-Handling Units} & \multicolumn{2}{c}{Extraction Units} \\
\cmidrule{2-12}          & \multicolumn{1}{c|}{\multirow{2}[4]{*}{Name}} & \multicolumn{1}{p{2.5em}|}{N } & \multicolumn{1}{p{4.285em}|}{Cool. Power} & \multicolumn{1}{p{4.285em}|}{Heat. Power} & \multicolumn{1}{p{2.5em}|}{N } & \multicolumn{1}{p{4em}|}{Flow Rate} & \multicolumn{1}{p{2.5em}|}{N } & \multicolumn{1}{p{4.285em}|}{Cool. Power} & \multicolumn{1}{p{4.285em}|}{Heat. Power} & \multicolumn{1}{p{2.5em}|}{N } & \multicolumn{1}{p{4em}}{Flow Rate} \\
\cmidrule{3-12}          &       &       & kW    & kW    &       & m3/h  &       & kW    & kW    &       & m3/h \\
    \midrule
    \multirow{7}[2]{*}{\begin{sideways}1\end{sideways}} & \multicolumn{1}{p{11.07em}|}{Drive Beam Injector U} & 1     & 700   & 900   & \multirow{2}[1]{*}{3} & \multirow{2}[1]{*}{90740} & \multirow{2}[1]{*}{} & \multirow{2}[1]{*}{} & \multirow{2}[1]{*}{} & \multirow{2}[1]{*}{} & \multirow{2}[1]{*}{} \\
          &       & 2     & 160   & 90    &       &       &       &       &       &       &  \\
          & \multicolumn{1}{p{11.07em}|}{Frequ. Multip. U} & 1     & 200   & 70    & 1     & 96720 &       &       &       &       &  \\
          &       & 1     & 90    & 40    & 1     & 48360 &       &       &       &       &  \\
          &       & 1     & 70    & 30    & 1     & 36270 &       &       &       &       &  \\
          &       & 1     & 250   & 240   & 1     & 60450 &       &       &       &       &  \\
          & \multicolumn{1}{p{11.07em}|}{Transfer Line - CR2 to J.P.} & 1     & 160   & 170   & 1     & 67170 &       &       &       &       &  \\
    \midrule
    \multirow{19}[2]{*}{\begin{sideways}2/3\end{sideways}} & \multicolumn{1}{p{11.07em}|}{Transfer Line - J.P. to S.P.} & 1     & 230   & 40    & 1     & 134330 & 1     & 230   & 40    & 1     & 134330 \\
          & \multicolumn{1}{p{11.07em}|}{Transfer Line - Loop} & 2     & 120   & 40    & 2     & 68660 & 2     & 120   & 40    & 2     & 68660 \\
          & \multicolumn{1}{p{11.07em}|}{Transfer Line -  e+} & 1     & 130   & 120   & 0     & 0     & 1     & 130   & 120   & 0     & 0 \\
          & \multicolumn{1}{p{11.07em}|}{Transfer Line -  e- } & 1     & 190   & 180   & 0     & 0     & 1     & 190   & 180   & 0     & 0 \\
          & \multicolumn{1}{p{11.07em}|}{Detectors Hall U} & 1     & 350   & 460   & 1     & 64180 & 1     & 350   & 460   & 1     & 64180 \\
          & \multicolumn{1}{p{11.07em}|}{Main Beam Dumps} & 2     & 40    & 0     &       &       & 2     & 40    & 0     &       &  \\
          & \multicolumn{1}{p{11.07em}|}{Drive Beam Dumps} & 8     & 10    & 0     &       &       &       &       &       &       &  \\
          & \multicolumn{1}{p{11.07em}|}{Drive Beam Turnaround} & 8     & 20    & 0     &       &       &       &       &       &       &  \\
          & \multicolumn{1}{p{11.07em}|}{UTRA} & 8     & 110   & 0     &       &       & 8     & 110   & 0     &       &  \\
          & \multicolumn{1}{p{11.07em}|}{UTRC} & 2     & 110   & 0     & 2     & 56700 & 2     & 110   & 0     & 2     & 56700 \\
          & \multicolumn{1}{p{11.07em}|}{Caverns 1.3 and 1.4} & 2     & 110   & 0     &       &       & 2     & 110   & 0     &       &  \\
          & \multicolumn{1}{p{11.07em}|}{Survey Cavern 2.1 and 3.1} & 2     & 0     & 0     &       &       & 2     & 0     & 0     &       &  \\
          & \multicolumn{1}{p{11.07em}|}{Additional Caverns 2.2 and 3.2} & 2     & 170   & 0     &       &       & 2     & 170   & 0     &       &  \\
          & \multicolumn{1}{p{11.07em}|}{Service Cavern} & 2     & 110   & 0     & 2     & 56700 & 2     & 110   & 0     & 2     & 56700 \\
          & \multicolumn{1}{p{11.07em}|}{BDS} & 4     & 130   & 0     &       &       & 4     & 130   & 0     &       &  \\
          & \multicolumn{1}{p{11.07em}|}{Main Beam Turnaround e+/e- and Tunnel BC2 e+/e-} & 2     & 40    & 0     &       &       & 2     & 40    & 0     &       &  \\
          & \multicolumn{1}{p{11.07em}|}{BC2 Caverns} & 2     & 30    & 0     &       &       & 2     & 30    & 0     &       &  \\
          & \multicolumn{1}{p{11.07em}|}{Tunnel Purge} & 2     & 470   & 470   & 2     & 60570 & 4     & 690   & 970   & 5     & 97240 \\
          & Lift Pressurized Area & 3     & 0     & 0     &       &       & 3     & 0     & 0     &       &  \\
    \midrule
    \multirow{7}[2]{*}{\begin{sideways}4\end{sideways}} & \multicolumn{1}{p{11.07em}|}{Main Beam Injector U} & 1     & 530   & 690   & \multirow{2}[1]{*}{3} & 68790 & 1     & 530   & 690   & \multirow{2}[1]{*}{3} & 68790 \\
          &       & 2     & 120   & 40    &       & 0     & 2     & 120   & 40    &       & 0 \\
          & Booster U & 2     & 110   & 20    & 3     & 60070 & 2     & 110   & 20    & 3     & 60070 \\
          & Transfer Line - Booster to J.P. & 1     & 160   & 80    & 1     & 67170 & 1     & 160   & 80    & 1     & 67170 \\
          & Pre Damping Ring U & 1     & 840   & 70    & 1     & 159190 & 1     & 840   & 70    & 1     & 159190 \\
          & Damping Rings e+ U & 1     & 610   & 450   & 1     & 203960 & 1     & 610   & 450   & 1     & 203960 \\
          & Damping Rings e- U & 1     & 350   & 80    & 1     & 203960 & 1     & 350   & 80    & 1     & 203960 \\
    \bottomrule
    \end{tabular}%

\end{table}%

\subsubsection{Surface Buildings}

The surface buildings will be ventilated by dedicated air handling units where heat loads or the size of the building requires it, several units shall be installed in the same building with each unit ventilating a specific area. Smoke extraction shall be ensured by dedicated smoke extractors located on the roof.

The operational parameters for the ventilation of the surface buildings are listed in Table~\ref{CEIS_Table4} for the run mode.

\begin{table}[!htbp]
  \centering
\caption{Ventilation infrastructure for the surface buildings excluding redundant units.}
\label{CEIS_Table4}
    \begin{tabular}{c|p{14.645em}|c|c|c|c|c|c|}
    \toprule
    \multirow{4}[8]{*}{\begin{sideways}Sector\end{sideways}} & \multicolumn{1}{c|}{\multirow{2}[4]{*}{Structure}} & \multicolumn{3}{c|}{Two Beam Machine} & \multicolumn{3}{c|}{Klystron Machine} \\
\cmidrule{3-8}          & \multicolumn{1}{c|}{} & \multicolumn{3}{c|}{Air-Handling Units} & \multicolumn{3}{c|}{Air-Handling Units} \\
\cmidrule{2-8}          & \multicolumn{1}{c|}{\multirow{2}[4]{*}{Name}} & N     & \multicolumn{1}{p{4.5em}|}{Cooling Power} & \multicolumn{1}{p{4.5em}|}{Heating Power} & N     & \multicolumn{1}{p{4.5em}|}{Cooling Power} & \multicolumn{1}{p{4.5em}|}{Heating Power} \\
\cmidrule{3-8}          & \multicolumn{1}{c|}{} &       & kW    & kW    &       & kW    & kW \\
    \midrule
    \multirow{3}[2]{*}{\begin{sideways}1\end{sideways}} & Drive Beam Injector S & 40    & 130   & 110   &       &       &  \\
          & RF Power Distribution & 3     & 450   & 450   &       &       &  \\
          & CR1 S and CR2 S & 8     & 110   & 10    &       &       &  \\
    \midrule
    \multirow{11}[2]{*}{\begin{sideways}2/3\end{sideways}} & Detectors Hall S & 1     & 340   & 220   & 1     & 340   & 220 \\
          & IP - Electricity & 1     & 240   & 240   & 1     & 240   & 240 \\
          & IP - Reception & 1     & 40    & 40    & 1     & 40    & 40 \\
          & IP - Workshop & 1     & 70    & 70    & 1     & 70    & 70 \\
          & IP - Service Office  & 1     & 180   & 180   & 1     & 180   & 180 \\
          & IP - Control & 1     & 100   & 100   & 1     & 100   & 100 \\
          & IP - Cryo & 4     & 140   & 140   & 4     & 140   & 140 \\
          & IP - Survey & 1     & 120   & 120   & 1     & 120   & 120 \\
          & IP - Gaz & 1     & 60    & 60    & 1     & 60    & 60 \\
          & IP - Site Access Control & 1     & 10    & 10    & 1     & 10    & 10 \\
          & Injection Hall & 2     & 110   & 20    & 2     & 110   & 20 \\
    \midrule
    \multirow{5}[2]{*}{\begin{sideways}4\end{sideways}} & Main Beam Injector S & 9     & 100   & 20    & 9     & 100   & 20 \\
          & Compton Ring & 1     & 140   & 140   & 1     & 140   & 140 \\
          & Target Halls (LINACs 1 and 2) & 2     & 20    & 20    & 2     & 20    & 20 \\
          & Booster S & 6     & 100   & 10    & 6     & 100   & 10 \\
          & Damping Rings e+/e- S and Pre Damping Ring S & 6     & 110   & 20    & 3     & 110   & 20 \\
    \midrule
    \multirow{5}[2]{*}{} & Building Shaft 2 + 3 - Eletricity & 2     & 30    & 30    & 2     & 30    & 30 \\
          & Building Shaft 2 + 3 - Workshop & 2     & 70    & 70    & 2     & 70    & 70 \\
          & Building Shaft 2 + 3 - Survey & 2     & 10    & 10    & 2     & 10    & 10 \\
          & Building Shaft 2 + 3 - Access Control & 2     & 10    & 10    & 2     & 10    & 10 \\
          & Building Shaft 2 + 3 - Shaft Access & 2     & 160   & 160   & 2     & 160   & 160 \\
    \bottomrule
    \end{tabular}%
\end{table}%

\begin{figure}[h!]
\centering
\includegraphics[width=\textwidth]{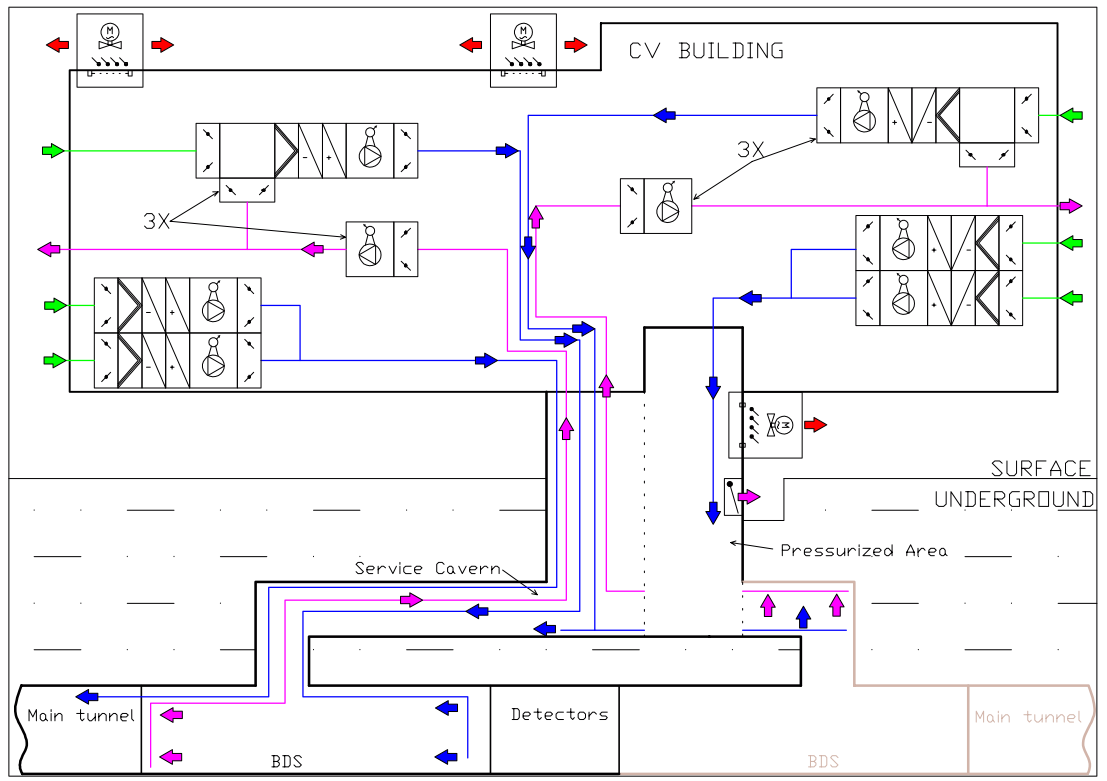}
\caption{\label{fig_CEIS_14a} Simplified P\&ID concerning the service cavern and adjacent areas.}
\end{figure}

A simplified P\&ID concerning the ventilation of the service cavern and adjacent areas is shown in Fig.~\ref{fig_CEIS_14a}{} 

\subsubsection{Safety}

Smoke extraction systems are present in all the facilities presenting a risk related to fire loads or, where it is necessary to ensure the safety of personnel. In the event of a fire, the Fire Service will be able to switch off or manually reconfigure the ventilation system. 

The concrete module for the lift and staircase in the shafts giving access to UTRCs will be kept over-pressured with respect to the surrounding underground areas thus allowing its use as a safe area in case of emergencies.

A pressure cascade is foreseen to prevent the migration of activated air from areas of high levels to areas with low levels of activation.

\section{Transport and Installation}
\label{sect:CEIS_Transport}
\subsection{Overview}

The transport and installation activities for the CLIC construction start from the unloading of components when they arrive at the CERN site.

The most notable issue for transport and handling is the installation of the underground equipment in both the Two-Beam module and klystron designs. This section will focus on the significant changes to the Two-Beam module design, briefly discussing the CDR proposals, and the challenges and proposed solutions for the klystron design. It is necessary to consider all the items that are to be transported and not only those that challenge the space constraints within the tunnel.

\subsection{Equipment to be Transported}

Below is a list of the key equipment that will need to be installed in the underground structures:

\begin{enumerate}
\item  Klystron modules
\item  Magnets
\item  Vacuum pipes
\item  Beam dumps
\item  Cooling and ventilation equipment
\item  Electrical cables and cable trays
\item  Racks
\end{enumerate}

Transport and handling solutions for all standard equipment are the same as the ones defined in detail in the CDR, it is therefore foreseen to use industrial off-the-shelf handling equipment. The most significant challenge for transport and handling for the 380\,GeV stage is the installation of the largest pieces of equipment, including, but not limited to, accelerating structures, klystron modules and magnets.

The transport and installation operations include:

\begin{enumerate}
\item  Unloading and transportation within and between surface buildings for the purposes of assembly, testing and storage.

\item  Transportation to the shaft access sites where items will, either, be installed in surface buildings or lowered to underground areas.

\item  Transportation and installation throughout the tunnels and underground structures.
\end{enumerate}

\subsection{Surface}

The capacities of the overhead travelling cranes are listed in Table\,\ref{CEIS_Table4a}.

\begin{table}[!h]
\caption{List of crane capacities}
\label{CEIS_Table4a}
\centering
\begin{tabular}{l c} \\ \toprule 
\bf{Building Type} & \bf{Crane load capacity (tonnes)} \\ \midrule
Detector Assembly & 2 x 80t + strand jacks (CMS approach) \\  
Cooling Tower and Pump Station & 3.2 \\
Cooling and Ventilation & 20 \\ 
Cryogenic Warm compressor & 20 \\ 
Cryogenic Surface Cold Box & 20 \\  
Workshop & 10 \\ 
Central Area Machine Cooling Towers & 5 \\
Shaft Access & 20 \\ 
Drive-Beam Injectors (only in Drive Beam option) & 5 x 5 (for 380\,GeV) \\ \bottomrule
\end{tabular}
\end{table}

\begin{figure}[h!]
\centering
\includegraphics[width=\textwidth]{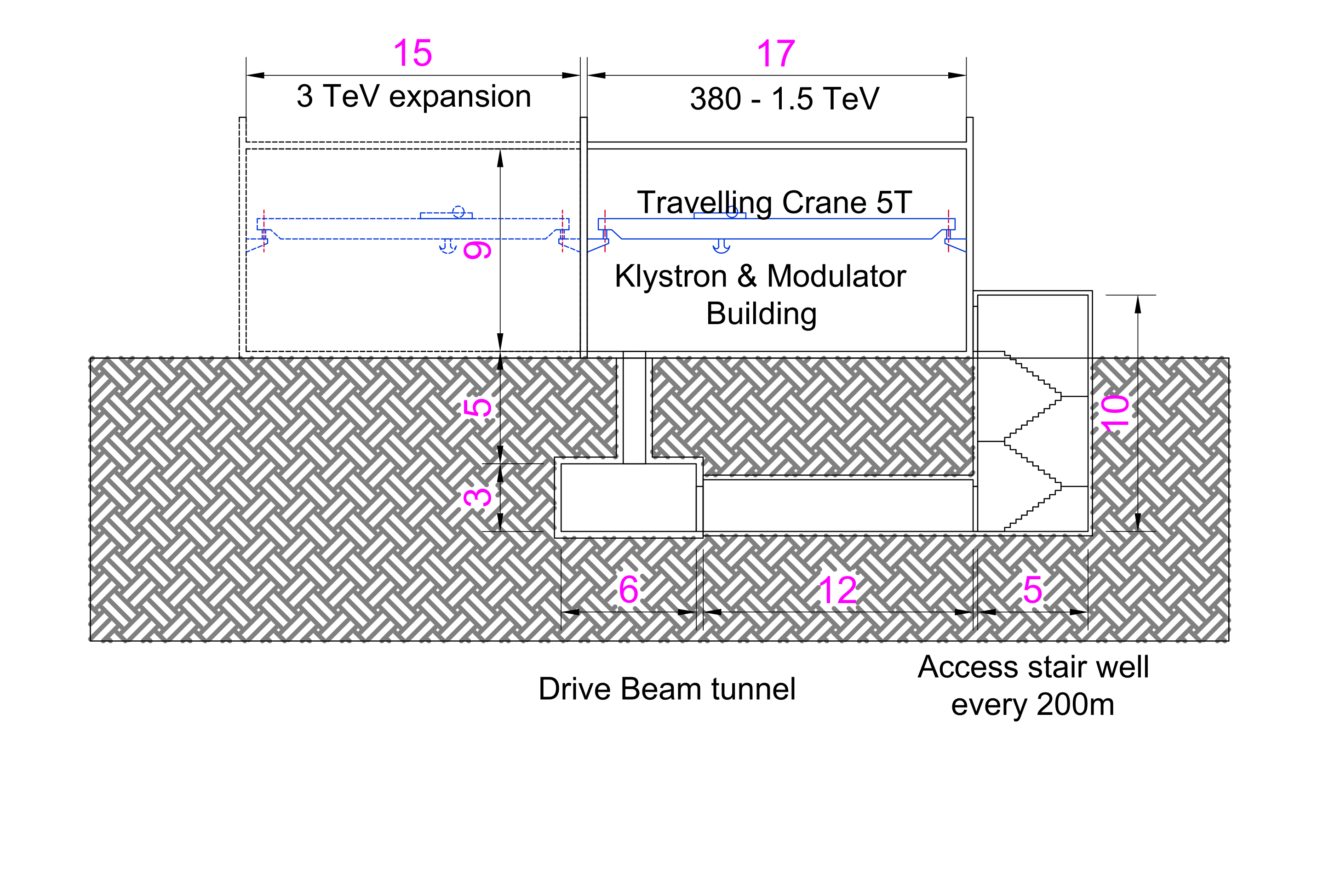}
\caption{\label{fig_CEIS_15} Cross section of the Drive-Beam Injectors building (Drive Beam option).}
\end{figure}

The most notable change for the klystron design is the removal of the Drive-Beam Injection complex. This will considerably reduce the number of cranes required and the number of items to be transported on the surface. The removal of the Drive-Beam building seen in Fig.\,\ref{fig_CEIS_15} will lead to the removal of the Drive-Beam Injection overhead cranes identified in Table\,\ref{CEIS_Table4a}.

\subsection{Shafts}

\begin{figure}[h!]
\centering
\includegraphics[width=0.7\textwidth]{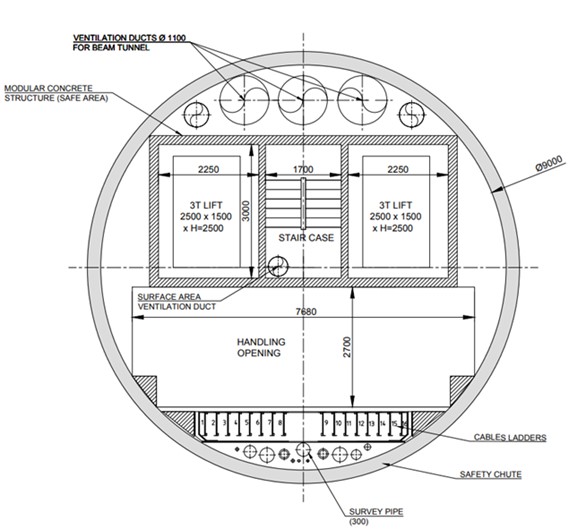}
\caption{\label{fig_CEIS_16} Cross-section of the access shaft.}
\end{figure}

In the 380\,GeV option three shafts provide underground access for the transportation of equipment. Considering the distances between the shafts (approximately 5\,km) it was decided that two 3 tonne lifts per shaft will be used as well as a handling opening (Fig.~\ref{fig_CEIS_16} This will significantly increase the rate of equipment transfer and provide a much-needed redundancy in case of failure or emergencies.

\subsection{Underground}

The main challenges for transport and handling throughout the main tunnel and ancillary structures are described.

As the Drive Beam option has undergone only minor civil engineering upgrades, the most notable change between the 3\,TeV and 380\,GeV study is the number of modules to be transported in the tunnel presented in Table\,\ref{CEIS_Table6}.

\begin{table}[!h]
\caption{Number of the modules for the 380\,GeV stage}
\label{CEIS_Table6}
\centering
\begin{tabular}{l r r} \hline 
\textbf{Type of equipment} & \textbf{Drive Beam} & \textbf{Klystron} \\ \hline \hline
Modules & 2,976 & 2,912 \\ \hline 
DB quadrupoles & 5,952 & - \\ \hline 
MB Short quadrupoles & 712 & 724 \\ \hline 
MB long quadrupoles & 436 & 452 \\ \hline \hline
\textbf{Total} & \textbf{10,076} & \textbf{4,088} \\ \hline 
\end{tabular}
\end{table}

\subsubsection{Drive Beam Option}

A conceptual design produced for the CDR \cite{Aicheler2012} of a combined trailer-crane vehicle unit was elaborated for the transport and installation of modules and magnets to achieve the highest rates of installation compatible with space, precision, interconnection and fragility constraints. In view of the narrow transport passage and the distances to travel, the trailer-crane transport vehicles will be equipped with an automated guidance system. Each vehicle will be able to simultaneously transport two modules for logistics reasons.

UTRCs are designed so that modules and other equipment can be taken from the lift and positioned close to the transport and installation vehicle loading area, the loading of equipment is achieved by the vehicles' own lifting equipment or by the 10\,tonne travelling crane installed in the UTRC. During the module installation phase it is planned to use the whole UTRC gallery floor for transport activities.

\subsubsection{Klystron Option}

It is anticipated that transport operations for the installation of equipment in the klystron option will be considered on two fronts; the klystron module installation and the accelerating structure installation as shown in Fig.~\ref{fig_CEIS_7} on the left and right sides of the tunnel cross section respectively.

\subsubsubsection{Main Beam Side}

The dimensions and weight of the modules and magnets that need to be transported requires the development of special equipment similar to that identified in the Drive Beam option. Specialized equipment will allow one to achieve the highest rate of installation compatible with the constraints identified in the CDR \cite{Aicheler2012}. 
These operations will not significantly differ from that of the Drive Beam option, it is foreseen that the transport vehicles will be equipped with an automated guidance system. A clear interconnection plane between the modules is necessary to allow each module to be lowered into position without interfering with adjacent modules, supports are installed before installation of the modules.

\subsubsubsection{Klystron Modules: Maintenance and Installation}

Two options were considered for the maintenance of the klystron modules, the first option requires a continuous monorail to be installed above the solenoids. To allow extraction of the solenoids, hoists are foreseen. It is essential that there is adequate clearance above the solenoids to extract them from the modulator tanks, therefore, the tunnel diameter and cross-section have been designed accordingly, see Fig.~\ref{fig_CEIS_7}. As the solenoids are expected to be extracted relatively frequently, in comparison to the other klystron components, it is recommended to have a fixed maintenance solution installed within the tunnel. 

On the other hand, as the monorail requires fixing before installation of the equipment can begin, the mounting accuracy of the solenoids would need to be defined beforehand. Therefore, the solenoid design must be adapted to include lifting points located directly below the hoists. The fact that the continuous rail is located throughout the entirety of the 380\,GeV tunnel means that obtaining the desired accuracy will be a significant challenge. Further integration studies will be required if this installation scenario for the Klystron option is selected. 

The preferred solution is the use of traditional transport means such as forklifts (jib instead of fork) or pallet trucks. Although this method of transportation is more flexible, it is currently limited by space constraints. For both of these maintenance approaches there is a requirement to ensure the necessary space needed for operation is available. It is essential that adequate space be provided to allow the transportation of the solenoids subsequent to extracting them with hoists, forklifts or other equipment.

More information and details of the transport systems will be provided after a detailed design of the tunnel services mounted on the ceiling has been done. Also clarification of the modulator tank and solenoid design as well as the necessity for the electrical racks is required.

\subsection{Cost Considerations}

The major difference between the two  380\,GeV options is the requirement for klystrons and modulators within the main tunnel. A smaller diameter tunnel is required for the Drive Beam option with a large amount of machine modules to be transported and installed, and, the requirement for a surface Drive Beam Injector complex with a 2\,km long building. Whereas, for the Klystron solution we are considering the Main Linac as two separate tunnels (no access from one side to another) with smaller machine equipment but equally challenging Klystron modules in the other part.

The greatest change in cost estimate for 380\,GeV in comparison with 3\,TeV comes from the reduction of the cost of operation (generally manpower), lifts and cranes (number of shafts) which is directly related to the length of the tunnel. As the installation rate is the same, based on the installation schedule, the number of special underground vehicles and expenses related to them will stay the same.

\section{Safety Systems during Operations}
\label{sect:CEIS_Safety}

The scale and complexity of a major scientific facility such as CLIC requires detailed consideration of the impacts that this could have on the health and safety of workers, visitors and those in the local community, as well as on the sustainability of the local environment. Through identification of the hazards at the earliest possible stage the risks can be mitigated by implementing appropriate control measures throughout the design process. Hazard controls can be implemented using standard best practices, or if necessary by performance based design (empirical methods or simulations, evaluated against key health, Safety and Environment performance criteria). A hazard register has been drawn up to categorise such hazards, and identify the measures needed to assure both personal and process safety; this register will be kept live throughout the lifecycle of the project, to encompass changes and add detail as it develops. It is to be noted that, while many of the hazards identified are common to both the Drive Beam and Klystron options for the 380\,GeV stage, there are some significant differences, and these have been highlighted for the relevant domains. No substantial additional hazards have been identified for the expansion to the 1.5\,TeV and 3\,TeV stages, with the same hazards as those for the 380\,GeV Drive Beam option expected to apply.

\subsection{Mechanical Hazards}

No significant mechanical hazards have been identified beyond those expected for a comparable scientific or industrial installation. Lifting and handling of large pieces of equipment must be considered, as well as the number of motorised vehicles throughout the site, both during technical stops and operation. The limited number of cryogenic components (i.e. the superconducting wiggler magnets and other superconducting components) may introduce oxygen deficiency hazards, and a further risk assessment is to be made as part of detailed design to determine the need for additional mitigation or compensatory measures. There will also be a danger of hot or cold surfaces due to thermal inertia; the magnets and klystron HV tanks will be water cooled, and should therefore maintain a safe surface temperature. All of these hazards can be mitigated through compliance with the applicable CERN Safety Rules and use of European harmonised standards for design and fabrication.

\subsection{Chemical Hazards}
\subsubsection{Klystron Option}

The most significant hazard identified is the large quantity of oil in the HV tanks for the klystron and modulator assemblies. There will be 800-1000\,litres of oil in each of the 2912 modulator assemblies. This is currently expected to be a highly refined mineral oil, capable of withstanding the high voltages. The mitigation strategies will consist of a retention basin for each HV tank, with sufficient capacity for the entire quantity of oil stored in the tank. Coupled with this, an oil level detector shall be used to indicate any drop in level. As a further measure, consideration shall be given to the potential leak path should oil be spilled outside of the retention  basins.

Another chemical hazard to be considered is that of lead used in shielding, most particularly around the klystrons for the shielding of X-rays. If embedded within the klystrons, care and appropriate personal protective equipment (PPE) will be required when handling the lead plates. However, if separate plates or blocks are used care must be taken so that the necessary procedures are followed for purchasing, shipping, storing and handling of the blocks to limit the dangers of lead poisoning or exposure to activated materials. 

\subsubsection{Drive Beam Option}

The klystrons for the Drive-Beam option will be located within the surface building, with each modulator assembly being immersed in 1500\,litres of oil, with the potential to use a lower flashpoint oil. However, the mitigation strategies will be the same as for the Klystron option. Similarly, the hazards and mitigation strategies associated with lead for the Klystron option also apply.

\subsection{Fire Safety}
\subsubsection{Drive Beam Option}

The CDR \cite{Aicheler2012} set out the fire strategy for the Drive Beam option to include the following measures:

\begin{itemize}
\item  Limiting the probability of onset of fire, with fire compartments every 439\,m in the main accelerator tunnels,

\item  Early fire detection and intervention,

\item  Safe evacuation of personnel,

\item  Limiting the propagation of fire and smoke along the facility, including hot smoke extraction in the main accelerator tunnels.
\end{itemize}

These remain valid for the current Drive Beam option design. 

\subsubsection{Klystron Option}

The strategy for the Klystron option has to be adjusted due to the requirement for large quantities of flammable electrical insulating oil within the tunnels, as this represents an additional and very significant fire load. The longitudinal fire compartment spacing will be the same as for the Drive Beam option, however the cross section will additionally be divided into three separate compartments, corresponding to the Klystron side, the Accelerating Structure side and the Service Compartment below. Sprinklers will be required in the Klystron side of the tunnels, covering the Klystrons, modulators and the oil retention basins. Fire detection will also be required in each the separate compartments of the tunnel, and hot smoke extraction will at minimum be required in the Klystron and Accelerating Structure compartments.

\subsection{Environmental Hazards}

The environmental considerations are considerable, across a wide variety of environmental domains. Important considerations identified include:

\begin{itemize}
\item  The volume of soil that will need to be excavated (and relocated) for the tunnels and shafts; the presence of natural hydrocarbons in the soil will also need to be taken into account,

\item  The use of surface cooling towers will require a study into the likely environmental impact and strategies (such as the disposal of rejected water),

\item  The effect of the construction on the water table, natural aquifers, and natural protected areas,

\item  The effect of the extremely large electrical infrastructure that will need to be installed, including many kilometres of electrical transmission cabling,

\item  Potential leakage of oil, and the dangers of this entering the surrounding environment or water sources,

\item  Identification and monitoring of ionising radiation effects on the surrounding environment,

\item  Greenhouse gas emissions from electrical, HVAC or detection equipment,

\item  Noise emissions of mobile or fixed sources,

\item  Electricity consumption.
\end{itemize}

A strategy for addressing these issues will be split across the project lifecycle phases. Initially, as part of early detailed design, the project will review and determine any potential additional environmental and socio-economic implications of those surface sites on the affected French and Swiss territories, and thus define any required additional mitigation measures that will be needed.  All measures shall be planned and taken to reduce the environmental impact.

Once the general layout and site master plans are determined, the project shall make an independent Environmental Impact Assessment (EIA) to provide assurance that all necessary measures to limit the impact of the project on the environment to an appropriate level have been taken. This assessment shall be based on the requirements of the applicable Host State legislation, i.e. \textit{Ordonnance relative \`{a} l'\'{e}tude de l'impact sur l'environnement (OEIE) and the Code de l'environnement, Livre I, Titre II, Chapitre II de la Partie L\'{e}gislative et R\'{e}glementaire.}

\subsection{Electrical Hazards}
\subsubsection{Klystron and Drive Beam Options}

The electrical hazards present for this project are considered standard for such an installation, but are nevertheless significant in their scale. Inside the tunnels there will be a number of high voltage systems, including the klystrons, modulators and magnets. The HV oil tank will be in place to insulate the klystrons and modulators, with pulse quality monitoring to identify any breakdown in the oil. NF C 18-510 compliant covers, and restriction of access to those with the appropriate electrical training, shall also be in place. An effective strategy for the earthing system will also be required for areas of high voltage and power.

The supply infrastructure potentially represents a large number of hazards, from those frequently found at CERN, such as uninterruptable power supplies, transformers and power converters, through to some more novel challenges, such as the dangers of having high voltage and current supply cables close to workers during installation and maintenance, and the need to have protected control systems across adjacent fire compartments. The CERN Electrical Safety Rules shall be followed throughout the design process; where exceptions are required, this shall be subject to an appropriate level of risk assessment to evaluate the residual risk, and determine the mitigation strategies required. 

\subsection{Biological Hazards}

The primary biological safety consideration identified for the project is the danger of Legionella in the cooling water circuits, and most particularly in the cooling towers. This hazard can be mitigated by following the regulations and standards for limiting the risk of Legionella bacteria and its dispersion in the atmosphere, and practices already in place for CERN's existing installations. 

\subsection{Non-ionising Radiation Hazards}

One significant hazard in this domain will be the Q-Switched, polarised laser, operating at a mean power per pulse of 1.5\,kW, which will be used in the e$^-$ production target. This will likely be a Class\,4 laser, and the safety system will therefore need to meet the appropriate IEC required control measures; its operation in the 780-880\,nm wavelength range indicates a particular danger to eyesight, which shall be compensated for accordingly and the appropriate interlock systems installed. 

RF will also be a significant hazard throughout the accelerator. RF components purchased from industry are required to be CE marked and to comply with the EU emission norms for industrial environments. RF equipment built or installed at CERN shall be ``leak-tight'', and tested against EU industrial norms for RF emissions in situ. In case of an RF-related accident (for example the breaking of a waveguide), the mismatch in reflected power shall be detected, and trigger a cut in the electrical supply. 

One comparatively novel hazard for CLIC is the potential presence of high flux density permanent, or hybrid magnets in the main accelerator. The extent of stray magnetic fields is not fully determined at this stage, but having permanent strong magnetic fields would mean that access to affected areas must be denied to those with pacemakers, and similar measures to those implemented in equivalent existing areas at CERN must apply, such as the requirement for non-magnetic tools.

\subsection{Workplace Hazards}

Many standard industrial hazards will need to be considered for those working in the CLIC tunnels and surface complexes.These include noise, lighting, air quality, and working in confined spaces, which can be satisfied by following existing CERN Safety practices and the Host States' regulations for workplaces. 

The machine heat loads and the current design for the cooling and ventilation in the Main-Beam tunnel predict an ambient temperature of 30-40$^\circ$C during beam operation, when there is no access. The ambient temperature will therefore be important for access and work during both the short technical stops during the runs and the longer annual technical stops. With acceptable cooling periods during short shutdowns still to be decided, an initial guideline of 28$^\circ$C is recommended for manual work within the tunnels (across the full range of relative humidity values) during extended access periods. At higher temperatures, where manual work is required, care must be taken to monitor the condition of workers, ensuring reduced working time and increased rest and hydration in proportion to any increased heat stress.

\subsection{Structural Safety}

All infrastructure shall be designed in accordance with the applicable Eurocodes to withstand the expected loads during construction and operation, but shall also consider accidental actions, such as seismic activity, fire, release of cryogens and the effect of radiation on the concrete matrix of the tunnels.

\subsection{Access Safety and Control Systems}

The CLIC access system is based on several protection layers comprising site access, building access and personnel protection with dedicated access points to regulate the personnel access to the supervised and controlled machine areas. Each access point has one or more personnel access booths and an optional material booth. This ensures identification and biometric authentication thus regulating access at the highest safety level required according to the operation modes which are managed remotely by the CCC. In the machine tunnel, the safety is ensured by access or beam “Important Element for Safety (EIS)”. The access system controls a number of independent beam zones divided into access sectors equipped with various safety elements. These access sectors are important for patrolling the machine and to minimise radiation exposure. Three access modes are foreseen; general, restricted with a safety token, and equipment test, these are automatically managed by the PPS or controlled by a human operator, locally or remotely from the CCC.

Each beam zone has its own independent access conditions, the absence of beam in each beam zone is guaranteed by at least two beam safety elements, with at least one passive element and one active element. These safety measures are activated and interlocked by the access system, and, the access status can make a zone unsafe for operation with beam, or forbid access to the Machine if the status of a safety element is unsafe. 

For this study, we have considered three protection layers: site, building and PPS for each of the five Main-Beam areas of the CLIC infrastructure.

\begin{enumerate}
\item Injector Linacs;
\item Sub-surface accelerator complex;
\item Transfer tunnels;
\item Main-Beam tunnel;
\item Experimental area.
\end{enumerate}

The access conditions for CLIC require further studies once the accessibility requirements of each machine segment have been defined. Each of the four shafts are equipped with access points, at the surface and underground, to regulate access from the shaft caverns to the main tunnel and experimental area. There are also several other access points required to regulate access to each sector of the Machine.

\section{Radiation Protection}
\label{sect:CEIS_RP}
For the mitigation of risks associated with ionising radiation, the existing CERN radiation protection rules and procedures are used. Risks resulting from ionising radiation must be analysed from a very early design phase onwards and mitigation approaches must be developed. Design constraints will ensure that the doses received by personnel working on the sites, as well as the public, will remain below regulatory limits under all operation conditions. A reliable and continuously operating radiation monitoring system will therefore be an important part of the system implementing risk control measures. 

Radiation protection is concerned with two aspects: protection of personnel operating and maintaining the installations and the potential radiological environmental impact. The potential radiological hazards to personnel working on the site are classified by the following sources, particle beam operation, activated solids, liquids and gases and parasitic X-ray emitters.

\subsection{Particle Beam Operation}

Radiation hazards will arise from the operation of the electron and positron beams. There is no difference expected in this respect between the Drive-Beam option and the Klystron option. The direct exposure to stray radiation is prevented by installing the high energy accelerator parts in a deep underground tunnel, which is inaccessible during beam operation. The central injector complex, where beams operate at lower energies, is closer to the surface and covered by about 5\,m of shielding. Access shafts and ducts to the tunnels are located and designed to effectively reduce the transmission of ionising radiation to acceptable levels in all accessible areas.

The radiation protection aspects remain unchanged with respect to the points addressed in the CDR \cite{Aicheler2012}. For both the Drive and Main Beams, an upper beam loss limit of 10${}^{-3}$ was determined. Monte Carlo simulations demonstrated that even with this pessimistic beam loss scenario, residual dose rates in the range of 10-100$\mu$\,Sv/h in the accelerator tunnel will allow hands-on maintenance after reasonable cool-down times.

\subsection{Activated Solids, Liquids and Gases}

The residual dose rate levels from activation in the Main and Drive-Beam tunnels are of lesser concern \cite{Aicheler2012}. The average dose rate levels will be relatively low in the $\mu$Sv/h range or below, and compatible with standard intervention procedures. Special attention needs to be paid to particular beam line elements and areas where beam losses will be concentrated.

The activation of the positron production target and its surroundings must be studied in detail. The design will need to take into account the potentially high activation of the target and the need for specific shielding and target handling.

Further radiation protection studies are required on: 

\begin{itemize}
\item  Activation in the various bending structures,  

\item  The Main-Beam collimation systems,

\item  The Drive-Beam PET structures. 
\end{itemize}

Activation of beam line elements and the resulting needs for material optimisation and specific handling procedures may be a technical and engineering challenge, but, are considered to be feasible considering the existing experience in the design of targets and collimators at CERN.

The beam dumps for the Drive and Main Beams require a careful technical design to contain and handle the activated water. More detailed studies on the activation of the post-collision absorbers, intermediate and final dumps and their respective shielding are required. Initial studies of the Main-Beam dump indicate a considerable activity inventory of several tens of TBq of Tritium after several years of operation \cite{Mereghetti2011}, which nevertheless can be technically handled.

The activation of water in the closed water cooling circuits needs to be evaluated in more detail. However, given the low residual dose rates from the limited beam losses in the Main-Beam tunnel, water activation in these circuits is not expected to be a major concern.

Air activation needs to be studied as it may be the major contributor to the radiological environmental impact. A good design approach will foresee recycling modes for the ventilation systems during beam operation which will effectively reduce the release of short-lived isotopes. Before access is granted, the tunnel air will need to be purged after a short waiting time to avoid undue exposure of personnel during the interventions.

\subsection{Parasitic X-ray emitters}

Klystrons are strong parasitic X-ray emitters. The design of these devices must include shielding against the generated X-rays so that they can be installed in areas accessible during their operation. Standard prescriptive regulations are an effective control measure for this specific hazard. 

\subsubsection{Klystron Option}

This option introduces a klystron gallery running parallel and close to the Main-Beam tunnel. The gallery will be closed during beam operation, but will remain accessible during certain operational conditions without beam. While the distances between klystrons and RF cavities are kept short, minimum shielding is required for radiation protection when the RF cavities in the Main-Beam tunnel are powered. The maximum dose rate in the gallery, will be compliant with a Supervised Radiation Area, therefore the radiation levels must not exceed 3$\mu$\,Sv/h.

The driving parameter for the shielding dimension is the expected dark current generated by the electric field in the accelerating structures and the level and spectrum of the produced X-rays. This source term cannot be easily determined and requires a study of the operational conditions taking into account the experiences gained during the development and testing of the prototype RF structures.

The input parameters for a first radiation transport study were based on data from CLIC-type RF structures operated at the X-band facilities at CERN, where average dark currents of about 10\,nA per structure were measured. The dark current was scaled to a full module, consisting of 8\,structures, and a scaled energy spectrum was used from tracking studies for a single structure.

Dephased RF conditioning of modules will be implemented to prevent the capture and subsequent acceleration of the dark current in adjacent modules. With this restriction, a 1.5\,m thick shielding wall is compatible with the targeted maximum dose rate level. Specific shielding arrangements still need to be made at the locations where ventilation ducts have to cross the shielding wall. Further safety margins could be obtained by optimising the concrete composition of the shielding wall or the use of local shielding at the loss points.

Further studies and optimisation during the technical design phase must be foreseen, using refined input parameters.

\printbibliography[heading=subbibintoc]
\endrefsection
\addtocontents{toc}{\vspace{\normalbaselineskip}}
\refsection 
\chapter{The CLIC Accelerator Implementation}
\label{Chapter:IMP}

\begin{figure}[h!]
\centering
\includegraphics[scale=0.4]{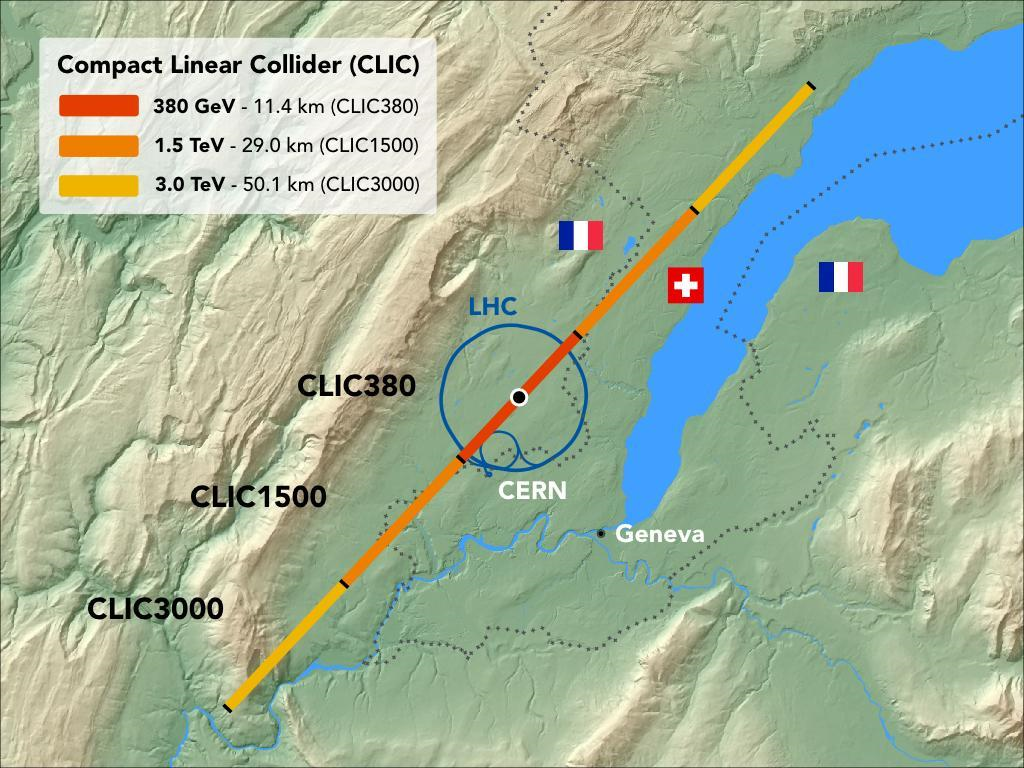}
\caption{\label{fig_IMP_1} The CLIC Main Linac footprints near CERN, showing the three implementation stages.(image credit: CLIC)}
\end{figure}

This chapter describes the implementation of the CLIC accelerator. Section~\ref{sect:IMP_Stages} recalls the staging scenario of CLIC. This is followed by a description of the schedule for the full CLIC programme in Section~\ref{sect:IMP_Sched}. Sections~\ref{sect:IMP_Cost} and \ref{sect:IMP_Power} cover the costs and power \& energy consumption of the accelerator. Section~\ref{sect:NextPhase} summarizes the main outstanding challenges for the project that need to be addressed before construction can start. 
\section{The CLIC Stages}
\label{sect:IMP_Stages}

The CLIC accelerator is foreseen to be built in three stages with centre-of-mass energies of 380\,GeV, 1.5\,TeV and 3\,TeV. The implementation of CLIC at CERN in these three stages is shown in Fig.~\ref{fig_IMP_1}. The key parameters for the three energy stages of CLIC are given in Table~\ref{t:scdup2} in Chapter~\ref{Chapter:HE_Design}. The concept of the staging implementation is illustrated in Figs.~\ref{f:scdup1} and~\ref{f:scdkl1} in the same chapter.

\begin{figure}[h!]
\centering
\includegraphics[scale=0.9]{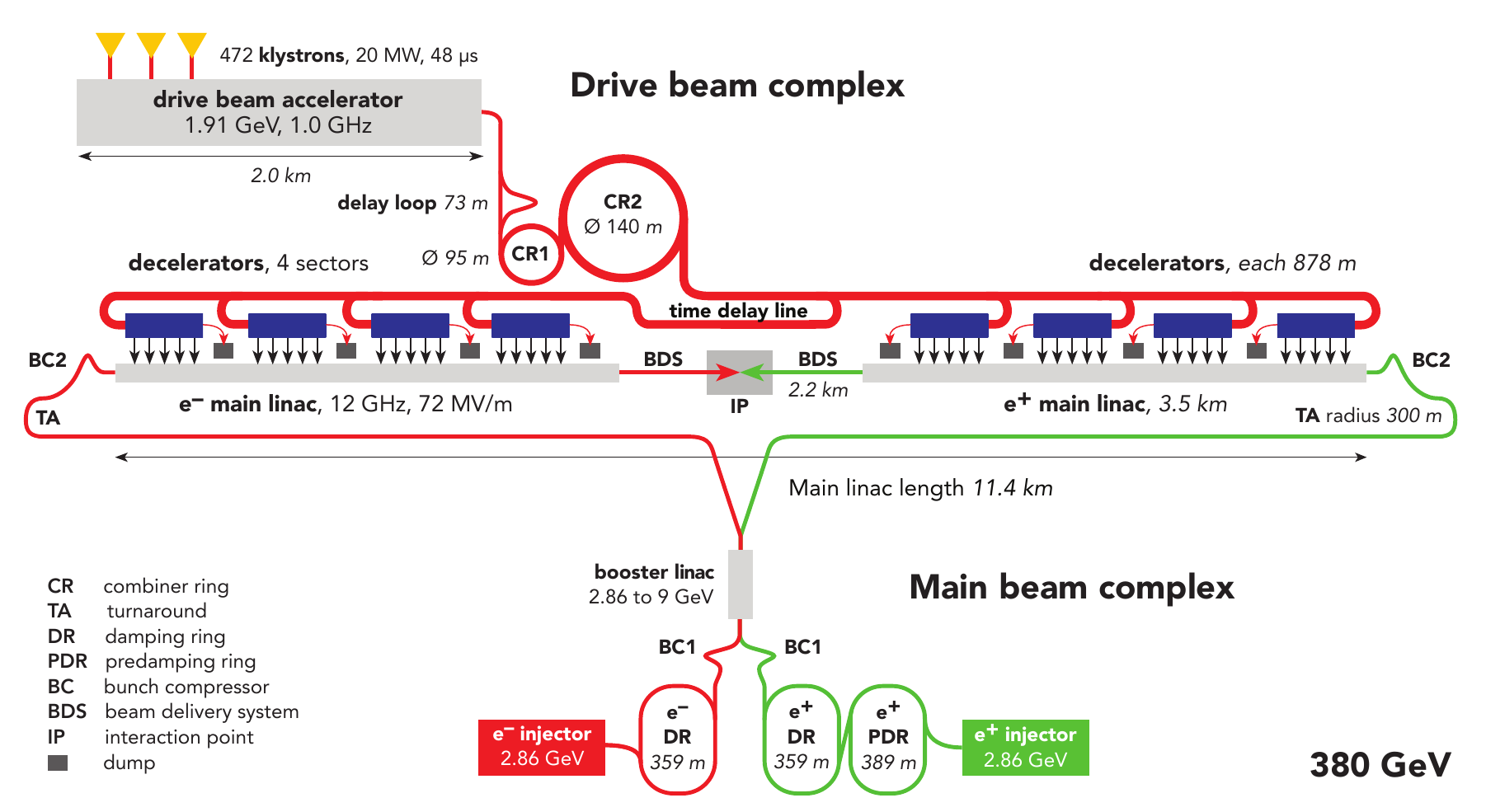}
\caption{\label{fig_IMP_3} Overview of the CLIC layout at $\sqrt{s}$~=~380\,GeV. Only one Drive-Beam complex is needed for this and for the 1.5\,GeV stage. In case of a Klystron-based implementation the Drive-Beam complex is replaced by X-band RF powering units installed along the Main Linac. (image credit: CLIC)}
\end{figure}

\begin{figure}[h!]
\centering
\includegraphics[scale=0.9]{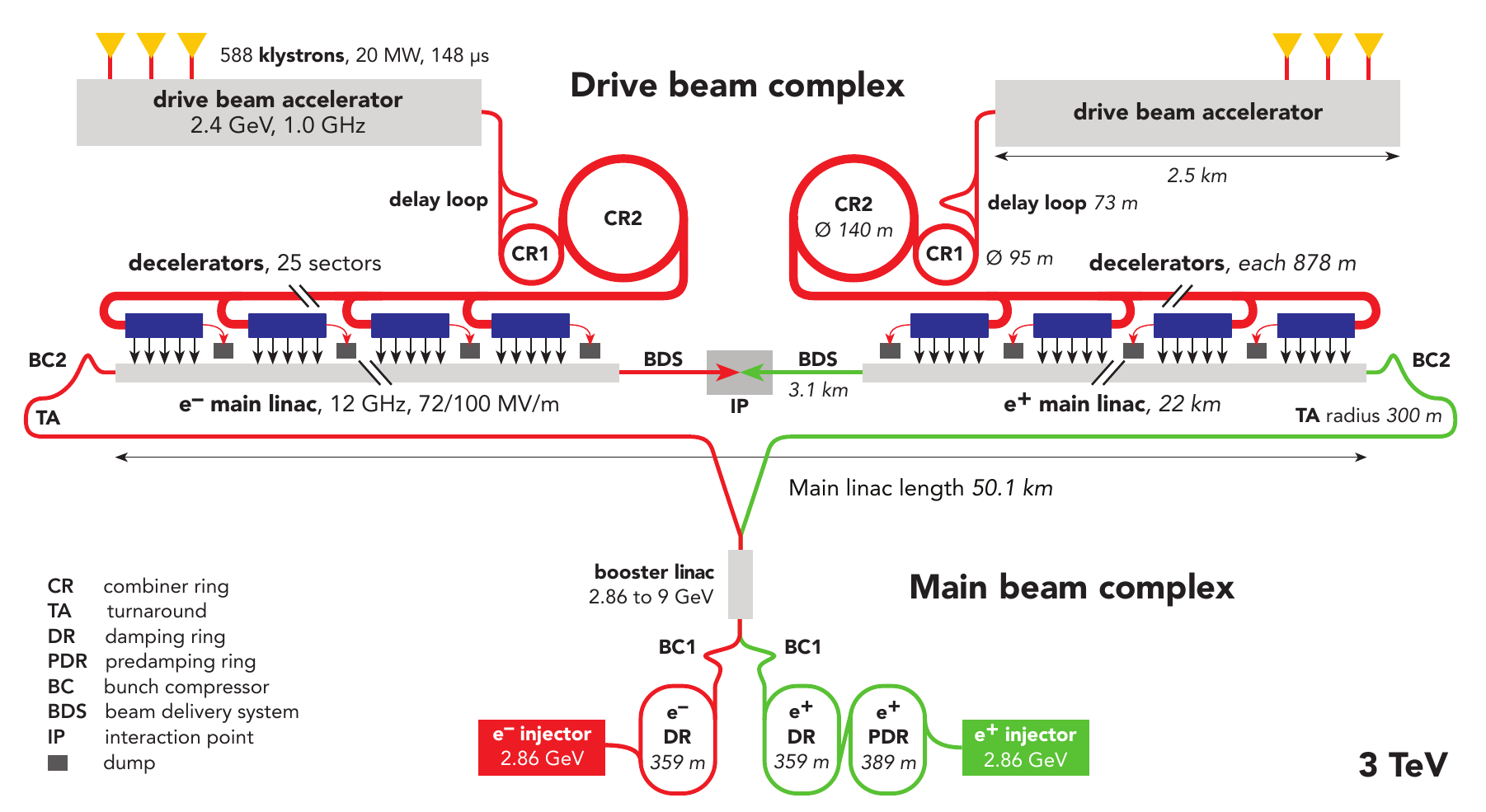}
\caption{\label{fig_IMP_4} Overview of the CLIC layout at $\sqrt{s}$ = 3 TeV. (image credit: CLIC)}
\end{figure}

The schematic layouts of the accelerator complex at 380\,GeV and 3\,TeV are shown in Figs.~\ref{fig_IMP_3} and \ref{fig_IMP_4}. The first and second stages use a single Drive-Beam complex to feed the two linacs, while in the third stage two Drive-Beam complexes are needed. 

\begin{figure}[h!]
\centering
\includegraphics[scale=0.3]{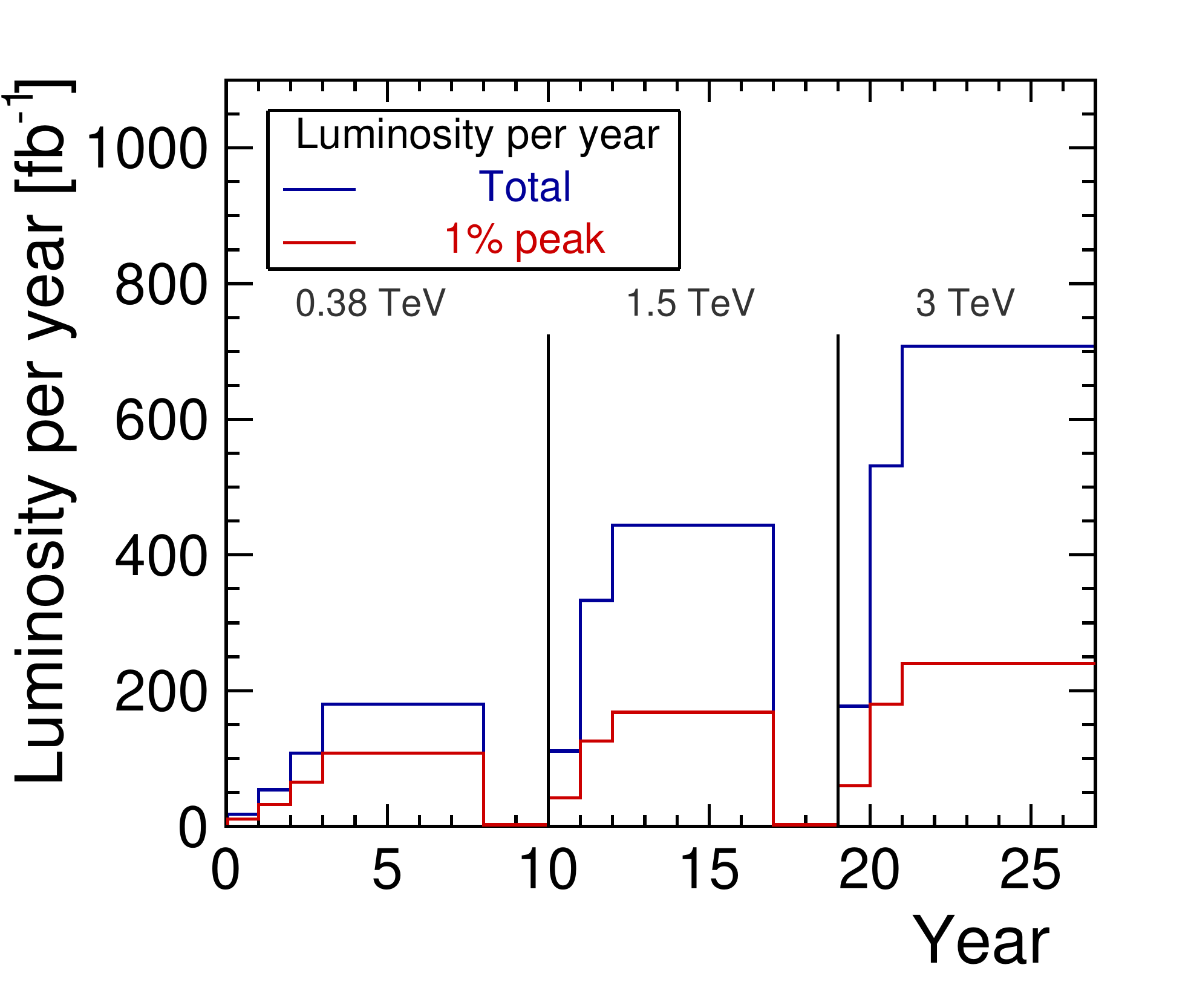}
\includegraphics[scale=0.3]{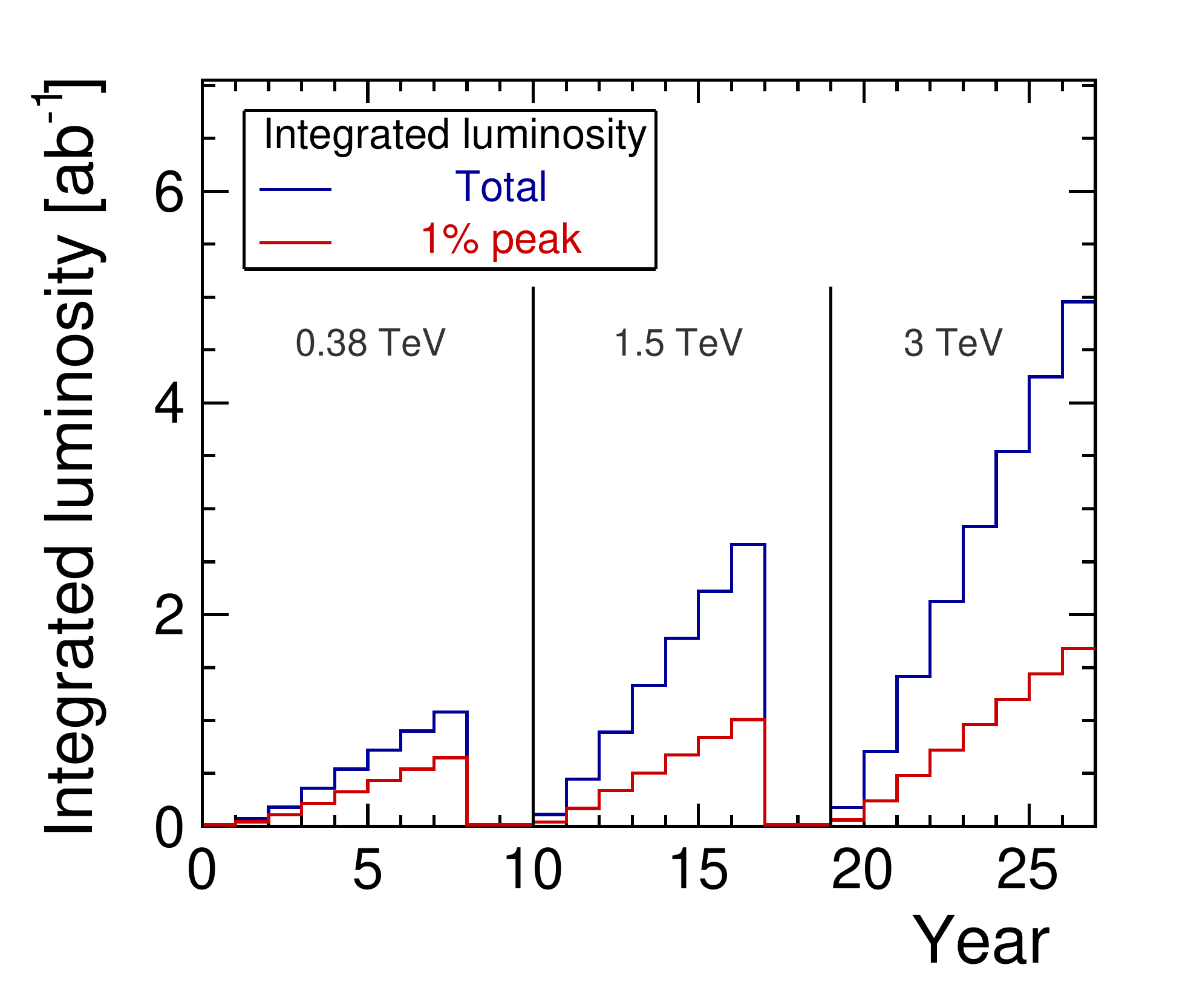}
\caption{\label{fig_IMP_5} 
(a) Luminosity and (b) integrated luminosity per year in the 
proposed staging scenario, for the total luminosity in blue and the luminosity at centre-of-mass energies above 99\% of the nominal centre-of-mass energy in red. Years are counted from the start of physics running.}
\end{figure}

Estimates of the integrated luminosities are based on an annual operational scenario~\cite{Bordry2018}. After completion of CLIC commissioning, it is estimated that 185 days per year will be available for operation, with an average accelerator availability of 75\%, thus yielding physics data-taking for 1.2~$\times$~10$^7$ seconds annually. The remaining time is shared between maintenance periods, technical stops and extended shut-downs as discussed in Section~\ref{sect:IMP_Power}. The yearly luminosity and the cumulative integrated luminosity for the three stages of the CLIC programme are shown in Fig.~\ref{fig_IMP_5}. A luminosity ramp-up of three years (10\%, 30\%, 60\%) is assumed for the first stage and two years (25\%, 75\%) for subsequent stages. Prior to data-taking at the first stage, commissioning of the individual systems and one full year of commissioning with beam are foreseen. 
These are part of the construction schedule.

\section{Schedule}
\label{sect:IMP_Sched}

The construction schedules presented here are based on the same methodologies as those used for the CLIC CDR~\cite{Lebrun2012}. Following input from equipment experts and the CERN civil engineering and infrastructure groups, small adjustments were made to the construction and installation rates used for the schedule estimates. The installation is followed by hardware commissioning, final alignment and commissioning with beam.

\subsection{Construction and Installation Rates}
The schedules shown below are based on the following set of time and rate estimates for the four main phases of the construction project.

\subsubsection{Civil Engineering Phase}

\begin{itemize}
\item  Site installation: 15\,weeks
\item  Shaft excavation and concrete: 1.2\,m/day
\item  Service and detector caverns: 300\,m$^3$/day 
\item  Excavation by tunnel-boring machine (TBM): 15-16\,m/day for 5.6m and 10m diameter tunnels.
\end{itemize}
                       
\subsubsection{General Infrastructure Phase}
\begin{itemize}
\item  Survey works -- 9 weeks/km/front
\item  General services -- 2 weeks/km/front
\item  Electrical installation and cabling: -- 7 weeks/km/front
\item  Cooling \& ventilation: ventilation ducts and pipes -- 7 weeks/km/front
\end{itemize}

\subsubsection{Machine Installation Phase}
Once most of the cabling work is completed, installation of the ground supports will start. The Two-Beam modules will then be transported, pre-aligned, and interconnected, working on two fronts of each sector. 

The modules will be transported to their final locations in the tunnel at a maximal rate of 200/550 (stage 1/stages 2-3) modules per month. This rate is compatible with the production rates for structures and modules. The interconnections will follow, starting at a maximal rate of 150~modules per month for the 380\,GeV stage and reaching a maximal rate of 400~modules interconnected per month for the higher energy stages. The Beam Delivery System will be installed partly in parallel but a dedicated period is foreseen after the Main Linac module installation. 

\subsubsection{Commissioning Phase}

This phase involves commissioning of the individual accelerator systems without beam; it will take about eight months followed by final alignment of the machine. Furthermore, one year of commissioning with beam is foreseen before the CLIC running scenarios shown in Fig.~\ref{fig_IMP_5} starts.

\subsection{380\,GeV Drive-Beam Schedule}

\begin{figure}[h!]
\centering
\includegraphics[width=0.8\textwidth]{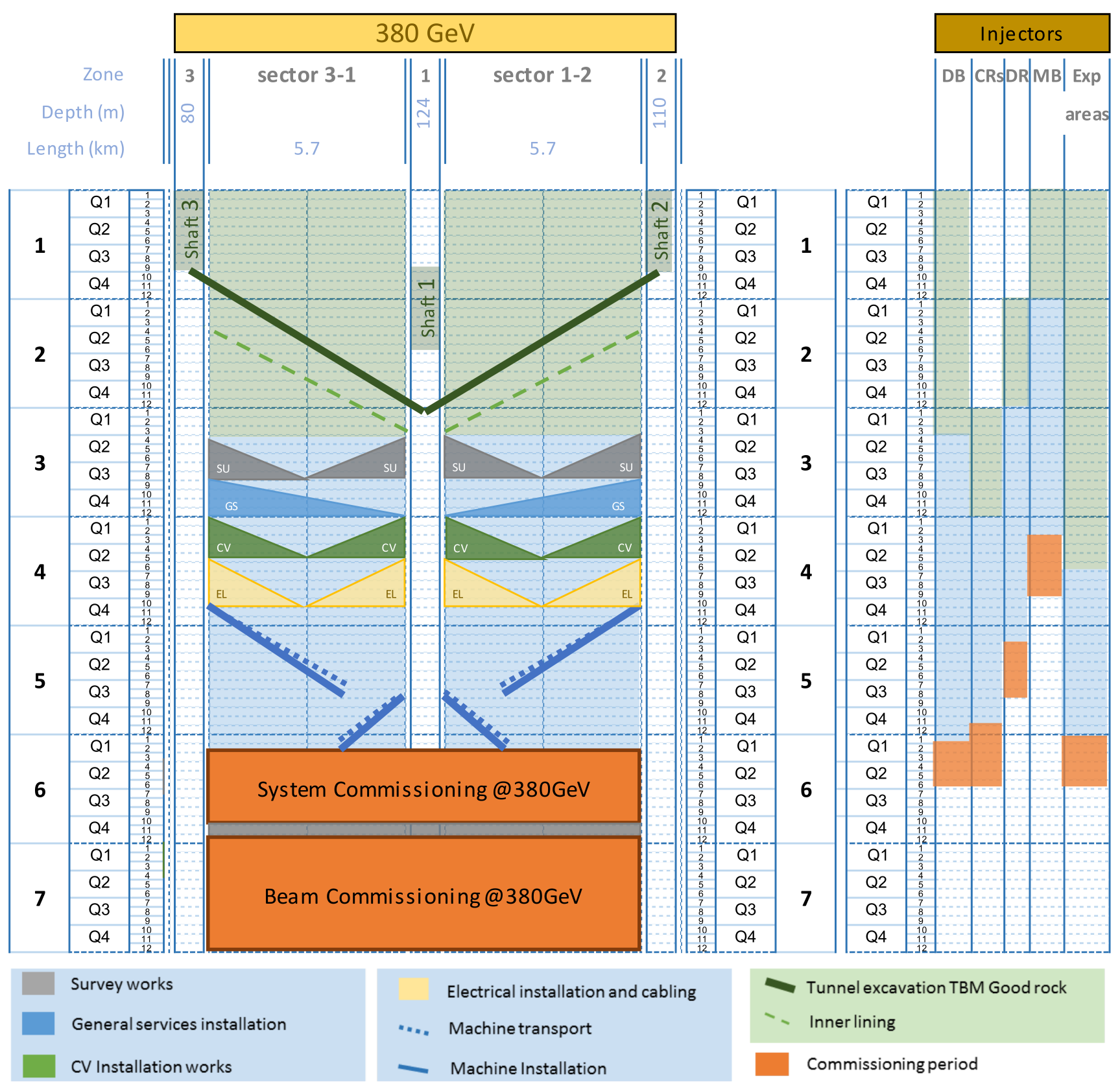}
\caption{\label{fig_IMP_6} Construction and commissioning schedule for the 380\,GeV drive-beam based CLIC facility. 
The vertical axis represents time in years. The abbreviations are introduced in Figure~\ref{f:scdup1}. (image credit: CLIC)}
\end{figure}

The schedule for the first stage of CLIC at 380\,GeV, based on the Drive-Beam design, is shown in Fig.~\ref{fig_IMP_6}. It consists of the following time-periods:

\begin{itemize}
\item  Slightly more than five years for the excavation and tunnel lining, the installation of the tunnel infrastructures, and the accelerator equipment transport and installation.
\item  Eight months for the system commissioning, followed by two months for final alignment.
\item  One year for the accelerator commissioning with beam.
\end{itemize}

In parallel, time and resources are allocated for the construction of the Drive-Beam surface building, the Combiner Rings, the Damping Rings, the Main-Beam building and experimental areas, and the corresponding system installation and commissioning, as shown in Fig.~\ref{fig_IMP_6}. 

\subsection{380\,GeV Klystron-driven Schedule}

\begin{figure}[h!]
\centering
\includegraphics[scale=0.6]{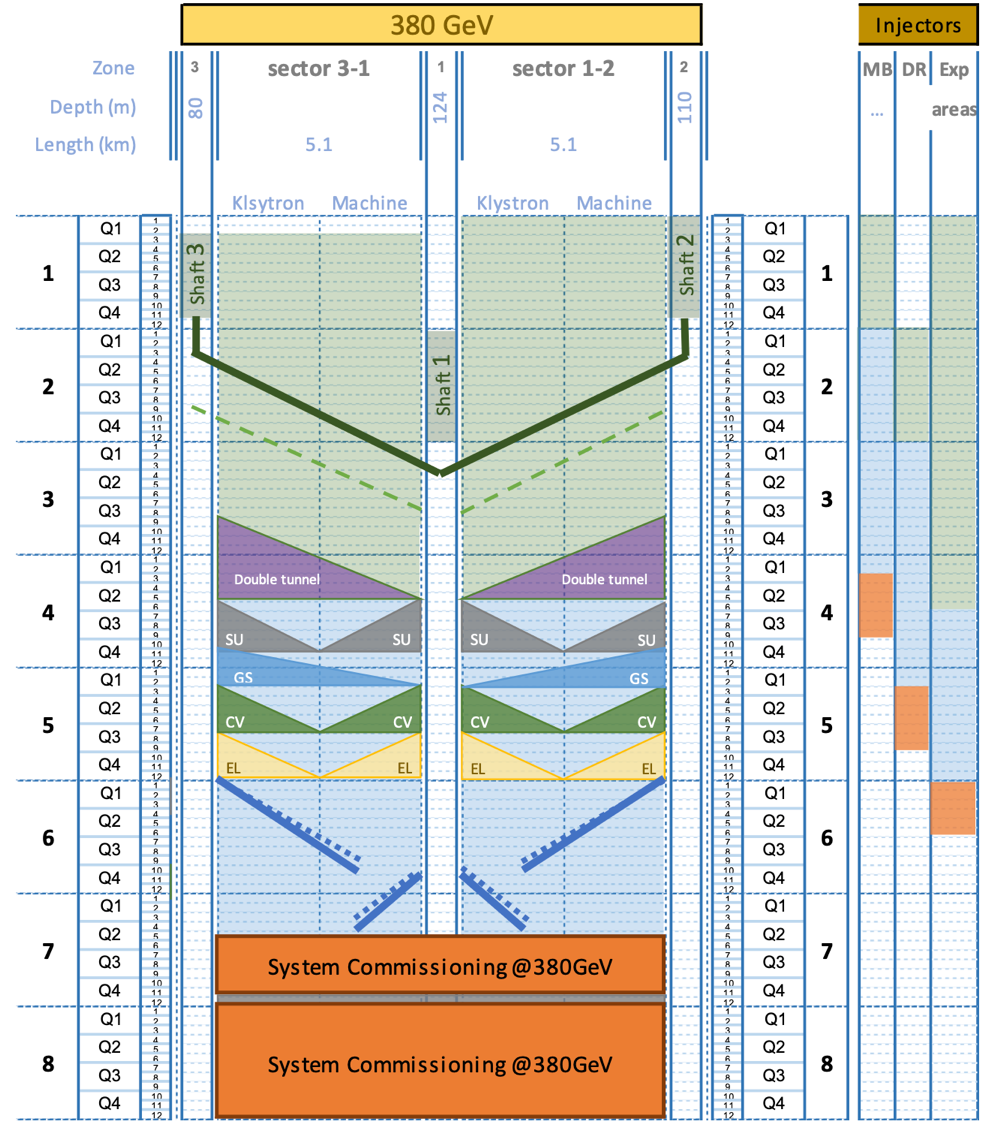}
\caption{\label{fig_IMP_7} The 380\,GeV Klystron-based CLIC construction schedule. (image credit: CLIC)}
\end{figure}

In this scheme the RF power is provided by X-band klystrons and modulators, installed underground all along the Main Linac. The total time for installation is slightly different from the Drive-Beam case. The surface buildings and installations are reduced to those exclusively needed for the Main Beam and experimental area, reducing the surface construction activities correspondingly.  The installation time in the main tunnel is longer, due to the RF units and the additional infrastructures required. Even though it is possible to work in parallel in the Main-Linac tunnel and in the klystron gallery, the overall transport, installation and handling logistics are more time consuming. The time needed for construction, installation and commissioning is eight years as shown in Fig.~\ref{fig_IMP_7}, compared to seven years for the Drive-Beam option at the same CLIC energy of 380\,GeV.

\subsection{Schedules for the Higher Energy Stages and the Complete Project}
\begin{figure}[h!]
\centering
\includegraphics[scale=0.41]{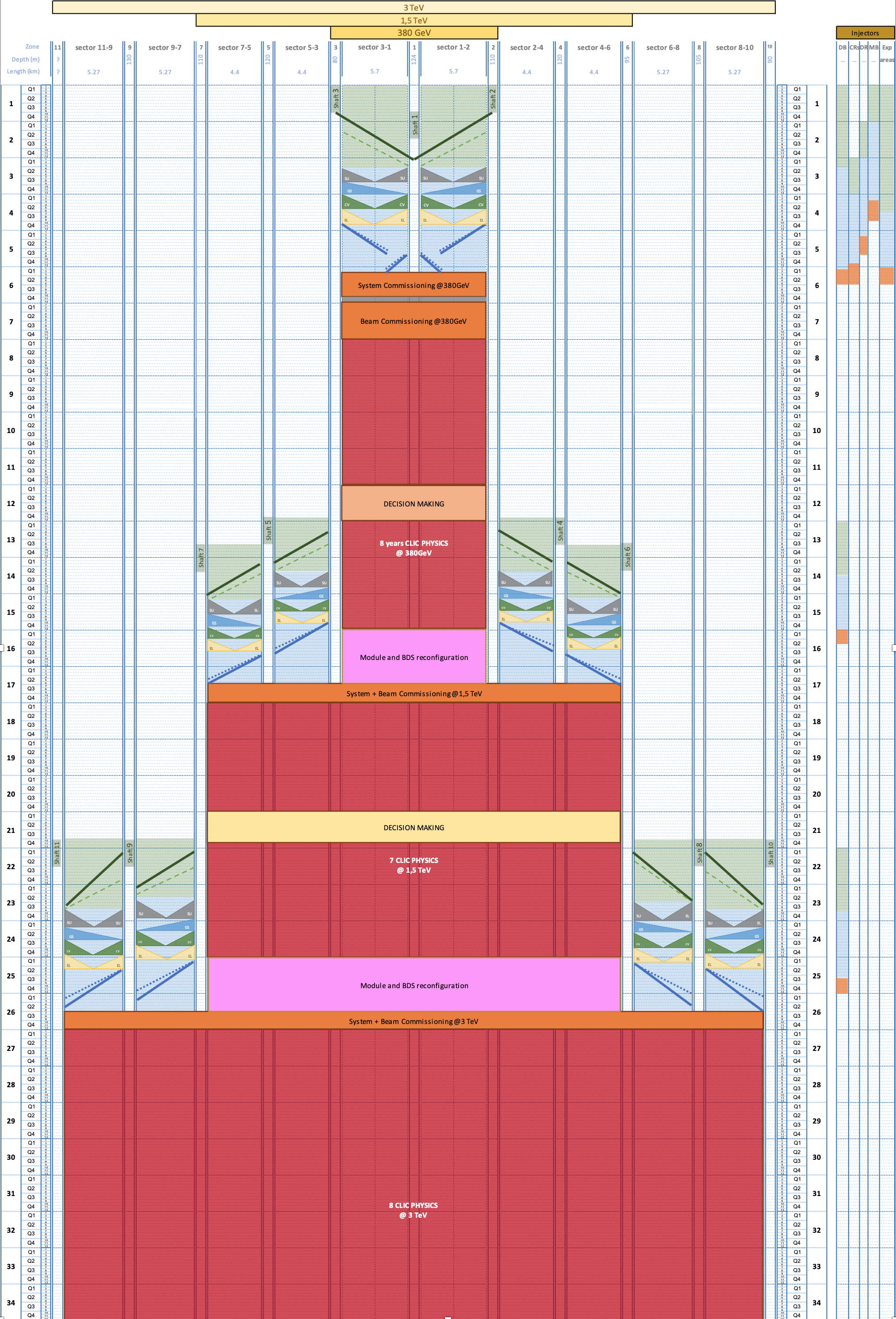}
\caption{\label{fig_IMP_8} The three stages of CLIC showing an overall project length of 34 years covering construction and operation. (image credit: CLIC)}
\end{figure}

In both cases discussed above, the 380\,GeV collider is designed to be extended to higher energies. Most of the construction and installation work can be carried out in parallel with the data-taking at 380\,GeV. However, it is estimated that a stop of two years in accelerator operation is needed between two energy stages. This time is needed to make the connection between the existing machine and its extensions, 
to reconfigure the modules for use at the next stage, to modify the Beam-Delivery System, to commission new equipment and to commission the new accelerator complex with beam. 

As the construction and installation of the 1.5\,TeV and subsequent 3\,TeV equipment cover periods of 4.5 years, the decision about the next higher energy stage needs to be taken after $\sim$4-5 years of data taking at the existing stage, based on physics results available at that time.  The corresponding scenario is shown in Fig.~\ref{fig_IMP_8} for the Drive-Beam based scenario. The overall upgrade schedule is very similar for the case in which the first stage will be powered by klystrons.

\subsection{Concluding Remarks on the Schedule}

\begin{figure}[h!]
\centering
\includegraphics[scale=0.5]{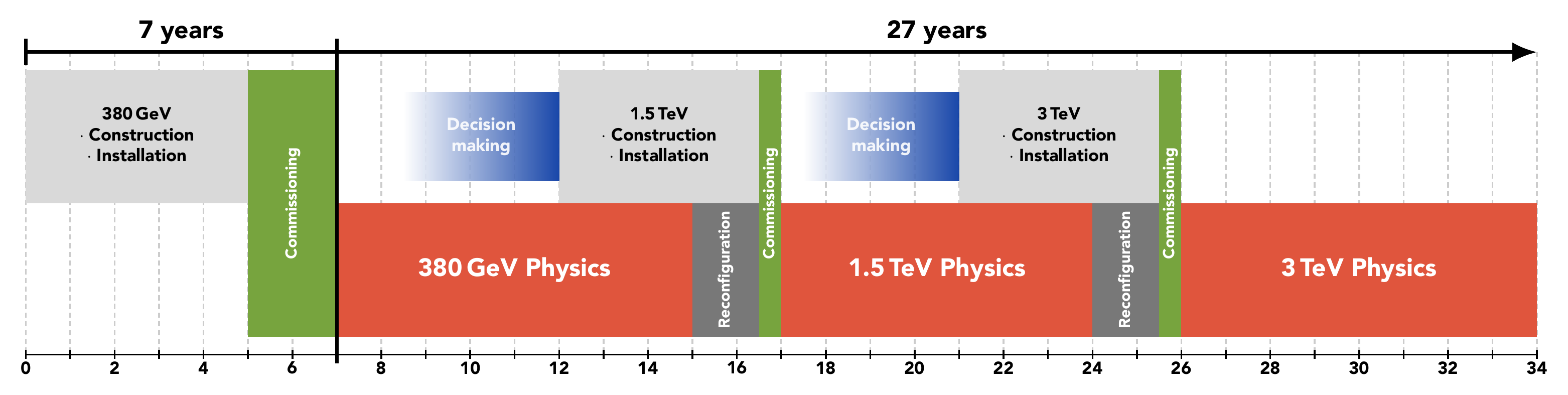}
\caption{\label{fig_IMP_9} Technology-driven CLIC schedule, showing the construction and commissioning period and the three stages for data taking. The time needed for reconfiguration (connection, hardware commissioning) between the stages is also indicated. (image credit: CLIC)}
\end{figure}

The schedule overview shown in Fig.~\ref{fig_IMP_9}, shows that the CLIC project covers 34 years from the start of construction. About 7 years are scheduled for initial construction and commissioning and a total of 27 years for data-taking at the three energy stages, which includes two 2-year intervals between the stages. 

The schedule for construction and installation shows that the CLIC project can be implemented well in time for first collisions in 2035, provided it can be launched by 2026. The most critical CLIC technology-specific items driving the schedule are the Main-Beam module production and installation, and the RF units. The other schedule drivers, such as the tunnelling, the buildings and the infrastructures are similar to other projects at CERN and elsewhere. 

\section{Cost Estimate}
\label{sect:IMP_Cost}
For the cost estimate of CLIC, the methodology used is the same as for previous CLIC cost estimates and estimates of other projects, such as the LHC experiments and the Reference Design Report and Technical Design Report of the International Linear Collider (ILC)~\cite{Phinney2007,Adolphsen:2013kya}. Previous CLIC cost estimates were reported in the CLIC CDR~\cite{Lebrun2012} for two different implementation options at 500\,GeV. An initial cost estimate for the first stage at 380\,GeV was presented together with the introduction of the corresponding CLIC energy staging scenario in~\cite{StagingBaseline}. Since then, many CLIC optimisation studies have been undertaken with a particular focus on cost reduction, as reported in Chapter~\ref{Chapter:Base_Design}. The resulting cost estimates, as well as the methodologies and assumptions used have been presented in November 2018 to a Cost Review Panel composed of international experts. After recommendations on minor issues by the review panel, the estimates have been updated accordingly. The resulting estimated cost of the 380\,GeV stage is presented, together with an estimate for upgrading to higher energies. 

\subsection{Scope and Method}
CLIC is assumed to be a CERN-hosted project, constructed and operated within a collaborative framework with participation and contributions from many international partners. Contributions from the partners are likely to take different forms (e.g. in kind, in cash, in personnel, from different countries, in different currencies or accounting systems). Therefore a``value and explicit labour'' methodology is applied. 

The value of a component or system is defined as the lowest reasonable estimate of the price of goods and services procured from industry on the world market in adequate quality and quantity and satisfying the specifications. Value is expressed in a given currency at a given time. Explicit labour is defined as the personnel provided for project construction by the central laboratory and the collaborating institutes, expressed in Full Time Equivalent (FTE) years. It does not include personnel in the industrial manufacturing premises, as this is included in the value estimate of the corresponding manufactured components. The personnel in industrial service contracts that are part of the accelerator construction, outside CERN or at CERN, are also accounted for in the value estimate of the corresponding items.

For the value estimate, a bottom-up approach is used, following the work breakdown structure of the project, starting from unit costs and quantities for components, and then moving up to technical systems, sub-domains and domains. This allows accounting for all aspects of the production process and the application of learning curves for large series. For each item of the two upper levels (domains and sub-domains, for example Main-Beam production, Injectors) of the work breakdown structure of the CLIC accelerator complex, coordinators were appointed with the mandate to collect costs from the technical groups performing design work, and to exchange information in an \textit{ad hoc} working group, meeting on a regular basis and gradually addressing all domains and sub-domains as refinements of the design progressed. For some parts (e.g. standard systems), cost scaling based on suitable estimators and scaling laws from similar items is used, implying that detailed knowledge on the work breakdown is not required, but rather estimators characterising the component. Unless the complete work breakdown is established and settled down to the component level, cost estimates are based on a hybrid of these two approaches; this is the case of the value estimate presented here. 

The basic value estimate concerns the construction of the 380\,GeV CLIC stage on a site close to CERN, where the 380\,GeV stage of CLIC constitutes a project in itself. As a consequence, large-series effects expected on unit costs -- learning curves and quantity rebates -- remain limited to the quantities required for the completion of the 380\,GeV stage. Estimates are provided both for the Drive-Beam and the Klystron-based options, together with the corresponding incremental value for upgrading to higher energies.

The value estimates given cover the project construction phase, from approval to start of commissioning with beam. They include all the domains of the CLIC complex from injectors to beam dumps, together with the corresponding civil engineering and infrastructures. Items such as specific tooling required for the production of the components, reception tests and pre-conditioning of the components, and commissioning (without beam) of the technical systems, are included.On the other hand, items such as R\&D, prototyping and pre-industrialisation costs, acquisition of land and underground rights-of-way, computing, and general laboratory infrastructures and services (e.g.\ offices, administration, purchasing and human resources management) are excluded. Spare parts are accounted for in the operations budget. The value estimate of procured items excludes VAT, duties and similar charges, taking into account the fiscal exemptions granted to CERN as an Intergovernmental Organisation.

A centralised repository for the value estimates is the CLIC Study Costing Tool~\cite{Jonghe2010}, developed and maintained by the CERN Advanced Information Systems (AIS) group. It presents an on-line, updated view of the value estimates, based on the corresponding work breakdown structures which can be entered at any level. The costing tool includes features for fixed and variable costs, currency conversion, escalation and uncertainty, as well as full traceability of input data and production of tabulated reports which can be exported for further processing.

\subsection{Uncertainty, Escalation and Currency Fluctuations}

The uncertainty objective for the final outcome is $\pm$25\%. To this aim, uncertainties on individual items are grouped in two categories. The first one, \textit{technical uncertainty}, relates to technological maturity and likelihood of evolution in design or configuration. This estimation was provided for each component or system by the expert in charge, according to three levels of relative standard deviation $\sigma^{\text{tech}}$: 0.1 for equipment of known technology, 0.2 for equipment requiring extrapolation from known technology and 0.3 for equipment requiring specific R\&D.  

The second category, \textit{commercial uncertainty}, relates to uncertainty in commercial procurement. A statistical analysis of the offers received for the procurement of accelerator components and technical systems for the LHC, in response to invitations to tender based on precise technical specifications, yielded a probability distribution function which could be modelled as exponential above the threshold of the lowest bid, with a relative standard deviation of about 0.5~\cite{Lebrun2010}. Sampling randomly from such a distribution to obtain a sample of \textit{n} bids, and buying from the lowest bidder among the sample, yields a distribution of prices for the procured components showing a relative standard deviation of $\sigma^{\text{comm}}$= 0.5/\textit{n}. This formula, requiring assessment of the number of valid offers (n) expected for each component, was used to characterise the uncertainty in commercial procurement of CLIC components.

In view of the period in which most of the work was performed, the value estimates presented in the following are expressed in Swiss Francs (CHF) as of December 2018, with the following average exchange rates: 1\,EUR~=~1.13\,CHF, 1\,CHF~=~1\,USD, 1\,CHF~=~114\,JPY. Only a small fraction (5\%) of the CLIC cost estimates rely on industrial quotes more than a year old. The escalation until end of 2018 is therefore negligible, however future index and currency changes can be more significant and will be tracked.

The basis for tracking price escalation is the application of economic indices such as those published on a periodic basis by national or international economic agencies. The CLIC value estimates are escalated according to indices published by the Swiss Office Federal de la Statistique, among which we have chosen two compound indices: the global "Construction" index, updated every semester, is applied to the value estimates of civil engineering, while the global "Arts \& Metiers et Industrie" index, updated monthly, is applied to all other technical systems. 

\subsection{Value Estimates and Cost Drivers}

\begin{figure}[h!]
\centering
\includegraphics[scale=0.8]{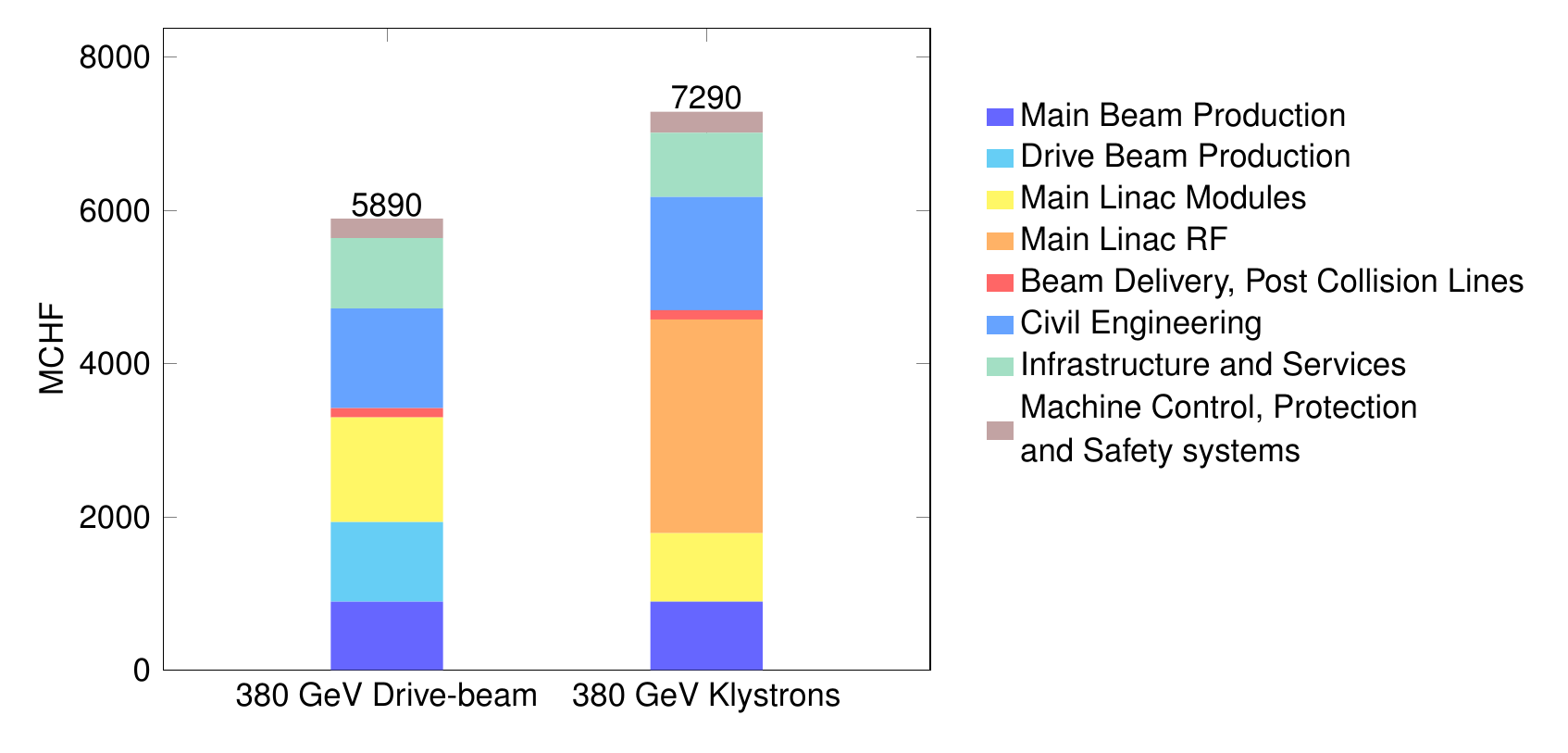}
\caption{\label{fig_IMP_10} Cost breakdown for the 380\,GeV stage of the CLIC accelerator, for the Drive-Beam option and for the Klystron option. (image credit: CLIC)}
\end{figure}

The breakdown of the resulting cost estimate up to the sub-domain level is presented in Table~\ref{Tab:Cost} for the 380\,GeV stage of the accelerator complex, both for the baseline design with a Drive Beam and for the Klystron-based option. Figure~\ref{fig_IMP_10} illustrates the sharing of cost between different parts of the accelerator complex. 
The injectors for the Main Beam and Drive Beam production are among the most expensive parts of the project, together with the Main Linac, and the civil engineering and services.

\begin{table}[ht]
\caption{Cost breakdown for the 380\,GeV stage of the CLIC accelerator, for the Drive-Beam baseline option and for the klystron option.}
\label{Tab:Cost}
\centering
\begin{tabular}{l l c c}
\toprule
\multirow{2}{*}{Domain} & \multirow{2}{*}{Sub-Domain} & \multicolumn{2}{c}{Cost [MCHF]} \\
 &  & {Drive-Beam} & {Klystron} \\ 
 \midrule
 \multirow{3}{*}{Main Beam Production} & Injectors & 175 & 175 \\
 & Damping Rings & 309 & 309 \\
& Beam Transport & 409 & 409 \\ \hline
\multirow{3}{*}{Drive Beam Production} & Injectors & 584 &  {---} \\
& Frequency Multiplication & 379 & {---}  \\
& Beam Transport & 76 &  {---} \\ \hline
\multirow{2}{*}{Main Linac Modules}  & Main Linac Modules & 1329 & 895 \\
 & Post decelerators  & 37 &  {---}  \\ \hline
Main Linac RF  & Main Linac Xband RF & {---} & 2788 \\ \hline
\multirow{3}{*}{\makecell[l]{Beam Delivery and \\ Post Collision Lines}}   & Beam Delivery Systems & 52 & 52 \\
 & Final focus, Exp. Area & 22 & 22 \\
 & Post-collision lines/dumps & 47 & 47 \\ \hline
Civil Engineering & Civil Engineering & 1300 & 1479 \\ \hline
\multirow{4}{*}{Infrastructure and Services}  & Electrical distribution  & 243 & 243 \\
 & Survey and Alignment & 194 & 147 \\
 & Cooling and ventilation  & 443 & 410 \\
 & Transport / installation & 38 & 36 \\ \hline
\multirow{4}{*}{\makecell[l]{Machine Control, Protection \\ and Safety systems}} & Safety system  & 72 & 114 \\
  & Machine Control Infrastructure & 146 & 131 \\
 & Machine Protection & 14 & 8 \\
 & Access Safety \& Control System & 23 & 23 \\ \midrule
\bfseries Total (rounded) & & \bfseries 5890 & \bfseries 7290 \\
\bottomrule
\end{tabular}
\end{table}

Combining the estimated technical uncertainties yields a total (1$\sigma$) error of 1270\,MCHF for the drive-beam based facility, and 1540\,MCHF when using klystrons. In addition, the commercial uncertainties, defined above, need to be included. They amount to 740\,MCHF and 940\,MCHF for the drive-beam and klystron-based options, respectively. The total uncertainty is obtained by adding technical and commercial uncertainties in quadrature. Finally, for the estimated error band around the cost estimate, the resulting total uncertainty is used on the positive side, while only the technical uncertainty is used on the negative side~\cite{Lebrun2012}. The cost estimate for the first stage of CLIC including a 1$\sigma$ overall uncertainty is therefore:
\begin{center}
\begin{tabular}{lc}
CLIC 380\,\text{GeV} Drive-Beam based:& $5890^{+1470}_{-1270}\, \text{MCHF}$;\\ \\
CLIC 380\,\text{GeV} Klystron based:& $7290^{+1800}_{-1540}\,\text{MCHF}$.
\end{tabular}
\end{center}

The difference between the Drive-Beam and Klystron-based values is mainly due to the current cost estimates for the X-band klystrons and corresponding modulators. The increased diameter of the Main-Linac tunnel, required to host the RF gallery in the klystron-based option, also contributes to the cost-difference. By reducing the X-band RF costs by 50\% in the klystron option, the overall cost of the two options becomes similar. To achieve such a reduction would require a dedicated development programme in collaboration with industry for X-band klystrons and associated modulators.
There is still room for possible gains through optimising the accelerating structure parameters, klystron design and luminosity performance. The cost of the Klystron option is more affected by the luminosity specification than the Drive-Beam option.

The cost composition and values of the 1.5\,TeV and 3\,TeV stages have also been estimated. The energy upgrade to 1.5\,TeV has a cost estimate of $\sim \text{5.1}\,\text{BCHF}$, including the upgrade of the Drive Beam RF power needed for the 1.5\,TeV stage. In the case of expanding from a klystron-based initial stage this energy upgrade will be 25\% more expensive. A further energy upgrade to 3\,TeV has a cost estimate of $\sim \text{7.3}\,\text{BCHF}$, including the construction of a second Drive-Beam complex. 

The CLIC technical cost drivers have been identified, together with potential cost mitigation alternatives. These will be addressed in the next phase of the CLIC project as discussed in Section~\ref{sect:NextPhase}.
In general, further cost reduction studies will require close collaboration with industry. Beyond technical developments, optimal purchase models need to be defined, optimising the allocation of risks and production responsibilities between industry, CERN and collaboration partners in each case. In particular, the module production and RF units have a potential for cost reduction. For a Klystron-based implementation, the cost reductions of the RF system are of crucial importance.

\subsection{Labour Estimates}
A first estimate of the explicit labour needed for construction of the CLIC accelerator complex was obtained~\cite{Lebrun2012} by assuming a fixed ratio between personnel and material expenditure for projects of similar nature and size. Scaling with respect to the LHC - a CERN-hosted collider project of similar size to CLIC - provides a good estimator. Data from the LHC indicate that some 7000\,FTE-years were needed for construction, for a material cost of 3690\,MCHF (December 2010), corresponding to about $1.9\,\text{FTE-year}/\text{MCHF}$.  About 40\% of this labour was scientific and engineering personnel, and the remaining 60\% worked on technical and project execution tasks.

In terms of complexity, the different CLIC sub-systems resemble the LHC case. Therefore, following the LHC approach outlined above, construction of the 380\,GeV stage of the CLIC accelerator complex would require 11500\,FTE-years of explicit labour. It is worth noting that this preliminary result is rather similar to the $1.8\,\text{FTE-year}/\text{MCHF}$ derived for the ILC~\cite{Adolphsen:2013kya}. Although the RF technology differs between ILC and CLIC, the main elements of the accelerator complex are similar in the two projects. 

\subsection{Operation Costs}
A preliminary estimate of the CLIC accelerator operation cost, with focus on the most relevant elements, is presented here. The material cost for operation is approximated by taking the cost for spare parts as a percentage of the hardware cost of the maintainable components. These annual replacement costs are estimated at the level of:
\begin{itemize}
\item 1\% for accelerator hardware parts (e.g.\ modules).
\item 3\% for the RF systems, taking the limited lifetime of these parts into account. 
\item 5\% for cooling, ventilation and electrical infrastructures etc. (includes contract labour and consumables)
\end{itemize}

These replacement/operation costs represent 116\,MCHF per year.

An important ingredient of the operation cost is the CLIC power consumption and the corresponding energy cost, which is discussed in Section~\ref{sect:IMP_Power} below. This is difficult to evaluate in CHF, as energy prices are likely to evolve. The expected energy consumption of the 380\,GeV CLIC accelerator, operating at nominal luminosity, corresponds to 2/3 of CERN's current total energy consumption. 

For personnel needed for the operation of CLIC, one can assume levels that are similar to large accelerator facilities operating today. Much experience was gained with operating Free Electron Laser linacs and light-sources with similar technologies. As CLIC is a normal-conducting accelerator operated at room temperature, one can assume that the complexity of the infrastructure, and therefore the maintenance efforts, compare favourably with other facilities. The maintenance programme for equipment in the klystron galleries is demanding, but is not expected to impact strongly on the overall personnel required for operation. The ILC project has made a detailed estimate of the personnel needed to operate ILC, yielding 640\,FTE. This number includes scientific/engineering (40\%), technical/junior level scientific staff (40\%) and administrate staff (20\%) for the operation phase \cite{Adolphsen:2013kya,Evans:2017rvt}. The difference between a 250\,GeV and a 500\,GeV ILC implementation was estimated to be 25\%. In the framework of CERN, these numbers would distribute across scientific/engineering/technical staff, technical service contracts, fellows and administrative staff.  The level of CLIC operational support required is expected to be similar to the ILC estimates.  

Given the considerations listed above, one can conclude that operating CLIC is well within the resources deployed for operation at CERN today. 
Operating CLIC concurrently with other programmes at CERN is also technically possible. This includes LHC, as both
accelerator complexes are independent. Building CLIC is not destructive with respect to the existing CERN accelerator complex. Electrical GRID connections are also independent. The most significant limitation will therefore be the resources, in particular personnel and overall energy consumption. 

\section{Power and Energy Consumption}
\label{sect:IMP_Power}

\begin{figure}[h!]
\centering
\includegraphics[width=\textwidth]{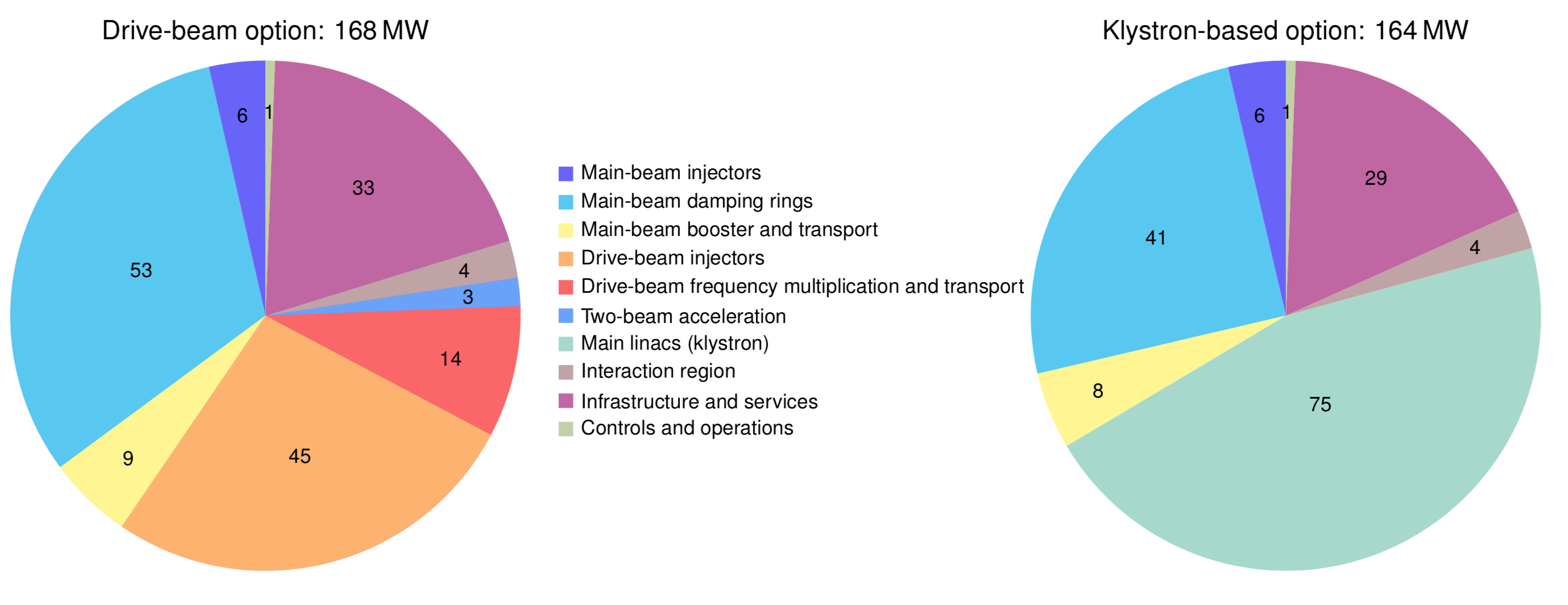}
\caption{\label{fig_IMP_11} Breakdown of power consumption between different domains of the CLIC accelerator in MW at a centre-of-mass energy of 380\,GeV, for the Drive-Beam option on the left and for the Klystron-option on the right. The contributions add up to a total of 168\,MW and 164\,MW in the two cases. (image credit: CLIC)}
\end{figure}

The nominal power consumption at the 380\,GeV stage has been estimated based on the detailed CLIC work breakdown structure. This yields for the Drive-Beam option a total of 168\,MW for all accelerator systems and services, taking into account network losses for transformation and distribution on site. The breakdown per domain in the CLIC complex (including experimental area and detector) and per technical system is shown in the left part of Fig.~\ref{fig_IMP_11}.  Most of the power is used in the Drive Beam and Main-Beam Injector complexes and comparatively little in the Main Linacs. Among the technical systems, the RF represents the major consumer. For the Klystron-based version the total power consumption is very similar at 164\,MW as shown in the right part of Fig.~\ref{fig_IMP_11}.

These numbers are significantly reduced compared to earlier estimates due to optimisation of the injectors for 380\,GeV, introducing optimised accelerating structures for this energy stage, significantly improving the RF efficiency, and consistently using the expected operational values instead of the full equipment capacity in the estimates.  For the 1.5\,TeV and 3.0\,TeV stages these improvements have not been studied in detail and the power estimates from the CDR are used~\cite{Lebrun2012}. 

\begin{table}[ht]
\caption{Estimated power consumption of CLIC at the three centre-of-mass energy stages and for different operation modes. The 380\,GeV numbers are for the drive beam option, whereas the estimates for the higher energy stages are from~\cite{Lebrun2012}.}
\label{Tab:Power}
\centering
\begin{tabular}{c c c c}
\toprule
Collision Energy [GeV]  & Running [MW] &  Standby [MW] & Off [MW] \\
\midrule
380     &  168 & 25   & 9 \\
1500    &  364  & 38  & 13 \\
3000    &  589  & 46  & 17 \\
\bottomrule
\end{tabular}
\end{table}

Table~\ref{Tab:Power} shows the nominal power consumption in three different operation modes of CLIC, including the ``running'' mode at the different energy stages, as well as the residual values for two operational modes corresponding to short (``standby'') and long (``off'') beam interruptions. Intermediate power consumption modes exist, for example when a part of the complex is being tested, or during transitional states as waiting for beam with RF on. The contribution of these transitional states to the annual energy consumption is dealt with by averaging between ``running'' and ``standby'' for certain periods, as described below.

\subsection{Energy Consumption}

\begin{figure}[h!]
\centering
\includegraphics[scale=0.6]{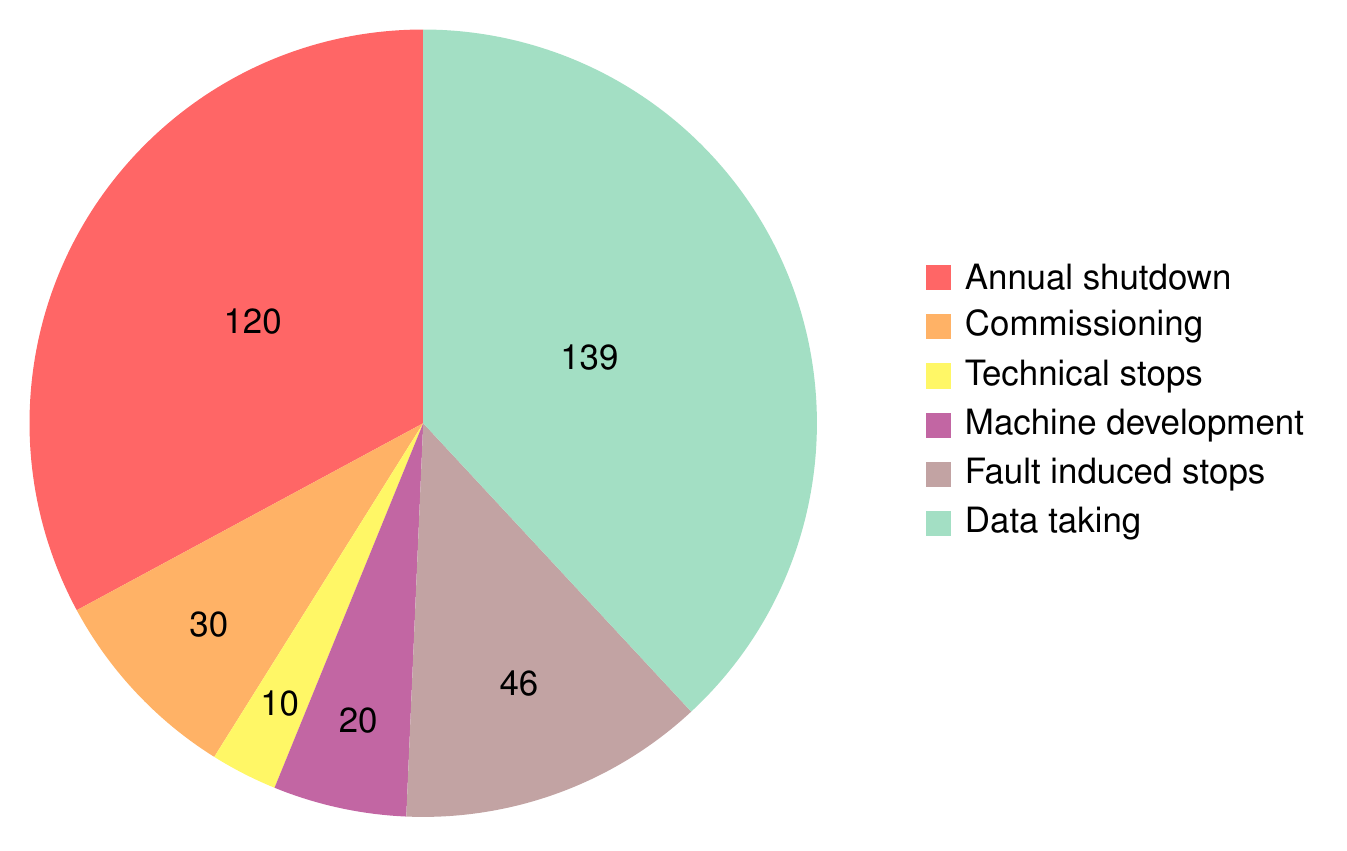}
\caption{\label{fig_IMP_12} Operation schedule in a "normal" year (days/year). (image credit: CLIC)}
\end{figure}

Estimating the yearly energy consumption from the power numbers requires an operational scenario, which is detailed in~\cite{Bordry2018} and depicted in Fig.~\ref{fig_IMP_12}. In any ``normal'' year, i.e.\ once CLIC has been fully commissioned, the scenario assumes 120 days of annual shutdown, 30 days for beam-commissioning, and 30 days of scheduled maintenance, including machine development and technical stops (typically 1 day per week, or 2 days every second week). This leaves 185 days of operation for physics, for which 75\% availability is assumed,
 i.e.\ 46 days of fault-induced stops. This results in 139 days, or 1.2~$\times$~10$^7$ seconds, per year for physics data taking. 

In terms of energy consumption the accelerator is assumed to be ``off'' during 120 days and ``running'' during 139 days. The power consumption during the remaining time, covering commissioning, technical stops, machine development and fault-induced stops is taken into account by estimating a 50/50 split between ``running'' and ``standby''. In addition, one has to take reduced operation into account in the first years at each energy stage to allow systematic tuning up of all parts of the accelerator complex. A luminosity ramp-up of three years (10\%, 30\%, 60\%) in the first stage and two years (25\%, 75\%) in subsequent CLIC stages is considered. For the energy consumption estimate we change the corresponding reduction in ``running'' time to a 50/50 mixture of the two states mentioned above, resulting in a corresponding energy consumption ramp-up.

\begin{figure}[h!]
\centering
\includegraphics[width=0.5\textwidth]{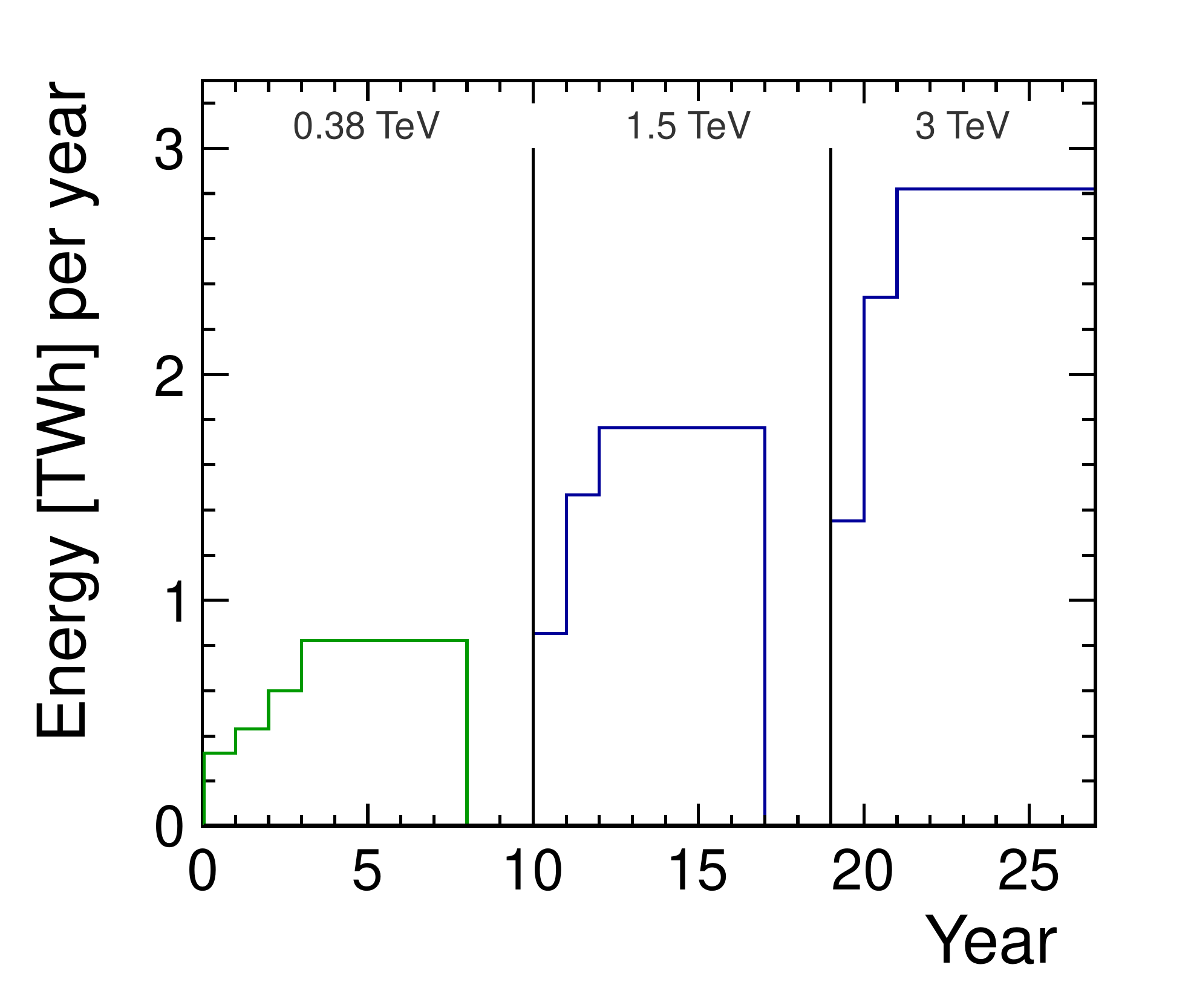}
\caption{\label{fig_IMP_13} Estimated yearly energy consumption of CLIC. The initial stage estimate is revised in detail (green), while numbers for the higher energy stages are from~\cite{Lebrun2012} (blue). (image credit: CLIC)}
\end{figure}

The evolution of the resulting electrical energy consumption over the years is illustrated in Fig.~\ref{fig_IMP_13}. For comparison, CERN's current energy consumption is  approximately 1.2\,TWh per year, of which the accelerator complex uses around 90\%.

\subsection{Power Reduction Studies and Future Prospects}
Since the CDR~\cite{Lebrun2012} in 2012 the CLIC collaboration has systematically explored power reduction and technical system optimisation across the complex. As a result the power estimate is reduced by around 35\% for the initial stage. The main contributors are:
\begin{itemize}
\item The accelerating structures were optimised for 380\,GeV and corresponding luminosity, impacting among others RF power needs and the machine length. The optimisation was done for cost but it was also shown that cost and power are strongly correlated.
\item The injector systems and drive beam facility were optimised to the 380 GeV parameters taking into account R\&D on various technical systems, for example reducing the number of drive beam klystrons to around 60\% of earlier designs.
\item High efficiency klystron studies have reached a maturity such that 70\% efficiency can be taken as the baseline.
\item Permanent magnets can partly replace electromagnets.
\item Nominal settings of RF systems, magnets, cooling have consistently been used analysing the power consumption when running at full luminosity, replacing earlier estimates which in some cases were based on maximum equipment capacity.
\end{itemize}

In summary, the estimate of the power consumption can be considered to be detailed and complete for the initial 380\,GeV stage. The estimates for the higher energy stages have not been scrutinised in order to include the saving measures listed above. Also for the initial stage further work can lead to additional savings. This concerns, in particular, the damping ring RF power, where further studies are needed before a revised baseline can be introduced. The total power consumption of the damping rings (53\,MW) is dominated by the RF system (45\,MW). In the present design, the power efficiency of the RF system is rather low due to high peak power requirements for compensation of transient beam loading effects.  Work is ongoing to improve the design and reduce the peak RF power requirements by introducing an optimum modulation of both phase and amplitude of the input RF signal.  This may result in a significant (up to a factor of 2) reduction of the damping ring RF system power consumption, potentially reducing the overall damping ring power consumption to around 30\,MW.

\subsection{Energy Cost Optimisation}

CLIC is a normal-conducting accelerator running at room temperature. Turning it on/off or into intermediate power states can, with appropriate thermo-mechanical considerations, be done relatively quickly. This means that CLIC could possibly be operated as a ``Peak Shaving Facility'' for the electrical network, matching not only seasonal, but also daily fluctuations of the demand. This particular feature constitutes a strong asset towards optimal energy management, a necessary approach in view of the large values of power consumption of the CLIC complex during nominal operation, in particular for stages 2 and 3. Furthermore, given the societal move towards renewable energies and the increased focus on energy recovery measures for any future facility, studies were launched addressing the following three issues: energy management and costs, operating with a large part of the power from renewable sources, and possible energy recovery measures. All of these issues require much more work in the next phase of the project, but a brief summary of the results obtained in initial case-studies~\cite{Fraunhofer} based on CLIC 380\,GeV  specifications is provide in the following three paragraphs. 

\begin{figure}[h!]
\centering
\includegraphics[scale=0.7]{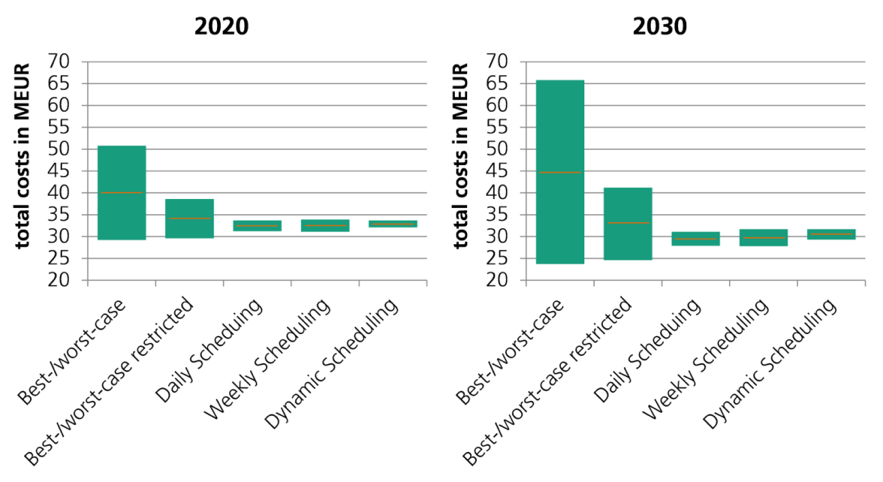}
\caption{\label{fig_fraunhofer} Relative energy cost by no scheduling, avoiding the winter months (restricted), daily, weekly and dynamic scheduling. As explained in the text the central values of the ranges shown should be considered the best estimates. The absolute cost scale will depend on prices, contracts and detailed assumption about running times, but the relative cost differences indicate that significant cost-reductions could be achieved by optimising the running schedule of CLIC to avoid high energy cost periods, also outside the winter shut-down periods. (image credit: Fraunhofer)}
\end{figure}

Cost reductions have been studied by optimising the CLIC running schedule, at a daily, weekly and yearly level. With the flexibility of the accelerator to go from one mode to another quickly it is possible to avoid the high cost energy periods. In order to minimize energy costs, optimization of operational strategies for the accelerator to achieve a fixed integrated luminosity per year have been investigated. For these investigations, the operation states of CLIC were described by a finite state machine (FSM). The general approach was to develop an operation strategy by dynamically shifting the operation states with high power demand to time slots with low energy prices. Three different operating strategies were developed, which generate optimized operating schedules for CLIC. Daily and Weekly Scheduling optimize the distribution of the various operation states of CLIC over the respective period using fixed state sequences with variable durations. Dynamic Scheduling distributes all operating states over the whole year without using predefined sequences, which leads to a high flexibility. The investigations are mainly based on the energy price curve for 2020 from the European energy spot market (EPEX SPOT) and a forecast for 2030 covering much more renewable energies. In addition, numerous boundary conditions and the requirement to reach a fixed integrated luminosity were considered. Figure~\ref{fig_fraunhofer} shows the main investigation results for the study. The best-/worst-case analysis provides the theoretical range of energy operation costs. The central value of the range should be considered to best estimate. Best-/worst-case restricted can be reached by avoiding the cold months with high energy prices and results already in a reduction of almost two thirds of the total energy cost savings. Furthermore, more than one third of the savings is attained by using optimized operational strategies. The simulation results for three different schedulers show similar cost distributions. Regarding the price curve of 2020, the minimal attainable total energy costs are 31\,million\,EUR for Daily/Weekly Scheduling and 32\,million\,EUR for Dynamic Scheduling. For the forecast of 2030, the lowest reachable energy price is even lower due to the higher fluctuation of the energy price. The actual running scenario, as well as prices and flexibility obtainable in future energy contract for CERN are very uncertain for the time period in question, but the possibility to adjust the running scenario and hence power consumption of CLIC to take advantage of low cost periods also outside the winter shut-downs seems worth pursuing. 

Given the flexibility on running and power consumption, it is also interesting to consider how effectively the accelerator can be powered by renewable energies. First of all, it is likely the overall energy landscape in Europe will shift over the next decades towards renewables, secondly the investment costs of such power sources are decreasing so one can consider moving investments in energy production into the construction costs, hence lowering the operation costs. By installing a portfolio of different renewable generators (different technologies, like wind and photovoltaic (PV), or different types of installations, like photovoltaic modules orientated into different directions) it becomes possible to partly level out the individual fluctuations of single generators in the aggregated generation curve. Combining the flexibility of the CLIC with this generation curve can increase the level of self-sufficiency. While it is possible to fully supply the annual electricity demand of the CLIC by installing local wind and PV generators (this could be e.g. achieved by 330\,MW-peak PV and 220\,MW-peak wind generators, at a cost of slightly more than 10\% of the CLIC 380\,GeV cost), self-sufficiency during all times can not be reached and only 54\% of the time CLIC could run independently from public electricity supply with the portfolio simulated. About 1/3 of the generated PV and wind energy will be available to export to the public grid even after adjusting the load schedule of CLIC. Because of the correlation between electricity price and (national) generation from wind and PV, own local generators can generally not step in during times of high energy prize. Large storage systems are still too expensive to shift power accordingly. Installing such large renewable power plants is a process of some years, so energy would not be available on short term. Besides the direct investment in the generation technology, many aspects of standards, regulations, land-use, landscape-protection etc. would have to be considered. One alternative to own renewable power plants could be the participation in projects of other investors to build large renewable power plants. Yet financial advantages from local own consumption (like lower grid fees) would get lost in that case. The conclusion is that on a medium to long-term time scale local renewable generation could cover a fraction of the total load, but self-sufficiency of CLIC based on local renewables is not a realistic scenario. 

Waste heat recovery also constitutes an interesting option in view of the large power rejected into water from CLIC. However, the use of waste heat to generate electricity is technically difficult due to the low temperature of the waste heat. The heat would have to be raised to a significantly higher level and more electricity would be consumed than can be generated again in the later process. A reasonable option is to use the waste heat to provide space heating. Also for this option, the temperature must be raised via a heat pump and thus additional electricity must be used. Another possibility would be the research of further innovative concepts for the use of waste heat with very low temperature (for example very low temperature ORCs, thermoelectric generators or the storage of heat in zeolites). The fact that the maximum energy need locally is during the winter, when it is favourable of energy cost reasons to not run the accelerator, also makes is more difficult today to envisage efficient large scale energy recovery strategies.

\section{CLIC Objectives for the Period 2020--2025}
\label{sect:NextPhase}

\begin{figure}[h!]
\centering
\includegraphics[width=\textwidth]{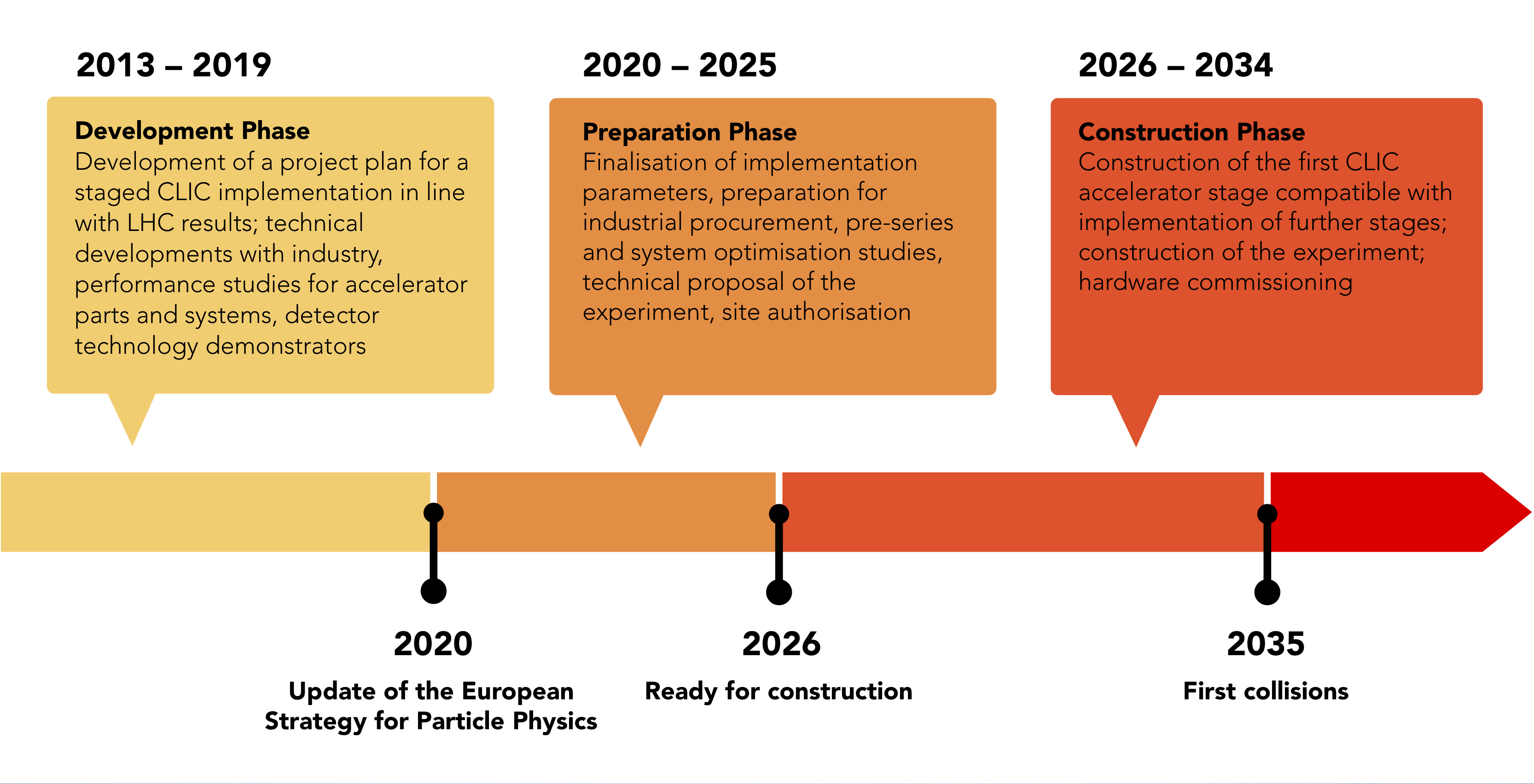}
\caption{Schematic view of the CLIC implementation schedule, with first collisions in 2035. (image credit: CLIC)}
\label{fig_IMP_14}
\end{figure}

The project implementation for CLIC foresees an initial five-year preparation phase prior to a construction start envisaged by 2026. 
The overall schedule towards first beams by 2035 is shown in Fig.~\ref{fig_IMP_14}. 
This leaves a 2-year margin in addition to the construction and commissioning period estimated in the technology-driven schedule shown in Figure~\ref{fig_IMP_9}.
In order to analyse the priorities for the preparation phase, the following project risks and mitigations have been considered:
\begin{itemize}
\item  Performance: The dominant performance risk is related to the luminosity.  Luminosity performance is based on technical performance and
reliability as well as design robustness and system redundancy. Risk mitigation implies further studies at design and technical level, including on variation of parameters such as temperatures, mechanical instabilities and vibrations, magnetic fields, etc. Most importantly, performance validations in normal-conducting Free Electron Laser (FEL) linacs and other compact linac systems will provide powerful demonstrations and new benchmarks for reliability, technical parameters, simulation and modelling tools on the timescale of 2020--2025.  
\item  Technical systems: 
The main technical risks are related to RF sources, the X-band components, and overall system integration for the Main Linac. 
Reliable, efficient and cost-effective klystrons, modulators and X-band structures are components which are crucial for the machine. 
Additional thermo-mechanical engineering studies of the Main-Linac tunnel, integrating all components, are important in order to further improve the understanding
of the mechanical and thermal stability needed for CLIC. 
In addition, further system tests (beyond what has been achieved with CTF3) of the high-power Drive Beam would be most desirable.
\item  Implementation: 
Principal risks are associated with the industrial production of large numbers of  modules and the civil engineering. 
Work during the preparation phase includes qualifying companies for industrial production and optimising the work distribution and 
component integration. The module installation and conditioning procedures need to be refined and further verified. 
Cost control is crucial and is an integral part of these studies. This requires work on optimising the 
risk sharing models between industry, CERN and collaborative partners for the most critical and costly components.
Detailed site-specific design work related to civil engineering and infrastructure needs to be performed.
\end{itemize}

\subsection{Accelerator Programme Overview}

\begin{table}[h]
\begin{center}
\caption{Main CLIC accelerator objectives and activities in the next phase.}
\label{tab:CLIC2025}
\small
\begin{tabular}{c c}
\toprule
\textbf{Activities} & \textbf{Purpose} \\
\midrule
\multicolumn{2}{c}{\textbf{Design and parameters}}\tabularnewline  
\begin{minipage}[t]{0.46\columnwidth}%
Beam dynamics studies, parameter optimisation, cost, power, system verifications in linacs and low emittance rings
\end{minipage} & %
\begin{minipage}[t]{0.46\columnwidth}%
Luminosity performance and reduction of risk, cost and power
\end{minipage} \tabularnewline %
\\
\multicolumn{2}{c}{\textbf{Main Linac modules}}\tabularnewline 
\begin{minipage}[t]{0.46\columnwidth}%
Construction of 10 prototype modules in qualified industries, Two-Beam and klystron versions, optimised design of the modules with their supporting infrastructure in the Main-Linac tunnel
\end{minipage} & %
\begin{minipage}[t]{0.46\columnwidth}%
Final technical design, qualification of industrial partners, production models, performance verification 
\end{minipage} \tabularnewline %
\\
\multicolumn{2}{c}{\textbf{Accelerating structures}}\tabularnewline
\begin{minipage}[t]{0.46\columnwidth}%
Production of $\sim50$ accelerating structures, including structures for the modules above
\end{minipage} & %
\begin{minipage}[t]{0.46\columnwidth}%
Industrialisation, manufacturing and cost optimisation, conditioning studies in test-stands  
\end{minipage} \tabularnewline %
\\
\multicolumn{2}{c}{\textbf{Operating X-band test-stands, high efficiency RF studies}}\tabularnewline  
\begin{minipage}[t]{0.46\columnwidth}%
Operation of X-band RF test-stands at CERN and in collaborating institutes for structure and component optimisation, further development of cost-optimised high efficiency klystrons 
\end{minipage} & %
\begin{minipage}[t]{0.46\columnwidth}%
Building experience and capacity for X-band components and structure testing, validation and optimisation of these components, cost reduction and increased industrial availability of high efficiency RF units 
\end{minipage} \tabularnewline %
\\
\multicolumn{2}{c}{\textbf{Other technical components}}\tabularnewline
\begin{minipage}[t]{0.46\columnwidth}%
Magnets, instrumentation, alignment, stability, vacuum   
\end{minipage} & %
\begin{minipage}[t]{0.46\columnwidth}%
Luminosity performance, costs and power, industrialisation 
\end{minipage} \tabularnewline %
\\
\multicolumn{2}{c}{\textbf{Drive beam studies}}\tabularnewline  
\begin{minipage}[t]{0.46\columnwidth}%
Drive beam front end optimisation and system tests to $\sim20\,\text{MeV}$
\end{minipage} & %
\begin{minipage}[t]{0.46\columnwidth}%
Verification of the most critical parts of the drive beam concept,  further development of industrial capabilities for L-band RF systems
\end{minipage} \tabularnewline %
\\
\multicolumn{2}{c}{\textbf{Civil Engineering, siting, infrastructure}}\tabularnewline  
\begin{minipage}[t]{0.46\columnwidth}%
Detailed site specific technical designs, site preparation, environmental impact study and corresponding procedures in preparation for construction
\end{minipage} & %
\begin{minipage}[t]{0.46\columnwidth}%
Preparation for civil engineering works, obtaining all needed permits, preparation of technical documentation, tenders and commercial documents
\end{minipage} \tabularnewline %
\\
\bottomrule
\end{tabular}
\end{center}
\end{table}

To address these issues the forthcoming preparation phase will comprise further design, technical and industrial developments,
with a focus on cost, power and risk reduction, in preparation for the Technical Design Report. System verifications in FEL linacs and low emittance rings will be increasingly important. The governance structure and the international collaboration agreements for the construction phase will be prepared during this time. 

Civil engineering and infrastructure preparation will become increasingly detailed during the preparation phase.
An environmental impact study and corresponding public enquiry will be needed as a prerequisite to authorisations for construction. 
Experience from the LEP and LHC projects indicates that approximately two years will be needed for such formal steps, as required by the procedures in the CERN host states. 

The key elements of the CLIC accelerator activities during the period 2020--2025 are summarised in Table~\ref{tab:CLIC2025}.

\subsection{Programme Implementation, Technology Demonstrators and Collaboration}
The design studies and technical work for CLIC are broadly shared among the CLIC collaboration partners. The CLIC (accelerator) collaboration currently comprises 53 institutes from 31 countries. 

The potential for collaborative projects is increasing with the current expansion in the field of Free Electron Laser (FEL) linacs and next-generation light sources. In particular, the increasing use of X-band technology, either as the main RF technology or for parts of the accelerators (deflectors, linearisers), is of high relevance for the next phase of CLIC. Construction, upgrades and operation of FEL linacs and conventional light sources, several of which are located at laboratories of CLIC collaboration partners, provide many opportunities for common design and component developments, and for acquiring crucial system test experience. Furthermore, the fact that there are significant resources invested in such accelerators world-wide, provides excellent opportunities for building up industrial capabilities and networks.

X-band RF systems and structure manufacturing used to be exclusively available in the US and Japan. However, today there are fourteen institutes capable of developing and testing X-band structures. All are working together on optimising the technology. 

The increasing number of qualified companies for accelerating structure manufacturing, together with the growing industrial availability
of RF systems, make it easier for new groups to engage in these technologies.
As a consequence, several smaller accelerators using X-band technology are in proposal or technical preparation phase. In this context it is important to mention the SPARC 1\,GeV X-band linac at INFN~\cite{Diomede:IPAC18}, a possible upgrade of CLARA at Daresbury~\cite{CLARA-upgrade}
and the CompactLight~\cite{DAuria:2062591} FEL study. The CompactLight design study is co-financed by the European Commission. It involves 24 partners preparing technical designs for compact FELs based on X-band linacs at energies ranging from 6\,GeV down
to small room-size systems for X-ray production through Inverse Compton Scattering (e.g.\ SmartLight~\cite{SmartLight}).
Furthermore, the implementation of a 3.5\,GeV X-band linac (eSPS) has been proposed~\cite{Akesson:2640784} at CERN. It aims at injecting electrons into the SPS for further acceleration, and suggests implementing the linac during the period 2019--2024. 

In summary, the growing use of CLIC technology and industrial capabilities allows for implementing
several of the CLIC project activities described in Table~\ref{tab:CLIC2025} in the form of collaborative projects together with the projects and technology partners mentioned above. 
Nevertheless, the principal ingredient to a successful preparation phase for CLIC,
and the ability to team up within such partner projects, is an increase in resources for CLIC at CERN in next phase of the project.

\subsection{Industrial Capabilities and Status}
The industrial capabilities to build CLIC exist already today. For the most part the requirements are demanding but equivalent to other projects in science and industry. This makes the series production of CLIC components well within reach on the timescale of the construction start-up.

Examples of main elements of the accelerator that are readily available in industry are civil engineering capabilities, production and installation of all infrastructure items, control-systems and the majority of magnets and vacuum systems. Some of these items need adapted designs but the industrial capabilities exists and have been demonstrated on relevant prototypes, not only within the CLIC study, but also related to other projects.

Other items have been subject to extensive development programmes, including industrial prototyping and subsequent tests, e.g. alignment and stabilisation systems, instrumentation and special magnets are prime examples. These items are industrially available but the final design needs to be further optimized for industrial manufacturability and costs. The mechanical, thermal and electrical integration of these parts at large scale in CLIC remains a challenge. 

\begin{figure}[h!]
\centering
\includegraphics[scale=0.4]{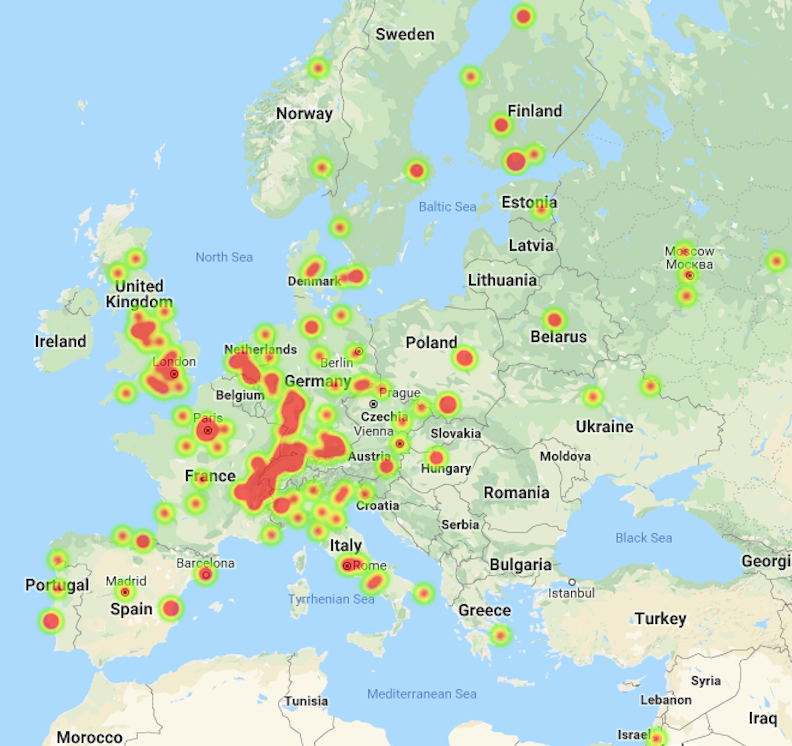}
\caption{Summary of CLIC X-band related collaboration agreements and contracts for the European region, with the colour intensity linked to contract-volume, resources engaged and publications within the subject. (image credit: CLIC)}
\label{fig_contracts}
\end{figure}

The two areas where significant efforts have been devoted to qualify vendors are RF power sources (klystrons, modulators and amplifiers), and X-band components (accelerating structures, power extraction units, RF network and distribution components). In these two areas the requirements of CLIC are demanding and these components are also cost-drivers for the project.
In the first case, dedicated prototyping efforts have been made with the few available companies world-wide producing RF power systems. This work involves design, efficiency optimisation, cost optimisation and pro-actively distributing prototyping contracts when possible to ensure that the most relevant companies can produce RF units for CLIC.
For the X-band components, an extensive design, manufacturing and qualification programme has been carried out, including detailed mechanical, thermal as well as RF power testing of industrially produced prototypes. The dissemination of these results, targeted collaboration with industries and the use of Open Hardware Licensing has facilitated the spread of the X-band technology around the world and the production of small series. While there are now numerous companies in the area of precision mechanics that can provide components according to the CLIC specifications, the procedures and steps needed to reliably produce large quantities still need to be developed. 

As an example of the broadness of the X-band development effort, on the academic, technological and industrial side Fig.~\ref{fig_contracts} show how the CLIC X-band technology studies are distributed across Europe. Similar maps can be made for other regions with clearly identifiable centres and collaborative partners or industrial contractors linked to these centres and European partners.

\printbibliography[heading=subbibintoc]
\endrefsection
\addtocontents{toc}{\vspace{1.2 cm}}
\refsection 
\chapter{Performance}
\label{Chapter:PERF}
\section{Introduction}
\label{sect:PERF_Intro}
CLIC requires excellent beam quality to achieve its luminosity goal. The assumed performances are based on extensive theoretical studies and experimental results. Development of components and their testing and a number of different beam facilities were instrumental for this.
Foremost the SLC~\cite{c:slc} has demonstrated the feasibility of the concept. Many lessons have been learned there and have been integrated into the design of CLIC. Light sources and FELs have advanced the design of low emittance rings and linacs, respectively. The experience gained in b-factories has also been important, and the construction and operation of test facilities specifically dedicated to linear colliders, such as CTF3 and ATF2, has been fundamental.

The SLC reached more than half of the luminosity goal. Its failure to reach two design parameters were the main cause; The repetition rate was 120\,Hz, instead of 180\,Hz and the bunch charge reached $3$--$4\times10^{10}$ particles instead of $7.2\times10^{10}$~\cite{c:Nan}. For otherwise unchanged parameters this would have led to a luminosity reduction by more than a factor of six. However the SLC achieved better than planned beta-functions and emittances at the collision point, mainly in the vertical plane, which allowed it to recover more of the luminosity goal.

In the SLC two main limitations existed for the bunch charge and have been addressed in CLIC, the beam stability in the damping rings and in the Main Linac:
\begin{itemize}
\item
The SLC damping ring showed a sawtooth instability at a bunch charge of $3\times10^{10}$, which made the extracted beam useless. Later upgrades of the damping ring design to lower its impedance managed to reduce the instability but the bunch charged remained below $4.5\times10^{10}$~\cite{c:Nan}. This effect is theoretically understood and the CLIC damping ring design takes it into account and profits from the enormous progress that has been made in the field of low emittance rings also thanks to the development of light sources and the b-factories.
\item
Wakefield effects in the main linac of the SLC amplified incoming beam jitter if the charge reached about $3\times10^{10}$. CLIC avoids such amplification by the use of strong focusing and a choice of bunch charge and length that is consistent with stable beam transport in the presence of the wakefields. This design choice is one of the main ingredients of the CLIC parameter optimisation and includes some margin.
\end{itemize}

While the SLC failed to reach the full bunch charge it provided important physics results.
It also successfully demonstrated a number of important key concepts and managed to address critical performance limitations.
In particular:
\begin{itemize}
\item  The strong beam-beam forces in CLIC leads to shrinking beam sizes during the collision and an increase in luminosity. This effect has been experimentally observed at the SLC, in which a luminosity enhancement by a factor two has been gained. The experiment agrees very well with the simulations~\cite{c:slc_bb}.
\item CLIC aims at 80\% electron polarisation at the IP. The SLC demonstrated a value of 78\%. This gives confidence that the CLIC goal can be met with the new, improved photo cathodes.
\item In CLIC, stable beam transport in the main linac requires BNS damping. This concept has been successfully tested in routine operation at the SLC.
\item Beam jitter is a key issue in CLIC. At the SLC during operation more than 50 feedback loops were developed in iterations and successfully implemented to stabilise the beam. In CLIC similar feedback systems are integrated from the beginning and profit from the SLC experience. Feedback systems were also a key ingredient in the successful operation of the CLIC Test facility CTF3, where further experience has been gained. 
\item Machine-induced background in the detector can limit the physics reach. In the SLC beam tails in the collimation system produced muons that perturbed the experiment. Muon spoilers resolved the problem and space for them is integrated in the CLIC design.
\end{itemize}

The CLIC parameters are more ambitious than the ones of SLC but also the understanding of the relevant physics and the technologies have improved significantly. Important examples are:
\begin{itemize}
\item
As discussed above, the SLC charge limitation is understood and avoided in the CLIC design.
\item
The novel Drive-Beam scheme and the Drive-Beam quality have been demonstrated in CTF3, as well as its use to produce high-power RF pulses and the Two-Beam acceleration concept.
\item
Modern light sources achieve CLIC-level, nano-metre normalised emittances in the vertical plane in routine operation. In the vertical plane they are three orders of magnitude below the SLC level.
\item
CLIC parameters require strong focusing at the IP. This focusing has been demonstrated at two test facilities, i.e. FFTB at SLAC and ATF2 at KEK. The achieved vertical beam sizes were close to the target, see Section~\ref{sect:PERF_BDS}. Also the super B-factory at KEK aims at beta functions that are slightly larger but this in a circular collider where the beam repeatedly passes through the system.
\item
The use of dispersion-free steering to maintain small emittances in a linac has successfully been tested in FACET, a test facility that used a part of the old SLC linac.
\item
Better understanding of the technical limitations and the improved CLIC structure design allow to reach higher gradients than the SLC. High gradient accelerating structures are in routine operation today and prove high gradient and reliability.
\item
The novel precision pre-alignment system of CLIC and sophisticated beam-based alignment and tuning ensure the preservation of the beam quality during transport. The alignment system is based on a concept developed for the LHC interaction regions with improved performance and prototypes have been built. The beam-based alignment has been simulated and experimentally has been verified.
\item Quadrupole jitter has been an important source of beam jitter in the SLC. For CLIC this has been addressed by designing the magnet supports to avoid resonances at low frequencies and by developing an active stabilisation system for the magnets, which demonstrated a reduction of the jitter to the the sub-nm regime.
\item
CLIC requires excellent relative timing at the femtosecond level over the collider complex. Modern FELs have developed the relevant technology.
\item
The impact of time varying magnetic fields are being explored and the necessary mitigation methods are being developed, see Section~\ref{sect:PERF_Stray}.
\item
High availability is key to achieve the luminosity goal. The very reliable routine operation of light sources, FELs, the b-factories and the LHC provide key concepts to address this issue.
\end{itemize}

In conclusion, the CLIC parameters are ambitious but are supported by experience in existing facilities. This permits a confidence that the goals can be met. In the following these performance benchmarks are presented in more detail.

\section{Drive Beam Generation, Power Production and Two-Beam Acceleration in the CLIC Test Facility CTF3}
\label{sect:PERF_DB}

\begin{figure}[ht]
\begin{center}
\includegraphics[width=10cm]{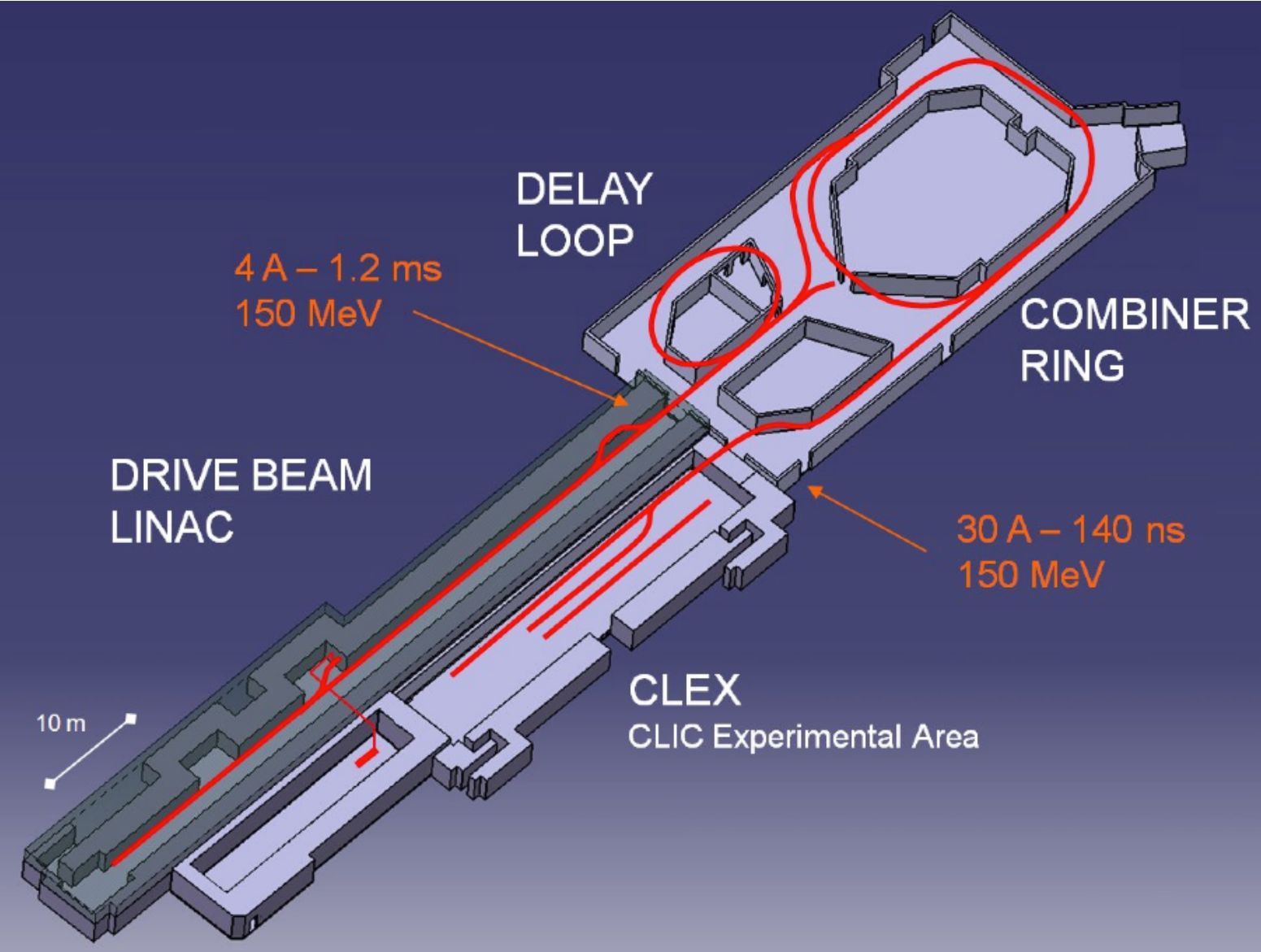}
\caption{Schematic CTF3 layout}
\label{fig:CTF3_Layout}
\end{center}
\end{figure}

The aim of the CLIC Test Facility CTF3 (see Fig.~\ref{fig:CTF3_Layout}), built at CERN by the CLIC International Collaboration, was to prove the main feasibility issues of the Two-Beam acceleration technology~\cite{Geschonke2002}. CTF3 consisted of a 150\,MeV electron linac followed by a 42\,m long Delay Loop (DL) and a 84\,m Combiner Ring (CR). The beam current from the linac was first doubled in the loop and then multiplied by a further factor of four in the ring, by interleaving bunches in transverse RF deflectors. The beam was then sent into the CLIC experimental area (CLEX) where it was decelerated to extract from it RF power at 12\,GHz. Such power was used to accelerate a probe beam, delivered by a 200\,MeV injector (Concept d'Acc\`el\`erateur Lin\'eaire pour Faisceaux d'Electrons Sondes, CALIFES) located in the same area. 

The main issues explored in CTF3 can be divided in two main areas~\cite{Aicheler2012}:

\begin{enumerate}
\item Drive-Beam generation: efficient generation of a high-current electron beam with the proper time structure to generate 12\,GHz RF power. In order to achieve this, CLIC relies on a novel technique: fully-loaded acceleration in normal conducting travelling wave structures followed by beam current and bunch frequency multiplication in a series of delay lines and rings by injection with RF deflectors. CTF3 used such method to produce a 28\,A electron beam with 12\,GHz bunch repetition frequency. The Drive Beam was then sent to the experimental area, CLEX.
\item RF power production and Two-Beam acceleration: in CLIC the needed 12\,GHz RF power is obtained by decelerating the high current Drive Beam in special resonant structures called PETS (Power Extraction and Transfer Structures). The power is then transferred to high gradient accelerating structures, operated at about 100\,MV/m. In the CTF3 experimental area (CLEX), the Drive Beam is decelerated in a string of PETS in the Test Beam Line, (TBL). The Drive Beam can alternatively be sent to another beam line (Two Beam Test Stand, TBTS, renamed later to Test Beam Module, TBM) where one or more PETS powered one or more structures, further accelerating a 200\,MeV electron beam provided by CALIFES.
\end{enumerate}

CTF3 was installed and commissioned in stages starting from 2003. The beam commissioning of the DL was completed in 2006. The CR and the connecting transfer line were installed and put into operation in 2007, while the transfer line to CLEX was installed in 2008. In 2009 this last beam line and the CLEX beam lines, including the CALIFES injector, were commissioned. During the autumn of 2009, recombination with the DL and CR together was achieved, yielding up to 28\,A of beam current. In 2010 the nominal power production from the PETS was obtained, and the first Two-Beam test was performed, reaching a measured gradient of 100\,MV/m. In 2011 a gradient of 145\,MV/m was reached and the PETS On-off mechanism was successfully tested. At the end of 2014 the TBTS was replaced by the Two-Beam Module, TBM, a 2\,m long fully representative unit of the CLIC Main Linac. In 2015 the Drive Beam was decelerated by 50\% of its initial energy in the TBL. Drive Beam stability and the overall performances of the facility were continually improved after the initial commissioning, until the final run in 2016. 

\subsection{Drive Beam Generation: Injector -- Beam Current and Time Structure}

\begin{figure}[ht]
\begin{center}
\includegraphics[width=8cm]{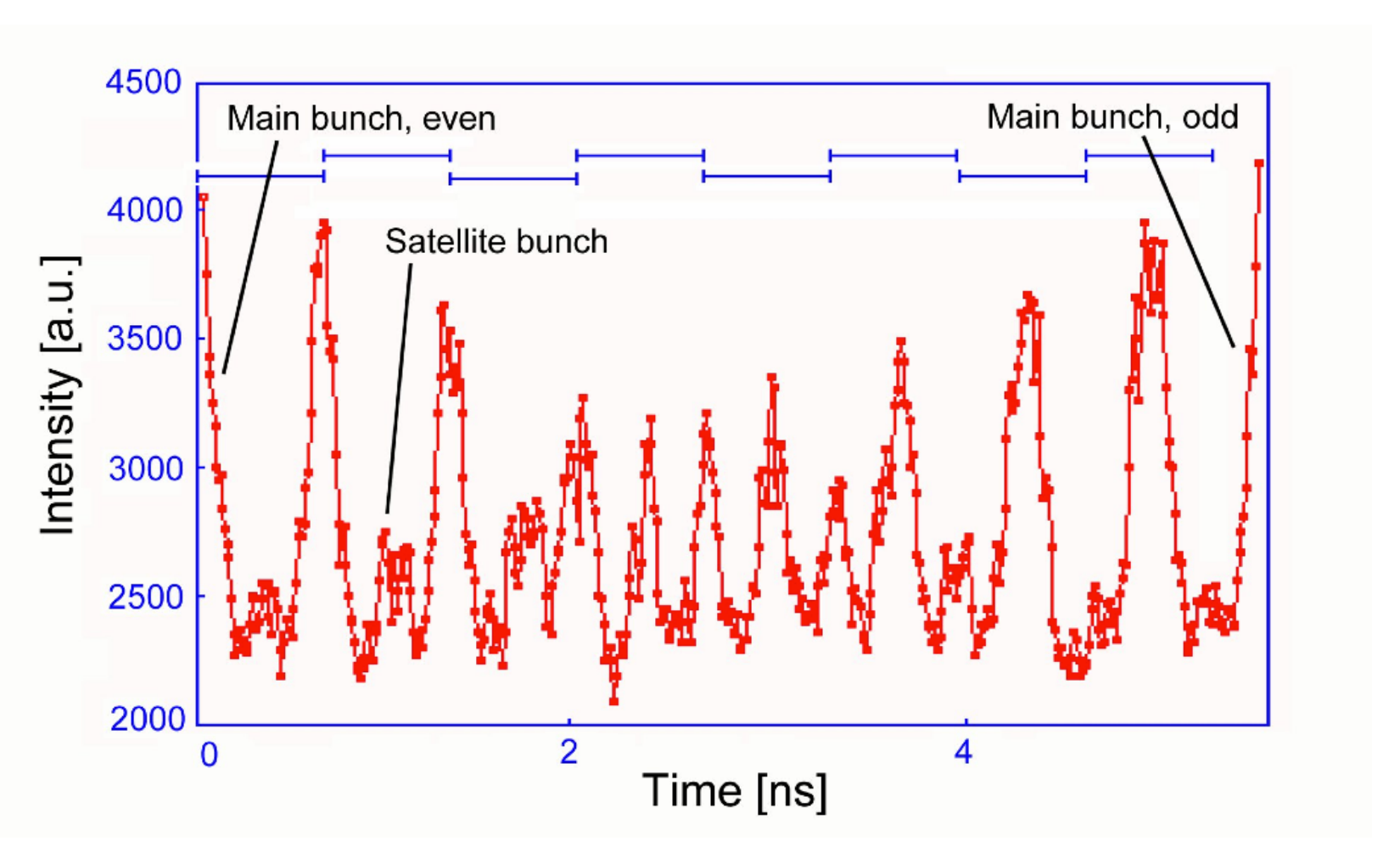}
\caption{Fast bunch phase switch, measured by a streak camera. At the top the 1.5 GHz periods are shown.}
\label{fig:Bunch_Switch}
\end{center}
\end{figure}

The CTF3 Drive-Beam Injector consisted of a high current thermionic gun, three 1.5\,GHz sub-harmonic bunchers (SHBs) and a 3\,GHz system composed of a pre-buncher and a buncher~\cite{Thiery2000}. The SHBs gave the first energy-time modulation to the beam and performed the phase coding by means of fast 180$^{\circ}$ RF phase switches. The 6-cell travelling wave (TW) SHBs had a nominal power of 40\,kW. Downstream, a 3\,GHz single-cell pre-buncher and a TW buncher were installed to create the final bucket structure and accelerate the beam up to 6\,MeV/c. The 2\,cm long pre-buncher nominal power was 100\,kW, while the half-meter long buncher was fed a maximum power of 40\,MW. Exhaustive simulations were performed using PARMELA to guide the optimization of the transverse emittance, the bunch length and the satellite population. The magnetic field distribution was optimized to keep the emittance at the exit of the injector below 50\,$\mu$m, as confirmed by measurements~\cite{Urschuetz2006}. A bunch length of 1\,mm was measured with a streak camera at the end of the linac ~\cite{Dabrowski2010}. Some particles form unwanted satellites in between the 1.5\,GHz main bunches; the measured fraction of the satellites is about 8\%, compared to the design figure of 7\%. Figure~\ref{fig:Bunch_Switch}, a projection of a streak camera image, shows the bunch population vs. time during the 180$^{\circ}$ phase switch. As can be seen from the figure, the measured switch time is less than 6\,ns (eight 1.5\,GHz periods), well below the 10\,ns target~\cite{Urschuetz2006b}.

\subsection{Drive Beam Generation: Linac --- Full Beam-Loading Acceleration}

Overall efficiency is paramount for linear colliders, and a very efficient RF energy transfer to the Drive Beam, obtained by means of full beam-loading operation, is one of the key ingredients in CLIC. The high pulse current in both CLIC and CTF3 (about 4\,A in both cases), accelerated in short travelling-wave RF structures with relatively low gradient, results in an extremely high transfer efficiency. No RF power is transmitted to the load when the beam is present and the resistive losses in the cavity walls are minimal, such that an overall efficiency of about 98\% is calculated in the CLIC case. However, an energy transient is present at the beginning of the pulse, and the first bunches may have twice the energy of the steady-state part; this mode of operation also strongly couples beam current fluctuations to the beam energy. 
 
The 3\,GHz TW accelerating structures designed and built for CTF3~\cite{Jensen:2002} work in the \(2 \pi / 3\) mode, have a length of 1.22\,m and operate at a loaded gradient (nominal current) of 6.5\,MV/m. The large average current also implies that transverse higher order modes (HOMs) must be damped in order to prevent beam instability and control emittance growth. A Slotted Iris - Constant Aperture structure (SICA), in which irises are radially slotted to guide dipole and quadrupole modes into SiC loads, was designed for the purpose. The selection of the damped modes is obtained through their field distribution, strongly damping the HOMs (Q typically below 20), while monopole modes are not influenced. In addition HOM detuning along the structure (by nose cones of variable geometry) is used; this improves their suppression and modulates the group velocity to control the gradient profile. The aperture is therefore kept constant along the structure, reducing short range wakefields. The RF is supplied by klystrons with power ranging from 35\,MW to 45\,MW, compressed by a factor two to provide 1.3 $\mu$s pulses with over 30\,MW at each structure input. The pulse compression system uses a programmed phase ramp to provide a constant RF power.

Beam commissioning started in June 2003. The design beam current and pulse length were rapidly reached. The beam was remarkably stable and no sign of beam break-up was observed at high current, thus proving for the first time operation under full beam loading~\cite{Corsini2004}. The measured normalized emittance at the end of the CTF3 linac was routinely about \(\epsilon_{x,y}\simeq 50 \mu\)m, confirming negligible wakefield effects as predicted by simulations. The energy spread of the initial beam transient ($\simeq$ 100 ns) could also be reduced to a few percent by partial RF filling of the structures at beam injection. The observation of the RF signals at the structure output couplers was particularly useful, allowing to easily adjust the beam-to-RF phase by maximizing the beam loading. The acceleration efficiency was demonstrated in a dedicated experiment~\cite{Urschuetz2006a}. 

\begin{figure}[ht]
\begin{center}
\includegraphics[width=7cm]{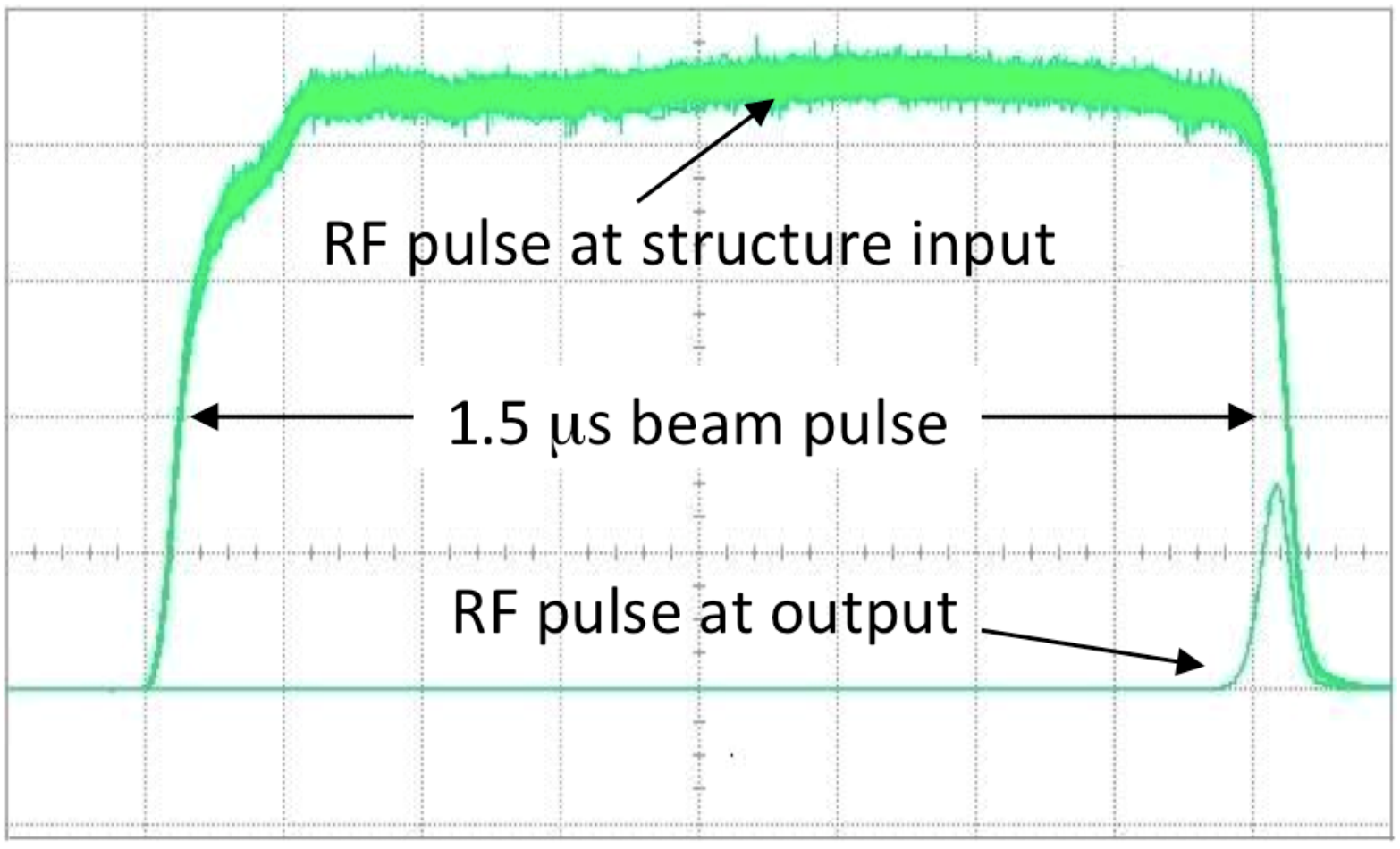}
\caption{RF power measured at the accelerating structure input and output with beam.}
\label{fig:Beam_Loading}
\end{center}
\end{figure}

After careful calibration of beam current and RF power measurements, the beam energy gain was calculated and compared to spectrometer measurements. Figure~\ref{fig:Beam_Loading} shows an example of the RF pulse measured at the structure input and output, showing that the RF power is almost fully absorbed by the beam. The measurements were in excellent agreement with the theoretical energy gain. Including the resistive losses, the obtained RF-to-beam transfer efficiency was 95.3\%. CTF3 was routinely operated over several years with fully loaded structures, successfully proving the stable, highly efficient acceleration of the Drive Beam.

\subsection{Drive Beam Generation: Delay Loop and CR -- Bunch Combination}

Beam recombination in CTF3 is done in two stages. First, using the Delay Loop (DL), a 1120\,ns long bunch train with a current of 4\,A is converted into 4 pulses of 140\,ns and 7.5\,A (not counting the charge contained in satellite bunches). Later, the pulses are interleaved in the Combiner Ring (CR) to produce a single 140\,ns long pulse with a maximum current of about 30\,A.  The first RF deflector, operating at 1.5\,GHz, sends odd and even phase-coded sub-pulses either straight to the CR or into the DL, whose length is equal to the sub-pulse length. The sub-pulses circulating in the DL come back in the deflector at half a wavelength distance, and their orbits are merged with the following ones to obtain 140\,ns long pulses with twice the initial current and twice the bunch repetition frequency. The pulses are combined again in the CR. A pair of RF deflectors is employed to create a time-dependent closed bump at injection, used to interleave bunches. 

The combination process must preserve transverse and longitudinal beam emittances: isochronous lattices, smooth linear optics, low impedance vacuum chambers and diagnostics, HOM free RF active elements are all needed to accomplish this task. 
A short bunch length is fundamental for efficient RF power production in the PETS. Bunch length preservation requires the use of isochronous optics (which implies $R_{56}=0$) in the DL, the CR and the transfer line connecting them. The isochronicity requirement is $| R_{56}| \le \pm1$ cm. The DL and CR are based on the use of three-dipole isochronous cells with three independent quadrupole families, whose tunability range fits well the requirements. Sextupoles can be used to control the second-order matrix term $T_{566}$. Bunch length control to down to less than 1\,mm R.M.S. was shown after the linac. In CLEX a bunch length of 2\,mm R.M.S. was estimated from RF power production and confirmed by direct streak camera measurements. Such length is consistent with required isochronicity conditions and entirely sufficient for CTF3 operation, avoiding also coherent synchrotron radiation, which would affect shorter bunches. 
Damping and detuning is used in the RF deflectors of the ring in order to minimize wakefields in the vertical plane~\cite{Alesini2011}. The lowest order horizontal dipole mode is the operational one and cannot be damped or detuned. However, the fill-time of the travelling wave deflectors is short enough to avoid turn-by-turn direct build-up. In order to avoid any residual amplification of the orbit errors by wakefields, the fractional tune of the CR is set to be about 0.6 in both planes. Also, the $\beta$-function in the deflectors is kept small to minimize amplification. The ring length must be
\( (N \pm 1/N_f) \lambda_{RF}\), where $N$ is an integer number, $N_f$ the combination factor (here 4), and $\lambda_{RF}$ is the RF wavelength. The fractional part $\lambda_{RF}/N_f$ can be determined precisely from a Fourier transform of the dedicated beam phase monitor signal, and when needed the ring length can be adjusted using a wiggler.

In the last experimental runs, a large fraction of beam time was dedicated to improvements of the Drive Beam performance, especially control of emittance growth, availability and stability. Apart from enabling a better exploitation of the beam by the users, this was an integral part of the CTF3 experimental program, aimed at demonstrating a beam quality close to the one required for CLIC. Emittance preservation requires good control of the optics, a very good closure of the DL and CR orbits, proper matching from the linac and control of spurious dispersion. The CTF3 Drive Beam has a R.M.S. energy spread of about 0.6\%, and the isochronous optics in DL, CR and transfer lines are strongly focusing. This leads to a large non-linear dispersion, which is the main source of emittance growth. The main contribution was coming from the DL, whose optics is constrained by building space limitations. A new more forgiving DL optics was developed and deployed, and tools to precisely measure Twiss parameters and dispersion in the different beam lines were put in place. Dispersion Free Steering and Dispersion Target Steering procedures were implemented and applied. While the beam combined in the CR met the CLIC emittance requirements (<150 $\mu$m) in both planes, the minimum horizontal emittance for the factor 8 recombined Drive Beam was about 250 $\mu$m (100 $\mu$m in vertical). Besides demonstrating the feasibility of the CLIC bunch combination principle, CTF3 has allowed the development of an optimized setting-up procedure of such a process, validating the special diagnostics needed~\cite{Corsini2013,Dabrowski2010}. 

\subsection{Drive Beam Generation: Beam Stability Issues and Phase Feed-Forward}

In CLIC, the Two-Beam Acceleration scheme puts tight constraints on the Drive-Beam current, energy and phase stability. In particular, both bunch charge and phase jitter contribute quadratically to the luminosity loss~\cite{Schulte2010}; a maximum variation of 0.75*10$^{-3}$ for the Drive-Beam current and about 2$^{\circ}$ at 12\,GHz for the Drive-Beam bunch phase after combination are allowed, for a maximum of 1\% to the luminosity loss per parameter and assuming a feed-forward system -- discussed later -- capable of reducing the phase jitter to 0.2$^{\circ}$ at 12\,GHz. During the first years of operation CTF3 suffered from relatively large beam jitters and drifts; dedicated studies discovered most of the sources, which were either removed or corrected by feedback systems. Dedicated tests aimed at demonstrating performances close to the CLIC requirements (including a feed-forward experiment, see later for a full description) were added to the CTF3 experimental program in its final years. The CLIC tolerances on the Drive-Beam Linac RF, for instance, were verified for the CTF3 klystrons, measuring the short-term RF stability over 500 consecutive RF pulses ($\simeq$ 10\,min).  The mean pulse-to-pulse phase jitter measured with respect to an external reference was 0.035$^{\circ}$ (3 GHz) and the relative pulse-to-pulse amplitude jitter has been 0.21\%, to be compared with the CLIC requirements of 0.05$^{\circ}$ R.M.S. phase jitter and 0.2\% R.M.S. amplitude jitter ~\cite{Sterbini2010}. 
The pulse-to-pulse current variations in the CTF3 linac were measured using inductive beam position monitors. The initial stability (\(\Delta I / I = 2 \times 10^{-3} \) ) was improved by better stabilizing the gun heater power supply and by a feedback, to obtain \(\Delta I / I =0.54 \times 10^{-3} \) ~\cite{Sterbini2010}, better than the required current stability for CLIC of \(\Delta I / I =0.75 \times 10^{-3} \). After further improvements, in 2016 the R.M.S. current jitter at the end of the Drive-Beam Linac was routinely better then \(2 \times 10^{-4} \), corresponding to the electronic noise floor of the BPMs (the observed jitter is the same with and without beam). Such performance is currently achieved for long periods, even tens of hours~\cite{Skowronski2016}. The stability of the combined beam current in CLEX, at the end of the experimental lines, was also largely improved and \(3 \times 10^{-3} \) R.M.S. was measured for periods longer then 5 hours~\cite{Malina2016}.
 
In CLIC a Drive-Beam phase feed-forward system is required to achieve a timing stability of 50\,fs R.M.S., or equivalently a phase stability (jitter) of 0.2\,degrees at 12\,GHz. This system poses a significant challenge in terms of bandwidth, resolution and latency and therefore a prototype of the system was designed, installed and tested in  CTF3. After one year of experience a phase jitter of $0.28 \pm 0.02$ was demonstrated, very close to the CLIC specifications~\cite{Roberts2016}. With additional fine-tuning in 2016, the final year of operation at CTF3, it was possible to further reduce the achieved phase jitter to below the CLIC requirement of 0.2$^{\circ}$ and to keep it at such level on timescales of about 10\,minutes~\cite{Roberts2018}.

\subsection{Two-Beam Acceleration: Power Generation, PETS On-Off and Deceleration}

In CTF3, PETS prototypes were tested with beam in TBL and in TBTS (and later TBM). The prototypes tested in both lines were very similar, differing mainly in length and some construction detail. In both cases the nominal CLIC parameters for power and pulse length were reached and exceeded. In particular during the 2010 run, RF power levels of about 300\,MW were reached at the nominal pulse length of 240\,ns, well above the nominal value for CLIC, 135\,MW~\cite{Skowronski2011}. Another milestone was the validation of the PETS On-Off concept. The PETS On-Off mechanism is required in CLIC in order to be able to switch on and off individual PETS whenever localized breakdowns threaten the normal machine operation. The system should also provide a gradual ramp-up of the generated power in order to reprocess either the main accelerating structure and/or the PETS itself. A prototype of the CLIC PETS On-Off mechanism, a compact external extension to PETS, consisting of two high power variable RF reflectors, was developed, manufactured and extensively tested with beam in the TBTS. The switching off of the PETS power production has been demonstrated with different beam and RF settings, up to 16\,A and 200\,MW~\cite{Syratchev2012}. The model describing the RF behaviour agreed with the measurements under all conditions. When the PETS is switched off, the accelerating structure sees less than 1\% of the extracted power, basically making breakdown events impossible. The likelihood of a breakdown in PETS is also reduced by a factor $10^2-10^3$ due to the  power attenuation to $\sim$ 25\% of the nominal value. 

The stability of the Drive Beam during deceleration was studied in the Test Beam Line (TBL), consisting of a FODO lattice with 14 consecutive PETS installed in the drift spaces. High precision BPMs and quadrupole active movers are used for precise beam alignment. In the TBL a 25\,A Drive Beam was decelerated from it initial energy, 135\,MeV, down to a minimum of 67\,MeV, reaching the 50\% deceleration milestone which was one of the initial goals of CTF3. On average the produced power was 90\,MW per PETS. The total peak RF power produced in less then 20\,m deceleration line reached 1.3\,GW. The beam energy loss and its energy spread was measured in a time-resolved spectrometer, and agreed very well with simulations. The dynamics of the Drive Beam while undergoing deceleration was studied in detail, and both transverse and longitudinal parameters were in agreement with the simulations~\cite{Lillestoel2010}. 

\subsection{Two-Beam Acceleration: Two-Beam Test Stand and Two-Beam Module}

The key purpose of CTF3 was to demonstrate the CLIC Two-Beam Acceleration scheme, including efficient power transfer to high-gradient structures and acceleration of a Probe Beam. The Probe Beam was provided by the 24\,m long injector linac, CALIFES~\cite{Peauger2008}, located in CLEX and delivering single bunches or bunch trains at 1.5\,GHz bunch repetition rate and energies up to 200\,MeV. The beam was generated and accelerated to $\sim$ 5\,MeV/c in a photo-injector and further accelerated in three 3\,GHz accelerating structures recuperated from the LEP Injector Linac (LIL). The accelerating structures and the photo-injector were powered by a single 3\,GHz klystron delivering 4\,MW RF pulses during 5.5\,$\mu$s to an RF pulse compressor. CALIFES was usually operated with bunch charges of $\sim$\,0.1\,nC and a normalized R.M.S. emittance of 10\,$\mu$m. 

\begin{figure}[ht]
\begin{center}
\includegraphics[width=7cm]{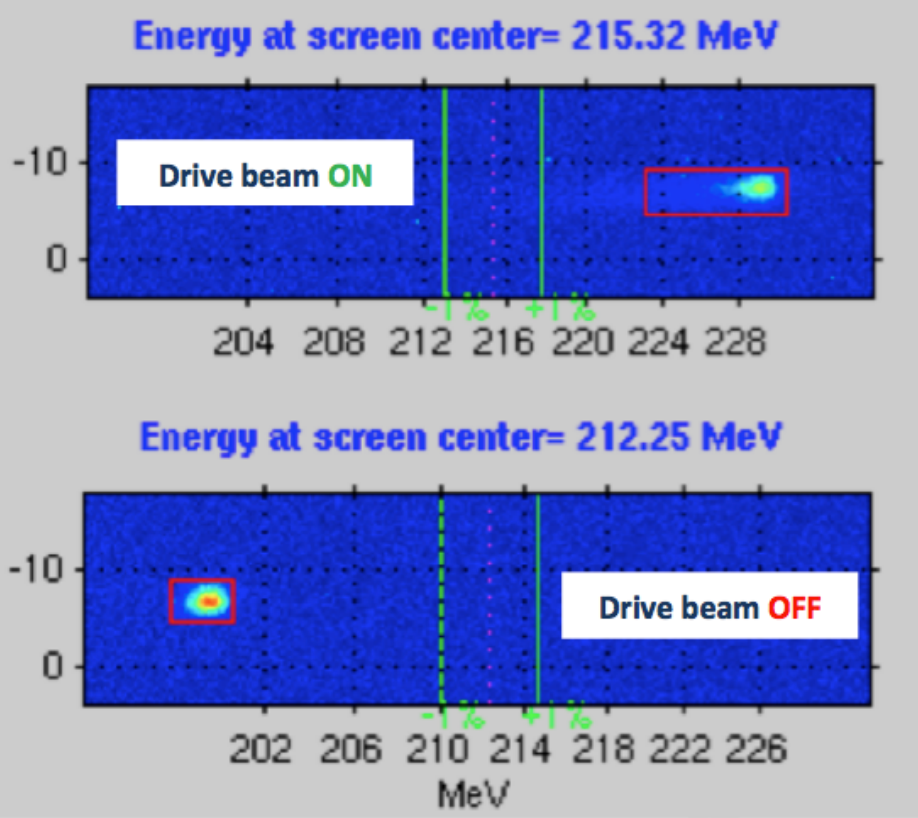}
\caption{Probe Beam observed in the TBTS spectrometer screen with the 12\,GHz RF power from the Drive Beam on (top) and off (bottom). The energy gain is about 32\,MeV, corresponding to a gradient of 145\,MV/m.}
\label{fig:Two_Beam}
\end{center}
\end{figure}

The Probe-Beam energy was measured in a spectrometer as a function of the Probe Beam 3\,GHz RF phase, phase-locked to the laser pulse timing. A phase scan was then used to adjust the relative timing between Probe and Drive Beam for maximum acceleration. The nominal CLIC accelerating gradient of 100\,MV/m corresponds to an energy gain of \(\Delta E = 21\)\,MeV. Energy gains of up to \(\Delta E = 32\)\,MeV were achieved with relatively low breakdown rates (a few 10$^{-3}$), while higher gradients, up to 165\,MV/m were achieved with higher rates~\cite{Skowronski2011}. Figure~\ref{fig:Two_Beam} shows an example of \(\Delta E = 32\)\,MeV Probe-Beam Acceleration measured on the spectrometer screen, corresponding to an accelerating gradient of 145\,MV/m. 

\begin{figure}[ht]
\begin{center}
\includegraphics[width=10cm]{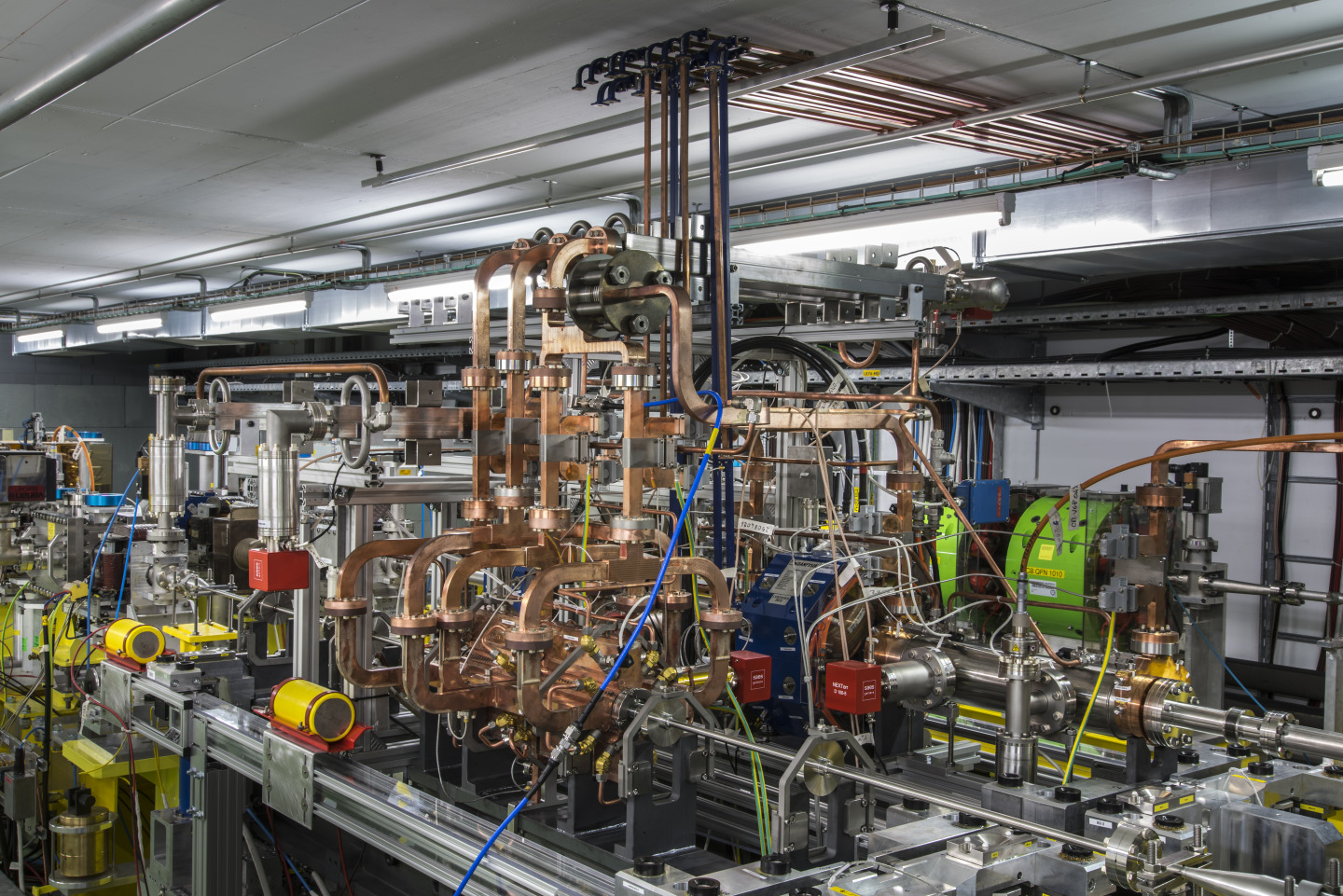}
\caption{The Two-Beam Module, TBM, a 2\,m long fully representative unit of the CLIC main linac, installed in the CLEX area of CTF3.}
\label{fig:Two_Beam_Module}
\end{center}
\end{figure}

The effect of breakdown kicks on the Probe Beam was also extensively studied~\cite{Palaia2013}. At the end of 2014 the TBTS was replaced by the Two-Beam Module, TBM, a 2\,m long fully representative unit of the CLIC Main Linac, including an active alignment system (Fig.~\ref{fig:Two_Beam_Module}). The nominal CLIC gradient/pulse length was again established, and extensive tests of the module functionality, including precision and reliability studies of the active alignment system, and measurements of transverse wakefield effects were carried out. The experimental results are still undergoing a full analysis, but are so far consistent with specifications and simulations.

\subsection{Conclusions on CTF3}
The CLIC Test Facility CTF3 provided a rich experimental programme, addressing various aspects of the accelerator technology needed for CLIC and solving the vast majority of issues related to Drive-Beam generation, power production and Two-Beam Acceleration. In particular, high-gradient acceleration beyond 10\,MV/m using X-band room temperature is now well established, as well as the production and use of a high-current Drive Beam as an efficient and reliable source of X-band RF power in the range of hundreds of MWs. CTF3 successfully completed its experimental program in December 2016 as planned, and stopped operation. 

\section{Drive-Beam Injector Performance}
\label{sect:PERF_DBI}
\subsection{Drive-Beam Electron Source}

\begin{figure}[!htb]
\begin{center}
\includegraphics[width = 0.95 \columnwidth]{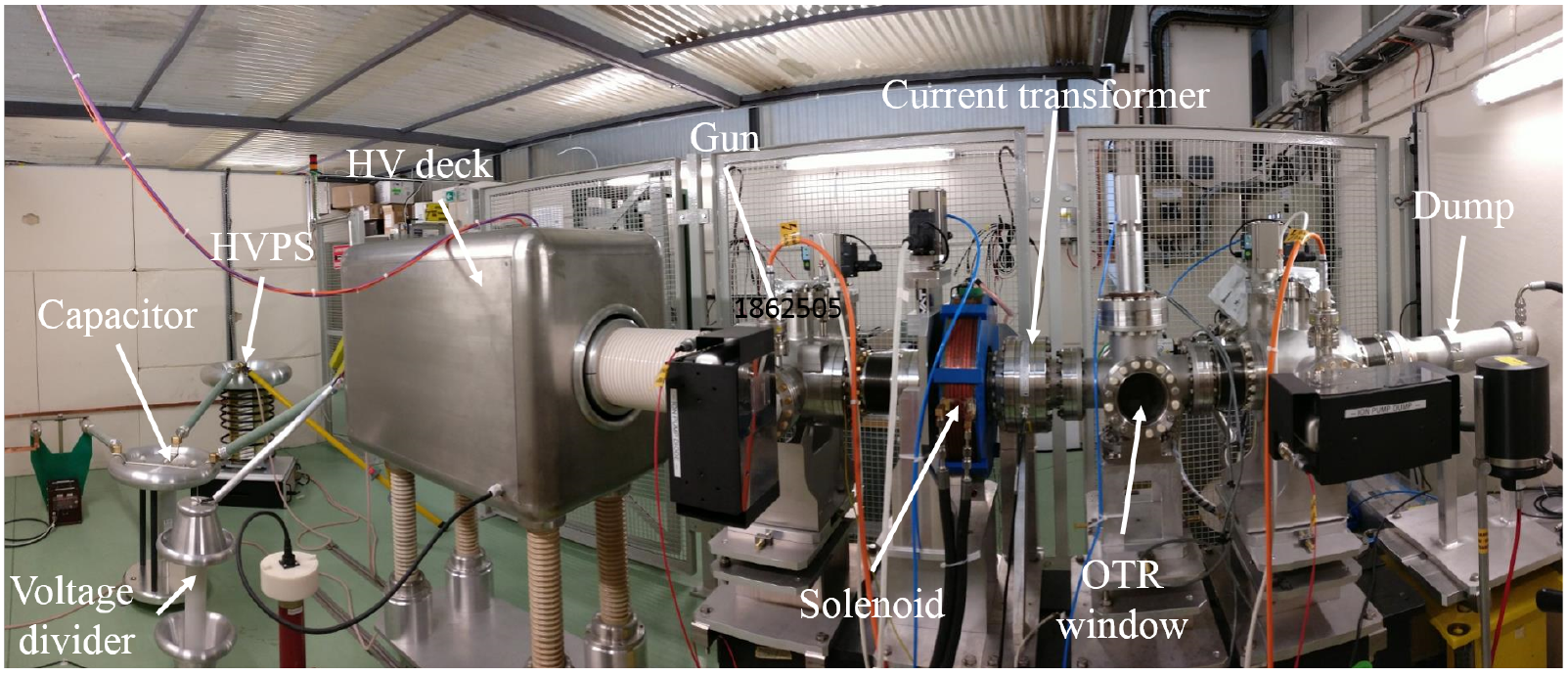}
\caption{Drive-Beam Electron Source test stand at CERN. The cathode is installed within a ceramic insulator while the anode is on ground potential. All drive electronics for the cathode is installed on a high voltage deck at a potential of 140\,kV. The beam can be diagnosed to measure its current, shape and emittance.}
\label{fig:DBI_Figure1}
\end{center}
\end{figure}

The CLIC Drive-Beam electron source has been designed taking into account the results from CTF3. The source is a scaled version of the CTF3 gun originally developed and contributed by SLAC and LAL. The current of 4.2\,A is very similar to the one used in CTF3 but the pulse duration is 140\,s and therefore a 100 times longer than in CTF3. The second very challenging specification is the required beam stability. The charge along the long pulse and pulse to pulse needs to be stable to 0.1\%. An electron gun based on a commercial thermionic cathode with a voltage of 140\,kV has been developed and a prototype was built~\cite{Doebert2014}. The cathode has a grid which allows beam current regulations. A test stand has been built at CERN to demonstrate the performance of the Drive-Beam electron source. The experimental setup can be seen in Fig.~\ref{fig:DBI_Figure1}. The source produced pulses of 140\,s duration with a current of 4.2\,A. The stability of the beam has been measured and a shot to shot reproducibility of 0.2\% has been achieved already. The emittance of the beam has been estimated to be below 10\,mm\,mrad by comparison with PIC simulations. 
Since the stored energy needed to use a DC high voltage with a 10$^{-4}$ stability is quite substantial, a pulsed high-voltage modulator based on a Marx-topology has been developed in collaboration with CEA ~\cite{Cassany2018}. The modulator demonstrated a stability of 0.1\%. 
The experiments with this source are ongoing, in particular to check long term stability and 50\,Hz operation, but the essential beam parameters have been already shown successfully.  Therefore the concept for the critical Drive-Beam electron source as described in the CDR can be considered as validated.

\subsection{Sub-harmonic Bunching System}

Apart from the source another critical item of the Drive-Beam Injector, the sub-harmonic buncher, has been investigated experimentally. The Drive-Beam Injector uses three sub-harmonic bunchers at 499.75\,MHz to pre-bunch the beam out of the DC source and switch the phase of the bunches by 180\,deg every 240\,ns to enable the beam for later recombination. The buncher and its RF power supply need therefore to have a bandwidth of about 10\%. In order to ease the requirements for the RF power source a buncher was developed which minimizes the power needs. The solution found is a two cell, backward-wave, travelling-wave structure. The detailed design of the buncher cavities can be found in~\cite{Shaker2013}. The innovative prototype was built out of aluminium because of weight and cost considerations. The cells were joined by electron beam welding. 
In parallel a solid state amplifier with an output power of 20\,kW was developed by RRCAT in India as a prototype for the necessary RF source. The amplifier architecture could be scaled up the 100\,kW needed for the third buncher in the CLIC injector scheme. The amplifier and the buncher were tested together at full output power, pulse length and duty factor successfully fulfilling as well the stability requirements. Finally a phase switching test has been performed checking the speed of the 180$^\circ$ phase shift. The switching time was limited somewhat by the amplifier to about 100\,ns compared to the required 15\,ns. The problem was investigated and could be mitigated by improving the amplifier circuit. The work to validate this concept is still ongoing.

\begin{figure}[!htb]
\begin{center}
\includegraphics[width = 0.65 \columnwidth]{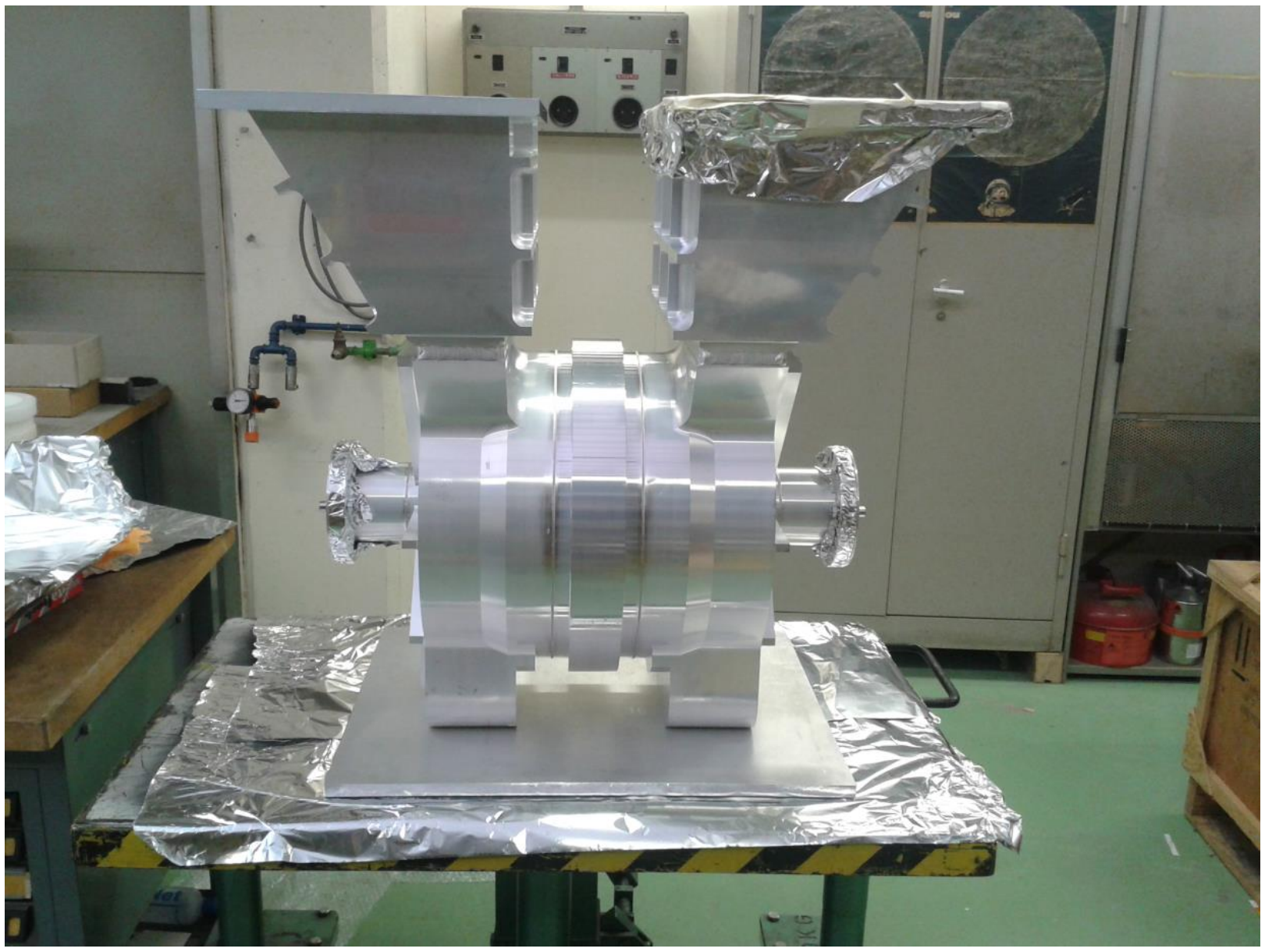}
\caption{Prototype of a 500 MHz sub-harmonic buncher for the CLIC Drive-Beam Injector. The cavity was made of aluminium.}
\label{fig:DBI_Figure2}
\end{center}
\end{figure}

\section{BDS Beam Dynamics, Experimental Studies in ATF2 and FFTB}
\label{sect:PERF_BDS}
\subsection{Achievements and Plans}
The Final Focus Systems (FFS) envisaged for future linear
colliders have always been regarded as major challenges
requiring experimental demonstrations.
Table~\ref{ffsexps} shows the main FFS parameters of the
different experimental projects~\cite{Plassard2018a,Balakin1995,White2014,Kuroda2016,Okugi2016,Tomas2009,Thrane2017}.

The first FFS experiment was FFTB~\cite{Balakin1995}, which used
a traditional chromaticity correction scheme. The achieved vertical
beam size was more than 50\% larger than the design. It was suspected
that this deviation was due to orbit jitter.

ATF2 was conceived
to demonstrate the compact chromaticity correction scheme
presented in~\cite{Raimondi2001} and it still operates with extended goals
to demonstrate CLIC chromaticity levels. It has occassionally reached
41\,nm vertical beam size, just 10\% above design, but using a relaxed optics
with ten times larger $\beta^*_x$. According to simulations the need to
enlarge $\beta^*_x$ may come from poor orbit control or magnetic aberrations~\cite{Patecki2016}.

The ATF2 Ultra-low $\beta^*$ proposal aims at reducing beam size below 30\,nm achieving similar chromaticity levels as the CLIC FFS. Two octupoles manufactured by CERN have been installed in order to cancel high order aberrations. Another option to increase the ATF2 chromaticity without reducing the IP beam size is to increase $L^*$. This modification could be considered once ATF2 has reached its main goals. Actually further R\&D proposals exist related to CLIC as the
study of crystal focusing~\cite{Cilento2017}.

SuperKEKB is an e$^+$e$^-$ circular collider aiming at reaching 60\,nm beam size at the IP with a traditional chromaticity correction system.
An alternative FFS design using the traditional chromaticity correction was
proposed for CLIC~\cite{Garcia2014}.
A pushed optics version of SuperKEKB FFS has been presented in~\cite{Thrane2017} in order to approach its chromaticity levels to CLIC and an IP beam size about 40\,nm. An experimental program
to test  CLIC FFS in SuperKEKB is being considered. 

\begin{table} [b]
\caption{FFS parameters for CLIC and related experimental projects~\cite{Plassard2018a,Balakin1995,White2014,Kuroda2016,Okugi2016,Tomas2009,Thrane2017}.}\label{ffsexps}
\centering
  \begin{tabular}{|lc|r|r|r|r|r|r|}\hline
    &      & CLIC &FFTB & \multicolumn{3}{c|}{ATF2} & SKEKB \\\cline{5-7}
   &      & 3 TeV &  & Nom. & UL $\beta^*$ & Long $L^*$ &  Low $\beta^*$\\ \hline
$L^*$       & [m]  & 6&0.4  &  1   &  1      &  2    & 0.9\\ 
$\beta^*_y$ & [mm] & 0.12&0.1  &  0.1 &  0.025  &  0.1 & 0.09\\
$\xi_y\approx L^*/\beta^*_y$ & [$10^3$]&50&4&10&40&20   & 10    \\ 
$\epsilon_y$& [pm] &0.003 & 22  & 12 & 12 & 12 & 13\\
$\sigma_y\ design$   &[nm]&1 &52&37 &23 & 37 & 34\\
$\sigma_y\ measured$ &[nm]&-&70$\pm$6& 41$\pm$2 & - & - & -\\
\hline
  \end{tabular}

\end{table}

\subsection{Ultra-low $\beta^*$ with octupoles in ATF2}
Reducing the $\beta$-functions at IP to such  small values reported in Tab.~\ref{ffsexps} imposes tight constraints on the machine imperfections. A clear example was the field quality of the quadrupole magnets installed in 2009 in the ATF2 beamline. Some of the measured~\cite{Masuzawa2011} multipolar components present in the Final Doublet magnets exceed the tolerances of both the ATF2 Nominal and Ultra-low $\beta^*$ lattices, as shown in~\cite{Okugi2014,MarinLacoma2012,Okugi2016}. Being the Ultra-low $\beta^*$ case the most severe, as expected. Possible solutions can be found by decreasing the horizontal $\beta$-function along the beamline, so the impact of the multipolar components is reduced accordingly. However this option deviates from the beam size aspect ratio required to test the FFS of the future linear colliders. A pair of octupole magnets would be required to effectively test the pushed optics at ATF2 without altering the $\beta^*$, as shown in~\cite{Marin2014}. Additionally it was found that the effect of the fringe fields of the FD quadrupoles would also preclude to reach a $\sigma_y^*\leq 30$\,nm. Nevertheless this detrimental effect could also be compensated thanks to the octupole magnets, as shown in~\cite{Patecki2014}. Finally the pair of octupole magnets would provide additional knobs to carry out the tuning of ATF2 under machine imperfections, as considered in~\cite{Marin2013}. Therefore a pair of octupole magnets were fabricated~\cite{Modena2016} at CERN following the specifications given in~\cite{Marin2014a} and installed in ATF2 by 2016. 
\begin{figure}[!htb]
\begin{center}
\includegraphics[width = 0.4 \columnwidth]{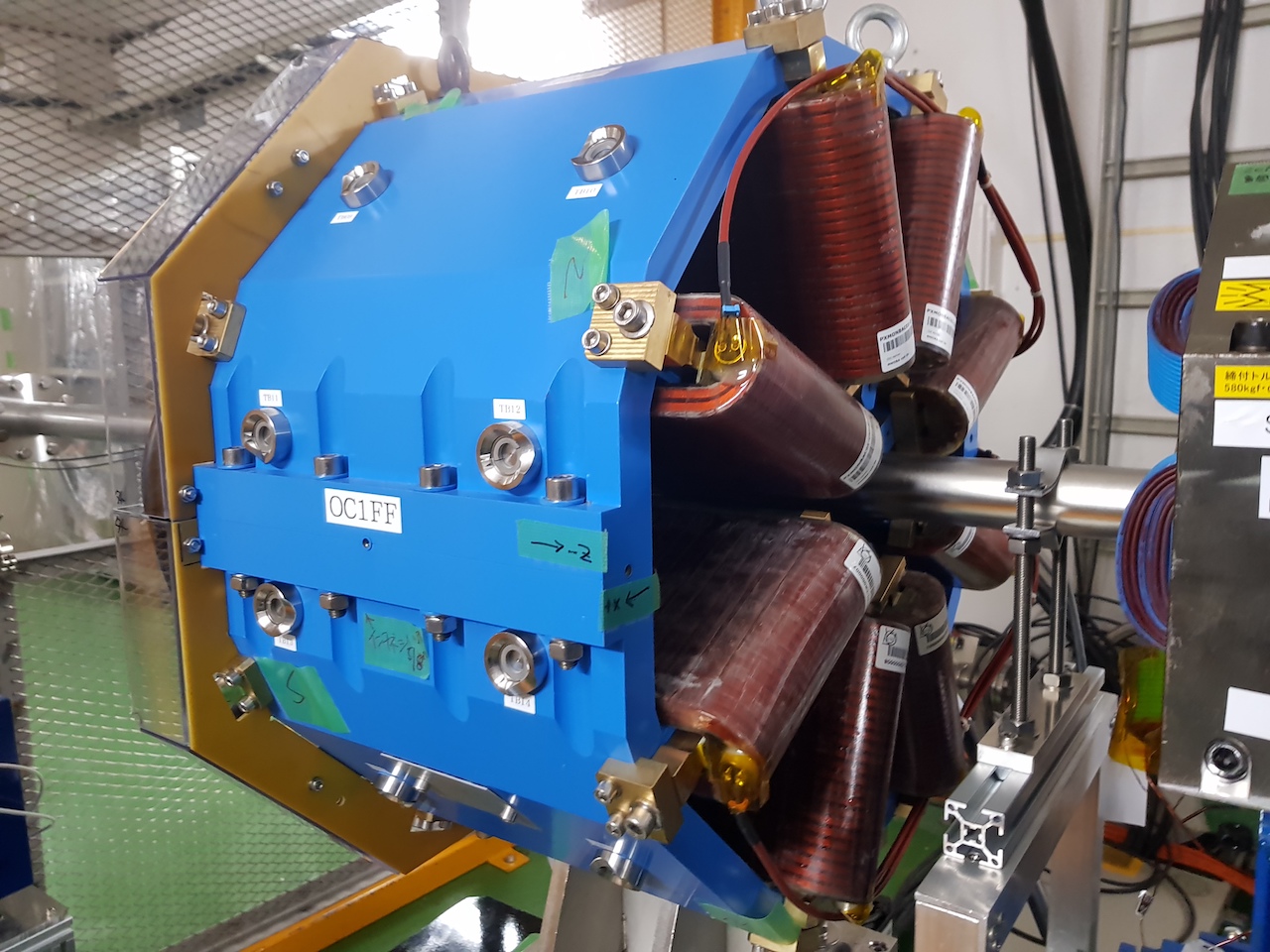}
\includegraphics[width = 0.4 \columnwidth]{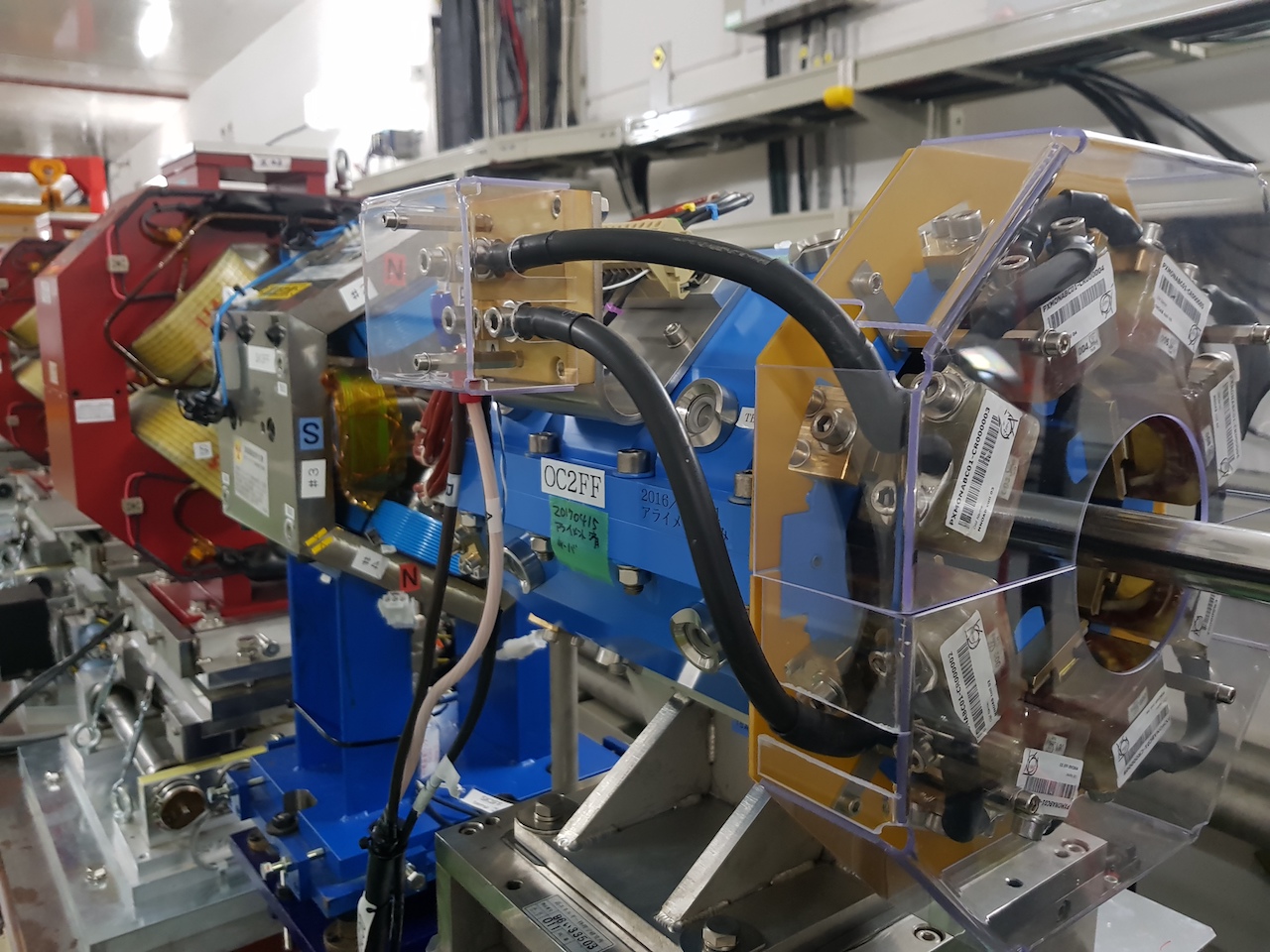}
\caption{Octupole magnets OCT1 (left) and OCT2 (right) installed at ATF2 beamline.}
\label{fig:octupoles}
\end{center}
\end{figure}
Presently the octupole magnets are in operation and provide additional tuning knobs for the ATF2 ultra-low $\beta^*$ tuning attempts carried out in December 2017 and February 2018, see more details in~\cite{Plassard2018}.

\subsection{Ground Motion in ATF2}
Ground motion (GM) is not a limiting factor in the ATF2 performance. We use ATF2 as a test bed for CLIC GM concepts. 14 G\"uralp 6T seismometers have been installed at ATF2. They have been used to regularly measure GM and magnet vibrations and to identify their main sources~\cite{Pfingstner2014a}.
Recently, these seismometers in combination with a transverse kicker have been used to demonstrate for the first time an orbit feed-forward system based on quadrupole motion measurements~\cite{Bett2018}. About 80\% of the quadrupole induced jitter was successfully canceled with the transverse kicker. 
Future studies will aim at using more than one kicker to further increase the efficiency of the feed-forward. 

\subsection{Wakefields}
ATF2 IP vertical beam size was observed to increase with bunch intensity by about a factor 2 in~\cite{Okugi2016,Kuroda2016,Patecki2016}.
This could partially be explained by the combination 
of wakefields and a large orbit jitter of the order of 0.4$\sigma_y$. Many wakefield sources were removed from the beamline, including cavity BPMs, observing a reduction of the beam size growth with intensity. However $\sigma_y=$41\,nm has not been reached again in ATF2 and the intensity dependence has not been evaluated in the same conditions.  

An attempt to study this intensity dependent effect has been done using the beam orbit~\cite{Korysko2018}. However 36 of the 46 BPMs in the machine (striplines and cavity BPMs) do not provide the required precision for such studies.
A Wakefield Free Steering (WFS) correction scheme has been tested on the machine in February 2018 with promising results. Its goal is to correct the orbit difference due to the charge using the steering magnets in ATF2. Further work will be done to improve the technique. WFS could be considered as integral part of the tuning procedure.

\section{Low Emittance Preservation (FACET / Elettra)}
\label{sect:PERF_FACET}
The proposed high luminosity of CLIC (and ILC) requires normalised vertical emittances of the order of few tens of nanometers at the interaction point in order to reach the target luminosity. Because of this ultra-low emittances, the preservation of the emittance throughout the entire accelerator from the damping rings to the interaction point has been regarded as a potential feasibility issue for the entire project. The main sources of emittance growths are misalignment of the accelerator components, which cause transverse deflections, introduce dispersions and might create short and long wakefields in the accelerator structures. The long train of bunches might suffer from long-range wakefields, which induce beam breakup and cause beam losses. Spurious dispersion in the quadrupoles causes emittance dilution which scales as the squared of the absolute misalignment of the beam BPMs, and the emittance dilution due to single-bunch transverse wakefields which scales as the square of the accelerator structure offset \cite{Raubenheimer2000}.  Other sources of the emittance growth include bending magnets where consistent emission of synchrotron radiation occurs, geometric and chromatic aberrations in nonlinear magnets, correlated energy spread due to longitudinal short-range wakefield effects in the accelerator structures, etc.
 
Most of these effects are excited by transverse installation misalignments of the elements, and therefore a tight pre-alignment of the components is required; in the case of the CLIC Main Linacs, this means one order of magnitude better than the available off-the-shelf solutions. The CLIC-motivated, EU-funded, PACMAN project \cite{Pacman2014} was launched to address this issue, and within its duration successfully reached the required precision and accuracy. Yet, excellent pre-alignment is not sufficient to preserve the ultra-low emittances, and dedicated beam-based techniques must be put in place to further mitigate the impact of the imperfections. In CLIC, this problem is even magnified by the sheer number of accelerator components: the Main Linacs of CLIC, at 380\,GeV centre-of-mass energy, are about 3\,km long and contain about 600 quadrupoles and as many BPMs (which, anyway, is just less than half the total number of components traversed by the electrons and by the positrons along their way from the damping rings through the interaction point). Such a  large number of components adds at least two complications: (1) it increases the number of potential sources of emittance-growth, (2) it makes local correction techniques impractical and excessively time consuming. Instead, in order to align such linacs in a robust and more practical way, global correction algorithms need to be used. Such algorithms have been studied and modelled intensively during the last decades \cite{Raubenheimer1991,Eliasson2008,Adli2008}. A global dispersion-free correction, in conjunction with even more sophisticated algorithms, have been proposed and foreseen as integral part of the linac tuning procedures for both CLIC and ILC.
 
Another important effect that might harm the beam is the long-range wakefields in the accelerator structures. In this case each bunch can affect later bunches within the same train and induce beam breakup. Analytic evaluations of the tolerance of the beam to such long-range wakefields kicks lead to key design choices for the CLIC main accelerating structure, which in fact features both waveguide damping and cell-to-cell detuning to provide strong broadband suppression of the long-range wakefields (see \cite{Grudiev2010} and CLIC CDR\,\cite{Aicheler2012}). Our tests at FACET aimed at a direct measurement of the long-range wakefield kicks \textit{with beam}, and ultimately validate this design.
 
In order to assess the criticality of the above-mentioned issues, a programme of experimental verifications was launched making use of the FACET linac at the SLAC National Accelerator Laboratory \cite{FACET} and of the linac of FERMI@Elettra free-electron laser in Trieste, Italy \cite{FERMI}. The goals of this measurement campaign were twofold; (1) demonstrate the effectiveness of global beam-based alignment algorithms, (2) measure the long-range wakefields in the CLIC structure and validate the design. Both goals have been successfully achieved.
 
\subsection{Beam-based Alignment}
\subsubsection{Experience at FACET}
The latest incarnation of the SLC linac, i.e. the FACET test facility at SLAC, was the ideal testbed for our linear collider beam-based techniques. The FACET facility made use of the first 2\,km of the Stanford Linear Collider, accelerating electrons (and positrons) from 1\,GeV to 20\,GeV beam energy, with several tens of correctors and BPMs. Here we report on the experimental proof-of-principle of emittance reduction techniques, in a long linac, by applying dispersion-free correction with automatic ``system identification'' algorithms.  The test was structured in two phases; first we deployed an automatic procedure to perform a system identification aimed at computing the response matrices from measurements, then we made use of such response matrices to perform tests of a dispersion-free steering (DFS) correction. We verified the combined effect of the system identification and dispersion-free correction by preparing extensive PLACET \cite{PLACET} simulations of a model of the FACET linac, prior to the measurement. Furthermore, PLACET was used in flight-simulator mode in order to develop and fine-tune all the scripts for the on-line operation. Detailed explanations of the procedure are presented in \cite{Latina2014}.
 
\begin{figure}[!htb]
\begin{centering}
\includegraphics[width=0.5\columnwidth]{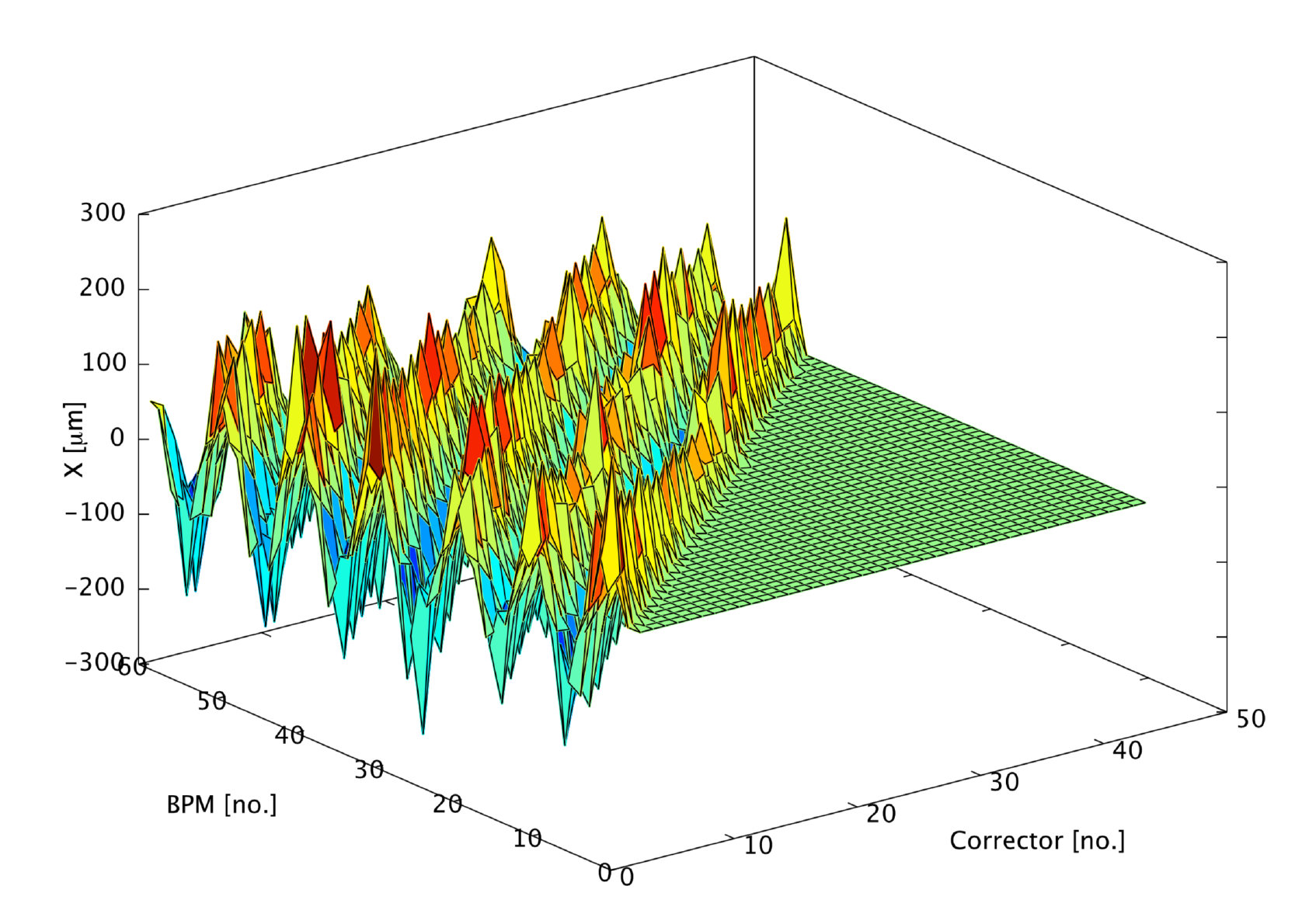}
\par\end{centering}
\caption{\label{facet.response} The orbit response matrix, $R$,
as measured by estimating the orbit difference from systematic alternating-sign
correctors kicks and applying the system identification algorithm.}
\end{figure}

The response of each BPMs to beam corrector excitations was measured using state-of-the-art system identification procedures \cite{Pfingstner2012} both for the nominal beam and for the dispersive test beam, required for the dispersion correction. The response matrices were obtained operating the linac in nominal conditions for the orbit correction, and changing the phase of a single klystron in the upstream section of interest from 0 to 90 degrees for the test beam. The selection of klystron and phase change was based on simulated evaluations. Figure~\ref{facet.response} shows the measured orbit response matrix.
 
\begin{figure}[!htb]
\begin{centering}
\begin{tabular}{ccc}
\includegraphics[width=0.25\textwidth]{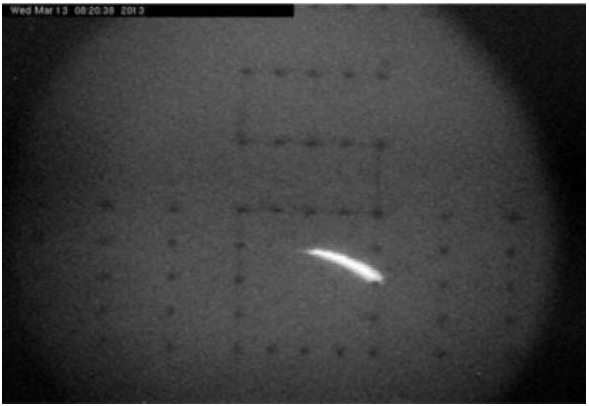} &
\includegraphics[width=0.25\textwidth]{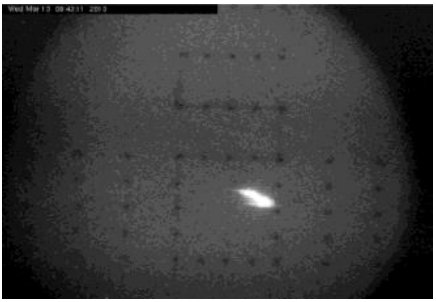} &
\includegraphics[width=0.25\textwidth]{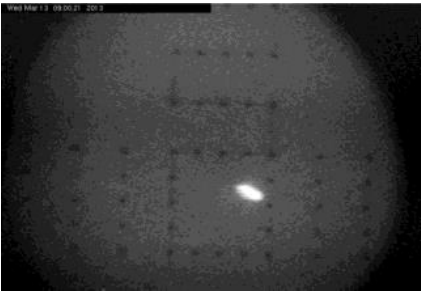} \\
\small{iteration 0} &
\small{iteration 1} &
\small{iteration 3}
\end{tabular}
\par\end{centering}
\caption{\label{facet.fig_screen} Phosphorous beam profile monitor measurements at the end of the FACET linac, before the dispersion correction, after
one iteration step, and after three iteration steps. Iteration zero is before the correction.}
\end{figure}

Figure~\ref{facet.fig_screen} shows one of the best results of the experiment, which demonstrated how the algorithm successfully corrected dispersion and, in so doing, helped to recover the transverse emittances. The horizontal dispersion was robustly corrected by a factor 3, and the vertical by a factor 4. It should be noted that during the correction the absolute orbit diverges from its initial course, as the dispersion is reduced. This is to be expected.
 
\subsubsection{Experience at FERMI@Elettra}
Spurred on by the success of the tests of beam-based alignment performed at FACET, we had the chance to test our beam-based alignment algorithms at other machines. The linac of the free-electron laser FERMI@Elettra at Sincrotrone Trieste, Italy, was probably the ideal one for us, given the presence of an X-band accelerator structure, used as RF lineariser, and of several C-band structures featuring strong transverse wakefields \cite{FERMI}. At FERMI@Elettra we took the chance to test a variant of dispersion-free steering, called wakefield-free steering (WFS), where the effect of the wakefields on the orbit is removed by means of dipole corrector magnets, rather than the dispersion. The results of this experiment have been published in \cite{Latina2014a}.
 
\subsection{Long-range Wakefields}
\begin{figure}[!bht]
\begin{centering}
\includegraphics[width=0.45\textwidth] {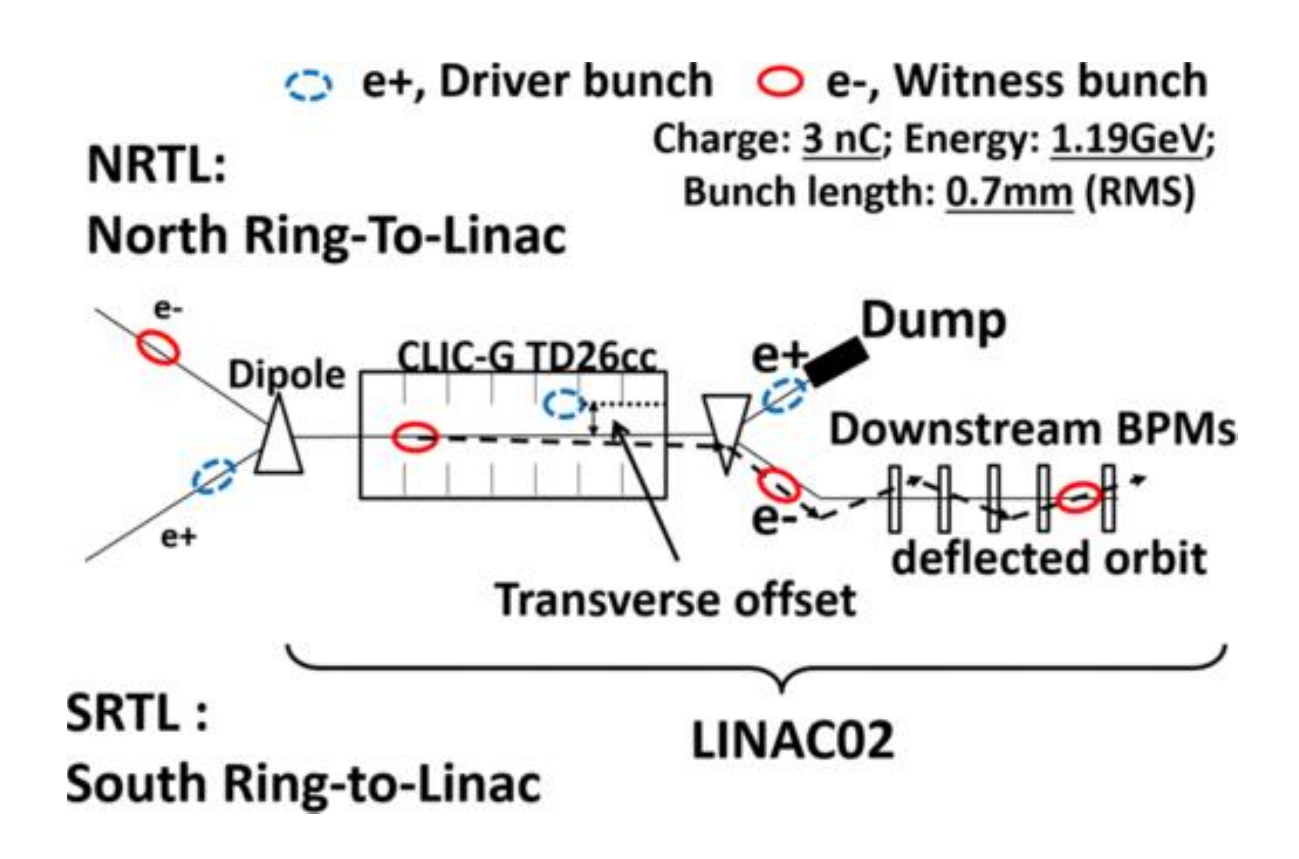}
\par\end{centering}
\caption{\label{facet.setup}Layout of the wakefields measurement experiment.}
\end{figure}

\begin{figure}[!htb]
\begin{centering}
\includegraphics[width=0.45\textwidth]{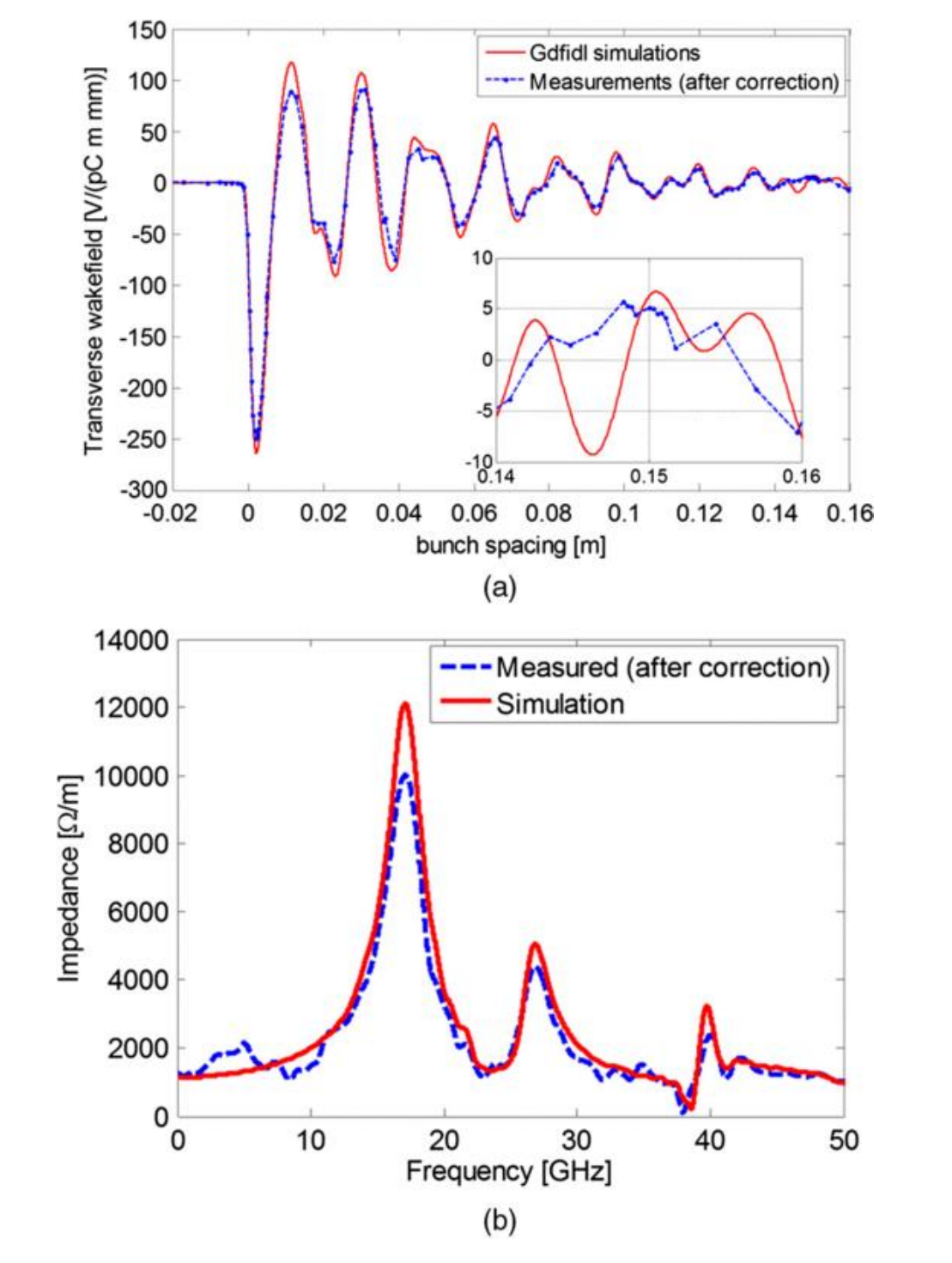}
\par\end{centering}
\caption{\label{facet.classe}(a) Wakefield plots compared with numerical simulations. (b) Spectrum of measured data versus numerical simulation.}
\end{figure}

The design of the CLIC main accelerator structure, heavily damping the high-order modes, was validated at FACET. The aim of this demonstration was twofold; validate that the beam dynamics requirements are met, and prove that the computer-simulated transverse wakefields are accurate \cite{GDFIDL}. This latter condition was essential, since the design may evolve in the future. The experimental demands were significant since the suppression must be measured at bunch distances from the peak of the transverse wakefield to the end of CLIC bunch train, which is spanning more than 4 orders of magnitude, namely from 1 mm to 50 m and over approximately 3 orders of magnitude of wakefield level. These requirements could only be met at the FACET facility at SLAC. The peculiarity of this facility, unique and so important for this type of measurement, was that it could simultaneously deliver positron and electron bunches. This gave us the possibility of exciting wakes with a positron bunch and then measuring the transverse kick directly on a following electron bunch at variable distance. The setup of the experiment and the main results are shown in Fig. \ref{facet.setup} and \ref{facet.classe}. The agreement between simulation and measurement is strikingly good. A detailed explanation of the experiment was presented in \cite{Zha2016}.
 
\subsection{Summary}
The campaign of experiments at FACET and FERMI@Elettra proved extremely fruitful:
\begin{enumerate}
\item It demonstrated the effectiveness and the portability of global system identification and beam-based algorithms
\item It validated the design of the CLIC accelerator structure with respect to the heavy damping of high-order modes
\item It validated the numerical wakefields calculation by 3-D electromagnetic codes, and increased our trust in these codes.
\end{enumerate}

\section{Performance of High-Gradient RF systems}
\label{sect:PERF_RF}
The performance of the CLIC Main-Linac RF system has been studied from a comprehensive range of perspectives and the results show that all major performance criteria and specifications can be achieved. The different performance categories are given in the list below. Highlights of developments which have occurred since the CDR are addressed individually in more detail below. In the last part of this section relevant performances of other linacs are described.
Performance criteria for the high-gradient RF system:
\begin{itemize}
  \item Accelerating gradient
	\begin{list}{\textbullet}{}
  		\item {3\,TeV structure -- 100\,MV/m loaded}
  		\item {380\,GeV structure -- 72\,MV/m loaded}
  		\item {The effect of beam loading}
  		\item {Conditioning, variability and long term operation}
	\end{list}
  \item Power production -- Two Beam
	\begin{list}{\textbullet}{}
  		\item {PETS power capability}
  		\item {Power modulation}
	\end{list}  
  \item Power production -- klystron-based
	\begin{list}{\textbullet}{}
  		\item {Modulator}
  		\item {Klystron}
        \item {Pulse compressor}
        \item {Waveguide components}
        \item {Control and operation}
        \end{list} 
  \item Integrated system operation        
\end{itemize}

The most challenging performance parameter is the accelerating gradient. The required 3\,TeV gradient is a factor three higher than the one in state-of-the-art operational linacs. But already by the time of the CDR, accelerating structures were operated at full pulse length and CLIC nominal breakdown rate in excess of 100\,MV/m (unloaded) showing the fundamental feasibility of the 3\,TeV design. 

In the intervening period, testing has continued in order to continue to address and improve accelerating gradient related performance issues including final performance, conditioning time and strategy, optimized recovery from breakdown and operational strategy, variation among structures and long term behavior~\cite{Wuensch2017,Degiovanni2016}. In addition the testing has also had the objective to determine and improve the performance of the complete high-power RF system including modulators, klystrons, pulse compressors, waveguide network, low-level RF etc. In order to address this broad range of issues the RF test stand infrastructure has been significantly increased. There are now three X-band test stands with a total of six testing slots currently under operation at CERN~\cite{Catalan2017,Catalan2014}. Testing in Nextef at KEK also continues.

\begin{figure}[ht]
\begin{center}
\includegraphics[width=15cm]{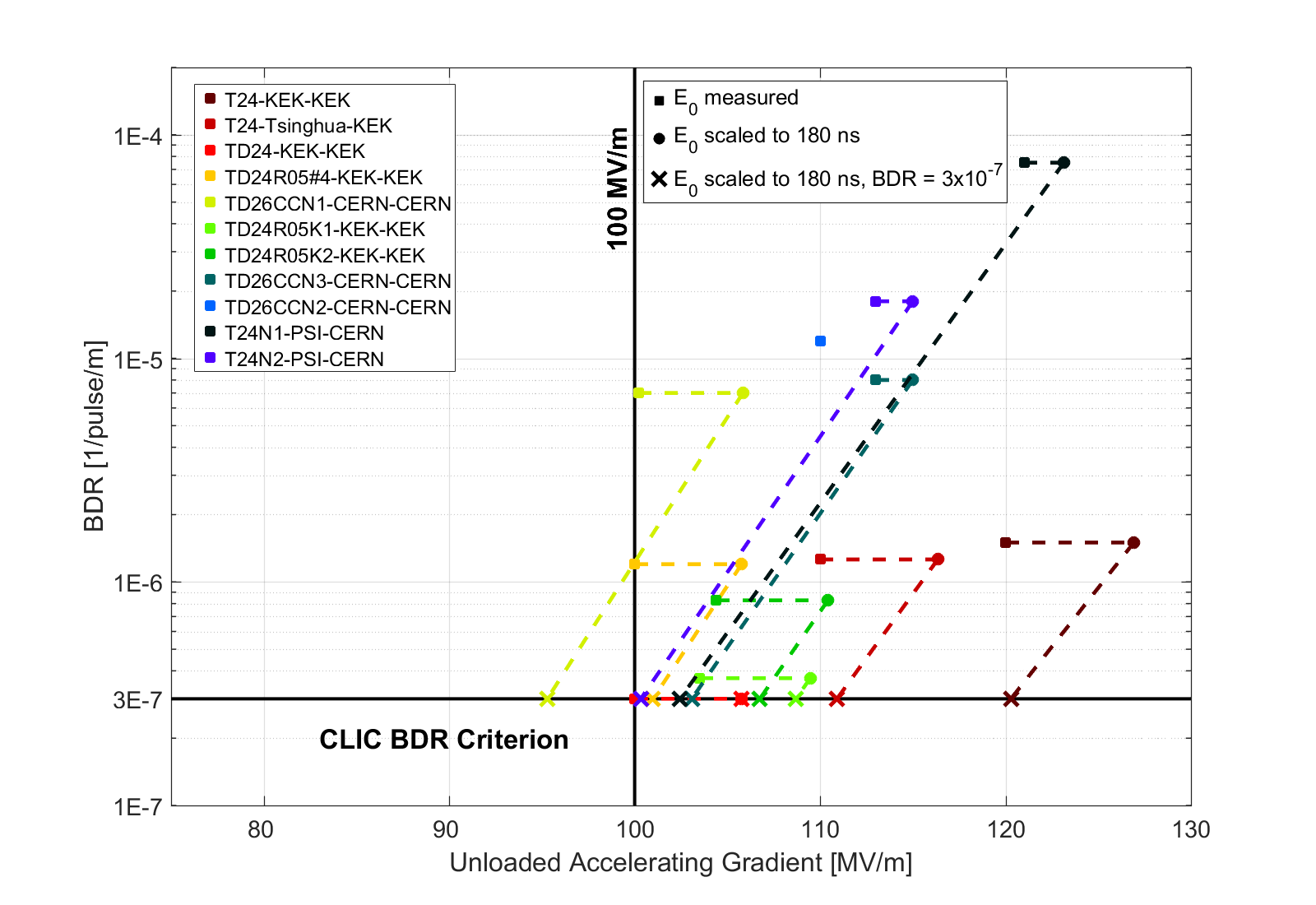}
\caption{A summary of the achieved performances of CLIC-G (3\,TeV) type test structures. The square points are measured operating conditions, the round points are scaled to a CLIC pulse length and the crosses are scaling to the CLIC breakdown rate.}
\label{fig:RF_Performances}
\end{center}
\end{figure}

The accelerating structures which have been operated during the past five years are all test versions of the CLIC-G 3\,TeV structure, but which incorporate varying degrees of complexity; without damping waveguides, with damping waveguides, with mode-launcher then compact coupler and finally including SiC absorbers and associated manifolds. The strategy to implement key features one at a time was to determine the individual effects on performance of each. The testing program has continued with the CLIC-G-based structures to complete this study and to be able to make direct comparisons to existing benchmark data. A summary plot of the performances of CLIC-G (3\,TeV) type structures is shown in Fig.~\ref{fig:RF_Performances}. 

One important observation given by the plot is that undamped structures can be consistently operated in excess of 110\,MV/m, while damped structures achieve a gradient of approximately 10\,MV/m lower with an average of around 105\,MV/m. The presence of the SiC damping material is confirmed to have no influence on performance.
For the same level of input power the average accelerating gradient with beam in CLIC is approximately 15\,MV/m lower than without due to beam loading. This means that the most conservative performance criterion for testing without beam is that the structures meet the gradient and breakdown rate specifications at the input power which corresponds to an unloaded gradient of 115\,MV/m. Thus test results so far show that on average undamped structures meet this performance goal, while the damped structures fall short by approximately 10\,MV/m.

Based on these high-gradient tests and post-mortem analysis of the structures, an improved version of the 3\,TeV structure CLIC-G* has been designed~\cite{Zha2016a}. It includes an improved cell outer wall profile to address the gradient loss due to the damping waveguides and the expectation is that the degradation will be significantly reduced. This also means that the CLIC-G* structures are expected to fulfil the gradient criterion including the effect of beam loading. CLIC-G* test structures will become available for testing in the coming year. 

There is additional evidence that the current results indicate that 100\,MV/m with loading are already being met. For constant input power, the beam loading will result in lower fields inside the structure and it is expected that this will result in a reduction of breakdown rate. In order to investigate this effect an experiment was carried out in CTF3 in which a high-gradient structure was measured with and without beam~\cite{Senes2017}. The overall results are that the effect is present but that there are long transients in behaviour when operating conditions are changed. The limited length and number of testing slots with beam meant that such transients tended to dominate the measurements, but indications are that this effect will be beneficial in CLIC.
In addition, the data points shown in Fig.~\ref{fig:RF_Performances} are those achieved at the end of a limited duration test run. The breakdown rates of the structures during constant gradient running decrease with time even when the gradient can no longer be increased [performance,  conditioning paper]. This effect is systematically seen in the tests. This means that the final operational data point shown as squares in the plot, has the trend to decrease in breakdown rate with time. This means for example that the TD26CCN3 would be expected to reach the required specifications, including the margin needed for beam loading, given sufficient testing time. Making such a demonstration of long-term behaviour possible is one of the major objectives of expanding the testing infrastructure.
The initial energy stage of CLIC will be 380\,GeV, not 3\,TeV. The accelerating structure for this energy stage has a target gradient of 72\,MV/m loaded gradient and consequently a different optimum design. Most importantly the lower energy results in the lower optimum gradient, and gives more room for margin in the high-gradient design of the structure. The 380 GeV structure has been designed using the field values in achieved tests, corresponding to the data points as they appear in Fig.~\ref{fig:RF_Performances} plus a margin of 10\%. Consequently it is expected that the 380\,GeV structure should easily exceed performance specifications. The first prototype 380\,GeV structures are expected towards the end of this year.
Another important aspect of the accelerating structure performance is the conditioning time need to achieve the operating gradient. Currently the full conditioning time is of the order of two months at 50\,Hz. Important insights have been gained into the mechanism of the conditioning process, in particular that the conditioning proceeds as the number of pulses [performance,  conditioning paper] and the pulses drive dislocation dynamics in a hardening process. These insights are being used to optimize the conditioning strategy and develop acceptance and conditioning strategies for large series production. A strategy with high-repetition rate conditioning, in the range of 200-400 Hz as being done in Xbox-3, will allow the operating gradient at short pulse length to be achieved within the order of a day and will validate acceptance of structures after fabrication. Further conditioning to full pulse length can be then done in-situ in parallel with linac commissioning.

Operation of accelerating structures at high-gradient requires power sources which produce power in the 50\,MW range for each accelerating structure. The feasibility of Two-Beam power production is addressed in Section~\ref{sect:PERF_DB}. The other option for power production in the 380\,GeV initial energy stage is to directly use 12\,GHz klystrons. The RF units in this option are based on similar components as used in the high-gradient test stands and have a similar configuration. The operation of the test stands, therefore, provides important performance verification for the klystron-based option.  
The Xbox-1 and 2 are each based on one 50\,MW CPI klystron followed by a pulse compressor giving a theoretical maximum pulse of 150\,MW, 200\,ns long. Currently the Xbox-1 and Xbox-2 test stands are powering a single accelerating structure so are regularly running at reduced power, with a compressed pulse in the range of 50\,MW. At these power levels the systems operate very stably for months with most of the down time occurring due to failures in the auxiliary systems like water cooling and interruption of electricity supply. A key priority for the testing program is to move to powering two accelerating structures which will require that the klystrons operate at approximately 120\,MW, so much closer to full specifications. In the longer term, operation with two combined klystrons must be demonstrated although the power levels present in this configuration have already been demonstrated in the Two-Beam Test Stand. Operation of two combined klystrons is now routine in Xbox-3.

In the meantime, a wide range of waveguide components are routinely running at required peak power levels, in particular those associated with one accelerating structure such as directional couplers, hybrids, pumping ports, terminating loads, etc. In fact these components are also run at approximately four times specified average power levels in Xbox-3. This is possible because Xbox-3 klystron modulator units are capable of operating at 400\,Hz, and when summed give specified power at 200\,Hz.

A number of other projects are operating, preparing and designing RF systems similar to CLIC, especially the klystron-based version. The most advanced is the SwissFEL. The specifications of the SwissFEL are somewhat different, primarily the  accelerating gradient is just under 30\,MV/m and the frequency is 5.7\,GHz, but the configuration of the RF system with solid state modulators, 50\,MW-range klystrons, BOC type pulse compressors and micron-precision accelerating structures is quite similar. The accelerating structure fabrication technology has been shown to be compatible with the successful test of two CLIC test structures (T24-type) built by PSI using the SwissFEL production line. In tests they have operated to over 115\,MV/m, and are included in Fig.~\ref{fig:RF_Performances}. Over 100, 2\,m long SwissFEL accelerating structures have been built and verified to a sub-micron tolerance on inner cell geometry. The CLIC cell geometry is more complicated due to the higher-order-mode damping waveguides, however the demonstration of sub-micron precision in a large series production is very important. At the time of this writing the linac is being commissioned and very high-stabilities of the modulators and resulting RF pulses are being achieved. The SXFEL at SINAP has a similar C-band linac and further similar experience is being gained there. 

X-band energy spread linearizer systems based on 50\,MW klystrons are under routine operation at PSI and at FERMI in Trieste, and are crucial for the performance of both machines. The systems only need to run at reduced specifications, approximately 30\,MW, but experience with long-term operation is being gained through their operation. The SXFEL at SINAP also has an operational X-band linearizer system, but based on the 6\,MW Toshiba klystron as used in Xbox-3. 

Upcoming X-band and high-gradient applications include; deflector systems at SINAP, DESY and PSI, a test stand followed by a linac energy upgrade, a test stand at the University of Melbourne, a test stand at the University of Valencia, the Smart*Light compact Inverse Compton Scattering source, a linearizer system at CLARA in Daresbury and a 1\,GeV linac-based test facility in Frascati. The operation of these systems will increase the commercial supplier base, operational experience, number of experts and likely produce further new ideas.

In addition to the electron linac applications listed above, there is also interest in high-gradient compact proton linacs for cancer therapy. The ADAM company is preparing to commercial supply such linac-based therapy machines. In parallel development is proceeding on higher performance linacs and a pair of structure has been built at CERN using the CLIC high-gradient design methodology and fabrication technology. The primary challenge is to adapt the high-gradient design developed for relativistic electrons to low velocity protons or carbon ions, v/c~=~0.38 in this case. The first of these structures is under testing and is now operating at over 60\,MV/m which means that surface fields are consistent with those seen in the CLIC test structures. This represents a significant gradient advance over proton linac structures.

\section{Damping Rings}

\begin{figure}[h]
\centering\includegraphics[width=.7\linewidth]{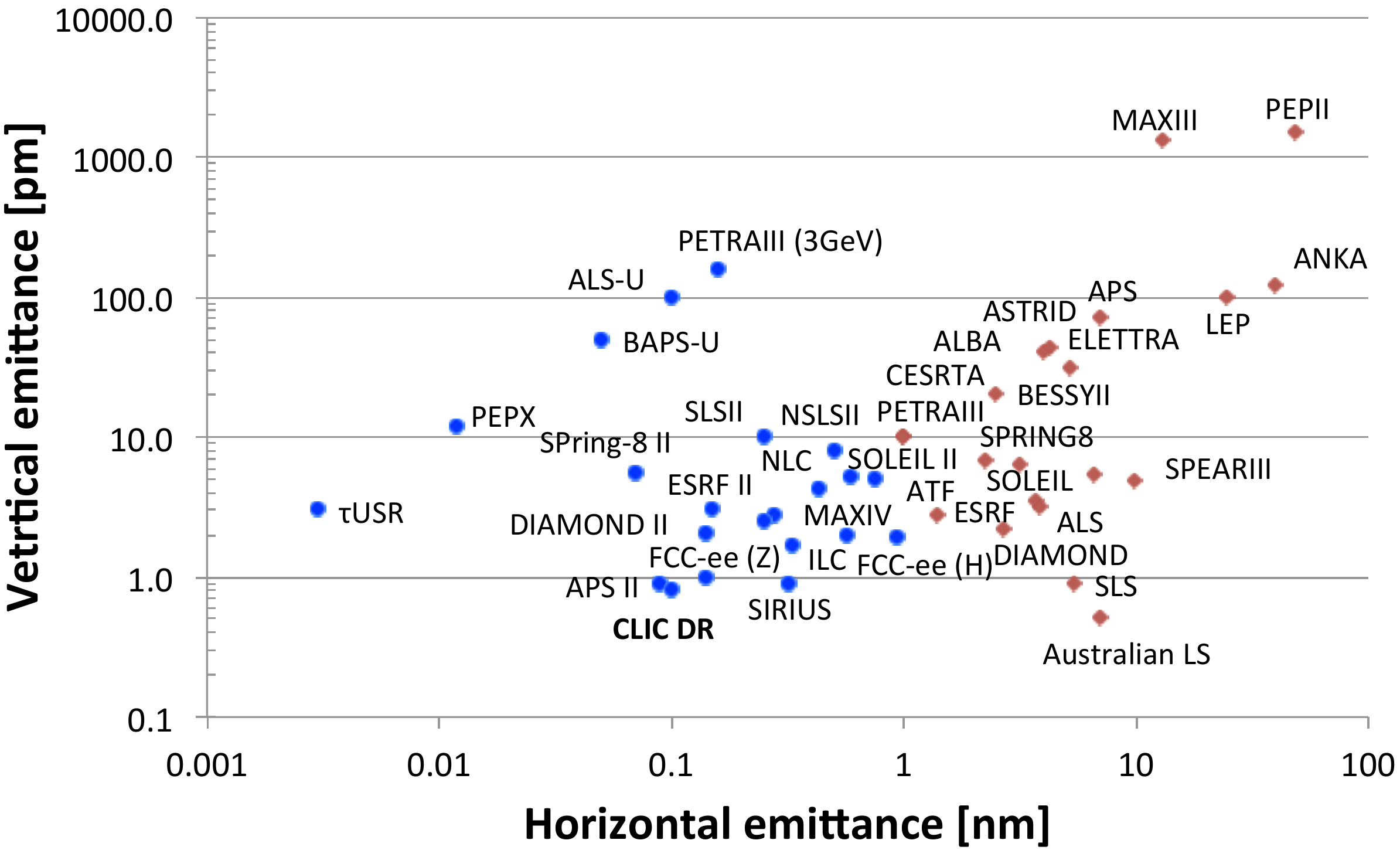}
\caption{Horizontal versus vertical geometrical emittance of low emittance rings in operation (red) and design (blue).}
\label{fig:emittance}
\end{figure}

The main challenge of the Damping Rings (DRs) is the generation of ultra-low emittances in all planes in order to achieve the high brightness beams required for the linear collider luminosity production. In Fig.~\ref{fig:emittance} \cite{LowemittanceWS}, the transverse geometrical emittances of low emittance rings in operation (red)  and design are presented. In recent years, it is apparent that, in particular, the new generation light sources are approaching and even superseding the horizontal emittance target of the CLIC DRs, whereas the vertical emittance has been reached in operating storage rings. What differentiates the CLIC DRs compared to other low emittance rings, is the necessity to reach these emittances simultaneously, and having a low longitudinal emittance, translated to bunch lengths, of below 2\,mm.

This high beam brightness drives the design choices of the DRs. The approach is to design a compact ring, attaining a sufficiently low emittance taking into consideration collective effects, such as Intra-Beam Scattering (IBS), e-cloud or even space-charge, which dominate the emittance evolution along the 20\,ms cycle. In this respect, all the design parameters including optics were carefully chosen and optimised in order to mitigate the impact of collective effects~\cite{Papaphilippou2012,Antoniou2012}.  In particular, an effort was made to reduce the effect of IBS on the "zero-current" emittance.  A further improvement can be made by replacing the arc bending magnets with ones including a variation in the longitudinal plane and considering wigglers with even higher field in the straight sections. 

The natural lattice choice for the arc optics are the Theoretical Minimum Emittance (TME) cells~\cite{Antoniou2014},  which can provide very small emittances, as well as high compactness. The horizontal emittance can be further reduced below the TME limit by considering dipole magnets with longitudinal variable bending field~\cite{Guo2002,Nagaoka2007,Papaphilippou2005,Papaphilippou2005a,Wang2009,Wang2011}. Based on an analytical evaluation~\cite{Papadopoulou2018} and magnet design considerations~\cite{Martinez2018}, an optimised trapezium bending radius profile is employed, i.e. with parabolic field evolution in the edges and constant in the centre.

\subsection{Optimization of the arc TME cell}
In the considered arc cell, the dipole length is $L=0.58$\,m and the maximum dipole field is $2.3$\,T. When the uniform dipoles of the current design are replaced by variable bends, the resulted emittance is lower than the required one. In this case, the subtraction of some TME cells from the existing arc is possible. Actually, the number of dipoles (i.e. total number of TME cells) can be reduced to such an extent that the required emittance is still achieved, thereby resulting in a shorter ring. 

As soon as the characteristics of the dipole are fixed, the drift space lengths are chosen~\cite{Papadopoulou2018}. Afterwards, it is necessary to find the phase advances that guarantee low chromaticities and small quadrupole strengths.  Keeping in mind that a modified TME cell  with a combined function dipole, i.e. having a small defocusing gradient, has a positive impact on the IBS effect~\cite{Braun2010} assists in choosing the proper phase advances.  The parametrization with the emittance showed that the quadrupole strengths for which $f_1<0$ and $f_2>0$,  can only be found for $\mu_{x}<0.5\cdot2\pi$.  Another reason that renders larger phase advances inappropriate is that they result in huge anharmonicities.  A good compromise for the horizontal phase advance is to be around $0.43\cdot2\pi$ and for the vertical phase advances to be always below $0.5\cdot2\pi$. After a detailed scanning of the cell characteristics, the horizontal and vertical phases advances of the TME cell are respectively chosen to be around $0.44\cdot2\pi$ and $0.1\cdot2\pi$, corresponding to an emittance reduction factor that is $\epsilon_r = 6.5$. Taking into account the fact that for the designed variable bend it is $F_{TME}=7.1$, emittances smaller than the ones of the current design with the  uniform dipoles, are reached. In this respect, it was possible to reduce the number of dipoles down to  $N_d=90$ for the case of the designed trapezium profile with 2.3\,T maximum field.

\subsection{Optimization of the FODO cell}

\begin{figure}[h]
\centering\includegraphics[width=.33\linewidth]{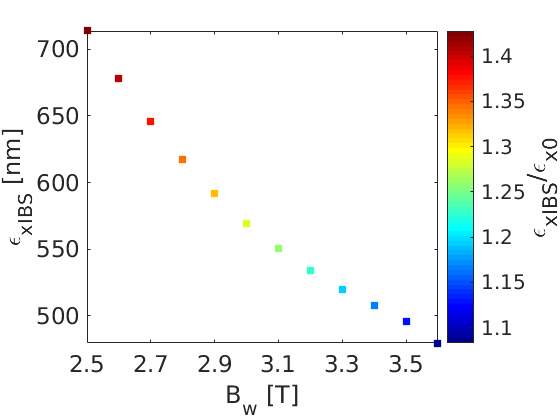}
\quad
\centering\includegraphics[width=.33\linewidth]{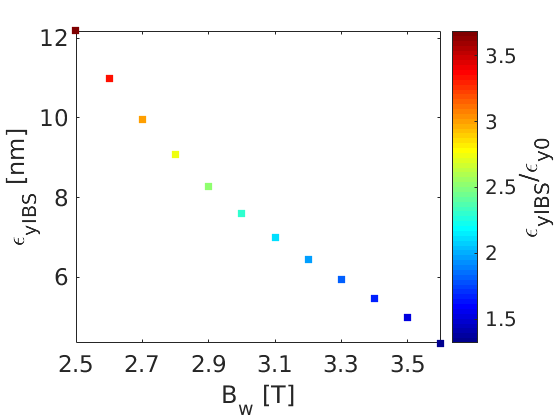}
\caption{{\small The dependence of the steady state emittances ($\epsilon_{xIBS}$ and $\epsilon_{yIBS}$) and their ratio with the corresponding equilibrium emittances ($\epsilon_{x0}$ and $\epsilon_{y0}$) on the wiggler peak field $B_w$, for the  trapezium dipole  profile.}}
\label{fig:ibs}
\end{figure}

The damping wigglers are necessary for achieving  low emittances within a fast damping time. Each FODO structure of the straight section accommodates two wigglers. The use of super-conducting technology is mandatory in order to have a high wiggler field and a relatively short period for obtaining low emittances and fast damping time. The super-conducting magnet wigglers used in the current design have a $B_w=2.5$\,T peak field and $\lambda_w=5$\,cm period~\cite{Schoerling2012}. A prototype wiggler with these characteristics was build by BINP and installed and tested in the ANKA storage ring, in particular for the cooling principle~\cite{Bernhard2016}.

It was further shown that by targeting  higher wiggler fields not only the emittance but also the IBS effect can be reduced~\cite{Antoniou2012,Fajardo2015}. Taking into account the optimization of the arc cells and the fact that the emittance with IBS is significantly lowered after increasing the wiggler's peak field, the FODO cells per straight section can be reduced from 13 down to 10. The plots in Fig.~\ref{fig:ibs} show the MADX results of the parametrization of the steady state transverse emittances including the IBS effect with the wiggler peak field $B_w$, starting from the  2.5\,T, i.e. the field of the existing wiggler design, for the 10 FODO cells. Clearly, the wiggler field increase corresponds to a significant reduction of the IBS effect, through the minimisation of the damping time. Regarding the fact that the required output emittance is 500 nm-rad, a new working point for the damping wigglers that complies with the technological restrictions  is proposed to be at 3.5\,T  and with a 49\,mm period length. This design necessitates a different wire technology, using Nb$_3$Sn material~\cite{Fajardo2015}. 

\subsection{Optical Functions and New Design Parameters}

\begin{figure}[h]
	\centering\includegraphics[width=.4\linewidth]{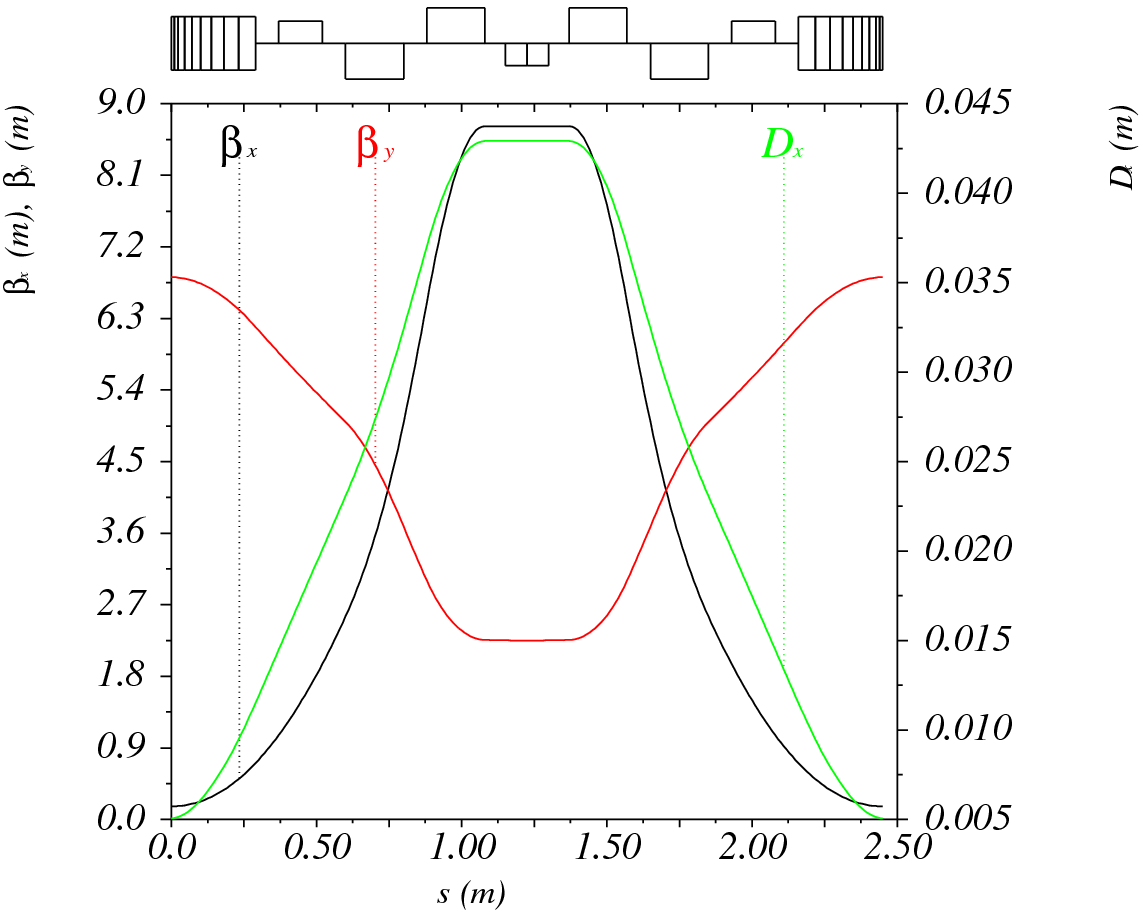}
	\quad
	\centering\includegraphics[width=.4\linewidth]{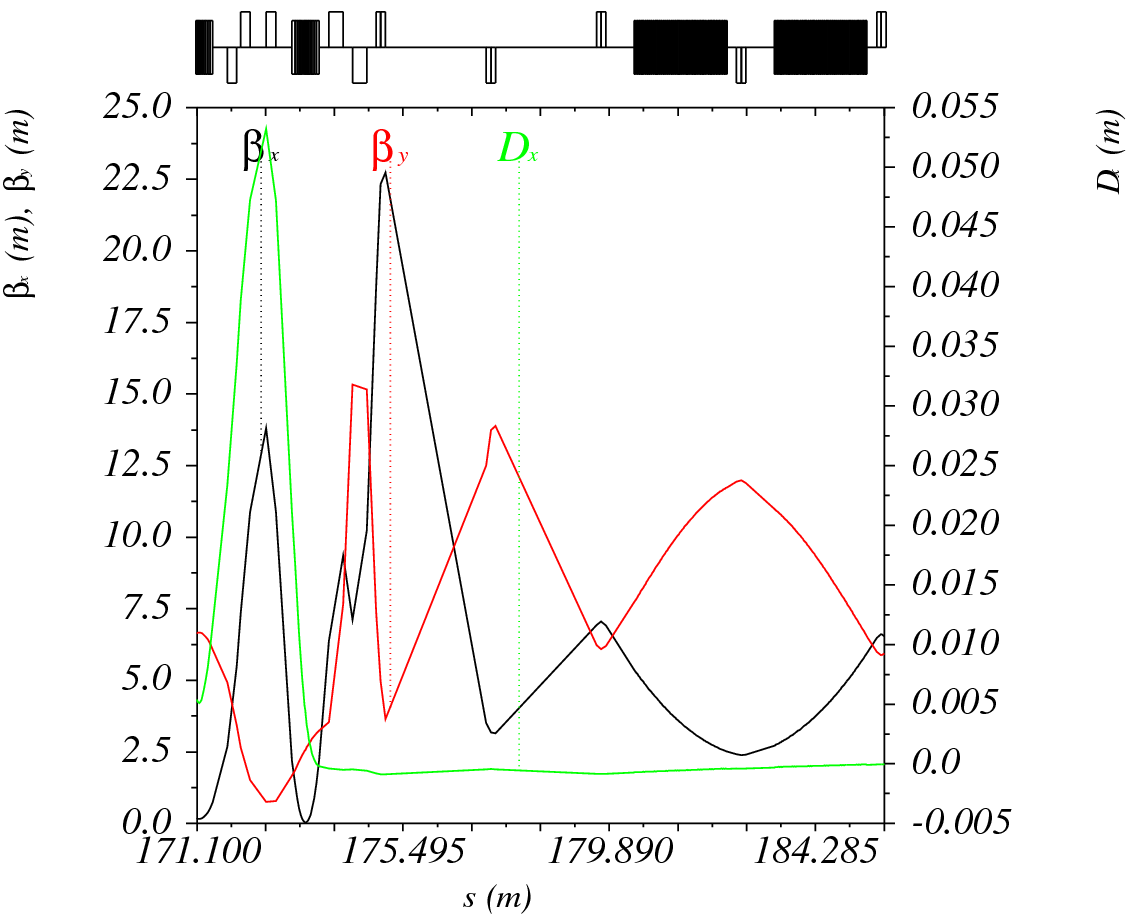}
	\caption{{\small Optical functions of the TME cell (left) and of the dispersion suppressor-beta matching section followed by the FODO cell (right), when using in the arcs the trapezium dipole profiles.}}
	\label{fig:tmecells}
\end{figure}

The final lattice, with a smaller number of dipoles and wigglers than the ones of the existing design, is the one designed for the re-baselining exercise. In Fig.~\ref{fig:tmecells} (left), the matched optics, i.e. horizontal dispersion, horizontal and vertical beta functions, are plotted for one arc TME cell. On the top part of the figure, a schematic layout of the cell is presented, showing the two doublets of quadrupoles and the sextupoles that are placed between the two mirror symmetric defocusing quadrupoles and between the dipole and the focusing quadrupoles. In Fig.~\ref{fig:tmecells} (right), the matched optics of the dispersion suppressor-beta matching section followed by the wiggler FODO cell, are presented.

The parameters of the original design and the alternative one are displayed in Table~\ref{tab:DRparams2}. Note that for the nominal bunch charge, all emittance targets are reached, including IBS, as calculated by the Bjorken-Mtingwa formalism through MADX.  Due to the larger bunch population of  $N_b=5.7\times10^9$ for the case of low energy CLIC and $V_{RF}=6.5$\,MV~\cite{Boland2016}, the final emittances are increased up to $536$\,nm and $6.5$\,nm in the horizontal and the vertical plane, respectively. One of the main advantages of the alternative design is that the damping ring becomes around 20~\% shorter. In addition, the damping times are significantly reduced and this is quite beneficial for all collective effects, including IBS.

\begin{table*}
	\begin{center}
		
		\caption{ Design parameters for the original and the improved designs of the CLIC DRs, for the case of $f_{RF}=2$~GHz and $N_b=5.7 \times 10^9$}
		\label{tab:DRparams2}
		
		\centering
		{\small \begin{tabular}{lcc}\hline\hline
				
				\multirow{1}{*}{Parameters, Symbol [Unit]}&\multirow{1}{*}{uniform}&\multirow{1}{*}{trapezium}\\ 
				\hline  \hline 
				Energy, $E$ [GeV]         		    				& 	\multicolumn{2}{c}{2.86}  \\     
				Bunch population, $N_b$	[$10^9$]      				&   \multicolumn{2}{c}{5.7}	  \\                          
				Circumference, $C$ [m]             					& 427.5  &  359.4 \\                
				Basic cell type  in the arc/LSS               	    & 	\multicolumn{2}{c}{TME/FODO} \\
				Number of arc cells/wigglers,	$N_d/N_{w}$			& 100/52 &  90/40\\
				RF Voltage,	$V_{RF}$ [MV]				        	& 4.50  & 6.50\\
				RF Stationary phase [$^o$]  						& 62.3  & 63.0\\
				Harmonic number,	$h$        						& 2850	 &	2398\\
				Momentum compaction,	$\alpha_c$ [$10^{-4}$]		& 1.3  &	1.2 \\
				Damping times, ($\tau_x, \tau_y, \tau_{l}$) [ms] 	&(1.98, 2.05, 1.04) &(1.15, 1.18, 0.60)\\
				Energy loss/turn, $U$ [MeV]							& 4.0  & 5.8 \\
				Quadrupole gradient strengths, ($k_1$, $k_2$) [T/m]	&(-26, 53)	 &  (-29, 56)\\
				Phase advances per arc cell, $(\mu_x, \mu_y)$ [$360^o$] & 0.408/0.050  &0.442/0.100 \\
				Horizontal and vertical tune, ($Q_x$, $Q_y$) 	   & (48.35, 10.40) &(45.61, 13.55)\\
				Horizontal and vertical chromaticity, ($\xi_x$, $\xi_y$) &(-113, -82) &(-169, -51)\\   
				\hline
				Dipole length, $L_d$ [m]               	       		& 	\multicolumn{2}{c}{0.58} \\
				Dipole field, ($B_{min}, B_{max}$) [T]		& (0.97, 0.97)   & (0.69, 2.32) \\
				Lengths and bending radii ratio, ($\lambda, \rho$) & (1, 1)   & (0.04, 0.29) \\
				Emittance reduction factor, $F_{TME}$				& 1  & 7.1 \\
				Normalized gradient in dipole  [m$^{-2}$ or T/m]	& 	\multicolumn{2}{c}{-1.1 or -10.5}  \\
				\hline      
				Wiggler peak field, $B_w$ [T]                       & 2.5  & 3.5\\  
				Wiggler length, $L_{w}$ [m]                         & \multicolumn{2}{c}{2} \\  
				Wiggler period, $\lambda_w$ [cm]					& 5.0  & 4.9 \\
				\hline
				& \multicolumn{2}{c}{without IBS} \\
				\hline		
				Normalized horiz. emittance, $\gamma\epsilon_{x}$ [nm-rad] & 312.2 &  438.46 \\
				Normalized vert. emittance, $\gamma\epsilon_{y}$ [nm-rad]  & 3.3  &  3.3    \\
				Energy spread (R.M.S.),	$\sigma_{\delta}$ [$\%$]		 		 & 0.11  & 0.13\\
				Bunch length (R.M.S.), $\sigma_s$ [mm]                   			 & 1.4   & 1.3\\
				Longitudinal emittance, $\epsilon_l$ [keVm] 			  		 & 4.4   &  4.6\\
				\hline
				& \multicolumn{2}{c}{with IBS} \\
				\hline			
				Normalized horiz. emittance, $\gamma\epsilon_{x}$ [nm-rad] & 478.9 &  535.9 \\
				Normalized vert. emittance, $\gamma\epsilon_{y}$ [nm-rad]  & 5.0 &  6.5    \\
				Energy spread (rms),	$\sigma_{\delta}$ [$\%$]		 		 & 0.11 & 0.13\\
				Bunch length (rms), $\sigma_s$ [mm]                   			 & 1.5  & 1.3\\
				Longitudinal emittance, $\epsilon_l$ [keVm] 			  		 & 4.7   &  4.8\\
				IBS factors hor./ver./long. 					   	& 1.53/1.52/1.08   &  1.22/1.96/1.05\\
				\hline\hline
			\end{tabular}
		}
		\vspace{-10pt} 
	\end{center}
\end{table*}

\section{Impact of Stray Magnetic Fields}
\label{sect:PERF_Stray}
A source of luminosity loss under investigation is the effect of external dynamic magnetic fields (stray fields).

\subsection{Tolerances}
\begin{figure}[!htb] 
  \subfigure[RTML long transfer line.]{%
    \includegraphics[width=0.45\textwidth]{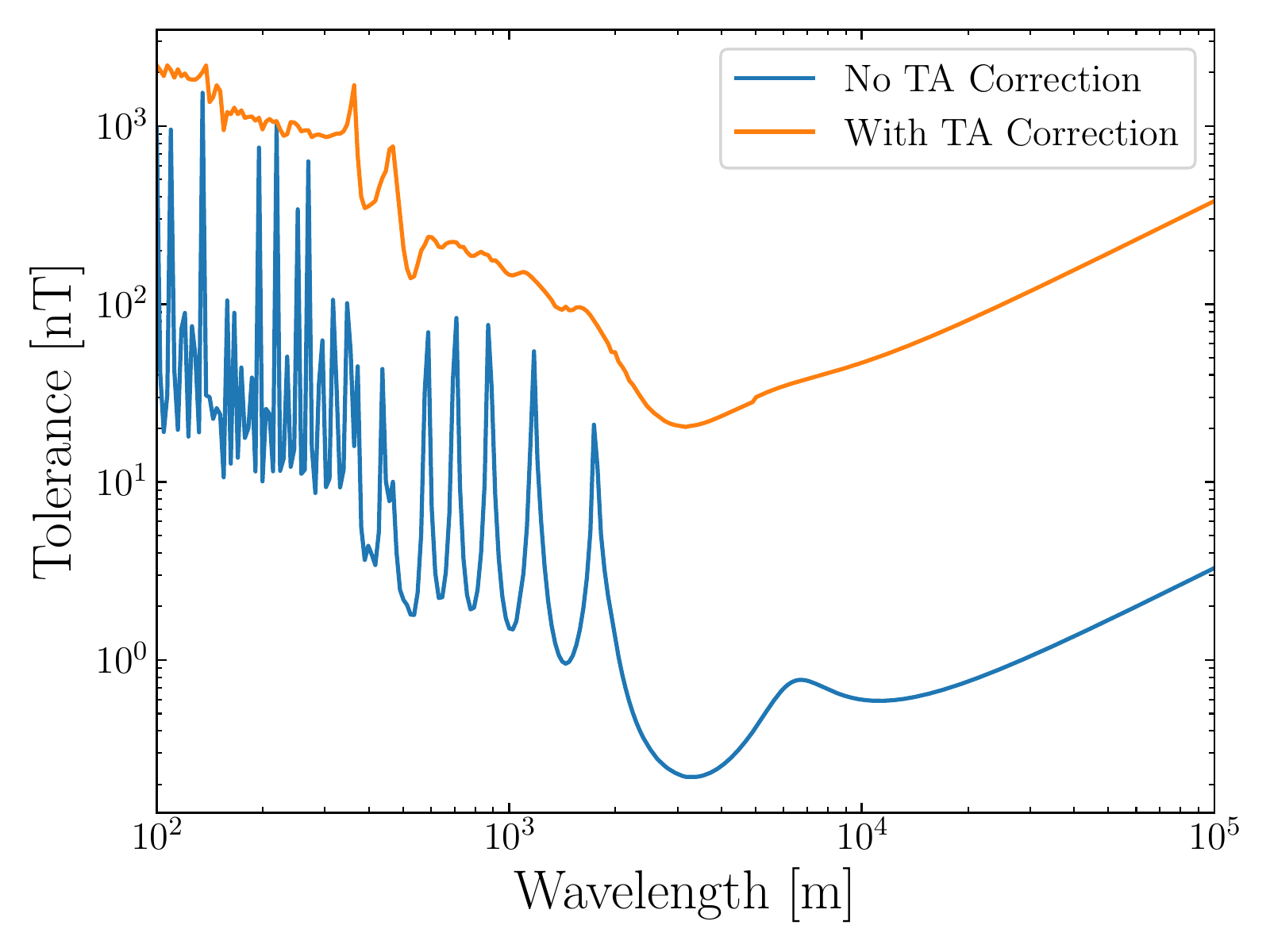} 
    \label{fig:StrayFields_rtml_tolerance}
  } 
  \quad 
  \subfigure[ML and ($L^*=6$\,m) BDS.]{%
    \includegraphics[width=0.45\textwidth]{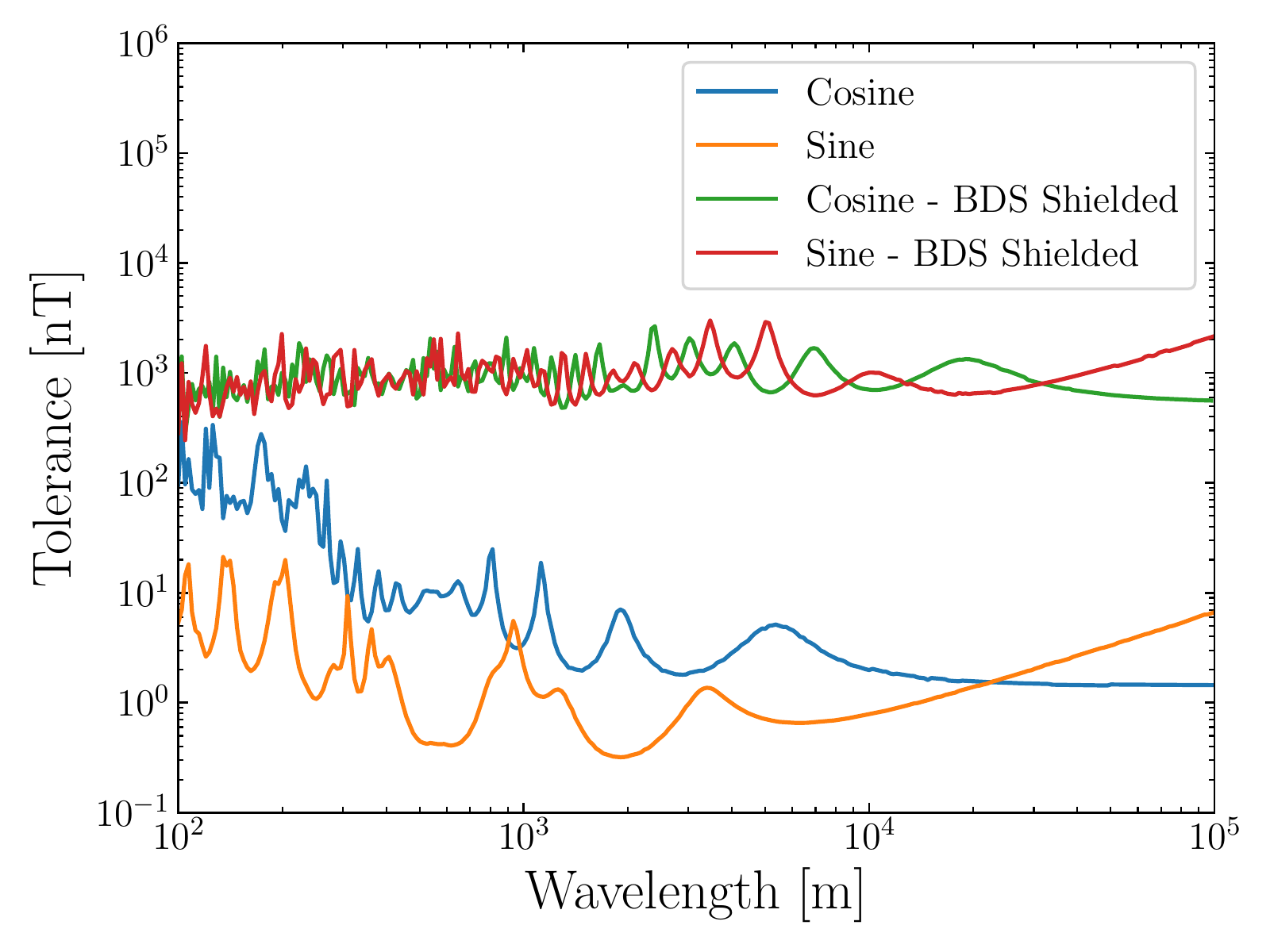}
    \label{fig:StrayFields_ml_bds_tolerance}
  } 
  \caption{Stray field tolerances for the 380 GeV CLIC design \cite{Gohil2019b}. Simulations were performed with the particle tracking code PLACET \cite{bib:HL:c:PLACET}. The tolerance was calculated as the stray field amplitude that corresponds to a 1.2 nm emittance growth (equivalent to a 2\,\% luminosity loss).} 
\end{figure}

Simulations of sinusoidally varying stray fields experienced directly by the beam have shown nT tolerances to remain within a 2\,\% luminosity loss budget \cite{Snuverink2010,Marin2017,Gohil2018,Gohil2019b}. Shown in Fig.\,\ref{fig:StrayFields_rtml_tolerance} is the tolerance for the long transfer line in the RTML, with and without a correction in the TA - discussed in Sec.\,\ref{sect:PERF_Stray_Mitigation}. The tight tolerance arises from the relatively weak focusing and long drifts in the transfer line. A minimum tolerance of approximately 0.1 nT corresponds to the stray field wavelength approaching the betatron wavelength of the lattice. 

Shown in Fig.\,\ref{fig:StrayFields_ml_bds_tolerance} is the tolerance for a stray field in the ML and BDS, along with the effect of shielding the BDS - discussed in Sec.\,\ref{sect:PERF_Stray_Mitigation}. Two types of stray fields have been considered, those which are symmetric about the IP (Cosine) and those which are anti-symmetric about the IP (Sine). Sine-like stray fields introduce an offset at the IP, which is not present for cosine-like stray fields, generally leading to tighter tolerances. The sensitivities in the tolerance for the ML and BDS come almost exclusively from the BDS, particularly in cases where the stray field wavelength resonates with the betatron motion at locations in the BDS where the beta function is large, such as the collimation section and bends before the final focus. The minimum tolerance observed is approximately 0.1 nT for sine-like stray fields and a tolerance of approximately 1 nT emerges for coherent (long wavelength, cosine-like) stray fields.

\subsection{Sources and Measurements}
Sources of stray fields can be classified as man-made or natural. Natural sources of dynamic stray fields typically have a coherence length of 100\,km or longer. The frequencies of most of the natural sources are below 1\,Hz. Such stray fields can be effectively mitigated with beam-based feedback systems. A review of natural sources above 1\,Hz \cite{Heilig2018} concludes that they occur infrequently (less than once a month) or typically have pT amplitudes and are therefore within the tolerances. Natural sources should not pose a serious problem for CLIC.

Man-made sources can either be an environmental source, which is a piece of equipment that produces a stray field, but is not an element of CLIC, or a technical source, which is an element of CLIC. Examples of environmental sources are the electrical grid and railways. Other running accelerators can also act as an environmental source, particularly on the CERN site where there are several other running experiments.

Technical sources, such as RF systems, vacuum pumps and power cables, pose the greatest risk. These are capable of producing stray fields across a wide frequency range. There is an on-going campaign to measure the power spectra of stray fields with an emphasis on technical sources.

\begin{figure}[!htb]
  \centering
    \includegraphics[width=0.48\textwidth]{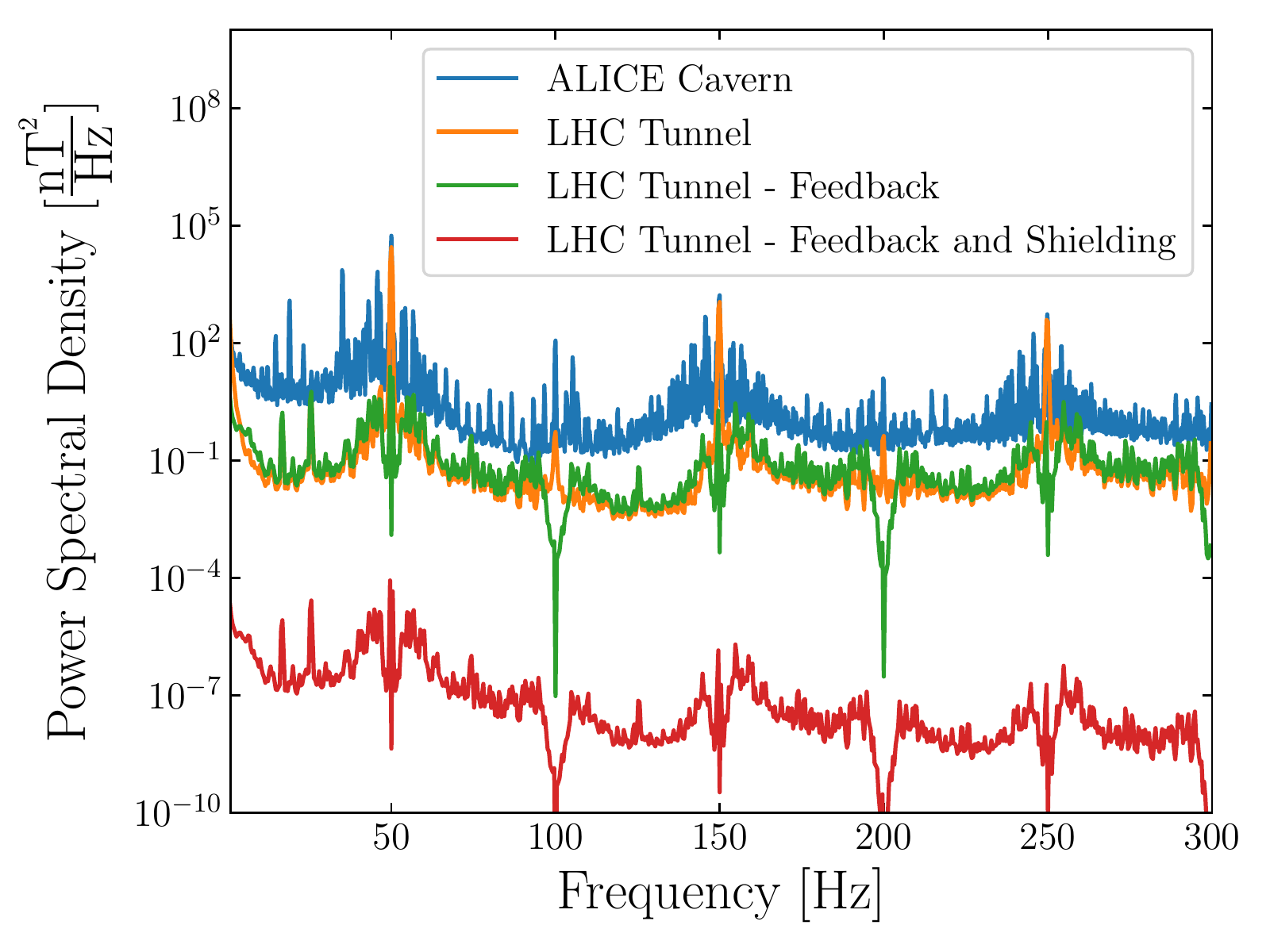}
    \hfill
    \includegraphics[width=0.48\textwidth]{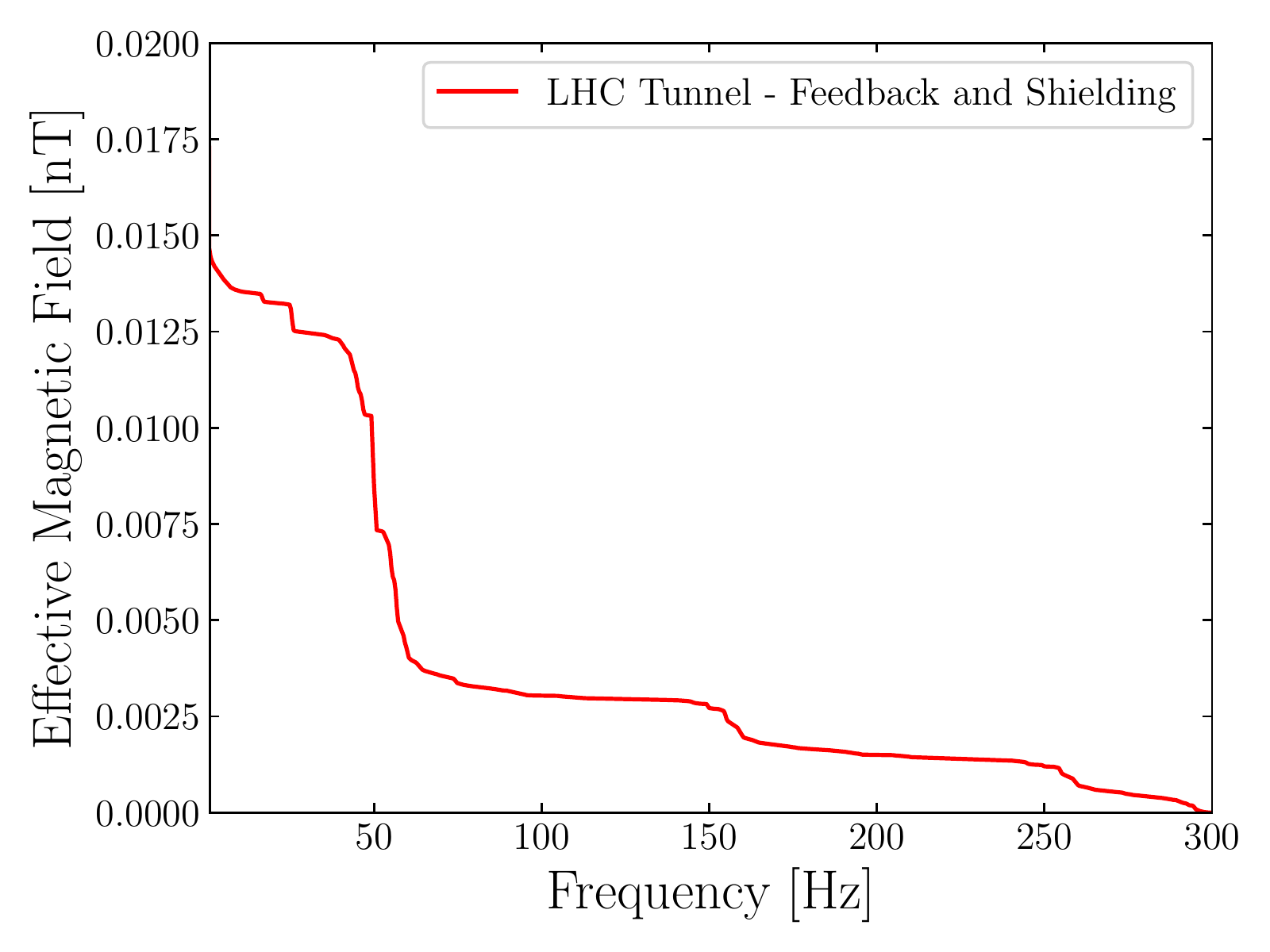}
    \caption{On the left is the average total power spectral density of the background magnetic field measured at 8 different locations in the LHC tunnel and ALICE detector cavern along with the effective reduction in power due to the currently designed beam-based feedback system and a 1 mm mu-metal coating (with a relative magnetic permeability of 10,000) around a beam pipe of radius 1 cm. On the right is the square root of the integrated power spectral density, representing the magnetic field in the LHC tunnel after mitigation.}
    \label{fig:StrayFields_lhc_measurement}
\end{figure}

To characterise a background magnetic field, the power spectrum in the LHC tunnel and ALICE detector cavern was measured. This is shown in Fig.\,\ref{fig:StrayFields_lhc_measurement} along the effect of different mitigation techniques - discussed in Sec.\,\ref{sect:PERF_Stray_Mitigation}. Although this measurement doesn't replicate the exact magnetic environment of CLIC, it does represent the magnetic field that can be expected underground and in the presence of technical equipment and environmental sources. There is a significantly larger magnetic field in the ALICE detector cavern, highlighting that the detector can act as a technical source. 

\subsection{Mitigation}
\label{sect:PERF_Stray_Mitigation}
The sensitivity to stray fields arises in the RTML transfer line and BDS. The ML was the most robust section with respect to stray fields, with tolerances on the order of $\mu$T. There will also be an additional benefit in the ML of magnetic shielding provided by the copper walls that form each cavity, suggesting that an effective mitigation strategy can be obtained by targeting just the RTML transfer line and BDS. Two mitigation techniques are anticipated: a beam-based feedback system that will act across the entire accelerator, and a metallic coating of the beam pipe for magnetic shielding. The coating will only be applied to particular sections. Calculations of tolerances with different mitigation techniques are presented in \cite{Gohil2019b}. 

A beam-based feedback system is modelled as an effective reduction in the power spectrum of the magnetic field. This is a simplification as the beam-based feedback does not act directly on the magnetic field, but its effect is equivalently modelled as a suppressed power spectrum. The effective reduction in power that the currently designed feedback system would provide on the magnetic field in the LHC tunnel is shown in Fig.\,\ref{fig:StrayFields_lhc_measurement}. The largest contributions to the magnetic field are the harmonics of 50 Hz coming from the electrical grid. The feedback system mostly cures these harmonics, however there is some amplification of the frequencies in between due to the feedback system.

The effectiveness of a metallic coating with a material of high magnetic permeability, such as mu-metal, is under investigation. The effect of a 1 mm mu-metal coating on the magnetic field in LHC tunnel, shown in Fig.\,\ref{fig:StrayFields_lhc_measurement}, is to reduce the effective magnetic field experienced by the beam to within the 0.1 nT tolerance of CLIC. This technique is well suited for long drifts and would be an effective strategy for mitigation in the RTML transfer line. It is expected that a stray field would not be able to penetrate through a magnet to influence the beam, therefore shielding the drifts only would suffice. Fig.\,\ref{fig:StrayFields_lhc_measurement} shows the effectiveness of the mu-metal shield increases with frequency, which means frequencies above 300 Hz will be suppressed heavily.

The RTML transfer line will be parallel to the ML and the Drive-Beam Decelerators. Because of this, the optics of the RTML transfer line has been designed to reduce any effect stray fields associated with the decelerators could have. By ensuring a $90^\circ$ phase advance between decelerators, any periodic perturbation from the decelerators would cancel. A feedforward trajectory correction could also be applied in the RTML TA. Fig.\,\ref{fig:StrayFields_rtml_tolerance} shows the tolerance can be increased by roughly two orders of magnitude with a feedforward correction.

A mu-metal shield can also be applied to sections of the BDS. As shown in Fig.\,\ref{fig:StrayFields_ml_bds_tolerance}, the tolerance can be increased by two orders of magnitude by surrounding the BDS with a perfect shield. It may not be necessary to shield the entire BDS, an order of magnitude increase in the tolerance can be obtained by shielding just the bends in the BDS \cite{Gohil2019b}.

The exact design of the mitigation is still to be optimised and will be based upon the on-going measurement campaign to characterise a realistic power spectrum of magnetic fields for CLIC. However, it is expected that a combination of beam-based feedback and magnetic shielding will be able to effectively mitigate the effects of stray fields.

\printbibliography[heading=subbibintoc]
\endrefsection

\chapter{Conclusion}
\label{Chapter:Conclusion}

\section{Summary and Outlook}
\label{sec:summary}
In this document, the research and development on the Compact Linear Collider is summarised, with emphasis on recent studies and R\&D for the CLIC accelerator complex. CLIC is foreseen to be built and operated in stages. This report provides details of an updated staging scenario,
which is optimised for physics performance, and contains assumptions about commissioning and running time per year
which were recently harmonised with those of other future CERN  projects. 
For the first stage with a centre-of-mass energy of 380\,GeV an integrated luminosity of 1\,ab$^{-1}$ is foreseen. 
This is followed by operation at 1.5\,TeV with 2.5\,ab$^{-1}$, and by a third stage at 3\,TeV with 5\,ab$^{-1}$  integrated luminosity.
This CLIC physics programme spans over 25--30 years.
The updated baseline also specifies $\pm 80$\% electron polarisation, with the sharing between the two longitudinal polarisation states optimised for the best physics reach at each energy stage.

The construction and operation of CLIC is described, with the Two-Beam acceleration scheme as baseline scenario.
Normal-conducting high-gradient 12\,GHz accelerating structures are powered via a high-current drive beam. 
The accelerating structures will be operated in the range of 70 to 100\,MV/m, resulting in a total accelerator length of 11\,km for the 380\,GeV stage and 50\,km for 3\,TeV.
For the first energy stage, an alternative scenario is presented, with X-band klystrons powering the Main-Beam accelerating structures.
Details of an implementation of CLIC near CERN are described, generally with emphasis on the 380\,GeV stage. These include results on civil engineering studies, construction and upgrade schedules, electrical networks, cooling and ventilation, transport and safety aspects.  

Beam experiments and hardware tests described in this report demonstrate that the CLIC performance goals can be met.
For instance, accelerating gradients of up to 145\,MV/m are reached with the Two-Beam concept at CTF3,
and breakdown rates of the accelerating structures well below the limit of $3 \times 10^7\text{m}^{-1}$ are stably achieved at X-band test platforms.
High luminosities can be achieved by using nanometre beam sizes. This requires low-emittance beams as well as novel alignment and stabilisation techniques.
There is substantial progress in all of these domains: performances as needed for the CLIC damping rings are achieved by modern synchrotron light sources;
special alignment procedures for the main linac are now available; sub-nanometre stabilisation of the final focus quadrupoles is demonstrated.
In general, beam physics studies, technical developments and system tests for CLIC resulted in significant progress in recent years. Reductions in cost and energy consumption have been among the main objectives of these developments, resulting in a better energy efficiency of the 380\,GeV stage, with power around 165\,MW, together with a lower estimated cost, now around 6\, billion CHF.

CLIC accelerator technology has reached a mature state and is increasingly being put to use in accelerator projects around the globe.
A detector design concept exists, and technology demonstrators for the sub-detectors have been built.
A work-plan for the preparation phase towards building CLIC is outlined in this report.
The CLIC accelerator and detector can be ready for a construction start around 2026.
First collisions at the 380\,GeV energy stage would then take place towards 2035.
CLIC provides a broad set of precision Standard Model physics measurements, reaching well beyond the projections for HL-LHC.
In addition, searches for new physics at the energy frontier can be performed.
In summary, CLIC represents a compelling opportunity for the post-LHC era.

\chapter*{Acknowledgements}
This work benefited from services provided by the ILC Virtual Organisation, supported by the national resource providers of the EGI Federation. 
This research was done using resources provided by the Open Science Grid, which is supported by the National Science Foundation and the U.S. Department of Energy's Office of Science.
This work was supported by
the European Union's Horizon 2020 Research and Innovation programme under Grant Agreement No.\,654168 (AIDA-2020);
the European Union's Horizon 2020 Research and Innovation programme under Grant Agreement No.\,777431 (CompactLight);
the European Union's Horizon 2020 Marie Sklodowska-Curie Research and Innovation Staff Exchange programme under Grant Agreement No.\,645479 (E-JADE); 
the National Commission for Scientific and Technological Research (CONICYT), Chile;
the DFG cluster of excellence ``Origin and Structure of the Universe'', Germany;
the Federal Ministry of Education and Research (BMBF), Germany under Grant Agreement No.\,05H18VKRD1; 
the Israel Science Foundation (ISF);
the I-CORE Program, Israel;
the Israel Academy of Sciences;
the Programma per Giovani Ricercatori ``Rita Levi Montalcini'' of the Ministero dell'Istruzione, dell'Universit\`a e della Ricerca (MIUR), Italy; 
the Research Council of Norway;
the National Science Centre, Poland, HARMONIA project under contract UMO-2015/18/M/ST2/00518 and OPUS project under contract UMO-2017/25/B/ST2/00496;
the Polish Ministry of Science and Higher Education under contract No.\,3501/H2020/2016/2 and 3812/H2020/2017/2; 
the Ministry of Education, Science and Technological Development of the Republic of Serbia under contract No.\,OI171012;
the Spanish Ministry of Economy, Industry and Competitiveness under projects MINEICO/FEDER-UE, FPA2015-65652-C4-3-R, FPA2015-71292-C2-1-P and FPA2015-71956-REDT; 
the Generalitat Valenciana under grant PROMETEO/2018/060, Spain; 
the IFIC, IFCA, IFT and CIEMAT grants under the Centro de Excelencia Severo Ochoa and Maria de Maeztu programs, SEV-2014-0398, MDM-2017-0765, SEV-2016-059, MDM-2015-0509, Spain;
the Swedish Research Council;
the Swiss National Science Foundation FLARE and FORCE grants 147463, 141146, 135012, 131428, 125272 and 126838;
the Scientific and Technological Research Council of Turkey (TUBITAK) under grant number 118F333; 
the UK Science and Technology Facilities Council (STFC), United Kingdom;
and the U.S. Department of Energy, Office of Science under contract DE-AC02-06CH11357.



\end{document}